\documentclass[12pt,vi,twoside]{mitthesis}
\usepackage{amsmath}
\usepackage{amssymb}
\usepackage{lgrind,graphicx,axodraw,color,journals}
\usepackage{rotating} 
\usepackage[english,greek]{babel}
\newcommand{\href}[2]{#2}
\newcommand{\Tevatron}{Tevatron}
\newcommand{\Dzero}{D\O}
\newcommand{\DZero}{D\O}
\newcommand{\pbar}{\ensuremath{\bar p}}
\newcommand{\abs}[1]{\ensuremath{\left|#1\right|}}
\newcommand{\Ell}{\ensuremath{\mathcal L}}
\newcommand{\lumiUnits}{\ensuremath{\mbox{s}^{-1}\mbox{cm}^{-2}}}
\newcommand{\met}{\ensuremath{\slash\!\!\!\!E_T}}

\newcommand{\pslash}{\ensuremath{\slash\!\!\!p}}
\newcommand{\pmiss}{\ensuremath{\slash\!\!\!p}}
\newcommand{\pmissvector}{\ensuremath{\vec{\slash\!\!\!p}}}
\newcommand{\puncl}{\ensuremath{\vec{p}_{\rm{uncl}}}}

\newcommand{\detEta}{\ensuremath{\eta_{\rm det}}}
\newcommand{\Stntuple}{\texttt{Stntuple}}
\newcommand{\memo}[1]{.....{\sc #1}.....}
\newcommand{\poo}[2]{\ensuremath{p(#1\to #2)}}

\newcommand{\bra}[1]{\ensuremath{\left\langle#1\right|}}
\newcommand{\ket}[1]{\ensuremath{\left|#1\right\rangle}}
\newcommand{\highlight}[1]{#1}
\newcommand {\cdfSpecific}[1]{{}}
\newcommand{\nomenclature}[2]{\item[#1]{#2}}

\def \CdfSim {{\sc CDFsim}}
\def \Pythia {{\sc Pythia}}
\def \Alpgen {{\sc Alpgen}}
\def \MadEvent {{\sc MadEvent}}
\def \MadGraph {{\sc MadGraph}}
\def \Herwig {{\sc Herwig}}
\def \Vista {{\sc Vista}}
\def \Sleuth {{\sc Sleuth}}
\def \sumPt {\ensuremath{\sum{p_T}}}
\def \SumPt {\ensuremath{\sum{p_T}}}
\def \ttbar {\ensuremath{t\bar t}}
\def \st {\ensuremath{^{{\rm st}}}}
\def \nd {\ensuremath{^{{\rm nd}}}}
\def \rd {\ensuremath{^{{\rm rd}}}}
\def \th {\ensuremath{^{{\rm th}}}}
\def \scriptP {\ensuremath{{\cal P}}}
\def \tildeScriptP {\ensuremath{\tilde{\cal P}}}
\def \twiddleScriptP {\tildeScriptP}  
\def \pval {\ensuremath{p\mbox{\small -}{\rm val}}}
\def \pvalmin {\ensuremath{p\mbox{\small -}{\rm val}_{\rm min}}}
\def \tildePval {\ensuremath{\tilde{p}\mbox{\small -}{\rm val}}}

\def \pTmin {{17}}
\def \numberOfVistaDiscrepantDistributions {{\ensuremath{384}}}
\def \VistaApproximateDefiniteLuminosity {{1993}}

\begin{document}
\selectlanguage{english}
%
%
%
%
%
%
%
\title{Model Independent Search For New Physics At The Tevatron}

\author{Georgios Choudalakis}
\department{Department of Physics}
\degree{Doctor of Philosophy in Physics}
\degreemonth{April}
\degreeyear{2008}
\thesisdate{April 24, 2008}


\supervisor{Peter Fisher}{Professor of Physics}

\chairman{Thomas Greytak}{Associate Department Head for Education}

\maketitle



\cleardoublepage
\setcounter{savepage}{\thepage}
\begin{abstractpage}
%
%
%
The Standard Model of elementary particles can not be the final theory.  There are theoretical reasons to expect the appearance of new physics, possibly at the energy scale of few TeV.  Several possible theories of new physics have been proposed, each with unknown probability to be confirmed.  Instead of arbitrarily choosing to examine one of those theories, this thesis is about searching for any sign of new physics in a model-independent way.  This search is performed at the Collider Detector at Fermilab (CDF).  

The Standard Model prediction is implemented in all final states simultaneously, and an array of statistical probes is employed to search for significant discrepancies between data and prediction.  The probes are sensitive to overall population discrepancies, shape disagreements in distributions of kinematic quantities of final particles, excesses of events of large total transverse momentum, and local excesses of data expected from resonances due to new massive particles.

The result of this search, first in 1 fb$^{-1}$ and then in 2 fb$^{-1}$, is null, namely no considerable evidence of new physics was found.

\end{abstractpage}


\cleardoublepage

\begin{centering}
\vspace{4cm}
\begin{tabular}{p{2.4in}p{0.3in}p{2.4in}}
\greektext Stouc gone'ic mou, Qr'hsto kai Euaggel'ia, sthn adelf'h mou Sof'ia, kai ston Ep'ikouro.
& \ &
\latintext To my parents, Christos and Evangelia, my sister Sophia, and to Epicurus.
\end{tabular}
\end{centering}

\cleardoublepage

\section*{Acknowledgments}

I am indebted to my advisor, Bruce Knuteson.  He has been extremely supportive, and mentored me optimally from the first day.  I was given the opportunity to participate in many conferences, seminars and summer schools.  He offered me space to develop initiative and apply my own ideas.  For anyone who knows Bruce, he can only be a paradigm of perseverance and brightness.

It has been a big pleasure to work with Conor Henderson, our post-doc, on both hardware and analysis.  Conor has been to me a resourceful teacher, and effective project leader.  My classmate, Si Xie, who joined CDF later, has been a great person to work with, and I wish him the best as he may continue this project after I graduate.  With Khaldoun Makhoul, Si, Conor, Bruce, and Markus Klute for a while, we advanced Level3 and Event Builder to their best.  For this I also thank Ron Rechenmacher, who at times saved the day like deus ex machina.

Ray Culbertson contributed to this analysis both technically and mentally.  His office has the heaviest traffic in CDF, to which I contributed with my visits for questions, so my thanks are due.  The same for Stephen Mrenna, who has been very helpful as a theorist and event generator expert.

I thank for their support the CDF spokesmen, Rob Roser and Jaco Konigsberg;  the Physics Coordinator, Doug Glenzinski; our godparents, Louis Lyons, Andy Hocker, Guillelmo Gomez-Ceballos, and Michael Schmidt who passed away prematurely;  our reviewers, Al Goshaw, Sergey Klimenko and Mario Martinez-Perez;  our conveners, Ben Brau and Chris Hays.  They all worked very hard to bring this analysis to the community.

It is an honor to have my thesis evaluated by Physicists of the caliber of Jerry Friedman, Roman Jackiw and Peter Fisher.

I wish to thank many distinguished scientists at MIT for inviting me to their elite company.  I may name indicatively Wit Busza, Bolek Wyslouch, Christoph Paus, Bernd Surrow, Gabriella Sciolla, and Richard Yamamoto.
Finally, I warmly thank Steve Pavlon, the sweetest person I met in America.


\pagestyle{plain}
\tableofcontents
{\footnotesize
\newpage
\listoffigures
\newpage
\listoftables
}

\chapter{Introduction}
\label{chap:intro}
\section{The Standard Model}

Our current understanding of nature on its most fundamental level is encoded in the ``Standard Model'' of elementary particles.

\begin{figure}[t]
\centering
\includegraphics[width=6cm]{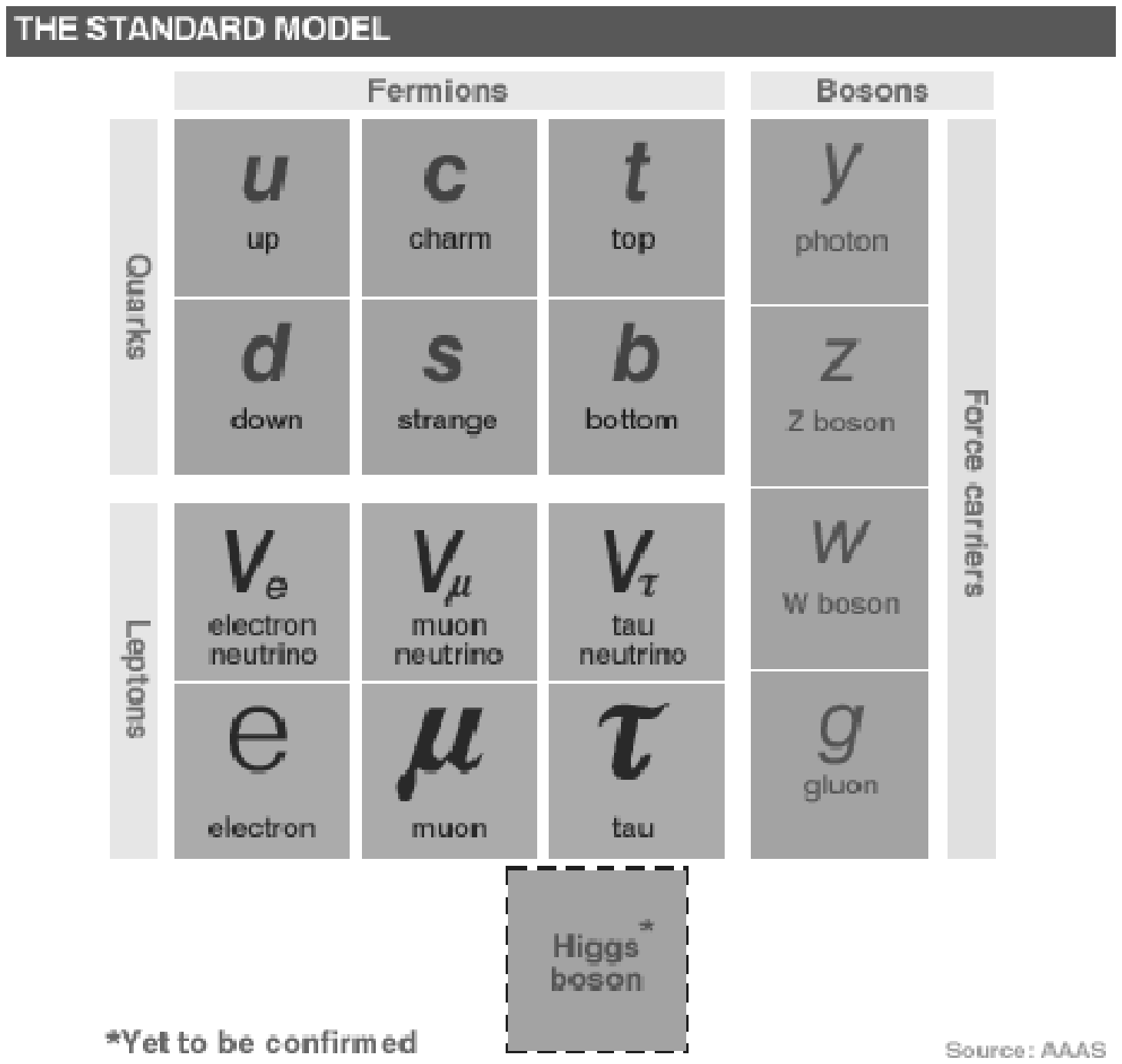}
\caption{Elementary particles in the Standard Model.}
\label{fig:buildingBlocks}
\end{figure}

The building blocks of matter are categorized into three families of fermions and four gauge bosons, shown in Fig.~\ref{fig:buildingBlocks}.

The Standard Model is a local gauge invariant quantum field theory, which describes electromagnetic, weak and strong interactions.  Interactions are introduced for free with the assumption that nature is symmetric under local gauge transformations of the $U(1)_Y \times SU(2)_L \times SU(3)_c$ group \cite{DynamicsOfSM}.  Electromagnetic and weak interactions are aspects of a unified electroweak interaction, which are distinguishable in result of electroweak symmetry breaking via the Higgs mechanism.  Elementary particles acquire bare mass by coupling to the same Higgs field that is responsible for the electroweak symmetry breaking.  The success of this model of electroweak interactions in describing experimental data from the last 35 years builds confidence in the existence of the Higgs boson, though it has not been directly observed as of today.

The Standard Model carries 26 free parameters, which are determined experimentally. Depending on how one counts, they are the 6 lepton masses, the 6 quark masses, 4 parameters from CKM plus 4 from PMNS matrix, the strong coupling $\alpha_s$, the QCD angle $\theta_{QCD}$, the electromagnetic coupling $\alpha$, Weinberg angle $\theta_w$, the vacuum expectation value ($\upsilon$) and the mass ($m_H$) of the Higgs.


The success of the Standard Model is certainly among the greatest achievements in physics.  At the same time, it is bound to not be the final theory.  Some reasons are explained in Section \ref{sec:SMlimitations}.

\subsection{Limitations}
\label{sec:SMlimitations}

The most obvious shortcoming of the Standard Model, as it stands, is that it does not describe gravity \cite{Roulet:2001se, PDBook}.  Its domain is limited to energies much smaller than Planck mass ($M_{Pl}$), where from dimensional analysis gravity is expected to be comparable to the other three known interactions.

Another nuisance is the presence of 26 free parameters.  Past successful theories have established in our minds some notion of scientific aesthetics, according to which the fundamental theory should be able to derive, from first principles, numbers such as the mass of the electron, or the amount of CP violation observed in systems like $K^0$ and $B^0$ mesons.  Otherwise one can not claim to understand those effects.  Grand Unification Theories try to address these issues by embedding the Standard Model into larger symmetry groups (Sec.~\ref{sec:GUTs}).

There is overwhelming evidence (from observations of the cosmic microwave background radiation, galaxy rotations, gravitational lensing, spectroscopy of clusters and super-novae) that dark matter and dark energy dominate the mass-energy density of the universe \cite{Olive:2003iq}.  Currently, the Standard Model fails to provide a good candidate for either.

Another puzzle is the so-called ``hierarchy problem'', namely why the electroweak symmetry is broken at energy $\lesssim$ 1 TeV, so much smaller than $M_{Pl}$, where gravity becomes significant.  Theories involving extra dimensions propose some answers (Sec.~\ref{sec:ExtraDimensions}).

\begin{figure}
\centering
\includegraphics[width=10cm]{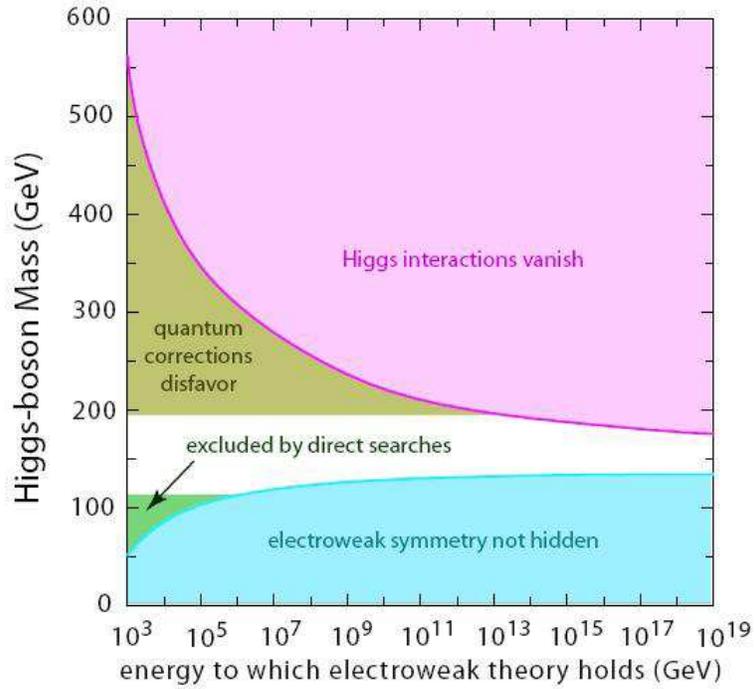}
\caption[Triviality and Stability limits on the Standard Model Higgs mass]{For any $m_H$ there is an energy scale at which the Standard Model stops making sense.  This happens in two ways \cite{0034-4885-70-7-R01}:  In one case the Higgs potential runs to $-\infty$ resulting in a trivial theory without Higgs interactions.  In the other case the Higgs potential has its minimum at 0, resulting in zero vacuum expectation value for Higgs, namely no electroweak symmetry breaking.}
\label{fig:SMvalidity}
\end{figure}

\begin{figure}
\centering
\begin{tabular}{cc}
\includegraphics[width=4cm]{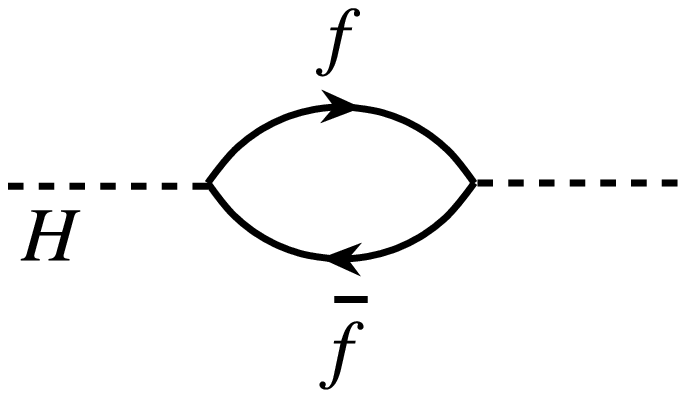} &
\includegraphics[width=4cm]{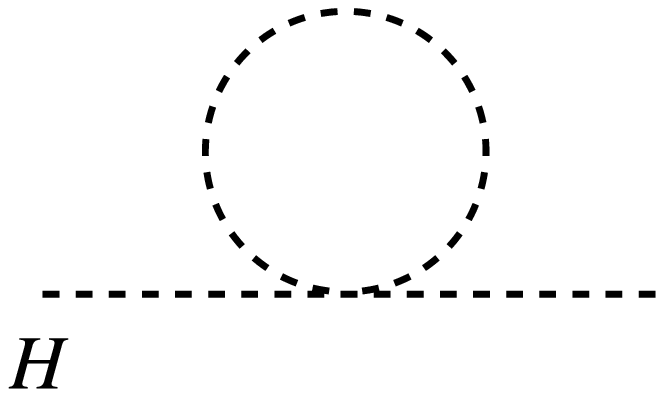} \\
(a) & (b)
\end{tabular}
\caption{Quantum corrections to the Higgs\index{SM!Higgs} $m_H^2$, through fermion loops (a) and Higgs's self-coupling (b).}
\label{fig:higgsLoopCorrections}
\end{figure}

Related to hierarchy is the the problem of ``naturalness'' in the Standard Model.  A small parameter in a theory is ``natural'' when setting it to zero increases some symmetry of the theory, therefore its smallness can be attributed to that very symmetry.  For instance, the masslessness of a vector field such as the photon can be related to the gauge invariance of the theory.  However, for a scalar field, such as the Standard Model Higgs, no symmetry is there to protect its mass from acquiring quadratically divergent corrections at the loop level (Fig.~\ref{fig:higgsLoopCorrections}), unless the theory is highly fine-tuned (Fig.~\ref{fig:SMvalidity}).  The required precision of fine-tuning depends on how far one wishes to extend the validity of the Standard Model.  If one wishes it account for loop corrections up to the Planck scale, while keeping the Higgs lighter than 1 TeV, as required by electroweak measurements, then the required fine-tuning is so precise that it seems unnatural (hence the connection between naturalness and hierarchy).  A solution to this can be either to abandon the concept of fundamental scalars, as in technicolor models (Sec.~\ref{sec:technicolor}), or to search for a theory where quadratic divergences cancel, as in Supersymmetry (Sec.~\ref{sec:SUSY}).

\section{Beyond the Standard Model}

Let me summarize the main proposals which address the limitations explained in Sec.~\ref{sec:SMlimitations}, and what observable implications each suggests.

\subsection{Grand Unification}
\label{sec:GUTs}
The motivation behind Grand Unification Theories (GUTs) \cite{Pati:1973uk,Georgi:1974sy} are questions such as ``why protons and electrons have exactly opposite charge'', or ``why have three generations of fermions and three interactions''.  These questions could become less thorny if instead of many we had just one symmetry group, which would make all particles look like components of just one particle, and all interactions like aspects of one force.  Such a theory wouldn't only satisfy common taste, but more importantly could derive from mathematical principles the values of some constants, such as $\sin{\theta_w}$, which would be a significant advancement in our understanding nature from a reductionist's point of view.

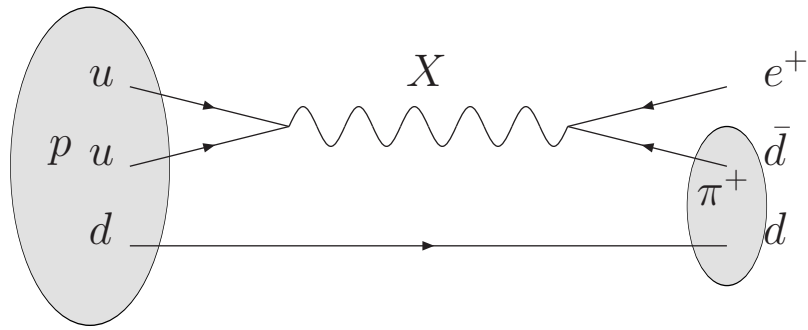
\begin{figure}
\begin{center}
\fcolorbox{white}{white}{
  \begin{picture}(315,120) (90,-180)
    \SetWidth{0.5}
    \SetColor{Black}
    \GOval(120,-120)(60,30)(0){0.882}
    \GOval(360,-135)(30,15)(0){0.882}
    \ArrowLine(135,-90)(195,-105)
    \Photon(195,-105)(300,-105){7.5}{5}
    \Text(375,-90)[lb]{\Large{\Black{$e^{+}$}}}
    \Text(375,-120)[lb]{\Large{\Black{$\bar{d}$}}}
    \ArrowLine(360,-120)(300,-105)
    \ArrowLine(360,-90)(300,-105)
    \ArrowLine(135,-150)(360,-150)
    \Text(375,-150)[lb]{\Large{\Black{$d$}}}
    \ArrowLine(135,-120)(195,-105)
    \Text(120,-90)[lb]{\Large{\Black{$u$}}}
    \Text(120,-150)[lb]{\Large{\Black{$d$}}}
    \Text(105,-120)[lb]{\Large{\Black{$p$}}}
    \Text(240,-90)[lb]{\Large{\Black{$X$}}}
    \Text(120,-120)[lb]{\Large{\Black{$u$}}}
    \Text(350,-135)[lb]{\Large{\Black{$\pi^{+}$}}}
  \end{picture}
}
\end{center}
\label{fig:protonDecay}
\caption{Diagram leading to proton decay in the context of $SU(5)$ Grand Unification.}
\end{figure}

Several Lie algebras have been studied; notably $SU(5)$, $SO(10)$, $E_6$ and more \cite{Roulet:2001se, PDBook}.  Phenomenology varies significantly depending on the assummed symmetry.  An effect predicted typically is proton decay, as new gauge bosons such as the one in Fig.~\ref{fig:protonDecay}, are predicted in breaking these hyper-symmetries at some large energy, typically $M_{\rm GUT}\simeq 10^{16}$ GeV.

\subsection{Supersymmetry}
\label{sec:SUSY}

\begin{table}
\centering
\begin{tabular}{|cc|cc|}
\hline
Particle &  Spin & Superpartner & Spin \\
\hline \hline
Photon & 1 & Photino & 1/2 \\
Gluon & 1 & Gluino & 1/2 \\
$W$ & 1 & Wino & 1/2 \\
$Z^0$ & 1 & Zino & 1/2 \\
$H$ & 0 & Higgsino & 1/2 \\
Graviton & 2 & Gravitino & 3/2 \\
\hline
Electron & 1/2 & Selectron & 0 \\
Muon & 1/2 & Smuon & 0 \\
Tau & 1/2 & Stau & 0 \\
Neutrino & 1/2 & Sneutrino & 0 \\
Quark & 1/2 & Squark & 0 \\
\hline
\end{tabular}
\caption{Ordinary particles and their superpartners.}
\label{tab:superpartners}
\end{table}

Supersymmetric theories take the approach of solving the problem of naturalness (Sec.~\ref{sec:SMlimitations}), by having a bosonic loop for each fermionic one,  thus canceling out the quadratically divergent loop corrections.

SUSY introduces boson partners to Standard Model fermions, and fermion partners to gauge bosons.  It introduces operators which transform fields into ``superpartners'' which differ from the original particles by half a unit of spin  \cite{Martin1997ns}.  The superpartners of gauge bosons are called ``gauginos'', those of leptons ``sleptons'' and those of quarks ``squarks'' (Table~\ref{tab:superpartners}).

SUSY can have additional favorable features, which increase interest in it.  With the extra assumption of a conserved multiplicative quantum number (R-parity), which is +1 for ordinary particles and -1 for superpartners, the lightest superpartner becomes stable, serving as a cold dark matter candidate \cite{AnIntroToTheScienceOfCosmology}.  Furthermore, a theory of local supersymmetry should lead to invariance under general coordinate transformations, which may be the road to incorporating General Relativity into the Standard Model.  Finally, SUSY can affect the running of couplings to make them exactly equal at some energy, in compliance with Grand Unification Theories.

If supersymmetry were exact, then each Standard Model particle would have a superpartner of equal mass.  Since this is not observed, SUSY has to be broken at some energy scale \cite{PDBook}.  It is non-trivial to construct models where SUSY is broken in ways that avoid contradicting observation, and simultaneously do not destroy its desirable features.

Higgs mass is predicted to be of order $10^2$ GeV$/c^2$, so for SUSY to secure it from divergences it has to be introduced at energy $\lesssim$ 1 TeV.  That happens to be also the energy scale where it needs to be introduced in order to equalize couplings at the scale of $10^{16}$ GeV, associated with Grand Unification.  These elements hint that, if SUSY is a correct theory, it may be within reach for current experiments.

Most SUSY signatures involve large missing energy accompanied by multiple leptons and jets.  Missing energy would be the effect of stable and elusive superpartners, while jets and leptons would result from long decay chains of unstable ones.

\subsection{Extra Dimensions}
\label{sec:ExtraDimensions}
Theories of extra dimensions are motivated by the hierarchy problem.

One hypothesis is that of large extra dimensions, where the known 4 dimensions, i.e. our ``brane'', are embedded in a manifold of higher dimensionality, and gravity only appears to be feeble because part of it is projected onto our brane, while the rest propagates in the extra dimensions, often referred to as ``the bulk''.  By adjusting the number of extra dimensions and their radius of curvature, one can make gravity appear significant at $M_{Pl}$ and still lower its natural scale down to the electroweak scale \cite{ArkaniHamed:1998rs}.  

Theories with universal extra dimensions exist too, where fermions and/or gauge bosons also propagate in the bulk \cite{Appelquist:2000nn}.

Other theories assume wrapped extra dimensions.  Hierarchy then emerges by exploiting the metric of the bulk space itself.  For example, with one wrapped extra dimension periodically bounded by two 3-dimensional branes, Einstein's equations result in an anti de Sitter metric, whose exponential factor makes gravity appear feeble on one of the 3-branes, where the Standard Model fields are supposed to be confined \cite{Randall:1999ee}.

\begin{figure}[t]
\centering
\begin{center}
\fcolorbox{white}{white}{
  \begin{picture}(193,268) (42,2)
    \SetWidth{0.5}
    \SetColor{Black}
    \ArrowLine(59,254)(105,224)
    \ArrowLine(104,223)(59,194)
    \ArrowLine(60,163)(106,133)
    \ArrowLine(105,132)(60,103)
    \ArrowLine(61,73)(107,43)
    \ArrowLine(106,42)(61,13)
    \Photon(105,224)(164,223){7.5}{3}
    \Photon(165,222)(195,251){7.5}{2}
    \Photon(165,226)(195,192){7.5}{2}
    \Photon(106,133)(165,132){7.5}{3}
    \ArrowLine(166,131)(196,164)
    \ArrowLine(196,101)(167,130)
    \Gluon(105,43)(163,43){7.5}{2.71}
    \Gluon(164,43)(194,74){7.5}{1.65}
    \Photon(163,43)(195,13){7.5}{2}
    \Text(42,254)[lb]{\Large{\Black{$q$}}}
    \Text(44,195)[lb]{\Large{\Black{$\bar{q}$}}}
    \Text(42,154)[lb]{\Large{\Black{$q$}}}
    \Text(44,95)[lb]{\Large{\Black{$\bar{q}$}}}
    \Text(43,61)[lb]{\Large{\Black{$q$}}}
    \Text(45,2)[lb]{\Large{\Black{$\bar{q}$}}}
    \Text(124,205)[lb]{\Large{\Black{$G$}}}
    \Text(130,113)[lb]{\Large{\Black{$G$}}}
    \Text(199,12)[lb]{\Large{\Black{$G$}}}
    \Text(131,25)[lb]{\Large{\Black{$g$}}}
    \Text(200,76)[lb]{\Large{\Black{$g$}}}
    \Text(202,249)[lb]{\Large{\Black{$\gamma$}}}
    \Text(203,193)[lb]{\Large{\Black{$\gamma$}}}
    \Text(201,163)[lb]{\Large{\Black{$e^{-}$}}}
    \Text(205,104)[lb]{\Large{\Black{$e^{+}$}}}
  \end{picture}
}
\end{center}
\caption{Possible signatures of graviton.}
\label{fig:graviton}
\end{figure}
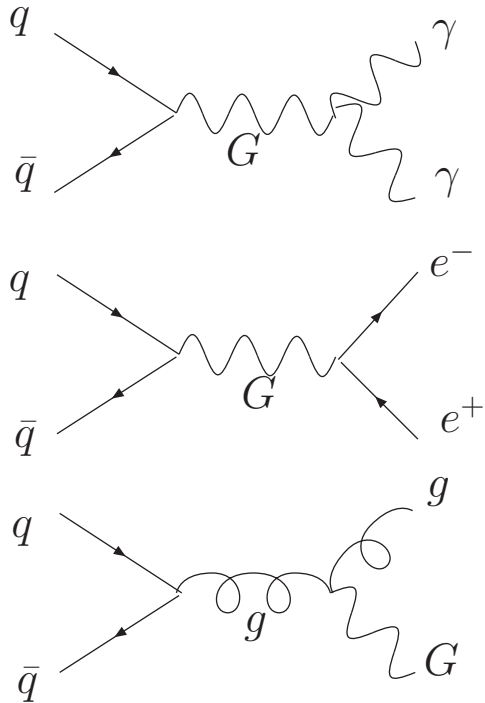

If at small distances gravity is not as feeble as suggested macroscopically by $M_{Pl}$, then collider experiments could reveal the coupling of gravitons.  For example, a signature could be $p\bar{p} \to g G_n$, i.e.\ mono-jet events with large missing energy due to the graviton $G_n$ escaping in the bulk (Fig.~\ref{fig:graviton}).  Another signature of the graviton could be the Standard-Model-forbidden $gg\to G_n\to\ell^+\ell^-$ \cite{PDBook}.  In the case of universal extra dimensions one may observe the Kaluza-Klein higher states of fermions and bosons, through $Z'\to t\bar{t}$ for instance.  

\subsection{Technicolor}
\label{sec:technicolor}
An alternative approach to electroweak symmetry breaking, which avoids the introduction of fundamental scalar fields, is new strong dynamics.  With the introduction of a new non-abelian gauge symmetry and additional fermions (``{\em technifermions}'') which have this new interaction, it becomes possible to form a technifermion condensate that can break the chiral symmetry of fermions, in a way analogous to QCD where the $q\bar q$ condensate breaks the approximate $SU(2)\times SU(2)$ symmetry down to $SU(2)_{isospin}$.  The breaking of global chiral symmetries implies the existence of Goldstone bosons, the ``technipions'' ($\pi_T$), in analogy with QCD pions.  Three of the Goldstone bosons are absorbed through the Higgs mechanism to become the longitudinal components of the $W$ and $Z$, which then acquire mass proportional to the technipion decay constant. 

Experimental signatures of technicolor are model dependent.  For example, they can be the resonance of a Standard Model gauge boson into an excited technivector meson, like a technirho ($\rho_T$), which subsequently decays into $W$ and $\pi_T$, with $\pi_T$ possibly decaying to regular quarks \cite{PDBook}.  For example, assuming that $\pi_T$ couples preferably to the third generation, such a process could be $\rho^\pm_T \rightarrow W^\pm \pi^0_T \rightarrow \ell^\pm \nu b \bar b$, or $\rho_T^0 \rightarrow W^+ \pi_T^- \rightarrow \ell^+ \nu_\ell b \bar c$.

\subsection{Compositeness}
Compositeness is the idea that the Higgs and possibly other bosons and fermions contain substructure.  Compositeness addresses the problem or naturalness similarly with technicolor, namely by avoiding the assumption of a fundamental scalar particle.

If quarks and leptons are not elementary, then they are predicted to have excited states ($q^\ast, \ell^\ast$).  For example, excited leptons could appear via $\ell^\ast \rightarrow \ell \gamma$ or $\ell^\ast \rightarrow W \nu.$.

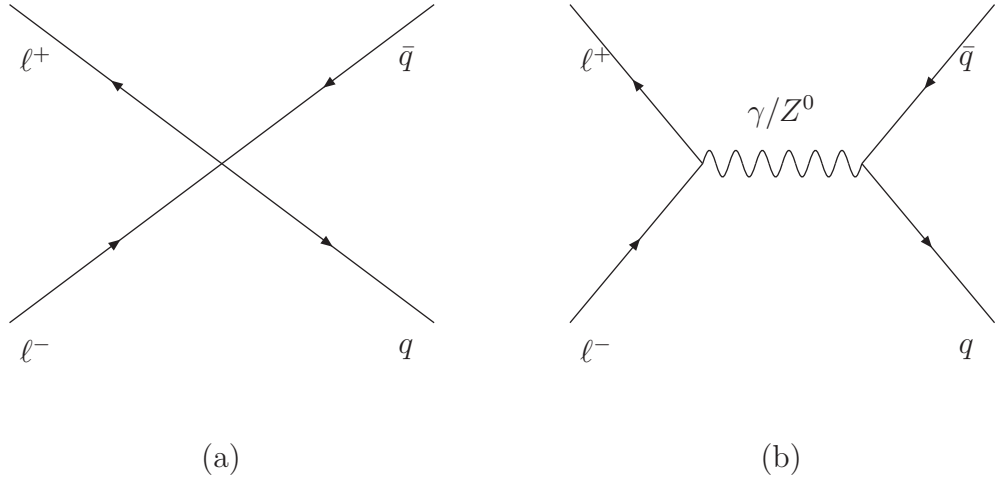
\begin{figure}
\centering
\begin{tabular}{cc}
\begin{picture}(200,200)(0,0)
\ArrowLine(20,40)(100,100)
\ArrowLine(100,100)(20,160)
\ArrowLine(180,160)(100,100)
\ArrowLine(100,100)(180,40)
\Text(30,30)[cc]{$\ell^-$}
\Text(30,140)[cc]{$\ell^+$}
\Text(170,30)[cc]{$q$}
\Text(170,140)[cc]{$\bar q$}
\end{picture} 
&
\begin{picture}(200,200)(0,0)
\ArrowLine(20,40)(70,100)
\ArrowLine(70,100)(20,160)
\ArrowLine(180,160)(130,100)
\ArrowLine(130,100)(180,40)
\Text(30,30)[cc]{$\ell^-$}
\Text(30,140)[cc]{$\ell^+$}
\Text(170,30)[cc]{$q$}
\Text(170,140)[cc]{$\bar q$}
\Photon(70,100)(130,100){5}{6}
\Text(100,120)[cc]{$\gamma / Z^0$}
\end{picture}
\\
(a) & (b)
\end{tabular}
\caption{(a) Contact interaction allowed in the case of compositeness. (b) Tree-level SM diagram with the same initial and final state, where the interaction is mediated by a gauge boson.}
\label{fig:contactInteraction}
\end{figure}

More importantly, if quarks and leptons have structure, new interactions should appear between them at the energy scale of their binding energy.  They would be contact interactions, allowing processes such as $\ell^+ \ell^- \rightarrow \ell^+ \ell^-$ and $\ell^+ \ell^- \rightarrow q \bar q$ to occur in ways additional to those of the SM (Fig.~\ref{fig:contactInteraction}) \cite{Chivukula2000mb, PDBook}.

\section{Current standpoint - Motivation}
\label{sec:intro-motivation}

In 1995, the discovery of the top quark was announced \cite{Abe1995hr}, leaving Higgs as the only unobserved Standard Model particle.  We now enter the Large Hadron Collider (LHC) era with some confidence that the Higgs will be observed to complete the Standard Model pantheon of particles.  At the same time, there is hope that even what has to lie beyond the Standard Model will be revealed soon.  If such a groundbreaking discovery is made, it will be different from the top quark or even a possible Higgs discovery, in the sense that it will signify the opening to a new continent of unexplored physics. 

Nature has proven its capacity to surprise us.  There are many ideas of what the new physics may be, but there is no need for any of them to be right.  So, especially in this historical time when we expect to overcome the current impasse, it makes sense to search for any sign of discrepancy between the data and the Standard Model, without introducing any bias in what it may look like.  This is the motivation behind performing a model-independent and global search.

\Tevatron\ stands at the current high energy frontier, producing $p\bar p$ collisions at energy 1.96 TeV and constantly increasing luminosity.  Although the size and reach of the \Tevatron\ are inferior to those of LHC, there is still a window of opportunity in the former, until the latter has collected data and understood systematic effects specific to it.  It would be undesirable to discover something at the LHC and then look back only to realize that it had been overlooked at the \Tevatron.  On the other hand, performing a global, model-independent analysis of the \Tevatron\ data has the potential of revealing evidence of new physics that can be cross-checked at the LHC.  This hope motivates the present work.

\chapter{Experimental apparatus}

\begin{figure}
\centering
\includegraphics[width=13cm,angle=0]{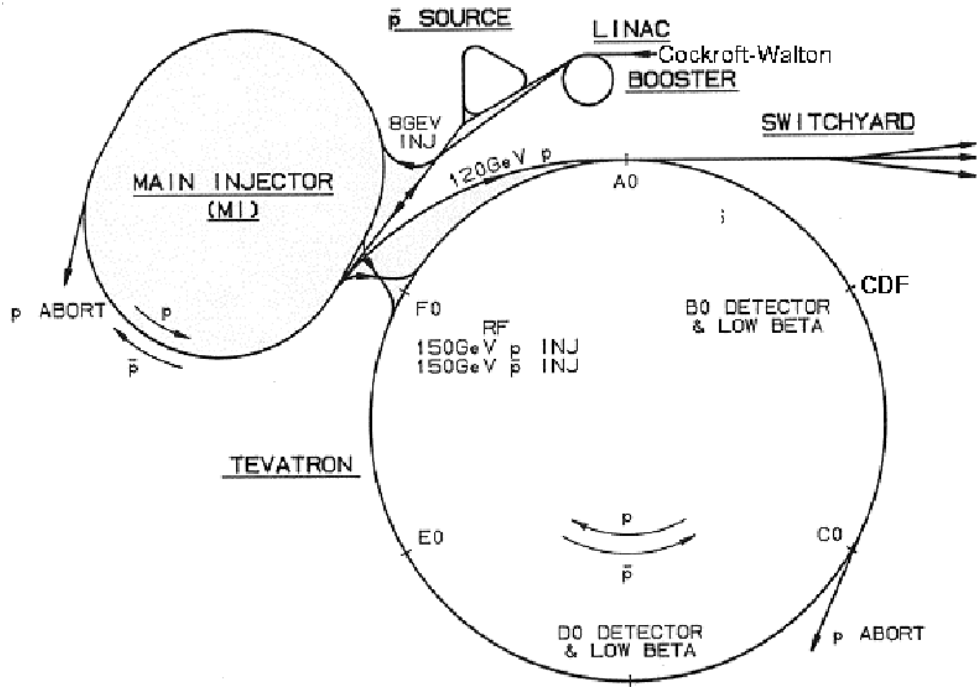} 
\caption{Sketch of the FNAL accelerator complex.}
\label{fig:tevatronSketch}
\end{figure}

The present search for new physics is performed in data collected with Collider Detector at Fermilab (CDF), a general scope detector for particles generated at high energy $p \bar p$ collisions produced by the \Tevatron\ accelerator.  \Tevatron\ and the Fermi National Accelerator Laboratory (FNAL) are shown in Fig.~\ref{fig:tevatronSketch}. 

This chapter describes the production of $p \bar p$ collisions and the CDF detector.  For the many acronyms used, please consult Appendix \ref{chapter:nomenclature}. 

\section{Beam Production}
Either due to CP violation or some other unknown reason, free protons outnumber antiprotons, which makes it easier to obtain the former, and use them to generate the latter.  In this section, the procedure leading to the production of the $p$ and \pbar\ beams is outlined.

\subsection{$p$ Source}
The production starts with storing hydrogen gas ($H_2$) in a {\bf Cockroft-Walton } chamber \cite{cockroft-walton}, in which a 750 kV DC voltage causes electric discharges which produce negative hydrogen ions ($H^-$). The $H^-$ are separated from the rest of the gas by use of a magnetic transport system and are channeled to the Linac.

The {\bf Linac} \cite{linac} is a 130 m long Alvarez linear accelerator that transfers the $H^-$ from the Cockroft-Walton to the Booster, accelerating them from 750 keV to 400 MeV.

The {\bf Booster} \cite{BoosterRookie} is a 475 m long synchrotron that accelerates the $H^-$ from 400 MeV to 8 GeV in just 67 ms, hence its name. One Linac load is 40 $\mu$s long and the rotation period of the beam in the Booster during injection is 2.22 $\mu$s, which means that in principle it could take $\frac{18 \times 2.22}{40}=99.9\%$ of the Linac's load in 18 turns. Operationally however, only 5 or 6 turns get used for maximum intensity, and the rest (66.7\%) of the Linac's load is dumped. 
At the entrance, the $H^-$ ions pass through a carbon foil, which strips off the electrons, transforming $H^-$ into $H^+$, viz.~protons. It is important that the $H^-$ pass through the carbon foil at their entrance to the ring, as they meet with the circulating $H^+$. This technique, named CEI, allows for higher beam brightness, avoiding limitations that would have otherwise followed from Liouville's theorem \cite{Hojvat:1979ya}.
A full Booster ``batch'' contains a maximum of $5\times 10^{12}$ protons at 8 GeV, coalesced into 84 bunches, ready to be delivered to the Main Injector.

\subsection{Main Injector}
The {\bf Main Injector} \cite{MI} is a 3.319 km long non-circular synchrotron, serving not only the  \Tevatron, but also providing protons for the production of the NuMI neutrino beam and the proton beam in the Fixed Target area. Its operations that relate to the \Tevatron\ are:

\begin{enumerate}
\item \pbar\ production: A single Booster batch is injected into the MI at 8 GeV. These protons are accelerated to 120 GeV and extracted in a single turn for delivery to the \pbar\ production target. The produced antiprotons will eventually return to the MI for acceleration to 150 GeV, before they are delivered to the \Tevatron.
\item Collider mode: Accelerate protons or antiprotons to 150 GeV and deliver them to the \Tevatron.
\item End of store: Accept 150 GeV antiprotons and decelerate them to 8 GeV for storage in the Recycler.
\end{enumerate}

\subsection{\pbar\ Source}
At the {\bf \pbar\ production area}, the 120 GeV protons coming from the MI are directed onto a nickel target \cite{pbarRookie}. Before the collision, the bunch undergoes some modulation called {\em RF bunch rotation}, so as to be shorter in time and, in agreement with Liouville's theorem, contain a wider spectrum of momenta. Its being more sudden maximizes the phase-space density of antiprotons produced as secondary products of the collision with the nickel target. First, the cone of particles produced at the collision is rendered parallel by means of a lithium lens \cite{lithium}. Then, a dipole magnet selects ~8 GeV antiprotons, as that is the standard MI injection energy, and directs them into the Debuncher.

At the {\bf Debuncher} \cite{pbarRookie}, which is a ``ring'' of rounded triangular shape, the 8 GeV antiprotons are subjected to a RF bunch rotation, this time in the reverse direction, so that their beam contains a narrower spectrum of momenta and, in agreement with Liouville's theorem, spans a longer time interval. This reduction in momentum spread is done to improve the Debuncher-to-Accumulator transfer, because of the limited momentum aperture of the Accumulator at injection. The Debuncher makes use of the time between MI cycles to reduce the beam transverse size and longitudinal momentum spread through betatron and momentum stochastic cooling respectively. This further improves the efficiency of the Debuncher-to-Accumulator transfer. 

The {\bf Accumulator} \cite{pbarRookie} is a rounded triangular ``ring'', similar to the Debuncher. The reason for that is that it also applies stochastic cooling to the \pbar\ beam, which requires linear segments along the ring to accommodate pickups and kickers. The main purpose of the Accumulator is to hold antiprotons until they are needed by the  \Tevatron. The antiprotons are stored in the Accumulator for hours or days, while they augment as more are produced at the nickel target. When a new pulse of antiprotons enters the Accumulator, it circulates along a trajectory of greater ``radius'' than the antiprotons that have already been cooled down. The RF decelerates the recently injected pulses of antiprotons from the injection energy to the edge of the stack tail. The stack tail momentum cooling system sweeps the beam deposited by the RF away from the edge of the tail and decelerates it towards the dense portion of the stack, known as the core. Additional cooling systems keep the antiprotons in the core at the desired momentum and minimize the transverse beam size.

There is yet another ring, the {\bf Recycler} \cite{recycler}, which has a role similar to that of the Accumulator. It is a 3.3 km long ring along the MI, being therefore much longer than the Accumulator, which means that if the Accumulator is getting full it can use the Recycler to hold some antiprotons too. Spread over a longer ring, the antiprotons in the Recycler are easier to maintain stable, since the beam is less dense and the dispersive forces weaker. In addition to being longer, the Recycler employs the electron cooling method to reduce the momentum spread of the antiprotons. Electron cooling is a more modern technique than stochastic cooling, in which a cold (small momentum spread) beam of electrons travels parallel to the hot antiproton beam, serving as a heat sink, where the heat of the antiproton beam is dumped, since the two beams interact electromagnetically and from thermodynamics it is known that heat goes from the hotter system to the cooler. Once the electron beam heats up, it is discarded for a new, cold electron beam to take over. The Recycler does not only accept antiprotons that the Accumulator can not hold, but also those that the \Tevatron\ does not need any more. Since antiprotons are so hard to produce, the Recycler keeps them to be reused in the next ``store'', hence its name. When the stored antiprotons reach adequate quantity, the {\bf \Tevatron} is ready to start $p \bar p$ collisions. 

\subsection{\Tevatron}

For over two decades, the \Tevatron\ \cite{RunIIhandbook, PDBook} has been the largest hadron collider, to be soon succeeded by the Large Hadron Collider (LHC) at CERN. It is a synchrotron accelerator with radius 1 km. Along its ring are 774 dipole and 216 quadrupole superconductive magnets, providing magnetic field of intensity 4.4 T. The magnets operate in superconductive state, with cooling from liquid helium.

The \Tevatron\ receives $p$ and \pbar\ bunches from the MI, where they have been accelerated from 8 to 150 GeV. The filling takes about 30 minutes, much longer than the acceleration period that is only 86 seconds. It accelerates the $p$ and the \pbar\ beam to the energy of $980$ GeV, producing head-on collisions at $\sqrt{s}=1.96$ TeV in the reference frame of CDF \cite{PDBook}. The proton and antiproton beams are both separated in 3 trains, each containing 12 bunches, therefore there are 36 $p$ and 36 \pbar\ bunches traveling in opposite directions at the same energy. Each bunch is about 18 ns (57 cm) long, which is the length of one RF bucket\footnote{A RF bucket is a slot defined by the RF electromagnetic waves, in which a bunch may be accommodated.} at the \Tevatron. The interval between successive bunch crossings is 396 ns (21 buckets), which is of course equal to the interval between successive bunches in a train. Successive trains are separated by longer (2621 ns or 139 buckets) intervals, called {\em abort gaps}. 

Each $p$ and \pbar\ bunch counts about $24 \times 10^{10}$ and $6 \times 10^{10}$ particles respectively. As of today, the beam's optical properties allow for instantaneous luminosity that is over $2\times 10^{32}~{\mbox{cm}}^{-2}{\mbox{s}}^{-1}$ at CDF, and about 15\% lower at \Dzero\ \cite{CDFvsD0lumi, VaiaCDFvsD0lumi}.


\section{The CDF detector}
\label{sec:CDFdetector}

\begin{figure}
\centering
\includegraphics[width=7cm,angle=0]{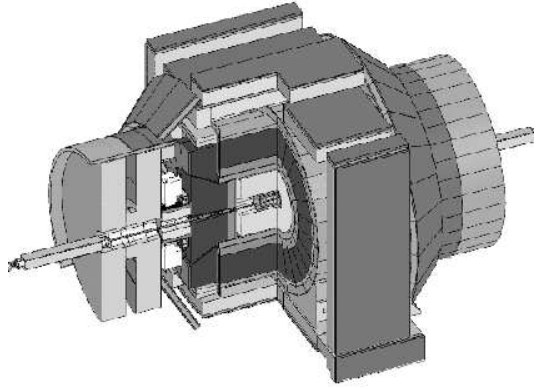} 
\caption{Cut-away view of the CDF detector.}
\label{fig:cdfCutAway}
\end{figure}

CDF is a $\sim$5,000 ton detector \cite{CDFTDR} enveloping the B0 collision point of the \Tevatron\ (Fit.~\ref{fig:tevatronSketch}). Externally, it looks forward-backward symmetric (Fig~\ref{fig:cdfCutAway}), mostly made of steel, of dimensions that are approximately $16~\mbox{m} \times 13~\mbox{m} \times 13~\mbox{m}$. It is underground, shielded behind tons of concrete, which keeps it somewhat insulated from environmental sources of noise and prevents potentially hazardous radiation from leaking into its immediate surroundings. A three story building houses in its basement the detector and its assembly site, while in the superjacent levels it accommodates the data acquisition devices and the Control Room, from where operations are managed.

\begin{figure}[t]
\centering
\includegraphics[width=15cm,angle=0]{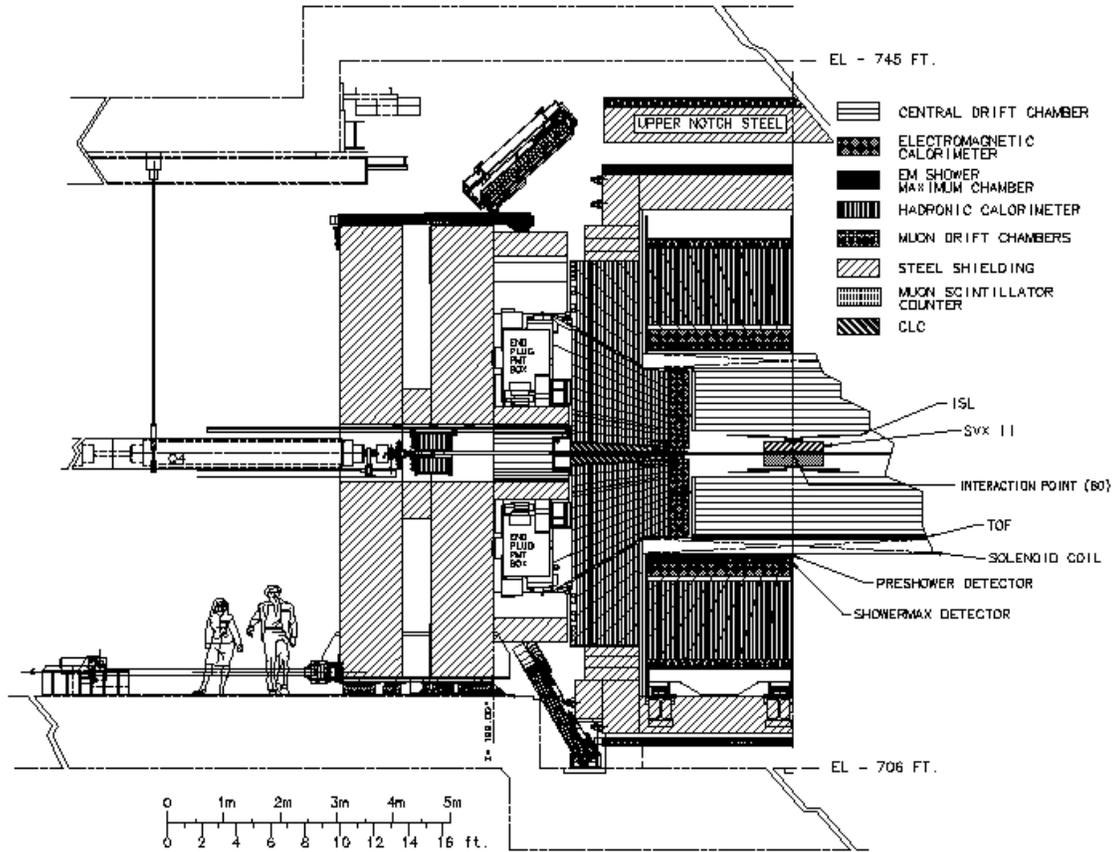} 
\caption{Transverse section of half of the CDF detector in Run~II.}
\label{fig:cdfII}
\end{figure}

The CDF detector allows for a broad range of physics searches, from heavy flavor physics to searches of exotic new phenomena. It combines a variety of features, i.e. tracking, timing, calorimetry and muon detection systems, all seamed together with powerful trigger and DAQ systems. 

By 1996, when the Run~I period of \Tevatron\ was over, about 90 $\mbox{pb}^{-1}$ of data had been collected, in which the long-sought $t$-quark had eventually been discovered \cite{Abe1995hr}. In preparation for the even more ambitious Run~II era, which started in 2001, CDF was decisively upgraded \cite{CDFTDR}, with new tracking and calorimetry capabilities and a much more efficient muon detection system. The DAQ system had to be upgraded too, to respond to the expected instantaneous luminosity of up to $5\times 10^{32}~{\mbox{cm}}^{-2}{\mbox{s}}^{-1}$. In the following sections, the current status of CDF will be described.

\subsection{Coordinate Systems}
Before describing the most important CDF components, it would be useful to present the established system of coordinates used at the experiment.

The Cartesian coordinate system has its axes starting at the detector's center, where the beams of $p$ and \pbar\ are supposed to collide. The $y$ axis is defined to point vertically up, and the $x$ to be perpendicular to the beam pipe and pointing in the direction away from the center of the \Tevatron\ ring. In terms of $\hat x$ and $\hat y$, $\hat z$ is $\hat x \times \hat y$, which approximately coincides with the direction in which the $p$ beam travels through the center of CDF.

The cylindrical coordinate system reflects the approximate axial symmetry of the tracker and the calorimeter around $\hat z$, which in cylindrical coordinates remains the same unit vector it was in Cartesian. The radial unit vector $\hat r$ at each point is perpendicular to and pointing away from the $z$ axis. The azimuthal angle $\phi$ is by definition $0$ on the semi-infinite $z - x$ plane that contains the positive $x$ axis and increases in the direction of $\hat \phi = \hat z \times \hat r$. 

Spherical coordinates are used more often than the above two systems. The reason is that, to the physical event occurring in a $p \bar p$ scattering, the cylindrical or any other symmetry of the surrounding detector is irrelevant. The dynamics of the event recognize one special axis, viz.~$z$, along which the $p$ and \pbar\ were traveling right before their collision. It is therefore convenient to define the angles of all outcoming particles with respect to $\hat z$. For any point in space, a radial unit vector $\hat r$ is defined to point in the direction away from the beginning of the coordinates. Also, a polar angle $\theta$ is defined, which is $0$ along the positive $z$ axis and increases in the direction of $\hat \theta = \hat r \times \frac{\hat r \times \hat z}{\sin\theta}$. Finally, the azimuthal angle $\phi$ is defined as in the cylindrical coordinates and increases along $\hat \phi = \hat \theta \times \hat r$.  

Since the $p$ and \pbar\ beams are unpolarized, $z$ has to be an axis of symmetry when examining a large set of events. In other words, based on the premise of isotropy of the universe which leaves $z$ as the only axis special to the scattering, there can be no law of physics that would cause a non-uniform $\phi$ distribution of the particles coming out of the scattering. 

It is common to not mention the polar angle $\theta$ per se, but instead a dimensionless quantity called ``pseudorapidity'', which is related to $\theta$ as
\begin{equation}
\eta = -\ln(\tan(\theta / 2)).
\end{equation}
$\eta$ is the $E \rightarrow \abs{\vec p}$ limit of the quantity called ``rapidity'', which is\footnote{The rapidity $y$ may not be confused with the Cartesian coordinate $y$.} 
\begin{equation}
y = \frac{1}{2} \ln{\frac{E+p_z}{E-p_z}},
\end{equation}
and has the beautiful property that for any pair of rapidities, the difference $\Delta y$ is invariant under Lorentz boosts along the $z$ axis.

\subsection{Tracking}
Tracking is crucial for particle identification; it has been so since the first experiments with wire and bubble chambers. Though technology has advanced, the principles remain:
\begin{itemize}
\item Only ionizing particles leave tracks, which distinguishes them from neutral ones.
\item The curvature of a track under the influence of Lorentz force in the presence of a magnetic field $\vec B$ is a measure of the transverse momentum ${\vec p}_T$ of the particle, namely of the projection of its momentum $\vec p$ on the plane transverse to $\vec B$.
\item The direction of the track can be used to estimate the direction ($\eta$,$\phi$) in which a particle is produced.
\item Being able to observe tracks improves our intuitive understanding of what particles are produced in an event. For example, the assembly of tracks within a cone is indicative of hadronic jet showers, while isolated tracks are more likely leptons\footnote{Even though $\tau$ is a lepton, it is common to include only electrons and muons in the term ``leptons'', because they are easier to identify than $\tau$ which often decays hadronically, so they consist more ``clear'' leptons in the experimental sense.}.
\item Extrapolating the tracks of an event down to their origin(s) indicates the position of the event. This can reveal the existence of displaced secondary vertices, indicative of the decay of a long-lived particle, such as a $B^0$ meson. It may also indicate the existence of multiple $p \bar p$ interactions in the same bunch crossing, by observation of multiple primary vertices in the same event.
\end{itemize}

\subsubsection{Silicon Detector}
\begin{figure}
\centering
\includegraphics[width=13cm,angle=0]{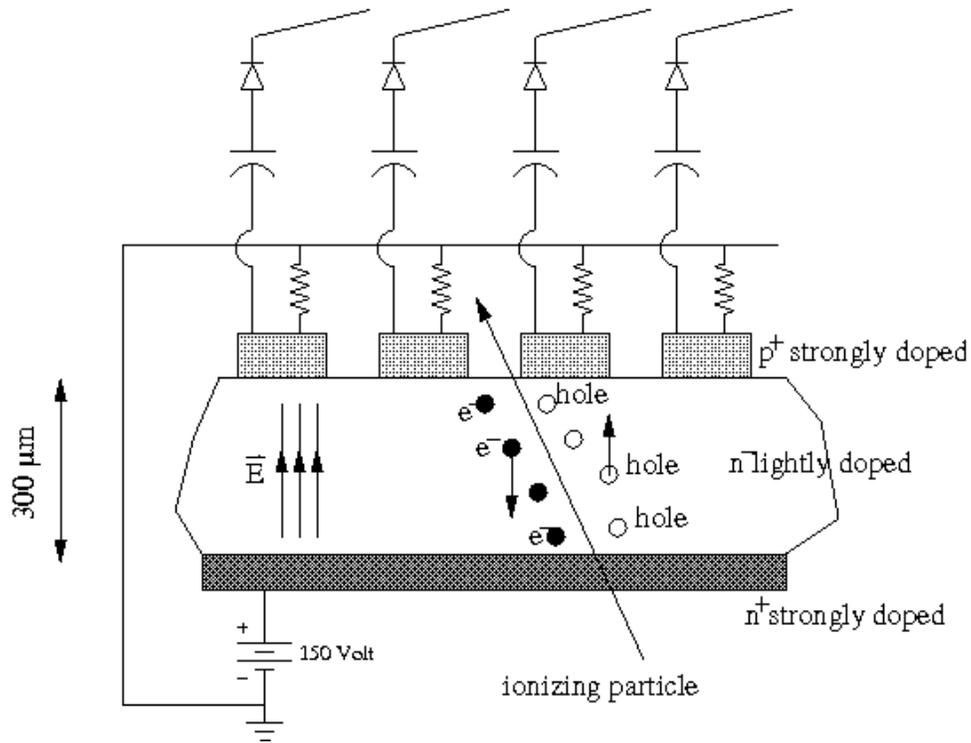} 
\caption[Schematic of a silicon particle sensor.]{Schematic of a silicon particle sensor. An array of finely spaced p-type silicon strips is implanted in an n-type silicon substrate, typically $300$ $\mu$m thick. The n-p contact is then reversely polarized, typically with a depletion voltage of 150 V. When an ionizing particle traverses the depletion zone it creates a localized stream of $e^-$-hole pairs, which are collected by the nearest strips, where after amplification they are detected as small current signals. There are variations in the design of silicon strips, such as double-sided strips where signals are read from both sides. The spatial resolution of the most advanced silicon strip can be as fine as 2 - 4 $\mu$m, limited mostly by diffusion \cite{PDBook,SiliconDetectors}.}
\label{fig:siliconSketch}
\end{figure}

The first tracking device particles pass through is the Silicon Detector. Silicon allows for a highly granular and radiation tolerant tracker that can survive as near as 1.5 cm from the collision point \cite{CDFTDR}. The operation principle of a silicon micro-strip is depicted in Fig.~\ref{fig:siliconSketch} \cite{PDBook, SiliconDetectors}.

\begin{figure}
\centering
\includegraphics[width=10cm,angle=0]{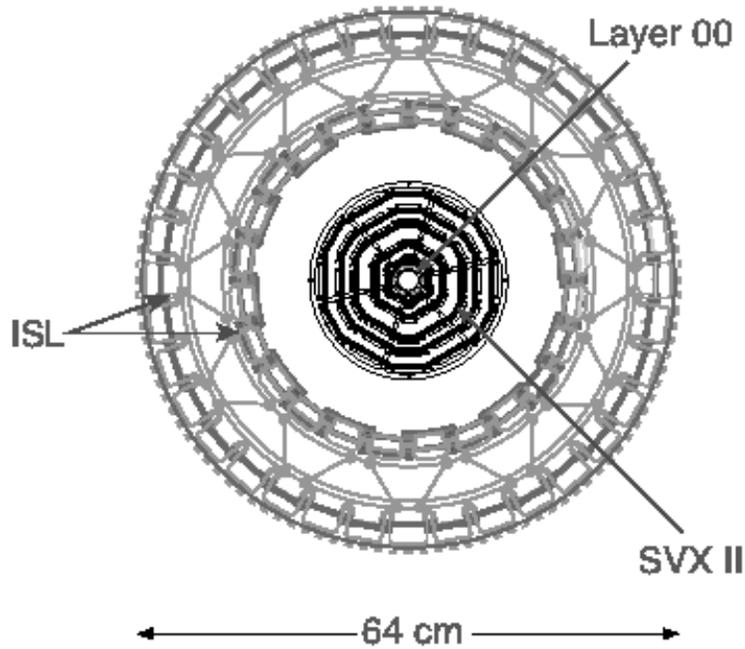} 
\caption[The CDF Silicon Detector (XY view)]{The CDF Silicon Detector (XY view) \cite{Boveia:2005kj}.}
\label{fig:SiliconXY}
\end{figure}

\begin{figure}
\centering
\includegraphics[width=15cm,angle=0]{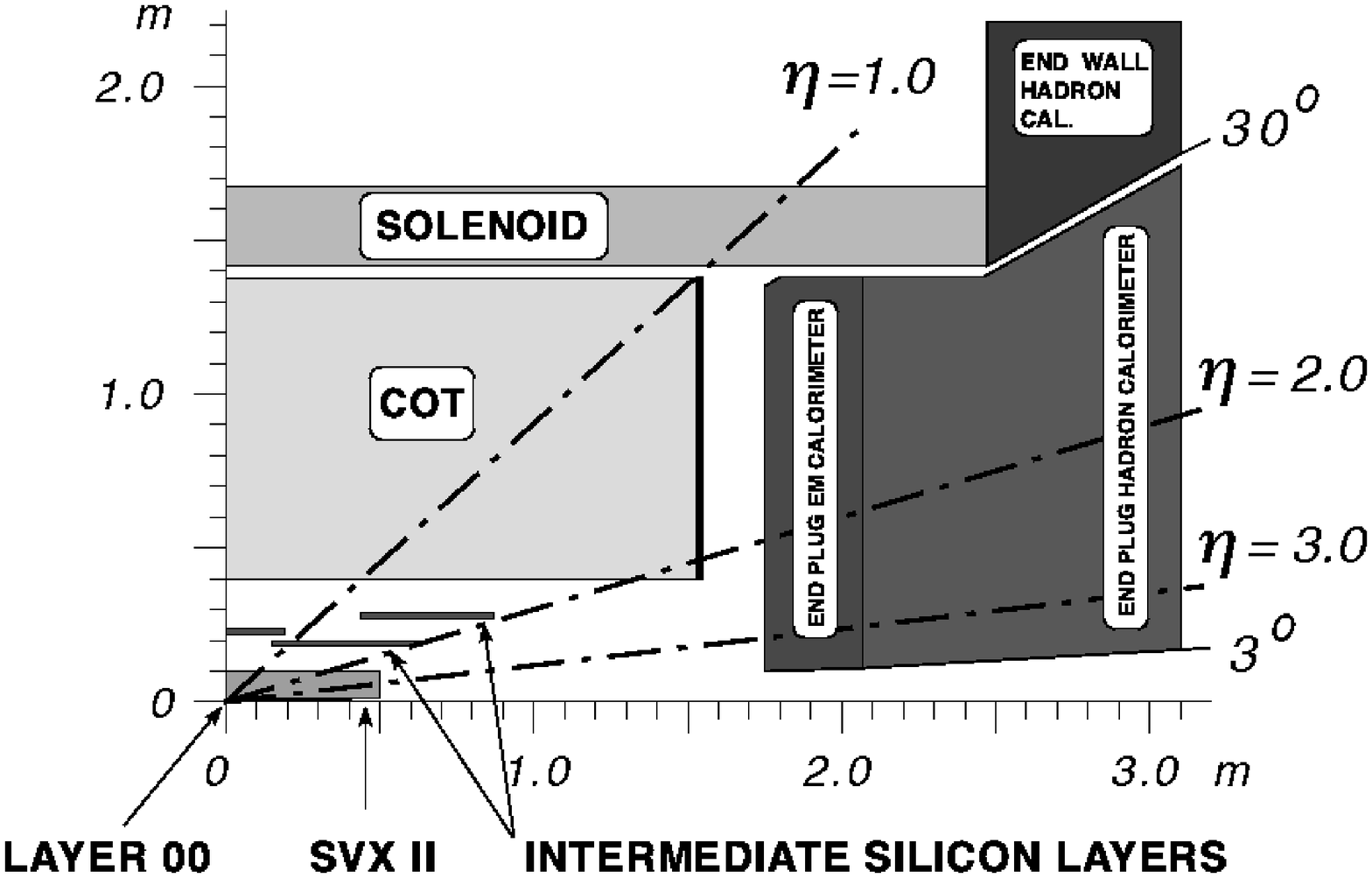} 
\caption[Schematic profile (RZ view) of the central part of the CDF detector.]{Schematic profile (RZ view) of the central part of the CDF detector \cite{SiliconCDF}. The Time Of Flight detector (not shown) is between the COT and the solenoid. The central electromagnetic and hadronic calorimeter are not depicted either.}
\label{fig:RZview}
\end{figure}

About 722,000 read-out channels come from the Silicon Detector \cite{Boveia:2005kj}, by far more than from any other CDF component. It is separated in three subsystems: L00, SVX and ISL (Fig.~\ref{fig:SiliconXY}, \ref{fig:RZview}).

L00 is a single layer of single-sided silicon built directly onto the beam pipe, at 1.5 cm radius. It provides precision position measurement before the particles undergo multiple scattering.

SVX is the heart of the Silicon Detector, consisting of 12 identical wedges in $\phi$. Each wedge contains 5 layers of double-sided silicon, oriented parallel to the beam pipe at radii from 2.5 to 10.6 cm. On one side, the silicon strips are aligned axially. The other side has $90^\circ$ stereo strips for 3 of the layers, and $1.2^\circ$ stereo strips for the remaining 2 layers. Obviously, the choice of aligning some strips non-axially was made to allow for three-dimensional track reconstruction.

The ISL envelops SVX. It carries $1.2^\circ$ stereo double-sided silicon in a single layer for intermediate radius measurement of central\footnote{Here and below the word ``central'' is used to describe objects with $\abs{\detEta}<1.0$; ``plug'' is used to describe objects with $1.0<\abs{\detEta}<2.5$.} tracks and in two layers for tracking in the region $1<\abs{\eta} < 2$, which is not completely covered by the COT (Fig.~\ref{fig:RZview}).

The silicon embedded strips are 8 $\mu$m wide \cite{SiliconCDF}, which brings the hit's spatial resolution down to about 12 $\mu$m. This resolution makes it possible to measure the impact parameter of a track to 40 $\mu$m, with 30 $\mu$m uncertainty due to the beam width. The $z_0$, namely the $z$-coordinate of the primary vertex, can be measured with 70 $\mu$m accuracy.

\subsubsection{Central Outer Tracker}
The COT \cite{COT1,COT2} is a cylindrical multi-wire open-cell drift chamber surrounding the Silicon Detector (Fig.~\ref{fig:RZview}). 

COT contains Argon-Ethane ($Ar-C_2H_6$) in a 1:1 mixture. When charged particles traverse the gaseous mixture they leave a trail of ionization electrons, which drift under the influence of an 1.9 kV/cm electric field. The latter is produced by field planes and homogenized by potential and shaper wires. After some time that depends on the distance they travel, the ionization electrons are collected by sense wires immersed in the gas producing a detectable\footnote{When an ionization electron approaches the 40$\mu$m thick sense wire it is accelerated by its rapidly increasing ($1/r$) electric field, producing an ``avalanche'' of secondary ionization electrons and thus enhancing the signal.} electric signal. The $r-\phi$ location of the track with respect to the sense wire is then estimated from the time it takes to detect the signal. The drift distance is less than 0.88 cm and is covered in less than 100 ns, which is less than the 396 ns between successive bunch crossings, therefore causes no pile-up of signals from different events. 

\begin{figure}
\centering
\includegraphics[width=8cm,angle=0]{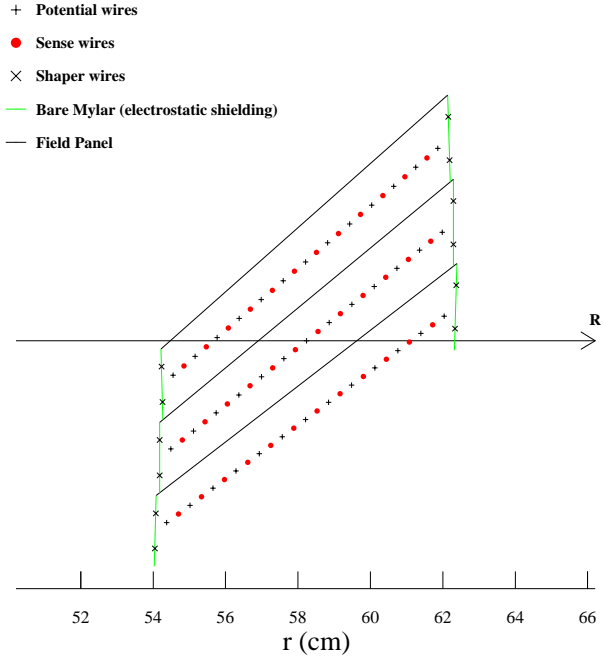} 
\caption[Three COT cells from the second superlayer (XY view).]{Three COT cells from the second superlayer (XY view). Their inclination with respect to the radial direction is equal to the Lorentz angle of $35^\circ$ (see text).}
\label{fig:COTcell}
\end{figure}

The field panels, shape, potential and sense wires are all grouped in electrostatically shielded cells (Fig.~\ref{fig:COTcell}). Each cell contains 12 sense, 13 potential and 4 shaper wires. Sense and potential wires alternate with successive sense wires being 7mm apart. Combining drift time information from several wires, the single hit resolution reduces to about 140 $\mu$m.

\begin{figure}
\centering
\includegraphics[width=12cm,angle=0]{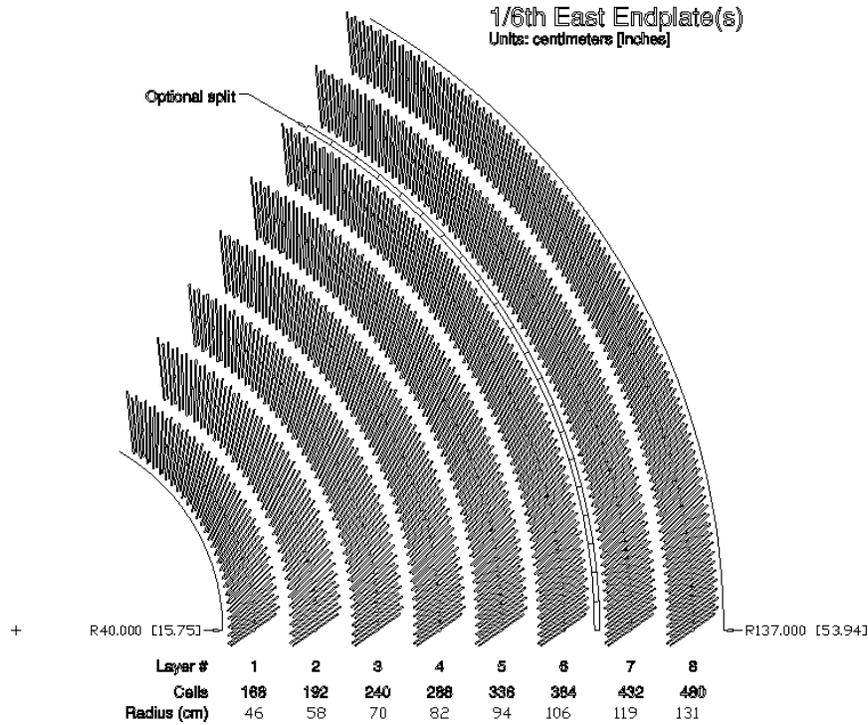} 
\caption[Part of the COT endplate (XY view).]{Part of the COT endplate (XY view). The wire-plane slots grouped into eight superlayers are shown.}
\label{fig:COTsuperlayers}
\end{figure}

Cells are arranged in 8 superlayers (Fig.~\ref{fig:COTsuperlayers}). The wires in the $1^{\mbox{st}}$ and $5^{\mbox{th}}$ superlayer are not oriented axially, but at a stereo angle of $+3^\circ$. Similarly, there is a stereo angle of $-3^\circ$ in superlayers 3 and 7. Like in the case of the Silicon Detector, the reason that 4 out of the 8 superlayers are oriented non-axially is to allow for tracking in the three dimensions\footnote{If all COT wires were parallel to the $z$ axis, then the $z$ coordinate of hits would be unknown.}. 

\begin{figure}
\centering
\includegraphics[width=12cm,angle=0]{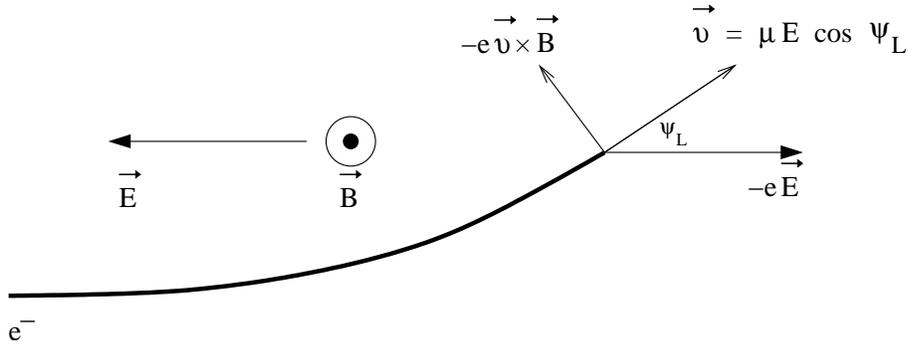} 
\caption[The trajectory of an ionization electron in the $\vec E$ and $\vec B$ field of the COT.]{The trajectory of an ionization electron in the $\vec E$ and $\vec B$ field of the COT. The condition $e E \sin\psi_L = e\upsilon B=e\mu E \cos\psi_L B$ determines the Lorentz angle $\psi_L = \arctan \mu B$.}
\label{fig:lorentzAngle}
\end{figure}

It was mentioned that ionization electrons drift under the influence of an electric field $\vec E$, but there is also a magnetic field $\vec B$ parallel to the $z$ axis. So, as the force $-e\vec E$ accelerates the electron, the force $-e\vec \upsilon \times \vec B$ turns it on the $x-y$ plane (Fig.~\ref{fig:lorentzAngle}). At any time the velocity of the electron in the medium can be parametrized as $\vec \upsilon = \mu \vec E$, where $\mu$ is the {\em mobility} of the medium. Assuming that the $\vec E$ field is homogeneous on the $x-y$ plane and the electron is non-relativistic, the equilibrium is at an angle $\psi$ with respect to $\vec E$ that is $\psi_L =\arctan \mu \abs{\vec B}$. $\psi_L$ is called the Lorentz angle and for the COT it is about $35^\circ$. The wires in the COT cells are then arranged along the direction determined by the Lorentz angle, to minimize the drift time and maximize the COT efficiency and resolution (Fig.~\ref{fig:COTcell}).

\subsubsection{Magnet}
A 1.4 T magnetic field is produced in the  $- \hat z$ direction by the superconductive solenoid surrounding the COT (Fig.~\ref{fig:RZview} and \ref{fig:cdfII}). 

The magnetic field is essential for the measurement of the transverse momentum ($p_T$) of ionizing particles. Greater magnetic field intensity and bigger tracking volume radius improve $p_T$ resolution, which on the other hand is limited by the spatial resolution of the tracker and multiple scattering \cite{PDBook}. At CDF, the $p_T$ resolution is $\delta(1/p_T)=\frac{0.15\%}{\mbox{{\small GeV/c}}}$.

\subsubsection{Track reconstruction} 
The Silicon Detector and the COT record a large number of hits in each event, viz.~discrete positions from which ionizing particles seem to have passed.
 But the hits alone do not suffice. In each event there are tens of charged particles, as well as false hits. What is needed is an algorithm to reconstruct tracks out of the thousands of hits of each event.

Every track is a helix that can be parametrized in terms of the variables in Table \ref{tab:helixParameters}. Essentially, tracking algorithms fit for those 5 parameters to best match the observed hits \cite{Tracking1, Tracking2}.

\begin{table}
\centering
\begin{tabular}{c p{13cm}}
$\theta$ & the polar angle at minimum approach, which refers to the point of the track closest to the $z$ axis. \\
$C$ & semi-curvature of the track (inverse of diameter), with the same sign as the particle's electric charge. \\
$z_0$ & $z$ coordinate at minimum approach. \\
$D$ & signed impact parameter: distance between helix and the $z$ axis at minimum approach. The sign of $D$ is given from its formal definition: $D = \mbox{sign}(q)(\sqrt{x_0^2+y_0^2}-\rho)$, where $q$ is the ionizing particle's charge, ($x_0$,$y_0$) is the center of the track's projection onto the $x-y$ plane, and $\rho$ is the radius of the same projection. Fig.~\ref{fig:signOfD} demonstrates combinations of positive and negative $D$ and $C$. \\
$\phi_0$ & Direction of track on $x-y$ plane at minimum approach, i.e.~the polar angle of the particle's $p_T$ at minimum approach.
\end{tabular}
\caption{The 5 parameters of a helical track.}
\label{tab:helixParameters}
\end{table}

\begin{figure}
\centering
\includegraphics[width=10cm,angle=0]{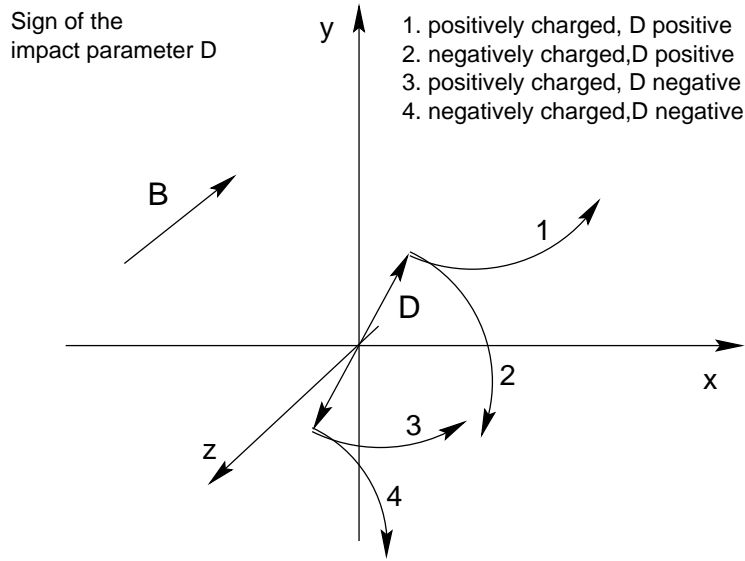} 
\caption{Combinations of positive and negative $D$ and $C$ (see Table \ref{tab:helixParameters}).}
\label{fig:signOfD}
\end{figure}




Tracking in the COT using the Segment Linking algorithm involves first reconstructing linear segments of the track in each of the eight superlayers \cite{Tracking2}. Then, the linear segments from the axial layers are linked to form a 2D track on the $x-y$ plane, starting the extrapolation with the outmost segment as seed. The $r-z$ projection of the track is attained by linking the segments from the stereo superlayers. Eventually, the track is characterized by the $\chi^2$ of the fit, and is only kept if that figure of merit is below threshold. 

\begin{figure}
\centering
\includegraphics[width=10cm,angle=0]{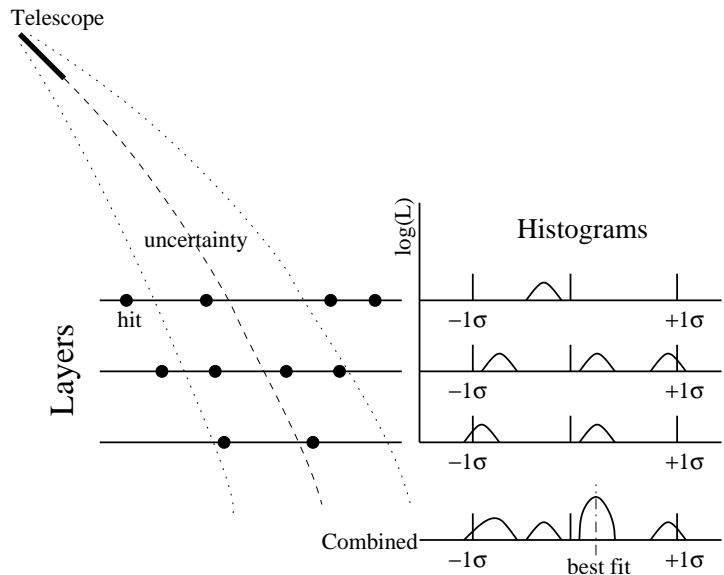} 
\caption{Schematic of the Histogram Tracking method.}
\label{fig:histogramMethod}
\end{figure}

An alternative is the Histogram Tracking algorithm \cite{Tracking2}. It starts with a coarse approximation of the final track, which is attained by extrapolating a segment of the track called ``telescope'', such as the outer superlayer segment. The extrapolated telescope corresponds to a helix whose parameters carry large uncertainty, therefore instead of a curve it can imagined as a tube, to visualize those uncertainties (Fig.~\ref{fig:histogramMethod}). In each layer the tube crosses there may be hits that fall inside the tube. For those hits, the likelihood is calculated to belong to the track. Each crossed layer is translated into a histogram of those likelihoods. Those histograms coming from different layers are then combined into a final one, and the track is reconstructed as the helix which maximizes the combined likelihood. Compared to the Segment Linking algorithm, this alternative is slower but more efficient in cases of missing and accurate in cases of spurious hits.

The Histogram Tracking algorithm is also applied in Silicon tracking, where the part of the track in the COT is used as the telescope. 

In Silicon tracking \cite{Tracking2}, the information of the $z$ of the primary vertex is used. That is known by combining hits from the stereo strips and extrapolating to the beam axis. This produces a variety of candidates, each of different likelihood, so in the end the primary vertex is at the most likely $z$.

The Stand-Alone algorithm for Silicon tracking uses information exclusively from silicon hits, therefore has the advantage of using the whole $\abs{\eta}<2$ acceptance of the Silicon Detector. It starts by finding hits in places where axial and stereo strips intersect. Then, triplets of aligned hits are identified. The information of the primary vertex is used to constrain the candidate helices. In the end the best fitting helix is kept.

The Outside-In algorithm \cite{TrackingOutsideIn} takes COT tracks and extrapolates them into the Silicon Detector, adding hits via a progressive fit. As each layer of silicon is encountered, a road size is established based on the error matrix of the track. Hits that are within the road are added to the track, and the track parameters and error matrix are refit with this new information. A new track candidate is generated for each hit in the road, and each of these new candidates are then extrapolated to the next layer in, where the process is repeated. As the extrapolation proceeds, the track error matrix is inflated to reflect the amount of scattering material encountered. At the end of this process, there may be many track candidates associated with the original COT track. The candidate that has hits in the largest number of silicon layers is chosen as the winner; if more than one candidate has the same number of hits, the $\chi^2$ of the fit in the silicon is used to decide.

The Inside-Out algorithm \cite{TrackingInsideOut} performs the reverse extrapolation: from the Silicon Detector to the COT. Its goal is to use the Stand-Alone silicon track to associate it with COT hits and improve the efficiency of reconstruction of tracks that do not cross more than 4 COT superlayers.

\subsection{Calorimetry}
CDF is equipped with sampling electromagnetic and hadronic calorimeters in the central and plug region, enhanced with shower maximum and preshower detectors for improved particle identification \cite{CDFTDR}. Central calorimeters cover $2\pi$ rads in $\phi$ (Fig.~\ref{fig:cdfCutAway}). The central electromagnetic calorimeter covers $\abs{\eta}<1.1$ and the hadronic $\abs{\eta}<1.3$. The plug calorimeters reach as far as $\abs{\eta}=3.6$. They are segmented in wedge-shaped towers pointing to the center of CDF. Each tower covers about 0.1 units of $\eta$ and $15^\circ$ in $\phi$ (Fig.~\ref{fig:cdfII}). For increased acceptance, the hadronic calorimeter has the endwall calorimeter, spanning $30^\circ < \abs{90^\circ-\theta} < 45^\circ$ (Fig.~\ref{fig:RZview}).

\subsubsection{Electromagnetic Calorimeter}
CEM and PEM comprise lead absorber sheets alternating with scintillator layers. Light produced at the scintillator is transfered by WLS fibers to two PMTs that correspond to each tower\footnote{Having two PMTs per tower allows for cross-check of the validity of signals, using time information and comparing the difference in the signal intensity in the two.}.

The CEM has a total maximum thickness of about 19 $X_0$, in 20-30 (varying with $\abs{\eta}$) layers of 3 mm lead and 5 mm scintillator. Its energy resolution, after \textit{in situ} calibration, is found to be $13.5\%/\sqrt{E_T}\oplus 2\%$. 

PEM contains 22 layers of lead, 4.5 mm each\footnote{The first layer is an exception, being 1 cm thick and read out separately to be used as a preshower detector.}, and its scintillator layers are 4 mm thick. Its total thickness is 21 $X_0$. Its resolution is $16\%/\sqrt{E_T}\oplus 1\%$.

In both CEM and PEM, there is a shower maximum detector, 6 $X_0$ into the calorimeter, where an electromagnetic shower statistically contains the biggest number of particles \cite{PDBook}. CES is a multi-wire proportional chamber with strip readout in the $z$ direction and wire along $\phi$. PES has scintillator strips that cross to form a 2-dimensional grid in each plug. 
With resolution of about 2 mm in the central and 1 mm in the plug, the showermax detectors facilitate the matching of tracks with calorimeter hits, improving $e^\pm/\gamma$ identification. Also, sampling the profile of the electromagnetic showers at 6 $X_0$ allows for improved $\gamma/\pi^0$ identification. 

Finally, between the solenoid and the first layer of the CEM lies a set of multi-wire proportional chambers, the CPR, which samples the electromagnetic showers at 1.075 $X_0$, viz.~the solenoid's thickness. This information greatly enhances $\gamma$ and soft $e^\pm$ identification \cite{CDFTDR}.

\subsubsection{Hadronic Calorimeter}
The hadronic calorimeter is similar to the electromagnetic, except that it uses iron for absorber instead of lead.
The CHA is 4.7 $\lambda_0$ thick, consisting of 32 2.5 cm iron layers alternating with 1 cm scintillator layers. Its energy resolution is $75\%/\sqrt{E_T}\oplus3\%$.

The WHA has similar energy resolution \cite{Abe:1988me}; $75\%/\sqrt{E_T}\oplus4\%$. It contains 15 layers of iron, 5 cm each, alternating with 1 cm layers of scintillator, adding up to 4.5 $\lambda_0$.

The PHA is thicker, containing 7 $\lambda_0$ in 23 layers of iron, 51 mm each, alternating with 6 mm layers of scintillator. Its energy resolution is $80\%/\sqrt{E_T}\oplus5\%$.

\subsection{Muon System}
\begin{figure}
\centering
\includegraphics[width=7cm,angle=0]{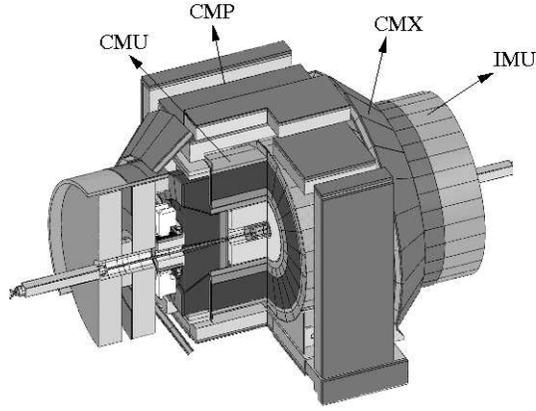} 
\caption{The muon detectors of CDF.}
\label{fig:muonDetectors}
\end{figure}

CDF is equipped with four muon detectors (Fig.~\ref{fig:muonDetectors}), which will be described in this section.

Muons weigh 200 times more than electrons, therefore radiate about $200^2=40,000$ times less by bremsstrahlung. They do not deposit much energy in the calorimeter, but rather traverse the whole detector almost unimpeded. This makes them easier to identify by installing wire chambers around the detector, beyond the calorimeter and even beyond extra absorbing material; muons are virtually the only ionizing particles that can reach there. 

Shielding the muon detectors behind absorber increases the detected muons' purity, but also enhances multiple scattering, which makes it harder to match the small track segment in the muon detector (called ``stub'') with the corresponding COT track. However this is not a very big problem, especially for high-$p_T$ muons, since the displacement due to multiple scattering is about $\frac{15\ \mbox{cm}}{p_T}$, for the $p_T$ is in GeV/c \cite{CDFTDR}. Furthermore, some low-$p_T$ muons can not reach the muon detectors, but that is not a problem either, since the threshold is lower than 2.2 GeV/c \cite{CDFTDR}, far lower than the $p_T$ of the muons considered in this analysis.

\subsubsection{Central Muon detector (CMU)}
\begin{figure}
\centering
\includegraphics[width=10cm,angle=0]{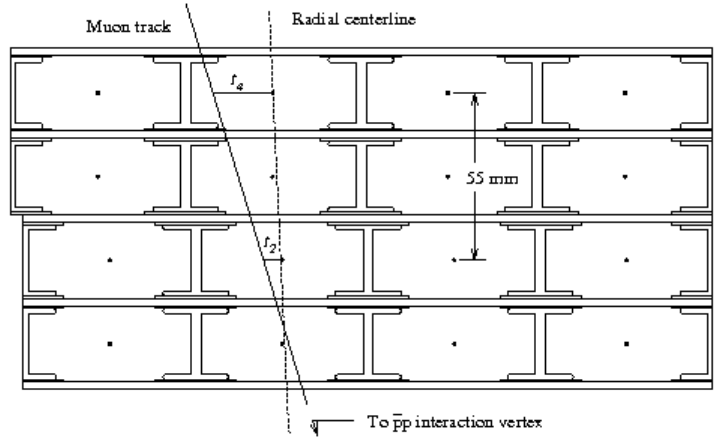} 
\caption{Cross section of a CMU chamber. Each vertical array is one stack.}
\label{fig:CMU}
\end{figure}

The CMU \cite{CDFTDR} surrounds the hadronic calorimeter, at radius 3.47 m, covering the $\abs{\eta}<0.6$ region. It consists of argon-ethane wire chamber cells operating in proportional mode, organized in stacks of four. Each wire chamber is $2.7\times6.4\times226\ \mbox{cm}^3$ with a resistive stainless steel wire along its biggest dimension, which is aligned parallel to the $z$ axis. In $\phi$ it is segmented in 24 wedges, each containing 4 stacks side by side, therefore each wedge contains a chamber of $4\times4=16$ cells (Fig.~\ref{fig:CMU}).

\begin{figure}
\centering
\includegraphics[width=5cm,angle=0]{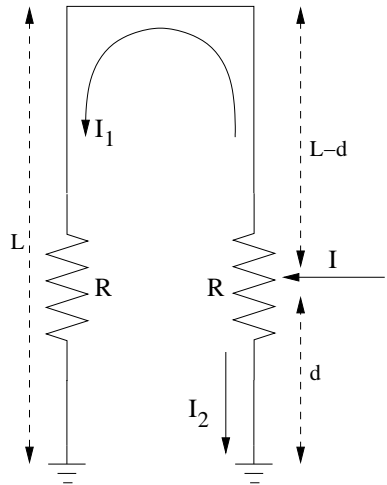} 
\caption[The principle of charge division method.]{The principle of charge division method. The ionization charge is collected at some position $d$ along the $z$ axis, and splits into two currents: $I_1$ and $I_2$. From Ohm's law, $I_1 R (1 + \frac{L-d}{L}) = I_2 R \frac{d}{L} \Rightarrow I_1 (2L-d) = I_2 d \Rightarrow d = \frac{2LI_1}{I_2 + I_1}$. With the approximation that all currents last for the same amount of time $\Delta t$, we can write $Q_i = I_i\mbox{const}_{\Delta t}$. Therefore, by measuring $Q_1$ and $Q_2$ one can determine $d=\frac{2LQ_1}{Q_1+Q_2}$.}
\label{fig:chargeDivision}
\end{figure}

The drift times ($<800$ ns) are used to measure the $r-\phi$ projection of the track. The $z$ coordinate of the track is extracted with about 10 cm precision, using the {\em charge division} method, whose principle is explained in Fig.~\ref{fig:chargeDivision}. To apply this method, every couple of $\phi-$adjacent cells have their wires ganged together at one end.

\subsubsection{Central Muon Upgrade detector (CMP)}
The CMP (Fig.~\ref{fig:muonDetectors}) is shielded behind about 7.8 $\lambda_0$, comprising the calorimeter, the magnet return yoke and extra steel absorber. Compared to the CMU, which was shielded behind only 5 hadronic interaction lengths, the CMP provides higher purity in muon identification \cite{CDFTDR}. Those reconstructed muons that have a stub in both the CMU and the CMP are called ``CMUP muons''.

The CMP is not azimuthally symmetric, but resembles a box surrounding the central region of the detector ($\abs{\eta} < 0.6$). It is made of wire chambers similar to those used for the CMU, but just bigger: $2.5\times15\times640\ \mbox{cm}^3$.

A bigger difference is that CMP contains scintillator counters in addition to wire chambers. The scintillator layers lie on the outer side of the chambers and provide timing information that is used to discard out-of-time muon candidates, which could not possibly be muons originating from the center of the detector. Furthermore, timing helps not have stubs from different bunch crossings piled up, given that the drift time in the CMP can be as large as 1.7 $\mu$s \cite{CDFTDR}. Eventually, the dimensions of the scintillator counters are $2.5\times30\times320\ \mbox{cm}^3$, so two silicon counters are needed to cover the $z$ dimension of the CMU, providing the very crude information of whether a muon stub has positive or the negative $z$ coordinate.

\subsubsection{CMX}
CMX \cite{CDFTDR} is very similar to CMP; it consists of same type wire chambers and silicon counters. It differs significantly in geometry though. It covers the region $0.6 < \abs{\eta} < 1$ and is shaped like a conic section on each side of the detector (Fig.~\ref{fig:muonDetectors}). The wire chambers are grouped in wedges, each $15^\circ$ in $\phi$. Each wedge contains 48 chambers, arranged in 8 layers. The lower $90^\circ$ of the CMX, which physically penetrate the floor supporting the detector, are called ``miniskirt'' for obvious reason (Fig.~\ref{fig:muonDetectors}). This part was not instrumented until past 2003.

\subsubsection{IMU}
IMU \cite{CDFTDR} covers the region $1 < \abs{\eta} < 1.5$ (Fig.~\ref{fig:muonDetectors}). It comprises silicon counters and wire chambers of dimensions $2.5\times8.4\times363\ \mbox{cm}^3$. In combination with ISL tracking, it provides muon reconstruction and momentum measurement in the $\abs{\eta} > 1$ region.

\subsection{Cerenkov Luminosity Counter}
CDF is equipped with the CLC \cite{CLC}, a detector dedicated to measuring instantaneous luminosity ($\Ell$). It consists of $2 \times 48$ Cerenkov counters placed in the far forward and backward region ($3.75 < \abs{\eta} < 4.75$). filled with isobutane at nearly atmospheric pressure.

The number of $p\bar p$ interactions ($n$) in a bunch crossing follows the Poisson distribution with mean $\mu = \sigma_{p\bar p}\Ell t_{BC}$, where $\sigma_{p \bar p}$ is the cross section of inelastic $p \bar p$ scattering and $t_{BC}$ is the time interval between bunch crossings.  

Bunch crossings with $n=0$ occur with probability $P_0(\mu)=e^{-\mu}$. By measuring the fraction of empty crossings $\mu$ can be measured\footnote{Of course it is necessary to correct the measured $\mu$ by dividing with the CLC acceptance $\epsilon$.} and therefore $\cal L$. 

An alternative method consists in measuring directly $\mu$ as $N/N_1$, where $N$ is the number of CLC counts of some bunch crossing, and $N_1$ is the average number of CLC counts in the case of single-interaction bunch crossings. $N_1$ can be measured at low $\Ell$, when $\mu \ll 1$.

The first method, of measuring empty crossings, has the advantage of not needing any information such as $N_1$, but at high $\Ell$ empty crossings become rare, making this method inefficient. On the other hand, the second method depends on the $N_1$ information, and $N/N_1$ in reality does not scale linearly with \Ell, as the CLC occupancy grows and is eventually saturated due to the finite number of counters, therefore correction for this non-linearity are required.

The uncertainty in the integrated luminosity measured with the CLC is 6\%, to which the biggest contribution comes from the uncertainty in $\sigma_{p\bar p}$ at 1.96 TeV.

\subsection{Data Acquisition}

CDF employs approximately $10^6$ readout channels. A bunch crossing at $\Ell \sim 2\times10^{32}$ \lumiUnits\ yields on average about 5 $p\bar p$ interactions. An event of such multiplicity takes about 200 kB of digitized information volume. It becomes then obvious that not every single bunch crossing can be read, as that would require the enormous bandwidth of $\sim$630 GB/s.

Apart from technically inevitable, it is also sensible to record only those events that pass some quality selection and would be of some interest\footnote{In an experiment of the broad scope of CDF it is not trivial to decide which events could be of some interest, since different analyses may see interest in different kinds of events. Furthermore, nobody is certain what the signature of physics beyond the Standard Model will be.}. For example, an event with leptons should be retained, while for multi-jet events it is enough to keep only a fraction of them, since they are so abundant in $p\bar p$ collisions.

\begin{figure}
\centering
\includegraphics[width=13cm,angle=0]{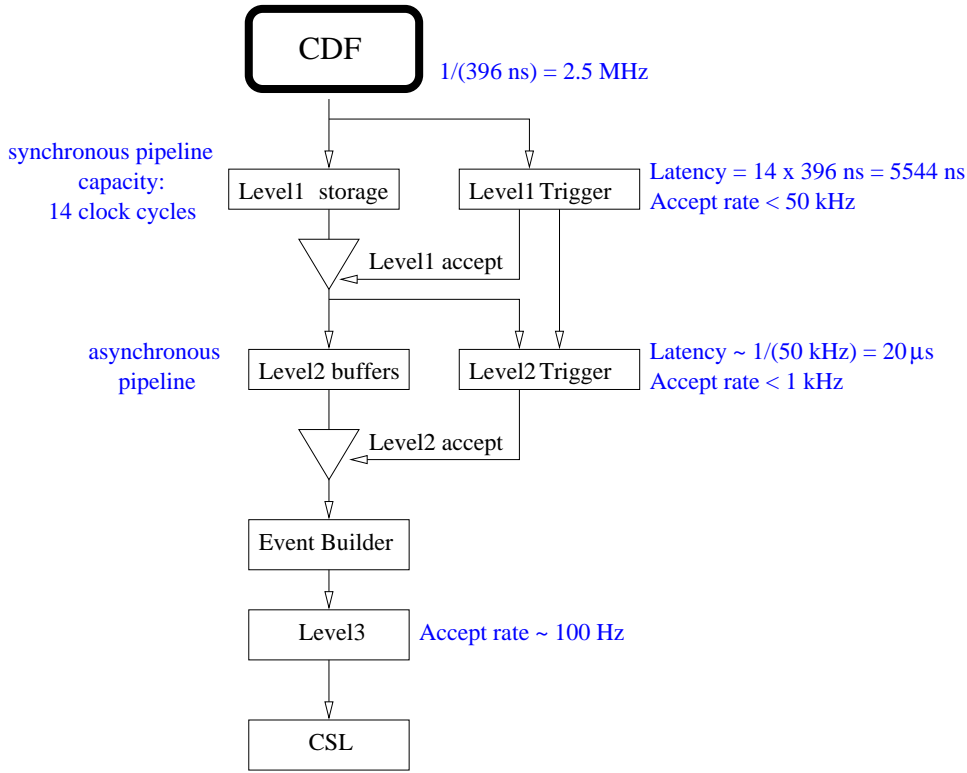}
\caption{Diagram of the CDF DAQ system.}
\label{fig:DAQ}
\end{figure}

The DAQ system \cite{CDFTDR} is responsible for selecting the best events as they occur. Fig.~\ref{fig:DAQ} provides an overview of the DAQ architecture.

\subsubsection{Level-1}

The frequency of 2.5 MHz at which bunches cross is too high to allow for full reconstruction of every event, so the first level of selection is based on fragments of information. This happens in Level-1; an accept/reject decision is made using ``primitives'', namely coarse information on COT tracks and stubs in the CMU, CMP and CMX \cite{CDFTDR}. Systems providing primitives are depicted in Fig.~\ref{fig:L1L2}. The XFT crudely reconstructs COT tracks on the $x-y$ plane. The XTRP extrapolates XFT tracks through the calorimeter and the muon system finding matching hits/towers.

Based on the primitives, several algorithms $-$ also called ``individual triggers'' $-$ contribute to the Level-1 decision. For example, effort is made to keep events with high-$p_T$ tracks, or leptons, or large missing transverse energy (\met) etc.

The latency of Level-1 is 5.5 $\mu$s, in which 14 bunch crossings occur. Therefore, all front-end electronics are equipped with buffers of enough capacity to contain information from 14 bunch crossings. Level-1 then works as a synchronous pipeline; by the time 14 events are pushed back into the buffer, at least one event has been examined and pulled from it, freeing a slot for the current event to be buffered.

Less than 2\% of the events pass Level-1, making its accept rate less than 50 kHz.

\subsubsection{Level-2}
Level-2 functions as an asynchronous pipeline, where events are processed in FIFO mode \cite{CDFTDR}. With no more than $50$ kHz input rate, it can afford up to 1/50 kHz = 20 $\mu$s to decide on each event\footnote{Actually, since up to 4 events can be kept in the Level-2 buffer, the latency can be even greater, without causing dead-time, provided that this is not the case for too many events.}.

\begin{figure}
\centering
\includegraphics[width=10cm,angle=0]{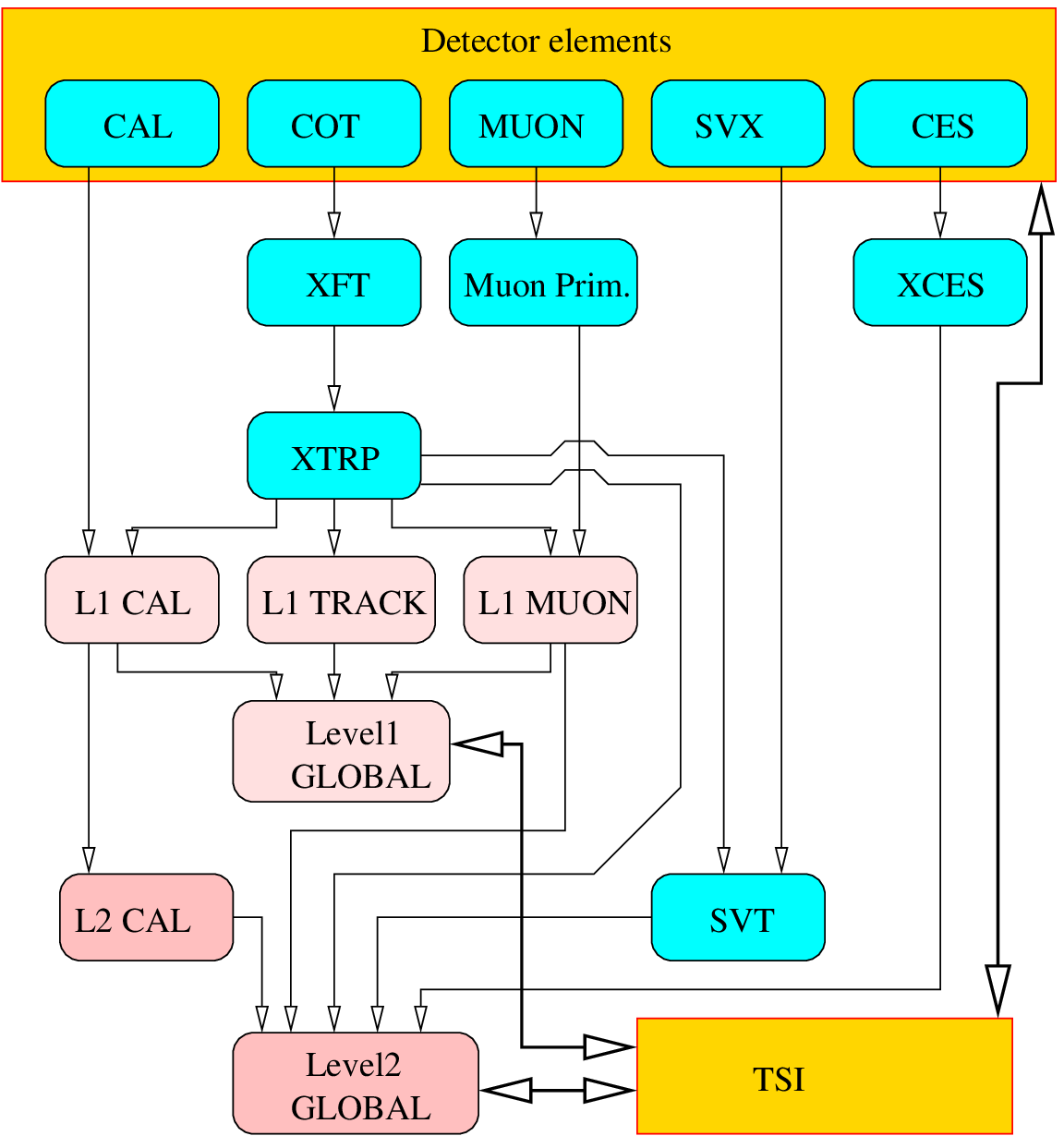}
\caption[Information flow within Level-1 and Level-2.]{Information flow within Level-1 and Level-2. XCES is a system that generates the stimulated showermax bitmap and finds matching tracks extrapolated by XTRP to define electron candidates. The SVT extrapolates XFT tracks into the SVX, providing the $D$ and $\phi_0$ information (Table \ref{tab:helixParameters}). The TSI coordinates the flow of information and interfaces to the CDF clock, which is used to know when a bunch crossing is occurring.}
\label{fig:L1L2}
\end{figure}

In its decision, Level-2 takes into account the primitives of Level-1, in addition to showermax information, as shown in Fig.~\ref{fig:L1L2}. 

The acceptance rate of Level-2 is less than 1 kHz. Effort is made to maintain this rate as close to 1 kHz as possible, by readjusting the trigger requirements as \Ell\ changes, making them stricter at high \Ell\ and looser at low \Ell.

\subsubsection{Event Builder}

In the case of a Level-2 accept, the whole detector is eventually read out. The EVB collects the fragments of the event and passes them to Level-3. Reading out the front-end electronics of the whole detector takes about 1 ms, which is why this step is only possible after having discarded over 99.96\% of the events. 

\begin{figure}
\centering
\includegraphics[width=10cm,angle=0]{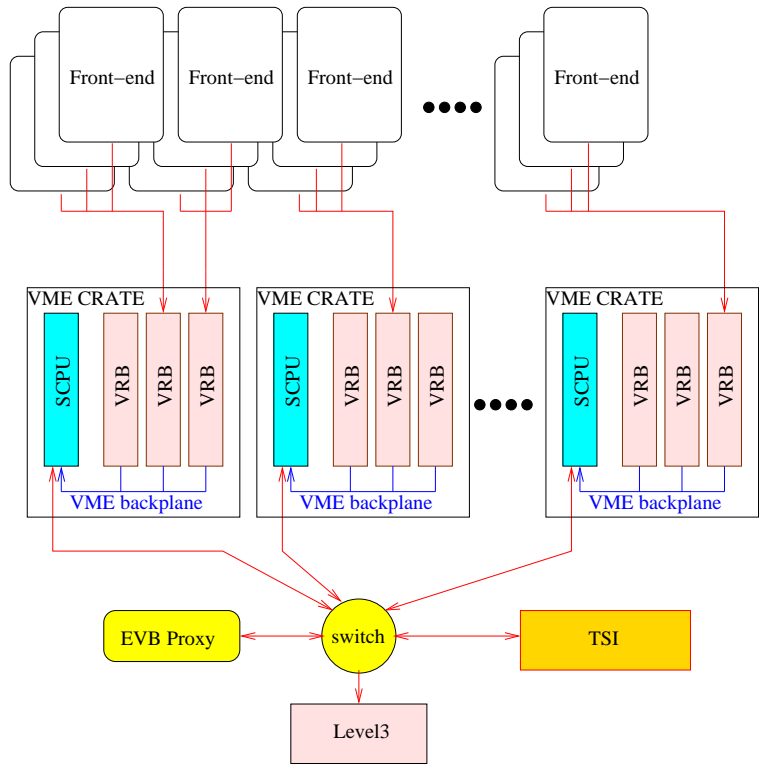}
\caption{Diagram of the Event Builder.}
\label{fig:EVB}
\end{figure}

EVB (Fig.~\ref{fig:EVB}) lies in 21 VME crates, each containing one Linux computer, referred to as SCPU \cite{EVBaces}. Each crate is dedicated to reading a different part of the detector. Apart from the SCPU, each crate contains a series of memory buffers, the VRBs. When the front-end crates are read, the information of the event is first stored in the VRBs. Each SCPU reads the VRBs of it own crate through the VME backplane of the crate, which in combination with the GigaBit Ethernet networking allows for the desired system speed. On reading the VRBs, a byte-count check is performed, as well as checks of the size of each buffer entry \cite{EVBchecks}. Though in principle EVB should not be discarding any events, it does so if information is missing or corrupted.

The function of the EVB is coordinated by the EVB Proxy, a process running on a dedicated Linux machine. All acknowledgement messages within the EVB are circulated through the EVB Proxy, and so does any information exchanged with the TSI and Level-3.

\subsubsection{Level-3}

Level-3 is the last stage of trigger selection \cite{EVBaces}. Receiving events from the EVB at $<1$ kHz, it is purely software implemented, performing three basic functions:
\begin{enumerate}
\item{Concatenates same-event fragments coming from the EVB into an event entry.}
\item{Imposes the final selection, taking into account the reconstructed objects information.}
\item{Submits passing events to the CSL for storage.}
\end{enumerate}

There is a whole cluster of 411 Linux computers counting 2.4 THz of CPU dedicated to Level-3. Though all computers are nearly identical, they are separated in three categories, depending on their task:
\begin{itemize}
\item{18 Converter nodes: They receive event fragments from the EVB and combine them to form self-contained event records which they pass to available Processor nodes.}
\item{384 Processor nodes: Upon reception of events from a Converter, they apply the Level-3 filter to either discard or pass them to an Output node, after some reformatting that reshapes the passing entries to their final format.}
\item{9 Output nodes: They receive the passing events from Processor nodes and propagate them to the CSL for storage.}
\end{itemize}

The Level-3 cluster is separated in 18 identical subsets, called ``subfarms''\footnote{A term appropriate for a subdivision of the whole Level-3 cluster, which is called ``farm'' in CDF jargon.}.  This way, data handling proceeds in 18 independent, parallel streams which share the load of incoming events. Each subfarm contains 1 Converter, 21 or 22 Processors, and shares an Output with another subfarm. On every Processor, 5 Level-3 filters run simultaneously, on hyper-threaded dual-core Intel CPUs. The Converter of each subfarm is allowed to only submit events to Processors of its own subfarm, and the Processors of each subfarm can only send events to the Output node serving it.

The operation of Level-3 is coordinated by the Level-3 Proxy application, running on a dedicated computer. The Proxy collects and sends acknowledgements from and to the computers of the cluster, and communicates with the EVB Proxy to indicate among other things which Converter is available to receive the next event.

Filtering is done by a program written in \texttt{C++}, the Level-3 filter executable, which applies criteria stored in a centralized database implemented in Oracle. In the database is stored the {\em trigger table}, which is a list of ``triggers''. Each trigger is structured to contain the following information:
\begin{enumerate}
\item{The prerequisite Level-1 and Level-2 triggers.}
\item{The \texttt{C++} reconstruction modules that should be used and in what order.}
\item{The specific selection criteria decided having some physics goal, for example a cut in some invariant mass in the event.}
\item{The name of the dataset in which to store the event if it passes the trigger selection.}
\end{enumerate}

The output rate of Level-3 is about 100 Hz. The events passing Level-3 are sent to the CSL for immediate storage. From there, they are shortly sent to the FCC for permanent storage on magnetic tape.

\subsection{Off-line production}
Data analysis is not performed on the raw data. Before the data on tape are usable, the off-line production process has to take place.

At production \cite{CDFTDR}, the raw data banks are unpacked and physics objects are reconstructed in full detail. This is similar to what is done at Level-3, but the off-line reconstruction is much more elaborate, applying the latest calibrations, since those reconstructed objects will be the final ones to be used for analysis.

Since passing Level-3, each event contains the information of the dataset(s) it belongs to. At the production, even further partitioning is made; datasets are collections of filesets, which are collections of files containing events. 

For the needs of each analysis, the raw data are taken from the appropriate dataset and are converted to a convenient format. Since ROOT \cite{ROOT} is the adopted analysis framework, the format varies between different architectures of ROOT Trees. For example, one is the ``topNtuple'', used mostly by collaborators doing $t$-quark analyses, but a more common format, used also in the present analysis, is the ``Standard Ntuple'' (\Stntuple).

\chapter{Data Analysis}
\label{chapter:1fb-1}

The analysis going into this thesis was conducted in two rounds: first with 1 fb$^{-1}$ of data, and then with 2 fb$^{-1}$.  The first round has been documented in \cite{prd1,prl1,susy07vista,susy07sleuth}.  An updated publication is currently being prepared for the second one.  This chapter is an adaptation of \cite{prd1}, while chapter \ref{chapter:2fb-1} presents material that will be in the publication of the second round.

\section{Strategy}

Sec.~\ref{sec:intro-motivation} motivates the goal of this analysis, viz.\ the model-independent search for new physics.  The method is to obtain a satisfactory description of the Standard Model expectation in channels where high-$p_T$ data are observed, and employ an array of probes to seek for statistically significant discrepancies between data and Standard Model background.

Crucial for model-independence is to not focus on channels sensitive to particular models, but examine data in as many channels as possible.  That introduces to this analysis over two million events (in 1 fb$^{-1}$), ranging from abundant QCD to rare electroweak ones.  Studying this large volume of qualitatively diverse data requires reducing the information content of each event to bare bones and characterizing each event in terms of physics objects that maintain the same meaning universally in any kind of event.  In each event, the 4-momenta of any reconstructed physics objects in its final state are recorded.  These objects can be leptons, photons, hadronic jets or missing energy. 

Another ingredient of model-independence is to not segregate the data into ``control'' and ``signal'' regions {\em a priori}, namely into regions where new physics is assumed to not exist or to exist respectively.  In most analyses control regions are predefined, to adjust correction factors, under the assumption that there is no new physics in those regions and that the extrapolation of correction factors from the control to the signal region is valid.  However, what is considered control region in one analysis is often signal region in some other, so, to be as generic as possible, one needs to treat all data as signal and control regions simultaneously, to address the question ``how well does the Standard Model implementation describe the data?''  If there is indeed detectable new physics, then it will be impossible to achieve good agreement between data and Standard Model simultaneously in all regions.  More in Sec.~\ref{sec:blindOrNot}. 

The Standard Model prediction is implemented in three steps:
\begin{enumerate}
\item{Monte Carlo generation and matching \cite{CKKW:Krauss:2002up} of samples simulating the Standard Model processes.}
\item{CDF detector simulation, which models the detector response to the MC generated events. For that, the {\sc Geant}-based package \CdfSim\ is used.}
\item{Fine-tuning of the outcome of \CdfSim\ to account for theoretical and experimental correction factors.}
\end{enumerate}

Structurally, the analysis contains four parts:
\begin{enumerate}
\item{The \Vista\ global fit\index{Vista@\Vista!global fit}, which adjusts and applies the correction model, providing the Standard Model background of the best possible global agreement with the data, exploiting the flexibility granted by the correction model.}
\item{The \Vista\ comparison\index{Vista@\Vista!comparison}, which examines the statistical significance of features in the bulk of all distributions and sorts the information in a comprehensive way.}
\item{The \Sleuth\ search, which focuses on the high-\sumPt\ tails searching for excesses of data.}
\item{The Bump Hunter search (present only for the second round of the analysis), which scans all mass variables for local excesses of data, potentially indicating a new resonance.}
\end{enumerate}
The above statistical probes are employed simultaneously, rather than sequentially.  So, an effect highlighted by \Sleuth\ prompts additional investigation of the discrepancy, usually resulting in a specific hypothesis explaining the discrepancy in terms of a detector effect or adjustment to the Standard Model prediction that is then fed back and tested for global consistency.

Statistical significance is a necessary but insufficient condition for discovery.  A statistically significant discrepancy could be attributed to inaccuracy in the Standard Model implementation, or in modeling the detector response.  These possibilities would need to be considered on a case-by-case basis.  In the event of a significant discrepancy, the breadth of view of this analysis can be exploited to evaluate the plausibility of it being a detector effect or a problem in the Standard Model implementation.

Forming hypotheses for the cause of specific discrepancies, implementing those hypotheses to assess their wider consequences, and testing global agreement after the implementation are emphasized as the crucial activities for the investigator throughout the process of data analysis.  This process is constrained by the requirement that all adjustments be physically motivated.  The investigation and resolution of discrepancies highlighted by the algorithms is {\em{the}} defining characteristic of this global analysis~\footnote{It is not possible to systematically simulate the process of constructing, implementing, and testing hypotheses motivated by particular discrepancies, since this process is carried out by individuals.  The statistical interpretation of this analysis is made bearing this process in mind.}.

This search for new physics terminates when either a compelling case for new physics is made, or there remain no statistically significant discrepancies on which a new physics case can be made. In the former case, to quantitatively assess the significance of the potential discovery, a full treatment of systematic uncertainties must be implemented.  In the latter case, it is sufficient to demonstrate that all observed effects are not in significant disagreement with an appropriate global Standard Model description.


\section{\Vista}
\label{sec:Vista}

This section describes \Vista: object identification, event selection, estimation of Standard Model backgrounds, simulation of the CDF detector response, development of a correction model, and results.


\subsection{Object identification}

Energetic and isolated electrons, muons, taus, photons, jets, and $b$-tagged jets with $\abs{\detEta}<2.5$ and $p_T>17$~GeV are identified according to CDF standard criteria.  The same criteria are used for all events.  The isolation criteria employed vary according to object, but roughly require less than 2~GeV of extra energy flow within a cone of $\Delta R = \sqrt{\Delta \eta^2 + \Delta \phi^2} = 0.4$ in $\eta$--$\phi$ space around each object.

Standard CDF criteria~\cite{WandZCrossSectionPaper:Abulencia:2005ix} are used to identify electrons ($e^\pm$) in the central and plug regions of the CDF detector.  Electrons are characterized by a narrow shower in the central or plug electromagnetic calorimeter and a matching isolated track in the central gas tracking chamber or a matching plug track in the silicon detector.  

Standard CDF muons ($\mu^\pm$) are identified using three separate subdetectors in the regions $\abs{\detEta}<0.6$, $0.6<\abs{\detEta}<1.0$, and $1.0<\abs{\detEta}<1.5$~\cite{WandZCrossSectionPaper:Abulencia:2005ix}.  Muons are characterized by a track in the central tracking chamber matched to a track segment in the central muon detectors, with energy consistent with minimum ionizing deposition in the electromagnetic and hadronic calorimeters along the muon trajectory.

Narrow central jets with a single charged track are identified as tau leptons ($\tau^\pm$) that have decayed hadronically~\cite{Ztautau}.  Taus are distinguished from electrons by requiring a substantial fraction of their energy to be deposited in the hadron calorimeter; taus are distinguished from muons by requiring no track segment in the muon detector coinciding with the extrapolated track of the tau.  Track and calorimeter isolation requirements are imposed.

Standard CDF criteria requiring the presence of a narrow electromagnetic cluster with no associated tracks are used to identify photons ($\gamma$) in the central and plug regions of the CDF detector~\cite{CdfPhotonId:Acosta:2004sn}.

Jets ($j$) are reconstructed using the JetClu~\cite{JetClu:Abe:1991ui} clustering algorithm with a cone of size $\Delta R = 0.4$, unless the event contains one or more jets with $p_T>200$~GeV and no leptons or photons, in which case cones of $\Delta R = 0.7$ are used.~\cdfSpecific{\footnote{Jet energies are corrected to level 7, using {\tt jetCorr04b}.}}  Jet energies are appropriately corrected to the parton level~\cite{jetEnergyScale:Bhatti:2005ai}.  Since uncertainties in the Standard Model prediction grow with increasing jet multiplicity, up to the four largest $p_T$ jets are used to characterize the event; any reconstructed jets with $p_T$-ordered ranking of five or greater are neglected and their energy is treated as unclustered, except in final states with small summed scalar transverse momentum containing only jets.

A secondary vertex $b$-tagging algorithm is used to identify jets likely resulting from the fragmentation of a bottom quark ($b$) produced in the hard scattering~\cite{CdfBtagging:Neu:2006rs}.

Momentum visible in the detector but not clustered into an electron, muon, tau, photon, jet, or $b$-tagged jet is referred to as unclustered momentum ({\tt uncl}).

Missing momentum ($\pmiss$) is calculated as the negative vector sum of the 4-vectors of all identified objects and unclustered momentum.  An event is said to contain a $\pmiss$ object if the transverse momentum of this object exceeds \pTmin~GeV, and if additional quality criteria discriminating against fake missing momentum due to jet mismeasurement are satisfied~\footnote{An additional quality criterion is applied to the significance of the missing transverse momentum $\vec{\pmiss}_T$ in an event, requiring that the energies of hadronic objects can not be adjusted within resolution to reduce the missing transverse momentum to less than 10~GeV.  The transverse components of all hadronic energy clusters $\vec{p}_{Ti}$ in the event are projected onto the unit missing transverse momentum vector $\hat{\pmiss}_T=\vec{\pmiss}_T/\abs{\vec{\pmiss}_T}$, and a ``conservative'' missing transverse momentum ${\pmiss_T}'=\pmiss_T - 2.5 \sqrt{ \sum_i{ \abs{ \vec{p}_{Ti} \cdot \hat{\pmiss}_T} } }$ is defined, where the sum is over hadronic energy clusters in the event, and the hadronic energy resolution of the CDF detector has been approximated as $100\% \sqrt{{p_T}_i}$, expressed in GeV.  An event is said to contain missing transverse momentum if $\pmiss_T>\pTmin$~GeV and ${\pmiss_T}'>10$~GeV.}.


\subsection{Event selection}
\label{sec:Vista:OfflineTrigger}

Events containing an energetic and isolated electron, muon, tau, photon, or jet are selected.  A set of three level online triggers requires:
\begin{itemize}
\item a central electron candidate with $p_T>18$~GeV passing level 3, with an associated track having $p_T>8$~GeV and an electromagnetic energy cluster with $p_T>16$~GeV at levels 1 and 2; or
\item a central muon candidate with $p_T>18$~GeV passing level 3, with an associated track having $p_T>15$~GeV and muon chamber track segments at levels 1 and 2; or
\item a central or plug photon candidate with $p_T>25$~GeV passing level 3, with hadronic to electromagnetic energy less than 1:8 and with energy surrounding the photon to the photon's energy less than 1:7 at levels 1 and 2; or
\item a central or plug jet with $p_T>20$~GeV passing level 3, with 15~GeV of transverse momentum required at levels 1 and 2, with corresponding prescales of 50 and 25, respectively; or
\item a central or plug jet with $p_T>100$~GeV passing level 3, with energy clusters of 20~GeV and 90~GeV required at levels 1 and 2; or
\item a central electron candidate with $p_T>4$~GeV and a central muon candidate with $p_T>4$~GeV passing level 3, with a muon segment, electromagnetic cluster, and two tracks with $p_T>4$~GeV required at levels 1 and 2; or
\item a central electron or muon candidate with $p_T>4$~GeV and a plug electron candidate with $p_T>8$~GeV, requiring a central muon segment and track or central electromagnetic energy cluster and track at levels 1 and 2, together with an isolated plug electromagnetic energy cluster; or
\item two central or plug electromagnetic clusters with $p_T>18$~GeV passing level 3, with hadronic to electromagnetic energy less than 1:8 at levels 1 and 2; or
\item two central tau candidates with $p_T>10$~GeV passing level 3, each with an associated track having $p_T>10$~GeV and a calorimeter cluster with $p_T>5$~GeV at levels 1 and 2.
\end{itemize}

Events satisfying one or more of these online triggers are recorded for further study.  
Offline event selection for this analysis uses a variety of further filters.
Single object requirements keep events containing:
\begin{itemize}
\item a central electron with $p_T>25$~GeV, or
\item a plug electron with $p_T>40$~GeV, or
\item a central muon with $p_T>25$~GeV, or
\item a central photon with $p_T>60$~GeV, or
\item a central jet or $b$-tagged jet with $p_T>200$~GeV, or
\item a central jet or $b$-tagged jet with $p_T>40$~GeV (prescaled by a factor of roughly $10^4$),
\end{itemize}
possibly with other objects present.  
Multiple object criteria select events containing:
\begin{itemize}
\item two electromagnetic objects (electron or photon) with $\abs{\eta}<2.5$ and $p_T>25$~GeV, or
\item two taus with $\abs{\eta}<1.0$ and $p_T>17$~GeV, or
\item a central electron or muon with $p_T>17$~GeV and a central or plug electron, central muon, or central tau with $p_T>17$~GeV, or
\item a central photon with $p_T>40$~GeV and a central electron or muon with $p_T>17$~GeV, or
\item a central or plug photon with $p_T>40$~GeV and a central tau with $p_T>40$~GeV, or
\item a central photon with $p_T>40$~GeV and a central $b$-jet with $p_T>25$~GeV, or
\item a central jet or $b$-tagged jet with $p_T>40$~GeV and a central tau with $p_T>17$~GeV (prescaled by a factor of roughly $10^3$), or
\item a central or plug photon with $p_T>40$~GeV and two central taus with $p_T>17$~GeV, or
\item a central or plug photon with $p_T>40$~GeV and two central $b$-tagged jets with $p_T>25$~GeV, or
\item a central or plug photon with $p_T>40$~GeV, a central tau with $p_T>25$~GeV, and a central $b$-tagged jet with $p_T>25$~GeV,
\end{itemize}
possibly with other objects present.  
Explicit online triggers feeding this offline selection are required.  The $p_T$ thresholds for these criteria are chosen to be sufficiently above the online trigger turn-on curves  that trigger efficiencies can be treated as roughly independent of object $p_T$.

Good run criteria are imposed, requiring the operation of all major subdetectors.  To reduce contributions from cosmic rays and events from beam halo, standard CDF cosmic ray and beam halo filters are applied~\cite{CdfCosmicFilter}.

These selections result in a sample of roughly two million high-$p_T$ data events in an integrated luminosity of 927~pb$^{-1}$.


\subsection{Event generation}
\label{sec:Vista:EventGeneration}

Standard Model backgrounds are estimated by generating a large sample of Monte Carlo events, using the \Pythia~\cite{Pythia:Sjostrand:2000wi}, \MadEvent~\cite{MadEvent:Maltoni:2002qb2}, and \Herwig~\cite{Herwig:Corcella:2002jc} generators.  \MadEvent\ performs a leading order matrix element calculation, and provides 4-vector information corresponding to the outgoing legs of the underlying Feynman diagrams, together with color flow information.  \Pythia\ 6.218 is used to handle showering and fragmentation.  The CTEQ5L~\cite{CTEQ5L:Lai:1999wy} parton distribution functions are used.  

\paragraph*{$\text{QCD jets}$.}
QCD dijet and multijet production are estimated using \Pythia.  Samples are generated with Tune A~\cite{RickFieldPythiaTunes} with lower cuts on $\hat{p}_T$, the transverse momentum of the scattered partons in the center of momentum frame of the incoming partons, of 0, 10, 18, 40, 60, 90, 120, 150, 200, 300, and 400~GeV.  These samples are combined to provide a complete estimation of QCD jet production, using the sample with greatest statistics in each range of $\hat{p}_T$.  
\paragraph*{$\gamma\text{+jets}$.}
The estimation of QCD single prompt photon production comes from \Pythia.  Five samples are generated with Tune A corresponding to lower cuts on $\hat{p}_T$ of 8, 12, 22, 45, and 80~GeV.  These samples are combined to provide a complete estimation of single prompt photon production in association with one or more jets, placing cuts on $\hat{p}_T$ to avoid double counting.  
\paragraph*{$\gamma\gamma\text{+jets}$.}
QCD diphoton production is estimated using \Pythia.  
\paragraph*{$V\text{+jets}$.}
The estimation of $V$+jets processes (with $V$ denoting $W$ or $Z$), where the $W$ or $Z$ decays to first or second generation leptons, comes from \MadEvent, with \Pythia\ employed for showering.  Tune AW~\cite{RickFieldPythiaTunes} is used within \Pythia\ for these samples.  The CKKW matching prescription~\cite{CKKW:Krauss:2002up} with a matching scale of 15~GeV is used to combine these samples and avoid double counting.  Additional statistics are generated on the high-$p_T$ tails using the MLM matching prescription~\cite{MrennaMatching:Mrenna:2003if}.  The factorization scale is set to the vector boson mass; the renormalization scale for each vertex is set to the $p_T$ of the jet.  $W$+4 jets are  generated inclusively in the number of jets; $Z$+3 jets are generated inclusively in the number of jets.  
\paragraph*{$VV\text{+jets}$.}
The estimation of $WW$, $WZ$, and $ZZ$ production with zero or more jets comes from \Pythia.  
\paragraph*{$V\gamma\text{+jets}$.}
The estimation of $W\gamma$ and $Z\gamma$ production comes from \MadEvent, with showering provided by \Pythia.  These samples are inclusive in the number of jets.
\paragraph*{$W(\rightarrow\tau\nu)\text{+jets}$.}
Estimation of $W\rightarrow\tau\nu$ with zero or more jets comes from \Pythia.
\paragraph*{$Z(\rightarrow\tau\tau)\text{+jets}$.}
Estimation of $Z\rightarrow\tau\tau$ with zero or more jets comes from \Pythia.  
\paragraph*{$t\bar{t}$.}
Top quark pair production is estimated using \Herwig\ assuming a top quark mass of 175~GeV and NNLO cross section of $6.77\pm 0.42$~pb~\cite{Kidonakis:2003qe}.
 
Remaining processes, including for example $Z(\rightarrow\nu\bar{\nu})\gamma$ and $Z(\rightarrow\ell^+\ell^-)b\bar{b}$, are generated by systematically looping over possible final state partons, using \MadGraph~\cite{MadGraph:Stelzer:1994ta} to determine all relevant diagrams, and using \MadEvent\ to perform a Monte Carlo integration over the final state phase space and to generate events.  The MLM matching prescription is employed to combine samples with different numbers of final state jets.

A higher statistics estimate of the high-$p_T$ tails is obtained by computing the thresholds in $\sum{p_T}$ corresponding to the top 10\% and 1\% of each process, where $\sum{p_T}$ denotes the scalar summed transverse momentum of all identified objects in an event.  Roughly ten times as many events are generated for the top 10\%, and roughly one hundred times as many events are generated for the top 1\%.  

\paragraph*{Cosmic rays.}
Backgrounds from cosmic ray or beam halo muons that interact with the hadronic or electromagnetic calorimeters, producing objects that look like a photon or jet, are estimated using a sample of data events containing fewer than three reconstructed tracks. This procedure is described in more detail in Appendix~\ref{sec:CorrectionModelDetails:CosmicRays}. 

\paragraph*{Minimum bias.}
Minimum bias events are overlaid according to run-dependent instantaneous luminosity in some of the Monte Carlo samples, including those used for inclusive $W$ and $Z$ production.  In all samples not containing overlaid minimum bias events, including those used to estimate QCD dijet production, additional unclustered momentum is added to events to mimic the effect of the majority of multiple interactions, in which a soft dijet event accompanies the rare hard scattering of interest.  A random number is drawn from a Gaussian centered at 0 with width 1.5 GeV for each of the $x$ and $y$ components of the added unclustered momentum.  Backgrounds due to two rare hard scatterings occurring in the same bunch crossing are estimated by forming overlaps of events, as described in Appendix~\ref{sec:Overlaps}.

Each generated Standard Model event is assigned a weight, calculated as the cross section for the process (in units of picobarns) divided by the number of events generated for that process, representing the number of such events expected in a data sample corresponding to an integrated luminosity of 1~pb$^{-1}$.  When multiplied by the integrated luminosity of the data sample used in this analysis, the weight gives the predicted number of such events in this analysis.


\subsection{Detector simulation}
\label{sec:Vista:DetectorSimulation}

The response of the CDF detector is simulated using a {\sc{geant}}-based detector simulation (\CdfSim)~\cite{Gerchtein:2003ba}, with {\sc{gflash}}~\cite{GFLASH:Grindhammer:1989zg} used to simulate shower development in the calorimeter.

In $p\bar{p}$ collisions there is an ordering of frequency with which objects of different types are produced:  many more jets ($j$) are produced than $b$-jets ($b$) or photons ($\gamma$), and many more of these are produced than charged leptons ($e$, $\mu$, $\tau$).  The CDF detectors and reconstruction algorithms have been designed so that the probability of misreconstructing a frequently produced object as an infrequently produced object is small.  The fraction of central jets that \CdfSim\ misreconstructs as photons, electrons, and muons is $\sim 10^{-3}$, $\sim 10^{-4}$, and $\sim 10^{-5}$, respectively.  Due to these small numbers, the use of \CdfSim\ to model these fake processes would require generating samples with prohibitively large statistics.  Instead, the modeling of a frequently produced object faking a less frequently produced object (specifically: $j$ faking $b$, $\gamma$, $e$, $\mu$, or $\tau$; or $b$ or $\gamma$ faking $e$, $\mu$, or $\tau$) is obtained by the application of a misidentification probability, a particular type of correction factor in the \Vista\ correction model, described in the next section.

In Monte Carlo samples passed through \CdfSim, reconstructed leptons and photons are required to match to a corresponding generator level object.  This procedure removes reconstructed leptons or photons that arise from a misreconstructed quark or gluon jet.


\subsection{Correction model}
\label{sec:Vista:CorrectionModel}

\begin{table*}
\footnotesize
\centering
\begin{minipage}{9in}
\begin{tabular}{lllllr}
{\bf Code } & {\bf Category } & {\bf Explanation } & {\bf Value } & {\bf Error } & {\bf Error(\%)} \\ \hline 
0001 & luminosity & CDF integrated luminosity & 927 & 20 & 2.2 \\ 
0002 & $k$-factor & cosmic $\gamma$ & 0.69 & 0.05 & 7.3 \\ 
0003 & $k$-factor & cosmic $j$ & 0.446 & 0.014 & 3.1 \\ 
0004 & $k$-factor & 1$\gamma$1$j$ photon+jet(s) & 0.95 & 0.04 & 4.2 \\ 
0005 & $k$-factor & 1$\gamma$2$j$ & 1.2 & 0.05 & 4.1 \\ 
0006 & $k$-factor & 1$\gamma$3$j$ & 1.48 & 0.07 & 4.7 \\ 
0007 & $k$-factor & 1$\gamma$4$j$+ & 1.97 & 0.16 & 8.1 \\ 
0008 & $k$-factor & 2$\gamma$0$j$ diphoton(+jets) & 1.81 & 0.08 & 4.4 \\ 
0009 & $k$-factor & 2$\gamma$1$j$ & 3.42 & 0.24 & 7.0 \\ 
0010 & $k$-factor & 2$\gamma$2$j$+ & 1.3 & 0.16 & 12.3 \\ 
0011 & $k$-factor & $W$0$j$ $W$ (+jets) & 1.453 & 0.027 & 1.9 \\ 
0012 & $k$-factor & $W$1$j$ & 1.06 & 0.03 & 2.8 \\ 
0013 & $k$-factor & $W$2$j$ & 1.02 & 0.03 & 2.9 \\ 
0014 & $k$-factor & $W$3$j$+ & 0.76 & 0.05 & 6.6 \\ 
0015 & $k$-factor & $Z$0$j$ $Z$ (+jets) & 1.419 & 0.024 & 1.7 \\ 
0016 & $k$-factor & $Z$1$j$ & 1.18 & 0.04 & 3.4 \\ 
0017 & $k$-factor & $Z$2$j$+ & 1.03 & 0.05 & 4.8 \\ 
0018 & $k$-factor & 2$j$ $\hat{p}_T<150$ & 0.96 & 0.022 & 2.3 \\ 
0019 & $k$-factor & 2$j$ $150<\hat{p}_T$ & 1.256 & 0.028 & 2.2 \\ 
0020 & $k$-factor & 3$j$ $\hat{p}_T<150$ & 0.921 & 0.021 & 2.3 \\ 
0021 & $k$-factor & 3$j$ $150<\hat{p}_T$ & 1.36 & 0.03 & 2.4 \\ 
0022 & $k$-factor & 4$j$ $\hat{p}_T<150$ & 0.989 & 0.025 & 2.5 \\ 
0023 & $k$-factor & 4$j$ $150<\hat{p}_T$ & 1.7 & 0.04 & 2.3 \\ 
0024 & $k$-factor & 5$j$+ & 1.25 & 0.05 & 4.0 \\ 
0025 & ID eff & \poo{e}{e} central & 0.986 & 0.006 & 0.6 \\ 
0026 & ID eff & \poo{e}{e} plug & 0.933 & 0.009 & 1.0 \\ 
0027 & ID eff & \poo{\mu}{\mu} $\abs{\eta}<0.6$ & 0.845 & 0.008 & 0.9 \\ 
0028 & ID eff & \poo{\mu}{\mu} $0.6<\abs{\eta}$ & 0.915 & 0.011 & 1.2 \\ 
0029 & ID eff & \poo{\gamma}{\gamma} central & 0.974 & 0.018 & 1.8 \\ 
0030 & ID eff & \poo{\gamma}{\gamma} plug & 0.913 & 0.018 & 2.0 \\ 
0031 & ID eff & \poo{b}{b} central & 1 & 0.04 & 4.0 \\ 
0032 & fake rate & \poo{e}{\gamma} plug & 0.045 & 0.012 & 27.0 \\ 
0033 & fake rate & \poo{q}{e} central & 9.71$\times 10^{-5}$ & 1.9$\times 10^{-6}$ & 2.0 \\ 
0034 & fake rate & \poo{q}{e} plug & 0.000876 & 1.8$\times 10^{-5}$ & 2.1 \\ 
0035 & fake rate & \poo{q}{\mu} & 1.157$\times 10^{-5}$ & 2.7$\times 10^{-7}$ & 2.3 \\ 
0036 & fake rate & \poo{j}{b} & 0.01684 & 0.00027 & 1.6 \\ 
0037 & fake rate & \poo{q}{\tau} $p_T<60$ & 0.00341 & 0.00012 & 3.5 \\ 
0038 & fake rate & \poo{q}{\tau} $60<p_T$ & 0.00038 & 4$\times 10^{-5}$ & 10.5 \\ 
0039 & fake rate & \poo{q}{\gamma} central & 0.000265 & 1.5$\times 10^{-5}$ & 5.7 \\ 
0040 & fake rate & \poo{q}{\gamma} plug & 0.00159 & 0.00013 & 8.2 \\ 
0041 & trigger & \poo{e}{\text{trig}} central, $p_T>25$ & 0.976 & 0.007 & 0.7 \\ 
0042 & trigger & \poo{e}{\text{trig}} plug, $p_T>25$ & 0.835 & 0.015 & 1.8 \\ 
0043 & trigger & \poo{\mu}{\text{trig}} $\abs{\eta}<0.6$, $p_T>25$ & 0.917 & 0.007 & 0.8 \\ 
0044 & trigger & \poo{\mu}{\text{trig}} $0.6<\abs{\eta}<1.0$, $p_T>25$ & 0.96 & 0.01 & 1.0 \\ 
\end{tabular}
\end{minipage}
\caption[The 44 factors introduced in the correction model.]{The 44 factors introduced in the correction model.  
All values are dimensionless with the exception of code {\tt 0001} (luminosity), which has units of pb$^{-1}$.  The values and uncertainties of these correction factors are valid within the context of this correction model.}
\label{tbl:CorrectionFactorDescriptionValuesSigmas}
\end{table*}

Unfortunately some numbers that can not be determined from first principles enter the comparison between data and the Standard Model prediction.  These numbers are referred to as ``correction factors''.  This correction model is applied to generated Monte Carlo events to obtain the Standard Model prediction across all final states.

Correction factors must be obtained from the data themselves.  These factors may be thought of as Bayesian nuisance parameters.  The actual values of the correction factors are not directly of interest.  Of interest is the comparison of data to Standard Model prediction, with correction factors adjusted to whatever they need to be, consistent with external constraints, to bring the Standard Model into closest agreement with the data.

The traditional prescription for determining these correction factors is to ``measure'' them in a ``control region'' in which no signal is expected.  This procedure encounters difficulty when the entire high-$p_T$ data sample is considered to be a signal region.  The approach adopted instead is to ask whether a consistent set of correction factors can be chosen so that the Standard Model prediction is in agreement with the CDF high-$p_T$ data.  

The correction model is obtained by an iterative procedure informed by observed inadequacies in modeling.  The process of correction model improvement, motivated by observed discrepancies, may allow a real signal to be artificially suppressed.  If adjusting correction factor values within allowed bounds removes a signal, then the case for the signal disappears, since it can be explained in terms of known physics.  This is true in any analysis.  The stronger the constraints on the correction model, the more difficult it is to artificially suppress a real signal.  By requiring a consistent interpretation of hundreds of final states, \Vista\ is less likely to mistakenly explain away new physics than analyses of more limited scope.

The 44 correction factors currently included in the correction model are shown in Table~\ref{tbl:CorrectionFactorDescriptionValuesSigmas}.  These factors can be classified into two categories: theoretical and experimental.  A more detailed description of each individual correction factor is provided in Appendix~\ref{sec:VistaCorrectionModel:CorrectionFactorValues}.

Theoretical correction factors reflect the practical difficulty of calculating accurately within the framework of the Standard Model.  These factors take the form of $k$-factors, so-called ``knowledge factors,'' representing the ratio of the unavailable all order cross section to the calculable leading order cross section.  Twenty-three $k$-factors are used for Standard Model processes including QCD multijet production, W+jets, Z+jets, and (di)photon+jets production.

Experimental correction factors include the integrated luminosity of the data, efficiencies associated with triggering on electrons and muons, efficiencies associated with the correct identification of physics objects, and fake rates associated with the mistaken identification of physics objects.  Obtaining an adequate description of object misidentification has required an understanding of the underlying physical mechanisms by which objects are misreconstructed, as described in Appendix~\ref{sec:MisidentificationMatrix}.

In the interest of simplicity, correction factors representing $k$-factors, efficiencies, and fake rates are generally taken to be constants, independent of kinematic quantities such as object $p_T$, with only five exceptions. The $p_T$ dependence of three fake rates is too large to be treated as approximately constant: the jet faking electron rate $\poo{j}{e}$ in the plug region of the CDF detector; the jet faking $b$-tagged jet rate $\poo{j}{b}$, which increases steadily with increasing $p_T$; and the jet faking tau rate $\poo{j}{\tau}$, which decreases steadily with increasing $p_T$.  Two other fake rates possess geometrical features in $\eta$--$\phi$ due to the construction of the CDF detector: the jet faking electron rate $\poo{j}{e}$ in the central region, because of the fiducial tower geometry of the electromagnetic calorimeter; and the jet faking muon rate $\poo{j}{\mu}$, due to the non-trivial fiducial geometry of the muon chambers.  After determining appropriate functional forms, a single overall multiplicative correction factor, determined by the global fit, is used

Correction factor values are obtained from a global fit to the data.  The procedure is outlined here, with further details relegated to Appendix~\ref{sec:CorrectionFactorFitDetails}.

Events are first partitioned into final states according to the number and types of objects present.  Each final state is then subdivided into bins according to each object's detector pseudorapidity ($\detEta$) and transverse momentum ($p_T$), as described in Appendix~\ref{sec:CorrectionFactorFitDetails:chi_k}.

Generated Monte Carlo events, adjusted by the correction model, provide the Standard Model prediction for each bin.  The Standard Model prediction in each bin is therefore a function of the correction factor values.  A figure of merit is defined to quantify global agreement between the data and the Standard Model prediction, and correction factor values are chosen to maximize this agreement, consistent with external experimental constraints.

Letting $\vec{s}$ represent a vector of correction factors, for the $k^\text{th}$ bin 
\begin{equation}
\chi^2_k(\vec{s})=\frac{(\text{Data}[k]-\text{SM}[k])^2}{\sqrt{\text{SM}[k]}^2 + \delta\text{SM}[k]^2},
\label{eq:chi_k}
\end{equation}
where $\text{Data}[k]$ is the number of data events observed in the $k^\text{th}$ bin, $\text{SM}[k]$ is the number of events predicted by the Standard Model in the $k^\text{th}$ bin, $\delta\text{SM}[k]$ is the Monte Carlo statistical uncertainty on the Standard Model prediction in the $k^\text{th}$ bin~\footnote{Given a set of Monte Carlo events with individual weights $w_j$, so that the total Standard Model prediction from these Monte Carlo events is $\text{SM}=\sum_j{w_j}$ events, the ``effective weight'' $w_{\text{eff}}$ of these events can be taken to be the weighted average of the weights: $w_{\text{eff}}=\frac{\sum_j{w_j w_j}}{\sum_j{w_j}}$.  The ``effective number of Monte Carlo events'' is $N_{\text{eff}}=\text{SM}/w_{\text{eff}}$, and the error on the Standard Model prediction is $\delta{\text{SM}}=\text{SM}/\sqrt{N_{\text{eff}}}$.}, and $\sqrt{\text{SM}[k]}$ is the statistical uncertainty on the expected data in the $k^\text{th}$ bin.  The Standard Model prediction $\text{SM}[k]$ in the $k^{\text{th}}$ bin is a function of $\vec{s}$. 

Relevant information external to the \Vista\ high-$p_T$ data sample provides additional constraints in this global fit.  The CDF luminosity counters measure the integrated luminosity of the sample described in this article to be 902~pb$^{-1} \pm 6\%$ by measuring the fraction of bunch crossings in which zero inelastic collisions occur~\cite{CdfLuminosityCountersCLC:Acosta:2002hx}.  The integrated luminosity of the sample measured by the luminosity counters enters in the form of a Gaussian constraint on the luminosity correction factor.  Higher order theoretical calculations exist for some Standard Model processes, providing constraints on corresponding $k$-factors, and some CDF experimental correction factors are also constrained from external information.  In total, 26 of the 44 correction factors are constrained.  The specific constraints employed are provided in Appendix~\ref{sec:CorrectionFactorFitDetails:chi_constraints}.

The overall function to be minimized takes the form
\begin{equation}
\chi^2(\vec{s}) = \left(\sum_{k\in\text{bins}}{\chi^2_k}(\vec{s})\right) + \chi^2_{\text{constraints}}(\vec{s}),
\label{eqn:chiSqd}
\end{equation}
where the sum in the first term is over bins in the CDF high-$p_T$ data sample with $\chi^2_k(\vec{s})$ defined in Eq.~\ref{eq:chi_k}, and the second term is the contribution from explicit constraints.

Minimization of $\chi^2(\vec{s})$ in Eq.~\ref{eqn:chiSqd} as a function of the vector of correction factors $\vec{s}$ results in a set of correction factor values $\vec{s}_0$ providing the best global agreement between the data and the Standard Model prediction.  The best fit correction factor values are shown in Table~\ref{tbl:CorrectionFactorDescriptionValuesSigmas}, together with absolute and fractional uncertainties.  The determined uncertainties are not used explicitly in the subsequent analysis, but rather provide information used implicitly to assist in appropriate adjustment to the correction model in light of observed discrepancies.  The uncertainties are verified by subdividing the data into thirds, performing separate fits on each third, and noting that the correction factor values obtained with each subset are consistent within quoted uncertainties.  Further details on the correlation matrix and other technical aspects of this global fit can be found in Appendix~\ref{sec:CorrectionFactorCovarianceMatrix}.

Although the correction factors are determined from a global fit, in practice the determination of many correction factors' values are dominated by one recognizable subsample.  The rate $\poo{j}{e}$ for a jet to fake an electron is determined largely by the number of events in the $ej$ final state, since the largest contribution to this final state is from dijet events with one jet misreconstructed as an electron.  Similarly, the rates $\poo{j}{b}$ and $\poo{j}{\tau}$ for a jet to fake a $b$-tagged jet and tau lepton are determined largely by the number of events in the $bj$ and $\tau j$ final states, respectively.  The determination of the fake rate $\poo{j}{\gamma}$, photon efficiency $\poo{\gamma}{\gamma}$, and $k$-factors for prompt photon production and prompt diphoton production are dominated by the $\gamma j$, $\gamma jj$, and $\gamma \gamma$ final states.  Additional knowledge incorporated in the determination of fake rates is described in Appendix~\ref{sec:MisidentificationMatrix}.

The global fit $\chi^2$ per number of bins is 288.1 / 133 + 27.9, where the last term is the contribution to the $\chi^2$ from the imposed constraints.  A $\chi^2$ per degree of freedom larger than unity is expected, since the limited set of correction factors in this correction model is not expected to provide a complete description of all features of the data.  Emphasis is placed on individual outlying discrepancies that may motivate a new physics claim, rather than overall goodness of fit. 

Corrections to object identification efficiencies are typically less than 10\%; fake rates are consistent with an understanding of the underlying physical mechanisms responsible; $k$-factors range from slightly less than unity to greater than two for some processes with multiple jets.  All values obtained are physically reasonable.  Further analysis is provided in Appendix~\ref{sec:VistaCorrectionModel:CorrectionFactorValues}.

\begin{table*}
{\mbox{
\tiny
\begin{minipage}{9in}
\begin{tabular}{l@{ }r@{ }r@{ $\pm$ }l@{ }l}
{\bf Final State} & {\bf Data} & \multicolumn{2}{c}{\bf Background} & \multicolumn{1}{c}{\bf $\sigma$} \\ \hline 
3j$\tau^\pm$ & $71$ & $113.7$ & $3.6$ & $-2.3$ \\ 
5j & $1661$ & $1902.9$ & $50.8$ & $-1.7$ \\ 
2j$\tau^\pm$ & $233$ & $296.5$ & $5.6$ & $-1.6$ \\ 
2j$2\tau^\pm$ & $6$ & $27$ & $4.6$ & $-1.4$ \\ 
b$e^\pm$j & $2207$ & $2015.4$ & $28.7$ & $+1.4$ \\ 
3j, high $\SumPt$ & $35436$ & $37294.6$ & $524.3$ & $-1.1$ \\ 
$e^\pm$3j$p\!\!\!/$ & $1954$ & $1751.6$ & $42$ & $+1.1$ \\ 
b$e^\pm$2j & $798$ & $695.3$ & $13.3$ & $+1.1$ \\ 
3j$p\!\!\!/$, low $\SumPt$ & $811$ & $967.5$ & $38.4$ & $-0.8$ \\ 
$e^\pm$$\mu^\pm$ & $26$ & $11.6$ & $1.5$ & $+0.8$ \\ 
$e^\pm$$\gamma$ & $636$ & $551.2$ & $11.2$ & $+0.7$ \\ 
$e^\pm$3j & $28656$ & $27281.5$ & $405.2$ & $+0.6$ \\ 
b5j & $131$ & $95$ & $4.7$ & $+0.5$ \\ 
j$2\tau^\pm$ & $50$ & $85.6$ & $8.2$ & $-0.4$ \\ 
j$\tau^\pm$$\tau^\mp$ & $74$ & $125$ & $13.6$ & $-0.4$ \\ 
b$p\!\!\!/$, low $\SumPt$ & $10$ & $29.5$ & $4.6$ & $-0.4$ \\ 
$e^\pm$j$\gamma$ & $286$ & $369.4$ & $21.1$ & $-0.3$ \\ 
$e^\pm$j$p\!\!\!/$$\tau^\mp$ & $29$ & $14.2$ & $1.8$ & $+0.2$ \\ 
2j, high $\SumPt$ & $96502$ & $92437.3$ & $1354.5$ & $+0.1$ \\ 
b$e^\pm$3j & $356$ & $298.6$ & $7.7$ & $+0.1$ \\ \hline
8j & $11$ & $6.1$ & $2.5$ & \\ 
7j & $57$ & $35.6$ & $4.9$ & \\ 
6j & $335$ & $298.4$ & $14.7$ & \\ 
4j, low $\SumPt$ & $39665$ & $40898.8$ & $649.2$ & \\ 
4j, high $\SumPt$ & $8241$ & $8403.7$ & $144.7$ & \\ 
4j$2\gamma$ & $38$ & $57.5$ & $11$ & \\ 
4j$\tau^\pm$ & $20$ & $36.9$ & $2.4$ & \\ 
4j$p\!\!\!/$, low $\SumPt$ & $516$ & $525.2$ & $34.5$ & \\ 
4j$\gamma$$p\!\!\!/$ & $28$ & $53.8$ & $11$ & \\ 
4j$\gamma$ & $3693$ & $3827.2$ & $112.1$ & \\ 
4j$\mu^\pm$ & $576$ & $568.2$ & $26.1$ & \\ 
4j$\mu^\pm$$p\!\!\!/$ & $232$ & $224.7$ & $8.5$ & \\ 
4j$\mu^\pm$$\mu^\mp$ & $17$ & $20.1$ & $2.5$ & \\ 
$3\gamma$ & $13$ & $24.2$ & $3$ & \\ 
3j, low $\SumPt$ & $75894$ & $75939.2$ & $1043.9$ & \\ 
3j$2\gamma$ & $145$ & $178.1$ & $7.4$ & \\ 
3j$p\!\!\!/$, high $\SumPt$ & $20$ & $30.9$ & $14.4$ & \\ 
3j$\gamma$$\tau^\pm$ & $13$ & $11$ & $2$ & \\ 
3j$\gamma$$p\!\!\!/$ & $83$ & $102.9$ & $11.1$ & \\ 
3j$\gamma$ & $11424$ & $11506.4$ & $190.6$ & \\ 
3j$\mu^\pm$$p\!\!\!/$ & $1114$ & $1118.7$ & $27.1$ & \\ 
3j$\mu^\pm$$\mu^\mp$ & $61$ & $84.5$ & $9.2$ & \\ 
3j$\mu^\pm$ & $2132$ & $2168.7$ & $64.2$ & \\ 
3bj, low $\SumPt$ & $14$ & $9.3$ & $1.9$ & \\ 
$2\tau^\pm$ & $316$ & $290.8$ & $24.2$ & \\ 
$2\gamma$$p\!\!\!/$ & $161$ & $176$ & $9.1$ & \\ 
$2\gamma$ & $8482$ & $8349.1$ & $84.1$ & \\ 
2j, low $\SumPt$ & $93408$ & $92789.5$ & $1138.2$ & \\ 
2j$2\gamma$ & $645$ & $612.6$ & $18.8$ & \\ 
2j$\tau^\pm$$\tau^\mp$ & $15$ & $25$ & $3.5$ & \\ 
2j$p\!\!\!/$, low $\SumPt$ & $74$ & $106$ & $7.8$ & \\ 
2j$p\!\!\!/$, high $\SumPt$ & $43$ & $37.7$ & $100.2$ & \\ 
2j$\gamma$ & $33684$ & $33259.9$ & $397.6$ & \\ 
2j$\gamma$$\tau^\pm$ & $48$ & $41.4$ & $3.4$ & \\ 
2j$\gamma$$p\!\!\!/$ & $403$ & $425.2$ & $29.7$ & \\ 
2j$\mu^\pm$$p\!\!\!/$ & $7287$ & $7320.5$ & $118.9$ & \\ 
2j$\mu^\pm$$\gamma$$p\!\!\!/$ & $13$ & $12.6$ & $2.7$ & \\ 
2j$\mu^\pm$$\gamma$ & $41$ & $35.7$ & $6.1$ & \\ 
2j$\mu^\pm$$\mu^\mp$ & $374$ & $394.2$ & $24.8$ & \\ 
\end{tabular}
\begin{tabular}{l@{ }r@{ }r@{ $\pm$ }l@{ }}
{\bf Final State} & {\bf Data} & \multicolumn{2}{c}{\bf Background} \\ \hline 
2j$\mu^\pm$ & $9513$ & $9362.3$ & $166.8$ \\ 
$2e^\pm$j & $13$ & $9.8$ & $2.2$ \\ 
$2e^\pm$$e^\mp$ & $12$ & $4.8$ & $1.2$ \\ 
$2e^\pm$ & $23$ & $36.1$ & $3.8$ \\ 
2b, low $\SumPt$ & $327$ & $335.8$ & $7$ \\ 
2b, high $\SumPt$ & $187$ & $173.1$ & $7.1$ \\ 
2b3j, high $\SumPt$ & $28$ & $33.5$ & $5.5$ \\ 
2b2j, low $\SumPt$ & $355$ & $326.3$ & $8.4$ \\ 
2b2j, high $\SumPt$ & $56$ & $80.2$ & $5$ \\ 
2b2j$\gamma$ & $16$ & $15.4$ & $3.6$ \\ 
2b$\gamma$ & $37$ & $31.7$ & $4.8$ \\ 
2bj, low $\SumPt$ & $415$ & $393.8$ & $9.1$ \\ 
2bj, high $\SumPt$ & $161$ & $195.8$ & $8.3$ \\ 
2bj$p\!\!\!/$, low $\SumPt$ & $28$ & $23.2$ & $2.6$ \\ 
2bj$\gamma$ & $25$ & $24.7$ & $4.3$ \\ 
2b$e^\pm$2j$p\!\!\!/$ & $15$ & $12.3$ & $1.6$ \\ 
2b$e^\pm$2j & $30$ & $30.5$ & $2.5$ \\ 
2b$e^\pm$j & $28$ & $29.1$ & $2.8$ \\ 
2b$e^\pm$ & $48$ & $45.2$ & $3.7$ \\ 
$\tau^\pm$$\tau^\mp$ & $498$ & $428.5$ & $22.7$ \\ 
$\gamma$$\tau^\pm$ & $177$ & $204.4$ & $5.4$ \\ 
$\gamma$$p\!\!\!/$ & $1952$ & $1945.8$ & $77.1$ \\ 
$\mu^\pm$$\tau^\pm$ & $18$ & $19.8$ & $2.3$ \\ 
$\mu^\pm$$\tau^\mp$ & $151$ & $179.1$ & $4.7$ \\ 
$\mu^\pm$$p\!\!\!/$ & $321351$ & $320500$ & $3475.5$ \\ 
$\mu^\pm$$p\!\!\!/$$\tau^\mp$ & $22$ & $25.8$ & $2.7$ \\ 
$\mu^\pm$$\gamma$ & $269$ & $285.5$ & $5.9$ \\ 
$\mu^\pm$$\gamma$$p\!\!\!/$ & $269$ & $282.2$ & $6.6$ \\ 
$\mu^\pm$$\mu^\mp$$p\!\!\!/$ & $49$ & $61.4$ & $3.5$ \\ 
$\mu^\pm$$\mu^\mp$$\gamma$ & $32$ & $29.9$ & $2.6$ \\ 
$\mu^\pm$$\mu^\mp$ & $10648$ & $10845.6$ & $96$ \\ 
j$2\gamma$ & $2196$ & $2200.3$ & $35.2$ \\ 
j$2\gamma$$p\!\!\!/$ & $38$ & $27.3$ & $3.2$ \\ 
j$\tau^\pm$ & $563$ & $585.7$ & $10.2$ \\ 
j$p\!\!\!/$, low $\SumPt$ & $4183$ & $4209.1$ & $56.1$ \\ 
j$\gamma$ & $49052$ & $48743$ & $546.3$ \\ 
j$\gamma$$\tau^\pm$ & $106$ & $104$ & $4.1$ \\ 
j$\gamma$$p\!\!\!/$ & $913$ & $965.2$ & $41.5$ \\ 
j$\mu^\pm$ & $33462$ & $34026.7$ & $510.1$ \\ 
j$\mu^\pm$$\tau^\mp$ & $29$ & $37.5$ & $4.5$ \\ 
j$\mu^\pm$$p\!\!\!/$$\tau^\mp$ & $10$ & $9.6$ & $2.1$ \\ 
j$\mu^\pm$$p\!\!\!/$ & $45728$ & $46316.4$ & $568.2$ \\ 
j$\mu^\pm$$\gamma$$p\!\!\!/$ & $78$ & $69.8$ & $9.9$ \\ 
j$\mu^\pm$$\gamma$ & $70$ & $98.4$ & $12.1$ \\ 
j$\mu^\pm$$\mu^\mp$ & $1977$ & $2093.3$ & $74.7$ \\ 
$e^\pm$4j & $7144$ & $6661.9$ & $147.2$ \\ 
$e^\pm$4j$p\!\!\!/$ & $403$ & $363$ & $9.9$ \\ 
$e^\pm$3j$\tau^\mp$ & $11$ & $7.6$ & $1.6$ \\ 
$e^\pm$3j$\gamma$ & $27$ & $21.7$ & $3.4$ \\ 
$e^\pm$$2\gamma$ & $47$ & $74.5$ & $5$ \\ 
$e^\pm$2j & $126665$ & $122457$ & $1672.6$ \\ 
$e^\pm$2j$\tau^\mp$ & $53$ & $37.3$ & $3.9$ \\ 
$e^\pm$2j$\tau^\pm$ & $20$ & $24.7$ & $2.3$ \\ 
$e^\pm$2j$p\!\!\!/$ & $12451$ & $12130.1$ & $159.4$ \\ 
$e^\pm$2j$\gamma$ & $101$ & $88.9$ & $6.1$ \\ 
$e^\pm$$\tau^\mp$ & $609$ & $555.9$ & $10.2$ \\ 
$e^\pm$$\tau^\pm$ & $225$ & $211.2$ & $4.7$ \\ 
$e^\pm$$p\!\!\!/$ & $476424$ & $479572$ & $5361.2$ \\ 
$e^\pm$$p\!\!\!/$$\tau^\mp$ & $48$ & $35$ & $2.7$ \\ 
\end{tabular}
\begin{tabular}{l@{ }r@{ }r@{ $\pm$ }l@{ }}
{\bf Final State} & {\bf Data} & \multicolumn{2}{c}{\bf Background} \\ \hline 
$e^\pm$$p\!\!\!/$$\tau^\pm$ & $20$ & $18.7$ & $1.9$ \\ 
$e^\pm$$\gamma$$p\!\!\!/$ & $141$ & $144.2$ & $6$ \\ 
$e^\pm$$\mu^\mp$$p\!\!\!/$ & $54$ & $42.6$ & $2.7$ \\ 
$e^\pm$$\mu^\pm$$p\!\!\!/$ & $13$ & $10.9$ & $1.3$ \\ 
$e^\pm$$\mu^\mp$ & $153$ & $127.6$ & $4.2$ \\ 
$e^\pm$j & $386880$ & $392614$ & $5031.8$ \\ 
$e^\pm$j$2\gamma$ & $14$ & $15.9$ & $2.9$ \\ 
$e^\pm$j$\tau^\pm$ & $79$ & $79.3$ & $2.9$ \\ 
$e^\pm$j$\tau^\mp$ & $162$ & $148.8$ & $7.6$ \\ 
$e^\pm$j$p\!\!\!/$ & $58648$ & $57391.7$ & $661.6$ \\ 
$e^\pm$j$\gamma$$p\!\!\!/$ & $52$ & $76.2$ & $9$ \\ 
$e^\pm$j$\mu^\mp$$p\!\!\!/$ & $22$ & $13.1$ & $1.7$ \\ 
$e^\pm$j$\mu^\mp$ & $28$ & $26.8$ & $2.3$ \\ 
$e^\pm$$e^\mp$4j & $103$ & $113.5$ & $5.9$ \\ 
$e^\pm$$e^\mp$3j & $456$ & $473$ & $14.6$ \\ 
$e^\pm$$e^\mp$2j$p\!\!\!/$ & $30$ & $39$ & $4.6$ \\ 
$e^\pm$$e^\mp$2j & $2149$ & $2152$ & $40.1$ \\ 
$e^\pm$$e^\mp$$\tau^\pm$ & $14$ & $11.1$ & $2$ \\ 
$e^\pm$$e^\mp$$p\!\!\!/$ & $491$ & $487.9$ & $12$ \\ 
$e^\pm$$e^\mp$$\gamma$ & $127$ & $132.3$ & $4.2$ \\ 
$e^\pm$$e^\mp$j & $10726$ & $10669.3$ & $123.5$ \\ 
$e^\pm$$e^\mp$j$p\!\!\!/$ & $157$ & $144$ & $11.2$ \\ 
$e^\pm$$e^\mp$j$\gamma$ & $26$ & $45.6$ & $4.7$ \\ 
$e^\pm$$e^\mp$ & $58344$ & $58575.6$ & $603.9$ \\ 
b6j & $24$ & $15.5$ & $2.3$ \\ 
b4j, low $\SumPt$ & $13$ & $9.2$ & $1.8$ \\ 
b4j, high $\SumPt$ & $464$ & $499.2$ & $12.4$ \\ 
b3j, low $\SumPt$ & $5354$ & $5285$ & $72.4$ \\ 
b3j, high $\SumPt$ & $1639$ & $1558.9$ & $24.1$ \\ 
b3j$p\!\!\!/$, low $\SumPt$ & $111$ & $116.8$ & $11.2$ \\ 
b3j$\gamma$ & $182$ & $194.1$ & $8.8$ \\ 
b3j$\mu^\pm$$p\!\!\!/$ & $37$ & $34.1$ & $2$ \\ 
b3j$\mu^\pm$ & $47$ & $52.2$ & $3$ \\ 
b$2\gamma$ & $15$ & $14.6$ & $2.1$ \\ 
b2j, low $\SumPt$ & $8812$ & $8576.2$ & $97.9$ \\ 
b2j, high $\SumPt$ & $4691$ & $4646.2$ & $57.7$ \\ 
b2j$p\!\!\!/$, low $\SumPt$ & $198$ & $209.2$ & $8.3$ \\ 
b2j$\gamma$ & $429$ & $425.1$ & $13.1$ \\ 
b2j$\mu^\pm$$p\!\!\!/$ & $46$ & $40.1$ & $2.7$ \\ 
b2j$\mu^\pm$ & $56$ & $60.6$ & $3.4$ \\ 
b$\tau^\pm$ & $19$ & $19.9$ & $2.2$ \\ 
b$\gamma$ & $976$ & $1034.8$ & $15.6$ \\ 
b$\gamma$$p\!\!\!/$ & $18$ & $16.7$ & $3.1$ \\ 
b$\mu^\pm$ & $303$ & $263.5$ & $7.9$ \\ 
b$\mu^\pm$$p\!\!\!/$ & $204$ & $218.1$ & $6.4$ \\ 
bj, low $\SumPt$ & $9060$ & $9275.7$ & $87.8$ \\ 
bj, high $\SumPt$ & $7236$ & $7030.8$ & $74$ \\ 
bj$2\gamma$ & $13$ & $17.6$ & $3.3$ \\ 
bj$\tau^\pm$ & $13$ & $12.9$ & $1.8$ \\ 
bj$p\!\!\!/$, low $\SumPt$ & $53$ & $60.4$ & $19.9$ \\ 
bj$\gamma$ & $937$ & $989.4$ & $20.6$ \\ 
bj$\gamma$$p\!\!\!/$ & $34$ & $30.5$ & $4$ \\ 
bj$\mu^\pm$$p\!\!\!/$ & $104$ & $112.6$ & $4.4$ \\ 
bj$\mu^\pm$ & $173$ & $141.4$ & $4.8$ \\ 
b$e^\pm$3j$p\!\!\!/$ & $68$ & $52.2$ & $2.2$ \\ 
b$e^\pm$2j$p\!\!\!/$ & $87$ & $65$ & $3.3$ \\ 
b$e^\pm$$p\!\!\!/$ & $330$ & $347.2$ & $6.9$ \\ 
b$e^\pm$j$p\!\!\!/$ & $211$ & $176.6$ & $5$ \\ 
b$e^\pm$$e^\mp$j & $22$ & $34.6$ & $2.6$ \\ 
\end{tabular}
\end{minipage}
}}
\caption[Subset of the populations comparison between data and Standard Model.]{A subset of the comparison between data and Standard Model prediction, showing the most discrepant final states and all final states populated with ten or more data events.  
Final states are labeled according to the number and types of objects present, and whether (high $\SumPt$) or not (low $\SumPt$) the summed scalar transverse momentum of all objects in the events exceeds 400~GeV.
Final states are ordered according to decreasing discrepancy between the total number of events expected, taking into account the error from Monte Carlo statistics and the total number observed in the data.  Final states exhibiting mild discrepancies are shown together with the significance of the discrepancy in units of standard deviations ($\sigma$) after accounting for a trials factor corresponding to the number of final states considered.  
Final states that do not exhibit even mild discrepancies are listed below the horizontal line in inverted alphabetical order.  Only Monte Carlo statistical uncertainties on the background prediction are included.
}
\label{tbl:VistaCdf}  
\end{table*}

\begin{figure}
\begin{tabular}{cc}
\includegraphics[width=3.2in]{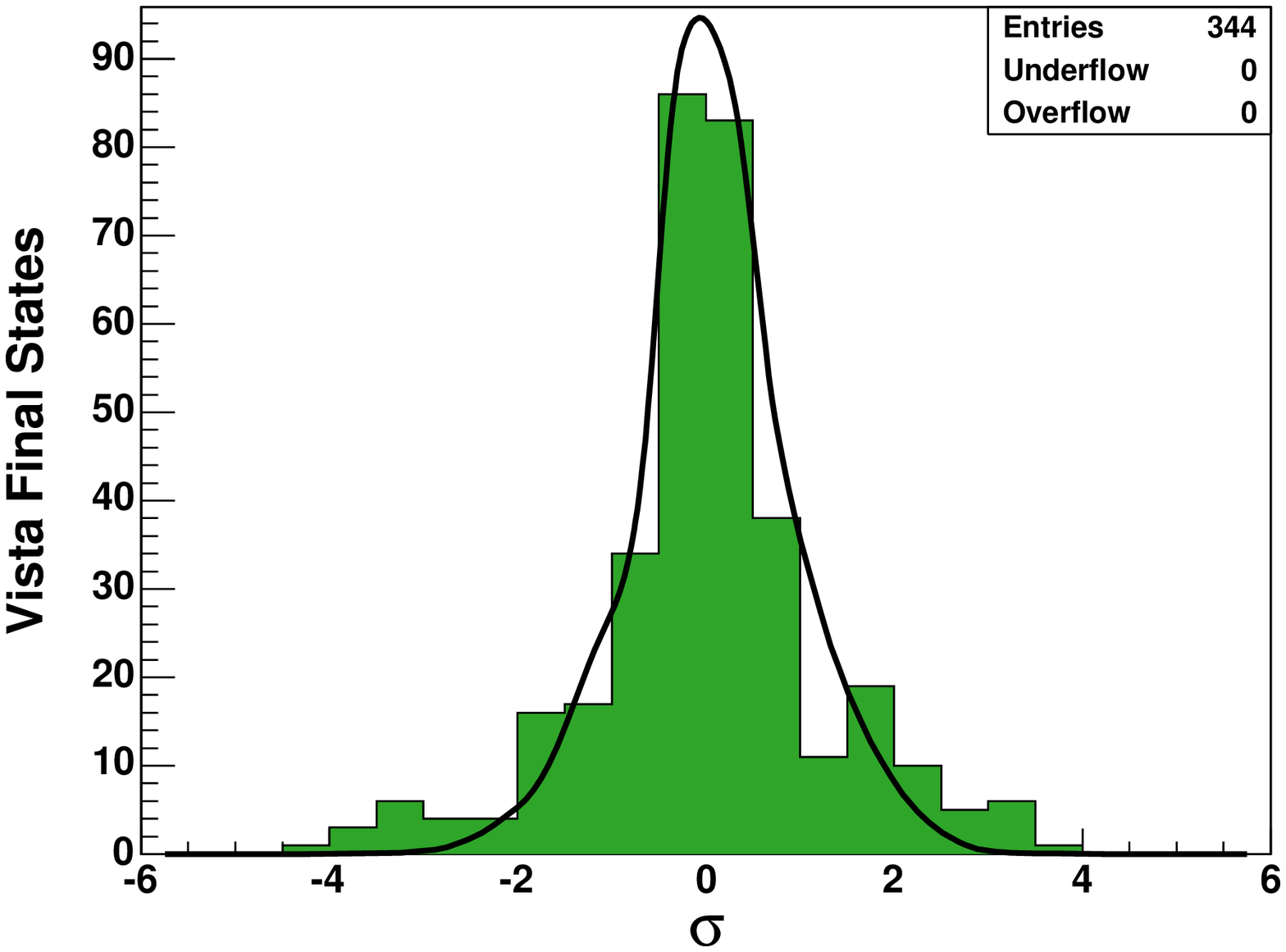} & \hspace{-0.4in}
\includegraphics[width=3.2in]{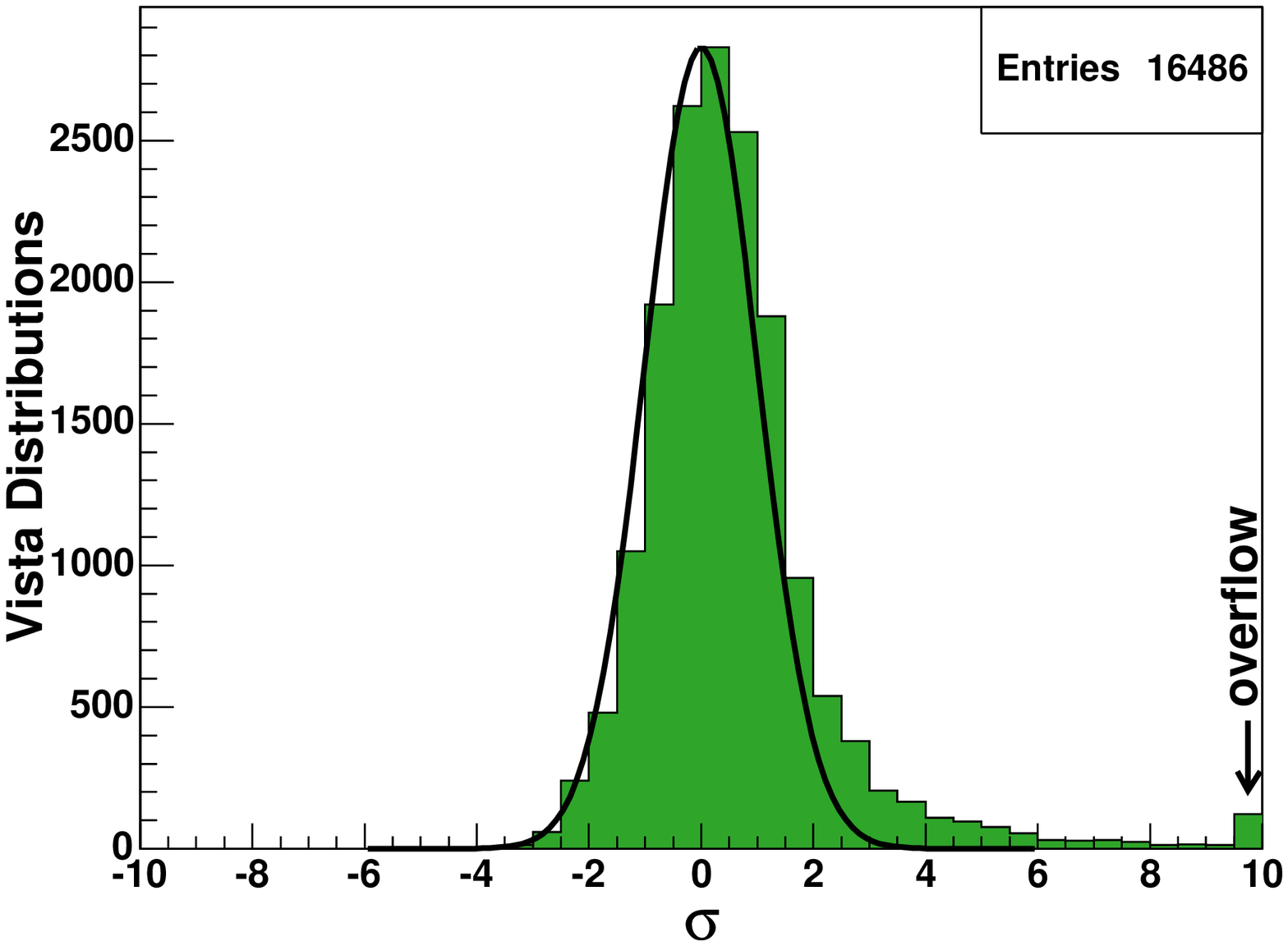} 
\end{tabular}
\caption[Distribution of observed discrepancy between data and the Standard Model prediction.]{Distribution of observed discrepancy between data and the Standard Model prediction, measured in units of standard deviation ($\sigma$), shown as the solid (green) histogram, before accounting for the trials factor.  The left pane shows the distribution of discrepancies between the total number of events observed and predicted in the 344 populated final states considered.  Negative values on the horizontal axis correspond to a deficit of data compared to Standard Model prediction; positive values indicate an excess of data compared to Standard Model prediction.  The right pane shows the distribution of discrepancies between the observed and predicted shapes in 16,486 kinematic distributions.  Distributions in which the shapes of data and Standard Model prediction are in relative disagreement correspond to large positive $\sigma$.  The solid (black) curves indicate expected distributions, if the data were truly drawn from the Standard Model background.  Interest is focused on the entries in the tails of the left distribution and the high tail of the right distribution.  
}
\label{fig:VistaSummaryCdf}
\end{figure}

\begin{figure}
\centering
\includegraphics[width=3.6in,angle=270]{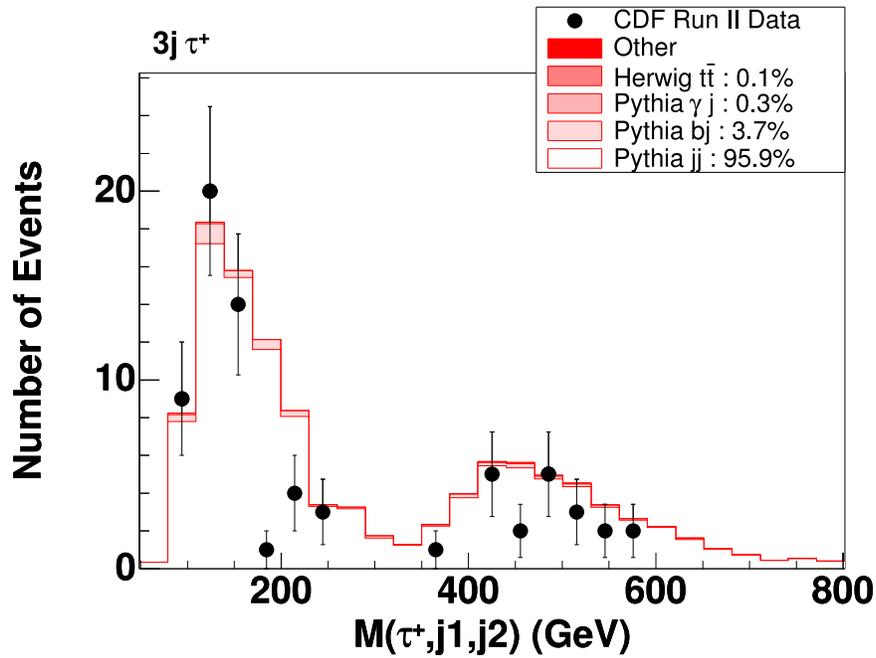}
\caption[The invariant mass of the tau lepton and two leading jets in the final state consisting of three jets and one positively or negatively charged tau.]{The invariant mass of the tau lepton and two leading jets in the final state consisting of three jets and one positively or negatively charged tau.  (The \Vista\ final state naming convention gives the tau lepton a positive charge.)  Data are shown as filled circles, with the Standard Model prediction shown as the shaded histogram.  This is the most discrepant kinematic distribution in the final state exhibiting the largest population discrepancy.}
\label{fig:3j1tau_mostDiscrepant}
\end{figure}

With the details of the correction model in place, the complete Standard Model prediction can be obtained.  For each Monte Carlo event after detector simulation, the event weight is multiplied by the value of the luminosity correction factor and the $k$-factor for the relevant Standard Model process.  The single Monte Carlo event can be misreconstructed in a number of ways, producing a set of Monte Carlo events derived from the original, with weights multiplied by the probability of each misreconstruction.  The weight of each resulting event is multiplied by the probability the event satisfies trigger criteria.  The resulting Standard Model prediction, corrected as just described, is referred to as ``the Standard Model prediction'' throughout the rest of this document, with ``corrected'' implied in all cases.

\subsection{Results}

\begin{figure}
\centering
\includegraphics[width=7cm,angle=0]{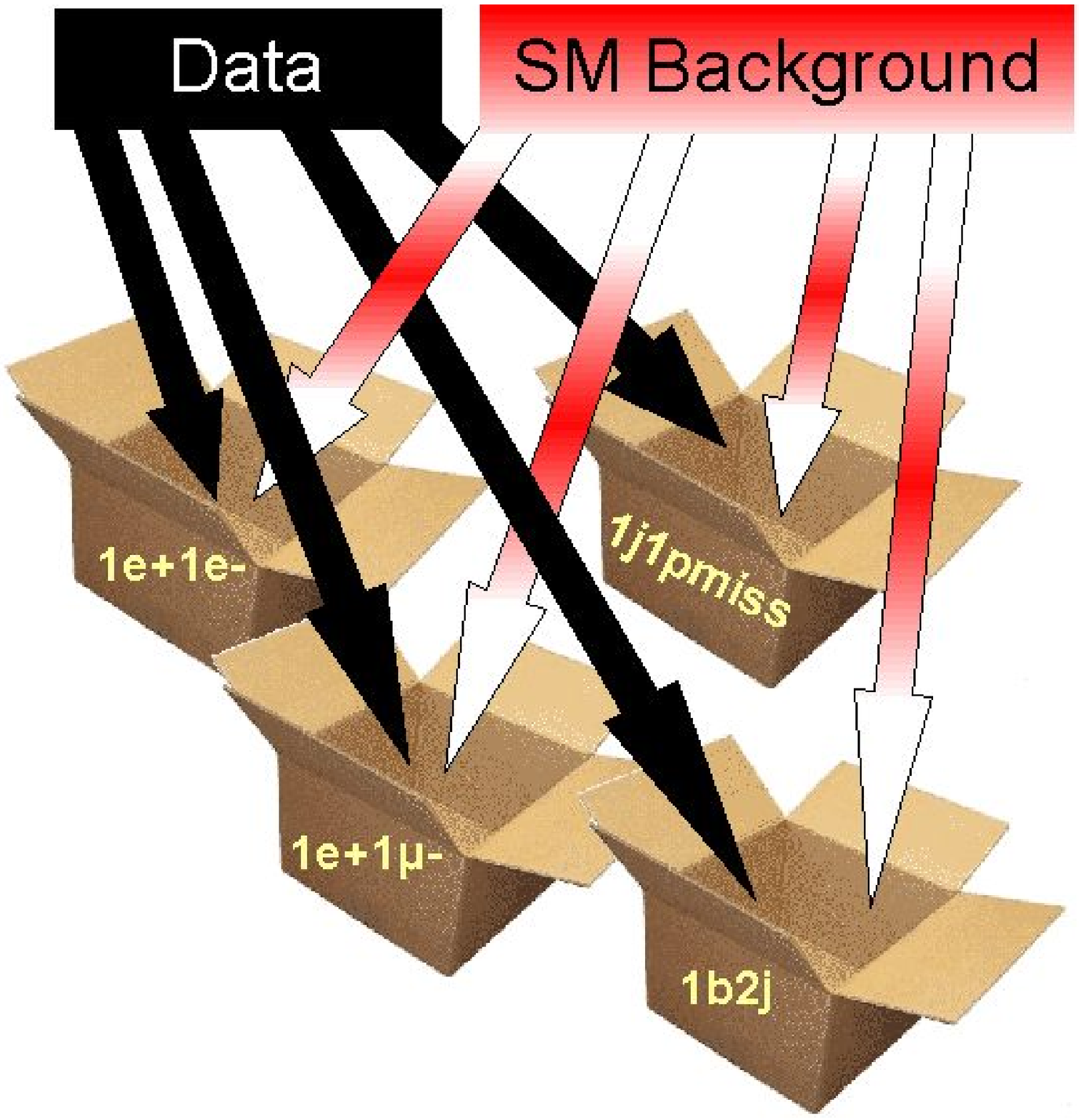} 
\caption[\Vista\ partitioning in final states.]{\Vista\ partitioning in final states.  Final states can be viewed as boxes, each containing events of one specific final configuration of objects.  Final states have not been prescribed, but are created automatically as new types of events appear.  In this way, every event, no matter how exotic, stays within the analysis, in the appropriate final state.}
\label{fig:partitioning}
\end{figure}

Data and Standard Model events are partitioned into exclusive final states, depending on the combinations of reconstructed final objects.  This partitioning is orthogonal, with each event ending up in one and only one final state, as shown schematically in Fig.~\ref{fig:partitioning}.  Data are compared to Standard Model prediction in each final state, considering the total number of events observed and predicted, and the shapes of relevant kinematic distributions.

In a data driven search, it is crucial to explicitly account for the {\em{trials factor}}, quantifying the number of places where we checked for an interesting signal.  Purely statistical fluctuations at the level of three or more standard deviations are expected to appear, simply because a large number of regions are considered.  A reasonably rigorous accounting of this trials factor is possible as long as the measures of interest and the regions to which these measures are applied are specified {\em{a priori}}, as is done here.  In this analysis a discrepancy at the level of $3\sigma$ or greater after accounting for the trials factor (typically corresponding to a discrepancy at the level of $5\sigma$ or greater before accounting for the trials factor) is considered ``significant.''  
It is worth noting that dedicated searches, checking only a small number of signal regions, typically do not account for any trials factor, simply because it is very difficult to quantify the effect of many people looking for new physics in different ways within the same experiment.  For that reason, instead of a mild $3\sigma$, a strong $5\sigma$ significance is considered necessary to discover something new in our field.  The assumption made silently is that if one observes a $5\sigma$ effect in just one attempt, then if one could include somehow the trials factor, the actual significance of the observation would turn out to be still greater than $3\sigma$, therefore convincing.  However, in cases where the ``new physics'' is well-expected (like $t\bar{t}$ or dibosons, which are processes within the Standard Model) ``discovery'' is claimed even with just $3\sigma$ \emph{without} considering the trials factor.  Certainly, for physics beyond the Standard Model, a $3\sigma$ sans trials factor should not be considered convincing proof of existence.

Discrepancy in the total number of events in a final state ($\text{fs}$) is measured by the Poisson probability $p_{\text{fs}}$ that the number of predicted events would fluctuate up to or above (or down to or below) the number of events observed.  
Since the expected population is known with some uncertainty, its probability density function is convoluted to obtain $p_{\rm fs}$.
To account for the trials factor due to the 344 \Vista\ final states examined, the quantity $p=1-(1-p_{\text{fs}})^{344}$ is calculated for each final state.  The result is the probability $p$ of observing a discrepancy corresponding to a probability less than $p_{\text{fs}}$ in the total sample studied.  This probability $p$ can then be converted into units of standard deviations by solving for $\sigma$ such that $\int_{\sigma}^{\infty}\, \frac{1}{\sqrt{2\pi}}e^{-\frac{x^2}{2}} dx = p$~\footnote{Final states for which $p>0.5$ after accounting for the trials factor are not even mildly interesting, and the corresponding $\sigma$ after accounting for the trials factor is not quoted.  For the mildly interesting final states with $p<0.5$ after accounting for the trials factor, $\sigma$ is quoted as positive if the number of observed data events exceeds the Standard Model prediction, and negative if the number of observed data events is less than the Standard Model prediction.}.  A final state exhibiting a population discrepancy greater than 3$\sigma$ after the trials factor is thus accounted for is considered significant.

Many kinematic distributions are considered in each final state, including the transverse momentum, pseudorapidity, detector pseudorapidity, and azimuthal angle of all objects, masses of individual jets and $b$-jets, invariant masses of all object combinations, transverse masses of object combinations including $\pmiss$, angular separation $\Delta\phi$ and $\Delta R$ of all object pairs, and several other more specialized variables.  A Kolmogorov-Smirnov (KS) test is used to quantify the difference in shape of each kinematic distribution between data and Standard Model prediction.  As with populations, a trials factor is assessed to account for the 16,486 distributions examined, and the resulting probability is converted into units of standard deviations.  A distribution with KS statistic greater than 0.02 and probability corresponding to greater than 3$\sigma$ after assessing the trials factor is considered significant.

Table~\ref{tbl:VistaCdf} shows a subset of the \Vista\ comparison of data to Standard Model prediction.  Shown are all final states containing ten or more data events, with the most discrepant final states in population heading the list.  After accounting for the trials factor, no final state has a statistically significant ($>3\sigma$) population discrepancy.  The most discrepant final state ($3j\,\tau^\pm$) contains 71 data events and $113.7\pm3.6$ events expected from the Standard Model.  The Poisson probability for $113.7\pm3.6$ expected events to result in 71 or fewer events observed in this final state is $2.8\times10^{-5}$, corresponding to an entry at $-4.03\sigma$ in Fig.~\ref{fig:VistaSummaryCdf}.  The probability for one or more of the 344 populated final states considered to display disagreement in population corresponding to a probability less than $2.8\times10^{-5}$ is 1\%.  The $3j\,\tau^\pm$ population discrepancy is thus not statistically significant.  The most discrepant kinematic distribution in this final state is the invariant mass of the tau lepton and the two highest transverse momentum jets, shown in Fig.~\ref{fig:3j1tau_mostDiscrepant}.

The six final states with largest population discrepancy are $3j\,\tau$, $5j$, $2j\,\tau$, $2j\,2\tau$, $b\,e\,j$, and the low-$p_T$ $3j$ final state, with $b\,e\,j$ being the only one of these six to exhibit an excess of data.  The $3j\,\tau$, $2j\,\tau$, and $2j\,2\tau$ final states appear to reflect an incomplete understanding of the rate of jets faking taus ($\poo{j}{\tau}$) as a function of the number of jets in the event, at the level of $\sim 30\%$ difference between the total number of observed and predicted events in the most populated of these final states.  The value of $\poo{j}{\tau}$ is primarily determined by the $j\,\tau$ final state.  Interestingly, although the underlying physical mechanism for $\poo{j}{e}$ is very similar to that for $\poo{j}{\tau}$, as discussed in Appendix~\ref{sec:MisidentificationMatrix}, a significant dependence on the presence of additional jets is not observed for $\poo{j}{e}$.

The $5j$ discrepancy results from a tension with the $e\,4j$ final state, whose dominant contribution comes from $5j$ production convoluted with $\poo{j}{e}$.  The low-$p_T$ $3j$ discrepancy results from a tension with the $e\,2j$ final state, whose dominant contribution comes from $3j$ production convoluted with $\poo{j}{e}$.  The $b\,e\,j$ final state is predominantly $3j$ production convoluted with $\poo{j}{b}$ and $\poo{j}{e}$; this discrepancy also arises from a tension with the low-$p_T$ $3j$ and $e\,2j$ final states.  The $b\,e\,j$ final state is the \Vista\ final state in which the largest excess of data over Standard Model prediction is seen.  The fraction of hypothetical similar CDF experiments that would produce a \Vista\ normalization excess as significant as the excess observed in this final state is $8\%$.  The $5j$, $b\,e\,j$, and low-$p_T$ $3j$ discrepancies correspond to a difference of $\sim 10\%$ between the total number of observed and predicted events in these final states.

Figure~\ref{fig:VistaSummaryCdf} summarizes in a histogram the measured discrepancies between data and the Standard Model prediction for CDF high-$p_T$ final state populations and kinematic distributions.  Values in this figure represent individual discrepancies, and do not account for the trials factor associated with examining many possibilities.

\begin{figure}
\centering
\includegraphics[width=3.6in,angle=270]{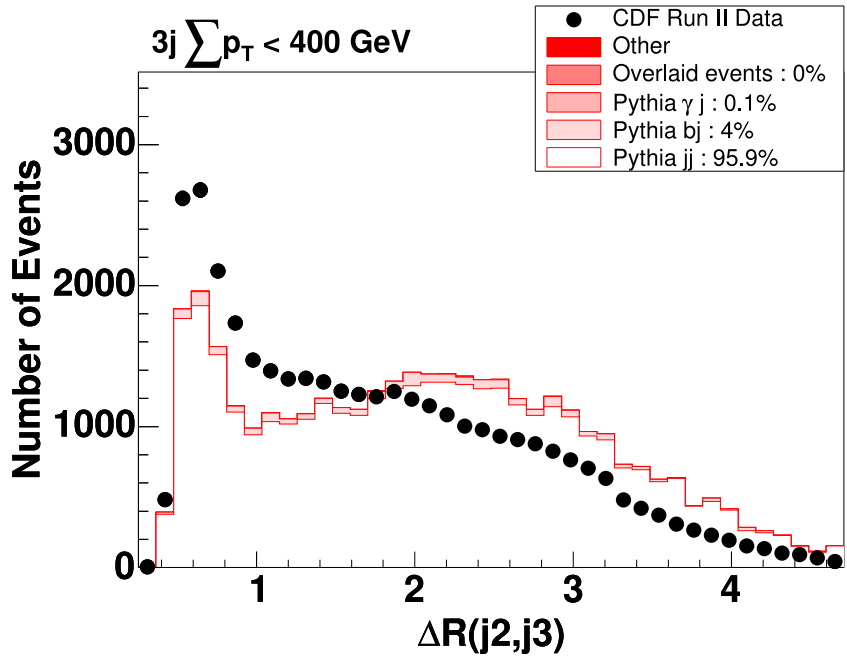}
\caption[A shape discrepancy highlighted by \Vista\ in the final state consisting of exactly three reconstructed jets with $\abs{\eta}<2.5$ and $p_T>17$~GeV, and with one of the jets satisfying $\abs{\eta}<1$ and $p_T>40$~GeV.]{A shape discrepancy highlighted by \Vista\ in the final state consisting of exactly three reconstructed jets with $\abs{\eta}<2.5$ and $p_T>17$~GeV, and with one of the jets satisfying $\abs{\eta}<1$ and $p_T>40$~GeV.  This distribution illustrates the effect underlying most of the \Vista\ shape discrepancies.  Filled circles show CDF data, with the shaded histogram showing the prediction of \Pythia.  The discrepancy is clearly statistically significant, with statistical error bars smaller than the size of the data points.  The vertical axis shows the number of events per bin, with the horizontal axis showing the angular separation ($\Delta R=\sqrt{\Delta\eta^2+\delta\phi^2}$) between the second and third jets, where the jets are ordered according to decreasing transverse momentum.  In the region $\Delta R(j_2,j_3)\gtrsim2$, populated primarily by initial state radiation, the Standard Model prediction can to some extent be adjusted.  The region $\Delta R(j_2,j_3)\lesssim2$ is dominated by final state radiation, the description of which is constrained by data from LEP\,1.}
\label{fig:3j_deltaR_j2j3}
\end{figure}

\begin{figure}
\centering
\includegraphics[width=3in,angle=270]{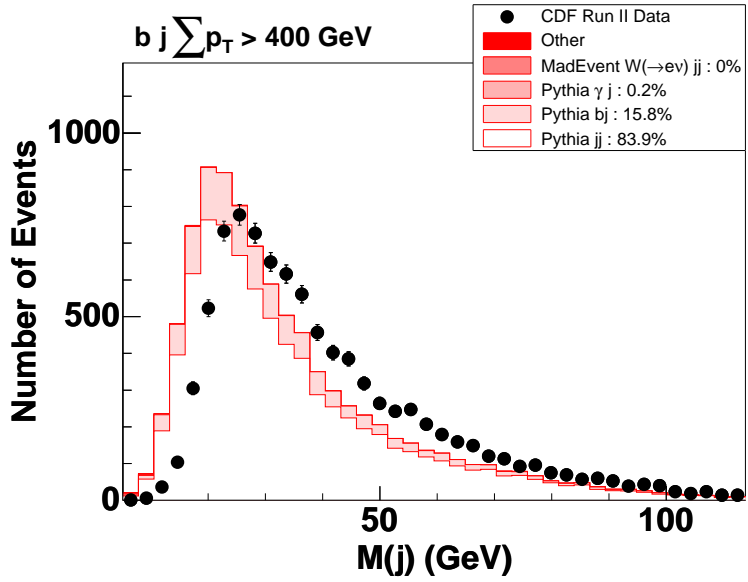}
\caption[The jet mass distribution in the $bj$ final state with $\SumPt>400$~GeV.]{The jet mass distribution in the $bj$ final state with $\SumPt>400$~GeV. The $3j$ $\Delta R(j_2,j_3)$ discrepancy illustrated in Fig.~\ref{fig:3j_deltaR_j2j3} manifests itself also by producing jets more massive in data than predicted by \Pythia's showering algorithm.  The mass of a jet is determined by treating energy deposited in each calorimeter tower as a massless 4-vector, summing the 4-vectors of all towers within the jet, and computing the mass of the resulting (massive) 4-vector.}
\label{fig:plots_1b1j_sumPt400+_mass_j}
\end{figure}

\begin{figure}
\centering
\includegraphics[width=3in,angle=270]{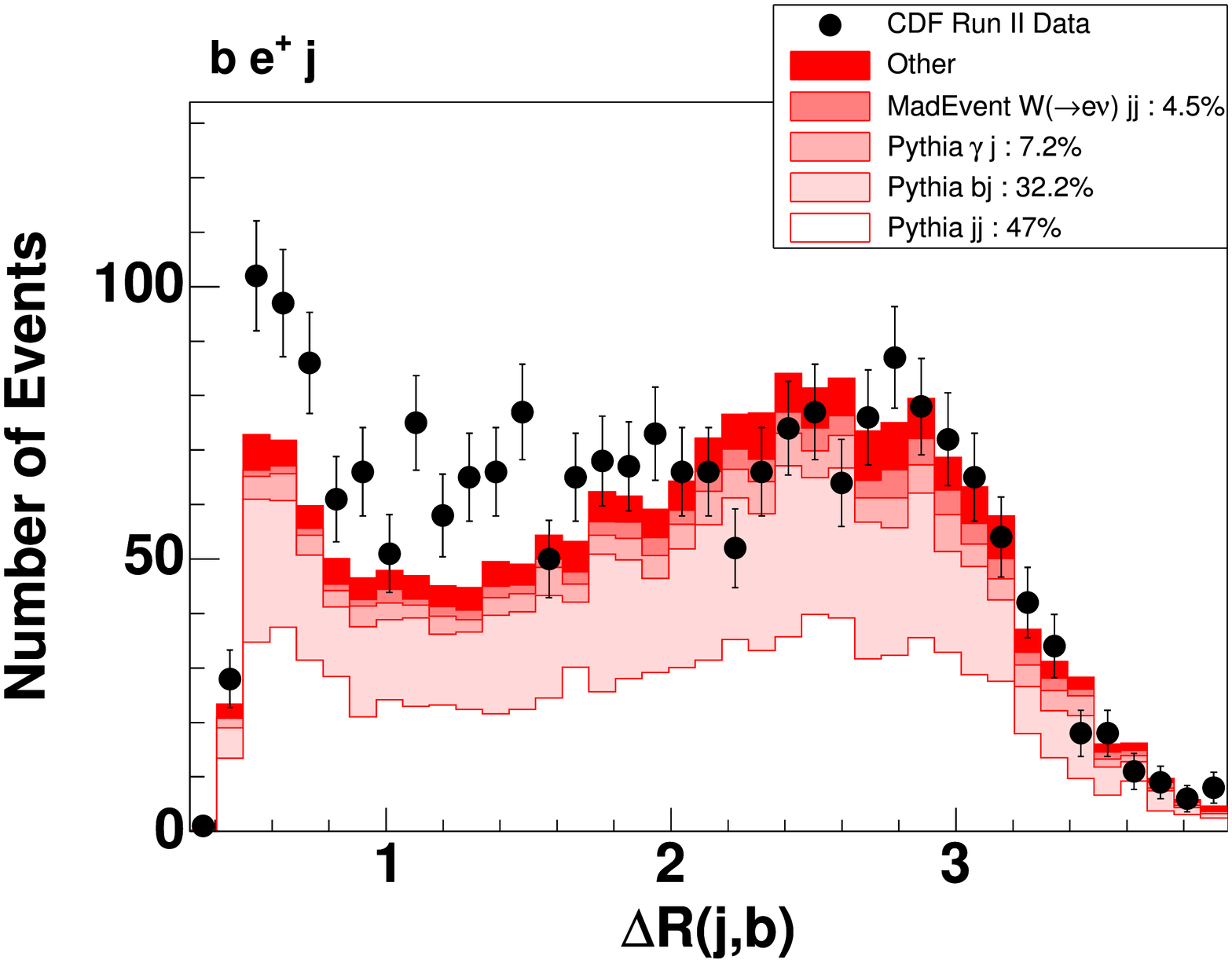}
\caption[The distribution of $\Delta R$ between the jet and $b$-tagged jet in the final state $b\,e\,j$.]{The distribution of $\Delta R$ between the jet and $b$-tagged jet in the final state $b\,e\,j$.  The primary Standard Model contribution to this final state is QCD three jet production with one jet misreconstructed as an electron.  The similarity to the $3j$ $\Delta R(j_2,j_3)$ discrepancy illustrated in Fig.~\ref{fig:3j_deltaR_j2j3} in the region $\Delta R(j,b)<2$ is clear.  Less clear is the underlying explanation for the difference with respect to Fig.~\ref{fig:3j_deltaR_j2j3} in the region $\Delta R(j,b)>2$.}
\label{fig:1b1e+1j_deltaR_jb}
\end{figure}

Of the 16,484 kinematic distributions considered, \numberOfVistaDiscrepantDistributions\ distributions are found to correspond to a discrepancy greater than 3$\sigma$ after accounting for the trials factor, entering with a KS probability of roughly $5\sigma$ or greater in Fig.~\ref{fig:VistaSummaryCdf}.  Of these \numberOfVistaDiscrepantDistributions\ discrepant distributions, 312 are attributed to modeling parton radiation, deriving from the $3j$ $\Delta R(j_2,j_3)$ discrepancy shown in Fig.~\ref{fig:3j_deltaR_j2j3}, with 186 of these 312 shape discrepancies pointing out that individual jet masses are larger in data than in the prediction, as shown in Fig.~\ref{fig:plots_1b1j_sumPt400+_mass_j}.  In the literature, that the same effect was observed (but not emphasized) by both CDF~\cite{Geer:CdfJetMass:Abe:1996nn,Geer:CdfJetMass:Abe:1997yb} and \DZero~\cite{D0JetMass:Abachi:1995zw} in Tevatron Run I.  The $3j$ $\Delta R(j_2,j_3)$ and jet mass discrepancies appear to be two different views of a single underlying discrepancy, noting that two sufficiently nearby distinct jets correspond to a pattern of calorimetric energy deposits similar to a single massive jet.  The underlying $3j$ $\Delta R(j_2,j_3)$ discrepancy is manifest in many other final states.  The final state $b\,e\,j$, arising primarily from QCD production of three jets with one misreconstructed as an electron, shows a similar discrepancy in $\Delta R(j,b)$ in Fig.~\ref{fig:1b1e+1j_deltaR_jb}.

While these discrepancies are clearly statistically significant, basing a new physics claim on them would be premature.  In the kinematic regime of the discrepancy, different algorithms to match exact leading order calculations with a parton shower lead to different predictions~\cite{MadEventAlpgenComparison:Alwall:2007fs}.  Newer predictions have not been systematically compared to LEP\,1 data, which provide constraints on parton showering reflected in \Pythia's tuning.  Further investigation into obtaining an adequate QCD-based description of this discrepancy continues.

An additional 59 discrepant distributions reflect an inadequate modeling of the overall transverse boost of the system.  The overall transverse boost of the primary physics objects in the event is attributed to two sources:  the intrinsic Fermi motion of the colliding partons within the proton, and soft or collinear radiation of the colliding partons as they approach collision.  Together these effects are here referred to as ``intrinsic $k_T$,'' representing an overall momentum kick to the hard scattering.  Further discussion appears in Appendix~\ref{sec:CorrectionModelDetails:IntrinsicKt}.

The remaining 13 discrepant distributions are seen to be due to the coarseness of the \Vista\ correction model.  Most of these discrepancies, which are at the level of 10\% or less when expressed as $({\text{data}}-{\text{theory}})/{\text{theory}}$, arise from modeling most fake rates as independent of transverse momentum. 

In summary, this global analysis of the bulk features of the high-$p_T$ data has not yielded a discrepancy motivating a new physics claim.  There are no statistically significant population discrepancies in the 344 populated final states considered, and although there are several statistically significant discrepancies among the 16,486 kinematic distributions investigated, the nature of these discrepancies makes it difficult to use them to support a new physics claim.

This global analysis of course can not conclude with certainty that there is no new physics hiding in the CDF data.  The \Vista\ population and shape statistics may be insensitive to a small excess of events appearing at large $\SumPt$ in a highly populated final state.  For such signals, different probes are required.  \Sleuth, and the Bump Hunter, which was added in the second round of this analysis, serve this purpose.


\section{\Sleuth}
\label{sec:Sleuth}

Taking a broad view of proposed models that might extend the Standard Model, something common is noted:  nearly all predict an excess of events at high $p_T$, concentrated in a particular final state.  This feature is exploited by \Sleuth~\cite{KnutesonThesis}.  \Sleuth\ is quasi model independent, where ``quasi'' refers to the assumption that the first sign of new physics will appear as an excess of events in some final state at large summed scalar transverse momentum ($\SumPt$).

The first version of \Sleuth\ was essentially developed by \DZero\ in Tevatron Run I~\cite{SleuthPRD1:Abbott:2000fb, SleuthPRD2:Abbott:2000gx, SleuthPRL:Abbott:2001ke}, and subsequently improved by H1 in HERA Run I~\cite{H1GeneralSearch:Aktas:2004pz}, with small modifications.  

\Sleuth\ relies on the following assumptions for new physics:
\begin{enumerate}
\item The data can be categorized into exclusive final states in such a way that any signature of new physics is apt to appear predominantly in one of these final states.  
\item New physics will appear with objects at high summed transverse momentum ($\SumPt$) relative to Standard Model and instrumental background.  
\item New physics will appear as an excess of data over Standard Model and instrumental background.  
\end{enumerate}
To the extent that the above are true, \Sleuth\ would be more sensitive to a new physics signal.

\subsection{Algorithm}
\label{sec:SleuthAlgorithm}
The \Sleuth\ algorithm consists of three steps, following the above three assumptions.

\subsubsection{Final states}
\label{sec:finalStateDefinitions}

In the first step of the algorithm, all events are placed into exclusive final states as in \Vista, with the following modifications.

\begin{itemize}

\item
Jets are identified as pairs, rather than individually, to reduce the total number of final states and to keep signal events with one additional radiated gluon within the same final state.  Final state names include ``$n$ $jj$'' if $n$ jet pairs are identified, with possibly one unpaired jet assumed to have originated from a radiated gluon.

\item
The present understanding of quark flavor suggests that $b$ quarks should be produced in pairs. Bottom quarks are identified as pairs, rather than individually, to increase the robustness of identification and to reduce the total number of final states.  Final state names include ``$n$ $bb$'' if $n$ $b$ pairs are identified.

\item
  Final states related through global charge conjugation are considered to be equivalent.  Thus $e^+e^-\gamma$ is a different final state than $e^+e^+\gamma$, but $e^+e^+\gamma$ and $e^-e^-\gamma$ together make up a single \Sleuth\ final state.  

\item
  Final states related through global interchange of the first and second generation are considered to be equivalent.  Thus $e^+\pmiss\gamma$ and $\mu^+\pmiss\gamma$ together make up a single \Sleuth\ final state.  The decision to treat third generation objects ($b$ quarks and $\tau$ leptons) differently from first and second generation objects reflects theoretical prejudice that the third generation may be special, and the experimental ability (in the case of $b$ quarks) and experimental challenge (in the case of $\tau$ leptons) in the identification of third generation objects. 

\end{itemize}

The symbol $\ell$ is used to denote electron or muon.  The symbol $W$ is used in naming final states containing one electron or muon, significant missing momentum, and perhaps other non-leptonic objects.  Thus the final states $e^+\pmiss\gamma$, $e^-\pmiss\gamma$, $\mu^+\pmiss\gamma$, and $\mu^-\pmiss\gamma$ are combined into the \Sleuth\ final state $W\gamma$.  A table showing the relationship between \Vista\ and \Sleuth\ final states is provided in Appendix~\ref{sec:Sleuth:Partitioning}.

\subsubsection{Summed Transverse Momentum Variable}

The second step of the algorithm considers a single variable in each exclusive final state:  the summed scalar transverse momentum of all objects in the event ($\sum{p_T}$).  Assuming momentum conservation in the plane transverse to the axis of the colliding beams,
\begin{equation}
\sum_i{\vec{p}_i} + \overrightarrow{\text{uncl}} + \vec{\pmiss} = \vec{0},
\end{equation}
where the sum over $i$ represents a sum over all identified objects in the event, the $i^\text{th}$ object has momentum $\vec{p}_i$, $\overrightarrow{\text{uncl}}$ denotes the vector sum of all momentum visible in the detector but not clustered into an identified object, $\vec{\pmiss}$ denotes the missing momentum, and the equation is a two-component vector equality for the components of the momentum along the two spatial directions transverse to the axis of the colliding beams.  The \Sleuth\ variable \SumPt\ is then defined by
\begin{equation}
\SumPt \equiv \sum_i{\abs{\vec{p}_i}} + \abs{\overrightarrow{\text{uncl}}} + \abs{\vec{\pmiss}},
\end{equation}
where only the momentum components transverse to the axis of the colliding beams are considered when computing magnitudes.

\subsubsection{Regions}
\label{sec:Sleuth:Regions}

The algorithm's third step involves searching for regions in which more events are seen in the data than expected from Standard Model and instrumental background.  This search is performed in the variable \SumPt\ defined in the second step of the algorithm, for each of the exclusive final states defined in the first step.  

The steps of the search can be sketched as follows.
\begin{itemize}
\item  In each final state, the regions considered are the one dimensional intervals in $\sum{p_T}$ extending from each data point up to infinity.  A region is required to contain at least three data events, as described in Appendix~\ref{sec:Sleuth:MinimumNumberOfEvents}.

\item 
In a particular final state, the data point with the $d^{\text{th}}$ largest value of $\SumPt$ defines an interval in the variable $\SumPt$ extending from this data point up to infinity.  This semi-infinite interval contains $d$ data events.  The Standard Model prediction in this interval, estimated from the \Vista\ comparison, integrates to $b$ predicted events.  In this final state, the interest of the $d^{\text{th}}$ region is defined as the Poisson probability $\pval = \sum_{i=d}^{\infty}\frac{b^i}{i!}e^{-b}$ that the Standard Model background $b$ would fluctuate up to or above the observed number of data events $d$ in this region.  The most interesting region in this final state is the one with smallest Poisson probability (\pvalmin).
\item For this final state, pseudo experiments are generated, with pseudo data pulled from the Standard Model background.  For each pseudo experiment, the interest of the most interesting region is calculated.  An ensemble of pseudo experiments determines the fraction $\scriptP$ of pseudo experiments in this final state in which the most interesting region is more interesting than the most interesting region in this final state observed in the data.  Namely, for each final state, \scriptP\ is the fraction of pseudo-data distributions, pulled from the Standard Model expectation, where \pvalmin\ was smaller than the \pvalmin\ observed in the actual data distribution.  If there is no new physics in this final state, $\scriptP$ is expected to be a random number pulled from a uniform distribution in the unit interval\footnote{
\label{footnote:scriptPdistribution} 
There is a small caveat, for final states with small expected population:  We require at least 3 data in a \sumPt\ tail.  If $d<3$, then $\pval=1$ by convention, i.e.\ the tail is totally uninteresting by definition.  Apart from $\pval=1$, the most uninteresting a tail can possibly be is to have exactly $d=3$ and as big a background $b$ as possible.
So, the largest \pval\ attainable for a final state with total background $b_{\rm tot}$, before we run into $\pval=1$, is $\pval_{\max}=\sum_{i=3}^{\infty}\frac{b_{\rm tot}^i}{i!}e^{-b_{\rm tot}}$.  I will show now that \scriptP\ can not assume values between $\pval_{\max}$ and 1, therefore its distribution is not exactly uniform, but has a gap: If the actual $\pvalmin$ were equal to $\pval_{\max}$, then the fraction of pseudo-data distributions which would have $\pvalmin > \pval_{\max}$ would be $\sum_{i=0}^{2}\frac{b_{\rm tot}^i}{i!}e^{-b_{\rm tot}}$, because they would be given $\pvalmin=1$ by convention.  The rest of the pseudo-data distributions would have $\pvalmin \le \pval_{\max}$, therefore $\scriptP=1-\sum_{i=0}^{2}\frac{b_{\rm tot}^i}{i!}e^{-b_{\rm tot}}=\pval_{\max}$.  For any actual $\pvalmin < \pval_{\max}$, \scriptP\ will be even smaller than $\pval_{\max}$, as it will be more challenging for a pseudo-data distribution to exceed that \pvalmin.  If $\pvalmin=1$, which has probability $\sum_{i=0}^{2}\frac{b_{\rm tot}^i}{i!}e^{-b_{\rm tot}}$, then all pseudo-data distributions would be at least as interesting, therefore $\scriptP=1$.  Therefore, the distribution of \scriptP\ has a Kronecker $\delta$ term at 1, multiplied by $\sum_{i=0}^{2}\frac{b_{\rm tot}^i}{i!}e^{-b_{\rm tot}}$, and the rest is spread at values $\scriptP \le \pval_{\max}$.  This gap in possible \scriptP\ values shrinks as $b_{\rm tot} \gg 3$, and practically vanishes for $b_{\rm tot} \gtrsim 10$.}. 
If there is new physics in this final state, $\scriptP$ is expected to be small.
\item  Looping over all final states, $\scriptP$ is computed for each final state.  The minimum of these values is denoted $\scriptP_{\text{min}}$.  Let ${\cal R}$ be the most interesting region in the final state with the smallest $\scriptP$.
\item The interest of the most interesting region ${\cal R}$ in the most interesting final state is defined as $\tildeScriptP = 1-\prod_a(1-\hat{p}_a)$, where the product is over all \Sleuth\ final states $a$, and $\hat{p}_a$ is the lesser of $\scriptP_{\text{min}}$ and the probability for the total number of events predicted by the Standard Model in the final state $a$ to fluctuate up to or above three data events. The quantity $\tildeScriptP$ is the fraction of hypothetical similar CDF experiments that would produce a final state with $\scriptP < \scriptP_{\text{min}}$\footnote{
\label{footnote:tildeScriptPdefinition}
This point deserves some explanation to become more obvious.  We have $N$ final states, and we want to find the probability that one or more of them would give a \scriptP\ smaller than the observed $\scriptP_{\min}$.  If the expectated distribution of \scriptP\ were exactly uniform for all $N$ final states, without the gap discussed in footnote \ref{footnote:scriptPdistribution}, then each final state would have equal probability $\scriptP_{\min}$ to give $\scriptP \le \scriptP_{\min}$.  In that simple case, we would just need to define $\tildeScriptP \equiv 1-\prod_a(1-\scriptP_{\min})=1-(1-\scriptP_{\min})^N$.  However, depending on the total background $b_{\rm tot}$, \scriptP\ is not distributed exactly uniformly for small final states, which gives rise to two possibilities:  If for a final state the gap starts at a $\pval_{\max} \ge \scriptP_{\min}$, then the probability that this final state would give $\scriptP \le \scriptP_{\min}$ is simply $\scriptP_{\min}$.  If, however, $b_{\rm tot}$ is such that $\pval_{\max} \le \scriptP_{\min}$, then $\scriptP_{\min}$ falls in the gap, and then that final state has probability $\sum_{i=3}^{\infty}\frac{b_{\rm tot}^i}{i!}e^{-b_{\rm tot}}$ to return $\scriptP \le \pval_{\max} < \scriptP_{\min}$, as explained in footnote \ref{footnote:scriptPdistribution}.  This complication necessitates the introduction of $\hat{p}_a$ in \tildeScriptP, to treat appropriately the two possible cases.}.
The range of $\tildeScriptP$ is the unit interval.  If the data are distributed according to our Standard Model prediction, $\tildeScriptP$ is expected to be a random number pulled from a uniform distribution in the unit interval, as was also demonstrated experimentally (see Fig.~\ref{fig:tildeScriptPdistribution}).  If new physics is present, $\tildeScriptP$ is expected to be small.
\end{itemize}

\begin{figure}
\begin{tabular}{cc}
\includegraphics[height=3.0in,angle=-90]{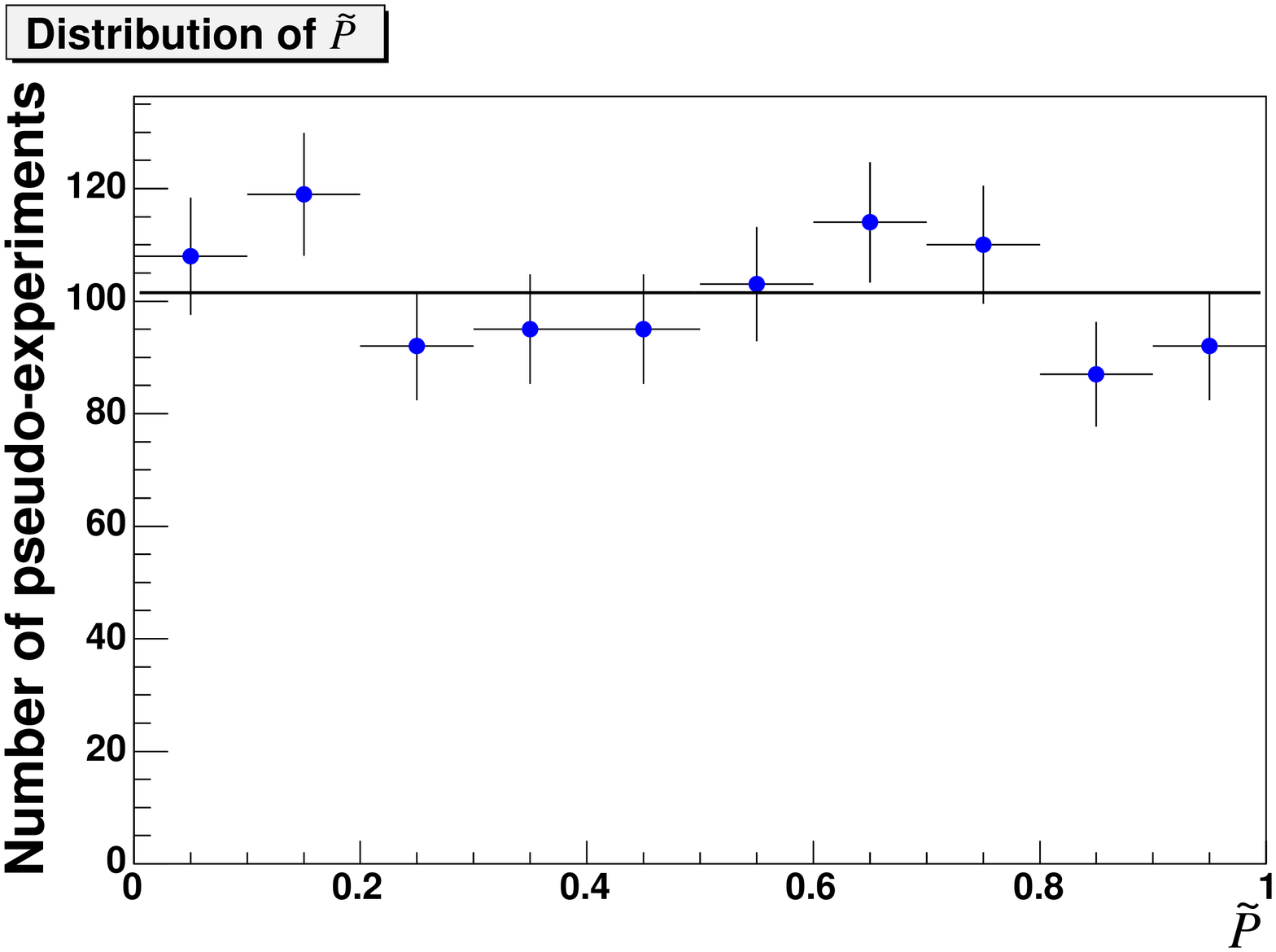} & \hspace{0in}
\includegraphics[height=3.0in,angle=-90]{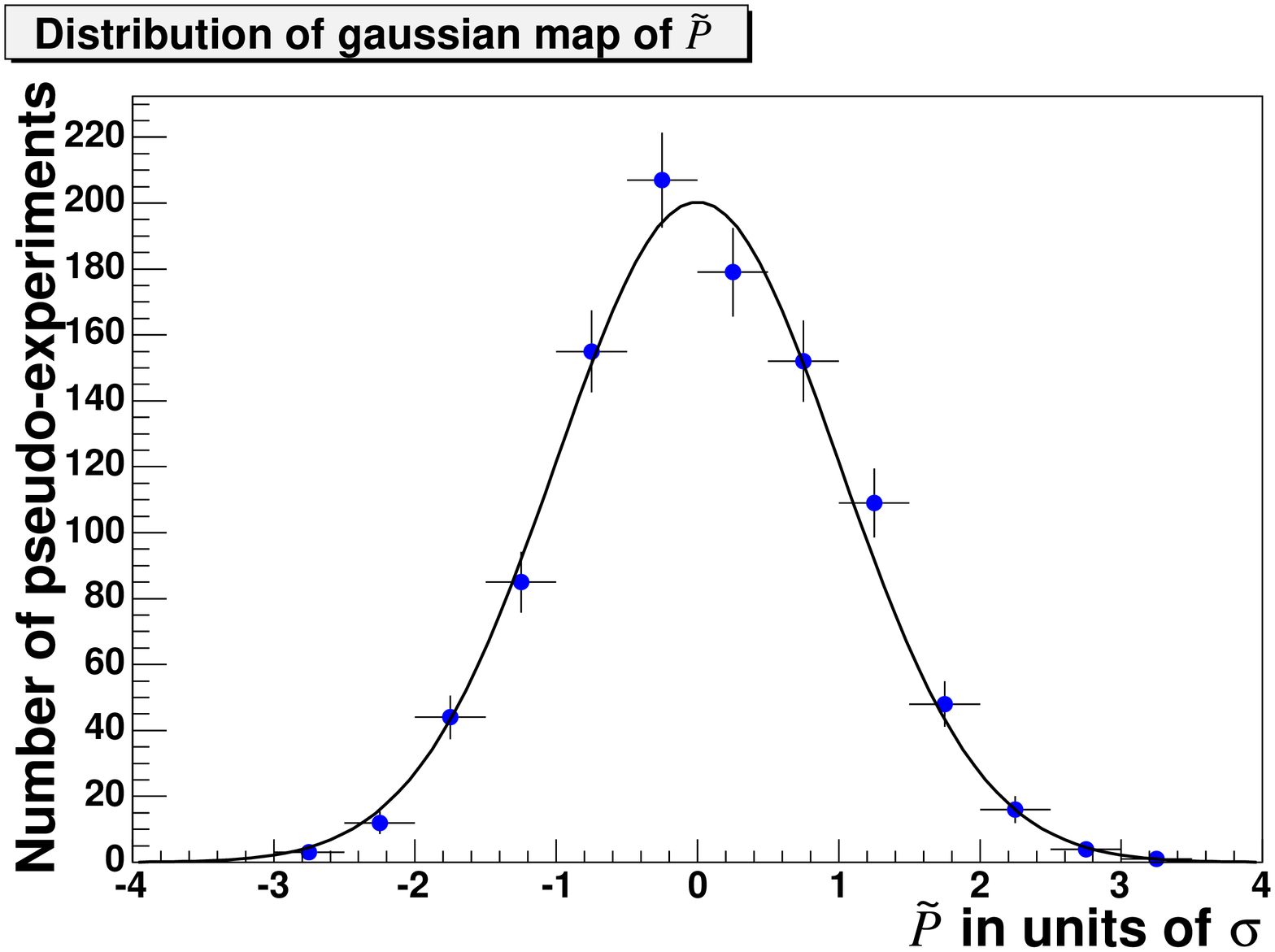} 
\end{tabular}
\caption[Distribution of expected values of \tildeScriptP\ in $\sim 1000$ pseudo-experiments, where pseudo-data are pulled from the Standard Model \SumPt\ distributions.]{Distribution of expected values of \tildeScriptP\ in $\sim 1000$ pseudo-experiments, where pseudo-data are pulled from the Standard Model \SumPt\ distributions.  On the right is shown the distribution of the same values of \tildeScriptP\ translated into standard deviations ($\sigma$) through the transformation: $\tildeScriptP = \int_{\sigma}^{\infty}\frac{1}{\sqrt{2\pi}}e^{-\frac{x^2}{2}}$.  The expected distribution is consistent with a uniform distribution in the interval $[0,1]$, represented by the black curve.}
\label{fig:tildeScriptPdistribution}
\end{figure}

An alternative statistic to \tildeScriptP\ was first implemented in this analysis.  The new measure of significance, \tildePval, is the probability that, in a pseudo-experiment, at least one \sumPt\ tail, in any final state, would have a \pval\ smaller than the smallest \pval\ found among all tails and all final states in the data.  In other words, \tildePval\ is the probability that in a pseudo-experiment some \sumPt\ tail would be more significant than the globally most significant tail found in the data.  The definition of \tildePval\ is
\begin{equation}
\tildePval \equiv 1 - \prod_{a}\left(1-\scriptP_{(a,\pvalmin)}\right),
\end{equation}
where $a$ denotes a final state, $\scriptP_{(a,\pvalmin)}$ is the probability for final state $a$ to have (in a pseudo-experiment) a \sumPt\ tail of $\pval \le \pvalmin$, and \pvalmin\ is the smallest \pval\ found among all tails in all final states using data.  Note that, unlike when defining \scriptP\ for a final state $a$, where \pvalmin\ was the smallest \pval\ within that final state, this \pvalmin\ going into $\scriptP_{(a,\pvalmin)}$ is the global smallest \pval.  Therefore, for a final state $a$, $\scriptP_{(a,\pvalmin)}$ is not the same as the \scriptP\ defined earlier for each final state, because there \scriptP\ was the probability for a final state to exceed in significance its own most interesting tail, while $\scriptP_{(a,\pvalmin)}$ is the probability for final state $a$ to exceed in significance the {\em globally} most interesting tail, which may or may not be within $a$.  

The qualitative difference between \tildePval\ and traditional \tildeScriptP\ is that \tildeScriptP\ focusses on fluctuations producing a smaller \scriptP\ than the $\scriptP_{\min}$ observed in the data, while \tildePval\ focusses on fluctuations producing a smaller \pval.  The \scriptP\ of a final state depends not only on the significance (\pvalmin) of the most interesting tail therein, but also on the total expected population of the final state where that tail is:  A \sumPt\ tail of numerically identical \pvalmin, but found in a final state with larger expected background, results into larger \scriptP, because bigger population means more pseudo-data, hence more \sumPt\ tails, hence more chances to have $\pval \le \pvalmin$.  So, \scriptP\ is not a measure of the significance of a tail {\em per se}, but rather of a whole \sumPt\ distribution.  Whether to use \tildeScriptP\ or \tildePval\ is a matter of preference.  \tildePval\ is more intuitive, because it quantifies the significance of \sumPt\ tails, which are fundamentally the features \Sleuth\ detects, while \tildeScriptP\ quantifies the significance of whole \sumPt\ distributions from the view-point of their own \sumPt\ excesses.  Since \tildeScriptP\ was invented first and has been part of \Sleuth\ since its conception, its use was continued in this analysis.

\subsubsection{Output}
\label{sec:Sleuth:Output}

The output of the algorithm is the most interesting region ${\cal R}$ observed in the final state with the smallest \scriptP, and a number $\tildeScriptP$ quantifying the interest of ${\cal R}$\footnote{If \Sleuth\ used \tildePval\ instead of \tildeScriptP, then the most interesting tail ${\cal R}$ would be the one with the globally smallest \pval.  That region may happen to be the same with the most interesting region within the final state with of smallest \scriptP, but it doesn't have to.}.  A reasonable threshold for discovery is $\tildeScriptP \lesssim 0.001$, which corresponds loosely to a local $5\sigma$ effect after the trials factor is accounted for\footnote{That is empirically confirmed in sensitivity tests (Sec.~\ref{sec:sleuthSensitivity}), where it was observed that the $\tildeScriptP$ discovery threshold is met approximately at the same time when $\pvalmin\simeq \int_{5}^{\infty}\frac{1}{\sqrt{2\pi}}e^{-\frac{x^2}{2}}$.}.

Although no integration over systematic errors is performed in computing $\tildeScriptP$, systematic uncertainties do affect the final \Sleuth\ result.  If \Sleuth\ highlights a discrepancy in a particular final state, explanations in terms of a correction to the background estimate are considered.  This process necessarily requires physics judgement.  A reasonable explanation of a \Sleuth\ discrepancy in terms of an inadequacy in the modeling of the detector response or Standard Model prediction that is consistent with external information is fed back into the \Vista\ correction model and tested for global consistency.  In this way, plausible explanations for discrepancies observed by \Sleuth\ are incorporated into the \Vista\ correction model.  This iteration continues until either all reasonable explanations for a significant \Sleuth\ discrepancy are exhausted, resulting in a possible new physics claim, or no significant \Sleuth\ discrepancy remains.


\begin{figure*}
\hspace{-2cm}
\begin{tabular}{cc}
\includegraphics[width=2.75in,angle=270]{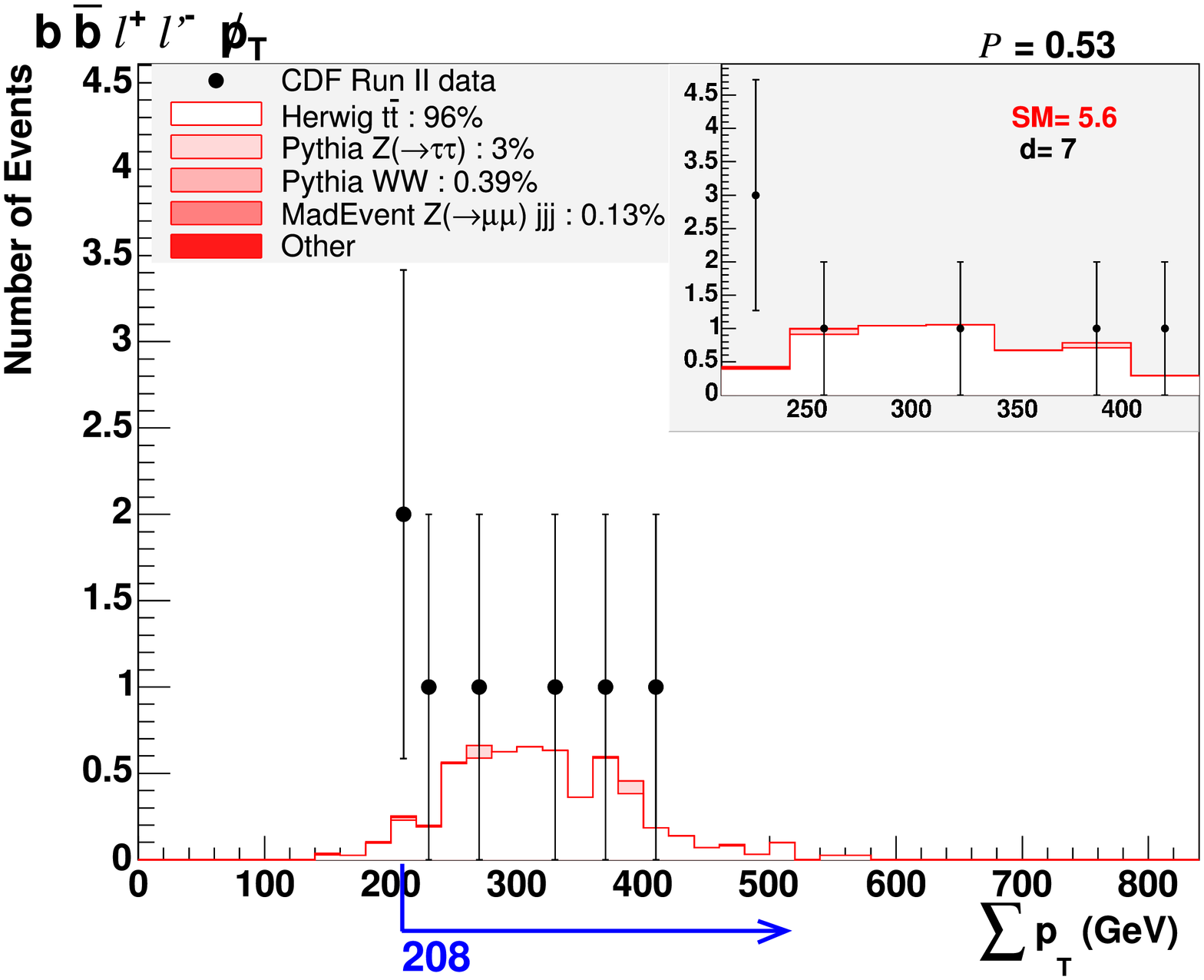} & 
\hspace{-1cm}\includegraphics[width=2.75in,angle=270]{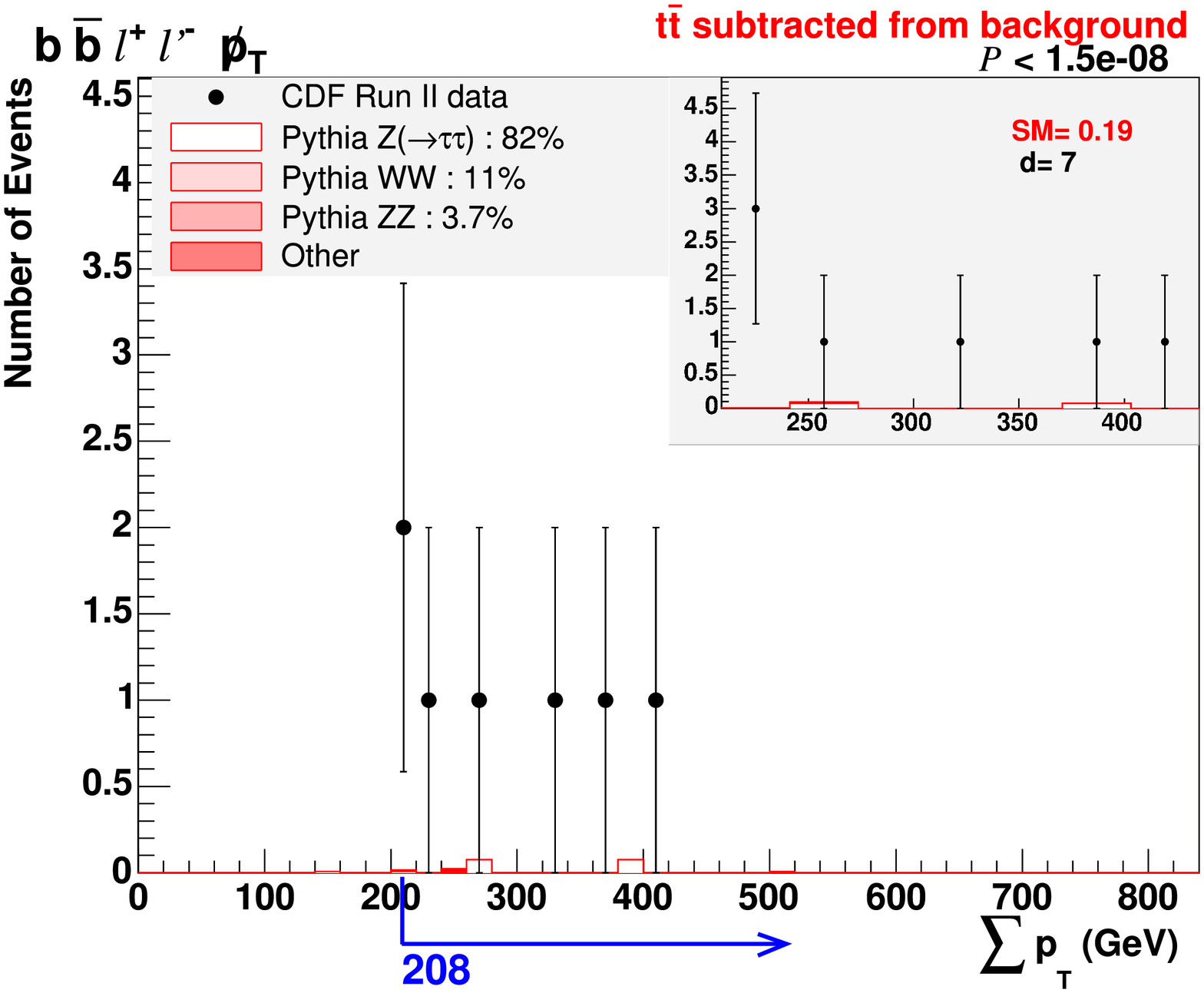} \\
\includegraphics[width=2.75in,angle=270]{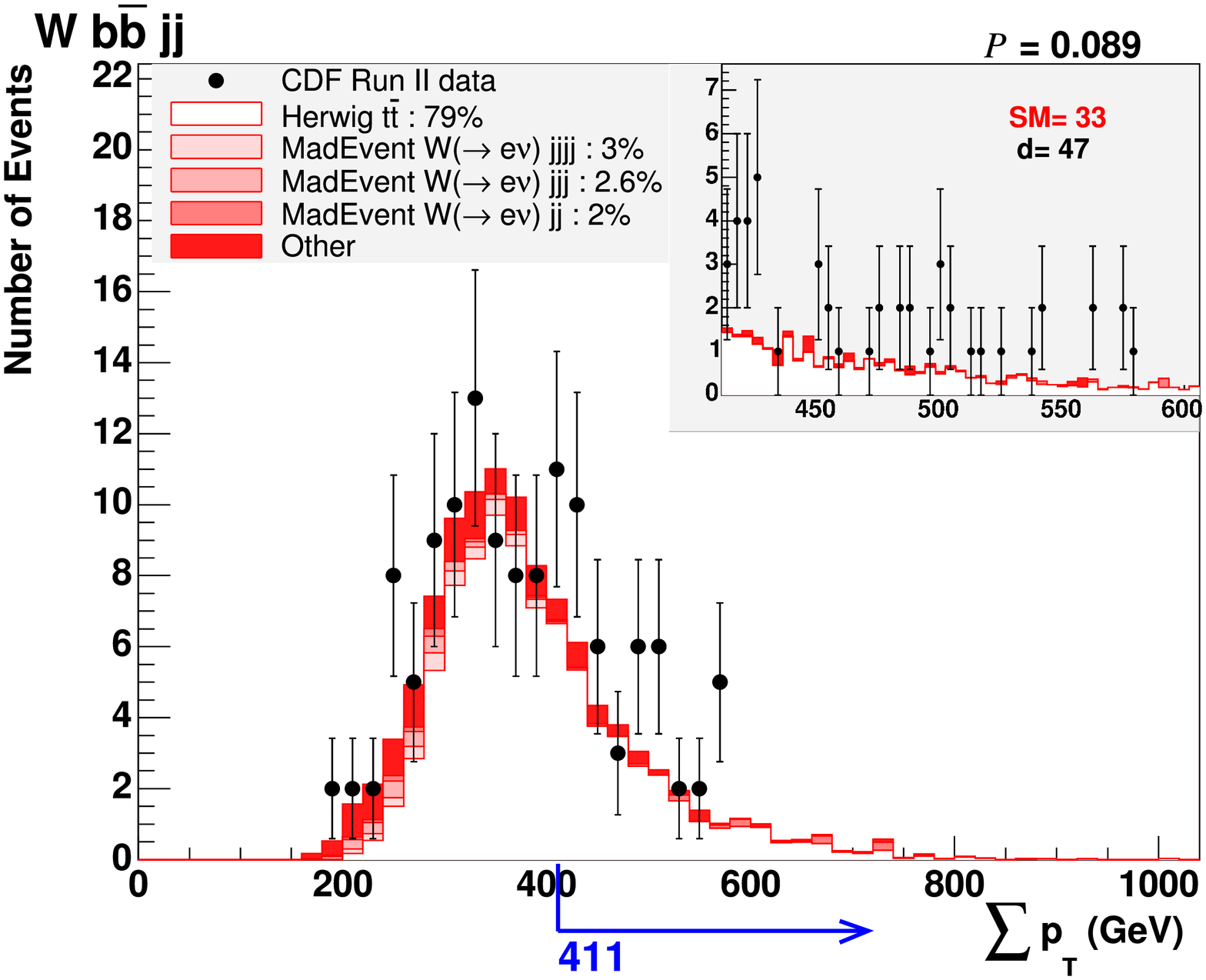} & 
\hspace{-1cm}\includegraphics[width=2.75in,angle=270]{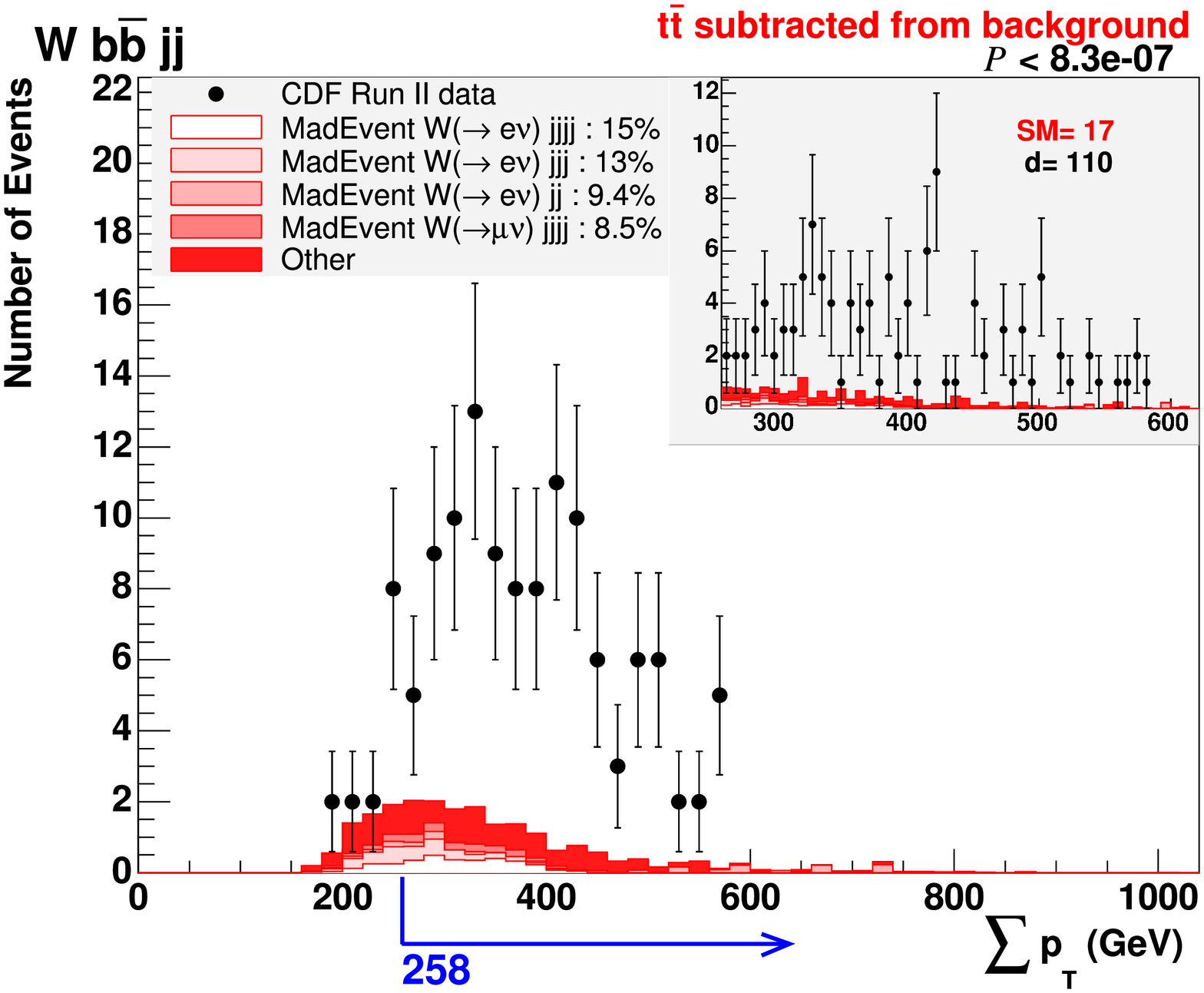} 
\end{tabular}
\caption[$\ttbar$ Sensitivity test]{(Top left)  The \Sleuth\ final state $b\bar{b}\ell^+\ell'^-\pmiss$, consisting of events with one electron and one muon of opposite sign, missing momentum, and two or three jets, one or two of which are $b$-tagged.  Data corresponding to 927~pb$^{-1}$ are shown as filled circles; the Standard Model prediction is shown as the shaded histogram.  (Top right)  The same final state with $t\bar t$ subtracted from the Standard Model prediction.  (Bottom row)  The \Sleuth\ final state $Wb\bar{b}jj$, with the Standard Model $t\bar t$ contribution included (lower left) and removed (lower right).  Significant discrepancies far surpassing \Sleuth's discovery threshold are observed in these final states with $t\bar{t}$ removed from the Standard Model background estimate.  
}
\label{fig:topless_SM_sensitivityTest}
\end{figure*}

\begin{figure}
\centering
\includegraphics[width=5.2in,angle=0]{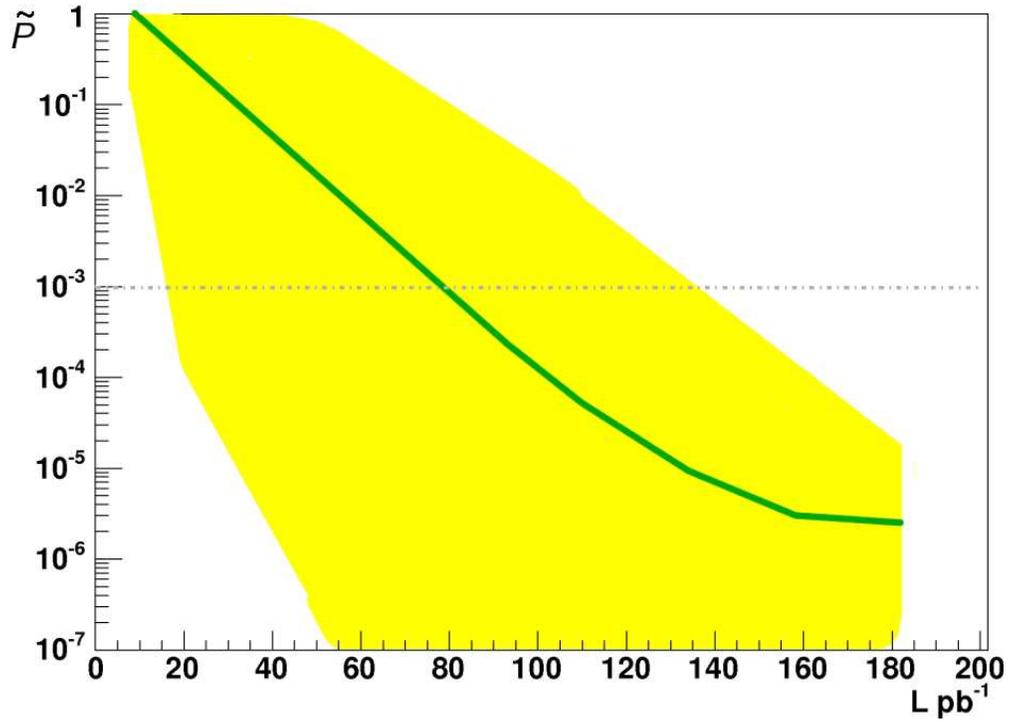}
\caption[\Sleuth's $\tildeScriptP$ as a function of assumed integrated luminosity, with $\ttbar$ removed.]{\Sleuth's $\tildeScriptP$ as a function of assumed integrated luminosity, with top quark pair production removed from the Standard Model background estimate.  The horizontal axis shows integrated luminosity, in units of pb$^{-1}$.  The vertical axis shows \Sleuth's $\tildeScriptP$.
  With Standard Model $t\bar{t}$ production omitted from the background estimate and actual data including $t\bar{t}$ production, \Sleuth's $\tildeScriptP$ decreases with increasing integrated luminosity, shown as the solid (green) line, crossing at roughly 80~pb$^{-1}$ the discovery threshold of $\tildeScriptP<0.001$, shown as the horizontal dashed (gray) line.  The shaded (yellow) band shows the range of values of $\tildeScriptP$ obtained in a number of trials, with the width of the band resulting from the statistical fluctuations of individual top quark events.}
\label{fig:tildeScriptPvsLuminosityTtbar}
\end{figure}

\begin{figure*}
\centering
\begin{tabular}{cc}
\includegraphics[width=2.3in,angle=270]{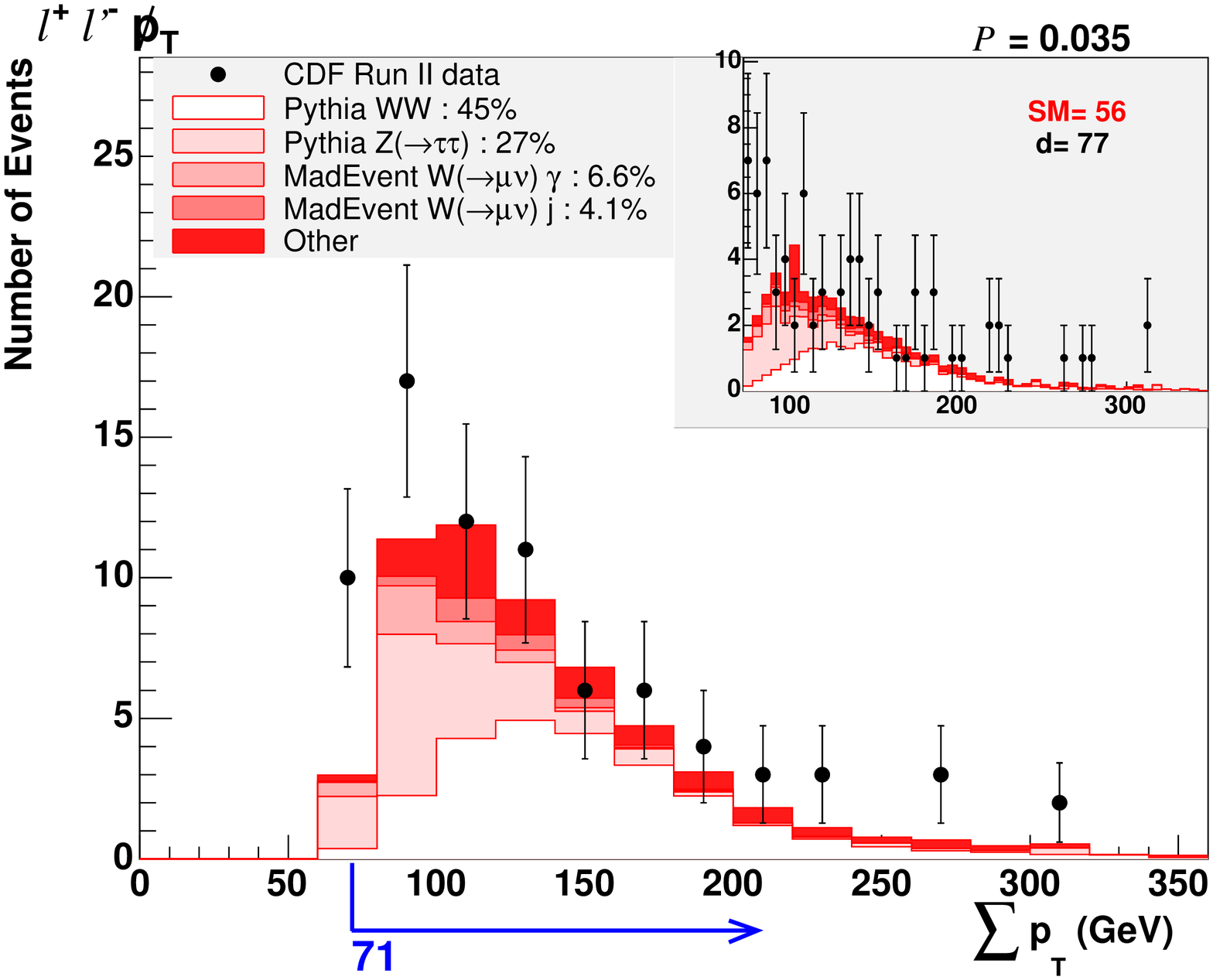} & \hspace*{0cm}
\includegraphics[width=2.3in,angle=270]{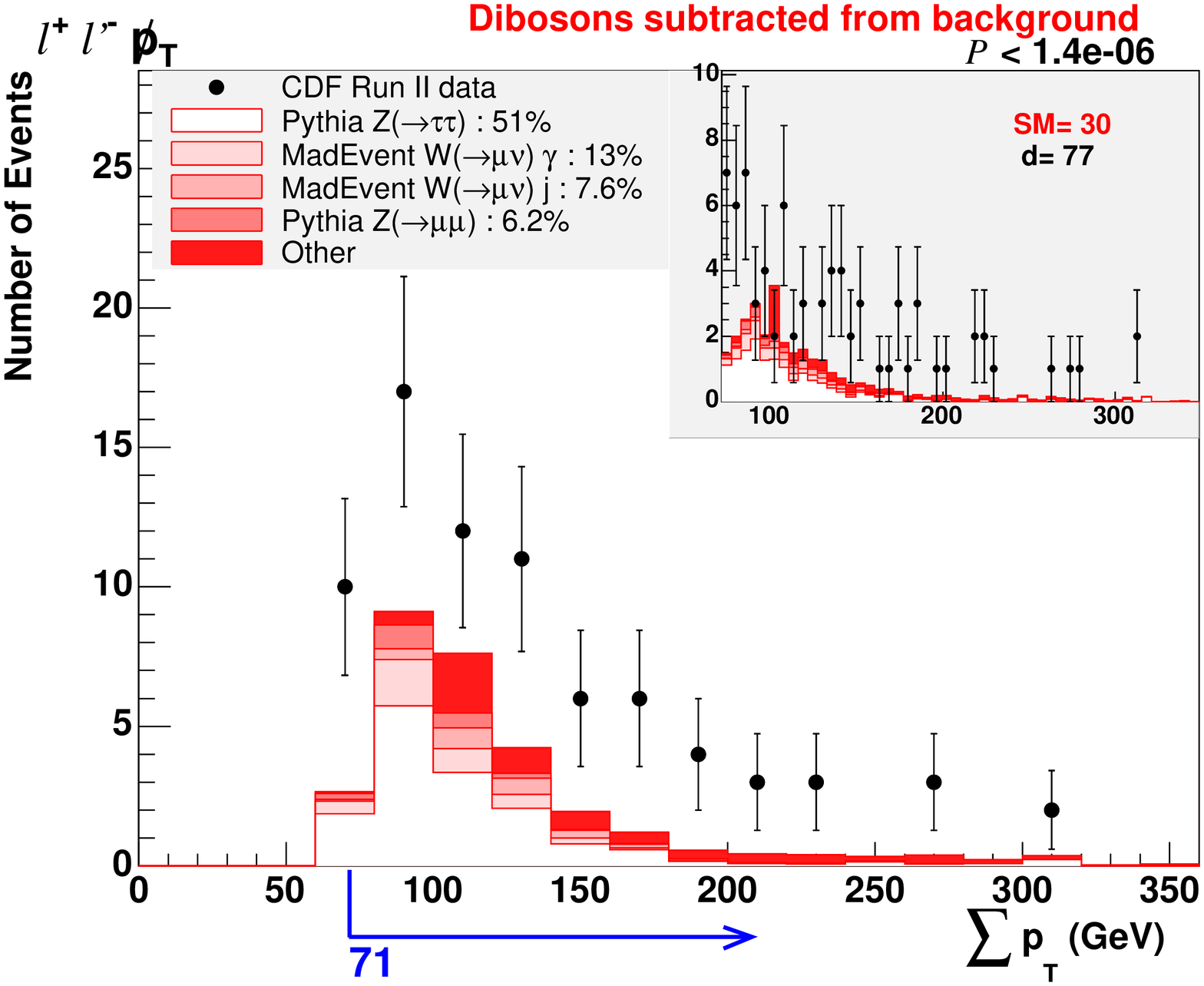} \\
\end{tabular}
\caption[\Sleuth\ diboson sensitivity test.]{(Left)  The final state $\ell^+{\ell'}^-\pmiss$, consisting of events with an electron and muon of opposite sign and missing transverse momentum, in 927~pb$^{-1}$ of CDF data.  (Right) The same final state with Standard Model $WW$, $WZ$, and $ZZ$ contributions subtracted, and with the correction factors re-fit in the absence of these contributions.  
\Sleuth\ finds the final state $\ell^+{\ell'}^-\pmiss$ to contain a discrepancy surpassing the discovery threshold of $\tildeScriptP<0.001$ with the processes $WW$, $WZ$, and $ZZ$ removed from the Standard Model background.}
\label{fig:VVless_SM_sensitivityTest}
\end{figure*}

\begin{table*}
\begin{minipage}{7in}
\begin{tabular}{cp{5cm}c}
  {\bf Model} & \multicolumn{1}{c}{{\bf Description}} & {\bf Sensitivity} \\
  \hline
  {1}
             & GMSB, $\Lambda=82.6$~GeV, $\tan{\beta}=15$, $\mu>0$, with one messenger of $M=2\Lambda$. 
  &  \raisebox{-0.5\height}{\includegraphics[width=7.5cm]{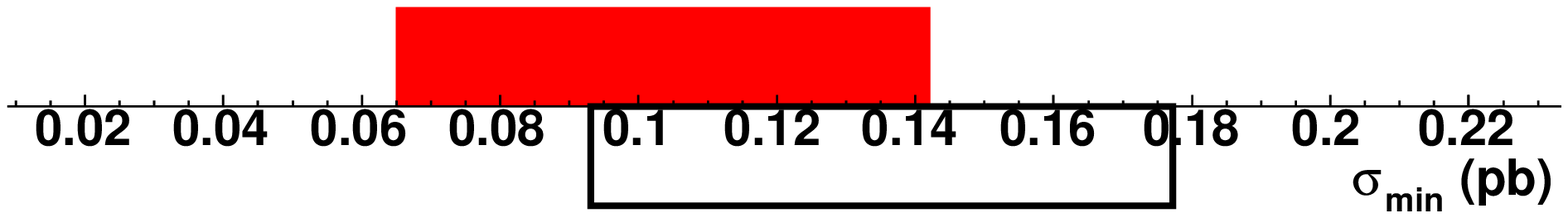}} \\ 
  {2}
             &  $Z'\to \ell^+\ell^-$, $m_{Z'}=250$~GeV, with standard model couplings to leptons.
  &   \raisebox{-0.5\height}{\includegraphics[width=7.5cm]{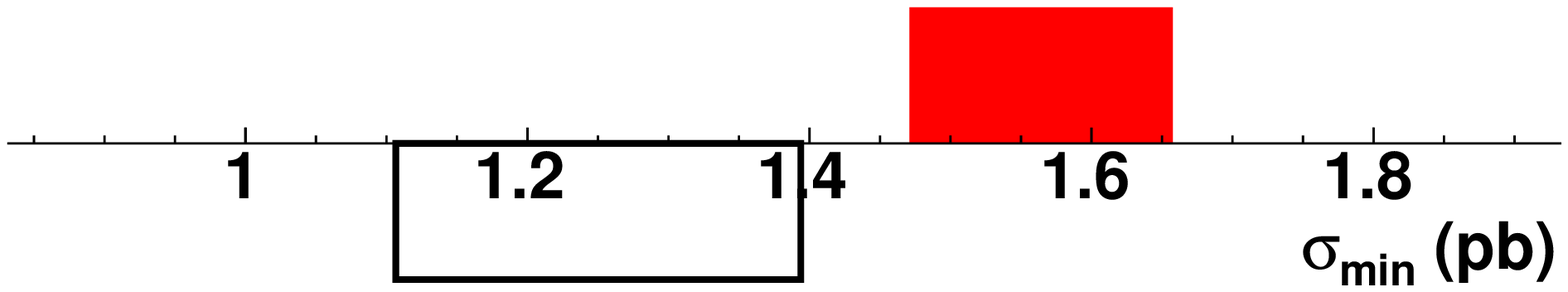}} \\ 
  {3}
             & $Z' \to q\bar{q}$, $m_{Z'}=700$~GeV, with standard model couplings to quarks.
  &   \raisebox{-0.5\height}{\includegraphics[width=7.5cm]{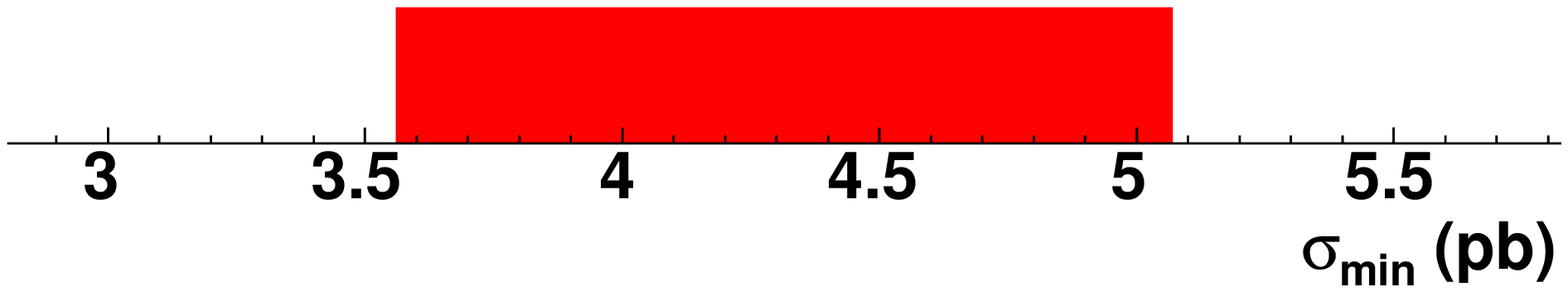}} \\ 
  {4}
             & $Z'\to q\bar{q}$, $m_{Z'}=1$~TeV, with standard model couplings to quarks.
  &   \raisebox{-0.5\height}{\includegraphics[width=7.5cm]{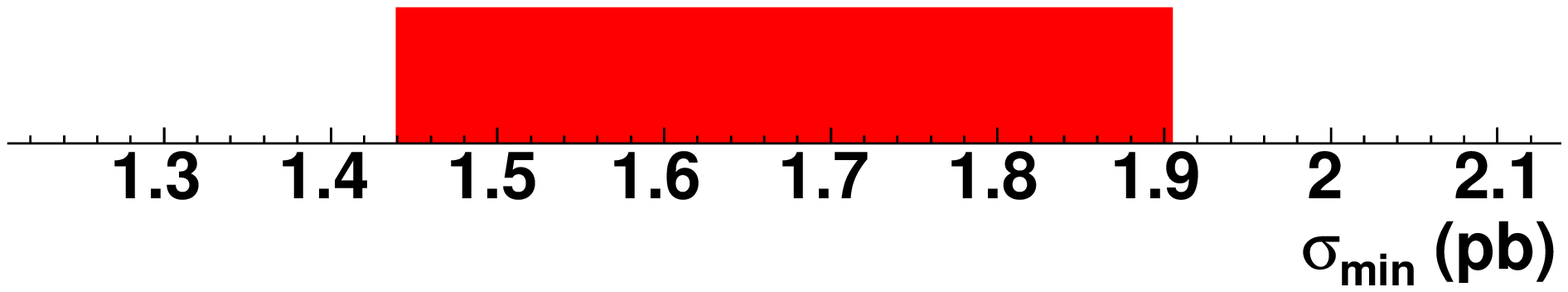}} \\ 
  {5}
             &  $Z'\to t\bar{t}$, $m_{Z'}=500$~GeV, with standard model couplings to $t\bar{t}$.
  &   \raisebox{-0.5\height}{\includegraphics[width=7.5cm]{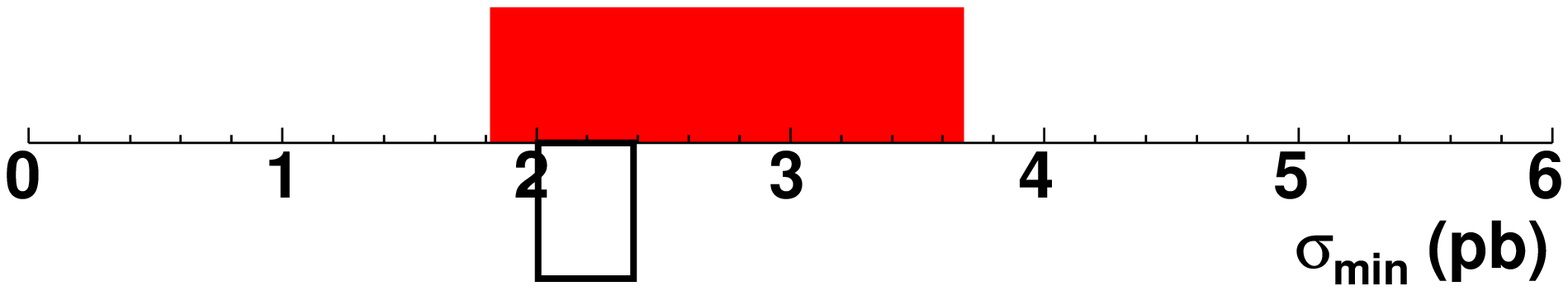}} \\ 
\end{tabular}
\end{minipage}
\caption[Summary of \Sleuth's sensitivity to several new physics models.]{Summary of \Sleuth's sensitivity to several new physics models, expressed in terms of the minimum production cross section needed for discovery with 927~pb$^{-1}$.  Where available, a comparison is made to the sensitivity of a dedicated search for this model.  The solid (red) box represents \Sleuth's sensitivity, and the open (white) box represents the sensitivity of the dedicated analysis. Systematic uncertainties are not included in the sensitivity calculation. The width of each box shows typical variation under fluctuation of data statistics.  In Models 3 and 4, there is no targeted analysis available for comparison.  
}
\label{tab:sensitivitySummary}
\end{table*}

\subsection{Sensitivity}
\label{sec:sleuthSensitivity}

Two important questions must be asked:
\begin{itemize}
\item Will \Sleuth\ find nothing if there is nothing to be found?
\item Will \Sleuth\ find something if there is something to be found?
\end{itemize}

If there is nothing to be found, \Sleuth\ will find nothing 999 times out of 1000, given a uniform distribution of $\twiddleScriptP$ and a discovery threshold of $\twiddleScriptP \lesssim 0.001$.  The uniform distribution of $\twiddleScriptP$ in the absence of new physics is illustrated in Fig.~\ref{fig:tildeScriptPdistribution}.  \Sleuth\ will of course return spurious signals if provided improperly modeled backgrounds.  The algorithm directly addresses the issue of whether an observed hint is due to a statistical fluctuation.  \Sleuth\ itself is unable to address systematic mismeasurement or incorrect modeling, but is useful in bringing these to attention.

The answer to the second question depends on the degree to which the new physics satisfies the three assumptions on which \Sleuth\ is based:  new physics will appear predominantly in one final state, at high summed scalar transverse momentum, and as an excess of data over Standard Model prediction.

\subsubsection{Known Standard Model processes} 

Consideration of specific Standard Model processes can provide intuition for \Sleuth's sensitivity to new physics.  This section tests \Sleuth's sensitivity to the production of top quark pairs, $W$ boson pairs, single top, and the Higgs boson.

\paragraph{Top quark pairs.}
Top quark pair production results in two $b$ jets and two $W$ bosons, each of which may decay leptonically or hadronically.  The $W$ branching ratios are such that this signal predominantly populates the \Sleuth\ final state $Wb\bar{b}jj$, where ``$W$'' denotes an electron or muon and significant missing momentum.  Although the final states $\ell^+\ell^-\pmiss b\bar{b}$ were important in verifying the top quark pair production hypothesis in the initial observation by CDF~\cite{CDFTopDiscovery:Abe:1995hr} and \DZero~\cite{D0TopDiscovery:Abachi:1995iq} in 1995, most of the statistical power came from the final state $Wb\bar{b}jj$.  The fully hadronic decay into $b\bar{b}\,4j$ has only convincingly been seen after integrating substantial Run II luminosity~\cite{TopAllHadronic:Aaltonen:2006xc}.  \Sleuth's first assumption that new physics will appear predominantly in one final state is thus reasonably well satisfied.  Since the top quark has a mass of $170.9\pm1.8$~GeV~\cite{TopQuarkMass:unknown:2007bx}, the production of two such objects leads to a signal at large \SumPt\ relative to the Standard Model background of $W$ bosons produced in association with jets, satisfying \Sleuth's second and third assumptions.  \Sleuth\ is expected to perform reasonably well on this example.  

To quantitatively test \Sleuth's sensitivity to top quark pair production, this process is removed from the Standard Model prediction, and the correction factors are re-obtained from a global fit assuming ignorance of $t\bar{t}$ production.  \Sleuth\ easily discovers $t\bar{t}$ production in 927~pb$^{-1}$ in the final states $b\bar{b}\ell^+\ell'^-\pmiss$ and $Wb\bar{b}jj$, shown in Fig.~\ref{fig:topless_SM_sensitivityTest}.  \Sleuth\ finds $\scriptP_{b\bar{b}\ell^+\ell'^-\pmiss}<1.5\times10^{-8}$ and $\scriptP_{Wb\bar{b}jj}<8.3\times10^{-7}$, far surpassing the discovery threshold of $\twiddleScriptP \lesssim 0.001$.

The test is repeated as a function of assumed integrated luminosity (Fig.~\ref{fig:tildeScriptPvsLuminosityTtbar}), and  \Sleuth\ is found to highlight the top quark signal at an integrated luminosity of roughly $80\pm60$~pb$^{-1}$, where the large variation arises from statistical fluctuations in the $t\bar{t}$ signal events.  Weaker constraints on the \Vista\ correction factors at lower integrated luminosity marginally increase the integrated luminosity required to claim a discovery.

\paragraph{$W$ boson pairs.}

The sensitivity to Standard Model $WW$ production is tested by removing this process from the Standard Model background prediction and allowing the \Vista\ correction factors to be re-fit.  \highlight{In 927~pb$^{-1}$ of Tevatron Run II data, \Sleuth\ identifies an excess in the final state $\ell^+{\ell'}^-\pmiss$, consisting of an electron and muon of opposite sign and missing momentum.}  This excess corresponds to $\twiddleScriptP < 2 \times 10 ^{-4}$, sufficient for the discovery of $WW$, as shown in Fig.~\ref{fig:VVless_SM_sensitivityTest}. 

\paragraph{Single top.}
Single top quarks are produced weakly, either through a $t$-channel process like $b u \to t d \to Wb+jet$, or through a $s$-channel, such as $u\bar{d}\to W^+ \to t \bar{b} \to Wb\bar{b}$.  Both of these final states are merged into \Sleuth's $Wb\bar{b}$ final state, satisfying \Sleuth's first assumption.
Single top production will appear as an excess of events, satisfying \Sleuth's third assumption.  \Sleuth's second assumption is not well satisfied for this example, since single top production does not lie at large \SumPt\ relative to other Standard Model processes.  \Sleuth\ is thus expected to be outperformed by a targeted search in this example.  

\paragraph{Higgs boson.}
Assuming a Standard Model Higgs boson of mass $m_h=115$~GeV, the dominant observable production mechanism is $p\bar{p}\rightarrow Wh$ and $p\bar{p}\rightarrow Zh$, populating the final states $Wb\bar{b}$, $\ell^+\ell^- b\bar{b}$, and $\pmiss\,b\bar{b}$.  The signal is thus spread over three \Sleuth\ final states.  Events in the last of these ($\pmiss\,b\bar{b}$) do not pass the \Vista\ event selection, which does not use $\pmiss$ as a trigger object.  \Sleuth's first assumption is thus poorly satisfied for this example.
The Standard Model Higgs boson signal will appear as an excess, but as in the case of single top production it does not appear at particularly large \SumPt\ relative to other Standard Model processes.  Since the Standard Model Higgs boson poorly satisfies \Sleuth's first and second assumptions, a targeted search for this specific signal is expected to outperform \Sleuth.  

\subsubsection{Specific models of new physics}
\label{sec:SleuthSensitivity:SpecificModels}

To build intuition for \Sleuth's sensitivity to new physics signals, several sensitivity tests are conducted for a variety of new physics possibilities.  Some of the new physics models chosen have already been considered by more specialized analyses within CDF, making possible a comparison between \Sleuth's sensitivity and the sensitivity of these previous analyses.  

\Sleuth's sensitivity can be compared to that of a dedicated search by determining the minimum new physics cross section $\sigma_\text{min}$ required for a discovery by each.  The discovery for \Sleuth\ occurs when $\tildeScriptP < 0.001$.  \highlight{In most \Sleuth\ regions satisfying the discovery threshold of $\tildeScriptP < 0.001$, the probability for the predicted number of events to fluctuate up to or above the number of events observed corresponds to greater than $5\sigma$.}  The discovery for the dedicated search occurs when the observed excess of data corresponds to a $5\sigma$ effect.  Smaller $\sigma_\text{min}$ corresponds to greater sensitivity.

The sensitivity tests are performed by first generating pseudo data from the Standard Model background prediction.  Signal events for the new physics model are generated, passed through the chain of CDF detector simulation and event reconstruction, and consecutively added to the pseudo data until \Sleuth\ finds $\tildeScriptP<0.001$.  The number of signal events needed to trigger discovery is used to calculate $\sigma_\text{min}$.

For each dedicated analysis to which \Sleuth\ is compared, the number of Standard Model events expected in 927~pb$^{-1}$ within the region targeted is used to calculate the number of signal events required in that region to produce a discrepancy corresponding to $5\sigma$.  Using the signal efficiency determined in the dedicated analysis, $\sigma_\text{min}$ is calculated.  The effect of systematic uncertainties is not included in \Sleuth, so it is also removed from the dedicated analyses.

The results of five such sensitivity tests are summarized in Table~\ref{tab:sensitivitySummary}.  \Sleuth\ is seen to perform comparably to targeted analyses on models satisfying the assumptions on which \Sleuth\ is based.  For models in which \Sleuth's simple use of $\SumPt$ can be improved upon by optimizing for a specific feature, a targeted search may be expected to achieve greater sensitivity.  One of the important features of \Sleuth\ is that it not only performs reasonably well, but that it does so broadly.  In Model 1, a search for a particular model point in a gauge mediated supersymmetry breaking (GMSB) scenario, \Sleuth\ gains an advantage by exploiting a final state not considered in the targeted analysis~\cite{Acosta:2004sb}.  In Model 2, a search for a $Z'$ decaying to lepton pairs, the targeted analysis~\cite{SamHarper:Abulencia:2006iv} exploits the narrow resonance in the $e^+e^-$ invariant mass.  In Models 3 and 4, which are searches for a hadronically decaying $Z'$ of different masses, there is no targeted analysis against which to compare.  In Model 5, a search for a $Z'\to t\bar{t}$ resonance, the signal appears at large summed scalar transverse momentum in a particular final state, resulting in comparable sensitivity between \Sleuth\ and the targeted analysis~\cite{JacoZprimettbar2}.

\begin{figure}
\centering
\begin{tabular}{c}
\includegraphics[width=2.5in,angle=270]{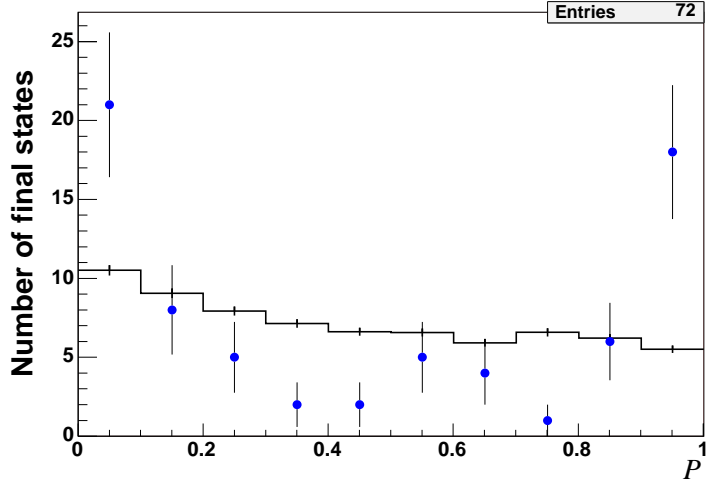} \\
\end{tabular}
\caption[Distribution of \scriptP.]{
{\em Blue points:} The \scriptP\ distribution observed in 927~pb$^{-1}$, with one entry for each of the 72 \Sleuth\ final states with at least 3 data.  There are 131 \Sleuth\ final states with non-zero background and less than 3 data, which are assigned $\scriptP=1$.
{\em Black histogram:} The expected \scriptP\ distribution from all 203 \Sleuth\ final states with non-zero background, if instead of actual data we use pseudo-data pulled from the expected \sumPt\ distribution of each final state, and omit the final states where pseudo-data are less than 3 and therefore have $\scriptP=1$.  As explained in Sec.~\ref{sec:Sleuth:Regions}, footnote~\ref{footnote:scriptPdistribution}, the \scriptP\ of final states with expected population $\lesssim 10$ is not uniformly distributed.  Of the 203 final states \Sleuth\ considers in 927~pb$^{-1}$, 150 have Standard Model background of less than 10 events, which causes the expected \scriptP\ distribution to slightly favor smaller values.
}
\label{fig:scriptPsPlots}
\end{figure}

\begin{figure*}
\centering
\begin{tabular}{cc}
\includegraphics[width=2.2in,angle=270]{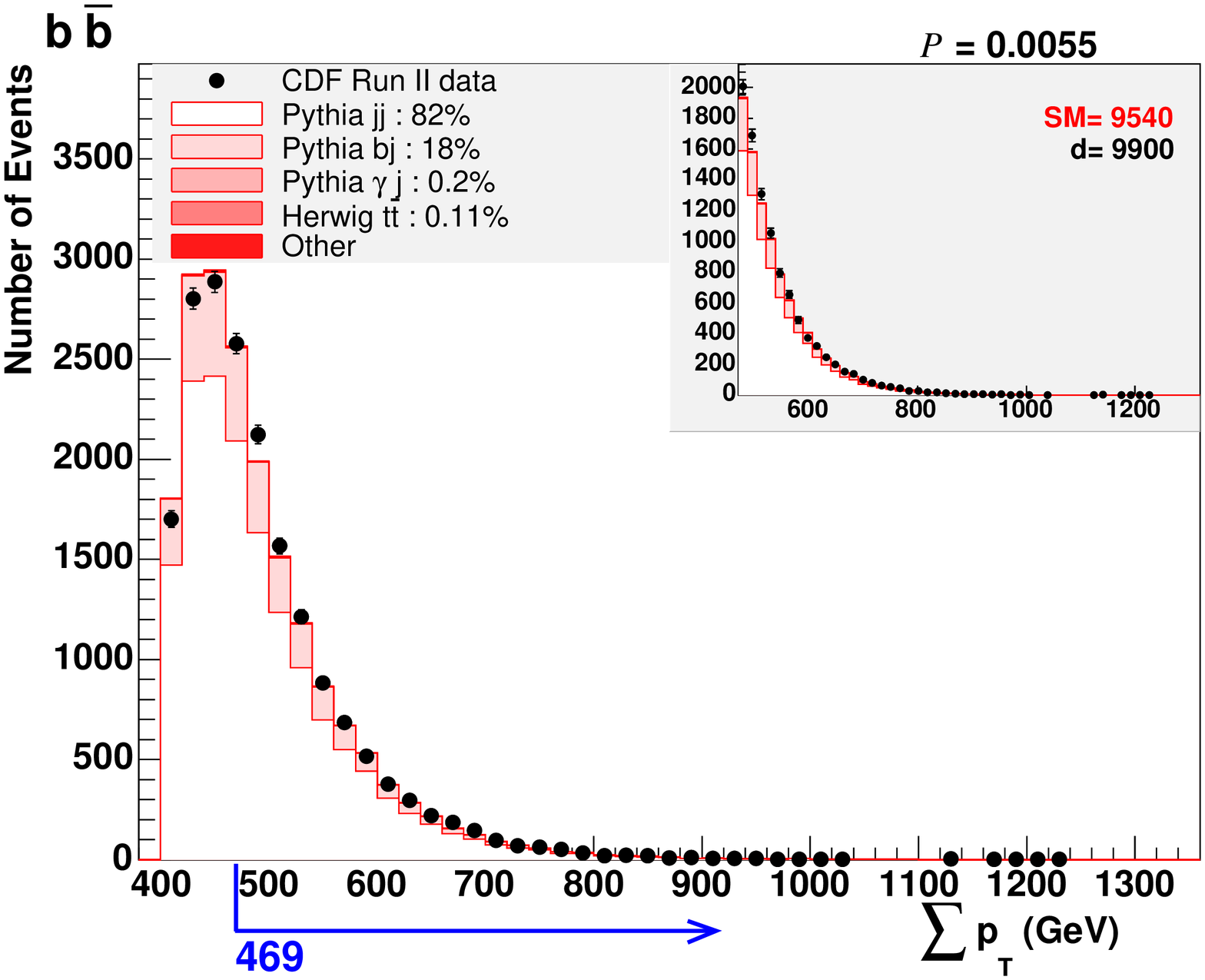} &
\includegraphics[width=2.2in,angle=270]{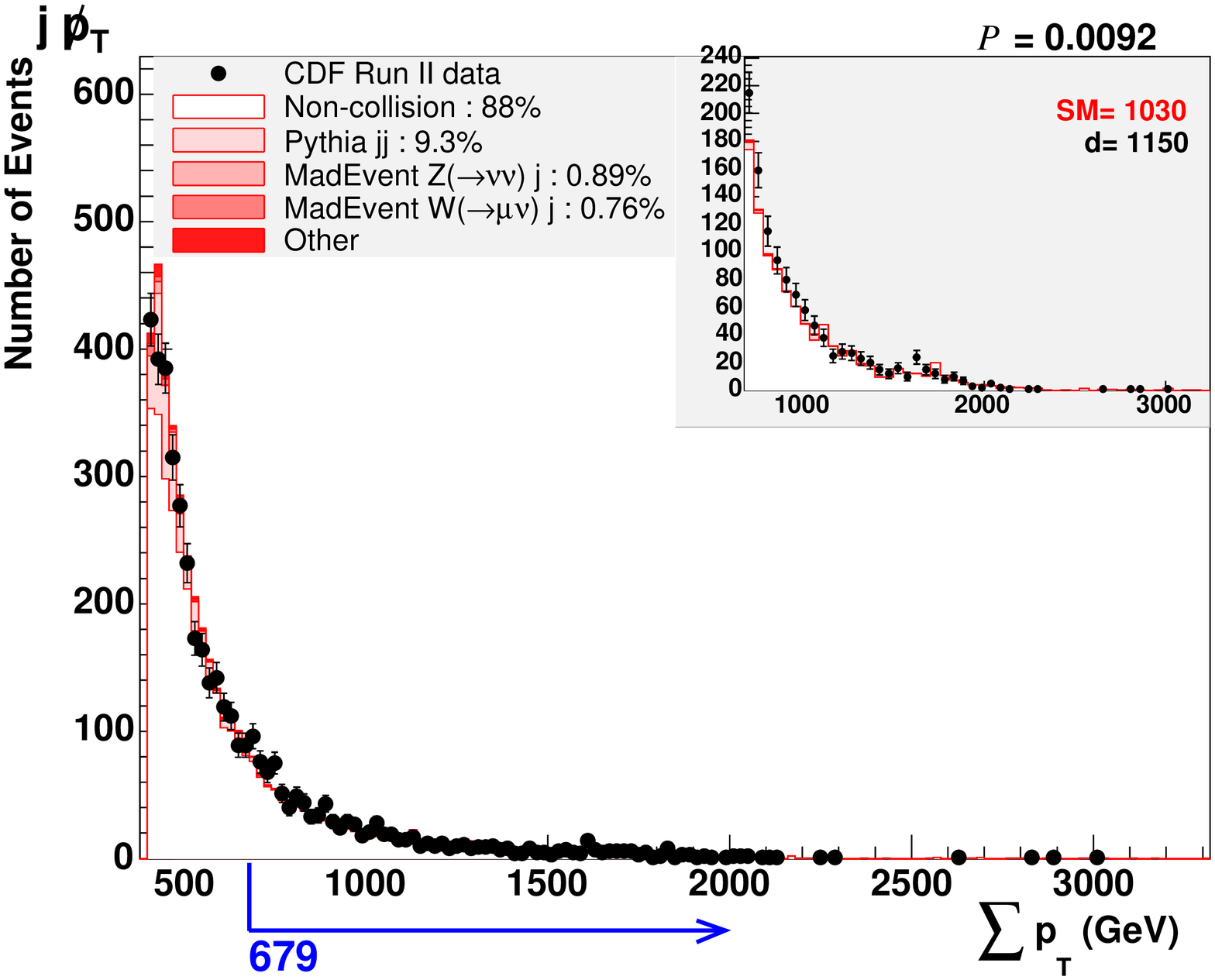} \\
\includegraphics[width=2.2in,angle=270]{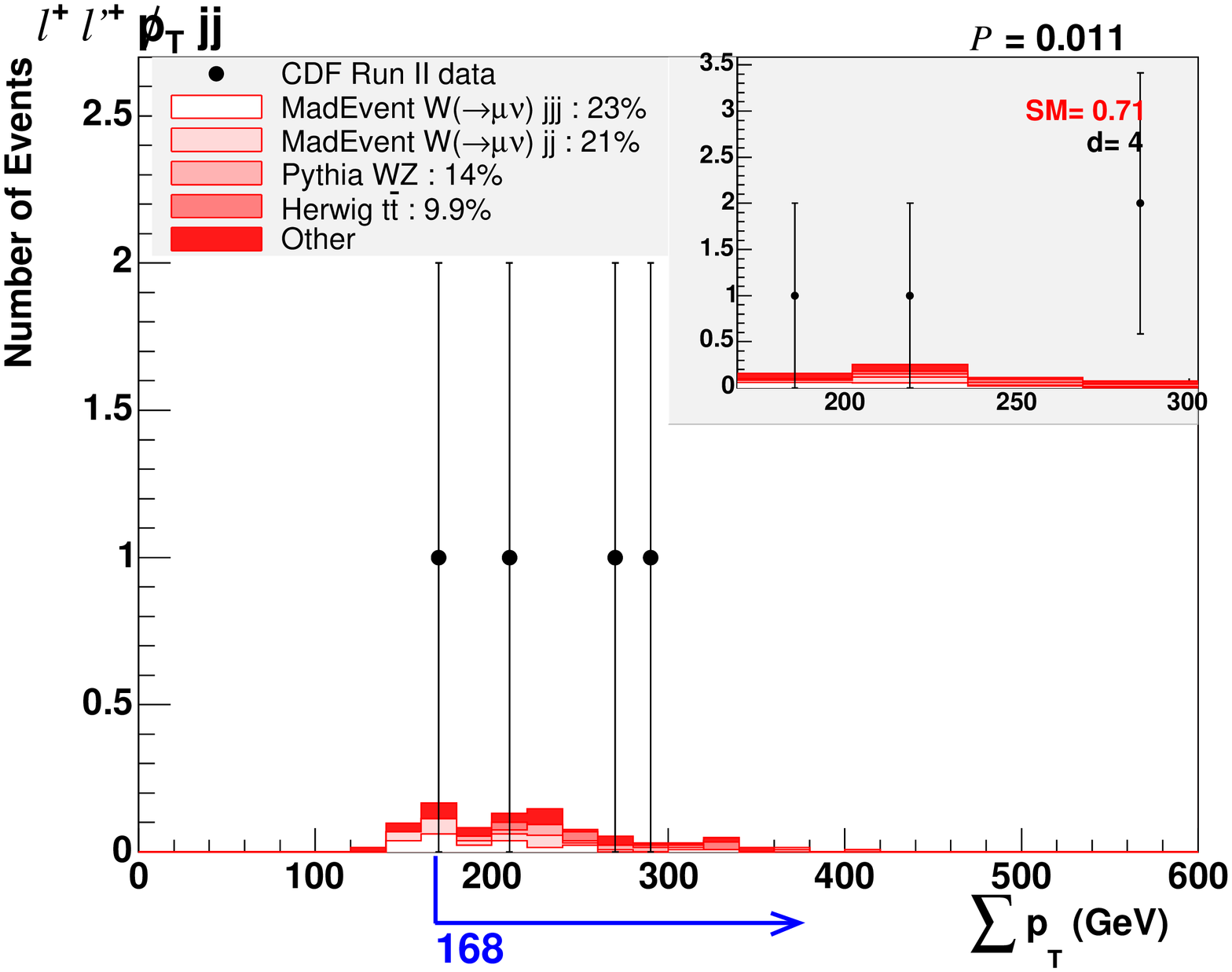} &
\includegraphics[width=2.2in,angle=270]{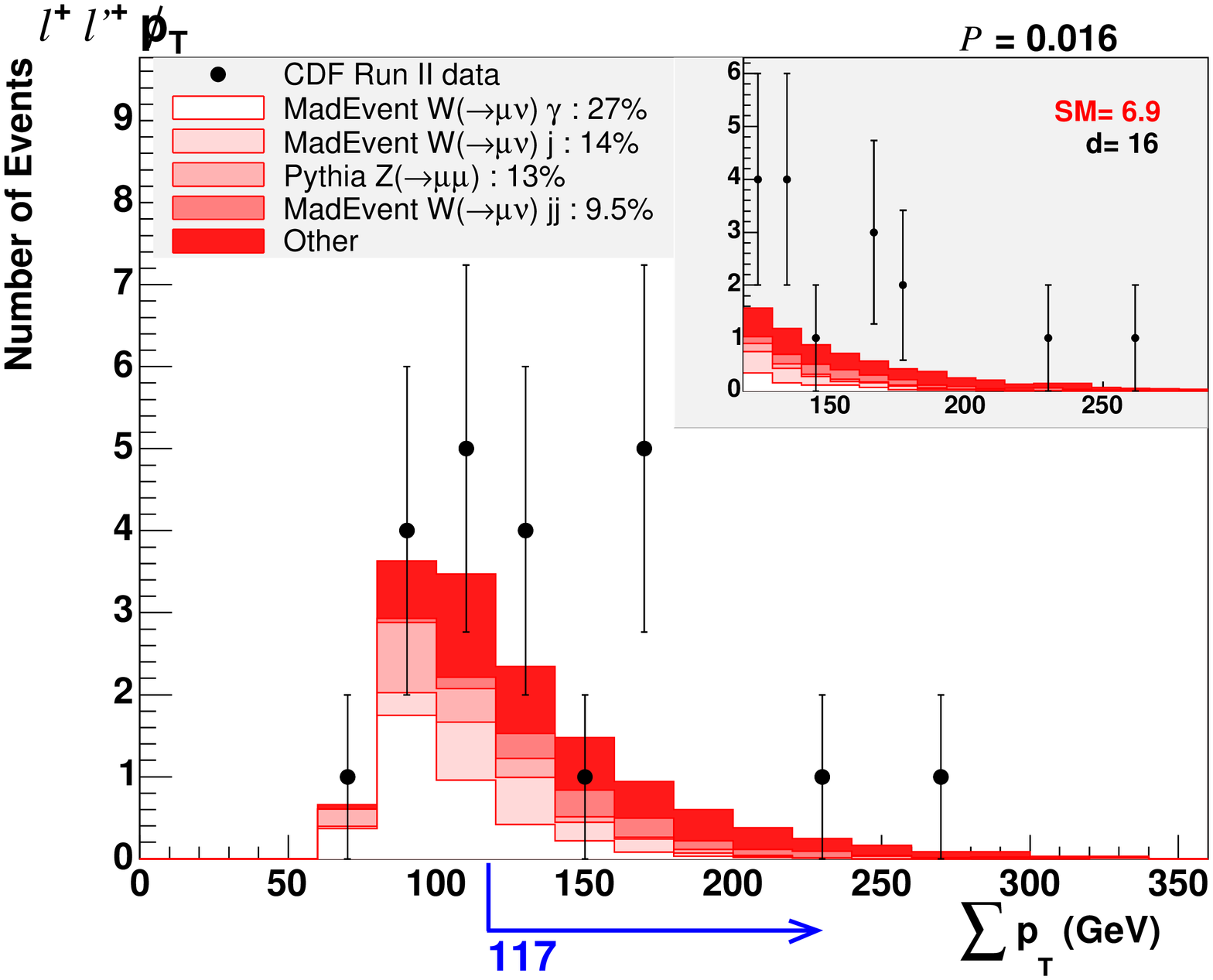} \\
\includegraphics[width=2.2in,angle=270]{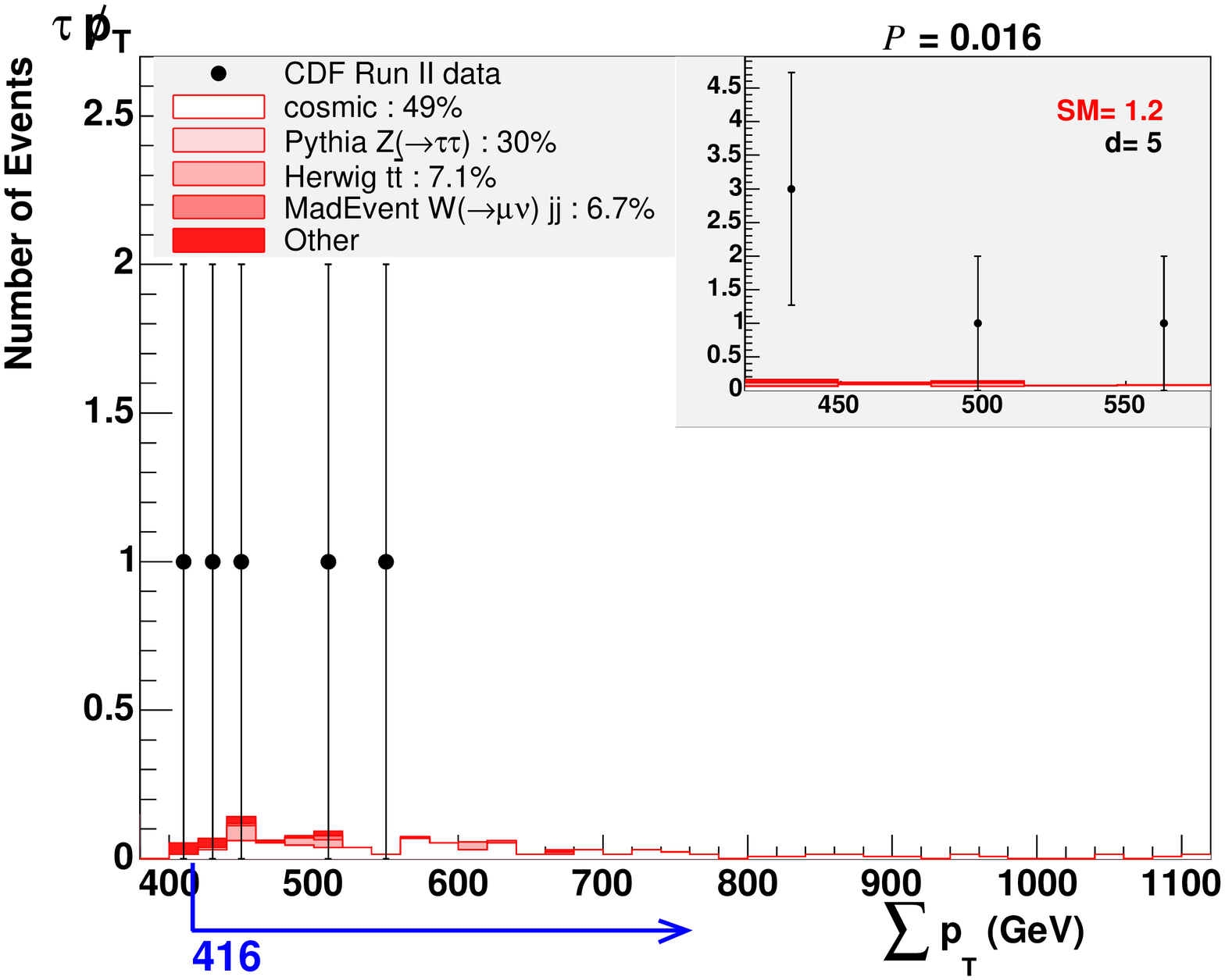} &
\includegraphics[width=2.2in,angle=270]{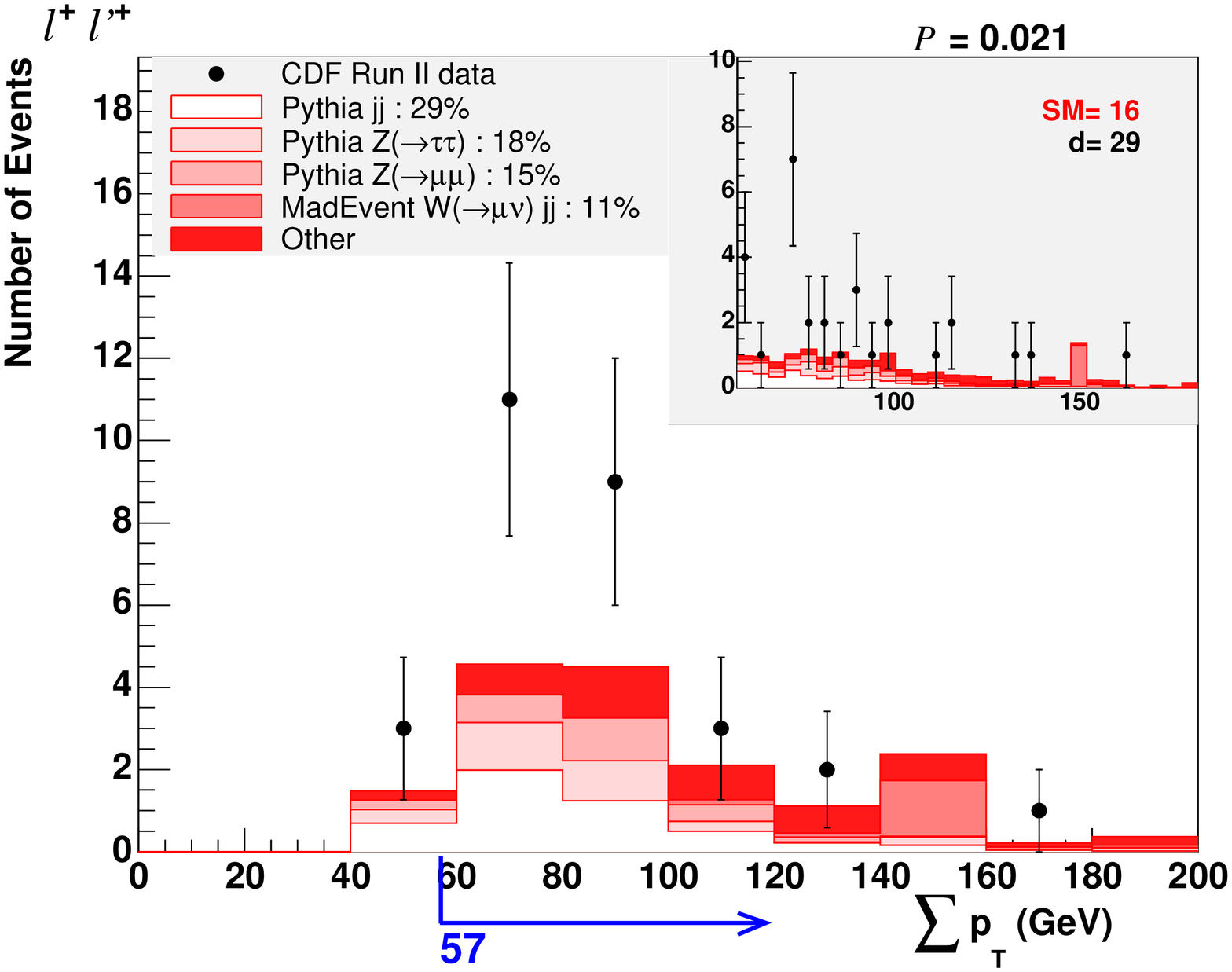} \\
\end{tabular}
\caption[The most interesting final states identified by \Sleuth.]{The most interesting final states identified by \Sleuth.  The region chosen by \Sleuth, extending up to infinity, is shown by the (blue) arrow just below the horizontal axis.  Data are shown as filled circles, and the Standard Model prediction is shown as the shaded histogram.  The \Sleuth\ final state is labeled in the upper left corner of each panel, with $\ell$ denoting $e$ or $\mu$, and $\ell^+\ell'^+$ denoting an electron and muon with the same electric charge.  The number at upper right in each panel shows \scriptP, defined in Sec.~\ref{sec:Sleuth:Regions}.  The inset in each panel shows an enlargement of the region selected by \Sleuth, together with the number of events (${\text{SM}}$) predicted by the Standard Model in this region, and the number of data events ($d$) observed in this region.
\label{fig:SleuthPlots}}
\end{figure*}


\subsection{Results}
\label{sec:Sleuth:Results}

The distribution of \scriptP\ for the final states considered by \Sleuth\ in the data is shown in Fig.~\ref{fig:scriptPsPlots}.  The concavity of this distribution reflects the degree to which the correction model described in Sec.~\ref{sec:Vista:CorrectionModel} has been tuned.  A crude correction model tends to produce a distribution that is concave upwards, as seen in this figure, while an overly tuned correction model produces a distribution that is concave downwards, with more final states than expected having $\scriptP$ near the midpoint of the unit interval.

The most interesting final states identified by \Sleuth\ are shown in Fig.~\ref{fig:SleuthPlots}, together with a quantitative measure (\scriptP) of the interest of the most interesting region in each final state, determined as described in Sec.~\ref{sec:Sleuth:Regions}.  The legends of Fig.~\ref{fig:SleuthPlots} show the primary contributing Standard Model processes in each of these final states, together with the fractional contribution of each.  The top six final states, which correspond to entries in the leftmost bin in Fig.~\ref{fig:scriptPsPlots}. span a range of populations, relevant physics objects, and important background contributions.  This picture is suggestive of statistical fluctuations, spread among unrelated final states.

The final state $b\bar{b}$, consisting of two or three reconstructed jets, one or two of which are $b$-tagged, heads the list.  These events enter the analysis by satisfying the \Vista\ offline selection requiring one or more jets or $b$-jets with $p_T>200$~GeV.  The definition of \Sleuth's \SumPt\ variable is such that all events in this final state consequently have $\SumPt>400$~GeV.  \Sleuth\ chooses the region $\SumPt>469$~GeV, which includes nearly $10^4$ data events.  \highlight{The Standard Model prediction in this region is sensitive to the $b$-tagging efficiency $\poo{b}{b}$ and the fake rate $\poo{j}{b}$, which have few strong constraints on their values for jets with $p_T>200$~GeV other than those imposed by other \Vista\ kinematic distributions within this and a few other related final states.}  For this region \Sleuth\ finds $\scriptP_{b\bar{b}}=0.0055$, which is unfortunately not statistically significant after accounting for the trials factor associated with looking in many different final states, as discussed below.

The final state $j\pmiss$, consisting of events with one reconstructed jet and significant missing transverse momentum, is the second final state identified by \Sleuth.  The primary background is due to non-collision processes, including cosmic rays and beam halo backgrounds, whose estimation is discussed in Appendix~\ref{sec:CorrectionModelDetails:CosmicRays}.  Since the hadronic energy is not required to be deposited in time with the beam crossing, \Sleuth's analysis of this final state is sensitive to particles with a lifetime between 1~ns and 1~$\mu$s that lodge temporarily in the hadronic calorimeter, complementing Ref.~\cite{Hugo:Abulencia:2006kk}.

The final states $\ell^+ {\ell'}^{+} \pmiss jj$, $\ell^+ {\ell'}^{+} \pmiss$, and $\ell^+ {\ell'}^{+}$ all contain an electron ($\ell$) and muon ($\ell'$) with identical reconstructed charge (either both positive or both negative).  The final states with and without missing transverse momentum are qualitatively different in terms of the Standard Model processes contributing to the background estimate, with the final state $\ell^+ {\ell'}^{-}$ composed mostly of dijets where one jet is misreconstructed as an electron and a second jet is misreconstructed as a muon; $Z\rightarrow \tau^+\tau^-$, where one tau decays to a muon and the other to a leading $\pi^0$, one of the two photons from which converts while traveling through the silicon support structure to result in an electron reconstructed with the same sign as the muon, as described in Appendix~\ref{sec:MisidentificationMatrix}; and $Z\rightarrow\mu^+\mu^-$, in which a photon is produced, converts, and is misreconstructed as an electron.  The final states containing missing transverse momentum are dominated by the production of $W(\rightarrow\mu\nu)$ in association with one or more jets, with one of the jets misreconstructed as an electron.  The muon is significantly more likely than the electron to have been produced in the hard interaction, since the fake rate $\poo{j}{\mu}$ is roughly an order of magnitude smaller than the fake rate $\poo{j}{e}$, as observed in Table~\ref{tbl:CorrectionFactorDescriptionValuesSigmas}.  The final state $\ell^+ {\ell'}^{-} \pmiss jj$, which contains two or three reconstructed jets in addition to the electron, muon, and missing transverse momentum, also has some contribution from $WZ$ and top quark pair production.

The final state $\tau\pmiss$ contains one reconstructed tau, significant missing transverse momentum, and one reconstructed jet with $p_T>200$~GeV.  This final state in principle also contains events with one reconstructed tau, significant missing transverse momentum, and zero reconstructed jets, but such events do not satisfy the offline selection criteria described in Sec.~\ref{sec:Vista:OfflineTrigger}.  Roughly half of the background is non-collision, in which two different cosmic ray muons (presumably from the same cosmic ray shower) leave two distinct energy deposits in the CDF hadronic calorimeter, one with $p_T>200$~GeV, and one with a single associated track from a $p\bar{p}$ collision occurring during the same bunch crossing.  Less than a single event is predicted from this non-collision source (using techniques described in Appendix~\ref{sec:CorrectionModelDetails:CosmicRays}) over the past five years of Tevatron running.

In these CDF data, \Sleuth\ finds $\twiddleScriptP = 0.46$.  The fraction of hypothetical similar CDF experiments (assuming a fixed Standard Model prediction, detector simulation, and correction model) that would exhibit a final state with \scriptP\ smaller than the smallest \scriptP\ observed in the CDF Run II data is approximately 46\%.  The actual value obtained for $\tildeScriptP$ is not of particular interest, except to note that this value is significantly greater than the threshold of $\lesssim 0.001$ required to claim an effect of statistical significance.  \Sleuth\ has not revealed a discrepancy of sufficient statistical significance to justify a new physics claim.\footnote{The alternative statistic, \tildePval, was found to be 22\%.  The region with the smallest \pval\ is in the final state $b\bar{b}$, which also has the smallest \scriptP.  Therefore, the most interesting region pointed by both statistics is the same: $\sumPt \ge 469$ in $b\bar{b}$.}

Systematics are incorporated into \Sleuth\ in the form of the flexibility in the \Vista\ correction model, as described previously.  This flexibility is significantly more important in practice than the uncertainties on particular correction factor values obtained from the fit.
The inclusion of additional systematic uncertainties would not qualitatively change the conclusion that \Sleuth\ has not revealed a discrepancy of sufficient statistical significance to justify a new physics claim.

\begin{figure}
\centering
\includegraphics[width=4in,angle=0]{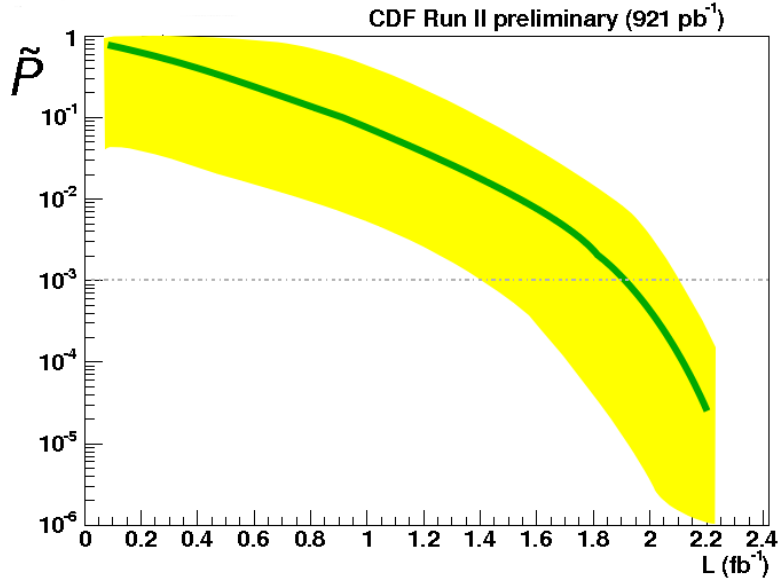}
\caption[Projection of \tildeScriptP\ towards lower and higher luminosities.]{Projection of \tildeScriptP\ towards lower and higher luminosities, starting from 927~pb$^{-1}$.  Values were obtained by scaling down or up both data and backgrounds.  The yellow band reflects uncertainty due to randomness in which of the present data events would have appeared in less data, or would recur in more.  The Standard Model implementation is assumed invariant in all except total populations.}
\label{fig:tildeScriptPvsLuminosity}
\end{figure}

Starting from the current result of \Sleuth\ in 927~pb$^{-1}$, a projection (Fig.~\ref{fig:tildeScriptPvsLuminosity}) shows that, if the dataset roughly doubles and nothing changes in the Standard Model implementation, then \tildeScriptP\ will likely be smaller than discovery threshold.  This implies that, either we are on the verge of a discovery that will happen with more data, or a doubling of data will likely enforce some more accurate modeling of Standard Model backgrounds, which will possibly increase \tildeScriptP\ away from its predicted small value.  This clue was the main motivation to repeat and improve this search with more data, as will be described in a later chapter.

\section{Summary of first round with 1~fb$^{-1}$}
\label{sec:Conclusions}

In the first round of this analysis, with 927~pb$^{-1}$,  a complete Standard Model background estimate has been obtained and compared with data in 344 populated exclusive final states and 16,486 relevant kinematic distributions.  Consideration of exclusive final state populations yields no statistically significant ($>3\sigma$) discrepancy after the trials factor is accounted for.  Quantifying the difference in shape of kinematic distributions using the Kolmogorov-Smirnov statistic, significant discrepancies are observed between data and Standard Model prediction.  These discrepancies are believed to arise from mismodeling of the parton shower and intrinsic $k_T$, and represent observables for which a QCD-based understanding is highly motivated.  None of the shape discrepancies highlighted motivates a new physics claim.

A further systematic search (\Sleuth) for regions of excess on the high-$\SumPt$ tails of exclusive final states has been performed, representing a quasi-model-independent search for new electroweak scale physics.  A measure of interest rigorously accounting for the trials factor associated with looking in many regions with few events is defined, and used to quantify the most interesting region observed in the CDF Run II data.  No region of excess on the high-$\SumPt$ tail of any of the \Sleuth\ exclusive final states surpasses the discovery threshold.


Although this result of course can not prove that no new physics is hiding in the studied data, this search is the most encompassing test of the Standard Model at the energy frontier.

\cdfSpecific{

\memo{OLD ATTEMPT}
\section{MC Production and Detector Simulation}

About $4.5\times10^{7}$ MC events of different weights compose the Standard Model background.  The weight\index{weight!uncorrected} of each MC event before the implementation of the correction model is equal to $\sigma_{{\rm proc}}(1{\rm pb}^{-1})/N_{{\rm MC}}$, where $\sigma_{{\rm proc}}$ is the cross section of the process and $N_{{\rm MC}}$ the population of events generated.  

The MC event generators used for the production of these events are \Pythia~\cite{Pythia}, \MadEvent~\cite{MadEvent}, \Alpgen~\cite{ALPGEN}, and \Herwig~\cite{Herwig}.  \MadEvent\ and \Alpgen\ perform an exact leading order matrix element calculation, and provide 4-vector information corresponding to the outgoing legs of the underlying Feynman diagrams, together with color flow information.  \Pythia\ 6.218 is used to handle showering and fragmentation.  The CTEQ5L~\cite{CTEQ5L} parton distribution functions are used.  

All MC events must pass through \CdfSim, which costs about 10 seconds per event.  Certain MC samples that the Collaboration has already generated and passed through \CdfSim\ are used.  These samples include:
\begin{itemize}
\item QCD jets: QCD dijet and multijet production are estimated using \Pythia.  Samples are generated with Tune A~\cite{RickFieldPythiaTunes} with lower cuts on $\hat{p}_T$, the transverse momentum of the scattered partons in the center of momentum frame of the incoming partons, of 0, 10, 18, 40, 60, 90, 120, 150, 200, 300, and 400~GeV.  These samples are combined to provide a complete estimation of QCD jet production, using the sample with greatest statistics in each range of $\hat{p}_T$.
\item $\gamma$+jets: The estimation of QCD single prompt photon production comes from \Pythia.  Five samples are generated with Tune A corresponding to lower cuts on $\hat{p}_T$ of 8, 12, 22, 45, and 80~GeV.  These samples are combined to provide a complete estimation of single prompt photon production in association with one or more jets, placing cuts on $\hat{p}_T$ to avoid double counting.  
\item $\gamma\gamma$+jets: QCD diphoton production is estimated using \Pythia.
\item $V$+jets: The estimation of $V$+jets processes (with $V$ denoting $W$ or $Z$), where the $W$ or $Z$ decays to first or second generation leptons, comes from \Alpgen, with \Pythia\ employed for showering.  Tune AW~\cite{RickFieldPythiaTunes} is used within \Pythia\ for these samples.
\item $VV$+jets: The estimation of $WW$, $WZ$, and $ZZ$ production with zero or more jets comes from \Pythia.  
\item $V\gamma$+jets: The estimation of $W\gamma$ and $Z\gamma$ production comes from \MadEvent, with showering provided by \Pythia.  These samples are inclusive in the number of jets.
\item $W\rightarrow\tau\nu$ and $Z\rightarrow\tau\tau$ with zero or more jets come from \Pythia.
\item $t\bar{t}$: Top quark pair production is estimated using \Herwig, assuming a top quark of mass 175~GeV and NNLO theoretical cross section $6.77\pm 0.42$~pb~\cite{Kidonakis:2003qe}.
\end{itemize}

Backgrounds from cosmic rays or beam halo that interacts with the hadronic or electromagnetic calorimeter, producing objects that look like a photon or jet, are estimated using a sample of data events containing fewer than three reconstructed tracks. This procedure is described in more detail in Sec.~\ref{sec:CosmicRays}. 

In each bunch crossing it can happen to have multiple $p\bar{p}$ collisions, appart from the one(s) that produce high-$p_T$ objects that make the event interesting.  This experimental complication needs to be modelled.  Minimum bias events\footnote{Minimum bias events can mean different things in different contexts.  Here, they are the kind of events we would be typically recording if we didn't have to impose online selection, but instead recorded all events produced.  Minimum bias events consist of low-$p_T$ jets.} are overlaid according to run-dependent instantaneous luminosity in some of the Monte Carlo samples, including those used for inclusive $W$ and $Z$ production.  In MC samples not containing overlaid minimum bias events, additional unclustered momentum is added to events to mimic the effect of the majority of multiple interactions, in which a soft dijet event accompanies the rare hard scattering of interest.  A random number is drawn from a Gaussian with mean 0 and standard deviation 1.5 GeV for each of the $x$ and $y$ components of the added unclustered momentum.  

Small backgrounds due to two rare hard scatterings occurring in the same bunch crossing are estimated by forming overlaps of events, as described in Appendix~\ref{sec:Overlaps}.

\section{Object Identification}

In \Stntuple\ format each event contains blocks\index{Stntuple@\Stntuple!blocks} of information.  These blocks contain candidate electrons, muons, jets etc., and additional information such as reconstructed tracks, secondary decay vertices, multiple interaction vertices etc.  One needs to loop through these blocks and keep any candidates that pass quality cuts.  A qualitative description of object ID criteria is given here, while exact criteria are listed in Appendix~\ref{sec:objectID}.

\paragraph{Central electrons} Energy deposited in the CEM and one associated COT track.  The CHA energy is required to be much less than the CEM in the same tower, to distinguish from charged hadrons.  The EM shower profile at the CES is required to be consistent with that of an electron.  Isolation\footnote{The calorimetric energy around a track.} has to be less than an upper limit, to reduce the influence from electrons that are produced inside jets\footnote{Or in processes such as $q\xrightarrow{\rm hadr.}\pi^0\to\gamma\gamma\to e^+e^-\gamma \to$ reconstructed $e^\pm$ if one of the two electrons is missed and the other is near the photon in the calorimeter.}.
\paragraph{Plug electrons} Similar to central electrons, except that the Phoenix algorithm\cite{phoenixElectrons} is required to have reconstructed a track associated with the PEM tower, and PES is used instead of CES.
\paragraph{Central photons} Similar to central electron, but the CEM tower must have no tracks of significant $p_T$ around it, or if there is one track near it it is required to be very soft. The CES EM shower profile is required to be consistent with photon.
\paragraph{Plug photons} Similar to central photon, but using PEM and PES.
\paragraph{Muons} A COT track is required, with upper cuts in the impact parameter ($d_0$) and in the track reconstruction $\chi^2$ to eliminate ``kinks'' resulting presumably from $\pi^\pm \rightarrow \mu^\pm\nu$. The calorimetric energy is required to be small, consistent with a MIP. Upper cut in isolation is applied to avoid muons produced within jets, mostly as $q\xrightarrow{\rm hadr.}\pi^\pm\to\mu^\pm\nu$. Hits aligned with the extrapolated track are required in the muon system. For muons of $\abs{\eta}<0.6$, hits are required both in CMU and CMP, hence the characterization ``CMUP'' for passing muons. CMX is used for muons of $0.6<\abs{\eta}<1.0$ and BMU for $1.0<\abs{\eta}<1.5$.
\paragraph{Central taus} Only one-prong taus are identified, to keep the object identification and fakes modeling simple and uniform across the three lepton generations.  Narrow central jets with a single charged track are identified as taus.  Electrons are discriminated against by requiring substantial fraction of the energy in the CHA; muons are vetoed by requiring to have no muon stub coinciding with the extrapolated track. Track and calorimeter isolation requirements are imposed.
\paragraph{$b$-jets} The identification of a jet originating from a $b$ quark is made using the SecVtx algorithm\cite{SecVtx}, which identifies the displaced secondary vertex\index{vertex!secondary} from the decay of the formed $B$ meson.
\paragraph{jets} Hadronic jets are reconstructed in $\abs{\eta}<2.5$ using calorimeter information with the JetClu clustering algorithm\cite{JetClu:Abe:1991ui}, using $\Delta R<0.4$.


\paragraph{Missing energy} The negative vector sum of the transverse momentum is used to calculate \pmissvector, as described in Sec.~\ref{sec:pmiss}.

\section{Event Selection Cuts}

Selection criteria are imposed to keep potentially interesting events.  Effort is made to keep high-$p_T$ events of as many final states as possible, with the requirement that they have at least one high-$pT$ object or specific combinations of objects.  Offline selection $p_T$ thresholds are safely higher than the turn-on region of online triggers, so that trigger efficiencies can be treated as roughly independent of object $p_T$.

Specifically, events enter the analysis if they have any of the following:
\begin{itemize}
\item a central $e$ of $p_T>25$~GeV
\item a plug $e$ of $p_T>40$~GeV
\item a central $\mu$ of $p_T>25$~GeV
\item a central $\gamma$ of $p_T>60$~GeV, or
\item a central jet (regular or $b$-tagged) of $p_T>200$~GeV, or
\item a central $b$-tagged jet of $p_T>60$~GeV (prescaled like the online trigger), or
\item a central jet or $b$-tagged jet of $p_T>40$~GeV (prescaled to one tenth of the online trigger prescale).


\item one $e$ plus another electromagnetic object ($e$ or $\gamma$) of $\abs{\eta}<2.5$ and $p_T>25$~GeV, or
\memo{bookmark}
\item a central or plug electron with $p_T>40$~GeV and a central tau with $p_T>17$~GeV, or
\item a central muon with $p_T>17$~GeV and a central or plug photon with $p_T>25$~GeV, or
\item a central muon with $p_T>25$~GeV and a central $b$-tagged jet with $p_T>17$~GeV, or
\item two taus with $\abs{\eta}<1.0$ and $p_T>25$~GeV, or
\item a central or plug photon with $p_T>40$~GeV and a central tau with $p_T>40$~GeV, or
\item one central photon with $p_T>25$~GeV and one other central or plug photon with $p_T>25$~GeV, or
\item a central photon with $p_T>40$~GeV and a central $b$-tagged jet with $p_T>25$~GeV, or\item a central jet or $b$-tagged jet with $p_T>40$~GeV and a central tau with $p_T>17$~GeV (prescaled by the online jet20 trigger), or
\item a central jet with $p_T>60$~GeV and a central $b$-tagged jet with $p_T>25$~GeV (prescaled by the online jet20 trigger), or
\item two central muons with $p_T>17$~GeV, or
\item one central electron and one central muon with $p_T>17$~GeV, or
\item one central electron with $p_T>20$~GeV and one central tau with $p_T>17$~GeV, or
\item one plug electron with $p_T>25$~GeV and one central muon with $p_T>17$~GeV, or
\item one central muon with $p_T>20$~GeV and one central tau with $p_T>17$~GeV.
\end{itemize}
Tri-object selections keep events containing
\begin{itemize}
\item a central or plug photon with $p_T>40$~GeV and two central taus with $p_T>17$~GeV, or
\item a central or plug photon with $p_T>40$~GeV and two central $b$-tagged jets with $p_T>25$~GeV, or
\item a central or plug photon with $p_T>40$~GeV, a central tau with $p_T>25$~GeV, and a central $b$-tagged jet with $p_T>25$~GeV.
\item one $b$-tagged jet with $p_T>90$~GeV and two other $b$-tagged jets with $p_T>60$~GeV
\item one central muon with $p_T>17$~GeV and two other central or plug muons with $p_T>17$~GeV.
\end{itemize}
Additional special selections keep events containing
\begin{itemize}
\item one central or plug electron with $p_T>40$~GeV, missing transverse momentum greater than $17$~GeV, and two or more jets or central $b$-tagged jets with $p_T>17$~GeV, or
\item one central muon with $p_T>25$~GeV, missing transverse momentum greater than $17$~GeV, and two or more jets or central $b$-tagged jets with $p_T>17$~GeV.
\end{itemize}

After the above selection cuts, some further clean-up cuts are imposed:
\begin{itemize}
\item{Duplicate events making it into more than one datasets are removed, so that only one instance of each event is kept.}
\item{Events are kept only from periods when all subdetectors are active.}
\item{Remove cosmic ray events by requiring at least one primary vertex and 3 tracks above 500~MeV.}
\item{Remove beam halo events, in the same way cosmics are removed, but by also exploiting geometric characteristics particular to beam halo events \cite{beamHaloCuts}.}
\end{itemize}

\section{Event Partitioning}
\label{sec:Partitioning}

Partition rules can be arbitrary; as long as both data and Standard Model background are partitioned alike, their comparison is legitimate. However, the partition rule influences the sensitivity to specific signatures. If one has a specific signature in mind, he tries to define a partition into bins that accumulate as much signal and less background as possible. In this anlysis, though, no specific signature is targeted, therefore the partition rule is chosen as generic and model-independent as possible.

Two different --- but related --- partition rules are applied, one before the \Vista\ comparison and one before the \Sleuth\ search.

\subsection{\Vista\ Partition Rule}
\label{sec:vistaPartitionRule}

\begin{figure}
\centering
\includegraphics[width=7cm,angle=0]{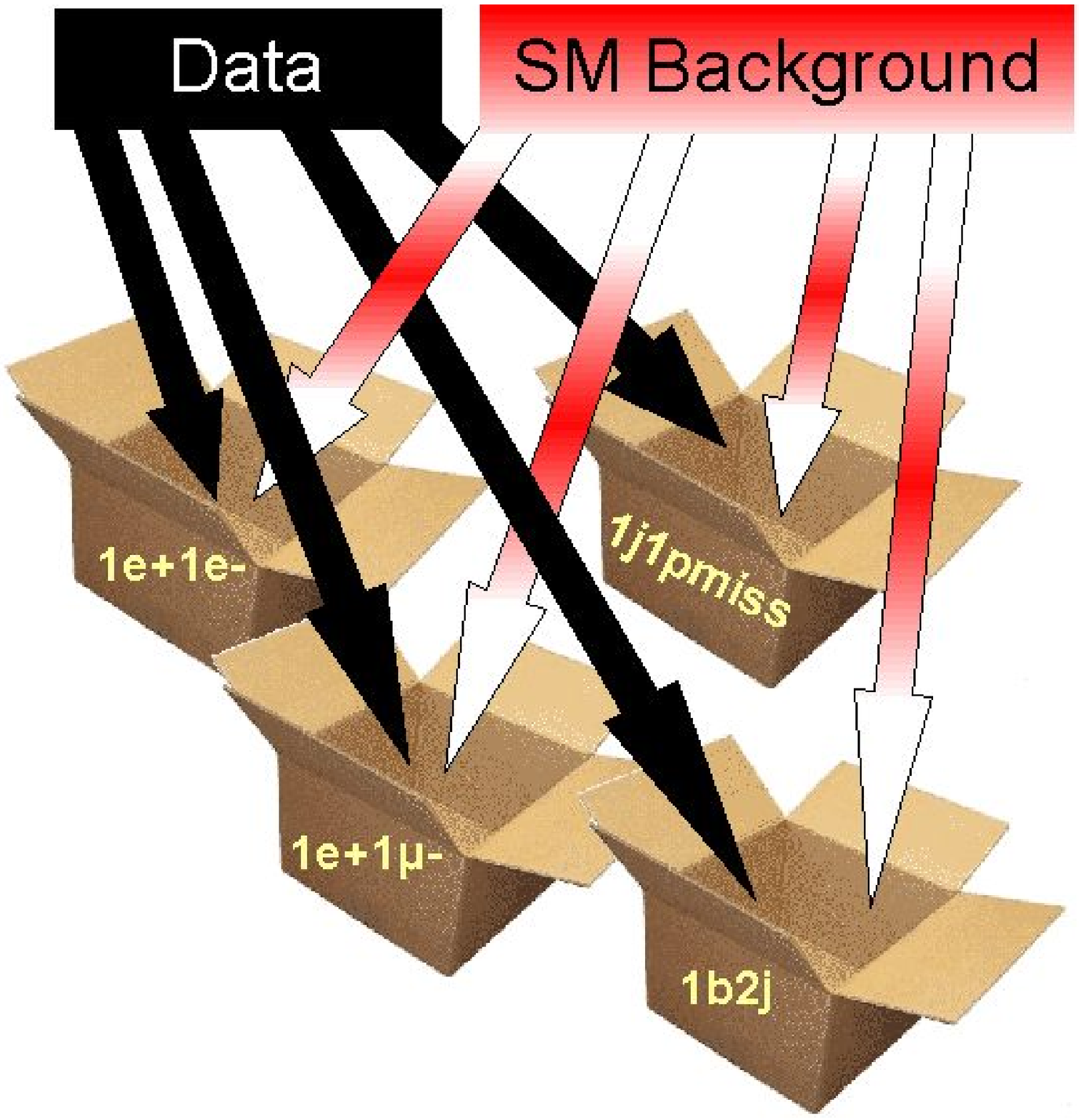} 
\caption{\Vista\ partitioning in final states.}
\label{fig:VistaPartitioning}
\end{figure}

For \Vista, the events are partitioned according to their reconstructed final objects. The idea is depicted in Fig.~\ref{fig:VistaPartitioning}. Each final state is exclusive and each event can belong to one and only one final state, characterized by the identity of its reconstructed final objects.

The \Vista\ final states are not predetermined, but are dynamically formed to accommodate the data. Namely, if a unique datum appears with an unprecedented combination of reconstructed final objects, then it will create its own final state, which will then be available to accomodate any further such events that may be found.\footnote{Exceptional treatment applies to events that have more than 4 jets: If an event has only jets and $\sumPt < 400$~GeV it is treated as described, like any other event. Those events are not considered by \Sleuth\ later. If however an event has more than 4 jets and at least one non-jet object, or it has more than 4 jets with $\sumPt > 400$~GeV, then the $k_T$ algorithm\cite{kTclustering} is used to cluster reconstructed jets until only 4 jets are left.}

The impartial and unbiased character of the \Vista\ partition rule, which discards no events and groups together no unlike events, gives \Vista\ much of its model-independence.

\subsection{\Sleuth\ Partition Rule}
\label{sec:sleuthPartitionRule}
For \Sleuth, a different partition rule is used, which is not as model-independent as the one used for \Vista, which makes \Sleuth\ a quasi model-independent search algorithm. It relies on \Vista\ partition, as it groups \Vista\ exclusive final states into \Sleuth\ final states. Each \Vista\ final state can be part of one and only one \Sleuth\ final state, therefore each event can belong to only one \Sleuth\ final state. So, the \Sleuth\ final states are non-overlapping sets of \Vista\ exclusive ones.

Model-dependence is introduced to improve \Sleuth's sensitivity to any new physics that would satisfy the following assumptions\index{Sleuth@\Sleuth!assumptions}:
\begin{itemize}
\item{Would appear as an excess of high-\sumPt\ data, which is what \Sleuth\ seeks.}
\item{Would treat equivalently the first two generations of leptons.}
\item{Would be symmetric under global charge conjugation.}
\item{Would produce jets in pairs.}
\item{Would conserve lepton number.}
\end{itemize}

The above assumptions are exploited by applying the following partition rules:
\begin{enumerate}
\item If two events transform into each other by a global $e\leftrightarrow\mu$ interchange, they are grouped in the same \Sleuth\ final state. E.g.\ $\mu^+\mu^-e^+j$ and $e^+e^-\mu^+j$.
\item If two events are global charge conjugates of each other, they are put in the same \Sleuth\ final state. E.g.\ $\mu^+\mu^+j$ and $\mu^-\mu^-j$.
\item If an event has $2N+1$ non $b$-tagged jets, where $N \geq 0$, then it is grouped in the same final state as events that have the same final objects, but $2N$ non $b$-tagged jets. The assumption is that the odd-numbered jet is due to radiation. E.g.\ $2j\pslash$ and $3j\pslash$.
\item If an event has $2N+1$ $b$-jets, where $N\geq 0$, and $M\geq1$ non-$b$ jets, then it is put in the same final state as events that have the same final objects, but $2N+2$ $b$-jets and $M-1$ non-$b$ jets. E.g.\ $3b3j$ and $4b2j$. The assumption is that one of the $M$ non-$b$ jets was actually a $b$-jet that was not tagged.
\item If there are three leptons and no \pslash, then the event is grouped together with the events having the same composition and also \pslash. The assumption is that a neutrino was produced but the \pslash\ was not resolved.
\item{If there is one lepton and no \pslash, then discard the event.}
\end{enumerate}

The result of the \Sleuth\ partition rules are shown in Table \ref{tab:sleuthFinalStatesContent}.
\begin{table}
\tiny
\begin{minipage}{7.0in}
\hspace{-0.0in}
\input{figures/analysis/sleuthFinalStateContentIndex}
\end{minipage}
\caption{Result of the \Sleuth\ parition rule. On the left column are the \Sleuth\ final states and on the rigth are the populated \Vista\ exclusive final states that were merged into each. Every \Vista\ final state written also implies its charge conjugate. }
\label{tab:sleuthFinalStatesContent}
\end{table}

\section{\Vista\ Algorithm}

\Vista\ is conceptually simple, though its practical implementation is more complicated. Its intermediate goal --- at the global fit stage --- is to adjust the parameters of the correction model, called correction factors, to obtain the Standard Model background that best matches the data. Its ultimate goal --- at the \Vista\ comparison stage --- is to compare the Standard Model background to the data globally and point out any discrepancies in the final state populations, or in kinematic variable distribution shapes.

\subsection{Correction Factors}
\label{sec:correctionFactors}
If the detector and the Standard Model were modeled at infinite precision, corrections to the output of \CdfSim\ would not be necessary. However, developing the correction model and applying it is an integral part of this analysis. The accuracy of the Standard Model implementation depends on the correction model used, namely the degree of sophistication introduced in fine-tuning the outcome of \CdfSim\ to represent the Standard Model more faithfully. 

In developing the correction model, one needs guidance from the data. Starting with no corrections, one observes significant and widespread discrepancies. Since the Standard Model is expected to describe the data well, that indicates that some corrections are needed to bring our implemented model closer to the Standard Model. After implementing the first set of corrections, another comparison may lead to another level of improvement by introducing richer corrections. As the correction model is developed, prudence is required to not introduce corrections that would {\em ad hoc} conceal potential signs of new physics. This takes intuition and understanding of the physical mechanisms that are not perfectly modeled. More on this issue is in Appendix \ref{sec:blindOrNot}.

\begin{table}
\label{tab:fudgeFactorDescriptions}
\begin{minipage}{6in}
\hspace{-0.5in}
\input{figures/analysis/fudgeFactorDescription}
\end{minipage}
\caption{Description of correction factors\index{correction!factors}.}
\end{table}

Table~\ref{tab:fudgeFactorDescriptions} lists all used correction factors. Each has a 4-digit reference code. They are of four kinds:
\begin{itemize}
\item the integrated luminosity,
\item $k$-factors representing the physical cross section of processes divided by their LO cross section,
\item identification probability scale factors and misidentification probabilities,
\item online trigger efficiency scale factors.
\end{itemize}
There are cases in which a correction factor is specific to a region of $p_T$ or $\abs{\eta}$. Typically, that happens because having a single correction factor for the whole $p_T$ and $\abs{\eta}$ range proves to be an oversimplification leading to discrepancies. In principle one could divide the $(p_T,\abs{\eta})$ space in an arbitrarily large number of regions, and define a different set of correction factors in each. That, however, would be the opposite of crude, i.e.\ too flexible, with enough degrees of freedom to fit almost anything, including potential signals of new physics. The current choice of correction factors was found to balance satisfactorily between the two extremes.

The set of correction factors of Table.~\ref{tab:fudgeFactorDescriptions} includes all the degrees of freedom the global fit can adjust (Sec.~\ref{sec:globalFit}). However, the Standard Model implementation can be adjusted in more ways, that are not subject to a fit, and are discussed in Sec.~\ref{sec:correctionModel}.

\subsection{Global Fit}
\label{sec:globalFit}

The global fit\index{Vista@\Vista!global fit} is a binned $\chi^2$ minimization procedure. Its free parameters ($\vec{s}$) are the correction factors (Sec.~\ref{sec:correctionFactors}), which are adjusted to minimize the quantity
\begin{equation}
\chi^2(\vec{s}) = \left(\sum_{k\in{\rm bins}}{\!\!\!\chi^2_k}(\vec{s})\right) + \chi^2_{{\rm constraints}}(\vec{s})\ .
\end{equation}

Each of the bins\index{Vista@\Vista!bins} used in defining $\chi^2$ contains events of specificied final reconstructed objects. However, what distinguishes it from a \Vista\ final state --- described in Sec.~\ref{sec:vistaPartitionRule} --- is that it is also characterized by specific object locations in $p_T$ and $\abs{\eta}$. The $(p_T,\abs{\eta})$ space is divided in large regions motivated by the detector geometry, with edges at $\abs{\eta}=\{0.6,1.0\}$ and $p_T=\{25,40,60,200\}$~GeV. So, events that have the same objects in the same $(p_T,\abs{\eta})$ regions are accounted in the same bin. Some sparcely populated bins are merged together to obtain better statistics.

The $\chi^2_{{\rm constraints}}$ term introduces to the global fit constraints from external information. For example, the CLC provides a measurement\cite{CLC} of the integrated luminosity of the data sample ($L$), therefore if the correction factor corresponding to $L$ (code {\tt 5001}) drifts far from the CLC measurement, a penalty is imposed to the $\chi^2$. For details on the constraints, see Appendix \ref{sec:CorrectionModelDetails}.

The term reflecting the agreement between data and Standard Model background in bin $k$ is
\begin{equation}
\chi^2_k(\vec{s})=\frac{({\rm Data}[k]-{\rm Standard Model}[k])^2}{\delta{\rm Standard Model}[k]^2 + \sqrt{{\rm Standard Model}[k]}^2}\ ,
\label{eq:chi_k}
\end{equation}
where ${\rm Data}[k]$ is the number of data and ${\rm Standard Model}[k]$ is the amount of Standard Model background expected in bin $k$. 
$\delta{\rm Standard Model}[k]$ is the statistical uncertainty in ${\rm Standard Model}[k]$ due to limited MC statistics.

$\chi^2_k$ depends on $\vec{s}$ because the term ${\rm Standard Model}[k]$ does, which is how by adjusting $\vec{s}$ the Standard Model comes to better agreement with the data and $\chi^2$ reaches minimum. The explicit relation between the Standard Model background expected in bin $k$ and $\vec{s}$ must take into account the different kinds of correction factors (luminosity, efficiencies, fake rates, $k$-factors) and which of them are relevant to the objects in the events of $k$ in the $(p_T,\abs{\eta})$ regions. Specifically, that relation is
\begin{eqnarray}
\label{eq:Standard Modelk}
{\rm Standard Model}[k] = {\rm Standard Model}[(k_1,k_2)] &=& \left(\int{{\cal L}\,dt}\right) \cdot P_{{\rm trigger}}[(k_1,k_2)] \cdot \nonumber \\
& & \textstyle \sum_{{k_2}'\in\,{\rm obj.\ comb.}} P_{{\rm ID}}[(k_1,{k_2}')][k_2] \cdot \nonumber \\
& & \textstyle \sum_{l\in\,{\rm proc.}} ({\rm kFactor}[l])\cdot ({\rm Standard Model}_0[(k_1,{k_2}')][l]) \ ,
\end{eqnarray}
where the bin index $k$ is analyzed into two indices $k_1$ and $k_2$. $k_1$ labels the $(p_T,\abs{\eta})$ regions in which there are energy clusters and $k_2$ represents the combination of reconstructed objects --- the two elements defining each bin $k$. 
${k_2}'$ is a dummy summation index representing different combinations of reconstructed objects.
$l$ traverses Standard Model background processes, such as dijet production, $W$+1~jet production etc.
The term ${\rm Standard Model}_0[(k_1,{k_2}')][l]$ is the sum of the uncorrected weights of all MC events generated by process $l$ and after passing through \CdfSim\ were reconstructed in the bin $(k_1,{k_2}')$.
$P_{{\rm ID}}[(k_1,{k_2}')][k_2]$ is the probability that an event produced with energy clusters in the detector regions labeled $k_1$ and with identified objects labeled ${k_2}'$ would be identified as having objects labeled $k_2$.
$P_{{\rm trigger}}[(k_1,k_2)]$ represents the probability that an event produced with energy clusters in the detector regions labeled by $k_1$ and contains identified objects labeled by $k_2$ would pass the online trigger.
The terms $\int{{\cal L}\,dt}$, $P_{{\rm trigger}}[(k_1,k_2)]$, $P_{{\rm ID}}[(k_1,{k_2}')][k_2]$ and ${\rm kFactor}[l]$ are the correction factors $\vec{s}$ and operate multiplicatively.

\subsubsection{Event weights}
\label{sec:eventWeights}

Each generated MC event of a process $l$ amounts to uncorrected weight $\tilde{w}_l = \frac{\sigma_l}{N_l \cdot (1{\rm pb})}$, where $\sigma_l$ is the LO cross section of $l$ and $N_l$ is the number of MC events generated of $l$. The term ${\rm Standard Model}_0[(k_1,{k_2}')][l]$ in eq.~\ref{eq:Standard Modelk} can be rewritten as
\begin{equation}
\label{eq:Standard Model0}
{\rm Standard Model}_0[(k_1,{k_2}')][l]=N_{[(k_1,{k_2}')][l]} \tilde{w}_l\ ,
\end{equation}
where $N_{[(k_1,{k_2}')][l]}$ is the number of MC events from process $l$ that after passing through \CdfSim\ were reconstructed as belonging to bin $(k_1,{k_2}')$. Using eq.~\ref{eq:Standard Model0}, eq.~\ref{eq:Standard Modelk} can be rewritten as a sum of as many terms as the total number ($N$) of generaged MC events:
\begin{eqnarray}
\label{eq:smk2}
{\rm Standard Model}[k](\vec{s})&=&\sum_{i=1}^{N} c_{k,\vec{s}}^{i}\,\tilde{w}_{l_i}\ ,\ {\rm with} \\
\label{eq:smk3}
c_{k,\vec{s}}^{i}=c_{(k_1,k_2),\vec{s}}^{i}&=&\int{{\cal L}\,dt}\cdot {\rm kFactor}[l_i] \cdot P_{{\rm trigger}}[(k_1,k_2)] \nonumber \\
 & & {} \cdot P_{{\rm ID}}[(k_1,{k_2}'_i)][k_2],
\end{eqnarray}
where $l_i$ is the process as an instance of which the MC event $i$ was generaged, and ${k_2}'_i$ denotes the combination of reconstructed objects in the MC event $i$ after passing through \CdfSim.

The corrected weight\index{weight!corrected} of the $i\th$ MC event, defined for a bin $k$ and a set of correction factors $\vec{s}$ as 
\begin{equation}
w_i = c_{k,\vec{s}}^{i}\,\tilde{w}_{l_i},
\label{eq:correctedWeight}
\end{equation}
can be used to rewrite eq.~\ref{eq:smk2} as ${\rm Standard Model}[k](\vec{s})=\sum_{i=1}^{N} w_i$, from which follows that the term $\delta{\rm Standard Model}[k]$ in eq.~\ref{eq:chi_k} is\footnote{Proof: $N_{\tiny{\rm eff}}=\frac{\sum w_i}{w_{\tiny{\rm eff}}}=\frac{\sum w_i}{{\sum w_i^2}\big/{\sum w_i}} \Rightarrow \sqrt{N_{\tiny{\rm eff}}}\,w_{\tiny{\rm eff}}=\sqrt{\sum w_i^2}$ }
\begin{equation}
\label{eq:deltaStandard Modelk}
\delta{\rm Standard Model}[k]=\delta\left(\sum w_i\right) = \sqrt{\sum{w_i^2}}\ .
\end{equation}

\subsubsection{Error matrix}
\label{sec:errorMatrix}

As a result of fitting all the correction factors simultaneously, the complete error matrix\index{error matrix} $(\Sigma)$ can be estimated, by approximating $\chi^2(\vec{s})$ with a multi-dimensional parabola around its minimum at $\vec{s}_0$:
\begin{eqnarray}
\chi^2(\vec{s}) &\simeq& \chi^2_{\min}+(\vec{s}-\vec{s}_0)^T \Sigma^{-1} (\vec{s}-\vec{s}_0) \Rightarrow \\
\Sigma^{-1}_{ij} &\simeq& \frac{1}{2}\frac{\partial^2\chi^2(\vec{s})}{\partial s_i \partial s_j}\Big|_{\vec{s}_0} \\
{} &\simeq& \frac{\chi^2(\vec{s}_0+\delta s_i\,\hat{i} + \delta s_j\,\hat{j}) - \chi^2(\vec{s}_0+\delta s_i\,\hat{i}) - \chi^2(\vec{s}_0 + \delta s_j\,\hat{j}) + \chi^2(\vec{s}_0)}{2\,\delta s_i\,\delta s_j}
\end{eqnarray}
where $\delta s_i\,\hat{i}$ and $\delta s_j\,\hat{j}$ are small displacements in the direction of the $i\th$ and $j\th$ correction factor respectively.

Then, $\Sigma$ is the inverse matrix of $\Sigma^{-1}$ and the variance of correction factor $s_i$ is
\begin{equation}
\sigma^2_i \simeq \Sigma_{ii} 
\end{equation}
and the correlation between $s_i$ and $s_j$ is 
\begin{equation}
\rho_{ij}=\rho_{ji}=\frac{{\rm cov}(s_i,s_j)}{\sigma_i \sigma_j} \simeq \frac{\Sigma_{ij}}{\sigma_i \sigma_j}
\end{equation}

\subsubsection{Pull-apart and Influence tables}

Visualizing the $\chi^2(\vec{s})$ in the multi-dimensional space of correction factors $\vec{s}$ is very difficult, but one can gain intuition about the global fit by knowing for the $k\th$ bin how much $\chi^2_k(\vec{s})$ changes with small variations of the $i\th$ correction factor $s_i$ around the minimum $\vec{s}_0$. The quantity of interest is 
\begin{equation}
{{\rm pull}}_{ki} \equiv \frac{\partial \chi^2_k(\vec{s})}{\partial s_i}\Big|_{\vec{s}_0}\,.
\end{equation}

The elements of the $N_{{\rm bins}}\times N_{{\rm final\ states}}$ matrix pull$_{ki}$ can be sorted to rigorously answer two questions:
\begin{itemize}
\item For the $k\th$ bin, how much is the total absolute pull $\sum_i \abs{{\rm pull}_{ki}}$ and which correction factors exert the absolutely biggest pulls? The answer for each bin is given in the {\em influence} table\index{influence table} (Table \ref{tab:}).

\memo{Complete references to tables, that will be in the results section, or in some appendix.}

\item For the $i\th$ correction factor $s_i$, how much in total does it pull apart the bins ($\sum_k \abs{{\rm pull}_{ki}}$), and which bins does it pull most? The answer for each $s_i$ is given in the {\em pull apart}\index{pull apart table} table (Table \ref{tab:}).

\memo{Complete references to tables, that will be in the results section, or in some appendix.}

\end{itemize}

\subsection{Correction Model}
\label{sec:correctionModel}

In this section, the implementation of the Standard Model will be described, i.e.~how after the global fit the Standard Model background expected in each \Vista\ final state is actually constructed. More details are in Appendix~\ref{sec:CorrectionModelDetails}.

The MC events available are coming from various sources, having some initial uncorrected weight, as described in Sec.~\ref{sec:eventWeights}. Each MC event, however, needs to enter the analysis with its corrected weight, representing the effect of the correction factors on it. The corrected weight is given by eq.~\ref{eq:correctedWeight}, which applies to any $k$ and $i$. In principle all MC events contribute to all bins, each time with a different weight. Therefore, it helps to think of the MC events not as events that are reweighted, stored in some bin and not used again\footnote{Which, apart from the reweighting step, is exactly how data events are treated.}, but rather as prototypes that are all reflected to all the bins with the appropriate weight each time (Fig.~\ref{fig:mirrorsVsBoxes}). When an event contributes to a final state by misreconstruction of one or more of its objects, the object's identity changes accordingly, along with the weight of the event.

\begin{figure}
\centering
\begin{tabular}{rl}
\includegraphics[width=7cm,angle=0]{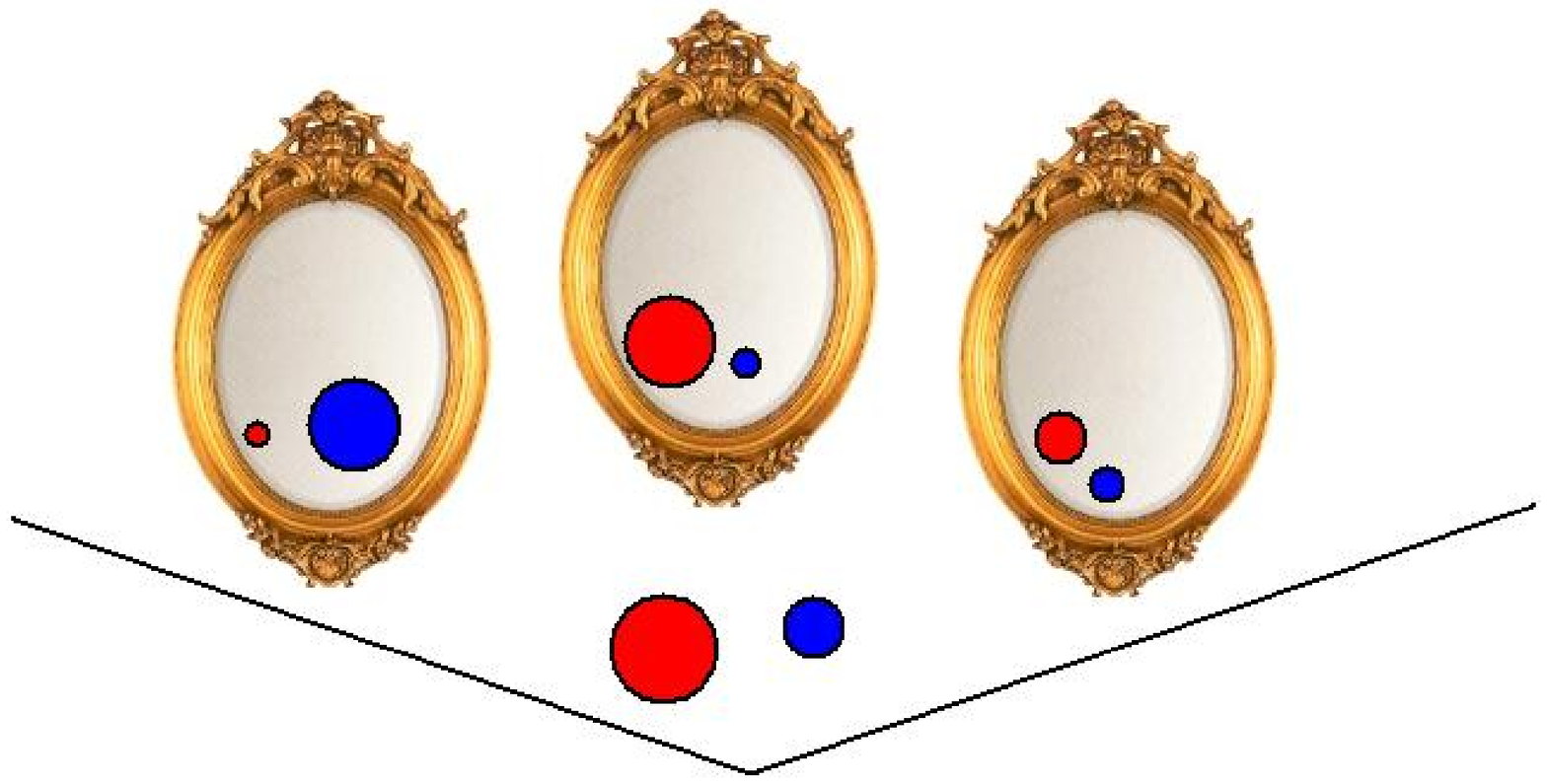} &
\includegraphics[width=7cm,angle=0]{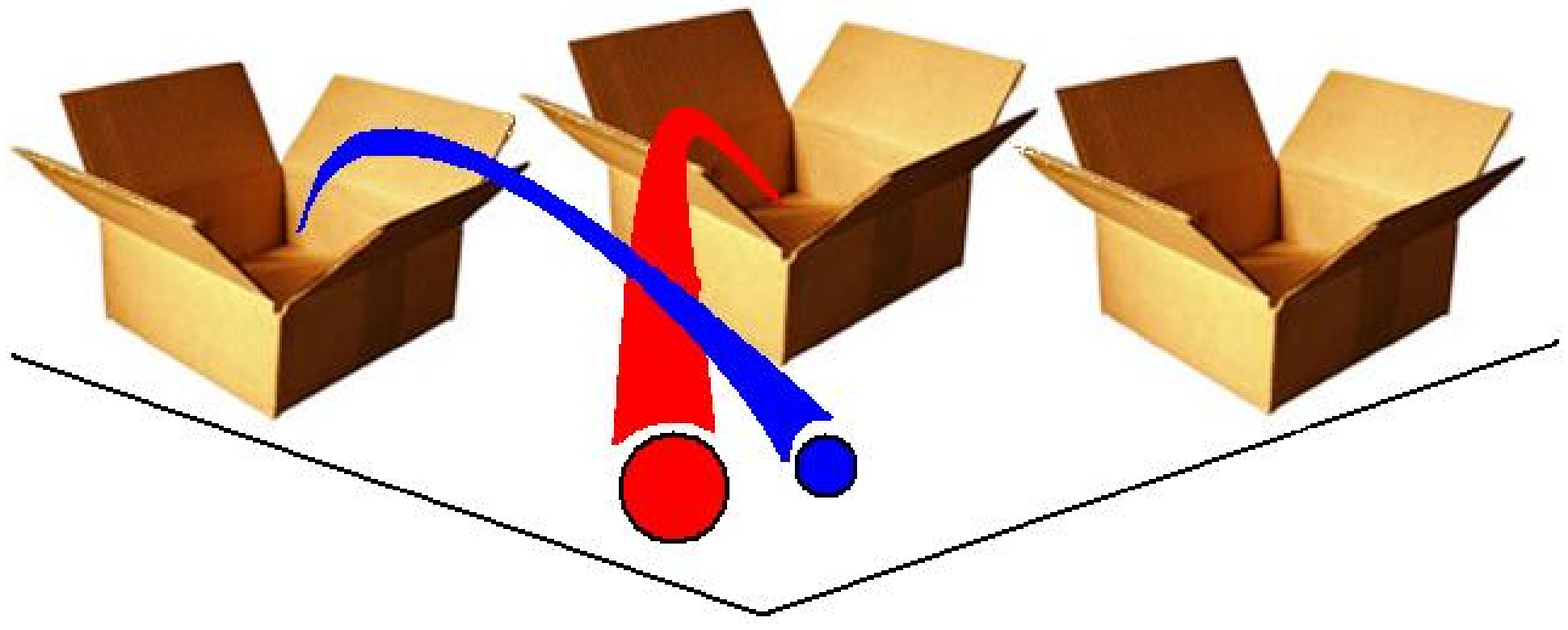}
\end{tabular}
\caption{The idea of MC events contributing in principle to all bins (mirrors) and not being used up, versus contributing to only one bin (box) each.}
\label{fig:mirrorsVsBoxes}
\end{figure}

Though it may seem unnatural to have events contributing to unrelated bins, with completely different reconstructed objects, this scheme is valid because for unnatural combinations of $i$ and $k$ the factor $c_{k,\vec{s}}^i$ is $\sim 0$, for the bin where the event would typically be expected $c_{k,\vec{s}}^i$ is $\sim \int{{\cal L}\,dt}$, and for bins where the event could only be found by some misreconstruction of probability $p_{\small{{\rm misID}}}$, $c_{k,\vec{s}}^i$ is $\sim \int{{\cal L}\,dt} \cdot p_{{\rm misID}}$. The exact $c_{k,\vec{s}}^i$ factor is found each time according to eq.~\ref{eq:smk3}.

\subsubsection{Treatment of fakes}
\label{sec:treatmentOfFakes}

When events pass through \CdfSim\ they are smeared by the limited detector resolution and sometimes may be misreconstructed, as would also happen in reality. By construction, the uncorrected weight of those ``fake'' events is proportional to the LO cross section of the generated process.

The LO cross sections for $p\bar{p}$ collisions at $1.96$~TeV are such that the generated objects are most often jets, less frequently $b$-jets and photons, and much less frequently leptons. So, if an event contained a generated high-$p_T$ quark that after \CdfSim\ was reconstructed as a lepton, the resulting fake\index{fakes} event would have a very big uncorrected weight, characteristic of jet production cross section, whose largeness would be propagated to the corrected weight too. That would affect final states with contributions from such fakes, because events with large weights appear like spikes and distort the actual shape of the Standard Model background distributions.\footnote{Practically, it would take too long to collect enough such fakes through \CdfSim\ because they occur rarely and each even takes about 10 seconds to pass through \CdfSim.} 

Instead of relying on \CdfSim\ for such fakes from abundant to rare objects, what is done is to reflect the abundant, non-faking events into the bin where misreconstruction could divert them, accounting in their corrected weights for the probability of that happening. Any such fakes that come out of \CdfSim\ are stripped away, to avoid spikes in the Standard Model background. This allows for well-populated Standard Model backgrounds in those final states containing fakes. 

The other category of fakes, namely where a rarely generated object fakes a frequently generated one, e.g.\ $\tau \to j$, is left for \CdfSim\ to take care of. The reason is obviously that it does not result in spikes, and generally is a small effect, since it is suppressed by both the fake rate and the small generation cross section. Not modeling it with correction factors saves a great number of correction factors and unnecessary complexity.\footnote{The only exception is the correction factor {\tt 5245}, which is used in addition to the \CdfSim\ fakes to better model the electron faking photon effect in the plug region.}

\section{\Vista\ Comparison}
\label{sec:VistaComparison}
Once the Standard Model background in each \Vista\ final state has been constructed, as described in Sec.\ \ref{sec:correctionModel} and \ref{sec:vistaPartitionRule}, the task of \Vista\ comparison is 
\begin{itemize}
\item to evaluate how well the populations of data and Standard Model agree in each final state,
\item to evaluate how well the shapes of all potentially interesting kinematic distributions of data and Standard Model agree,
\item to search for statistically significant localized excesses (``bumps'')\index{bumps} of data in mass distributions in all final states,
\item to sort and present the above information in a way that makes it easy to browse and understand.
\end{itemize}

\subsubsection{Populations Comparison}
\label{sec:populationsStatistic}

In testing the null hypothesis that the data population is consistent to the expected Standard Model population, the statistic used is the probability that the expected population $b=\sum w_i$ would randomly fluctuate to or beyond the observed data population $d$. Namely, if $n$ is a random number pulled from a Poisson distribution with mean $b$, the statistic is the probability that $n\ge d$ if $d>b$ or $n\le d$ if $d<b$. Actually, $b$ is determined with some uncertainty $\delta b$, which is partly from limited MC statistics, and partly from uncertainty in the correction factors. The latter part is ignored due to practical limitations and $\delta b=\sqrt{\sum w_i^2}$ is purely statistical uncertainty, like in eq.~\ref{eq:deltaStandard Modelk}. Assuming that the PDF of $b$ is Gaussian and $b>0$, the statistic $p$ is defined as
\begin{eqnarray}
\label{eq:vistaP}
p&=&\sum_{n=\alpha}^{\Omega} \int_0^\infty \frac{1}{K}\,\frac{1}{\sqrt{2\pi}\,\delta b}\,e^{-\frac{({b'}-b)^2}{2\,{\delta b}^2}}\,\frac{{b'}^n}{n!}\,e^{-b'}\,d{b'}\, , \\
\nonumber {\rm where}&{}&\left\{\alpha = 0\ ,\ \Omega = d\right\} \ \ {\rm\ if}\ d<b \\
\nonumber {}&{}&\left\{\alpha = d\ ,\ \Omega = \infty\right\} \ \ {\rm\ if}\ d\ge b \\
\nonumber {}&{}&K=\int_0^\infty \frac{1}{\sqrt{2\pi}\,\delta b}\,e^{-\frac{({b'}-b)^2}{2\,{\delta b}^2}} d{b'}\ .
\end{eqnarray}
The smaller the $p$, the more significant the discrepancy.

In the definition of the statistical significance of a population difference \abs{d-b}, the trials factor\index{trials factor} due to checking many final states needs to be accounted for, since this increases the probability that at least one final state would yield a $p$ as small as the observed one. For a given $p$, this probability is
\begin{equation}
\label{eq:vistaTildeP}
\tilde{p}=1-(1-p)^N\ ,
\end{equation}
where $N$ is the number of final states examined, which is about 350 in 1 fb$^{-1}$ of data. 

\subsubsection{Shapes Comparison}
\label{sec:shapesStatistic}

For each final state there is a number of relevant kinematic variables that can be defined, depending on the objects in the final state. For example, some of them are the $\eta$, $\phi$ and $p_T$ of all the objects, the invariant mass of the system of groups of objects, or more specific variables, indicative of some work hypothesis.

The distributions of all relevant variables are automatically made for the data and the background in each final state. Then, for each variable, the shapes of the distributions of data and background are compared, using KS test, which, loosely speaking,\footnote{It actually returns the probability that the maximum difference between the cumulative distributions of the two compared distributions would randomly appear to be bigger than observed, assuming the null hypothesis that the two distributions are just statistical variations of the same distribution.} returns the probability that the two distributions are consistent. The smaller the returned probability, the bigger the shape discrepancy.

Given that approximately $17\times 10^3$ kinematic variables are plotted and examined in this way, the statistical significance of each discrepancy needs to account for the trials factor\index{trials factor}, as described in eq.~\ref{eq:vistaTildeP}.

\subsubsection{Bump statistic}
\label{sec:bumpStatistic}

A bump\index{bumps} statistic measures the extent to which a localized excess of data exists in a mass window.  For each invariant mass distribution in each final state, windows of widths varying from 5 to 100~GeV are defined, in steps of 5~GeV.  Within each window, the Poisson probability for the Standard Model prediction to fluctuate up to or above the number of events observed is calculated.  For each mass distribution, the window for which this probability is smallest is noted.  

The statistical significance of this most interesting window must be reduced by a trials factor\index{trials factor} equal to the number of windows considered.  The counting is obscured by the overlaps of windows of different size, but reasonably estimated to be twice the number of 5 GeV windows considered.\footnote{A window is not considered if it contains zero Standard Model events.}  An additional trials factor is applied to account for the total number of mass distributions ($\sim 4\times10^3$) considered in all final states.  

\subsubsection{Information Organization}

The volume of the \Vista\ output\index{Vista@\Vista!output} is large, as it contains hundreds of final states and thousands of kinematic distributions, probed with the statistics described above.

The organization of this output is such that the biggest discrepancies between data and Standard Model background appear first. This proves particularly helpful at the stage of development of the correction model, as it reveals what requires better modelling.

Employing an Internet-based interface, the final states are listed in order of decreasing population discrepancy. For each final state, the most discrepant shapes are listed in order of decreasing significance, and the same is done for any bumps found in it. The distributions of all kinematic variables are available following a hyperlink corresponding to each final state.

\section{\Sleuth\ Algorithm}

\Sleuth\ uses the same Standard Model background and data as \Vista, thus could be viewed as another probe of \Vista. It stands, though, as an algorithm on its own because it uses the quasi model-independent partition rule of Sec.~\ref{sec:sleuthPartitionRule} and, instead of analytic calculation, it uses multiple trials to estimate the significance of the effects it targets.

\Sleuth\ considers a single variable\index{Sleuth@\Sleuth!\SumPt} in each exclusive final state:  the summed scalar transverse momentum of all objects in the event \SumPt:
\begin{equation}
\SumPt \equiv \sum_i{\abs{\vec{p}_{Ti}}} + \abs{\puncl} + \abs{\pmissvector},
\end{equation}
where only the momentum components transverse to the axis of the colliding beams are considered when computing magnitudes.

For each \Sleuth\ final state, the \sumPt\ disribution is made for data and Standard Model background. Then \Sleuth\ searches for regions in which more events are seen in the data than expected. The steps of the search can be sketched as follows:
\begin{itemize}
\item In each final state, the regions\index{Sleuth@\Sleuth!regions} considered are the semi-infinite intervals in \sumPt\ extending from each data point up to infinity.
\item In each specific final state, the region defined by the $d\th$ highest \sumPt\ datum contains $d$ data events.  The Standard Model prediction in this region integrates to $b$ predicted events.  In this final state, the interestingness of the $d\th$ region is defined as the Poisson probability (or ``$p$-value'') $p_d = \sum_{n=d}^{\infty}\frac{b^n}{n!}e^{-b}$ that the Standard Model background $b$ would fluctuate up to or above the observed number of data events $d$ in this region.  The most interesting region in the final state is the one with smallest $p_d$.
\item For this final state, pseudo experiments are generated, with pseudo data pulled from the Standard Model background.  For each pseudo experiment, the interestingness of the most interesting region is calculated.  The fraction $\scriptP$ (also written ``scriptP'')\index{Sleuth@\Sleuth!\scriptP} of pseudo experiments in this final state in which the most interesting region is more interesting than the most interesting region in this final state observed in the data is calculated.  If the null hypothesis holds, namely the data is consistent with the Standard Model background, $\scriptP$ is expected to be a random number pulled from a uniform distribution in the unit interval.
\item The above are repeated to determine $\scriptP$ for each final state.  The smallest of these values is denoted by $\scriptP_{\rm min}$.
\item $\twiddleScriptP = 1-\prod_{a=1}^{N_{\rm FS}}(1-\hat{p}_a)$ is defined (also written ``tildeScriptP'')\index{Sleuth@\Sleuth!\tildeScriptP}, where the product is over all $N_{\rm FS}$ \Sleuth\ final states $a$, and $\hat{p}_a$ is the lesser of $\scriptP_{\rm min}$ and the probability for the total number of events predicted by the Standard Model in the final state $a$ to fluctuate up to or above three data events, as discussed in Appendix~\ref{sec:sleuth:details}.\footnote{In the simple case where in all final states the background is more likely to fluctuate up to three than $\scriptP_{\rm min}$, $\twiddleScriptP = 1 - (1-\scriptP_{\rm min})^{N_{\rm FS}}$, in analogy with eq.~\ref{eq:vistaTildeP}.}
$\twiddleScriptP$ represents the fraction of hypothetical experiments, following the Standard Model implementation, that would produce by chance a region in any final state as interesting as the most interesting region observed in the most interesting final state in the data. 
\end{itemize}

The output\index{Sleuth@\Sleuth!output} is the most interesting region ${\cal R}$ observed in the data and a number $\twiddleScriptP$ that quantifies the interestingness of ${\cal R}$.  $\twiddleScriptP$ is pulled from a uniform distribution on the unit interval if the null hypothesis holds, i.e.\ the data come from the Standard Model background alone, and is expected to be small if the data contain a hint of new physics (Appendix~\ref{sec:sleuth:details}).

In addition to \tildeScriptP\ and the most interesting region, all \Sleuth\ final states are listed as part of the output, each with its \sumPt\ distribution, the most interesting region in it and the corresponding \scriptP. Like in \Vista, the list of \Sleuth\ final states is ordered in increasing \scriptP, namely from the most interesting to the least.

A reasonable threshold for discovery\index{Sleuth@\Sleuth!discovery threshold} is $\twiddleScriptP \lesssim 0.001$, which would mean that less than one in a thousand CDF experiments following just the Standard Model implementation would by fluctuation yield anything as significant as the most significant effect observed in \Sleuth.   As shown in Appendix~\ref{sec:sleuth:details}, the significance --- before accounting for the trials factor --- of the individual effect that would result in $\twiddleScriptP \lesssim 0.001$ corresponds loosely to the $5\sigma$ significance that is commonly recognized in the field as ``discovery level''.

}
\chapter{Update with 2 fb$^{-1}$}
\label{chapter:2fb-1}

This analysis was conducted in two rounds: first with 1 fb$^{-1}$ of data, and then with 2 fb$^{-1}$.  The first round was presented in Chapter \ref{chapter:1fb-1}.  This chapter summarizes the second round.

\section{Overview}

Four separate statistics are employed to search for evidence of new physics.  These statistics are
\begin{itemize}
\item a difference between the number of observed and predicted events in individual exclusive final states;
\item a difference in distribution shape between data and Standard Model prediction in a variety of kinematic variables;
\item an excess of data in the large \SumPt\ tail of exclusive final states; and
\item a local excess (bump) in some invariant mass distribution, reflecting possibly a new resonance.
\end{itemize}
The next sections discus these statistics: Sec.~\ref{sec:Vista2} is about the normalization and shape statistics, Sec.~\ref{sec:Sleuth2} about the \SumPt\ statistic, and Sec.~\ref{sec:BumpHunter} about the mass bump statistic.  Conclusions are provided in Sec.~\ref{sec:Conclusions2}.


\section{\Vista}
\label{sec:Vista2}

Conceptually, \Vista\ in the second round of analysis is the same as in the first.

\subsection{Object identification}
\label{sec:particleID2}
The particle identification criteria used in this analysis are the same as in the first round, except for the following changes:

\begin{itemize}
\item Changed previously suboptimal conversion filter to the standard one.  In the previous version, we required each lepton candidate to not have within $\Delta R<0.4$ another track of opposite sign.  The neighbor track was counted only if it had $p_T>2$~GeV.  In this version, we make no transverse momentum requirement on the candidate neighbor tracks.  This change reduces significantly the rate for jets and photons to fake electrons, since both fakings involve conversions.
\item For plug electrons we now require the presence of a good quality PES cluster\footnote{Variables $PES~5x9~U$ and $PES~5x9~V$ need to be defined and less than 0.65.}, and that the PHX track matches to the electromagnetic cluster to within $\Delta R < 0.01$.  This reduces the rate of jets faking electrons in the region $\abs{\detEta}>1$.
\item For CMUP muons, we require CMU the distance between a stub and the track extrapolation ($\Delta X$) to be less than 7 cm, instead of 3 cm.  This follows a change in the standard muon identification criteria used by the experiment.
\item For taus, the momentum is now taken from the calorimeter $E_T$ rather than visible momentum (track momentum plus $\pi^0$s). The minimum seed track $p_T$ requirement has been increased to 10.5 GeV, reflecting a change in online trigger criteria.  We also added an additional muon veto cut requiring that the calorimter $E_T$ over seed track $p_T$ be greater than 0.5, inconsistent with a minimum ionizing particle.
\item For plug photons, we apply the fiducial cut $\abs{\detEta}>1.2$.  
\end{itemize}

Tables with identification criteria for all objects can be found in Appendix~\ref{sec:particle_id_detail}.

\subsection{Event selection}

The following criteria are used to keep events of interest.  Single-object criteria accept events containing:
\begin{itemize}
\item a central electron with $p_T>25$~GeV, or
\item a plug electron with $p_T>40$~GeV, or
\item a central muon with $p_T>25$~GeV, or
\item a central photon with $p_T>60$~GeV, or
\item a central or plug photon with $p_T>300$~GeV, or
\item a central jet or $b$-tagged jet with $p_T>200$~GeV, or
\item a central $b$-tagged jet with $p_T>60$~GeV (prescaled by the online jet20 trigger), or
\item a central jet or $b$-tagged jet with $p_T>40$~GeV (prescaled by 10 in addition to the online jet20 trigger prescale).
\end{itemize}
Di-object criteria keep events containing:
\begin{itemize}
\item one electron plus one electron or photon with $\abs{\eta}<2.5$ and $p_T>25$~GeV, or
\item a central or plug electron with $p_T>40$~GeV and a central tau with $p_T>17$~GeV, or
\item a central muon with $p_T>17$~GeV and a central or plug photon with $p_T>25$~GeV, or
\item a central muon with $p_T>25$~GeV and a central $b$-tagged jet with $p_T>17$~GeV, or
\item two taus with $\abs{\eta}<1.0$ and $p_T>25$~GeV, or
\item a central or plug photon with $p_T>40$~GeV and a central tau with $p_T>40$~GeV, or
\item one central photon with $p_T>25$~GeV and one other central or plug photon with $p_T>25$~GeV, or
\item a central photon with $p_T>40$~GeV and a central $b$-tagged jet with $p_T>25$~GeV, or
\item a central jet or $b$-tagged jet with $p_T>40$~GeV and a central tau with $p_T>17$~GeV (prescaled by the online jet20 trigger), or
\item a central jet with $p_T>60$~GeV and a central $b$-tagged jet with $p_T>25$~GeV (prescaled by the online jet20 trigger), or
\item two central muons with $p_T>17$~GeV, or
\item one central electron and one central muon with $p_T>17$~GeV, or
\item one central electron with $p_T>20$~GeV and one central tau with $p_T>17$~GeV, or
\item one plug electron with $p_T>25$~GeV and one central muon with $p_T>17$~GeV, or
\item one central muon with $p_T>20$~GeV and one central tau with $p_T>17$~GeV.
\end{itemize}
Tri-object criteria keep events containing:
\begin{itemize}
\item a central or plug photon with $p_T>40$~GeV and two central taus with $p_T>17$~GeV, or
\item a central or plug photon with $p_T>40$~GeV and two central $b$-tagged jets with $p_T>25$~GeV, or
\item a central or plug photon with $p_T>40$~GeV, a central tau with $p_T>25$~GeV, and a central $b$-tagged jet with $p_T>25$~GeV, or
\item one $b$-tagged jet with $p_T>90$~GeV and two more $b$-tagged jets with $p_T>60$~GeV, or
\item one central muon with $p_T>17$~GeV and two other central or plug muons with $p_T>17$~GeV.
\end{itemize}
Additional special criteria accept events containing:
\begin{itemize}
\item one central or plug electron with $p_T>40$~GeV, missing transverse momentum greater than $17$~GeV, and two or more jets or central $b$-tagged jets with $p_T>17$~GeV, or
\item one central muon with $p_T>25$~GeV, missing transverse momentum greater than $17$~GeV, and two or more jets or central $b$-tagged jets with $p_T>17$~GeV.
\end{itemize}

The above criteria are set by the requirements that the corresponding Standard Model prediction can be generated with enough Monte Carlo event to have weights $\lesssim1$, and that trigger efficiencies can be treated as roughly independent of object $p_T$, while keeping as many potentially interesting events as possible.

Explicit online trigger paths are no longer required.  CDF specific details are provided in Sec.~\ref{sec:trigger2}.

\subsection{Event generation}
\label{sec:eventGeneration2}

Here are summarized changes made to our Monte Carlo event generation since the first round of analysis.

\begin{itemize}
\item A number of electroweak samples changed to use the newest (Gen6) \CdfSim\ version.  They include (the Stntuple sample names are given in parentheses):
 Pythia $W\rightarrow e \nu$~(we0sfe, we0sge, we0she),
 Pythia $W\rightarrow \mu \nu$~(we0s8m, we0s9m),
 Pythia $W\rightarrow \tau \nu$~(we0s9t, we0sat),
 Pythia $Z\rightarrow ee$~(ze1s6d, ze1sad, ze0scd, ze0sdd, ze0sed, ze0see),
 Pythia $Z\rightarrow \mu\mu$~(ze1s9m, ze0sbm, ze0scm, ze0sdm, ze0sem),
 Pythia $Z\rightarrow \tau\tau$~(ze0s8t, ze0sat),
 Pythia WW (we0sbd, we0sgd),
 Pythia WZ (we0scd),
 Baur $W(\rightarrow  e\nu)+\gamma$~(re0s28, re0s48),
 Baur $W(\rightarrow  \mu\nu)+\gamma$~(re0s29, re0s49),
 Baur $W(\rightarrow  \tau\nu)+\gamma$~(re0s1a, re0s4a).
\item A low mass Drell-Yan sample was added with $M_Z$ going down to 10 GeV (zx0sde, zx0sdm)
\item We switched from using the Mrenna matched W+jets sample to the standard Top Group Alpgen W+jets samples: 
$W(\rightarrow e\nu)$+jets (ptopw0, ptopw1, ptop2w, ptop3w, ptop4w),
$W(\rightarrow\mu\nu)$+jets (ptopw5, ptopw6, ptop7w, ptop8w, ptop9w),
$W(\rightarrow \tau\nu)$+jets (utopw0, utopw1, utop2w, utop3w, utop4w).
\item We switched from using MadEvent W+bbar to the standard Top Group W+bbar sample: 
$W(\rightarrow e\nu)$+bb+jets (btop0w, btop1w, btop2w), 
$W(\rightarrow \mu\nu)$+bb+jets (btop5w, btop6w, btop7w).
\end{itemize}

Table~\ref{tbl:SumWeight} summarizes the contributions from each Monte Carlo sample.

\begin{table}
\tiny
\hspace{-1.5cm}
\begin{minipage}{6.5in}
\begin{verbatim}
Dataset           Process                   Weights * Number  =  Total weight  |  Dataset          Process                   Weights * Number  =  Total weight
-----------       ----------------          -------   ------     ------------  |  -----------      ----------------          -------   ------     ------------
pyth_jj_000       Pythia jj 0<pT<10         1700           1         1720.13   |  alpgen_muvmj     Alpgen W(-> mu v) j           0.3    281072        83565.2
pyth_jj_010       Pythia jj 10<pT<18         330          74        24368.4    |  ut0s2w           Alpgen W(-> tau v)+jets       0.29     5230         1537.91
pyth_pj_008       Pythia j gamma 8<pT<12      86           5          430.59   |  mad_vtvt-a       MadEvent Z(->vv) gamma        0.27      136           37.22
mrenna_mu+mu-     MadEvent Z(-> mu mu)        29         220         6478.07   |  mad_veve-a       MadEvent Z(->vv) gamma        0.27      135           36.91
pyth_jj_090       Pythia jj 90<pT<120         22        2064        45825.1    |  we0s9t           Pythia W(-> tau v)            0.26    66024        17105.3
pyth_pj_012       Pythia j gamma 12<pT<22     21        1970        41809.2    |  ut0sw1           Alpgen W(-> tau v)+jets       0.24    27785         6634.6
pyth_jj_018       Pythia jj 18<pT<40          18       24807       444683      |  pyth_pp          Pythia gamma gamma            0.23    25552         5783.53
alpgen_eve        Alpgen W(->e v)             12        5807        68133.2    |  ze1s6d           Pythia Z(->ee)                0.22   484676       106397
mad_vtvt-j        MadEvent Z(->vv) j          11           3           32.25   |  mad_e+e-b-b      MadEvent Z(->ee) bb           0.22     1028          224
mad_veve-j        MadEvent Z(->vv) j          11           3           32.01   |  alpgen_evejj     Alpgen W(->e v) jj            0.21   175665        37470.9
mrenna_e+e-       MadEvent Z(->ee)            10        5965        60080.4    |  re0s28           Baur W(->ev) gamma            0.21    22074         4698.99
alpgen_muvm       Alpgen W(-> mu v)            9.9      4483        44217.8    |  alpgen_muvmjj    Alpgen W(-> mu v) jj          0.2    112546        22201.2
pyth_jj_120       Pythia jj 120<pT<150         8.2      3291        27109.7    |  ztopcz           Pythia ZZ                     0.19      588          110.12
pyth_bj_010       Pythia bj 10<pT<18           7.7        96          739.91   |  stelzer_Zaj      stelzer_Zaj                   0.18     1592          288.68
pyth_jj_060       Pythia jj 60<pT<90           6.7     25300       170628      |  mad_aajj         MadEvent jj gamma gamma       0.18     7825         1406.23
mrenna_mu+mu-j    MadEvent Z(-> mu mu) j       6.6      3209        21131      |  mad_mu+mu-b-b    MadEvent Z(-> mu mu) bb       0.17      624          109.01
pyth_jj_040       Pythia jj 40<pT<60           5       87760       440765      |  mad_e+e-jj       MadEvent Z(->ee) jj           0.17      775          134.36
pyth_jj_200       Pythia jj 200<pT<300         3.4     73024       249462      |  re0s29           Baur W(-> mu v) gamma         0.17    19972         3454.5
mad_veve-a_f      MadEvent Z(->vv) gamma       3.4        13           44.3    |  re0s1a           Baur W(-> tau v) gamma        0.17     2823          467.2
ut0sw0            Alpgen W(-> tau v)+jets      3.2       643         2065.45   |  pyth_jj_300      Pythia jj 300<pT<400          0.14   103800        14883.7
pyth_pj_022       Pythia j gamma 22<pT<45      3       31039        93761.3    |  mad_aaa_f        MadEvent gamma gamma gamma    0.14       52            7.4
pyth_jj_150       Pythia jj 150<pT<200         2.7     59183       162311      |  cosmic_j_hi      Cosmic (jet100)               0.12    36674         4487.59
we0sfe            Pythia W(->e v)              2.4    381263       921806      |  pyth_bj_040      Pythia bj 40<pT<60            0.12   160713        18850
cosmic_j_lo       Cosmic (jet20)               2.3       122          277.24   |  mrenna_e+e-jjj   MadEvent Z(->ee) jjj          0.11    23995         2667.48
cosmic_ph         Cosmic (photon_25_iso)       1.9      2694         4989.53   |  ze0s8t           Pythia Z(-> tau tau)          0.093   15030         1400.09
pyth_pj_080       Pythia j gamma 80<pT         1.5     18378        28063.9    |  pyth_bj_200      Pythia bj 200<pT<300          0.081  254807        20679.8
mrenna_e+e-j      MadEvent Z(->ee) j           1.5     28104        40784.4    |  hewk03           MadEvent Z(->ee) gamma        0.081   70476         5709.75
pyth_pj_045       Pythia j gamma 45<pT<80      1.4     82466       117398      |  mad_aaa          MadEvent gamma gamma gamma    0.08       73            5.82
mrenna_mu+mu-jj   MadEvent Z(-> mu mu) jj      1.3      4146         5503.42   |  wenubb0p         Alpgen W(->e v) bb            0.075   41459         3105.72
mad_veve-j_f      MadEvent Z(->vv) j           1.2         6            7.23   |  wmnubb0p         Alpgen W(-> mu v) bb          0.075   26067         1952.18
pyth_bj_018       Pythia bj 18<pT<40           1.2     15659        18163.1    |  zx0sem           Pythia Z(-> mu mu) (m_Z<20)   0.075      45            3.37
mad_e+e-          MadEvent Z(->ee)             1         520          540.85   |  zx0see           Pythia Z(->ee) (m_Z<20)       0.074      73            5.37
stelzer_l+l-j     stelzer_l+l-j                0.92      668          615.62   |  wenubb1p         Alpgen W(->e v) bb j          0.072   14111         1021.01
mrenna_e+e-jj     MadEvent Z(->ee) jj          0.92    11258        10307.9    |  wmnubb1p         Alpgen W(-> mu v) bb j        0.072    8426          609.28
mad_mu+mu-        MadEvent Z(-> mu mu)         0.89       82           72.79   |  hewk04           MadEvent Z(-> mu mu) gamma    0.072    2032          145.67
pyth_bj_060       Pythia bj 60<pT<90           0.87    10723         9348.26   |  overlay          Overlaid events               0.071   11118          794.91
mad_vtvt-a_f      MadEvent Z(->vv) gamma       0.84       38           32      |  pyth_jj_400      Pythia jj 400<pT              0.068   13153          894.13
pyth_bj_090       Pythia bj 90<pT<120          0.84     2374         1990.9    |  alpgen_evejjj    Alpgen W(->e v) jjj           0.068   92857         6284.23
mad_vtvt-j_f      MadEvent Z(->vv) j           0.81        5            4.07   |  alpgen_muvmjjj   Alpgen W(-> mu v) jjj         0.066   55704         3693.97
stelzer_Waj       MadEvent W(->l v)j gamma     0.69     1637         1124.5    |  ttop0z           Herwig ttbar                  0.065   30518         1983.29
pyth_bj_120       Pythia bj 120<pT<150         0.67     2848         1903.72   |  ut0s3w           Alpgen W(-> tau v)+jets       0.063    4458          281.47
mad_aaj           MadEvent j gamma gamma       0.52      559          289.03   |  ze0sat           Pythia Z(-> tau tau)          0.063   22882         1438.37
we0s8m            Pythia W(-> mu v)            0.49  1.2908e+06    630955      |  wmnubb2p         Alpgen W(-> mu v) bb jj       0.054    3529          189.63
pyth_bj_150       Pythia bj 150<pT<200         0.45    28272        12593.5    |  wenubb2p         Alpgen W(->e v) bb jj         0.054    6075          325.2
mrenna_mu+mu-jjj  MadEvent Z(-> mu mu) jjj     0.44     3435         1498.33   |  we0scd           Pythia WZ                     0.053    2890          154.28
mad_e+e-j         MadEvent Z(->ee) j           0.39      735          286.71   |  we0sbd           Pythia WW                     0.048    2839          136.4
alpgen_evej       Alpgen W(->e v) j            0.35   398558       140544      |  we0sgd           Pythia WW                     0.048    2567          122.92
we0sat            Pythia W(-> tau v)           0.35    49543        17155.3    |  alpgen_evejjjj   Alpgen W(->e v) jjjj          0.027   41696         1123.15
mad_mu+mu-j       MadEvent Z(-> mu mu) j       0.34      494          166.35   |  alpgen_muvmjjjj  Alpgen W(-> mu v) jjjj        0.024   27099          662.36
mad_mu+mu-jj      MadEvent Z(-> mu mu) jj      0.32     1681          532      |  ut0s4w           Alpgen W(-> tau v)+jets       0.022    2551           56.84
ze1s9m            Pythia Z(-> mu mu)           0.3    370995       110522      |  Total:                                                           4.36864e+06
\end{verbatim}\end{minipage}
\caption[The number of events contributing from each Standard Model process.]{The number of events contributing from each Standard Model process, ordered according to decreasing effective weight of individual Monte Carlo events.  The data set names are shown in the leftmost column, with the corresponding process shown in the second column.  The typical weight of individual events from each process is shown in the third column, and the ``effective'' number of events from each process contributing to the background estimate is shown in the fourth column.  The weight from each process is totaled in the rightmost column, and the total weight is provided at bottom.  The total weight is equal to the roughly four million events included in this analysis.}
\label{tbl:SumWeight}
\end{table}

\begin{table}
\centering
\footnotesize
\vspace{-1cm}
\begin{tabular}{lllllll}
{\bf Code } & {\bf Category } & {\bf Explanation } & {\bf Value } & {\bf Error } & {\bf Error(\%) } \\ \hline 
0001 & luminosity & CDF integrated luminosity & 1990 & 50 & 2.6 \\ 
0002 & $k$-factor & cosmic\_ph & 0.83 & 0.05 & 6.0 \\ 
0003 & $k$-factor & cosmic\_j & 0.192 & 0.006 & 3.1 \\ 
0004 & $k$-factor & 1$\gamma$1j photon+jet(s) & 0.92 & 0.04 & 4.4 \\ 
0005 & $k$-factor & 1$\gamma$2j & 1.26 & 0.05 & 4.0 \\ 
0006 & $k$-factor & 1$\gamma$3j & 1.61 & 0.08 & 5.0 \\ 
0007 & $k$-factor & 1$\gamma$4j+ & 1.94 & 0.16 & 8.3 \\ 
0008 & $k$-factor & 2$\gamma$0j diphoton(+jets) & 1.6 & 0.08 & 5.0 \\ 
0009 & $k$-factor & 2$\gamma$1j & 2.99 & 0.17 & 5.7 \\ 
0010 & $k$-factor & 2$\gamma$2j+ & 1.2 & 0.09 & 7.5 \\ 
0011 & $k$-factor & W0j W (+jets) & 1.38 & 0.03 & 2.2 \\ 
0012 & $k$-factor & W1j & 1.33 & 0.03 & 2.3 \\ 
0013 & $k$-factor & W2j & 1.99 & 0.05 & 2.5 \\ 
0014 & $k$-factor & W3j+ & 2.11 & 0.09 & 4.3 \\ 
0015 & $k$-factor & Z0j Z (+jets) & 1.39 & 0.028 & 2.0 \\ 
0016 & $k$-factor & Z1j & 1.23 & 0.04 & 3.2 \\ 
0017 & $k$-factor & Z2j+ & 1.02 & 0.04 & 3.9 \\ 
0018 & $k$-factor & 2j $\hat{p}_T$$<$150 dijet & 1.003 & 0.027 & 2.7 \\ 
0019 & $k$-factor & 2j 150$<$$\hat{p}_T$ & 1.34 & 0.03 & 2.2 \\ 
0020 & $k$-factor & 3j $\hat{p}_T$$<$150 multijet & 0.941 & 0.025 & 2.7 \\ 
0021 & $k$-factor & 3j 150$<$$\hat{p}_T$ & 1.48 & 0.04 & 2.7 \\ 
0022 & $k$-factor & 4j $\hat{p}_T$$<$150 & 1.06 & 0.03 & 2.8 \\ 
0023 & $k$-factor & 4j 150$<$$\hat{p}_T$ & 1.93 & 0.06 & 3.1 \\ 
0024 & $k$-factor & 5j low & 1.33 & 0.05 & 3.8 \\ 
0025 & $k$-factor & 1b2j 150$<$$\hat{p}_T$ & 2.22 & 0.11 & 5.0 \\ 
0026 & $k$-factor & 1b3j 150$<$$\hat{p}_T$ & 2.98 & 0.15 & 5.0 \\ 
0027 & misId & p(e$\rightarrow$e) central & 0.978 & 0.006 & 0.6 \\ 
0028 & misId & p(e$\rightarrow$e) plug & 0.966 & 0.007 & 0.7 \\ 
0029 & misId & p($\mu$$\rightarrow$$\mu$) CMUP+CMX & 0.888 & 0.007 & 0.8 \\ 
0030 & misId & p($\gamma$$\rightarrow$$\gamma$) central & 0.949 & 0.018 & 1.9 \\ 
0031 & misId & p($\gamma$$\rightarrow$$\gamma$) plug & 0.859 & 0.016 & 1.9 \\ 
0032 & misId & p(b$\rightarrow$b) central & 0.978 & 0.021 & 2.1 \\ 
0033 & misId & p($\gamma$$\rightarrow$e) plug & 0.06 & 0.003 & 5.0 \\ 
0034 & misId & p(q$\rightarrow$e) central & 7.09$\times 10^{-5}$ & 1.9$\times 10^{-6}$ & 2.7 \\ 
0035 & misId & p(q$\rightarrow$e) plug & 0.000766 & 1.2$\times 10^{-5}$ & 1.6 \\ 
0036 & misId & p(q$\rightarrow$$\mu$) & 1.14$\times 10^{-5}$ & 6$\times 10^{-7}$ & 5.2 \\ 
0037 & misId & p(b$\rightarrow$$\mu$) & 3.3$\times 10^{-5}$ & 1.1$\times 10^{-5}$ & 33.0 \\ 
0038 & misId & p(j$\rightarrow$b) 25$<$$p_T$ & 0.0183 & 0.0002 & 1.1 \\ 
0039 & misId & p(q$\rightarrow$$\tau$) & 0.0052 & 0.0001 & 1.9 \\ 
0040 & misId & p(q$\rightarrow$$\gamma$) central & 0.000266 & 1.4$\times 10^{-5}$ & 5.3 \\ 
0041 & misId & p(q$\rightarrow$$\gamma$) plug & 0.00048 & 6$\times 10^{-5}$ & 12.6 \\ 
0042 & trigger & p(e$\rightarrow$trig) plug, $p_T$$>$25 & 0.86 & 0.007 & 0.8 \\ 
0043 & trigger & p($\mu$$\rightarrow$trig) CMUP+CMX, $p_T$$>$25 & 0.916 & 0.004 & 0.4 \\ 
\end{tabular}
\caption[The correction factors of \Vista\ correction model.]{The correction factors of \Vista\ correction model.  The best fit values ({\tt Value}) are given in the 4\th\ column.  Correction factor errors ({\tt Error}) resulting from the fit are shown in the 5\th\ column.  The fractional error ({\tt Error(\%)}) is listed in the 6\th\ column.  All values are dimensionless except for the first one, which represents integrated luminosity and has units of pb$^{-1}$.  These values and uncertainties are valid within the context of this correction model.}
\label{tbl:CorrectionFactorDescriptionValuesSigmas}
\end{table}

Specific modifications to the correction model implemented since the first round are described here.

\begin{itemize}
 \item  The integrated luminosity of the data sample considered has increased from 927 to 1990~pb$^{-1}$.  The integrated luminosity correction factor has been adjusted accordingly.
 \item  Events from more recent data have been included in the high-$p_T$ jet and photon non-collision backgrounds.  For events with $\sumPt > 400$ GeV and at least two jets of $p_T > 10$ GeV and no objects of other kinds, we require the $p_T$ of the jet with the second largest $p_T$ to be greater than 75 GeV.  This cut is to clean multijet samples of events where the second jet comes from the underlying event but the first jet is due to a cosmic ray.  Such events are not modeled well by our cosmic background, which comprises events required to have less than three tracks; this requirement reduces the fraction of such cosmic + jet(s) events relative to the data sample, where more than three tracks are required. As a result of these changes, the {\tt cosmic\_ph} and {\tt cosmic\_j} correction factors have been readjusted. 
 \item  It was recognized that in the previous version of the analysis we had been using a suboptimal filter for conversion electrons. This filter has been updated and now yields a substantially reduced rate for jets faking electrons via fragmentation to a leading $\pi^0$.
 \item  In order to accommodate the ditau trigger, which in recent data requires a seed track with $p_T>10$~GeV, and recognizing our concentration on the identification of single-prong taus, the track requirement for taus has been increased to 10.5~GeV.  The fake rate $\poo{j}{\tau}$ and its dependence on $p_T$ have been adjusted accordingly.
 \item  In order to address questions regarding the fake rate $\poo{j}{\tau}$ and its consistent simultaneous application to many final states, the measurement of tau $p_T$ is now based on the energy deposited in the calorimeter.
 \item  In order to address questions regarding the fake rate $\poo{j}{\tau}$ and its consistent simultaneous application to multijet final states with large and small \sumPt, a monotonically decreasing dependence of the fake rate $\poo{j}{\tau}$ on the generated summed scalar transverse momentum has been imposed.
 \item  In the implementation of the fake rates $\poo{j}{e}$, $\poo{j}{\mu}$, and $\poo{j}{\tau}$, jets from a parent $u$ or $\bar{d}$ quark now only fake positively charged $\mu$ and $\tau$ leptons (and positrons rather than electrons at a ratio of 2:1), and jets from a parent $\bar{u}$ or $d$ quark now only fake negatively charged $\mu$ and $\tau$ leptons (and electrons rather than positrons at a ratio of 2:1).
 \item  The ditau trigger, which turned on roughly 300~pb$^{-1}$ into Run II, has now been live for a greater fraction of the total integrated dataset.  The effective ditau trigger effeciency has been adjusted accordingly.
 \item  A fake rate $\poo{b}{\mu}$ has been introduced.
 \item  The $p_T$ dependence of the fake rate $\poo{j}{b}$ and $\poo{j}{tau}$ has been adjusted.
 \item  The \detEta\ and $\phi$ dependence of the fake rate $\poo{j}{e}$ and $\poo{j}{ph}$ has been adjusted to take into account more geometric features of the detector including the calorimeter cracks at \detEta\ of 0 and 1.1.
 \item  The efficiency for reconstructing a jet as a non-$b$-tagged jet has been reduced from 1 to 1-$\poo{j}{b}$. 
 \item  Separate $k$-factors have been introduced for heavy flavor multijet production for the high-$p_T$ sample.  Specifically, a new $k$-factor has been introduced for events with at least one heavy flavor jet and three jets in total, with $\hat{p_T} > 150~GeV$. Another $k$-factor has been introduced for events with at least one heavy flavor jet and four or more jets in total, with $\hat{p_T} > 150~GeV$. They are listed in the table of correction factors \ref{tbl:CorrectionFactorDescriptionValuesSigmas} as 1b2j and 1b3j.
\end{itemize}

\subsection{Results}

\begin{table*}
\tiny
\hspace{-0.5cm}
\begin{minipage}{9in}
\begin{tabular}{l@{ }r@{ }r@{ $\pm$ }l@{ }l}
{\bf Final State} & {\bf Data} & \multicolumn{2}{c}{\bf Background} & {\bf $\sigma$} \\ \hline 
b$e^\pm$$p\!\!\!/$ & $690$ & $817.7$ & $9.2$ & $-2.7$ \\ 
$\gamma$$\tau^\pm$ & $1371$ & $1217.6$ & $13.3$ & $+2.2$ \\ 
$\mu^\pm$$\tau^\pm$ & $63$ & $35.2$ & $2.8$ & $+1.7$ \\ 
b2j$p\!\!\!/$ high-$\Sigma p_T$ & $255$ & $327.2$ & $8.9$ & $-1.7$ \\ 
2j$\tau^\pm$ low-$\Sigma p_T$ & $574$ & $670.3$ & $8.6$ & $-1.5$ \\ 
3j$\tau^\pm$ low-$\Sigma p_T$ & $148$ & $199.8$ & $5.2$ & $-1.4$ \\ 
$e^\pm$$p\!\!\!/$$\tau^\pm$ & $36$ & $17.2$ & $1.7$ & $+1.4$ \\ 
2j$\tau^\pm$$\tau^\mp$ & $33$ & $62.1$ & $4.3$ & $-1.3$ \\ 
$e^\pm$j & $741710$ & $764832$ & $6447.2$ & $-1.3$ \\ 
j$2\tau^\pm$ & $105$ & $150.8$ & $6.3$ & $-1.2$ \\ 
$e^\pm$2j & $256946$ & $249148$ & $2201.5$ & $+1.2$ \\ 
2bj low-$\Sigma p_T$ & $279$ & $352.5$ & $11.9$ & $-1.1$ \\ 
j$\tau^\pm$ low-$\Sigma p_T$ & $1385$ & $1525.8$ & $15$ & $-1.1$ \\ 
2b2j low-$\Sigma p_T$ & $108$ & $153.5$ & $6.8$ & $-1$ \\ 
b$\mu^\pm$$p\!\!\!/$ & $528$ & $613.5$ & $8.7$ & $-0.9$ \\ 
$\mu^\pm$$\gamma$$p\!\!\!/$ & $523$ & $611$ & $12.1$ & $-0.8$ \\ 
2b$\gamma$ & $108$ & $70.5$ & $7.9$ & $+0.1$ \\ 
8j & $14$ & $13.1$ & $4.4$ & $0$ \\ 
7j & $103$ & $97.8$ & $12.2$ & $0$ \\ 
6j & $653$ & $659.7$ & $37.3$ & $0$ \\ 
5j & $3157$ & $3178.7$ & $67.1$ & $0$ \\ 
4j high-$\Sigma p_T$ & $88546$ & $89096.6$ & $935.2$ & $0$ \\ 
4j low-$\Sigma p_T$ & $14872$ & $14809.6$ & $186.3$ & $0$ \\ 
4j$2\gamma$ & $46$ & $46.4$ & $3.9$ & $0$ \\ 
4j$\tau^\pm$ high-$\Sigma p_T$ & $29$ & $26.6$ & $1.7$ & $0$ \\ 
4j$\tau^\pm$ low-$\Sigma p_T$ & $43$ & $63.1$ & $3.3$ & $0$ \\ 
4j$p\!\!\!/$ high-$\Sigma p_T$ & $1064$ & $1012$ & $62.9$ & $0$ \\ 
4j$\gamma$$\tau^\pm$ & $19$ & $10.8$ & $2$ & $0$ \\ 
4j$\gamma$$p\!\!\!/$ & $62$ & $104.2$ & $22.4$ & $0$ \\ 
4j$\gamma$ & $7962$ & $8271.2$ & $245.1$ & $0$ \\ 
4j$\mu^\pm$$p\!\!\!/$ & $574$ & $590.5$ & $13.6$ & $0$ \\ 
4j$\mu^\pm$$\mu^\mp$ & $38$ & $48.4$ & $6.2$ & $0$ \\ 
4j$\mu^\pm$ & $1363$ & $1350.1$ & $37.7$ & $0$ \\ 
3j high-$\Sigma p_T$ & $159926$ & $159143$ & $1061.9$ & $0$ \\ 
3j low-$\Sigma p_T$ & $62681$ & $64213.1$ & $496$ & $0$ \\ 
3j$2\gamma$ & $151$ & $177.5$ & $7.1$ & $0$ \\ 
3j$\tau^\pm$ high-$\Sigma p_T$ & $68$ & $76.9$ & $3$ & $0$ \\ 
3j$p\!\!\!/$ high-$\Sigma p_T$ & $1706$ & $1899.4$ & $77.6$ & $0$ \\ 
3j$p\!\!\!/$ low-$\Sigma p_T$ & $42$ & $36.2$ & $5.7$ & $0$ \\ 
3j$\gamma$$\tau^\pm$ & $39$ & $37.8$ & $3.6$ & $0$ \\ 
3j$\gamma$$p\!\!\!/$ & $204$ & $249.8$ & $24.4$ & $0$ \\ 
3j$\gamma$ & $24639$ & $24899.4$ & $372.4$ & $0$ \\ 
3j$\mu^\pm$$p\!\!\!/$ & $2884$ & $2971.5$ & $52.1$ & $0$ \\ 
3j$\mu^\pm$$\gamma$$p\!\!\!/$ & $10$ & $3.6$ & $1.9$ & $0$ \\ 
3j$\mu^\pm$$\gamma$ & $15$ & $7.9$ & $2.9$ & $0$ \\ 
3j$\mu^\pm$$\mu^\mp$ & $175$ & $177.8$ & $16.2$ & $0$ \\ 
3j$\mu^\pm$ & $5032$ & $4989.5$ & $108.9$ & $0$ \\ 
3b2j & $23$ & $28.9$ & $4.7$ & $0$ \\ 
3bj & $82$ & $82.6$ & $5.7$ & $0$ \\ 
3b & $67$ & $85.6$ & $7.7$ & $0$ \\ 
$2\tau^\pm$ & $498$ & $512.7$ & $14.2$ & $0$ \\ 
$2\gamma$$p\!\!\!/$ & $128$ & $107.2$ & $6.9$ & $0$ \\ 
$2\gamma$ & $5548$ & $5562.8$ & $40.5$ & $0$ \\ 
2j high-$\Sigma p_T$ & $190773$ & $190842$ & $781.2$ & $0$ \\ 
2j low-$\Sigma p_T$ & $165984$ & $162530$ & $1581$ & $0$ \\ 
2j$2\tau^\pm$ & $22$ & $40.6$ & $3.2$ & $0$ \\ 
2j$2\gamma$$p\!\!\!/$ & $11$ & $8$ & $2.4$ & $0$ \\ 
2j$2\gamma$ & $580$ & $581$ & $13.7$ & $0$ \\ 
2j$\tau^\pm$ high-$\Sigma p_T$ & $96$ & $114.6$ & $3.3$ & $0$ \\ 
\end{tabular}
\begin{tabular}{l@{ }r@{ }r@{ $\pm$ }l@{ }l}
{\bf Final State} & {\bf Data} & \multicolumn{2}{c}{\bf Background} & {\bf $\sigma$} \\ \hline 
2j$p\!\!\!/$ high-$\Sigma p_T$ & $87$ & $80.9$ & $6.8$ & $0$ \\ 
2j$p\!\!\!/$ low-$\Sigma p_T$ & $114$ & $79.5$ & $100.8$ & $0$ \\ 
2j$p\!\!\!/$$\tau^\pm$ & $18$ & $13.2$ & $2.2$ & $0$ \\ 
2j$\gamma$$\tau^\pm$ & $142$ & $144.6$ & $5.7$ & $0$ \\ 
2j$\gamma$$p\!\!\!/$ & $908$ & $980.3$ & $63.7$ & $0$ \\ 
2j$\gamma$ & $71364$ & $73021.4$ & $595.9$ & $0$ \\ 
2j$\mu^\pm$$\tau^\mp$ & $16$ & $19.3$ & $2.2$ & $0$ \\ 
2j$\mu^\pm$$p\!\!\!/$ & $17927$ & $18340.6$ & $201.9$ & $0$ \\ 
2j$\mu^\pm$$\gamma$$p\!\!\!/$ & $31$ & $27.7$ & $7.7$ & $0$ \\ 
2j$\mu^\pm$$\gamma$ & $57$ & $58.2$ & $13$ & $0$ \\ 
2j$\mu^\pm$$\mu^\mp$$p\!\!\!/$ & $11$ & $7.8$ & $2.7$ & $0$ \\ 
2j$\mu^\pm$$\mu^\mp$ & $956$ & $924.9$ & $61.2$ & $0$ \\ 
2j$\mu^\pm$ & $22461$ & $23111.4$ & $366.6$ & $0$ \\ 
$2e^\pm$j & $14$ & $13.8$ & $2.3$ & $0$ \\ 
$2e^\pm$$e^\mp$ & $20$ & $17.5$ & $1.7$ & $0$ \\ 
$2e^\pm$ & $32$ & $49.2$ & $3.4$ & $0$ \\ 
2b high-$\Sigma p_T$ & $666$ & $689$ & $9.4$ & $0$ \\ 
2b low-$\Sigma p_T$ & $323$ & $313.2$ & $10.3$ & $0$ \\ 
2b3j low-$\Sigma p_T$ & $53$ & $57.4$ & $6.5$ & $0$ \\ 
2b2j high-$\Sigma p_T$ & $718$ & $803.3$ & $12.7$ & $0$ \\ 
2b2j$p\!\!\!/$ high-$\Sigma p_T$ & $15$ & $21.8$ & $2.8$ & $0$ \\ 
2b2j$\gamma$ & $32$ & $39.7$ & $6.2$ & $0$ \\ 
2b2j$\mu^\pm$$p\!\!\!/$ & $14$ & $17.3$ & $1.9$ & $0$ \\ 
2b2j$\mu^\pm$ & $22$ & $21.8$ & $2$ & $0$ \\ 
2b$\mu^\pm$$p\!\!\!/$ & $11$ & $14.4$ & $2.1$ & $0$ \\ 
2bj high-$\Sigma p_T$ & $891$ & $967.1$ & $13.2$ & $0$ \\ 
2bj$p\!\!\!/$ high-$\Sigma p_T$ & $25$ & $31.3$ & $3.1$ & $0$ \\ 
2bj$\gamma$ & $71$ & $54.5$ & $7.1$ & $0$ \\ 
2bj$\mu^\pm$$p\!\!\!/$ & $12$ & $10.7$ & $1.9$ & $0$ \\ 
2b$e^\pm$2j$p\!\!\!/$ & $30$ & $27.3$ & $2.2$ & $0$ \\ 
2b$e^\pm$2j & $72$ & $66.5$ & $2.9$ & $0$ \\ 
2b$e^\pm$$p\!\!\!/$ & $22$ & $19.1$ & $2.2$ & $0$ \\ 
2b$e^\pm$j$p\!\!\!/$ & $19$ & $19.4$ & $2.2$ & $0$ \\ 
2b$e^\pm$j & $63$ & $63$ & $3.4$ & $0$ \\ 
2b$e^\pm$ & $96$ & $92.1$ & $4.1$ & $0$ \\ 
$\tau^\pm$$\tau^\mp$ & $856$ & $872.5$ & $19$ & $0$ \\ 
$\gamma$$p\!\!\!/$ & $3793$ & $3770.7$ & $127.3$ & $0$ \\ 
$\mu^\pm$$\tau^\mp$ & $381$ & $440.9$ & $7.3$ & $0$ \\ 
$\mu^\pm$$p\!\!\!/$$\tau^\mp$ & $60$ & $75.7$ & $3.4$ & $0$ \\ 
$\mu^\pm$$p\!\!\!/$$\tau^\pm$ & $15$ & $12$ & $2$ & $0$ \\ 
$\mu^\pm$$p\!\!\!/$ & $734290$ & $734296$ & $4897.8$ & $0$ \\ 
$\mu^\pm$$\gamma$ & $475$ & $469.8$ & $12.5$ & $0$ \\ 
$\mu^\pm$$\mu^\mp$$p\!\!\!/$ & $169$ & $198.5$ & $8.2$ & $0$ \\ 
$\mu^\pm$$\mu^\mp$$\gamma$ & $83$ & $60$ & $3.1$ & $0$ \\ 
$\mu^\pm$$\mu^\mp$ & $25283$ & $25178.5$ & $86.5$ & $0$ \\ 
j$2\gamma$$p\!\!\!/$ & $36$ & $30.4$ & $4.2$ & $0$ \\ 
j$2\gamma$ & $1822$ & $1813.2$ & $27.4$ & $0$ \\ 
j$\tau^\pm$ high-$\Sigma p_T$ & $52$ & $56.2$ & $2.5$ & $0$ \\ 
j$\tau^\pm$$\tau^\mp$ & $203$ & $252.2$ & $8.7$ & $0$ \\ 
j$p\!\!\!/$ high-$\Sigma p_T$ & $4432$ & $4431.7$ & $45.2$ & $0$ \\ 
j$\gamma$$\tau^\pm$ & $526$ & $476$ & $9.3$ & $0$ \\ 
j$\gamma$$p\!\!\!/$ & $1882$ & $1791.9$ & $72.3$ & $0$ \\ 
j$\gamma$ & $103319$ & $102124$ & $570.6$ & $0$ \\ 
j$\mu^\pm$$\tau^\mp$ & $71$ & $98$ & $3.9$ & $0$ \\ 
j$\mu^\pm$$\tau^\pm$ & $15$ & $12$ & $2$ & $0$ \\ 
j$\mu^\pm$$p\!\!\!/$$\tau^\mp$ & $26$ & $30.8$ & $2.6$ & $0$ \\ 
j$\mu^\pm$$p\!\!\!/$ & $109081$ & $108323$ & $707.7$ & $0$ \\ 
j$\mu^\pm$$\gamma$$p\!\!\!/$ & $171$ & $171.1$ & $31$ & $0$ \\ 
j$\mu^\pm$$\gamma$ & $152$ & $190$ & $39.3$ & $0$ \\ 
\end{tabular}
\begin{tabular}{l@{ }r@{ }r@{ $\pm$ }l@{ }l}
{\bf Final State} & {\bf Data} & \multicolumn{2}{c}{\bf Background} & {\bf $\sigma$} \\ \hline 
j$\mu^\pm$$\mu^\mp$$p\!\!\!/$ & $32$ & $32.2$ & $10.9$ & $0$ \\ 
j$\mu^\pm$$\mu^\mp$$\gamma$ & $14$ & $11.5$ & $2.6$ & $0$ \\ 
j$\mu^\pm$$\mu^\mp$ & $4852$ & $4271.2$ & $185.4$ & $0$ \\ 
j$\mu^\pm$ & $77689$ & $76987.5$ & $930.2$ & $0$ \\ 
$e^\pm$4j$p\!\!\!/$ & $903$ & $830.6$ & $13.2$ & $0$ \\ 
$e^\pm$4j$\gamma$ & $25$ & $29.2$ & $3.6$ & $0$ \\ 
$e^\pm$4j & $15750$ & $16740.4$ & $390.5$ & $0$ \\ 
$e^\pm$3j$\tau^\mp$ & $15$ & $21.1$ & $2.2$ & $0$ \\ 
$e^\pm$3j$p\!\!\!/$ & $4054$ & $4077.2$ & $63.6$ & $0$ \\ 
$e^\pm$3j$\gamma$ & $108$ & $79.3$ & $5$ & $0$ \\ 
$e^\pm$3j & $60725$ & $60409.3$ & $723.3$ & $0$ \\ 
$e^\pm$$2\gamma$ & $41$ & $34.2$ & $2.6$ & $0$ \\ 
$e^\pm$2j$\tau^\pm$ & $37$ & $47.2$ & $2.2$ & $0$ \\ 
$e^\pm$2j$\tau^\mp$ & $109$ & $95.9$ & $6.8$ & $0$ \\ 
$e^\pm$2j$p\!\!\!/$ & $25725$ & $25403.1$ & $209.4$ & $0$ \\ 
$e^\pm$2j$\gamma$$p\!\!\!/$ & $30$ & $31.8$ & $4.8$ & $0$ \\ 
$e^\pm$2j$\gamma$ & $398$ & $342.8$ & $15.7$ & $0$ \\ 
$e^\pm$2j$\mu^\mp$$p\!\!\!/$ & $22$ & $14.8$ & $1.9$ & $0$ \\ 
$e^\pm$2j$\mu^\mp$ & $23$ & $15.8$ & $2$ & $0$ \\ 
$e^\pm$$\tau^\pm$ & $437$ & $387$ & $5.3$ & $0$ \\ 
$e^\pm$$\tau^\mp$ & $1333$ & $1266$ & $12.3$ & $0$ \\ 
$e^\pm$$p\!\!\!/$$\tau^\mp$ & $109$ & $106.1$ & $2.7$ & $0$ \\ 
$e^\pm$$p\!\!\!/$ & $960826$ & $956579$ & $3077.7$ & $0$ \\ 
$e^\pm$$\gamma$$p\!\!\!/$ & $497$ & $496.8$ & $10.3$ & $0$ \\ 
$e^\pm$$\gamma$ & $3578$ & $3589.9$ & $24.1$ & $0$ \\ 
$e^\pm$$\mu^\pm$$p\!\!\!/$ & $31$ & $29.9$ & $1.6$ & $0$ \\ 
$e^\pm$$\mu^\mp$$p\!\!\!/$ & $109$ & $99.4$ & $2.4$ & $0$ \\ 
$e^\pm$$\mu^\pm$ & $45$ & $28.5$ & $1.8$ & $0$ \\ 
$e^\pm$$\mu^\mp$ & $350$ & $313$ & $5.4$ & $0$ \\ 
$e^\pm$j$2\gamma$ & $13$ & $16.1$ & $3.9$ & $0$ \\ 
$e^\pm$j$\tau^\mp$ & $386$ & $418$ & $18.9$ & $0$ \\ 
$e^\pm$j$\tau^\pm$ & $160$ & $162.8$ & $3.5$ & $0$ \\ 
$e^\pm$j$p\!\!\!/$$\tau^\mp$ & $48$ & $44.6$ & $3.3$ & $0$ \\ 
$e^\pm$j$p\!\!\!/$$\tau^\pm$ & $11$ & $8.3$ & $1.5$ & $0$ \\ 
$e^\pm$j$p\!\!\!/$ & $121431$ & $121023$ & $747.6$ & $0$ \\ 
$e^\pm$j$\gamma$$p\!\!\!/$ & $159$ & $192.6$ & $10.9$ & $0$ \\ 
$e^\pm$j$\gamma$ & $1389$ & $1368.9$ & $38.9$ & $0$ \\ 
$e^\pm$j$\mu^\mp$$p\!\!\!/$ & $42$ & $33$ & $2.9$ & $0$ \\ 
$e^\pm$j$\mu^\pm$$p\!\!\!/$ & $16$ & $9.2$ & $1.9$ & $0$ \\ 
$e^\pm$j$\mu^\mp$ & $62$ & $63.8$ & $3.2$ & $0$ \\ 
$e^\pm$j$\mu^\pm$ & $13$ & $8.2$ & $2$ & $0$ \\ 
$e^\pm$$e^\mp$4j & $148$ & $159.1$ & $7$ & $0$ \\ 
$e^\pm$$e^\mp$3j & $717$ & $743.6$ & $24.4$ & $0$ \\ 
$e^\pm$$e^\mp$2j$p\!\!\!/$ & $32$ & $41.4$ & $5.6$ & $0$ \\ 
$e^\pm$$e^\mp$2j$\gamma$ & $10$ & $11.4$ & $2.9$ & $0$ \\ 
$e^\pm$$e^\mp$2j & $3638$ & $3566.8$ & $72$ & $0$ \\ 
$e^\pm$$e^\mp$$\tau^\pm$ & $18$ & $16.1$ & $1.7$ & $0$ \\ 
$e^\pm$$e^\mp$$p\!\!\!/$ & $822$ & $831.8$ & $13.6$ & $0$ \\ 
$e^\pm$$e^\mp$$\gamma$ & $191$ & $221.9$ & $5.1$ & $0$ \\ 
$e^\pm$$e^\mp$j$p\!\!\!/$ & $155$ & $170.8$ & $12.4$ & $0$ \\ 
$e^\pm$$e^\mp$j$\gamma$ & $48$ & $45$ & $3.9$ & $0$ \\ 
$e^\pm$$e^\mp$j & $17903$ & $18258.2$ & $204.4$ & $0$ \\ 
$e^\pm$$e^\mp$ & $98901$ & $99086.9$ & $147.8$ & $0$ \\ 
b6j & $51$ & $42.3$ & $3.8$ & $0$ \\ 
b5j & $237$ & $192.5$ & $7.1$ & $0$ \\ 
b4j high-$\Sigma p_T$ & $26$ & $23.4$ & $2.6$ & $0$ \\ 
b4j low-$\Sigma p_T$ & $836$ & $821.7$ & $15.9$ & $0$ \\ 
b3j high-$\Sigma p_T$ & $12081$ & $12071$ & $84.1$ & $0$ \\ 
b3j low-$\Sigma p_T$ & $2974$ & $2873$ & $31$ & $0$ \\ 
\end{tabular}
\end{minipage}
\caption[A subset of the populations comparison between Tevatron Run II data and Standard Model prediction.]{A subset of the comparison between Tevatron Run II data and Standard Model prediction.  
}
\label{tbl:VistaCdf2}  
\end{table*}

\begin{figure*}
\centering
\begin{tabular}{cc}
\includegraphics[width=3in]{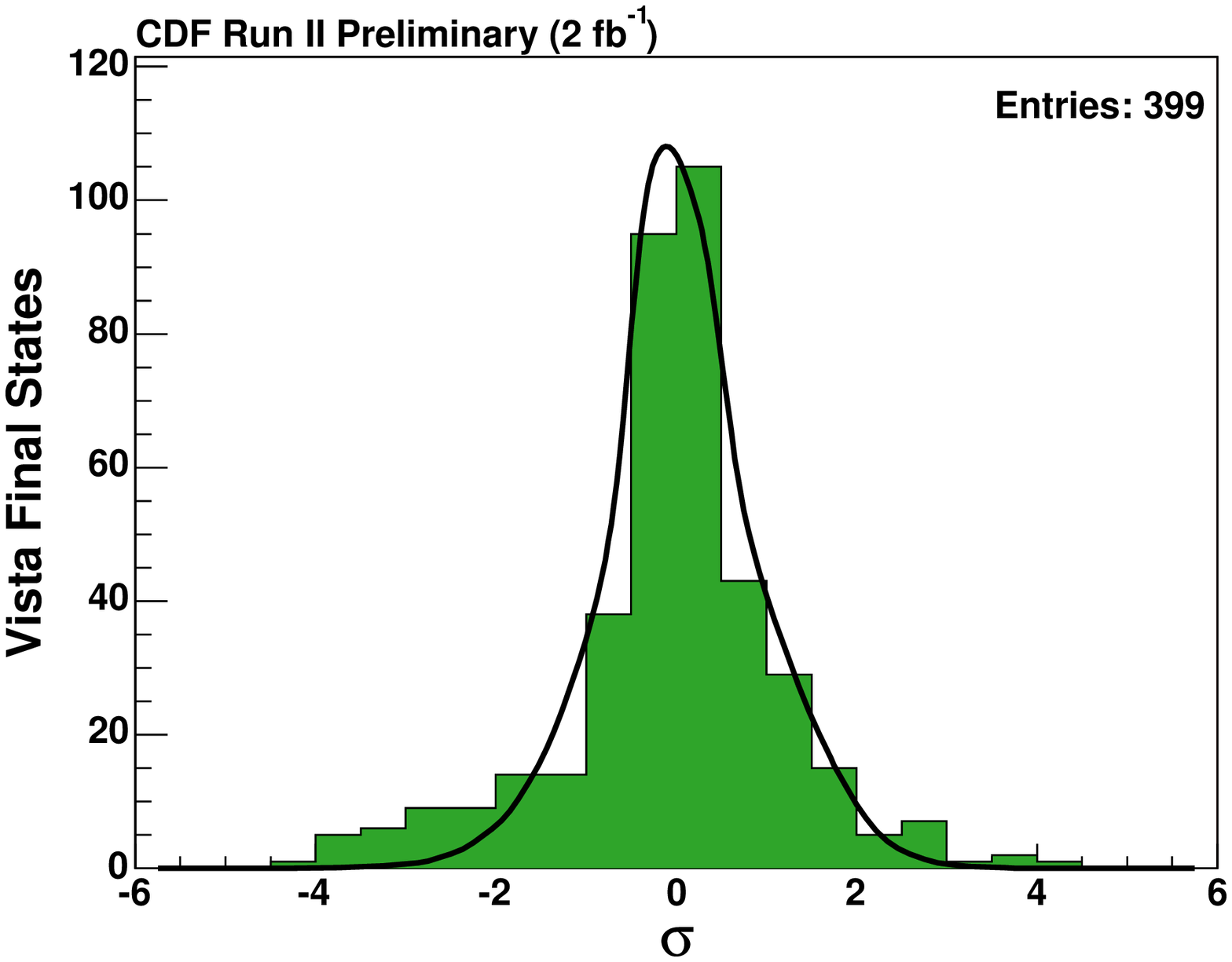} &
\includegraphics[width=3in]{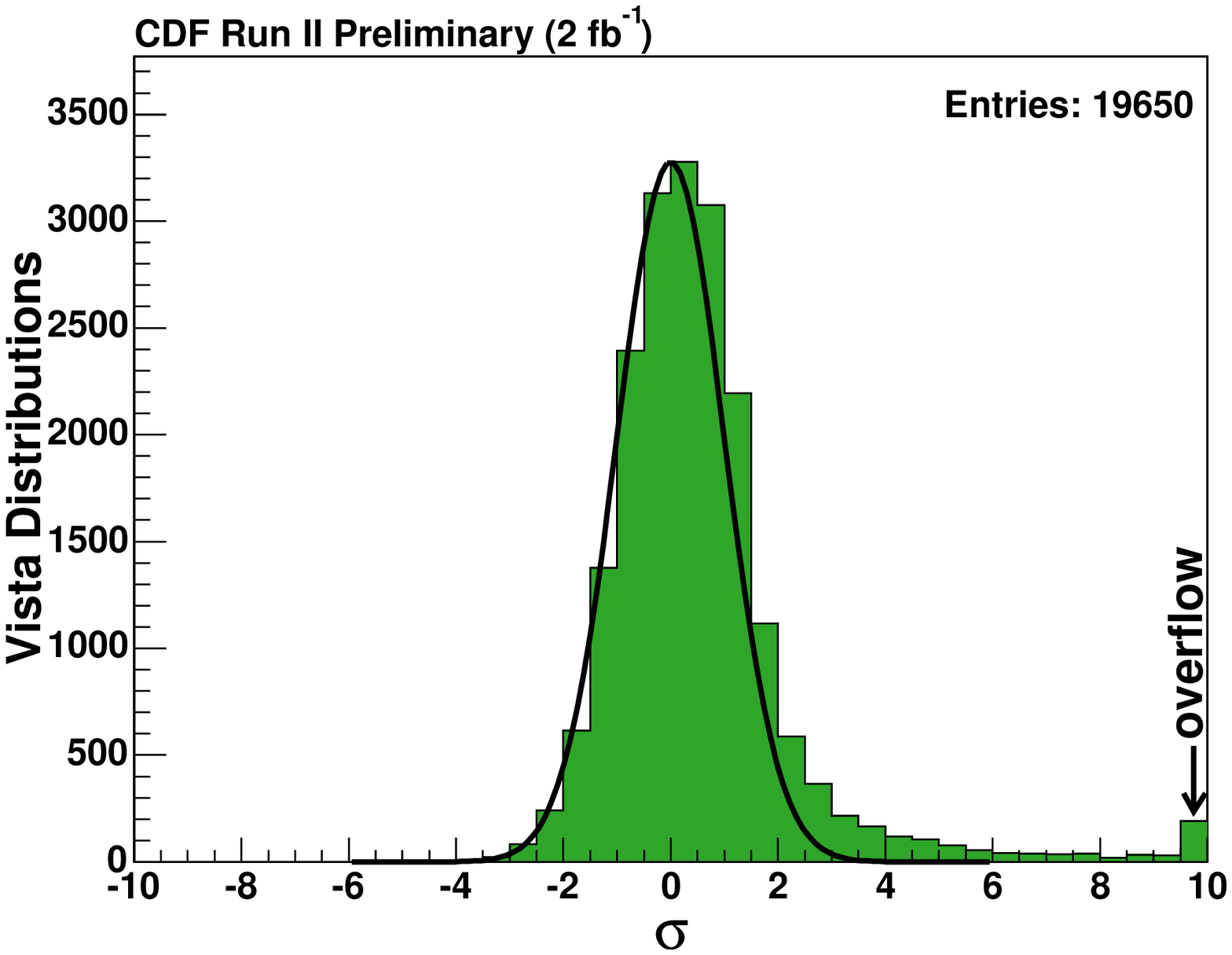} \\
\end{tabular}
\caption[Distribution of discrepancy between data and Standard Model prediction.]{Distribution of discrepancy (before accounting trials factor) between data and Standard Model prediction, measured in units of standard deviation ($\sigma$).  The left pane shows the distribution of discrepancies between the total number of events observed and predicted in the final states considered.  Final states with data excess populate the right tail, while those with data deficit populate the left tail.  The right pane shows the distribution of discrepancies between the observed and predicted shapes of roughly $2\times10^4$ kinematic distributions.  Distributions in agreement correspond to small or negative $\sigma$, and distributions in disagreement correspond to large positive $\sigma$.  Interest is in the entries in both tails of the distribution on the left, and in the right tail of the distribution on the right.\label{fig:VistaSummaryCdf2}}
\end{figure*}

The global fit $\chi^2$, described in Sec.~\ref{sec:Vista:CorrectionModel}, was in the second round $784.43$, from 335 bins, plus a 28.4 from external constraints. It is obviously a very large $\chi^2$, even more unlikely than it was in the first round of the analysis, indicating that deviations from the fit are clearly non-statistical, but due to systematic imperfections in our Standard Model implementation.  Higher statistics exacerbate systematic imperfections.

Table~\ref{tbl:VistaCdf2} shows the comparison of CDF Run II 2~fb$^{-1}$ data to Standard Model prediction.  All events have been partitioned in exclusive final states.  The number of events observed is compared to the number expected from the Standard Model, taking into account the uncertainty due to finite Monte Carlo statistics, and the trials factor due to examining 399 final states.  The final states are ordered in decreasing discrepancy.

No final state is found to have a population discrepancy that is considered significant after accounting for the trials factor.  The largest population discrepancy is a 2.7$\sigma$ deficit (including trials factor) observed in final state $be^\pm\pmiss$.
Fig.~\ref{fig:VistaSummaryCdf2} summarizes in a histogram the distribution of discrepancies observed in final state populations. Qualitatively, shape discrepancies give us the same information we had in the first round of the analysis.



Discrepant distributions are flagged using the Kolmogorov-Smirnov (KS) statistic.~\footnote{The KS statistic is defined in terms of the cumulative distributions of two populations.  Given a particular distribution, such as the invariant mass {\tt mass(j1,j2)} of the two jets in the {\tt 1e+2j1pmiss} final state, the Standard Model prediction and the data are both normalized to unit integral, and the cumulative distributions are drawn.  The maximal separation of the two cumulative distributions is the KS statistic, a number between 0 and 1.  This statistic can be translated into a probability for the data to have been pulled from the Standard Model distribution, with the translation depending only on the value of the statistic and the number of data events.  This KS probability $\text{KS}_p$ can then be converted into units of standard deviations $\text{KS}_\sigma$ by solving $\int_{-\infty}^{\text{KS}_\sigma}\, \frac{1}{\sqrt{2\pi}}e^{-\frac{x^2}{2}} dx = \text{KS}_p$.
}
Fig.~\ref{fig:VistaSummaryCdf2} shows a histogram of the disagreement seen in all kinematic distributions. 19,650 distributions are considered in 2~fb$^{-1}$, and 559 are found to have a significant disagreement. However, as in the first round with 1.0~fb$^{-1}$, no indication of new physics is found amongst these discrepant distributions; all are attributed to the ``3-jet effect'', difficulties with intrinsic $k_T$ or residual coarseness of the correction model.

\subsubsection{Evolution of the \Vista\ Global Comparison since 1~fb$^{-1}$}

\begin{table*}
\hspace{-0cm}
\begin{minipage}{9in}
\tiny
\begin{tabular}{l@{ }r@{ }r@{ $\pm$ }l@{ }l}
{\bf Final State} & {\bf Data} & \multicolumn{2}{c}{\bf Background} & {\bf $\sigma$} \\ \hline 
4j$\tau^\pm$$\tau^\mp$ & $1$ & $2.4$ & $1.5$ & $0$ \\ 
4j$\mu^\pm$$\tau^\mp$ & $1$ & $0.7$ & $1.1$ & $0$ \\ 
4j$\mu^\pm$$\mu^\mp$$p\!\!\!/$ & $1$ & $1.1$ & $1.5$ & $0$ \\ 
4b4j & $1$ & $0$ & $1$ & $0$ \\ 
4b2j & $1$ & $0.9$ & $1.3$ & $0$ \\ 
4b & $3$ & $1.3$ & $1.3$ & $0$ \\ 
3j$2\tau^\pm$ & $6$ & $8.1$ & $1.8$ & $0$ \\ 
3j$2\gamma$$p\!\!\!/$ & $1$ & $2$ & $1.6$ & $0$ \\ 
3j$\mu^\pm$$\tau^\pm$ & $3$ & $0.8$ & $1.2$ & $0$ \\ 
3b4j & $2$ & $2.9$ & $1.5$ & $0$ \\ 
3b3j & $8$ & $8.2$ & $2$ & $0$ \\ 
3b2j$p\!\!\!/$ & $1$ & $0.7$ & $1.2$ & $0$ \\ 
3b2j & $23$ & $27.2$ & $4.8$ & $0$ \\ 
3b$\gamma$ & $1$ & $0.4$ & $1.2$ & $0$ \\ 
3bj$p\!\!\!/$ & $1$ & $3$ & $1.6$ & $0$ \\ 
3bj$\gamma$ & $1$ & $1.8$ & $1.5$ & $0$ \\ 
3bj$\mu^\pm$$p\!\!\!/$ & $1$ & $1.1$ & $1.2$ & $0$ \\ 
3b$e^\pm$$p\!\!\!/$ & $1$ & $0.4$ & $1.2$ & $0$ \\ 
$2\mu^\pm$$p\!\!\!/$ & $3$ & $0.8$ & $1.1$ & $0$ \\ 
2j$\gamma$$p\!\!\!/$$\tau^\pm$ & $1$ & $1.1$ & $1.3$ & $0$ \\ 
2j$\mu^\pm$$p\!\!\!/$$\tau^\pm$ & $1$ & $1.7$ & $1.4$ & $0$ \\ 
2b6j & $2$ & $0.3$ & $1.2$ & $0$ \\ 
\end{tabular}
\begin{tabular}{l@{ }r@{ }r@{ $\pm$ }l@{ }l}
{\bf Final State} & {\bf Data} & \multicolumn{2}{c}{\bf Background} & {\bf $\sigma$} \\ \hline 
2b$\mu^\pm$$\mu^\mp$ & $2$ & $1.1$ & $1.2$ & $0$ \\ 
2bj$\tau^\pm$ & $1$ & $0.8$ & $1.2$ & $0$ \\ 
2bj$\mu^\pm$$\mu^\mp$ & $3$ & $0.3$ & $1.1$ & $0$ \\ 
2b$e^\pm$$p\!\!\!/$$\tau^\mp$ & $1$ & $0.2$ & $1.1$ & $0$ \\ 
2b$e^\pm$$\mu^\mp$$p\!\!\!/$ & $1$ & $2.2$ & $1.3$ & $0$ \\ 
$\gamma$$2\tau^\pm$ & $2$ & $0.1$ & $1.1$ & $0$ \\ 
j$2\gamma$$\tau^\pm$ & $2$ & $1.8$ & $1.4$ & $0$ \\ 
j$2\mu^\pm$$p\!\!\!/$ & $1$ & $0.6$ & $1.2$ & $0$ \\ 
j$\mu^\pm$$2\gamma$$p\!\!\!/$ & $1$ & $0.1$ & $1.1$ & $0$ \\ 
j$\mu^\pm$$\gamma$$\tau^\mp$ & $1$ & $0.1$ & $1.1$ & $0$ \\ 
$e^\pm$4j$\tau^\pm$ & $2$ & $3.1$ & $1.2$ & $0$ \\ 
$e^\pm$4j$\mu^\mp$$p\!\!\!/$ & $1$ & $0.6$ & $1.2$ & $0$ \\ 
$e^\pm$4j$\mu^\pm$$p\!\!\!/$ & $1$ & $0$ & $1$ & $0$ \\ 
$e^\pm$4j$\mu^\mp$ & $1$ & $0.7$ & $1.2$ & $0$ \\ 
$e^\pm$3j$\mu^\mp$ & $4$ & $3$ & $1.4$ & $0$ \\ 
$e^\pm$$\gamma$$\tau^\pm$ & $1$ & $0.9$ & $1.1$ & $0$ \\ 
$e^\pm$$\gamma$$p\!\!\!/$$\tau^\mp$ & $1$ & $0.5$ & $1.2$ & $0$ \\ 
$e^\pm$$\mu^\pm$$\mu^\mp$$p\!\!\!/$ & $1$ & $0.6$ & $1.1$ & $0$ \\ 
$e^\pm$j$2\gamma$$p\!\!\!/$ & $1$ & $0.2$ & $1.1$ & $0$ \\ 
$e^\pm$j$\mu^\pm$$\mu^\mp$ & $1$ & $0.8$ & $1.1$ & $0$ \\ 
$e^\pm$$e^\mp$2j$\mu^\pm$$p\!\!\!/$ & $1$ & $0$ & $1$ & $0$ \\ 
$e^\pm$$e^\mp$j$\mu^\pm$$p\!\!\!/$ & $1$ & $0.2$ & $1$ & $0$ \\ 
\end{tabular}
\begin{tabular}{l@{ }r@{ }r@{ $\pm$ }l@{ }l}
{\bf Final State} & {\bf Data} & \multicolumn{2}{c}{\bf Background} & {\bf $\sigma$} \\ \hline 
b6j$p\!\!\!/$ & $1$ & $0.1$ & $1.1$ & $0$ \\ 
b4j$p\!\!\!/$$\ 400+$ & $3$ & $1.6$ & $1.4$ & $0$ \\ 
b3j$\mu^\pm$$\tau^\pm$ & $1$ & $0.1$ & $1$ & $0$ \\ 
b2j$\tau^\pm$$\tau^\mp$ & $1$ & $0.1$ & $1.1$ & $0$ \\ 
b2j$\mu^\pm$$\gamma$ & $1$ & $0.9$ & $1.3$ & $0$ \\ 
b$\tau^\pm$$\tau^\mp$ & $2$ & $1.6$ & $1.3$ & $0$ \\ 
b$\mu^\pm$$p\!\!\!/$$\tau^\mp$ & $1$ & $1.1$ & $1.3$ & $0$ \\ 
b$\mu^\pm$$\gamma$ & $1$ & $0.7$ & $1.2$ & $0$ \\ 
b$\mu^\pm$$\mu^\mp$$p\!\!\!/$ & $3$ & $0.7$ & $1.3$ & $0$ \\ 
bj$\tau^\pm$$\tau^\mp$ & $1$ & $0.6$ & $1.2$ & $0$ \\ 
bj$\mu^\pm$$\tau^\mp$ & $1$ & $0.5$ & $1.2$ & $0$ \\ 
b$e^\pm$3j$\gamma$ & $1$ & $1.4$ & $1.2$ & $0$ \\ 
b$e^\pm$3j$\mu^\mp$$p\!\!\!/$ & $1$ & $0.8$ & $1.2$ & $0$ \\ 
b$e^\pm$$2\gamma$ & $2$ & $0.2$ & $1.1$ & $0$ \\ 
b$e^\pm$2j$\tau^\mp$ & $2$ & $1.6$ & $1.2$ & $0$ \\ 
b$e^\pm$2j$p\!\!\!/$$\tau^\mp$ & $2$ & $0.9$ & $1.2$ & $0$ \\ 
b$e^\pm$2j$\gamma$$p\!\!\!/$ & $3$ & $0.4$ & $1.2$ & $0$ \\ 
b$e^\pm$$\tau^\pm$ & $1$ & $1.2$ & $1.1$ & $0$ \\ 
b$e^\pm$$\gamma$$p\!\!\!/$ & $3$ & $2.7$ & $1.5$ & $0$ \\ 
b$e^\pm$j$p\!\!\!/$$\tau^\mp$ & $1$ & $1.5$ & $1.3$ & $0$ \\ 
\end{tabular}
\end{minipage}
\caption{New \Vista\ final states which appeared in the analysis of 2~fb$^{-1}$.}
\label{tbl:VistaNewFinalStates}  
\end{table*}

Table~\ref{tbl:VistaNewFinalStates} displays the \Vista\ final states which newly appeared in the present analysis.
A large number involve b-jets; this is a result of changes in our offline event selection criteria, which now accept more events containing b-tagged jets (previously events with a leading b-jet with $p_T<200$ were prescaled offline by a factor of 10; we also introduced a new tri-b offline selection).

There are also 11 final states which were populated in the 1.0~fb$^{-1}$ analysis, but are not now:
{\tt 
1b1e+3j1tau-		
1b3j2ph			
1e+			
1e+1e-1ph1tau+		
1e+3j2ph		
1j1pmiss2tau+		
1j3ph			
2b2ph			
3j1mu+1pmiss1tau+	
3j1pmiss1tau+1tau-	
1b1e+3j1ph1pmiss
}
These events were generally found to contain an object (usually a $\tau$ or plug photon) which now fails our tighter identification requirements.

A final reason for the increase of \Vista\ final states from 344 in 1.0~fb$^{-1}$ to 399, is that jet-tau final states have been divided into high-$p_T$ and low-$p_T$ states. 

The 3j$\tau^\pm$ and 2j$\tau^\pm$ final states remain among the `top ten' most discrepant states, but their significance has decreased compared to the first round.  The improvement in agreement was achieved after slight changes in modeling jets faking taus in events with large activity.
Other final states from the first round's top ten now exhibit zero discrepancy (after accounting for the trials factor).  We attribute this to a combination of general improvements in modeling and statistical fluctutations.


\section{\Sleuth}
\label{sec:Sleuth2}

\Sleuth\ algorithm was not modified in the second round.

\subsection{Results}
\label{sec:Sleuth:Results2}

\begin{figure*}
\begin{tabular}{cc}
\includegraphics[width=2.2in,angle=270]{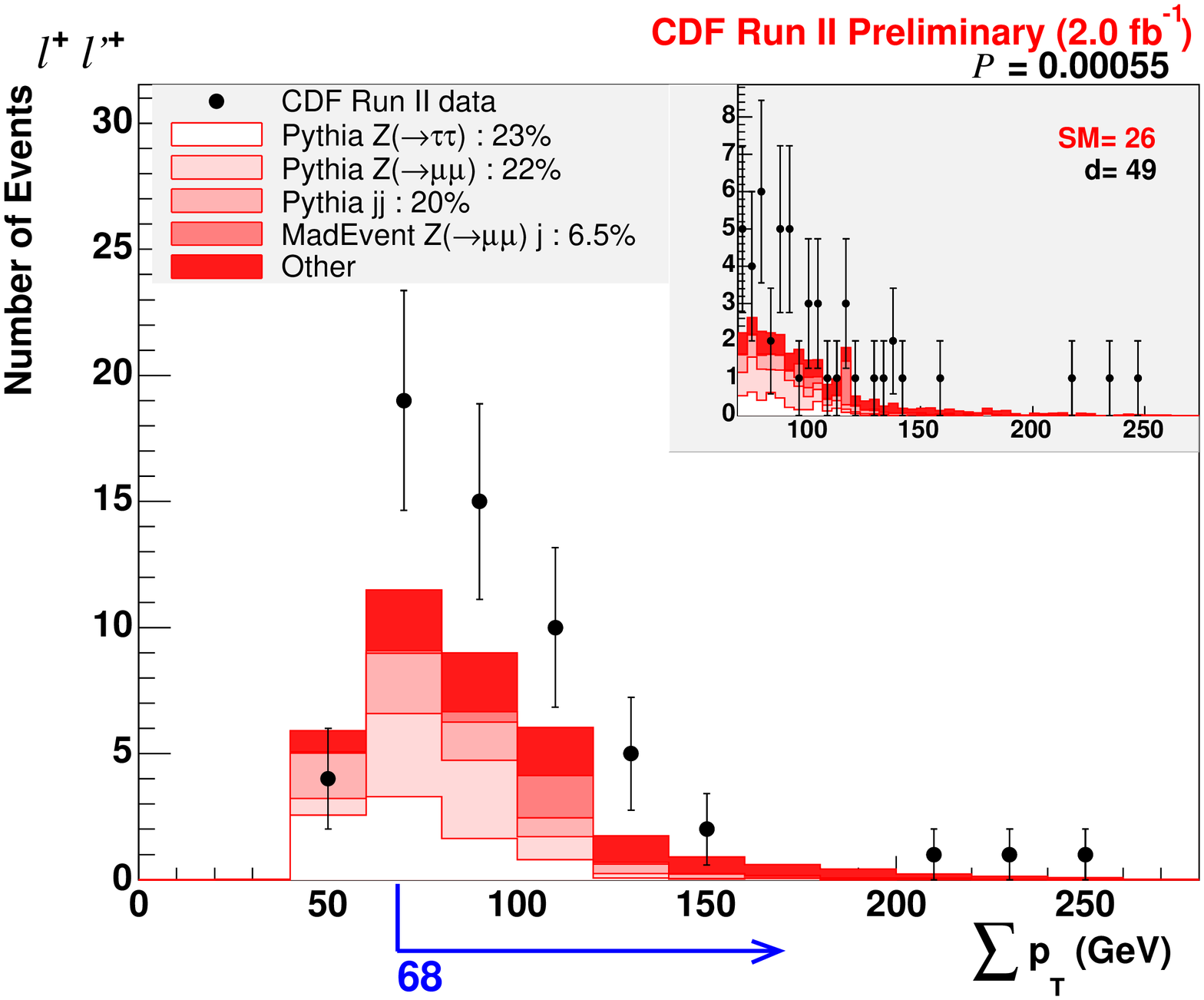}
&
\includegraphics[width=2.2in,angle=270]{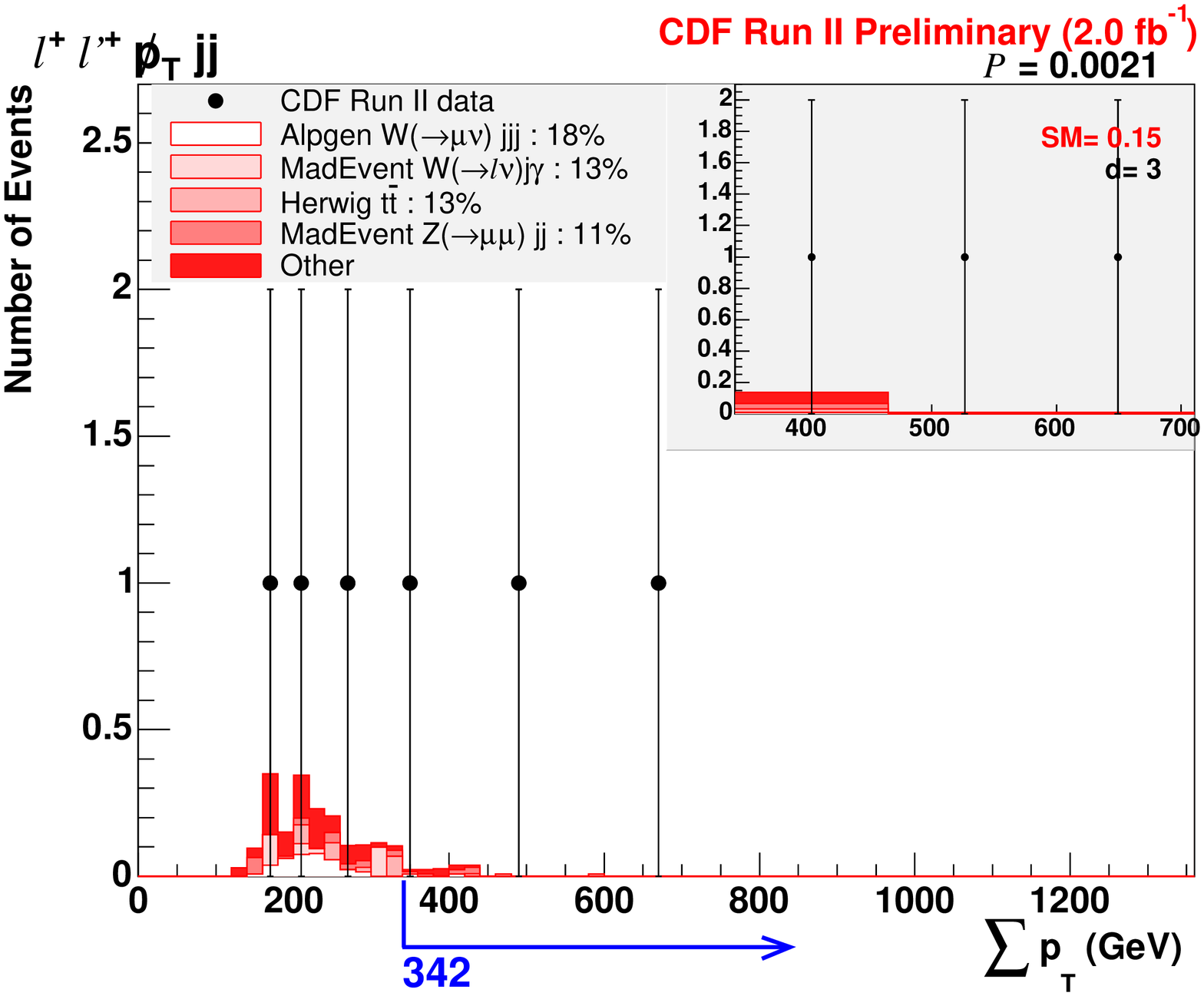}
\\
\includegraphics[width=2.2in,angle=270]{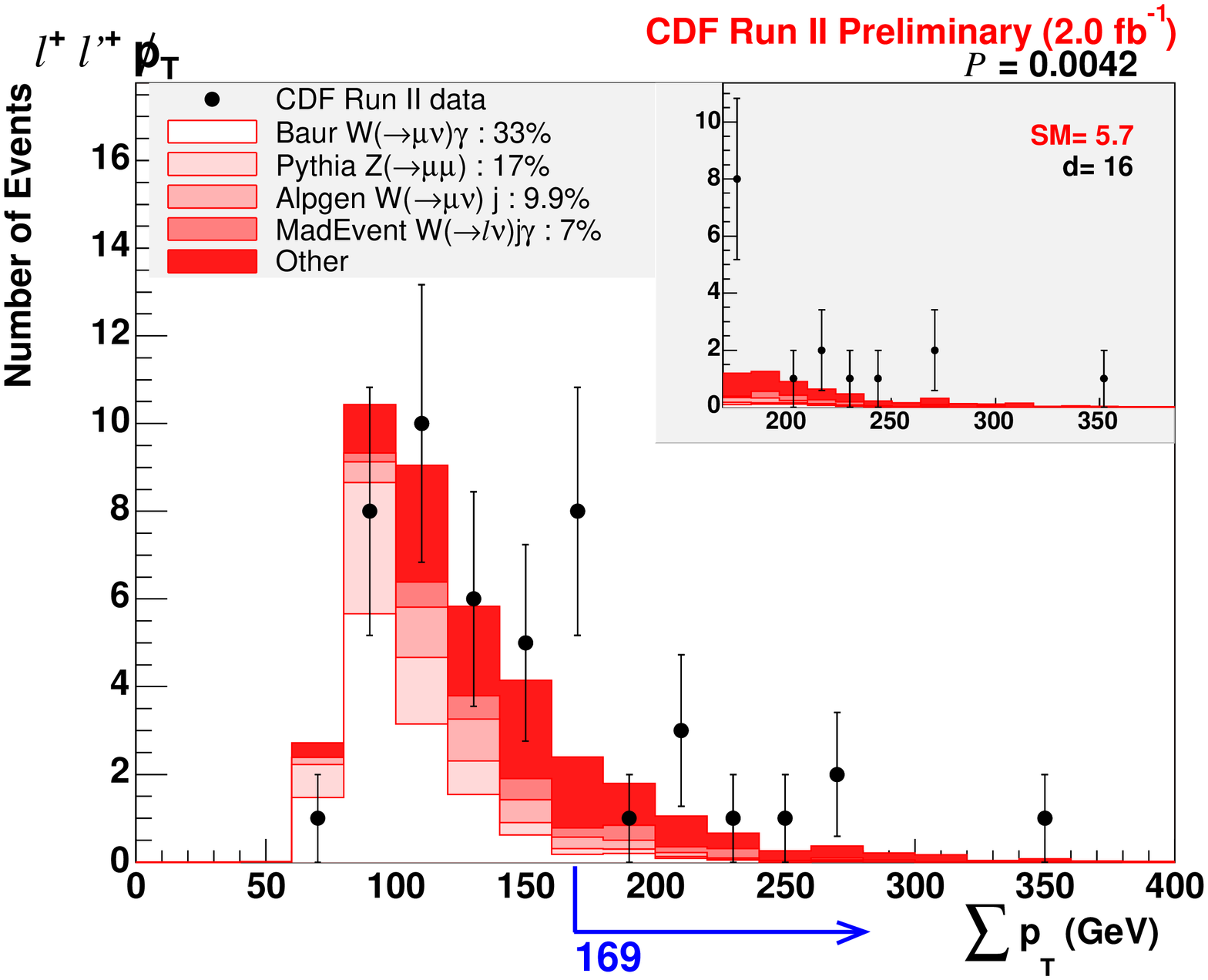}
&
\includegraphics[width=2.2in,angle=270]{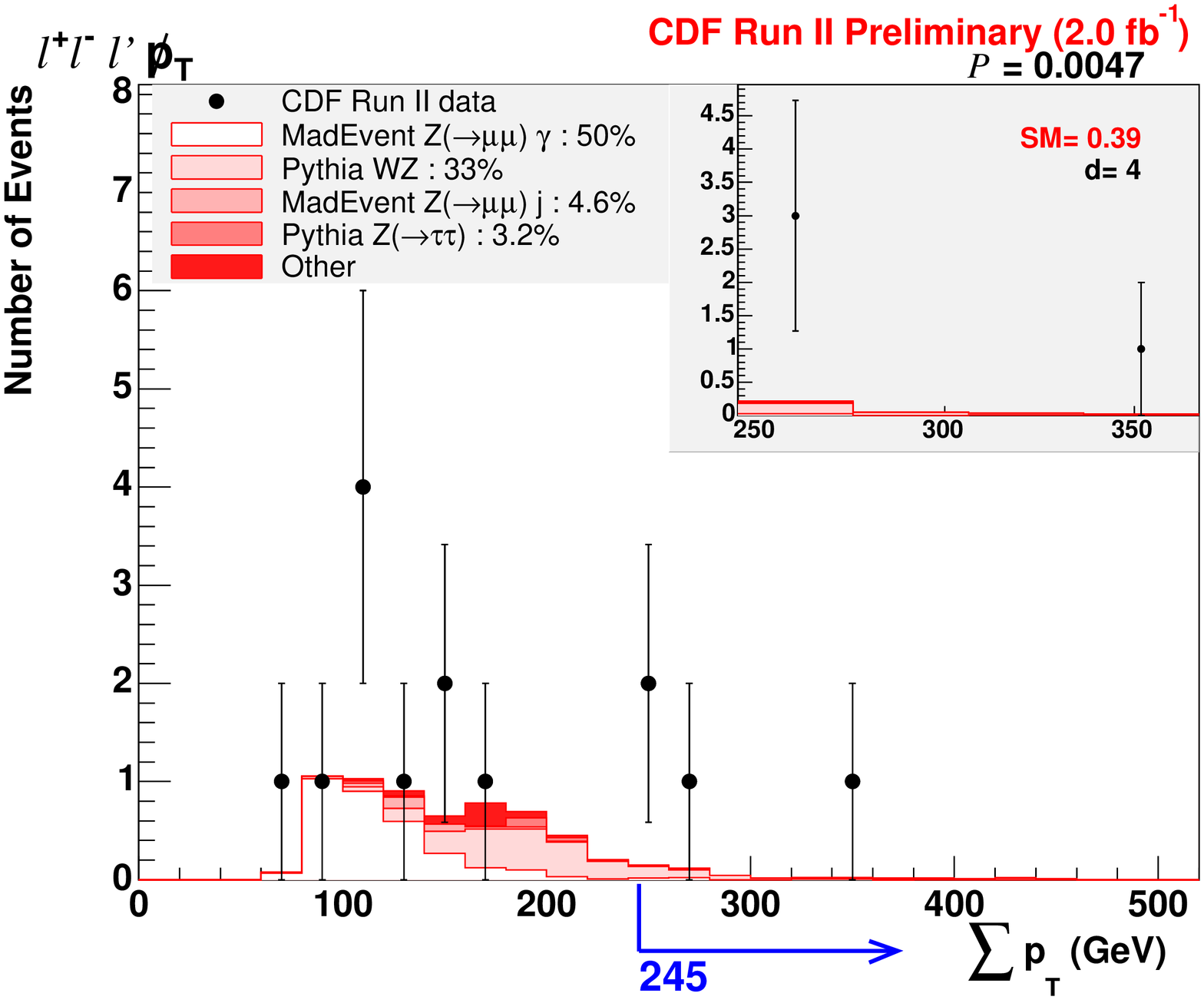}
\\
\includegraphics[width=2.2in,angle=270]{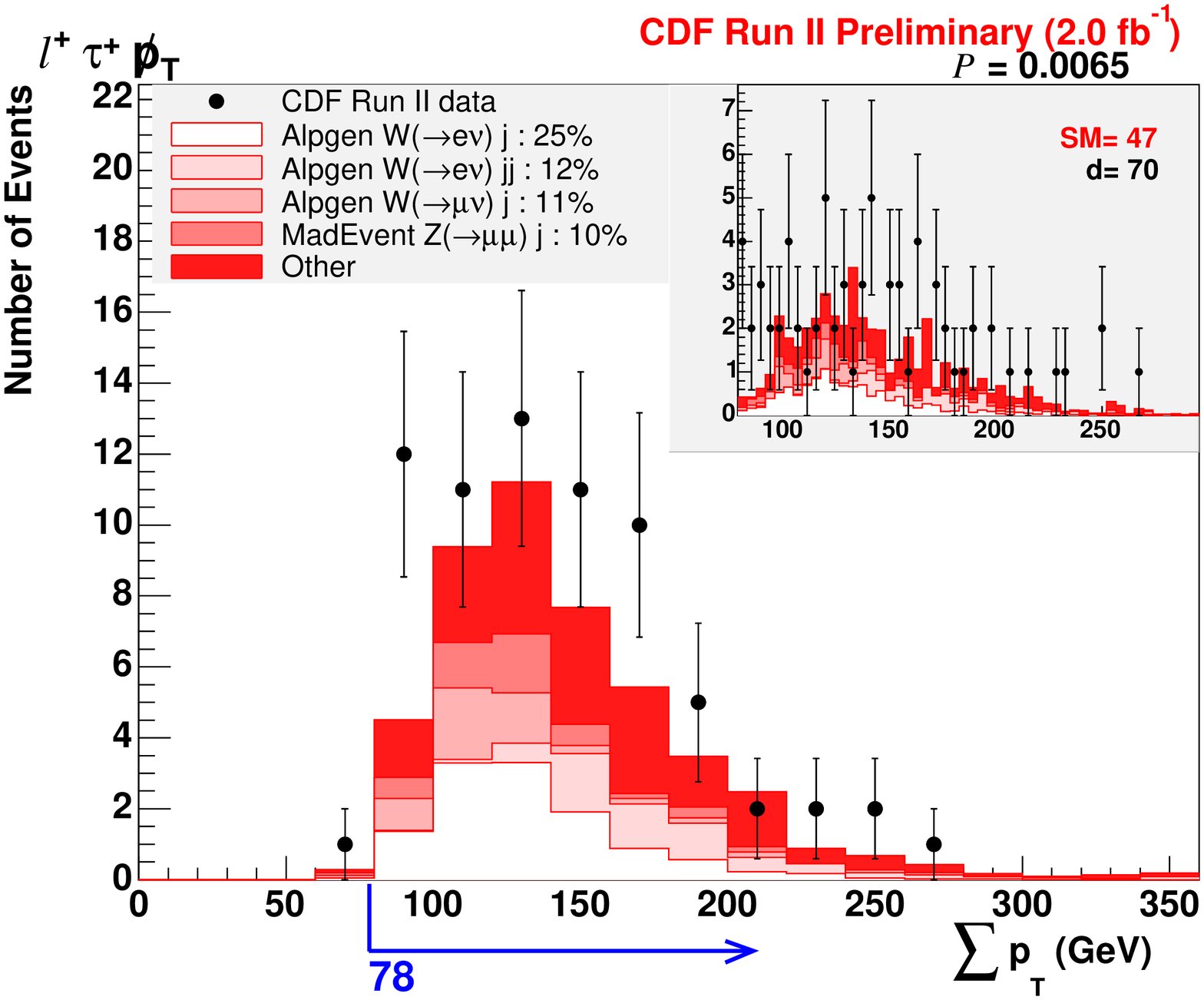}
&
\includegraphics[width=2.2in,angle=270]{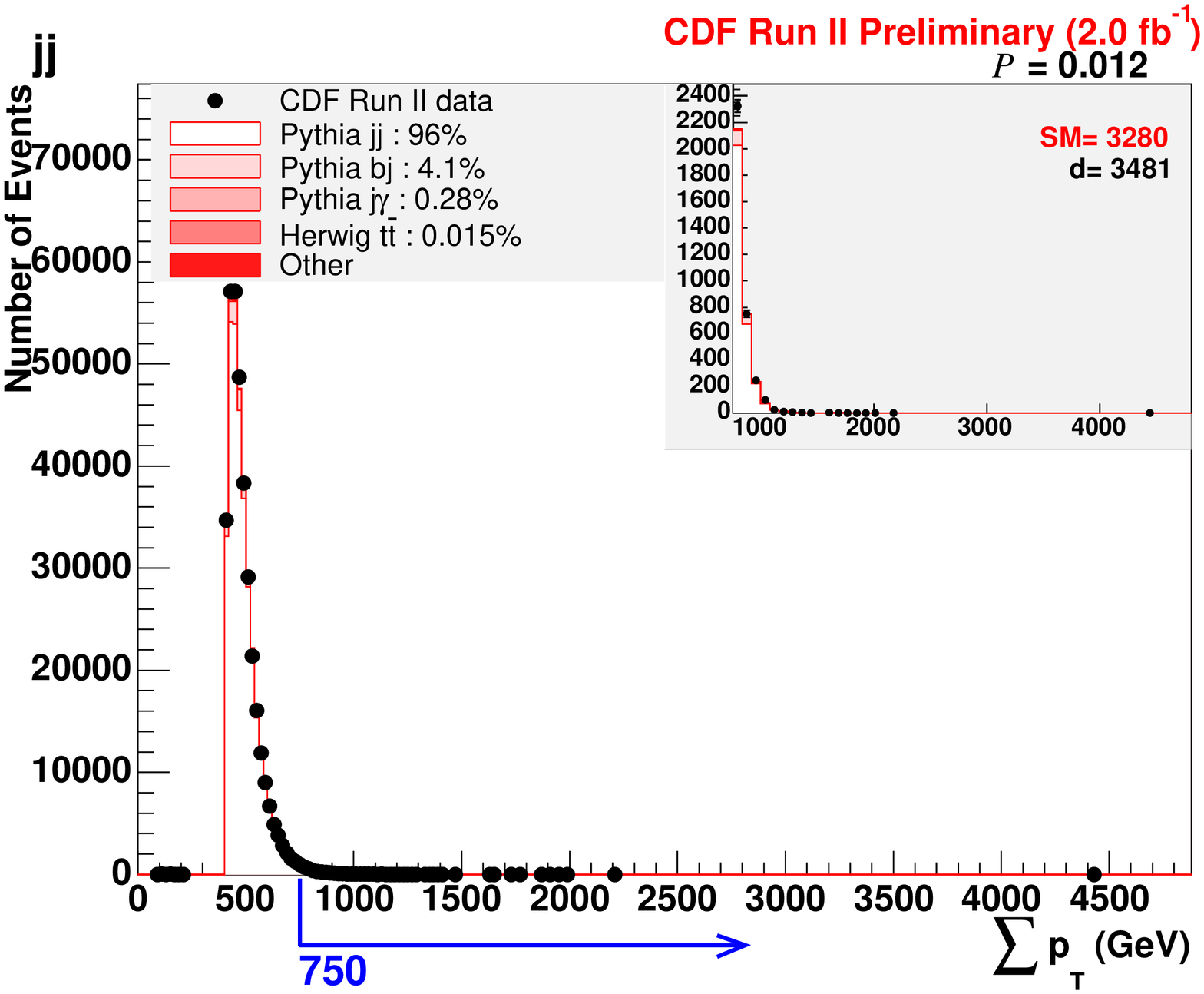}
\\
\end{tabular}
\caption{The most interesting final states identified by \Sleuth\ in 2~fb$^{-1}$.\label{fig:SleuthPlots2}}
\end{figure*}

\begin{figure}
\centering
\includegraphics[width=2.5in,angle=270]{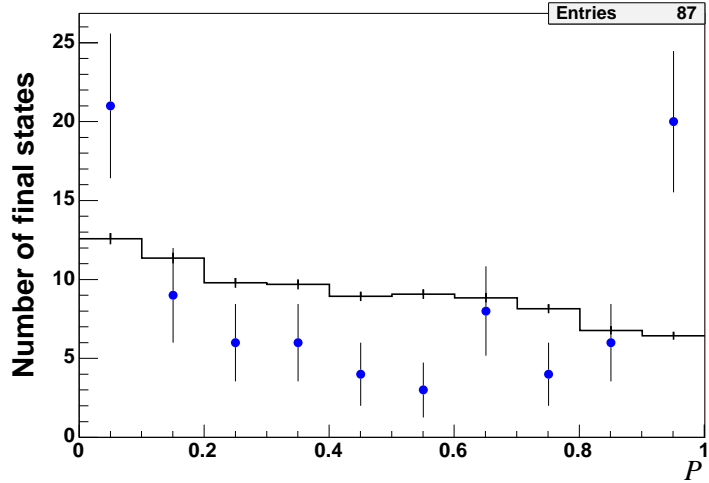} \\
\caption[\scriptP\ distribution]{
{\em Blue points:} The \scriptP\ distribution observed in 1990~pb$^{-1}$, with one entry for each of the 87 \Sleuth\ final states with at least 3 data.  There are 153 \Sleuth\ final states with non-zero background and less than 3 data, which are assigned $\scriptP=1$.
{\em Black histogram:} The expected \scriptP\ distribution from all 240 \Sleuth\ final states with non-zero background, if instead of actual data we use pseudo-data pulled from the expected \sumPt\ distribution of each final state, and omit the final states where pseudo-data are less than 3 and therefore have $\scriptP=1$.  As explained in Sec.~\ref{sec:Sleuth:Regions}, footnote~\ref{footnote:scriptPdistribution}, the \scriptP\ of final states with expected population $\lesssim 10$ is not uniformly distributed.  Of the 240 final states \Sleuth\ considers in 1990~pb$^{-1}$, 171 have Standard Model background of less than 10 events, which causes the expected \scriptP\ distribution to slightly favor smaller values.
}
\label{fig:scriptPsPlots2}
\end{figure}

The most interesting final states highlighted by \Sleuth\ are shown in Fig.~\ref{fig:SleuthPlots2}.  The region chosen by \Sleuth\ is shown by the (blue) arrow, extending up to infinity.
CDF Run II data are shown as filled circles; Standard Model prediction is shown as a histogram.  \Sleuth\ final state labels are in the upper left corner of each panel.  The number at upper right in each panel is \scriptP, the fraction of hypothetical similar experiments in which something as interesting as the region shown would be seen in this final state.  The inset in each panel shows an enlargement of the region selected by \Sleuth, together with the number of events (${\text{SM}}$) predicted by the Standard Model in this region, and the number of data events ($d$) observed in that region.

The distribution of \scriptP\ for the final states considered by \Sleuth\ in the CDF Run II data is shown in Fig.~\ref{fig:scriptPsPlots2}. 

In these CDF data, \Sleuth\ finds 
$
\twiddleScriptP = 0.085
$.
This is sufficiently far above the \Sleuth\ discovery threshold of \tildeScriptP$<0.001$ that no discovery claim can be made on the basis of \Sleuth\ for 2~fb$^{-1}$.

\subsubsection{Study of Same-Sign \Sleuth\ States}

The top \Sleuth\ final states appear a common trend to involve same-sign leptons.
We first consider the 2\nd\ and 3\rd\ \Sleuth\ final states, which both contain same-sign electron and muon, significant missing energy, and varying numbers of jets. The relevant \Vista\ final states are:

\begin{center}
\begin{tabular}{ccc}
Final State & data & background \\\hline
$e^+\mu^+\pmiss$ & 31 & 29.9 $\pm$ 1.6 \\
$e^+j\mu^+\pmiss$ & 16 & 9.2 $\pm$ 1.9 \\
$e^+2j\mu^+\pmiss$ & 6 & 1.7 $\pm$ 1.2 \\
$e^+3j\mu^+\pmiss$ & 0 & 0.26 $\pm$ 0.07 \\
\end{tabular}
\end{center}


The primary backgrounds for all these final states are similar, although the relative proportions vary with the number of reconstructed jets.
The three main backgrounds are: $W(\rightarrow\mu\nu)$+jets, with a jet faking the electron; $Z(\rightarrow\mu^+ \mu^-)$+jets, where 1 $\mu$ is not reconstructed, creating missing energy, and a jet fakes the electron; and $W\gamma$(+jets), where the photon fakes the electron.

All these processes involve real muons -- there is no significant Standard Model contribution to these final states from fake muons.  Therefore we can discard any explanation for the excess in data which involves charge assignment to muons faked by jets.

We can be confident that the charge-sign of a real muon is well-measured by the CDF tracking system.  The curvature resolution of the chamber is $\sigma_C = 3.6\times10^{-6}$~cm$^{-1}$.  The curvature corresponding to a track with momentum of 100~GeV/$c$ is $2.1 \times 10^{-5}$~cm$^{-1}$. The sign of the curvature of such a track, and hence the charge of such a particle, is thus typically determined with a significance of better than five standard deviations \cite{CDF8643}.
\Vista\ supports this conclusion, since we reconstruct $\sim$25,000 $\mu^{+} \mu^-$ events but only a single $\mu^+ \mu^+$ event (and even then, the $\mu^+ \mu^+$ invariant mass is $\sim$150~GeV, making it unlikely to be a $Z$ decay with wrong charge-reconstruction).

\begin{figure}
\centering
\includegraphics[width=0.4\columnwidth,angle=270]{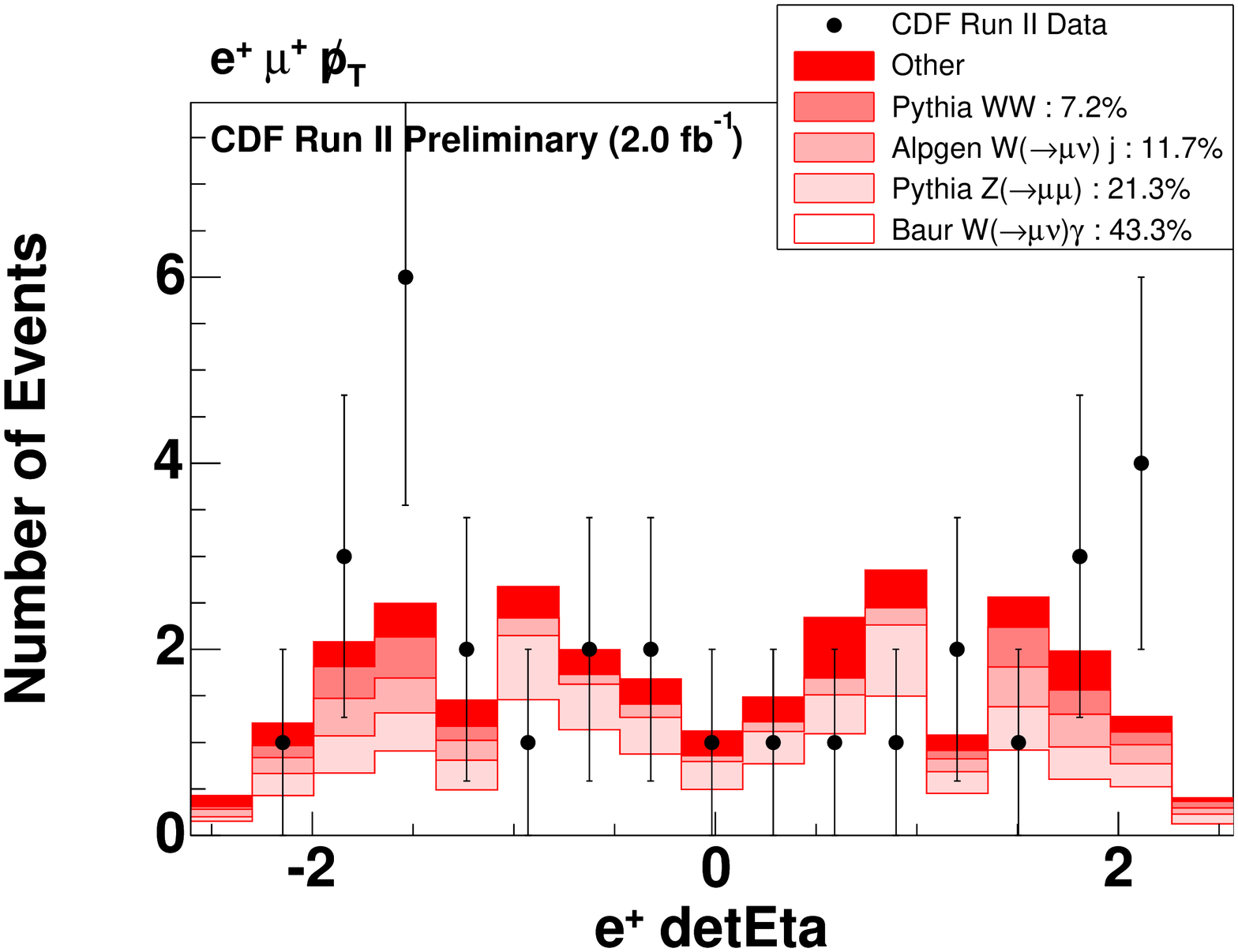}
\caption{Detector $\eta$ distribution for the electron in {\tt 1e+1mu+1pmiss}.}
\label{fig:1e+1mu+1pmiss_edeteta}
\end{figure}

\begin{figure}
\centering
\includegraphics[width=0.4\columnwidth,angle=270]{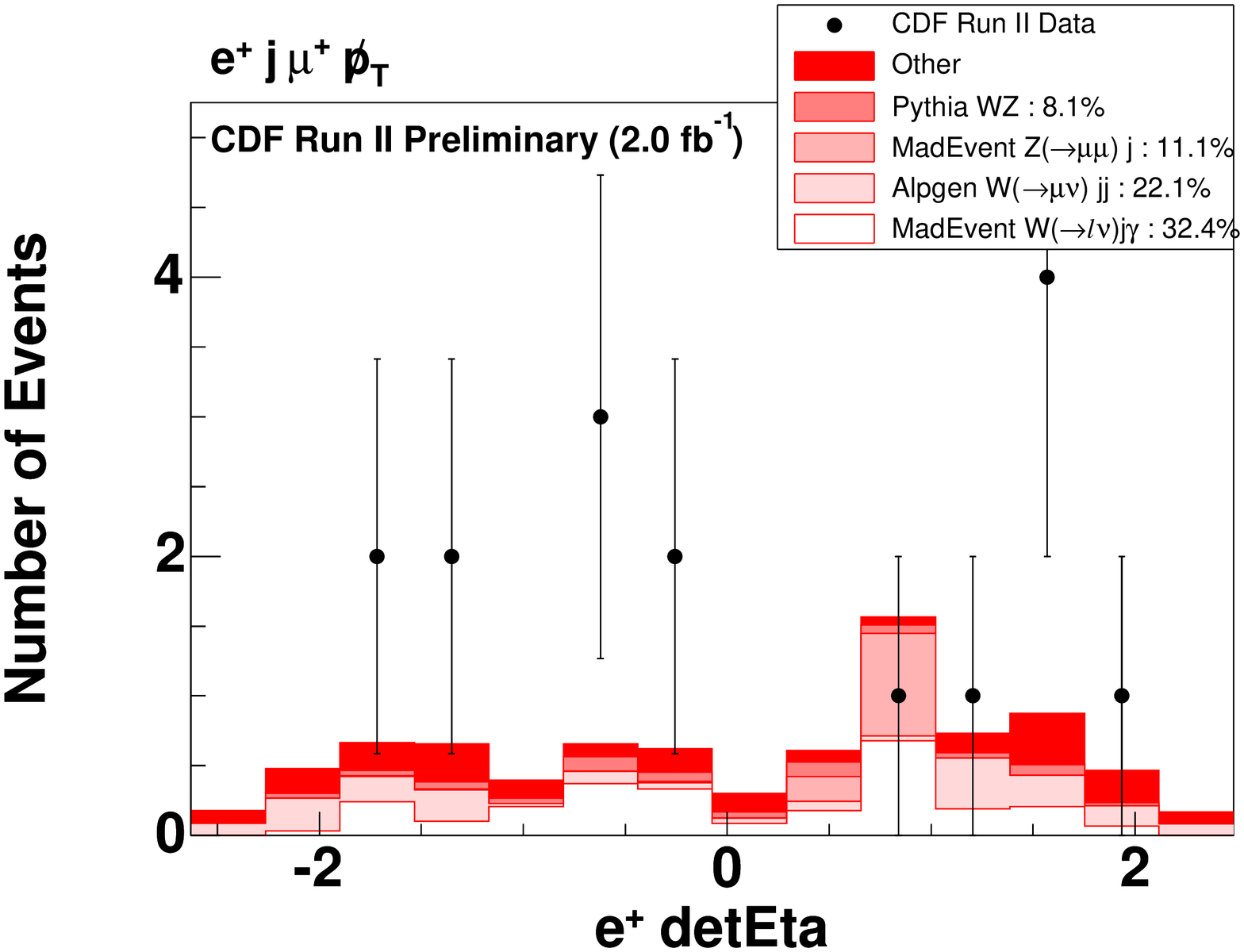}
\caption{Detector $\eta$ distribution for the electron in {\tt 1e+1j1mu+1pmiss}.}
\label{fig:1e+1j1mu+1pmiss_edeteta}
\end{figure}

We can assume the muon charge is correct therefore, and focus on the electron.  This is a fake electron from a jet.  This fake rate is well-determined from the electron+jet(s) events, and similarly the $k$-factors for the boson+jets processes are well-determined from other final states.  We expect the contribution from these processes to these particular final states to therefore be accurate.
Indeed, the most populous state {\tt 1e+1mu+1pmiss} is well described, and the mild excesses seen by \Sleuth\ arise from the {\tt 1e+1j1mu+1pmiss} and {\tt 1e+2j1mu+1pmiss} final states.  Examination of the kinematic distributions from thse final states yields nothing further (the electron $\detEta$ distributions for these final states are shown in Figs.~\ref{fig:1e+1mu+1pmiss_edeteta} and \ref{fig:1e+1j1mu+1pmiss_edeteta}), so, following the above reasoning and given that the effect is not statistically very signficant, we ascribe the presence of these two states towards the top of \Sleuth's list as likely just due to a fluctuation.


The 1\st\ \Sleuth\ final state {\tt 1e+1mu+} also has same sign electron and muon, but no missing energy, and 0 or 1 jets.
The potentially relevant \Vista\ final states are:
\begin{center}
\begin{tabular}{ccc}
Final State & data & background \\ \hline
$e^+\mu^+$ & 45 & 28.5 $\pm$ 1.8 \\
$e^+j\mu^+$ & 13 & 8.2 $\pm$ 2 \\
$e^+2j\mu^+$ & 2 & 2.6 $\pm$ 1.6 \\
$e^+3j\mu^+$ & 2 & 0.6 $\pm$ 1.2 \\
\end{tabular}
\end{center}
So only the data excess in $e^+\mu^+$ needs any potential investigation for evidence of Standard Model background mismodeling.
The largest background is from $Z\rightarrow(\mu^+\mu^-)$+jets, with one muon lost and a jet faking an electron. As explained earlier, this process is well-constrained and cannot explain the excess in data.

\begin{figure}
\centering
\includegraphics[width=0.5\columnwidth,angle=270]{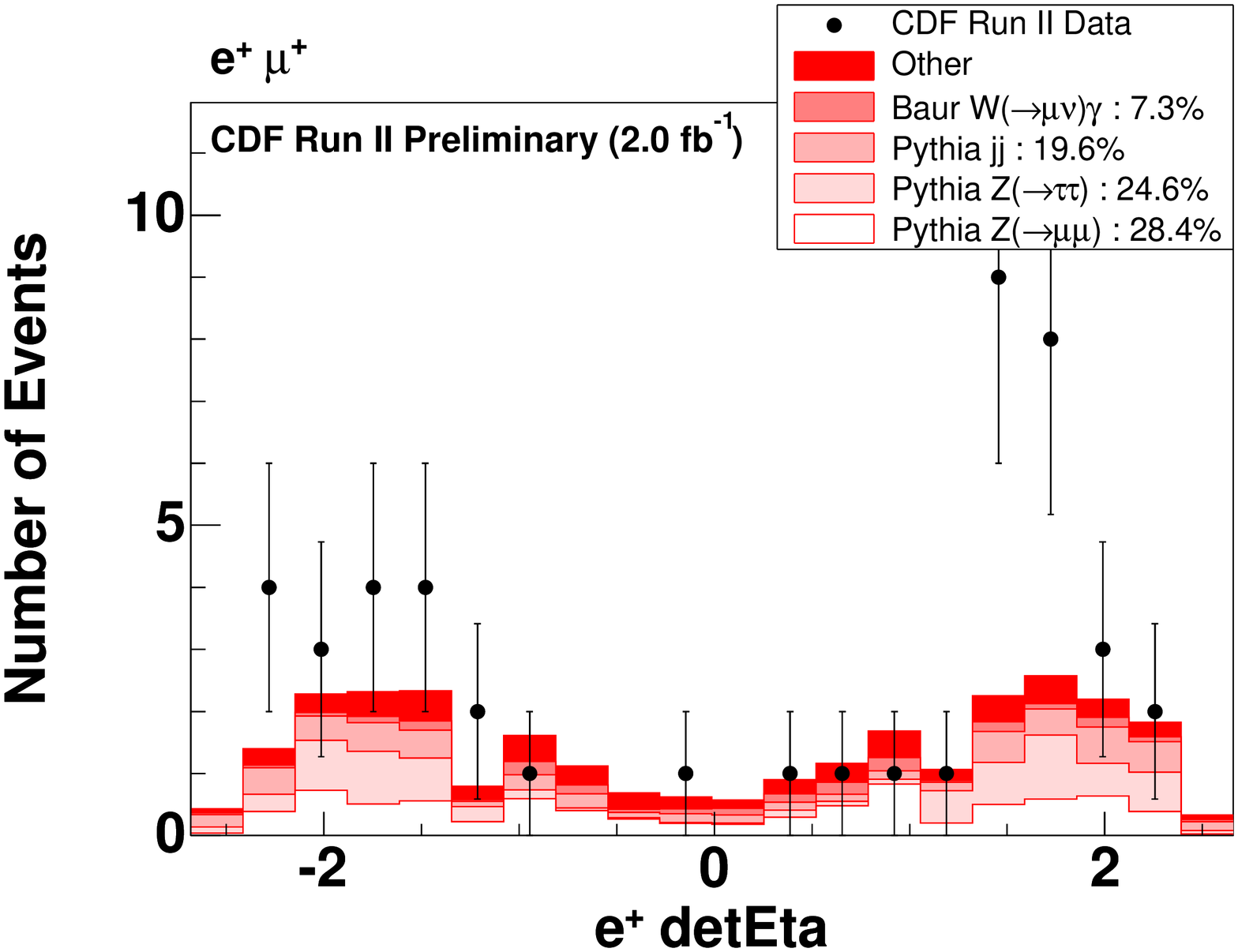}
\caption{$\detEta$ distribution for the electron in {\tt 1e+mu+}}
\label{fig:1e+1mu+_edeteta}
\end{figure}

\begin{figure}
\centering
\includegraphics[width=0.5\columnwidth,angle=270]{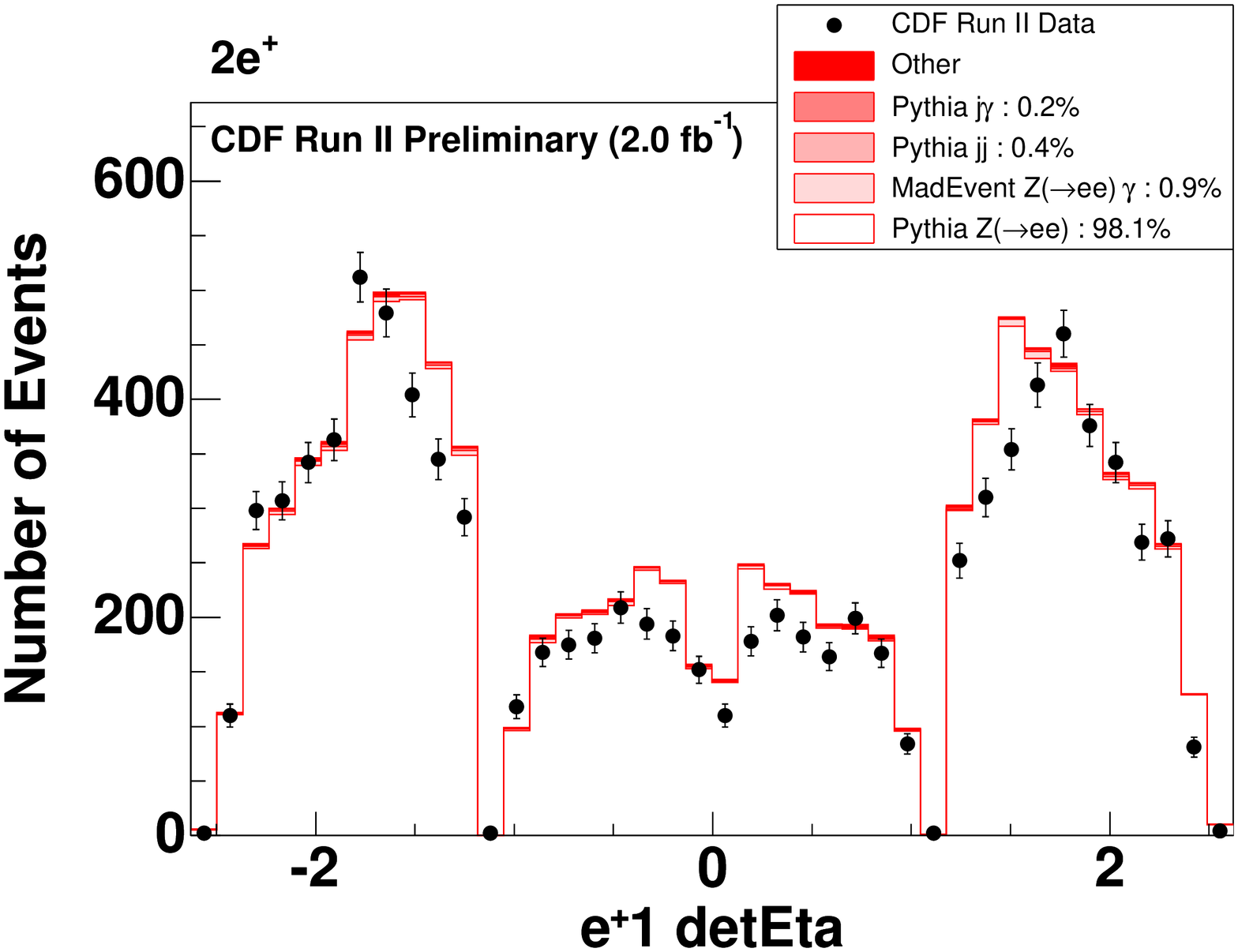}
\caption{$\detEta$ distribution for the electron in {\tt 2e+}}
\label{fig:2e+_edeteta}
\end{figure}

The next largest background is $Z\rightarrow\tau^+\tau^-$, with one $\tau$ decaying to an electron and the other to a muon. As discussed above, we trust the muon charge, so the electron must be reconstructed with the wrong charge.
For central electrons, this occurs at a rate on the order of 1 in $10^{-4}$, through electron bremstrahlung to a photon with an asymmetric conversion that half the time results in an opposite charge electron, and therefore is too small to play a role here.  For plug electrons, however, the track charge has a false-reconstruction rate of order 10\% \cite{CDF8614}.  Fig.~\ref{fig:1e+1mu+_edeteta} shows the $\detEta$ of the electron, and we indeed observe that the $Z\rightarrow \tau \tau$ contribution is almost entirely in the plug.  However, Fig.~\ref{fig:2e+_edeteta}, which shows electron $\detEta$ for the {\tt 2e+} final state (dominated by real electrons from $Z$ with phoenix track charge mis-assignment), demonstrates that this charge misidentification is quite well modeled -- there is certainly no room for the factor of two increase that would be needed to explain the data excess.
The only other large background is from QCD dijet events where both electron and muon are fakes. Both of these total fake rates are very well constrained from the electron+jets(s) and muon+jet(s) final states, so the only possible flexibility is in the charge assignment to the fakes, which would shift background events between the {\tt 1e+1mu+} and {\tt 1e+1mu-} final states.
However, with our current modeling, this process contributes an approximately equal number of expected events ($\sim5$) to each of these states. It is implausible to argue that the combination of QCD Feynman diagrams and faking mechanisms could be such as to significantly {\it anti-correlate} the fake electron and muon charge signs, so this cannot contribute to the data excess.
In conclusion, after examining the possibilities and reminding ourselves that the similar final states but with additional jets are actually well described, we have no explanation for this excess other than a statistical fluctuation.

The 5\th\ most discrepant state in \Sleuth\ is $\ell^+ \tau^+$. Since \Sleuth\ combines electrons and muons, the relevant \Vista\ final states are:
\begin{center}
\begin{tabular}{ccc}
Final State & data & background \\ \hline 
$e^+\pmiss \tau^+$ & 36 & 17.2 $\pm$ 1.7 \\
$e^+j\pmiss \tau^+$ & 11 & 8.3 $\pm$ 1.5 \\
$\mu^+\pmiss \tau^+$ & 15 & 12 $\pm$ 2 \\
$j\mu^+\pmiss \tau^+$ & 8 & 9.4 $\pm$ 3.1 \\
\end{tabular}
\end{center}

One sees that the excess comes only from $e^+\pmiss \tau^+$. This is actually among most discrepant final states in \Vista, with a significance of 1.4$\sigma$ after accounting for the trials factor.
The primary background is $W\rightarrow e \nu$+jet, where the jet ends up faking a $\tau$ with the same charge as the electron. This is rarer than the other case where the fake $\tau$ has opposite sign to the electron.
However, we appear to be modeling this process quite well, because it equally applies in the case when the $W$ decays to muon and neutrino, and \Vista\ predicts those final states correctly. We believe the excess in $e^+\pmiss \tau^+$ is therefore likely just a fluctuation.

In conclusion, although the top \Sleuth\ states all involve same-sign leptons, we find no explanation that can simultaneously account for all.  More data would help us see to what extent this is mismodeling, and to what statistical fluctuation.

\subsubsection{Evolution of the Top \Sleuth\ Final States from 1~fb$^{-1}$}


The {\tt 1bb} final state which was at the top of the list of \Sleuth\ discrepancies has now gone down the list.  The reason is that the region selected previously had been selected based on a relatively small excess in a particular region of \sumPt. Doubling the data caused that region to exceed the upper limit of 10,000 events. This upper limit is designed to reject excesses found in regions of high statistics where even a small systematic error would cause \Sleuth\ to give a large discrepancy. 

The discrepancy in the $j\pmiss$ final state, which is dominated by cosmic events, has been corrected by the additional quality criteria cuts on the cosmic background.

The 3\rd, 4\th\ and 6\th\ most discrepant \Sleuth\ final states from the first round were same sign dilepton final states.  These final states have become more discrepant in this round of the analysis as discussed in Sec.~\ref{sec:Sleuth:Results2}.

The 5\th\ most discrepant \Sleuth\ final state from the first round of the analysis was the $\pmiss\tau^+$. Then, we a major background contribution was missing, $W(\rightarrow \tau\nu)+jets$, which has been added.  

The remaining discrepancies were all corrected either by improving the background modelling, or were simply fluctuations.

\subsection{Sensitivity}
\label{sec:Sleuth:Sensitivity2}

For the 2~fb$^{-1}$ analysis, we have performed an additional test of the sensitivity of \Sleuth\ to Standard Model single top production.


\begin{table}[hbt]
\centering
\begin{tabular}{lll}
\centering
Final State & Events & Acceptance (\%) \\ \hline
$W jj$ & 5149 & 5.1 \\
$W b\bar{b}$ & 3231 & 3.2 \\
$W             $ & 1977 & 2.0 \\
$W 4j$ & 298 & 0.3 \\
$W b\bar{b} jj$ & 219 & 0.2 \\
$b \bar{b} \pmiss$ & 128 & 0.1 \\
$jj$ & 109 & 0.1 \\
$b \bar{b}$ & 96 & 0.1 \\
$jj \pmiss$ & 59 & 0.1 \\
$b \bar{b} 2j$ & 41 & 0.0 \\
\end{tabular}
\caption[Partitioning of events in Single Top into \Sleuth\ final states]{Partitioning of events in Single Top into \Sleuth\ final states. The most populous final states are shown. The offline selection filter accepts \% of the pseudo-signal events. The acceptance is shown for each individual final state.}
\label{tbl:modelPartitioning_model13}
\end{table}

\begin{table}[hbt]
\centering
\begin{tabular}{r|c|l}
 cost & Final state & \multicolumn{1}{c}{\tildeScriptP} \\ \hline
3600  &  $W b\bar{b}$  &  $=$0.0009669    \\
4800  &  $W b\bar{b}$  &  $=$0.0003004    \\
3800  &  $W b\bar{b}$  &  $=$0.0002808    \\
3600  &  $W jj$  &  $=$0.0008754    \\
3600  &  $W b\bar{b}$  &  $=$0.0002843    \\
3800  &  $W b\bar{b}$  &  $=$0.0007113    \\
5000  &  $W b\bar{b}$  &  $=$0.0007072    \\
3800  &  $W b\bar{b}$  &  $=$0.0003327    \\
5400  &  $W b\bar{b}$  &  $=$0.0003309    \\
2800  &  $W b\bar{b}$  &  $=$0.0004739    \\
\end{tabular}
\caption[Summary of ``discoveries'' for single top.]{Summary of ``discoveries'' for single top. Cost is the number of pseudo-signal events required to obtain $\tildeScriptP < 0.001$. The second column contains the final state in which the most interesting region is found at the point of discovery. The third column contains \tildeScriptP\ at discovery.}
\label{tbl:discoveries_model13}
\end{table}
\begin{figure}
\hspace{-1.5cm}
\begin{tabular}{cc}
\includegraphics[width=0.42\columnwidth,angle=270]{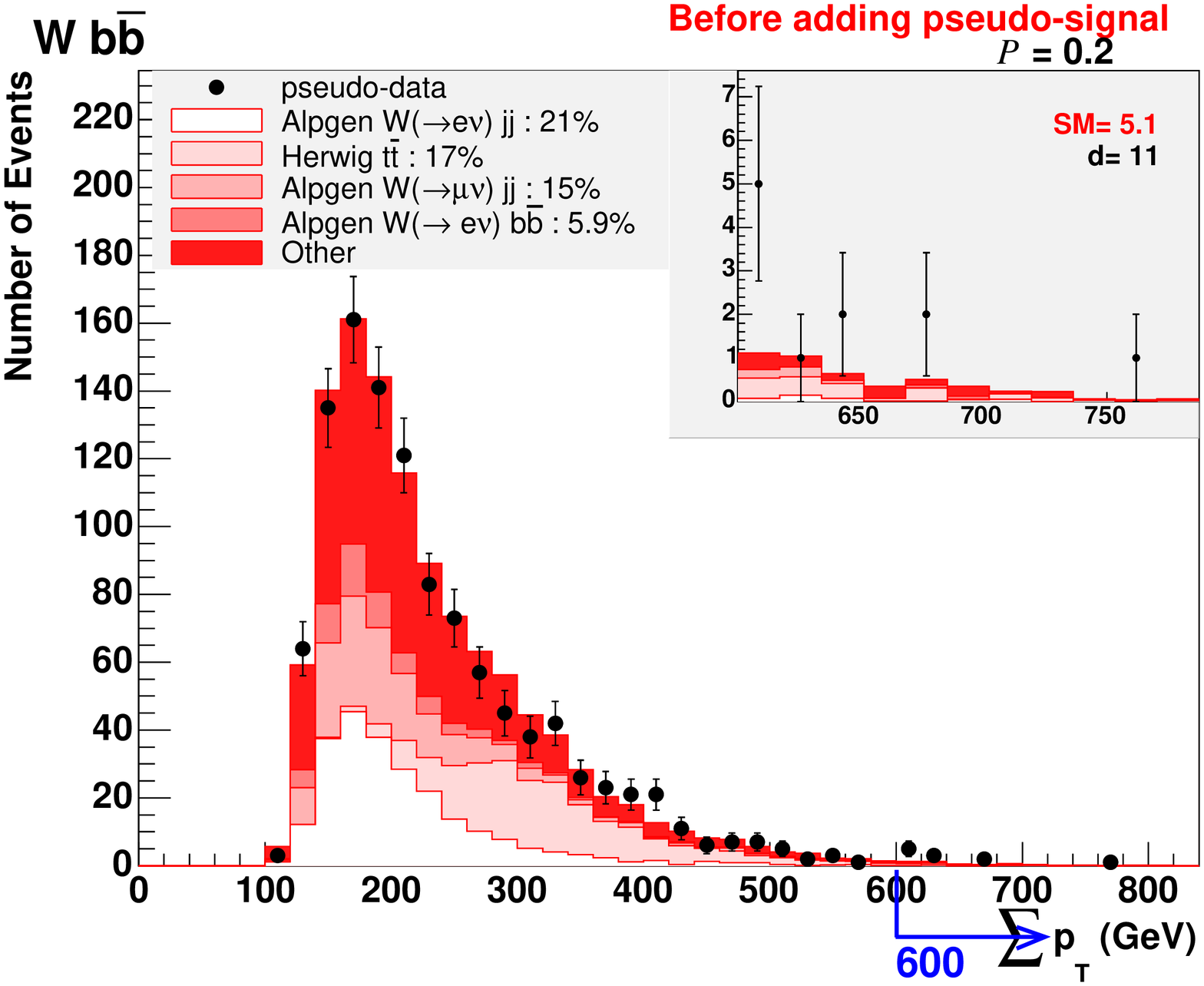} &
\includegraphics[width=0.42\columnwidth,angle=270]{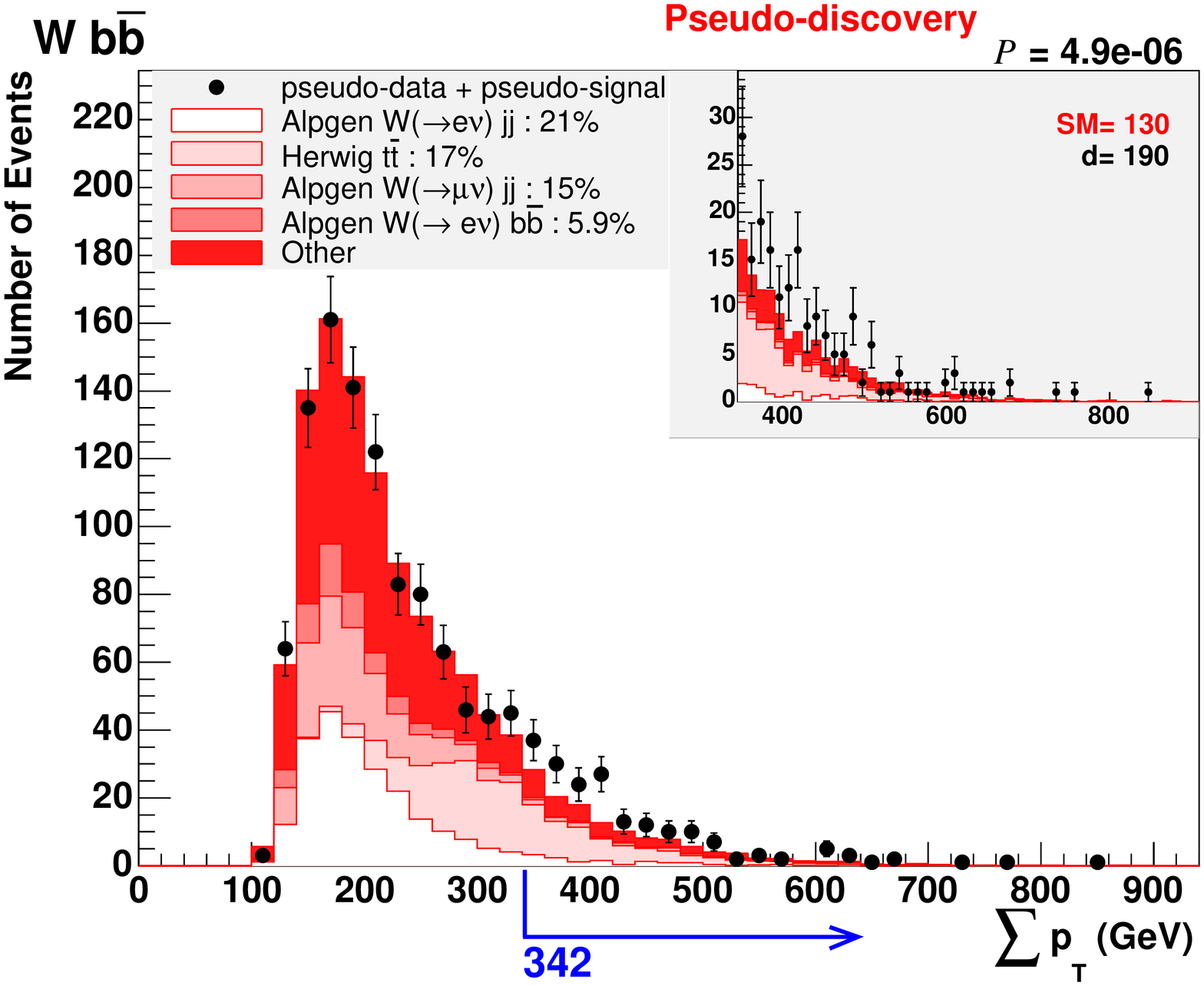} 
\end{tabular}
\caption[Pseudo-discovery of single top quark]{{\em (Left)}  The final state in which single top first appears, as it is before the addition of any pseudo signal.  {\em (Right)}  The same final state, after the addition of pseudo signal required for its discovery by \Sleuth.  For this discovery, 3600 pseudo signal events yields $\tilde{\cal P} = 0.0009669$.}
\label{fig:pseudoDiscovery_model13}
\end{figure}

This sensitivity test is performed by injecting `signal' single top events into pseudo-data generated from the background.
Single top events are obtained from the CDF Top Group Monte Carlo samples {\tt stop00} and {\tt stop01} ($s$-channel and $t$-channel production respectively), run through our standard event reconstruction. 
The acceptance for the signal events into \Sleuth\ final states is shown in Table~\ref{tbl:modelPartitioning_model13}.

Signal events are added to the pseudo-data in chunks, until \Sleuth's discovery threshold of \tildeScriptP$< 0.001$ is reached.  To account for random fluctuations, ten such trials are performed and the final result is averaged from all trials.  Table~\ref{tbl:discoveries_model13} summarizes the result of each trial.

As expected, \Sleuth's `golden' final state for discovering single top is $Wb\bar{b}$.  The $\sim4$\% acceptance into this final state is consistent with the numbers obtained for dedicated single top searches \cite{CDF8286}.
Note that due to the definition of final states in \Sleuth, $Wb\bar{b}$ contains events with 2 or 3 jets, with at least 1 $b$-tag. This merges somewhat the standard single top separation into distinct 2-jet and 3-jet bins, and this is why the $t\bar{t}$ background contribution is relatively large.

\begin{figure}
\centering
\includegraphics[width=0.6\columnwidth]{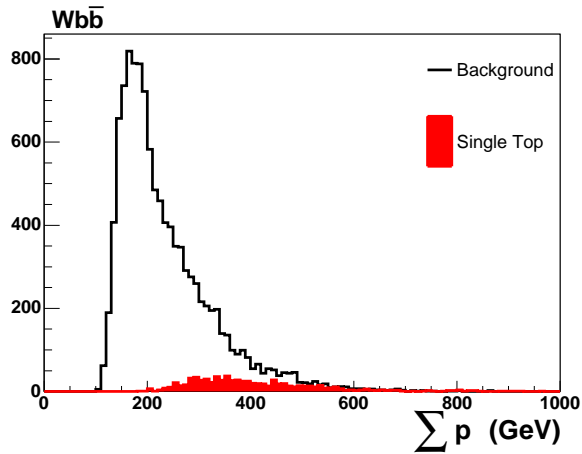}
\caption{Relative \SumPt\ distributions from single top signal and combined background prediction.}
\label{fig:singletop_bkg_sumpt}
\end{figure}

An example `discovery' is illustrated in Fig.~\ref{fig:pseudoDiscovery_model13}.
This shows the combined background prediction in the absence of signal, and the \SumPt\ distribution after adding sufficient signal to trigger \Sleuth's discovery threshold. Fig.~\ref{fig:singletop_bkg_sumpt} illustrates the \SumPt\ distribution from single top signal relative to the combined background prediction.

The result of this sensitivity test is that \Sleuth\ would be expected to discover single top at the $5\sigma$~level in 2~fb$^{-1}$ if it had a cross-section of $5.9\pm1.1$~pb.
The Standard Model expected cross-section is 2.86~pb (combined $s$- and $t$-channel). A naive extrapolation therefore leads to an expected luminosity for \Sleuth\ discovery of $2.0\times(5.9/2.86)^2=8.5\pm3.1$~fb$^{-1}$.

This conclusion seems perhaps surprising given the effort devoted to sophisticated tools such as Matrix Elements and Neural Networks for dedicated single top searches.
The apparent sensitivity of \Sleuth\ stems from the fact that it treats the background as being absolutely fixed. Any addition is therefore considered pure signal, allowing `discovery' of single top with relatively few extra events. 
In practice this is unrealistic, since \SumPt\ alone would find it hard to distinguish between single top production and excess $W$+heavy flavour relative to \Alpgen\ predictions, which have a large uncertainty.
In a realistic test, we would probably have to introduce a separate $k$-factor for $W$+heavy flavour, which would swallow up much of the single top signal, since there is no other populous final state that could constrain the W+heavy flavor $k$-factor independently of possible single top contributions.
For the dedicated single top searches, the total backgrounds are generally allowed to float, and more sophisticated purely `shape-based' variables are used to discriminate signal from background.


\section{Bump Hunter}
\label{sec:BumpHunter}

The bump hunter is a new feature added in the second round of this analysis, to enhance the sensitivity of the search to new physics involving narrow mass resonances. 

\subsection{Strategy}
\label{sec:BumpHunterStrategy}

The idea is to scan the spectrum of most mass variables with a sliding window.  The window needs to vary in width to follow the changing detector resolution.  As the window drifts accross a mass distribution, it evaluates the probability that the amount of data therein, or even more, could have emerged by fluctuation from the predicted population.  The window where this probability is smallest contains the most interesting local excess of data.

In each final state there are typically several mass variables to scan.  On average there are $5036/399 \simeq 13$.  They include masses of all combinations of reconstructed objects, such as pairs, triplets, or bigger ensembles.  

The width of the sliding window equals two times the characteristic mass resolution for the given combination of objects and at the given mass value.
Mass uncertainty results from uncertainties about the specific energies and momenta of all objects involved.
It is possible to have combinations of four-momenta that result in the same mass, but different mass uncertainties.
For example, if a $Z^0$ decays to $e^+e^-$, the mass of that pair will always be close to the nominal $m_Z \simeq 91$ GeV, though its resolution will depend on the boost of the decaying $Z^0$.
Obviously, each event has a different mass uncertainty, so we need to estimate the characteristic mass resolution at each value of mass and for each mass variable.  That characteristic mass resolution will be representative of the mass resolution of the events there.
To estimate it, we assume that all objects in the ensemble have equal momentum, negligible mass, and their momenta balance on a plane\footnote{If the (equal) momenta are two, to balance they have to be back-to-back.  If they are three, they have to be on the same plane, each separated by $120^\circ$ from its first neighbors.  If we have $N\ge 4$ equal, balancing momenta in 3 dimensions, then their angular configurations can be significantly more complicated, as there are many possible arrangements that satisfy the condition of ballance.  To avoid such complexity, we choose to constrain all $N$ vectors in one plane, and assume the solution where all vectors have separation $\frac{2\pi}{N}$ from their nearest neighbors.}.
Then, we assign to each involved individual energy the appropriate uncertainty, depending on what object it belongs to, since different objects are measured with different energy resolutions.
For electrons and photons, the uncertainty is assumed to be $\Delta E_{\rm EM} = 0.14 \sqrt{E} + 0.015 E$, determined by the electromagnetic calorimeter.
For jets and $\tau$s it is taken to be $\Delta E_{\rm HAD} = E \sqrt{ 0.457/E + 20.3/E^2 + 0.00834 }$, determined by the hadronic calorimeter.
For (beam constrained) muons it is $\Delta E_{\mu} = 0.0005 E^2$, determined by the COT track curvature resolution.
In cases of transverse mass involving \pmiss, we assume roughly $\Delta E_{MET} = 3\sqrt{\pmiss}$.
We propagate those $\Delta E$s corresponding to the members of the ensemble into the system's total mass.
For example, if we want to find the characteristic mass resolution for a $(e^+,\mu^-,j)$ triplet at system mass 90 GeV, we have $m = \sqrt{(E_e+E_\mu+E_j)^2-(\vec{p_e}+\vec{p_\mu}+\vec{p_j})^2}$.
We assume $E_e=E_\mu=E_j\equiv E$ and the planar configuration with zero net momentum, to obtain that $m=3E$, hence $E=30$ GeV for each object.
We use the above formulas for the three different $\Delta E$s, keeping in mind the different resolutions for electrons, muons and jets, and then we propagate those uncorrelated uncertainties to the mass, to find $\Delta m = \sqrt{(\Delta E_e)^2 + (\Delta E_\mu)^2 + (\Delta E_j)^2}=6.57$ GeV.

The step size by which the window drifts equals half a characteristic mass resolution, therefore it varies along the mass spectrum, as the width does too.  That way there are no gaps left between consecutive windows.  Instead, consecutive windows partly overlap.

Each window comes with two sidebands, extending on each side as far as the window's width. The region of the spectrum that is scanned is slightly narrower than the whole spectrum's span (defined as the interval between the highest-mass and the lowest-mass event in both data and background), so that all considered windows have sidebands lying within the spectrum.

As the window drifts along a mass spectrum, its \pval\ is calculated at each location.  That is defined as the Poisson probability that the Standard Model events expected in the window ($b$) would fluctuate up to or above the observed data ($d$), i.e.\ $\pval=\sum_{i=d}^{\infty}\frac{b^i}{i!}e^{-b}$.

A window qualifies as a bump if it satisfies the following criteria: 
\begin{itemize}
\item The central region must contain at least 5 data events.
\item Both sidebands must be less discrepant than the central region, i.e.\ both must have larger \pval.
\item If the background in a sideband is non-zero, then it must have $\pval > \int_{5}^{\infty}\frac{1}{\sqrt{2\pi}}e^{-x^2/2}\, dx$, namely it must not exhibit a significant ($5\sigma$) discrepancy.  If the background is zero, then it must have less than 5 data\footnote{This special treatment of the zero-background case is to be able to spot excesses of data that may be isolated at mass values where there is no Standard Model background at all.  If we had, for example, 6 events in the central window and 1 event in the sideband, we wouldn't like this band to disqualify due to having a discrepant sideband.}.
\item The above criteria need to hold even when we consider the possible effects of low Monte Carlo statistics in the background.  This is explained next.
\end{itemize}

\begin{figure}
\centering
\begin{tabular}{cc}
\includegraphics[angle=-90,width=0.45\columnwidth]{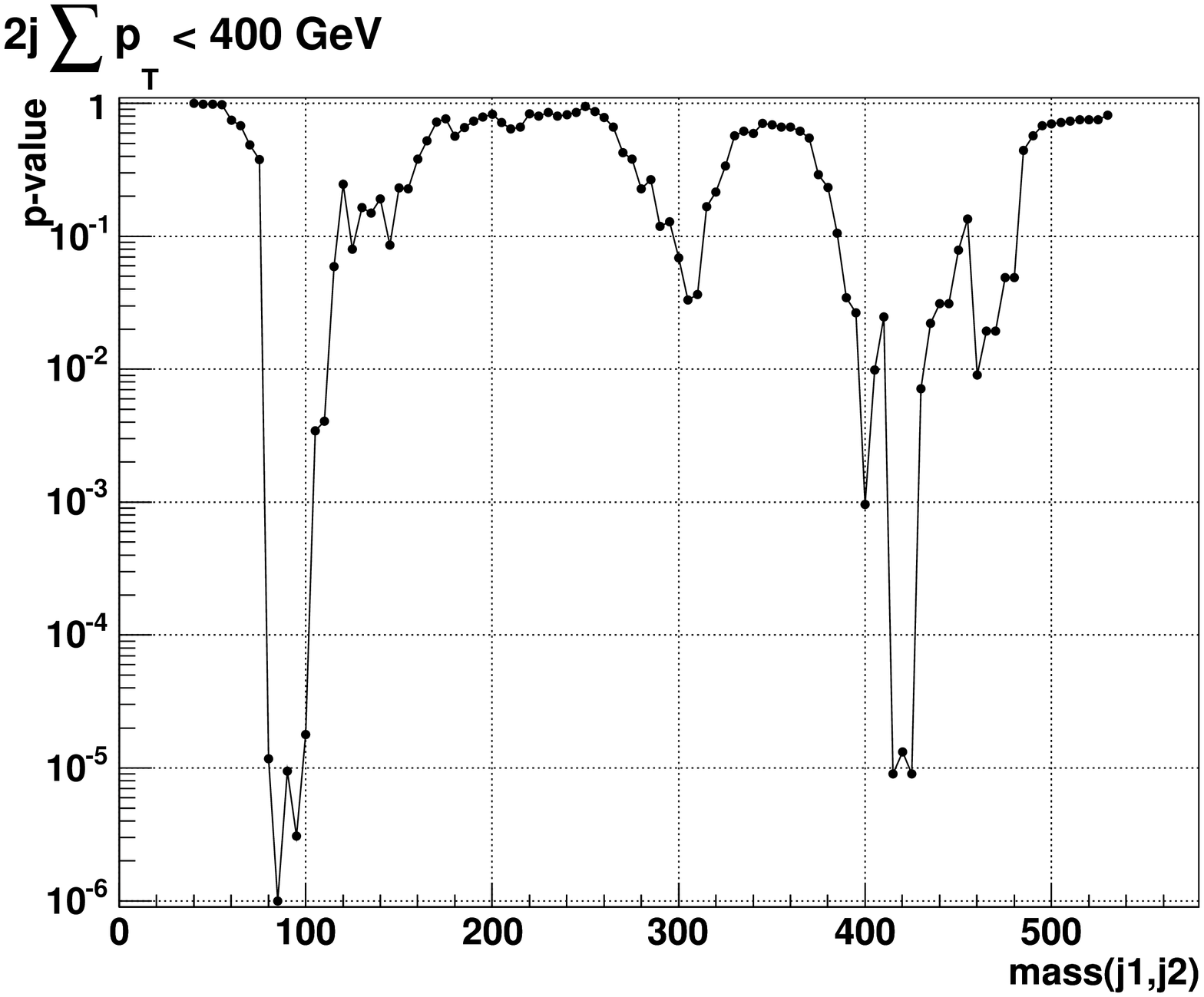} &
\includegraphics[angle=-90,width=0.45\columnwidth]{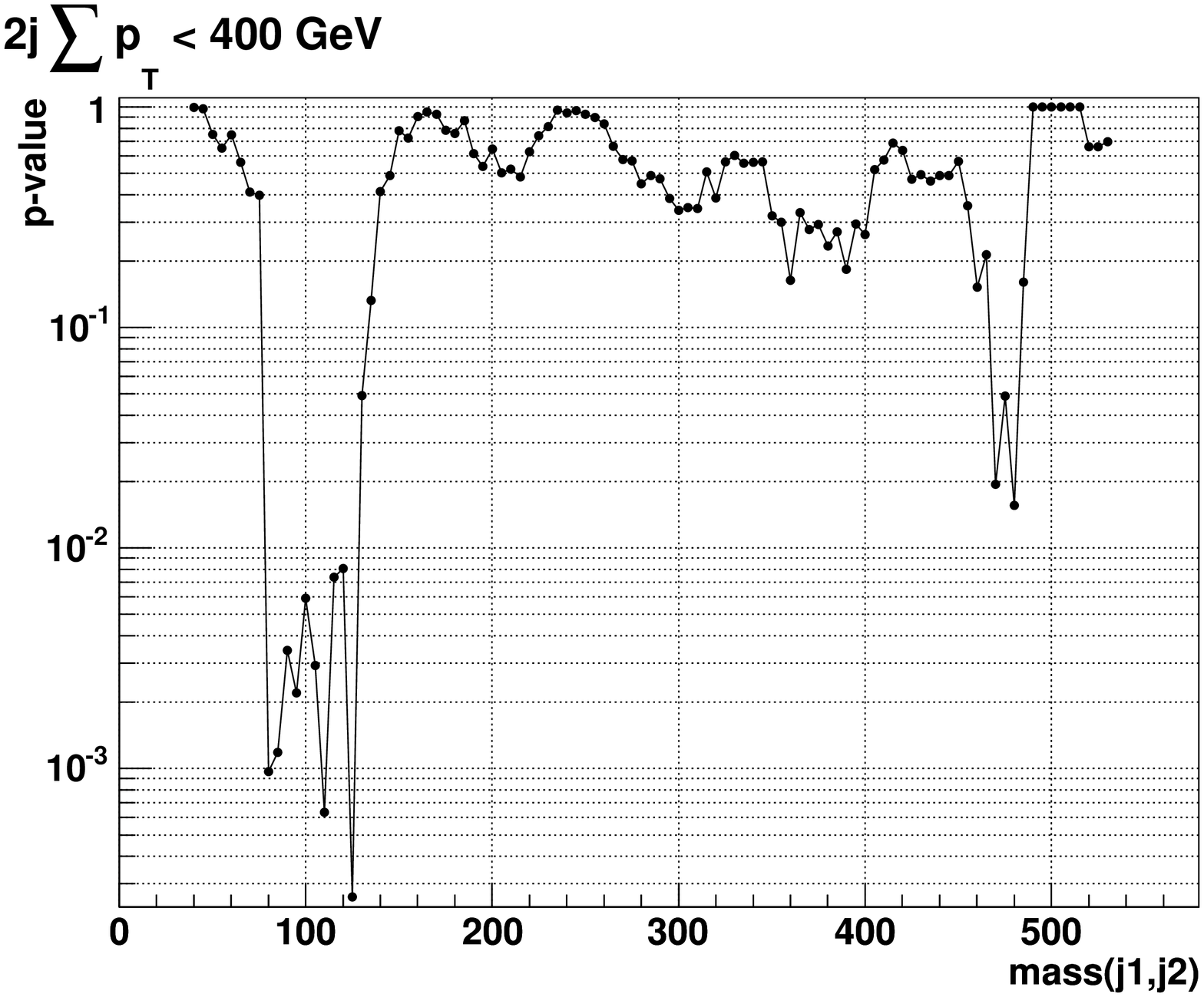}
\end{tabular}
\caption[\pval\ of all bumps in  mass($j1,j2$) in final state $2j\;\SumPt<400$~GeV]{{\em (Left)} The \pval\ of each bump candidate, as a function of the location of each window's center, along mass($j1,j2$) in final state $2j\;\SumPt<400$~GeV.  Bump candidates failing quality criteria have \pval=1. The most significant bump has $\pval \sim 10^{-6}$, which translates to $P_a\sim 3\times10^{-5}$ and $P_b \sim 0.15$, therefore all local excesses are insignificant. {\rm(Right)} For demonstration, we apply the conservative anti-spike treatment to all bump candidates.  The result of anti-spike treatment is to have larger $p$-values and the reduction of significance is greater in regions like around 400 GeV, where Monte Carlo statistics are poorer, therefore spikes contribute more.}
\label{fig:bumpsReport2j0-400_withSpikeTreatment}
\end{figure}

It can happen to have a great excess of data in the central window, and simultaneously non-discrepant sidebands, but realize that the sidebands contain only a couple of very large-weight events in the Standard Model background.  These large-weight events are called ``spikes'', and are the result of limited Monte Carlo statistics.  That bump would potentially pass all quality criteria, and appear to be statistically significant, but it would be prudent to treat conservatively the presence of spikes in the sidebands, and consider that these Monte Carlo events could easily have been in the central window instead.  In that case, the \pval\ of the central window would be larger (less significant) and the sidebands would have a higher probability to be discrepant, hence the bump could disqualify.  Since limited Monte Carlo statistics are a practical limitation, we have to be conservative and eliminate, if necessary, this bump.  To do that, we first need to define what we consider as a spike in each sideband, and reevaluate the \pval\ and the quality of the bump, assuming the spikes from both sidebands transfered into the central window.  To define the weight of spikes in a sideband, we look for outliers among the Monte Carlo events, namely for events with significantly larger weights than the average weight of the events in the sideband.  We find the average weight and the standard deviation of weights in the sideband, including in the calculation all Monte Carlo events therein.  If there is an event whose weight lies beyond 3 standard deviations from the average, then we gradually reduce its weight.  As we reduce it, we reevaluate the average and standard deviation of weights.  If along its path towards smaller weight it meets another event of same weight, then their weights are bound to be equal from then on, and keep being gradually reduced together.  To visualize this process, imagine the axis of weights as an horizontal stretched string, and the weight of each event represented by the position of a tiny bead along this string; the larger the weight, the farther on the right the bead is located.  If there are significant outliers, namely beads very far on the right, we start pushing the rightmost bead slowly from right to left, to bring it closer to the others.  On its way, the rightmost bead drags with it any beads it meets, since beads can not pass through each other.  We stop this reduction of weights when they are all within 3 standard deviations from their average.  Then, we compare the total initial weight to the total final weight in the sideband.  The difference is weight attributed to spikes.  If this difference turns out to be smaller than the largest single weight in the sideband, then we define the latter as spike instead.  For the sake of saving time, we do not apply the anti-spike treatment described above, unless the p-value of a qualifying bump candidate is smaller than $\int_{5}^{\infty}\frac{1}{\sqrt{2\pi}}e^{-x^2/2}\, dx$, since it is not crucial to be conservative, when a bump is not significant to begin with.  A demonstration of the effect of the anti-spike treatment is shown in Fig.~\ref{fig:bumpsReport2j0-400_withSpikeTreatment}.

When a variable's spectrum is scanned from one end to the other, the qualifying bump with the smallest \pval\ is the most interesting within that variable.  Its statistical significance is quantified on first level by its \pval; the smaller the more significant.  It is crucial, though, to account for the trials factor due to examining many windows within that spectrum.  We need, therefore, to estimate the probability that a qualifying bump candidate of such a small (or smaller) \pval\ would appear anywhere along the spectrum, if instead of the actual data we had pseudo-data pulled from the Standard Model distribution.  We denote this probability $P_a$, and it can be estimated either experimentally (by producing many sets of pseudo-data and scanning them for more interesting bumps), or using a semi-analytic calculation. 

The semi-analytic method, whose goal is to save the enormous time-cost of using Monte Carlo to experimentally evaluate $P_a$ for all mass variables, proceeds as follows:  For each window and its sidebands, we estimate with Monte Carlo the probability that it would satisfy quality criteria ($P(Q)$), if the data populations in the center and in the sidebands were pulled randomly from the respective expected populations therein.  Let's denote the \pval\ of the most interesting bump in the actual data \pvalmin.  Denote the probability that a window would have $\pval \le \pvalmin$ as $P(S)$.  The probability that a window would qualify and simultaneously have $\pval \le \pvalmin$ is $P(Q \land S) \simeq P(Q) \; P(S)$, where we assumed that $Q$ and $S$ are independent.  This is not generally true, but holds approximately in most cases.  In fact $P(Q|S) = \frac{P(Q\land S)}{P(S)} \ge P(Q)$, because if $S$ is true then we have a significant excess of data in the central window, which makes it somewhat less likely for the sidebands to exhibit a bigger discrepancy than the center, hence it's more likely that quality standards ($Q$) will also be met.  So, $P(Q)\;P(S)\le P(Q|S)\;P(S)=P(Q\land S)$, i.e.\ we slightly underestimate $P(Q\land S)$ by approximating it with $P(Q)\;P(S)$.
$P(S)$ is approximately $pval_{\min}$, but that is exactly correct only as long as there is an {\em integer} number of data that, given the background in the window, would result in a \pval\ of {\em exactly} \pvalmin.  If that is not the case, then $P(S) \le pval_{\min}$, because to exceed in significance the most interesting bump, this window would need to exhibit a \pval\ not just equal to \pvalmin, but smaller.  For example, if $\pvalmin=0.01$ and the background is $b=3.2$, then to exceed \pvalmin\ in significance we need the data to be at least $d=9$.  If $d=8$ then $\pval=0.016 > 0.01$.  However, if $d=9$ then $\pval=0.0057$, which means that the true $P(S)$ in this example would be $0.0057$ instead of $0.01$.  This difference becomes negligible for large backgrounds, where one event more or less changes \pval\ negligibly.

We find, as described, $P(Q \land S)$ for all windows considered along the spectrum, and set $P_a=1-\prod(1-P(Q\land S))$, namely the probability that at least one window would qualify and surpass in significance the most interesting bump in the actual data.  Here, another assumption is implicit: that windows are independent.  

\begin{figure}
\centering
\includegraphics[angle=-90,width=0.8\columnwidth]{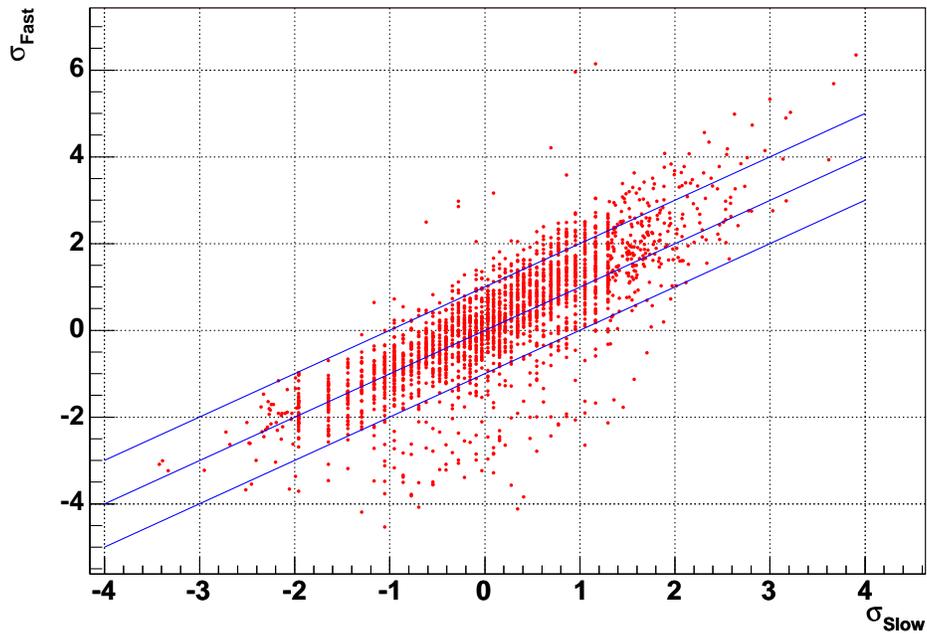}
\caption[Comparison of fast versus slow method to estimate $P_a$.]{Comparison of fast versus slow method to estimate $P_a$.  Each point corresponds to a mass variable with at least one qualifying bump in pseudo-data.  The three lines indicate the locus where the fast estimate of $P_a$ is equal to, or $\pm1\sigma$ away from the slow estimate of $P_a$.  Slow $P_a$ can be only a rational number, since it is the fraction of two integers, namely the number of pseudo-data distributions with a more interesting bump and the total number of tried pseudo-data distributions.  That is why the slow $P_a$ appears to assume a discrete spectrum of values.}
\label{fig:compareFastVsSlow}
\end{figure}
\begin{figure}
\begin{tabular}{cc}
\includegraphics[angle=-90,width=0.45\columnwidth]{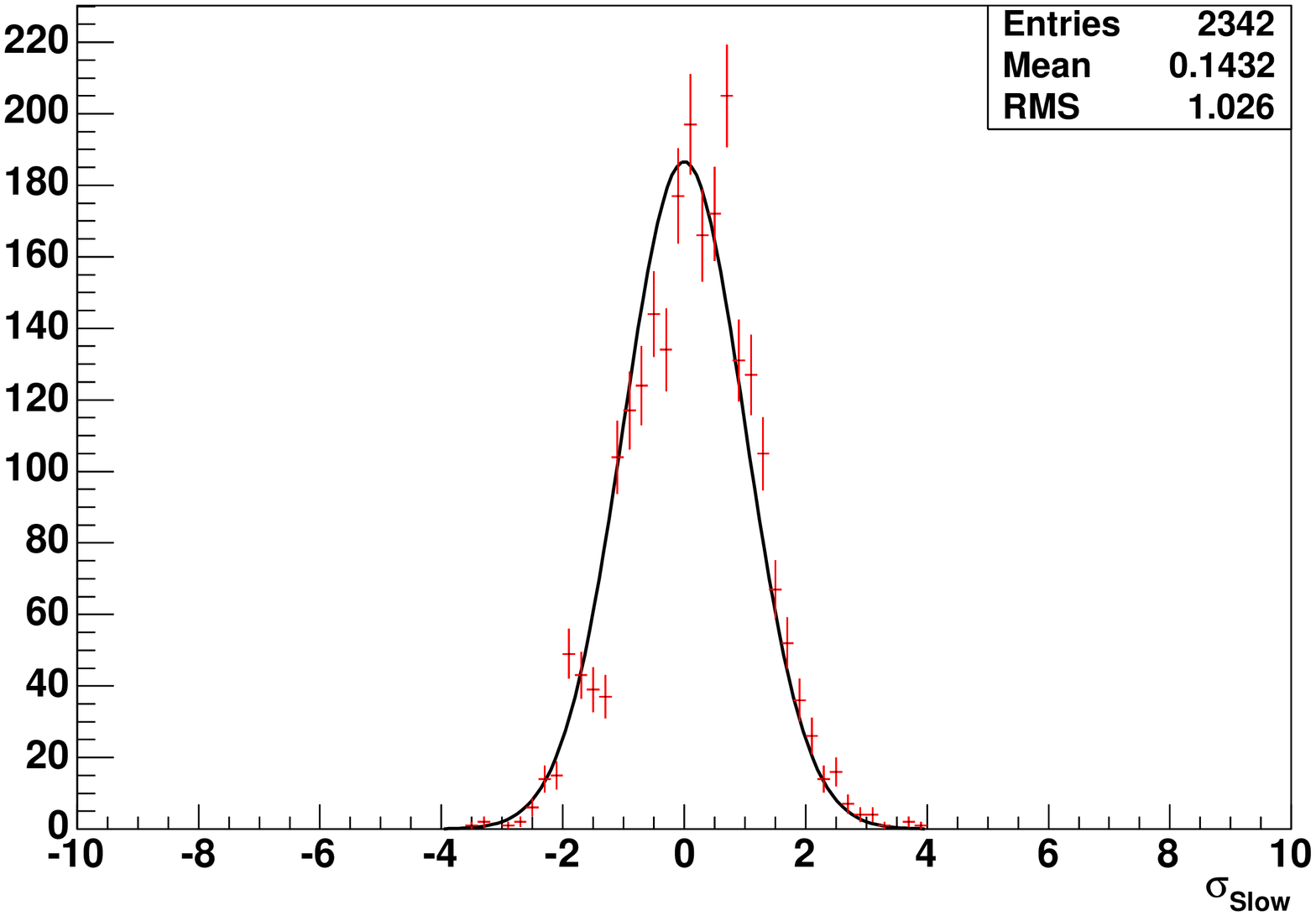} &
\includegraphics[angle=-90,width=0.45\columnwidth]{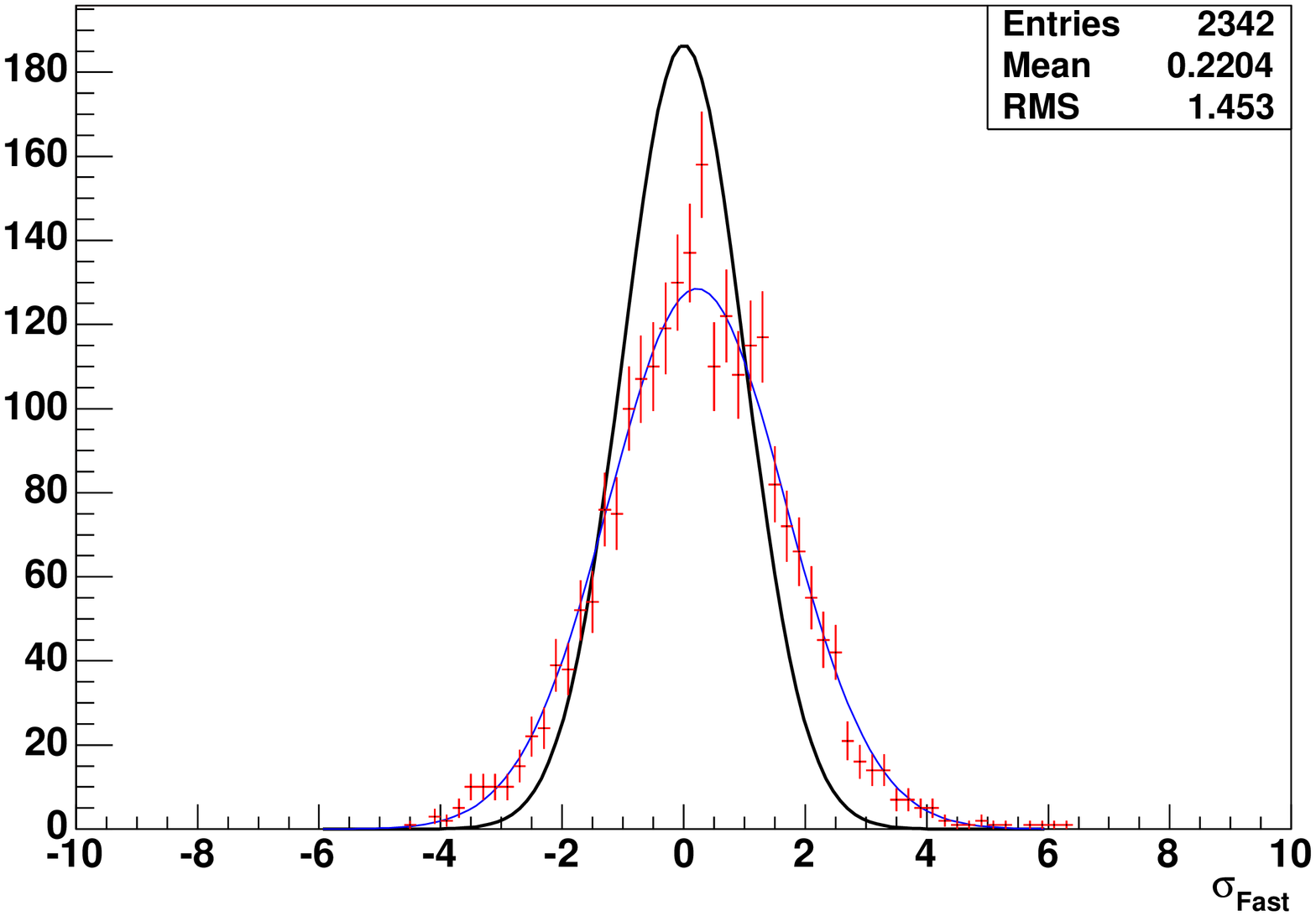}
\end{tabular}
\caption[Expected distribution of the fast and the slow estimator of $P_a$]{Expected distribution of the fast and the slow estimator of $P_a$, when applied on pseudo-data.  The slow estimator {\em (left)} is distributed according to a normal distribution (except for some recurrent values which reflect that the slow estimator can only be a rational number), while the fast one {\em (right)} follows a Gaussian probability density function with mean 0.2204 and standard deviation 1.453.  In the right plot, the Normal distribution has been drawn for comparison.}
\label{fig:expectedDistributionsOfPa}
\end{figure}

A comparison between the semi-analytic (fast) and the experimental method to estimate $P_a$ is shown in Fig.~\ref{fig:compareFastVsSlow}.  Pseudo-data were pulled from all mass distributions, and then both the slow and the fast methods were used to estimate $P_a$.  The comparison shows that, for pseudo-data, the fast method returns a $P_a$ which is, when translated into units of standard deviation, within about 1$\sigma$ from the $P_a$ determined by the slow method.  This difference reflects on the expected distributions of $P_a$ from all mass variables when using the two methods.  While the slow method returns a $P_a$ with uniform expected distribution, the fast method's $P_a$ is distributed as shown in Fig.~\ref{fig:expectedDistributionsOfPa}.

The slow method does not rely on any approximation, therefore its answer is more representative of the true $P_a$.  It is only limited by the number of pseudo-data sets that we can generate.  Its disadvantage is that even when applied on just one mass variable to estimate the significance of its most interesting bump, it can take prohibitively long.  How long depends on the number of expected events in the final state where the mass variable belongs, but more importantly on the smallness of \pvalmin.  For really significant bumps ($\pvalmin \lessapprox 10^{-8}$) it may take millions of sets of pseudo-data to start resolving $P_a$ experimentally.  The slow method returns the best estimate of $P_a$ it could obtain within the amount of time it was allowed to run.  If during this amount of time it is clear at 95\% confidence level that $P_a$ is either greater or smaller than what corresponds to a 5$\sigma$ effect ($\int_{5}^{\infty}\frac{1}{\sqrt{2\pi}}e^{-x^2/2}\, dx=2.87\times10^{-7}$), then the slow method returns the estimated value of $P_a$ at that time, since the conclusion is clear and additional accuracy would be of no use.
Due to its great time cost, we employ the slow method only if the fast (semi-analytic) method has returned a significant enough $P_a$, i.e. smaller than what corresponds to a $4.5\sigma$ effect.  The final significance of a bump is not quantified by $P_a$, but by $P_b$ (defined later), which includes the whole trials factor.  For $P_a$ equivalent to $4.5\sigma$, $P_b$ is 2.1$\sigma$, safely away from the discovery threshold of 3$\sigma$ in $P_b$, which corresponds to $P_a$ of 5$\sigma$.  This is mentioned to explain that the slow and more accurate estimator for $P_a$ is employed not just beyond the discovery threshold, but safely earlier, when a bump starts being mildly significant.

Since $P_a$ encompasses the trials factor from examining multiple windows within the mass variable, it characterizes the significance of the mass viariable in terms of its most interesting bump. The next question is what the probability is that in a pseudo-experiment, where data are pulled from the Standard Model epxectation, any mass variable would appear with a $P_a$ smaller than the actual $P_a$ of the mass variable. We denote this probability as $P_b$. We estimate it assuming all mass viariables are statistically independent trials, therefore $P_b = 1 - (1-P_a)^N$, where $N$ is the total population of scanned mass variables from all \Vista\ final states.

In summary, for each mass variable the most interesting bump is the one with the smallest \pval, and with all trials factor accounted for, its significance is approximately given by $P_b$. Then $P_b$ is converted to units of standard deviations, and if it corresponds to a 3$\sigma$ effect or more, then we consider it a discrepancy worth pursuing.

\subsection{Results}
\label{sec:BumpHunter:Results}

\begin{figure}
\centering
\includegraphics[angle=0,width=0.8\columnwidth]{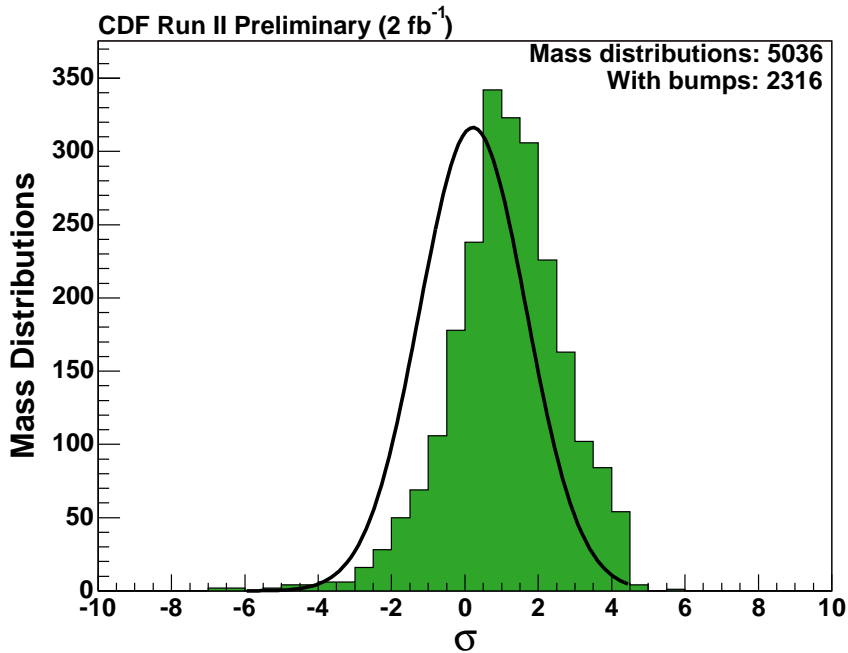}
\caption[Significance of the most interesting bump in each mass variable.]{Significance of the most interesting bump in each mass variable.  Each entry corresponds to one mass distribution found to contain at least one bump satisfying quality criteria.  The quantity distributed is $P_a$, transformed to units of standard deviation ($\sigma$), using the formula $P_a=\int_{\sigma}^{\infty}\frac{1}{\sqrt{2\pi}}e^{-x^2/2}\;dx$.  Large $P_a$ translates to a small number of $\sigma$ and signifies an insignificant effect.  The discovery threshold corresponds to 5$\sigma$.  The entries under 4.5$\sigma$ have been estimated using the semi-analytic (fast) method, which yields values distributed according to the black curve when applied on pseudo-data agreeing with the Standard Model background.  Values above 4.5$\sigma$ are estimated using the slow, more accurate method.   Therefore, values of $P_a$ corresponding to more than 4.5$\sigma$ can be translated directly to significance, since their expected values follow the Normal distribution.  About 5000 mass distributions are considered, which means that to have an effect of significance 3$\sigma$ after trials factor, it needs to have a significance of 5$\sigma$ or more in this scale of Pa.  Only one mass distribution has its most significant bump exceed this discovery threshold.  More details in the text.}
\label{fig:vistaSummary_bump}
\end{figure}

The summary of the most interesting bump in each mass variable is shown in Fig.~\ref{fig:vistaSummary_bump}.

\begin{figure}
\centering
\begin{tabular}{c}
\includegraphics[angle=-90,width=0.5\columnwidth]{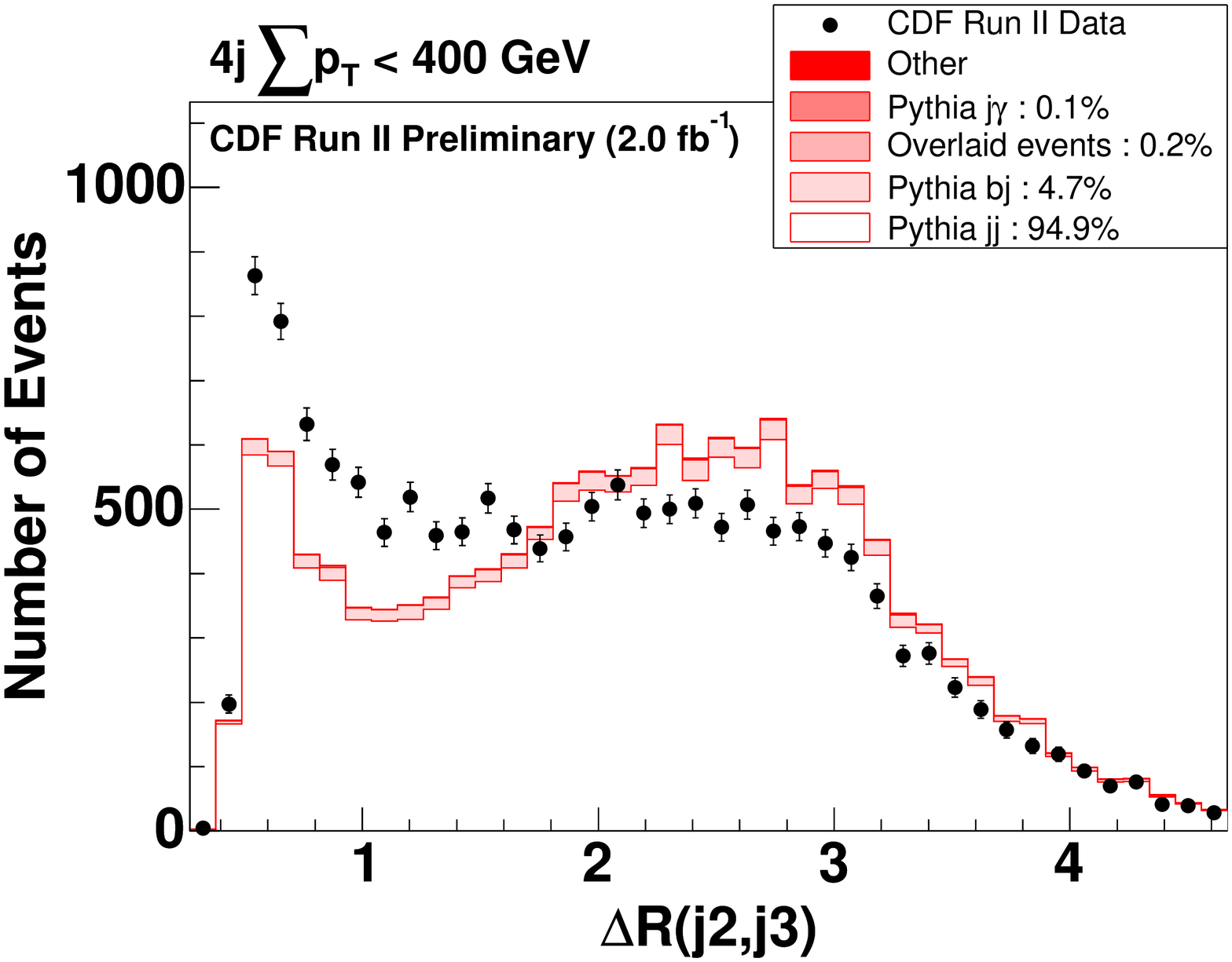}\\
\includegraphics[angle=-90,width=0.5\columnwidth]{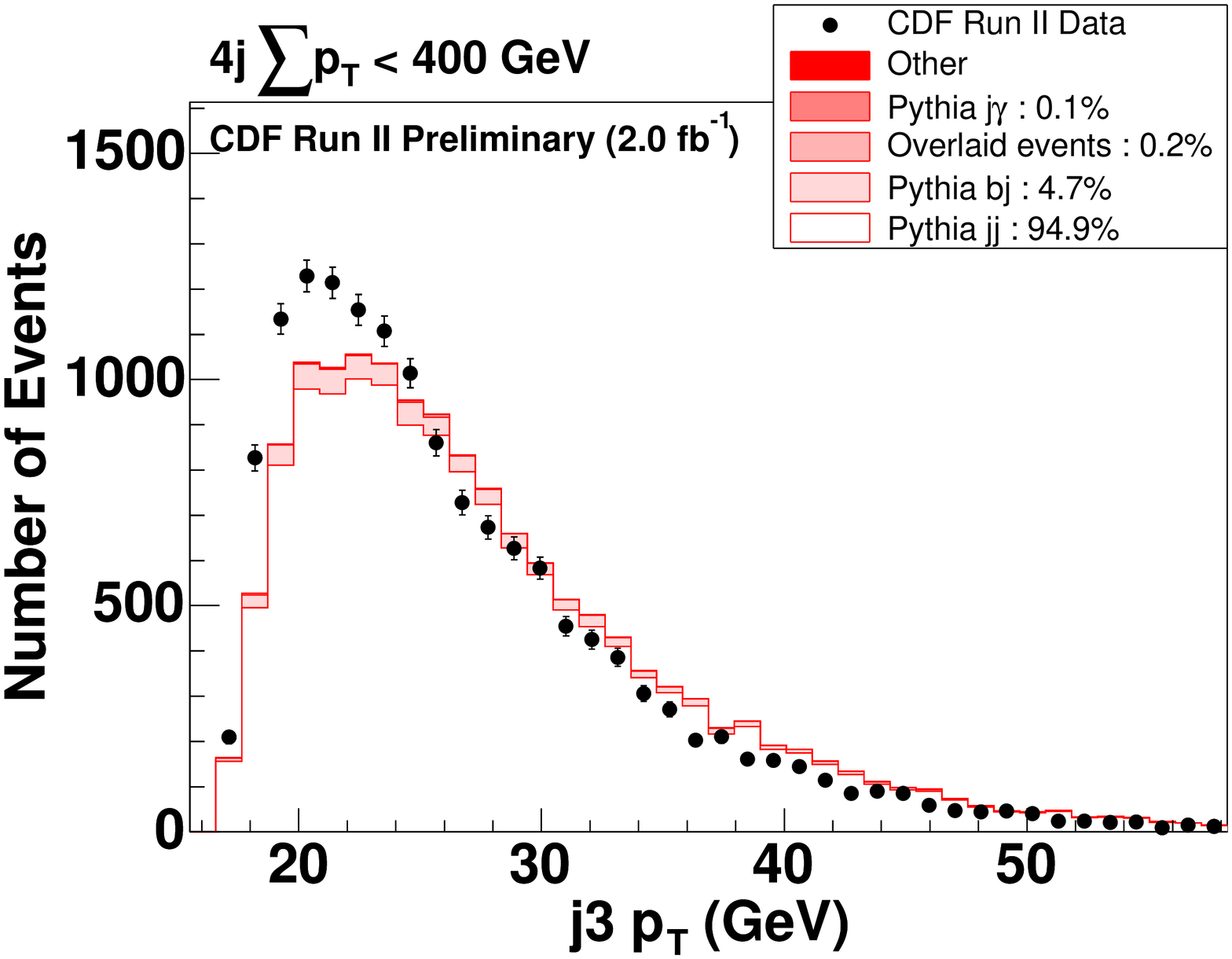}\\
\includegraphics[angle=-90,width=0.5\columnwidth]{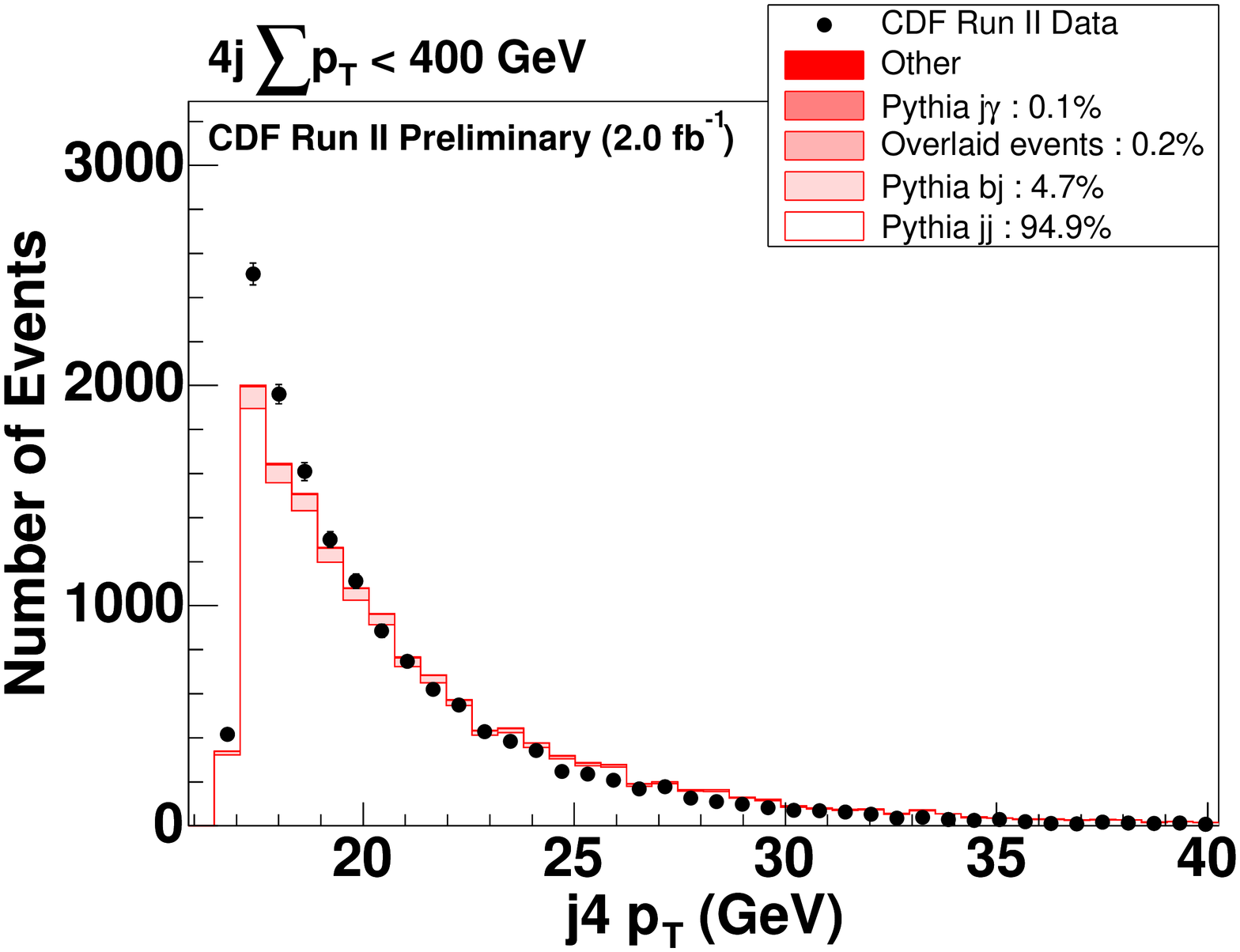}
\end{tabular}
\caption[Interpretation of the only significant mass bump found]{{\em(Upper)} The ``3-jet'' effect appearing in the angular separation between the second ($j2$) and the third ($j3$) leading jets, in final state $4j\;\sumPt < 400$ GeV.  There is an excess of soft final state radiated jets emitted at small angles. The {\em lower two} distributions from the same final state demonstrate exactly this excess, which is not present in the $p_T$ of the first and second leading jets.}
\label{fig:4j_sumPt0-400:3jEffect}
\end{figure}

\begin{figure}
\centering
\begin{tabular}{cc}
\includegraphics[angle=-90,width=0.45\columnwidth]{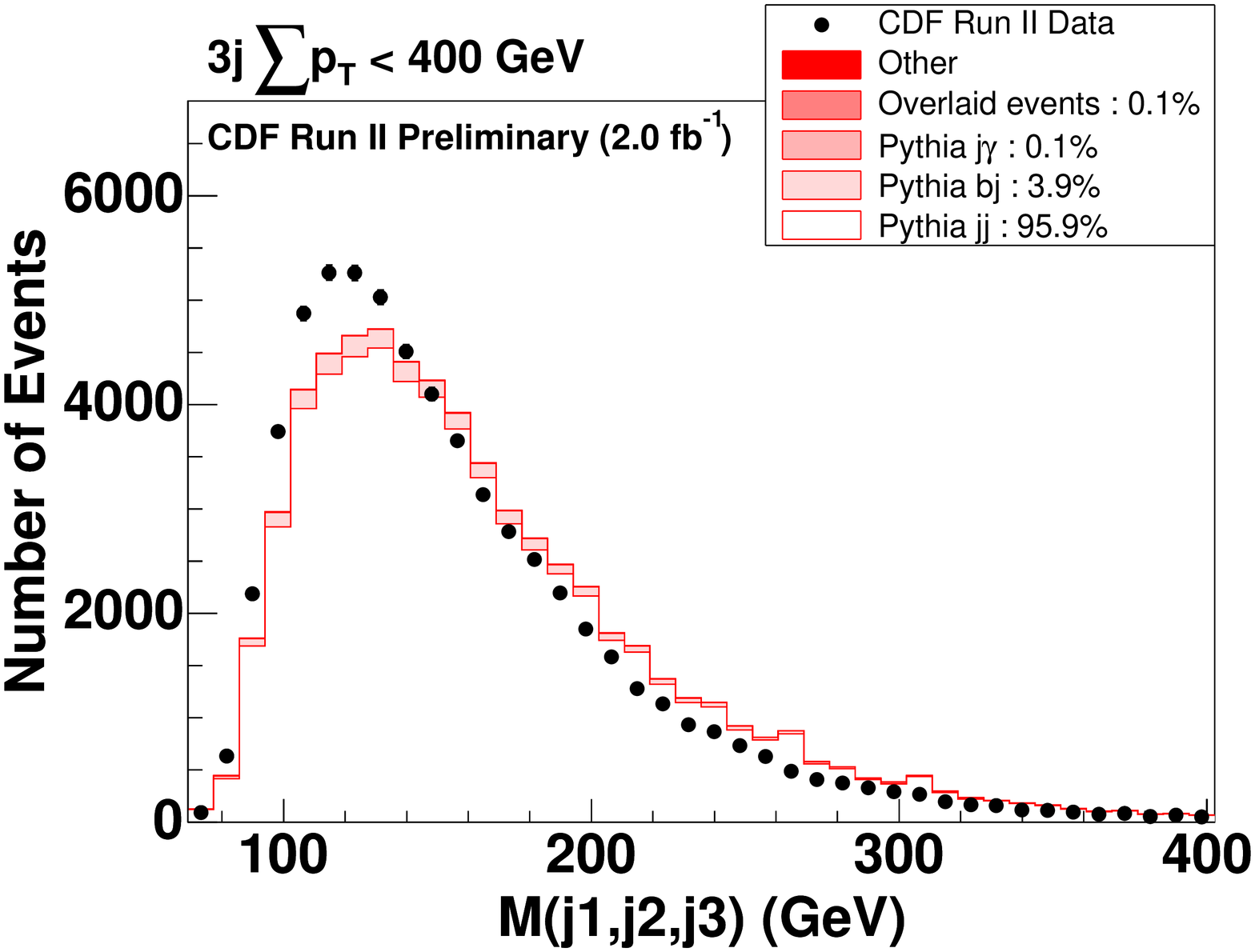} &
\includegraphics[angle=-90,width=0.45\columnwidth]{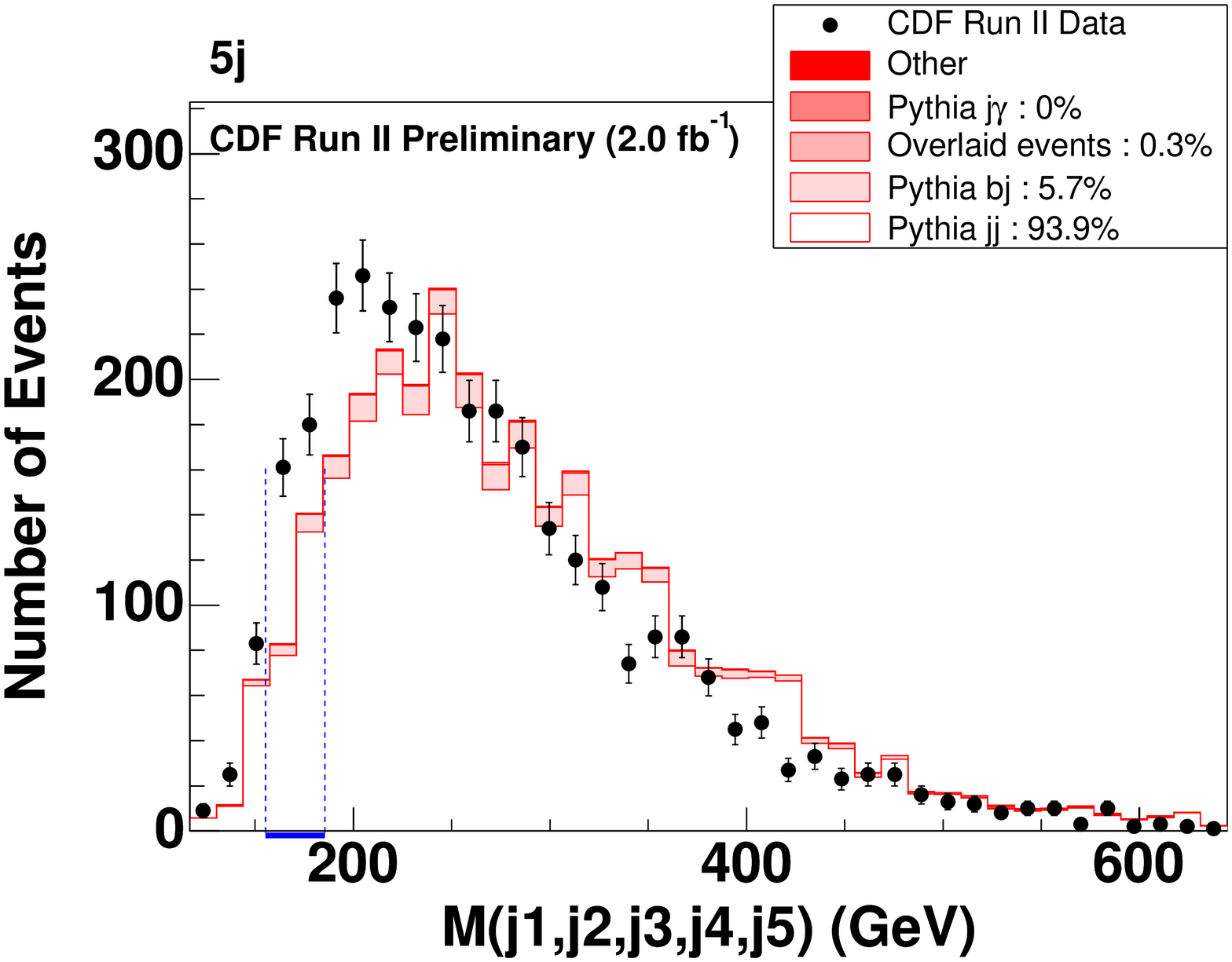}
\end{tabular}
\caption[The ``3-jet'' effect appearing in the mass of all jets in the final state with three {\em (left)} and five {\em (right)} jets.]{The ``3-jet'' effect appearing in the mass of all jets in the final state with three {\em (left)} and five {\em (right)} jets.  The excess is similar to the one identified as a bump in the $4j\;\sumPt<400$ GeV final state.  The difference in the case of $3j\;\sumPt<400$ GeV is that the excess is wide and the sidebands are discrepant, making this bump candidate disqualify, while in the case of $5j$ the excess satisfies bump quality criteria, but has $P_b$ corresponding to only 1.5$\sigma$.}
\label{fig:3jand5jmasses}
\end{figure}

The only mass variable with its most significant bump exceeding the discovery threshold is the mass of all four jets in the final states with four jets of $\sumPt < 400$ GeV, shown in Fig.~\ref{fig:4j_sumPt0-400:mass(j1,j2,j3,j4)}.  This is attributed to the ``3-jet'' effect, the main cause of all shape discrepancies in this analysis.  Fig.~\ref{fig:4j_sumPt0-400:3jEffect} shows another instance of the same effect in that final state.  The same effect is observed in final states of different jet multiplicity, as shown in Fig.~\ref{fig:3jand5jmasses}.

\begin{figure}
\centering
\includegraphics[angle=-90,width=0.8\columnwidth]{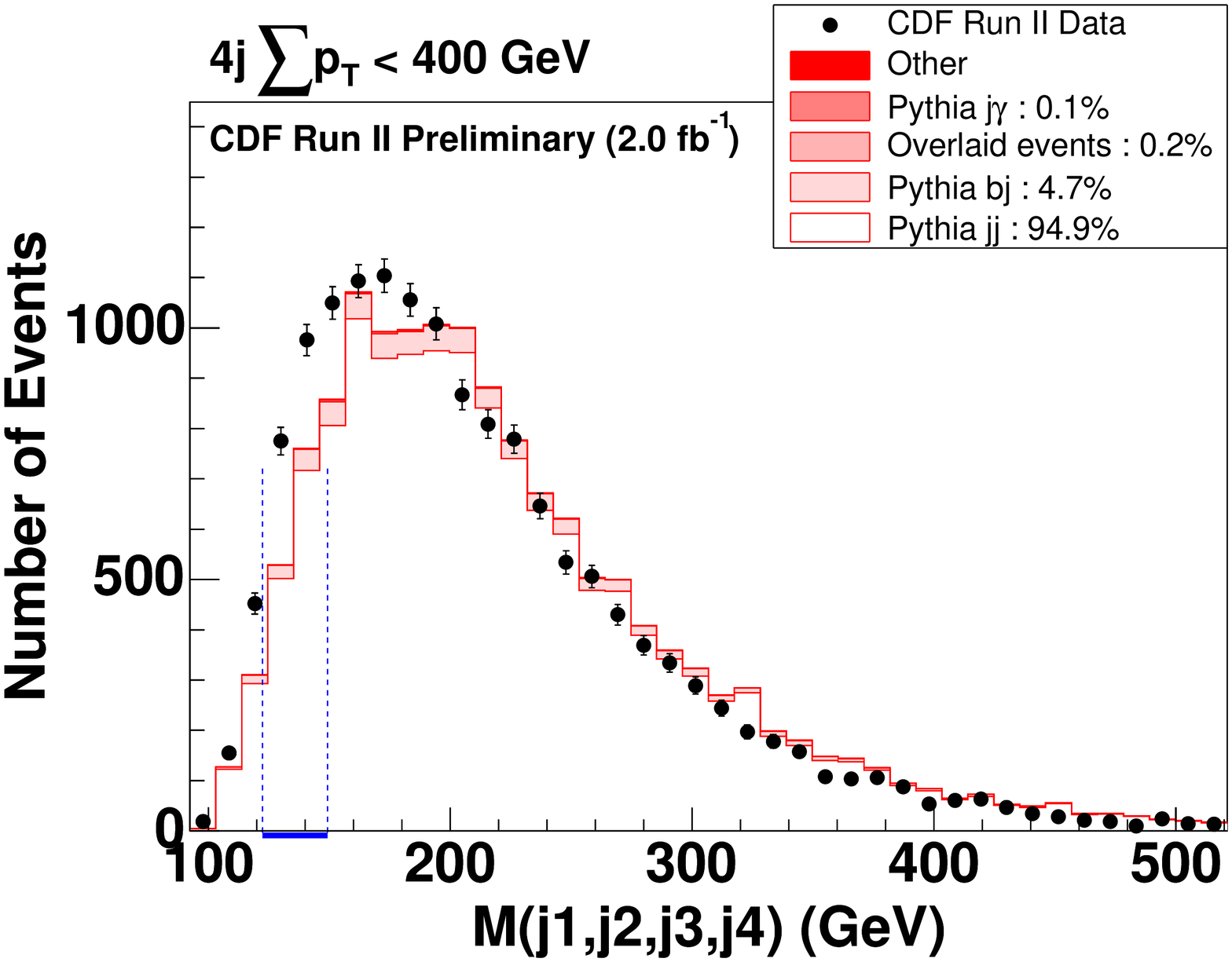}
\caption{The most significant bump found in the $4j\;\SumPt < 400$ GeV final state, indicated by the blue lines.  Its $P_b$ translates to 4.1$\sigma$.}
\label{fig:4j_sumPt0-400:mass(j1,j2,j3,j4)}
\end{figure}

Although no discovery-level bumps were found in other mass variables, it is interesting to present the most interesting bumps found in some mass distributions.

\begin{figure}
\centering
\begin{tabular}{c}
\includegraphics[angle=-90,width=0.5\columnwidth]{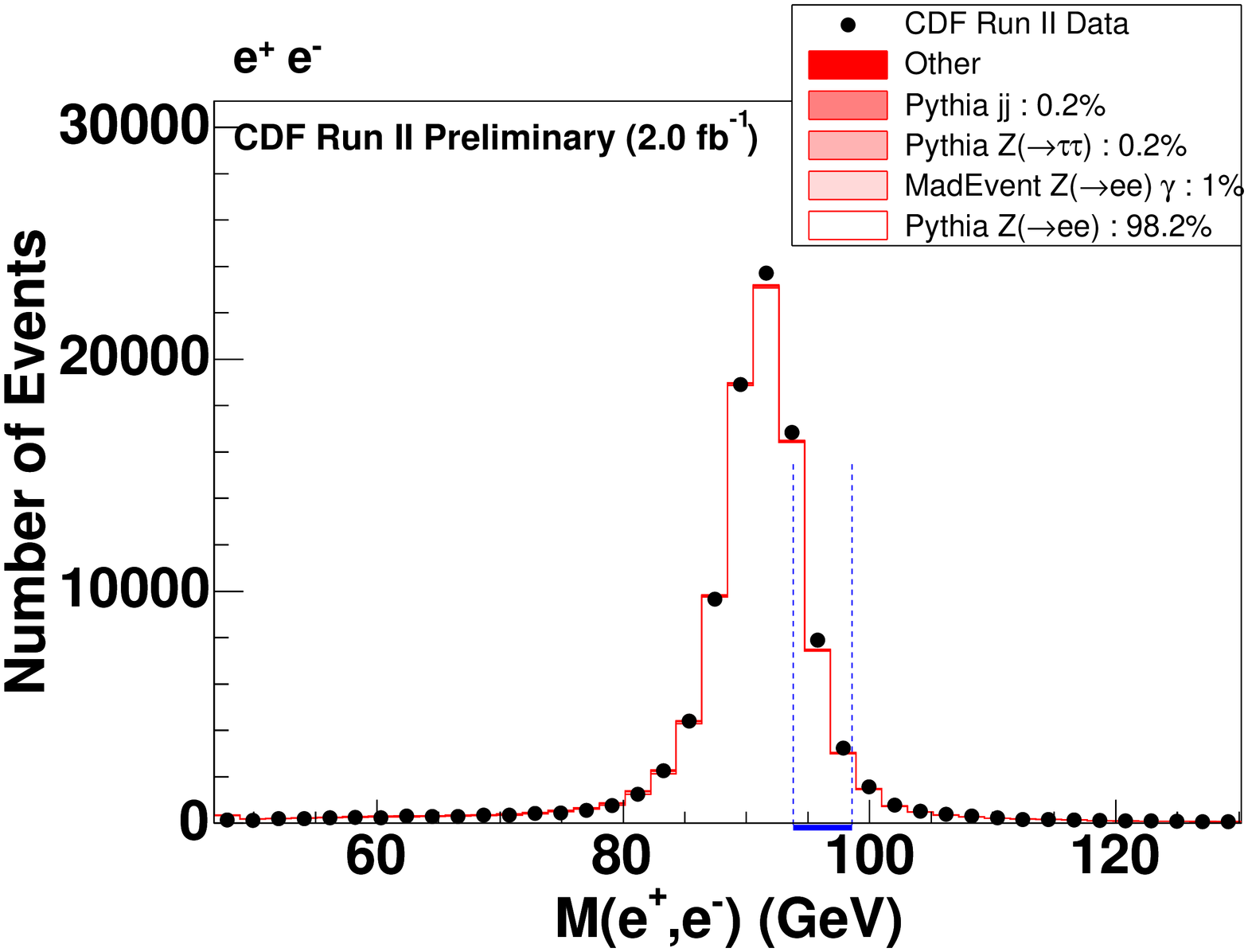}\\
\includegraphics[angle=-90,width=0.5\columnwidth]{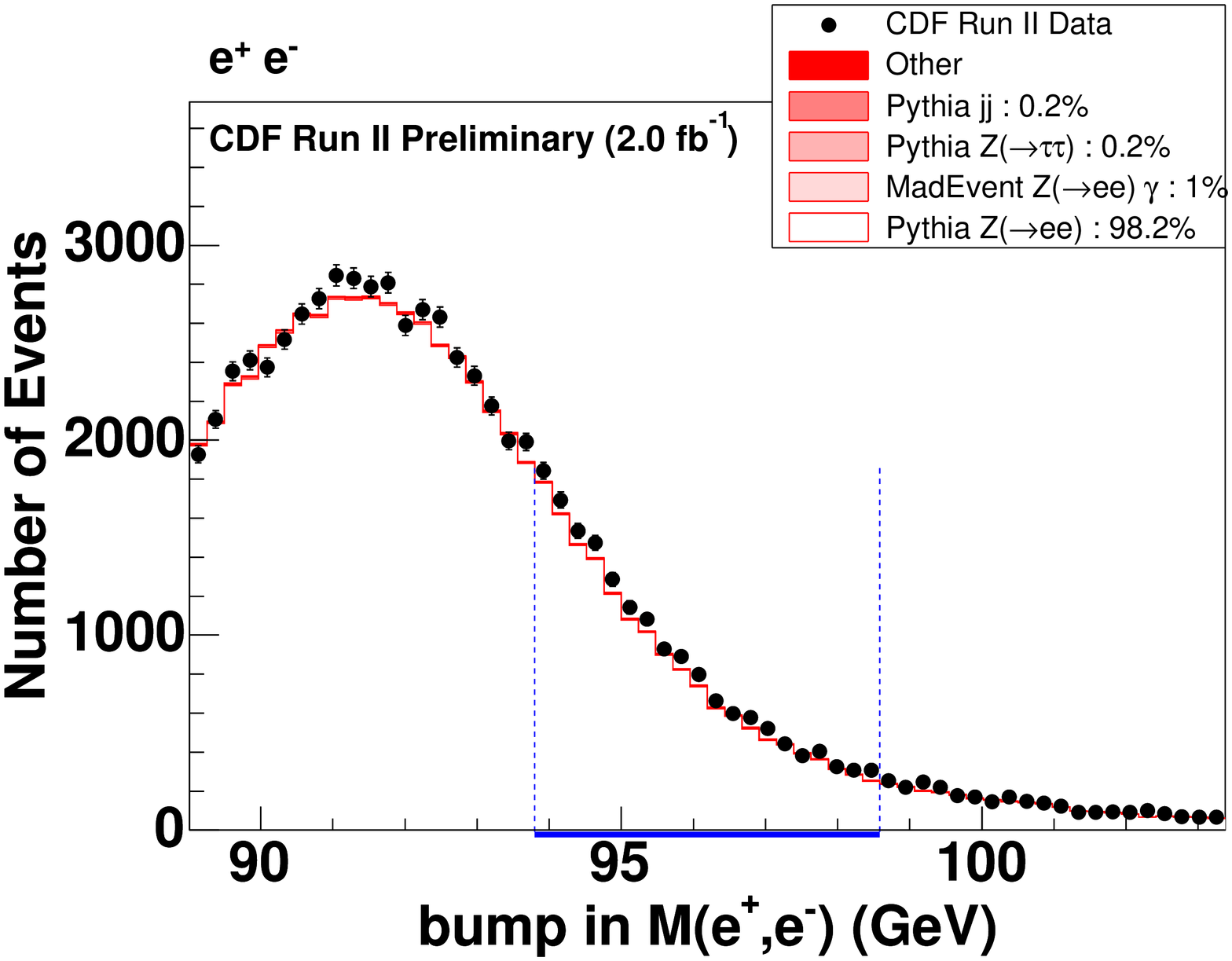}\\
\includegraphics[angle=-90,width=0.5\columnwidth]{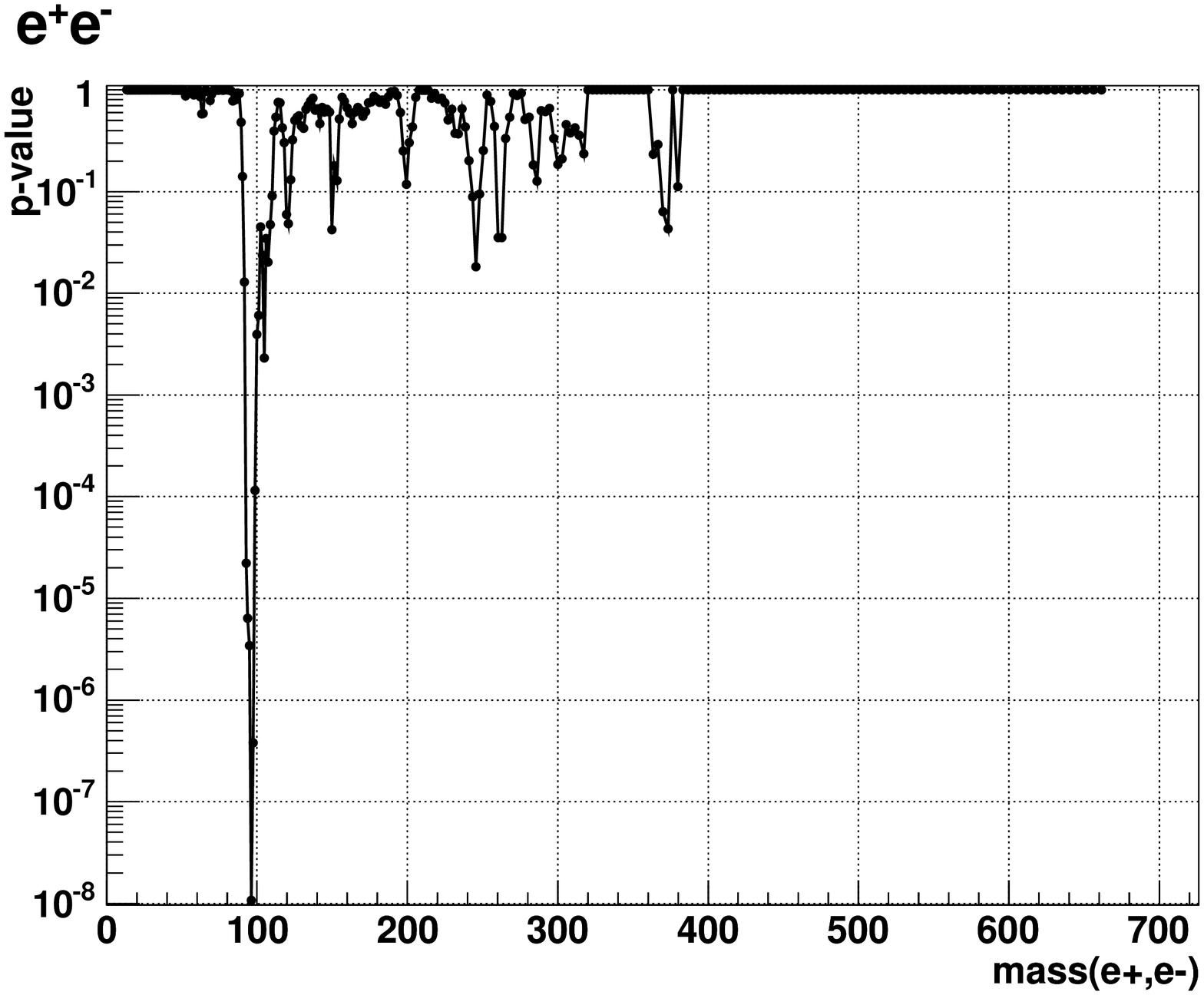}
\end{tabular}
\caption[Bumps found in $e^+e^-$.]{{\em (Upper two)} The most interesting bump found in final state $e^+e^-$.  {\em (Bottom)} The \pval\ of all bumps accross the mass spectrum of the two leptons.  Apart from this discrepancy at the $Z$-peak, which corresponds to a 2.7$\sigma$ effect after trials factor and reflects only a tiny mismodeling in a region with very high statistics, no other significant bump was found.}
\label{fig:bump1e+1e-}
\end{figure}

In the mass of the $(e^+,e^-)$ pair in the final state with two opposite sign electrons ($e^+e^-$) the most significant bump corresponds to a 2.7$\sigma$ effect, which is though exactly at the $Z$ boson resonance.  The number of expected events there is so high, that even the slightest systematic mismodeling would appear as very statistically significant.  From Fig.~\ref{fig:bump1e+1e-} it is clear that this ``bump'' is not due to new physics, but a tiny systematic mismodelling of the $Z$-peak, with no visible effect anywhere else.

\begin{figure}
\centering
\begin{tabular}{c}
\includegraphics[angle=-90,width=0.5\columnwidth]{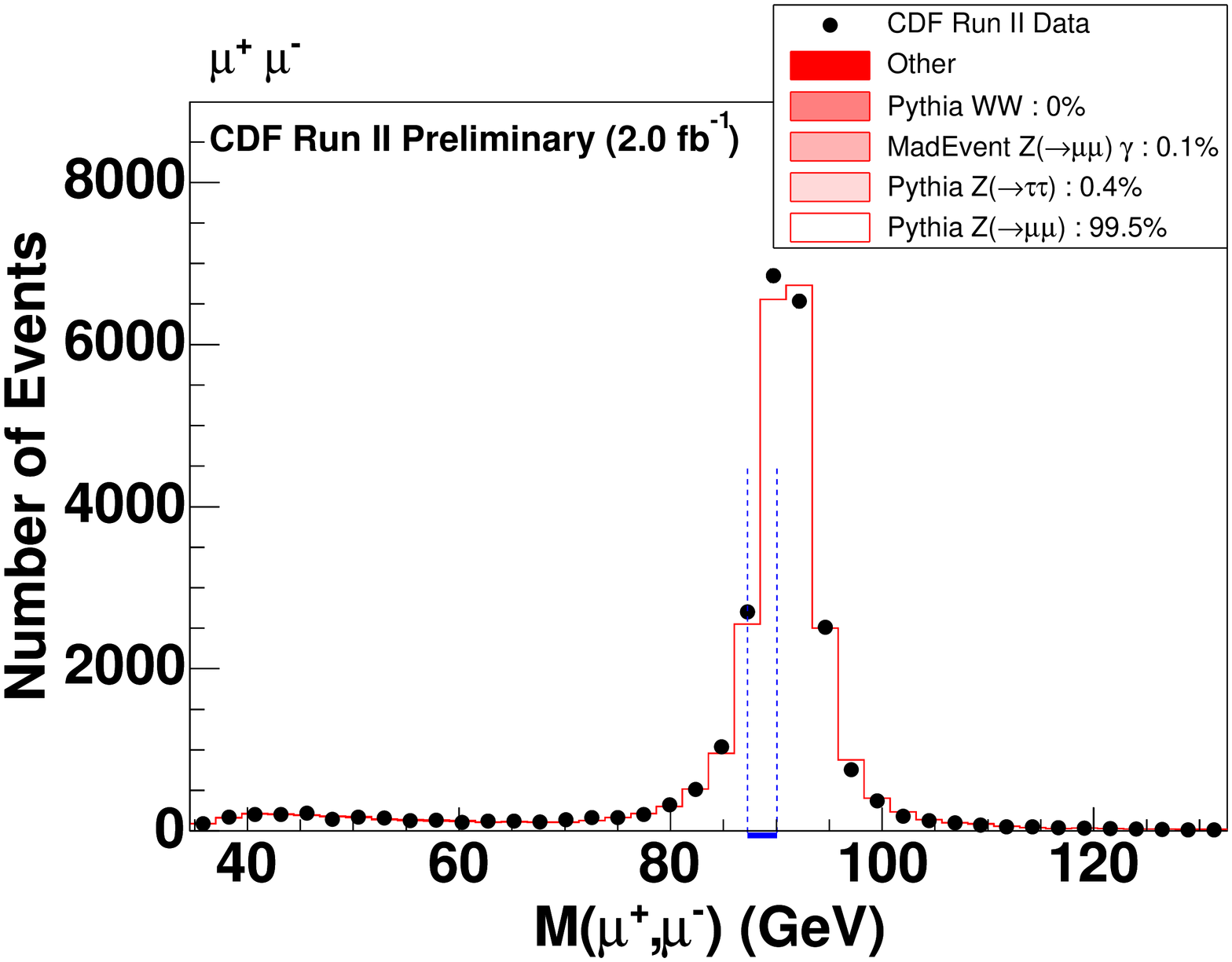}\\
\includegraphics[angle=-90,width=0.5\columnwidth]{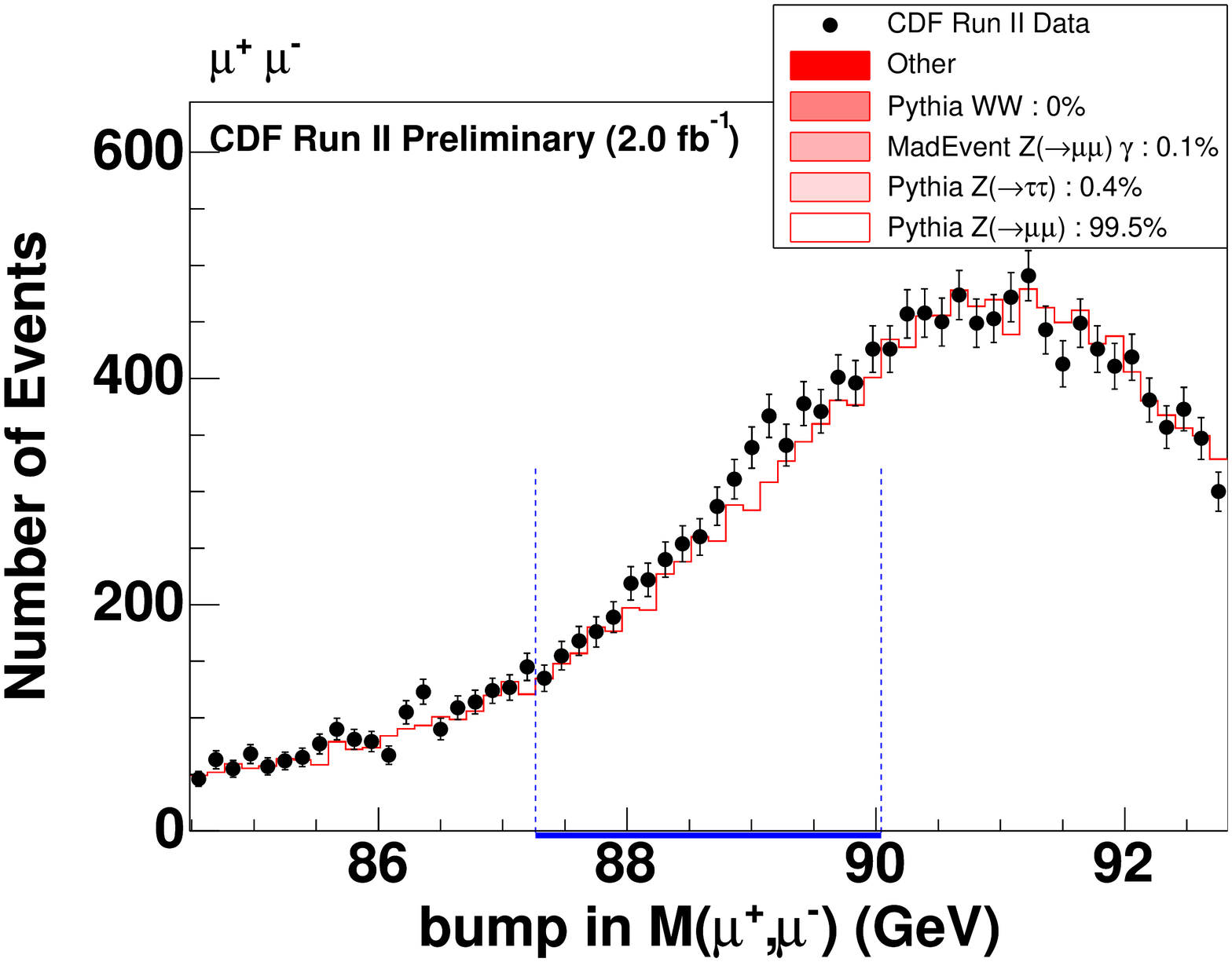}\\
\includegraphics[angle=-90,width=0.5\columnwidth]{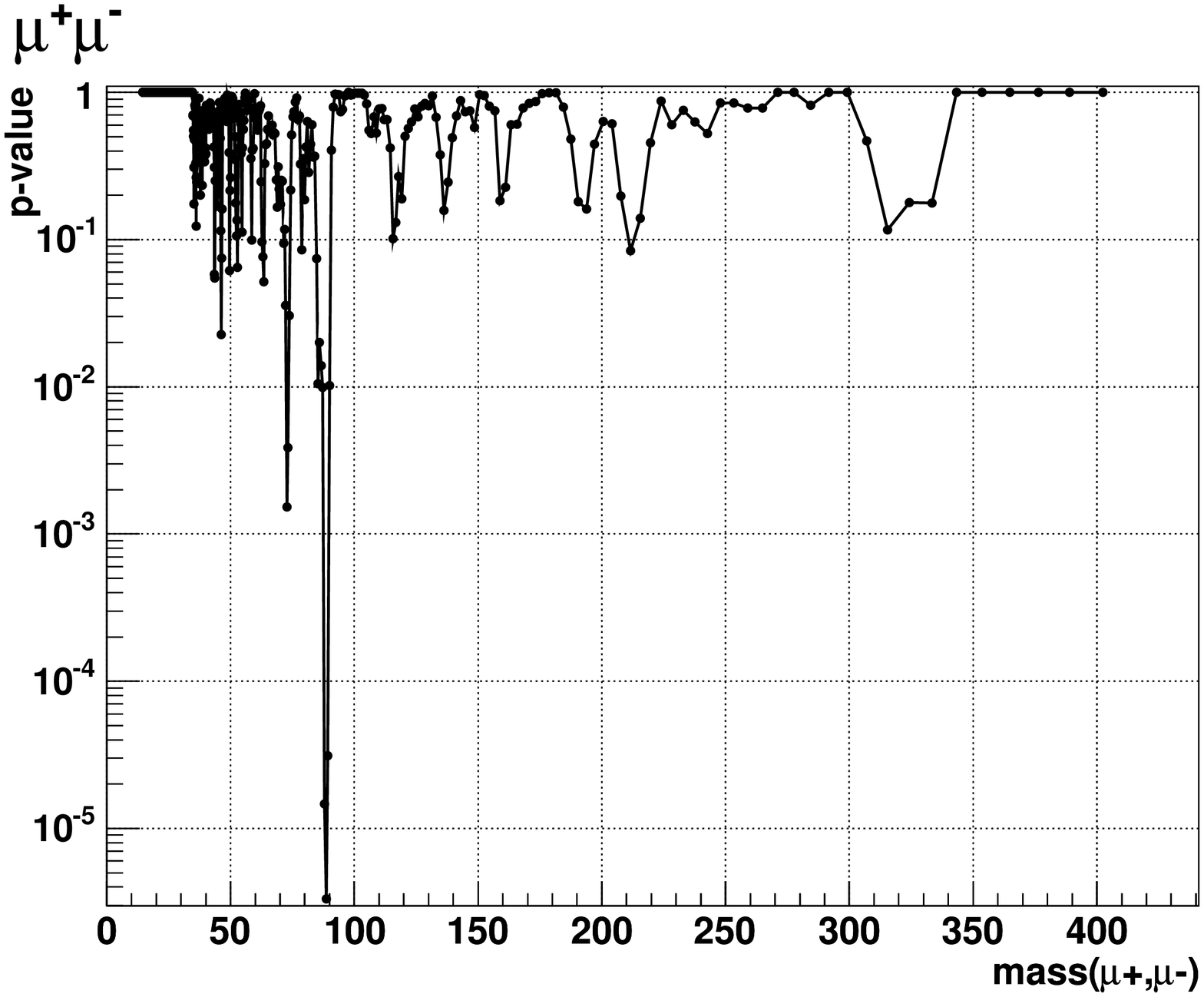}
\end{tabular}
\caption[Bumps found in $\mu^+\mu^-$]{{\em (Upper two)} The most interesting bump found in final state $\mu^+\mu^-$.  {\em (Bottom)} The \pval\ of all bumps accross the mass spectrum of the two leptons.  Even the most significant bump, at the $Z$-peak, has $P_b \simeq 0.74$, therefore is completely insignificant.}
\label{fig:bump1mu+1mu-}
\end{figure}

The mass of the two muons in the $\mu^+\mu^-$ final state does not have any significant bump either, not even of the mundane kind found in $e^+e^-$.  That is shown in Fig.~\ref{fig:bump1mu+1mu-}.

\begin{figure}
\centering
\begin{tabular}{c}
\includegraphics[angle=-90,width=0.5\columnwidth]{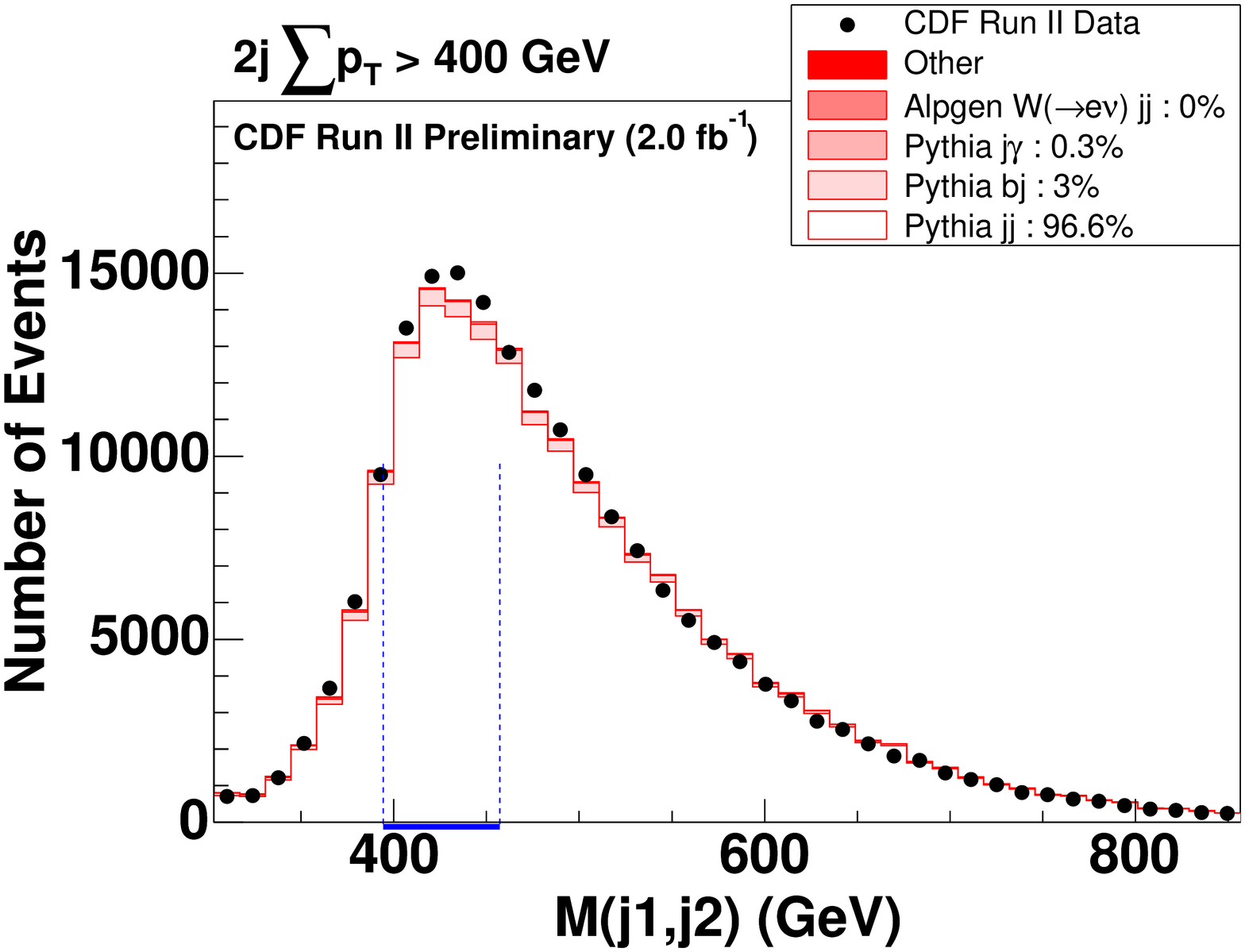}\\
\includegraphics[angle=-90,width=0.5\columnwidth]{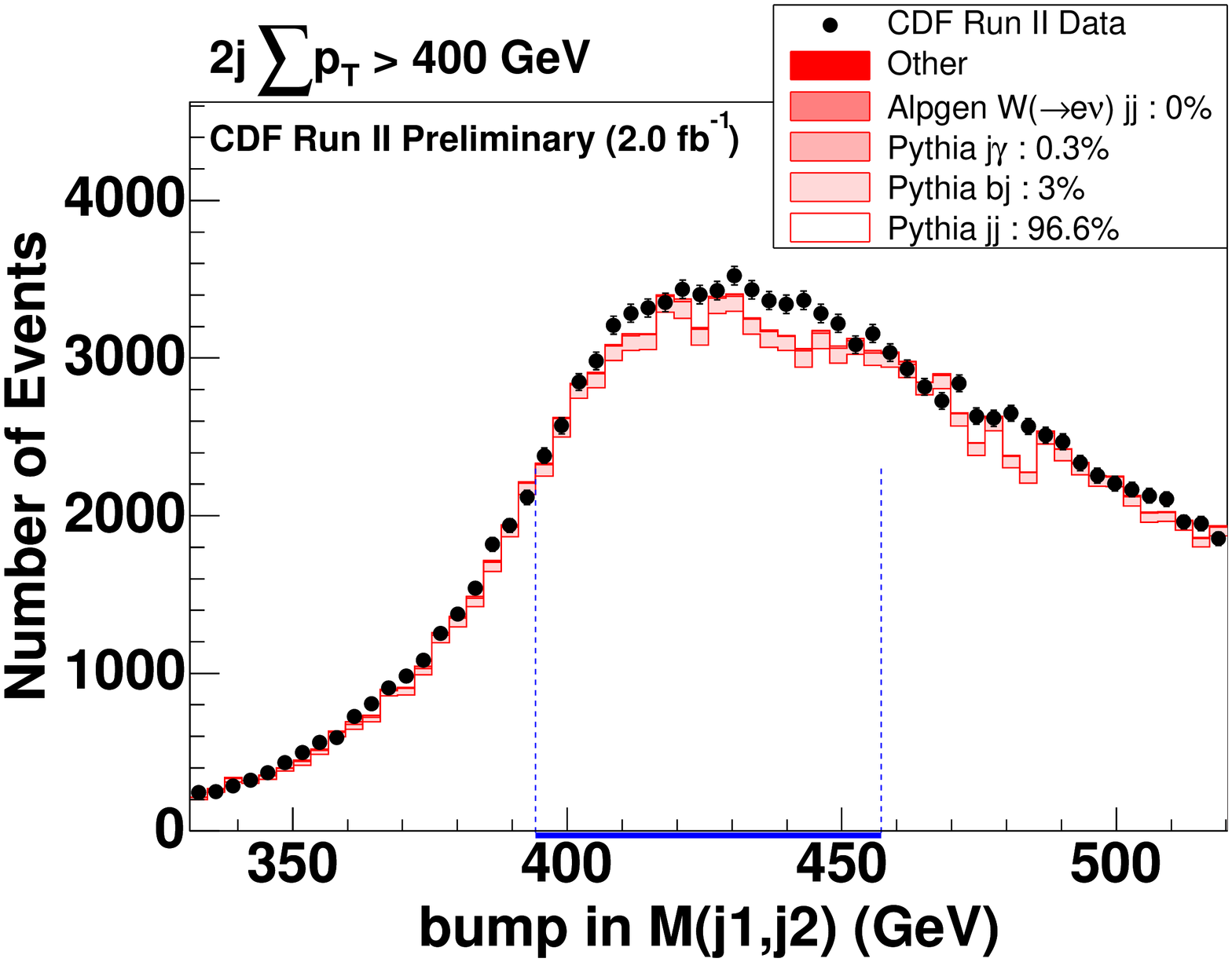}\\
\includegraphics[angle=-90,width=0.5\columnwidth]{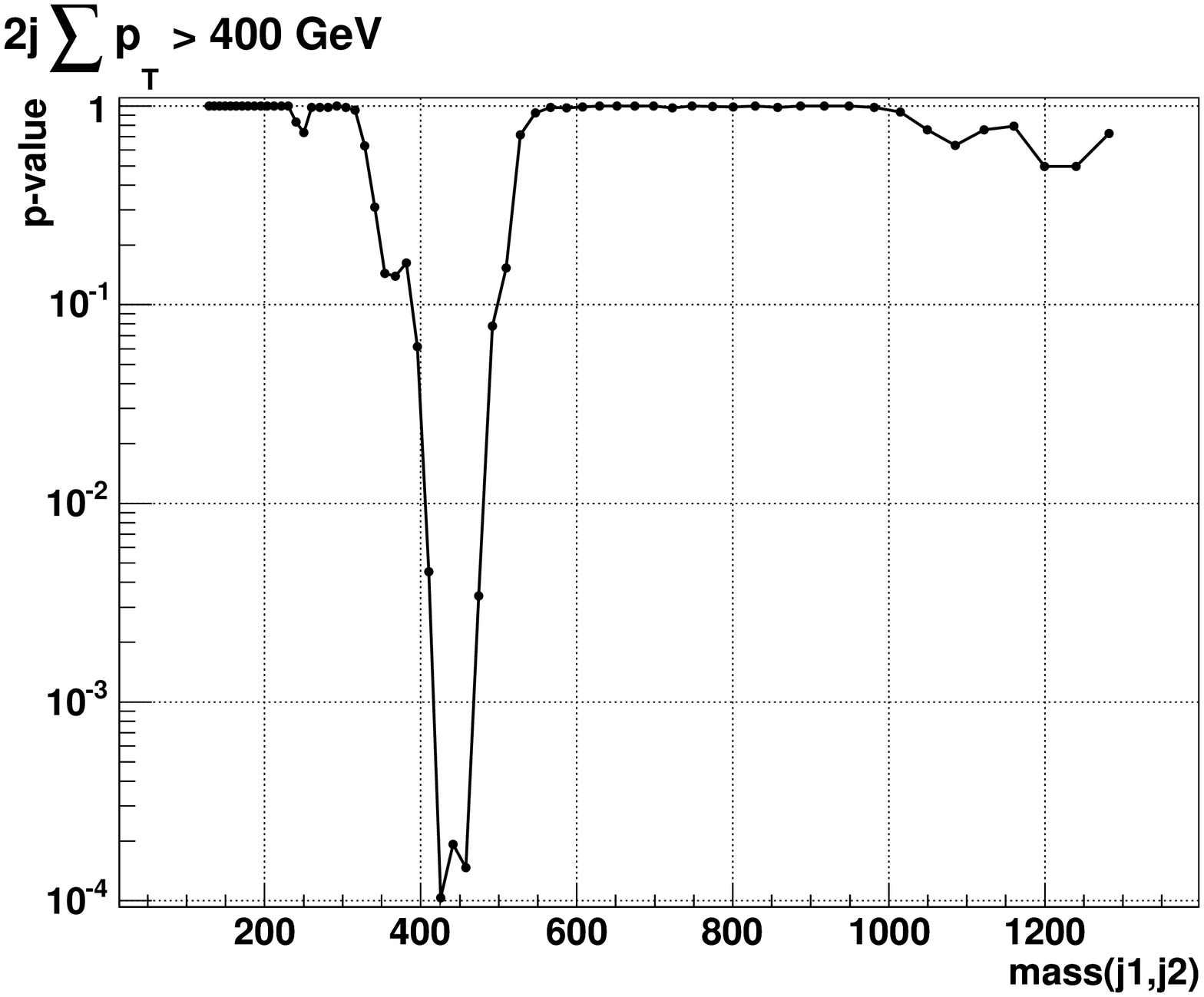}
\end{tabular}
\caption[Dijet bumps found]{{\em (Upper two)} The most interesting bump found in final state $2j\;\sumPt > 400$ GeV.  {\em (Bottom)} The \pval\ of all bumps accross the di-jet mass spectrum.  Even the most significant bump, yields $P_b \simeq 0.99$, therefore is completely insignificant.}
\label{fig:bump2j_sumPt400+}
\end{figure}

Another potentially interesting mass variable is the dijet mass in the final state with two high \sumPt\ jets.  That is shown in Fig.~\ref{fig:bump2j_sumPt400+}.  Unfortunately, no high-mass di-jet resonance was observed.

\subsection{Sensitivity}
\label{sec:BumpHunterSensitivity}

To test the sensitivity of the Bump Hunter, we generate some specific new physics signal, pass it through the full CDF detector simulation, and inject it gradually on top of pseudo-data pulled from the Standard Model background, until the Bump Hunter identifies a discovery-level bump.

\subsubsection{120 GeV Higgs in association with $W$}

The pseudo-signal use for this test contains a Standard Model Higgs of mass 120 GeV, allowed to decay to $b\bar{b}$, which has branching ratio 68\% \cite{Djouadi:1997yw}.  The associated $W$ decays to $e$ or $\mu$ or $\tau$ plus neutrino, with total branching ratio $\sim \frac{1}{3}$.

\begin{figure}
\centering
\includegraphics[angle=-90,width=0.7\columnwidth]{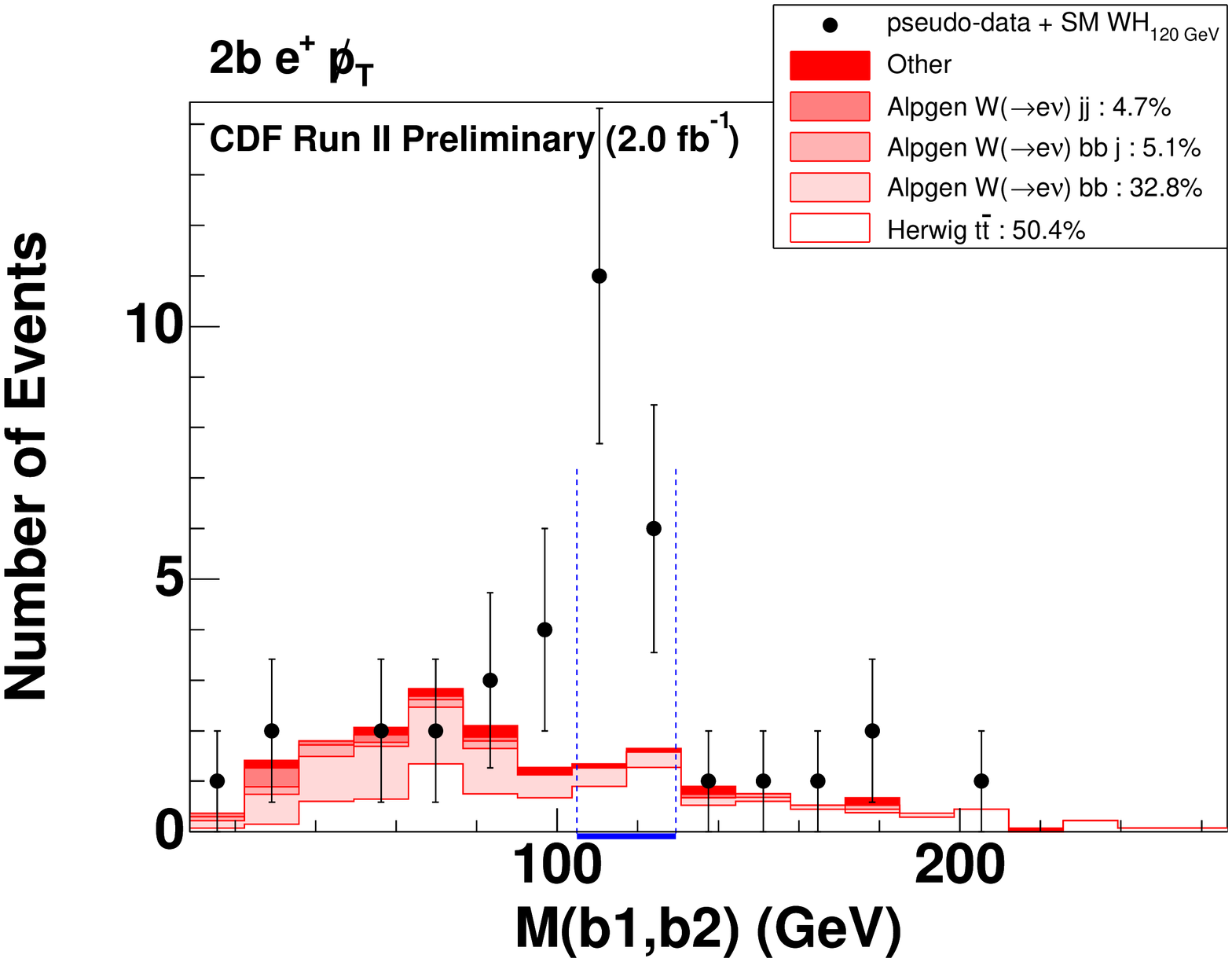}
\caption[Example of a pseudo-discovery of the Standard Model Higgs boson ($m_H=120$ GeV)]{Example of a pseudo-discovery of the Standard Model Higgs boson ($m_H=120$ GeV), produced in association with a $W$ boson. Out of 7000 generated $WH(\to \ell\nu b\bar{b})$ events, 15 populate the $2be^+\pmiss$ final state.  They cause this local excess which is identified by the Bump Hunter algorithm and its significance is estimated a 3.4$\sigma$ after trials factor.}
\label{fig:bumpSensitivityModel14}
\end{figure}

About 6500 signal events are required to obtain the first bump beyond discovery threshold.  Events passing selection criteria are distributed in several final states, and 15 of them make it to the $2be^+\pmiss$ final state, producing the bump in Fig.~\ref{fig:bumpSensitivityModel14}.
Compensating for the branching ratio, we find that the required cross section of $WH_{120 {\rm GeV}}$ to have this 5$\sigma$ level discovery would be about 14.4 pb, which is $\sim$90 times larger than the predicted Standard Model cross section.

\subsubsection{$Z'\to\ell^+\ell^-$ at mass 250 GeV}

\begin{figure}
\centering
\includegraphics[angle=-90,width=0.7\columnwidth]{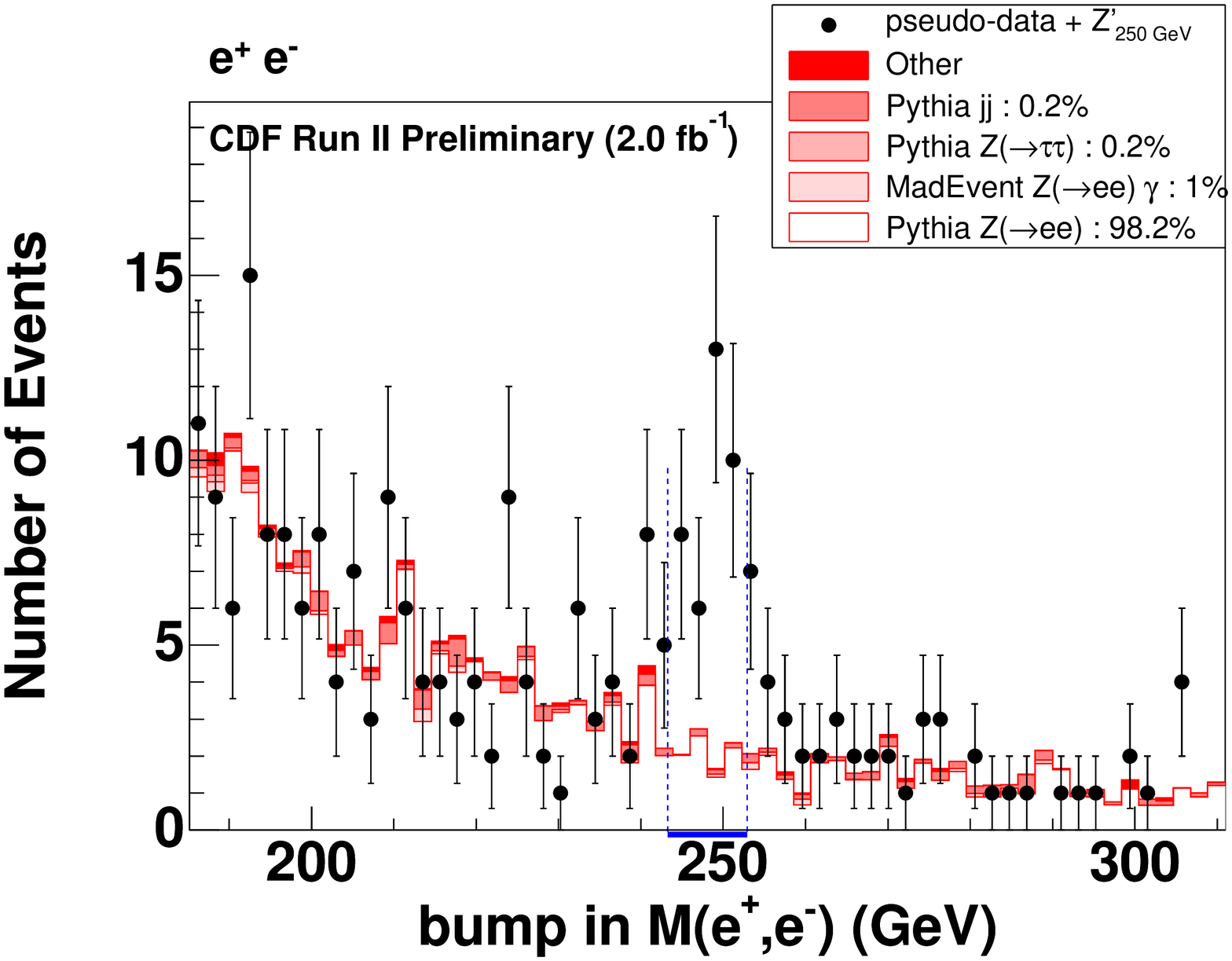}
\caption[Example of a pseudo-discovery of a 250 GeV $Z'$ decaying to charged leptons.]{Example of a pseudo-discovery of a 250 GeV $Z'$ decaying to charged leptons.  Out of 700 generated events, 55 populate the $e^+e^-$ final state, where the most significant bump appears.  The significance of this bump is estimated at 3.7$\sigma$ after trials factor.}
\label{fig:bumpSensitivityModel02}
\end{figure}

Pseudo-signal of a 250 GeV $Z'$ boson was generated, where $Z'$ may decay to $\ell^+\ell^-$, where $\ell$ can be $e$, $\mu$, or $\tau$.  The first discovery-level bump caused after injecting about 700 events of this pseudo-signal.  55 events end up in the 1e+1e- final state, and form the bump shown in Fig.~\ref{fig:bumpSensitivityModel02}.

With 700 injected events the significance found is 3.7$\sigma$, which is higher than the discovery threshold of 3$\sigma$.  That is because the pseudo-signal is injected in bunches of 100 events, so the actual requirement is between 600 and 700 events.  Dividing this number of generated events by our integrated luminosity shows that we would need the cross section times branching ratio of this signal to be approximately 0.325 pb.

\subsubsection{$Z'\to t\bar{t}$ at mass 500 GeV}

\begin{figure}
\centering
\begin{tabular}{cc}
\includegraphics[angle=-90,width=0.45\columnwidth]{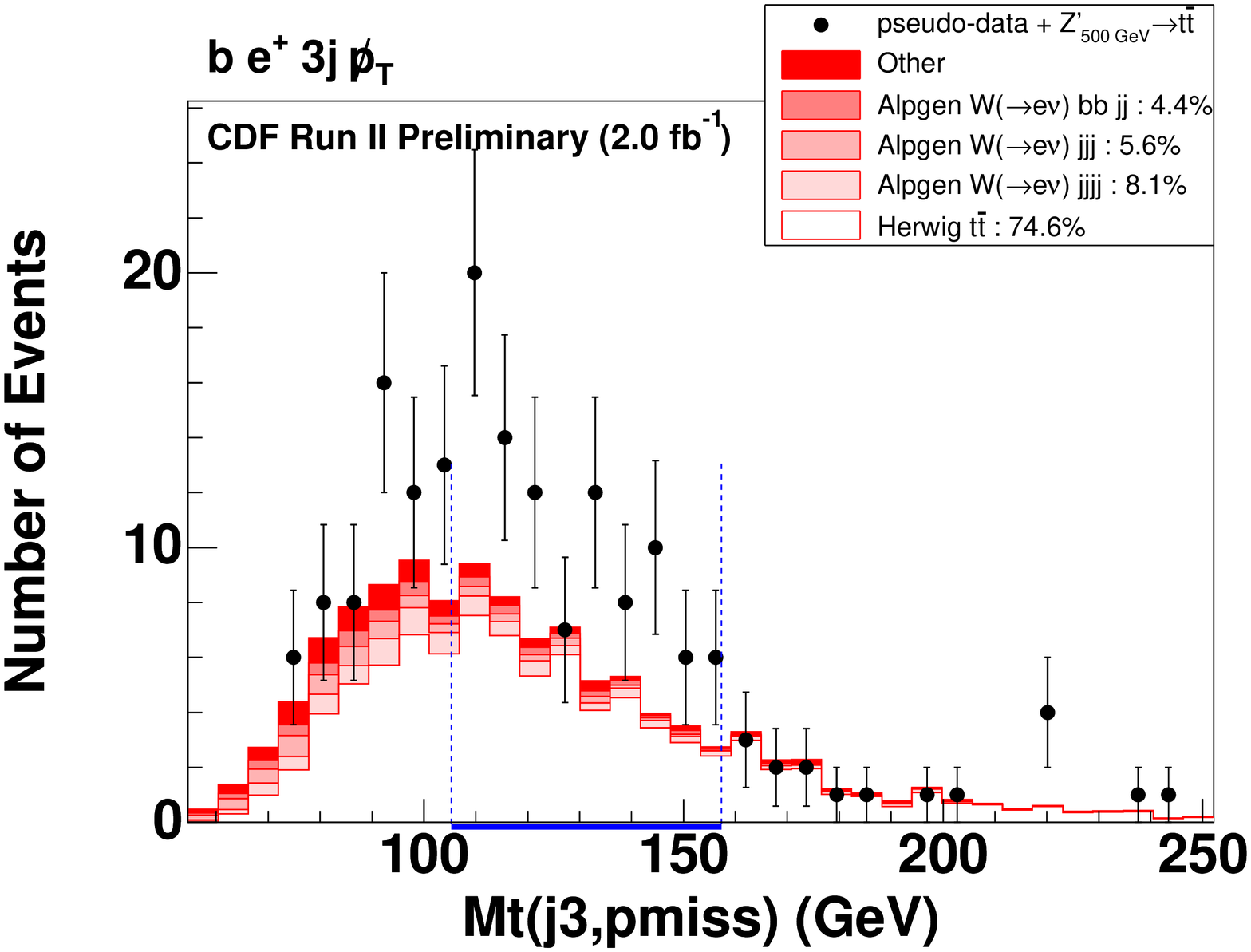} &
\includegraphics[angle=-90,width=0.45\columnwidth]{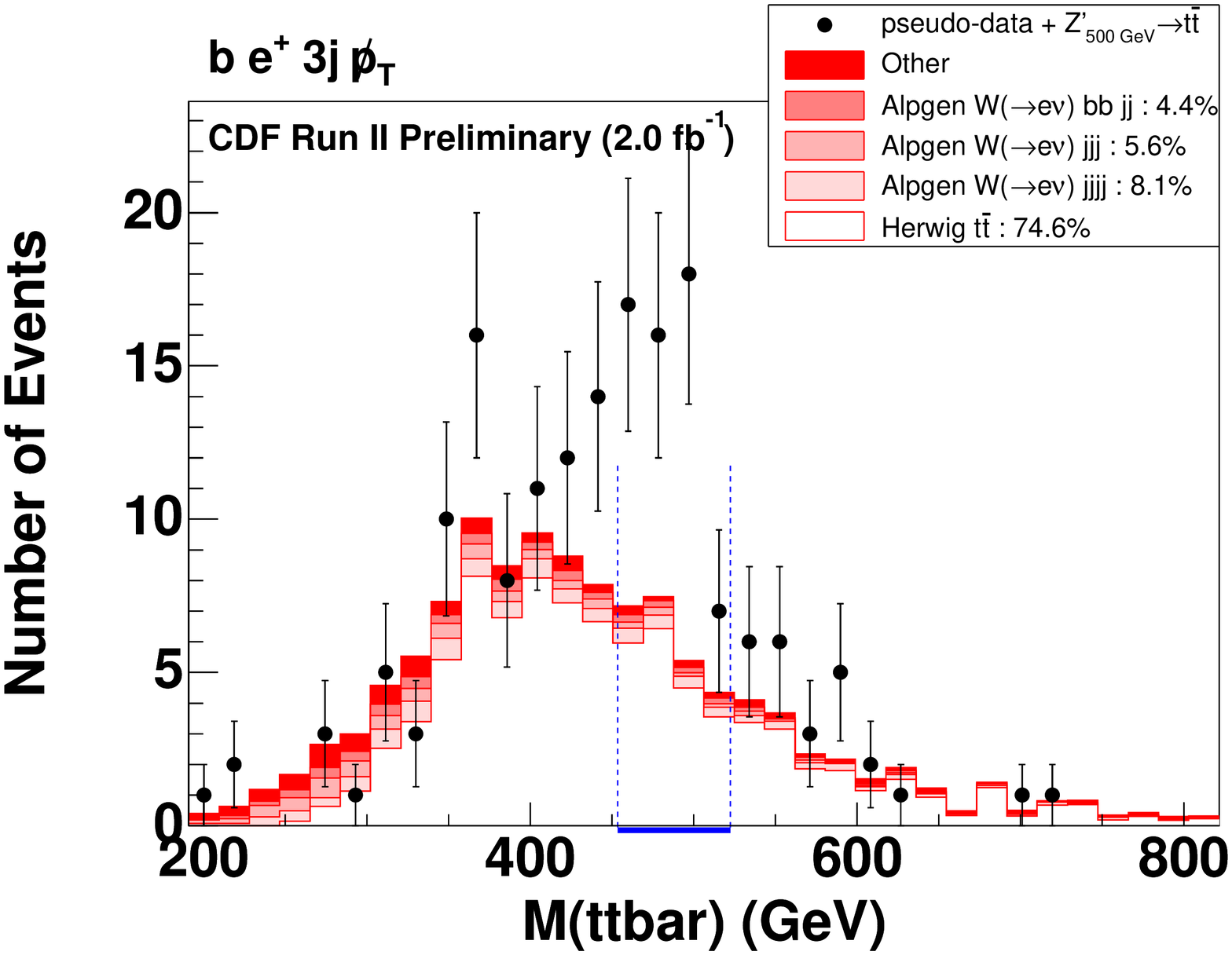}
\end{tabular}
\caption[Pseudo-discovery of $Z'_{500 GeV}\to t\bar{t}$]{{\em (Left)} Most significant bump after injecting 5000 $Z'_{500 GeV}\to t\bar{t}$ events.  47 signal events make it to the $be^+3j\pmiss$ final state, which cause this bump of significance 3.2$\sigma$ after trials factor in transverse mass of the third highest $p_T$ jet and \pmiss.  {\em (Right)} Most significant bump after injecting a different 4600 $Z'_{500 GeV}\to t\bar{t}$ events, on a background that was allowed to fluctuate anew.  41 signal events make it to the $be^+3j\pmiss$ final state, which cause this bump of significance 3.3$\sigma$ after trials factor in $m_{t\bar{t}}$, consistent naturally with the mass of the introduced $Z'$.}
\label{fig:bumpSensitivityModel11_mTj3pmiss}
\end{figure}

For this test we generated $Z'$ events of mass 500 GeV, where the heavy boson decays to a $t\bar{t}$ pair.  Injecting 5000 such events causes simultaneously two significant bumps in the $be^+3j\pmiss$ final state; one is in the transverse mass between \pmiss\ and the second highest $p_T$ jet ($j2$), with significance 3$\sigma$; the other is in the transverse mass of the third highest $p_T$ ($j3$) and \pmiss, with significance 3.2$\sigma$.  The latter is shown in Fig.~\ref{fig:bumpSensitivityModel11_mTj3pmiss}.

In another instance, after injecting 4600 different pseudo-signal events, a 3.3$\sigma$ effect after trials factor was created in the same final state ($be^+3j\pmiss$), but this time in the variable $m_{t\bar{t}}$, where one would more easily interpret the excess as due to resonant production of $t\bar{t}$.  That is shown in Fig.~\ref{fig:bumpSensitivityModel11_mTj3pmiss} as well.

%
With discovery cost of approximately 4800 events, the required cross section is approximately 2.4 pb.


\section{Summary of second round with 2 fb$^{-1}$}
\label{sec:Conclusions2}

\Vista\ and \Sleuth\ search for outliers, representing significant discrepancies between data and Standard Model prediction.  Unfortunately, the result obtained is that no signficant outliers have been found either in the total number of events in the \Vista\ exclusive final states, or in \Sleuth's search of the \SumPt\ tails.  Disregarding effects from tuning corrections to the data, \Sleuth's $\tildeScriptP$ provides a rigorous statistical calculation of the likelihood that the most discrepant \Sleuth\ final state seen would have arisen purely by chance from the Standard Model prediction and correction model constructed within \Vista.

\Vista's correction model does not explicitly include some sources of systematic uncertainty, including those associated with parton distribution functions and showering parameters in the event generators used; these sources of uncertainty are included implicitly, in that they would be considered if necessary in the event of a possible discovery.  Other uncertainties related to the modeling of the CDF detector response and object identification criteria are determined as part of \Vista\ but are not included in the calculation of $\tildeScriptP$.  For the correction model used, \Sleuth\ finds $\twiddleScriptP = 0.085$.

The Bump Hunter, a new algorithm for identification of mass resonances, did not find any significant mass bumps either, except for one that is attributed to \Pythia\ not modeling perfectly parton showering.

Although the \Vista\ correction model could presumably be improved further to show even better agreement with Standard Model prediction, finding $\tildeScriptP \gg 0.001$ indicates that even the most discrepant \SumPt\ tail is not of statistical interest.  The correction model used is thus good enough (even without considering effect of systematic uncertainties on the \Sleuth\ final states) to conclude this search for outliers using \Vista\ and \Sleuth\ in 2~fb$^{-1}$.

This analysis does not prove that there is no new hint of physics buried in these data; merely that this search does not find any.

\chapter{Grand Summary and Conclusion}

This thesis presents the first model-independent search for new physics of such scope.

The Standard Model was implemented using a simplified set of corrections.

New physics was sought that would cause significant discrepancies in (a)  populations of exclusive final states, (b) shapes of kinematic distributions, (c) mass spectra, and (d) high-\sumPt\ events.

The search was first conducted in 1 fb$^{-1}$ of CDF II data, revealing no ground on which to support a discovery claim.  It was then repeated in 2 fb$^{-1}$ of data, improved and enhanced with the Bump Hunter, an algorithm to locate narrow resonances due to new massive particles.

Unfortunately and surprisingly, even with 2 fb$^{-1}$ the result was null, in the sense that no new physics could be claimed with the findings.  The discrepancies seen were attributed mainly to the difficulty in modeling soft radiated parton showers with \Pythia.  This issue was suspected to be problematic, but no other analysis had illustrated so clearly its repercussion.

Although no single analysis can guarantee that new physics is nowhere in the data, it is highly informative that in a search of this scope nothing exploitable was found.  This is complemented and consented by the numerous searches, dedicated to specific signals, which so far have failed too to reveal what lies beyond the Standard Model.

Even with a null result, the value of this technique is great in providing an overview of all data, even those nobody ever considers.  It can make a big difference at the later stages of the LHC, or in any experiment where there is a proliferation of data, and a fairly accurate theoretical prediction analogous to what our event generators and detector simulation provide.

\appendix
\chapter{Correction Model Details}
\label{sec:CorrectionModelDetails}

Some aspects of the correction model are fixed, rather than dynamically adjusted by the global fit, which is viewed as just a tool to provide reasonable values for some parameters of the correction model.  Not every parameter needs to be determined by a fit, as long as it is reasonable or estimated beforehand, through a MC study for instance.

Implementation details of the correction model will be described in this chapter in some extra detail.

\section{Fake rate physics}
\label{sec:MisidentificationMatrix}

The following facts begin to build a unified understanding of fake rates for electrons, muons, taus, and photons.  This understanding is woven throughout the correction model, and significantly informs and constrains the \Vista\ correction process.  Explicit constraints derived from these studies are provided in Appendix~\ref{sec:CorrectionFactorFitDetails}.  The underlying physical mechanisms for these fakes lead to simple and well justified relations among them.

\begin{table}
\centering\mbox{
\begin{tabular}{c|rrrrrrrrr}
 & $e^+$ & $e^-$ & $\mu^+$ & $\mu^-$ & $\tau^+$ & $\tau^-$ & $\gamma$ & $j$ & $b$ \\ \hline 
$e^+$  & 62228 & 33 & 0 & 0 & 182 & 0 & 2435 & 28140 & 0  \\
$e^-$  & 24 & 62324 & 0 & 0 & 0 & 192 & 2455 & 28023 & 1  \\
$\mu^+$  & 0 & 0 & 50491 & 0 & 6 & 0 & 0 & 606 & 0  \\
$\mu^-$  & 0 & 1 & 0 & 50294 & 0 & 6 & 0 & 577 & 0  \\
$\gamma$  & 1393 & 1327 & 0 & 0 & 1 & 1 & 67679 & 21468 & 0  \\
$\pi^0$  & 1204 & 1228 & 0 & 0 & 5 & 8 & 58010 & 33370 & 0  \\
$\pi^+$  & 266 & 0 & 115 & 0 & 41887 & 6 & 95 & 54189 & 37  \\
$\pi^-$  & 1 & 361 & 0 & 88 & 13 & 41355 & 148 & 54692 & 44  \\
$K^+$  & 156 & 1 & 273 & 0 & 42725 & 7 & 37 & 52317 & 24  \\
$K^-$  & 1 & 248 & 0 & 165 & 28 & 41562 & 115 & 53917 & 22  \\
$B^+$  & 100 & 0 & 77 & 1 & 100 & 10 & 40 & 66062 & 25861  \\
$B^-$  & 2 & 85 & 3 & 68 & 11 & 99 & 45 & 66414 & 25621  \\
$B^0$  & 88 & 27 & 87 & 17 & 77 & 32 & 21 & 65866 & 25046  \\
$\bar{B^0}$  & 17 & 79 & 11 & 71 & 41 & 77 & 21 & 66034 & 25103  \\
$D^+$  & 126 & 6 & 62 & 0 & 1485 & 67 & 207 & 79596 & 11620  \\
$D^-$  & 4 & 134 & 3 & 74 & 64 & 1400 & 234 & 79977 & 11554  \\
$D^0$  & 60 & 13 & 27 & 2 & 312 & 1053 & 248 & 88821 & 5487  \\
$\bar{D^0}$  & 15 & 46 & 5 & 28 & 1027 & 253 & 237 & 89025 & 5480  \\
$K^0_L$  & 1 & 4 & 0 & 0 & 71 & 60 & 202 & 96089 & 26  \\
$K^0_S$  & 26 & 31 & 2 & 1 & 170 & 525 & 9715 & 76196 & 0  \\
$\tau^+$  & 1711 & 13 & 1449 & 0 & 4167 & 2 & 673 & 50866 & 607  \\
$\tau^-$  & 12 & 1716 & 0 & 1474 & 6 & 3940 & 621 & 51125 & 580  \\
$u$  & 8 & 10 & 1 & 0 & 446 & 31 & 247 & 94074 & 26  \\
$d$  & 3 & 4 & 0 & 0 & 64 & 308 & 191 & 94322 & 22  \\
$g$  & 2 & 0 & 0 & 0 & 17 & 14 & 12 & 81865 & 99  \\
\end{tabular}}
\caption[Central single particle misidentification matrix.]{Central single particle misidentification matrix.  Using a single particle gun, $10^5$ particles of each type shown at the left of the table are shot with $p_T=25$~GeV into the central CDF detector, uniformly distributed in $\theta$ and in $\phi$.  The resulting reconstructed object types are shown at the top of the table, labeling the table columns.  Thus the rightmost element of this matrix in the fourth row from the bottom shows $\poo{\tau^-}{b}$, the number of negatively charged tau leptons (out of $10^5$) reconstructed as a $b$-tagged jet.
\label{tbl:misId_cdfSim_central}}
\end{table}

\begin{figure}
\centering
\begin{minipage}{3.5in}
\includegraphics[width=3.2in]{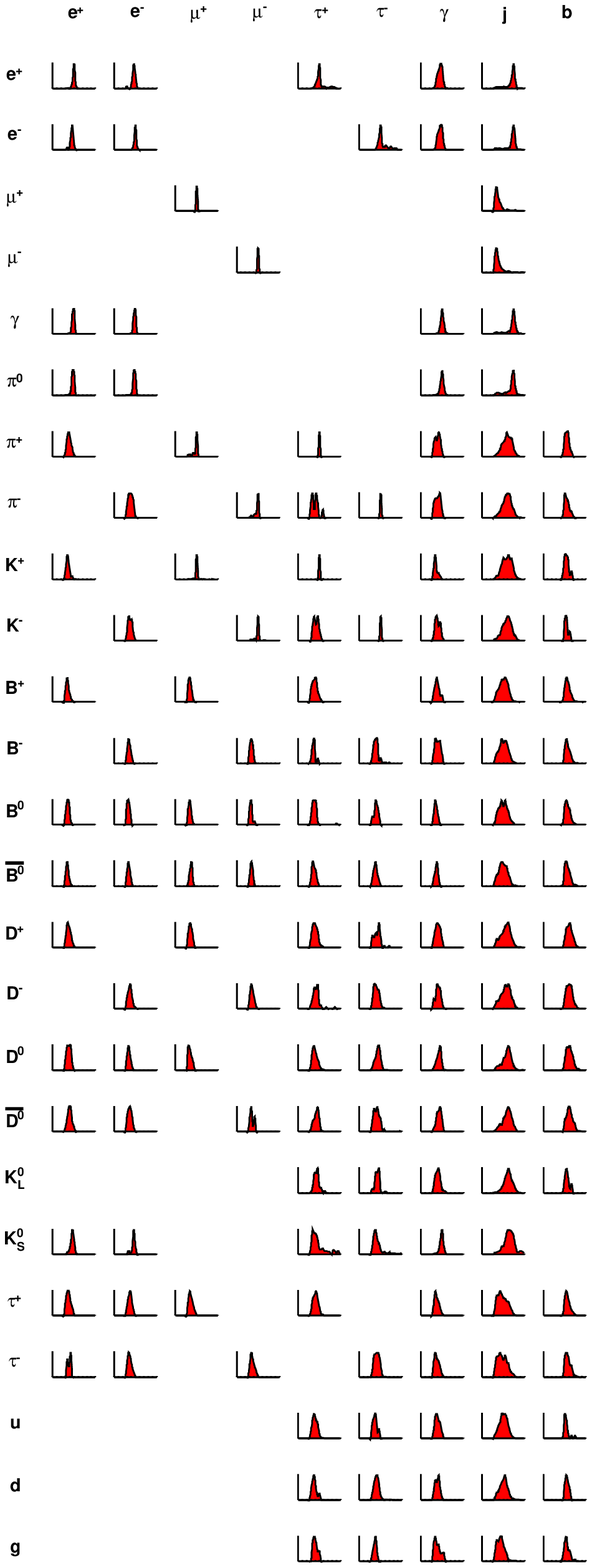}
\end{minipage}\hfill
\begin{minipage}{2.5in}
\caption[Transverse momentum distribution of reconstructed objects from a single particle gun shooting into the central CDF detector.]{Transverse momentum distribution of reconstructed objects (labeling columns) arising from single particles (labeling rows) with $p_T=25$~GeV shot from a single particle gun into the central CDF detector.  The area under each histogram is equal to the number of events in the corresponding misidentification matrix element of Table~\ref{tbl:misId_cdfSim_central}, with the vertical axis of each histogram scaled to the peak of each distribution.  A different vertical scale is used for each histogram, and histograms with fewer than ten events are not shown.  The horizontal axis ranges from 0 to 50~GeV.}
\end{minipage}
\label{fig:misId_cdfSim_central_pt}
\end{figure}


Table~\ref{tbl:misId_cdfSim_central} shows the response of the CDF detector simulation, reconstruction, and object identification algorithms to single particles.  Using a single particle gun, $10^5$ particles of each type shown at the left of the table are shot with $p_T=25$~GeV into the CDF detector, uniformly distributed in $\theta$ and in $\phi$.  The resulting reconstructed object types are shown at the top of the table, labeling the columns.  The first four entries on the diagonal at upper left show the efficiency for reconstructing electrons and muons~\footnote{The electron and muon efficiencies shown in this table are different from the correction factors {\tt 0025} and {\tt 0027} in Table~\ref{tbl:CorrectionFactorDescriptionValuesSigmas}, which show the ratio of the object efficiencies in the data to the object identification efficiencies in \CdfSim.}.  The fraction of electrons misidentified as photons (top row, seventh column) is seen to be roughly equal to the fraction of photons identified as electrons or positrons (fifth row, first and second columns), and measures the number of radiation lengths in the innermost regions of the CDF tracker.  The fraction of $B$ mesons identified as electrons or muons, primarily through semileptonic decay, are shown in the four left columns, eleventh through fourteenth rows.  Other entries provide similarly useful information, most easily comprehensible from simple physics.

The transverse momenta of the objects reconstructed from single particles are displayed in Fig.~\ref{fig:misId_cdfSim_central_pt}.  \cdfSpecific{Table~\ref{tbl:misId_cdfSim_50GeV_central} shows a similar study with $10^4$ particles at $p_T=50$~GeV.}  The relative resolutions for the measurement of electron and muon momenta are shown in the first four histograms on the diagonal at upper left.  The histograms in the left column, sixth through eighth rows, show that single neutral pions misreconstructed as electrons have their momenta well measured, while single charged pions misreconstructed as electrons have their momenta systematically undermeasured, as discussed below.  The histogram in the top row, second column from the right, shows that electrons misreconstructed as jets have their energies systematically overmeasured.  Other histograms in Fig.~\ref{fig:misId_cdfSim_central_pt} contain similarly relevant information, easily overlooked without the benefit of this study, but understandable from basic physics considerations once the effect has been brought to attention.


\begin{figure*}
\hspace{-1cm}
\begin{tabular}{cc}
\includegraphics[width=2.3in,angle=270]{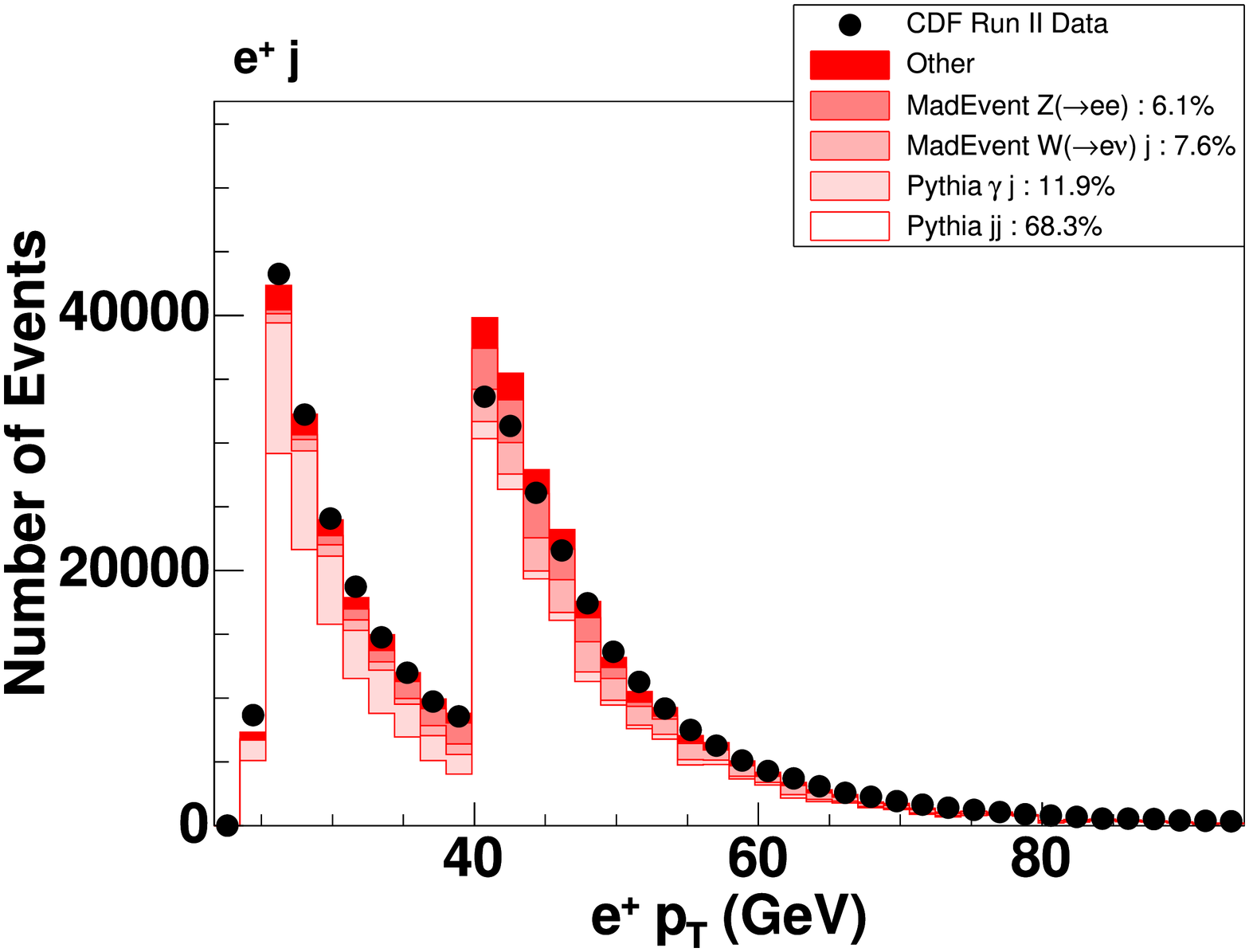} & \includegraphics[width=2.3in,angle=270]{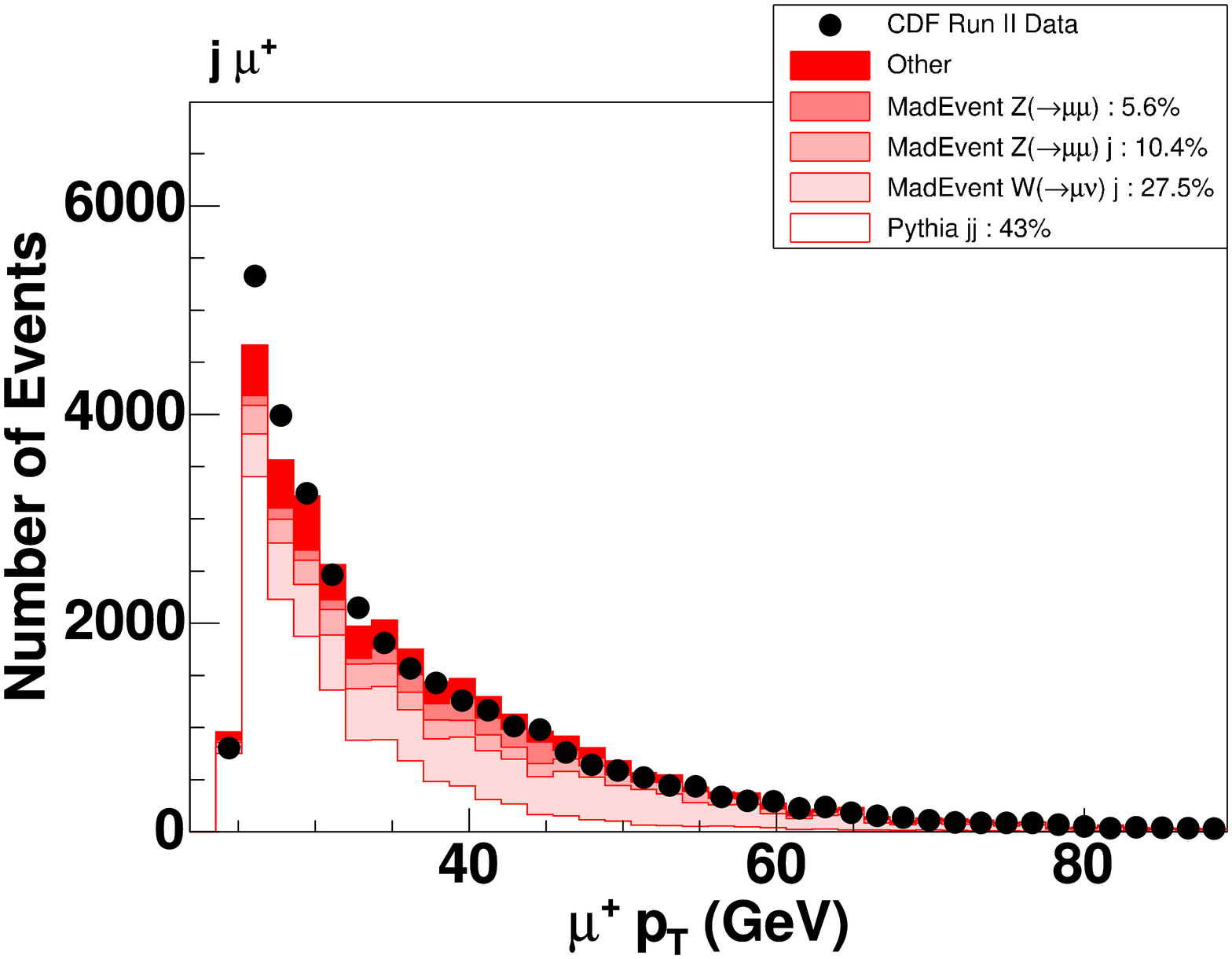} \\
\includegraphics[width=2.3in,angle=270]{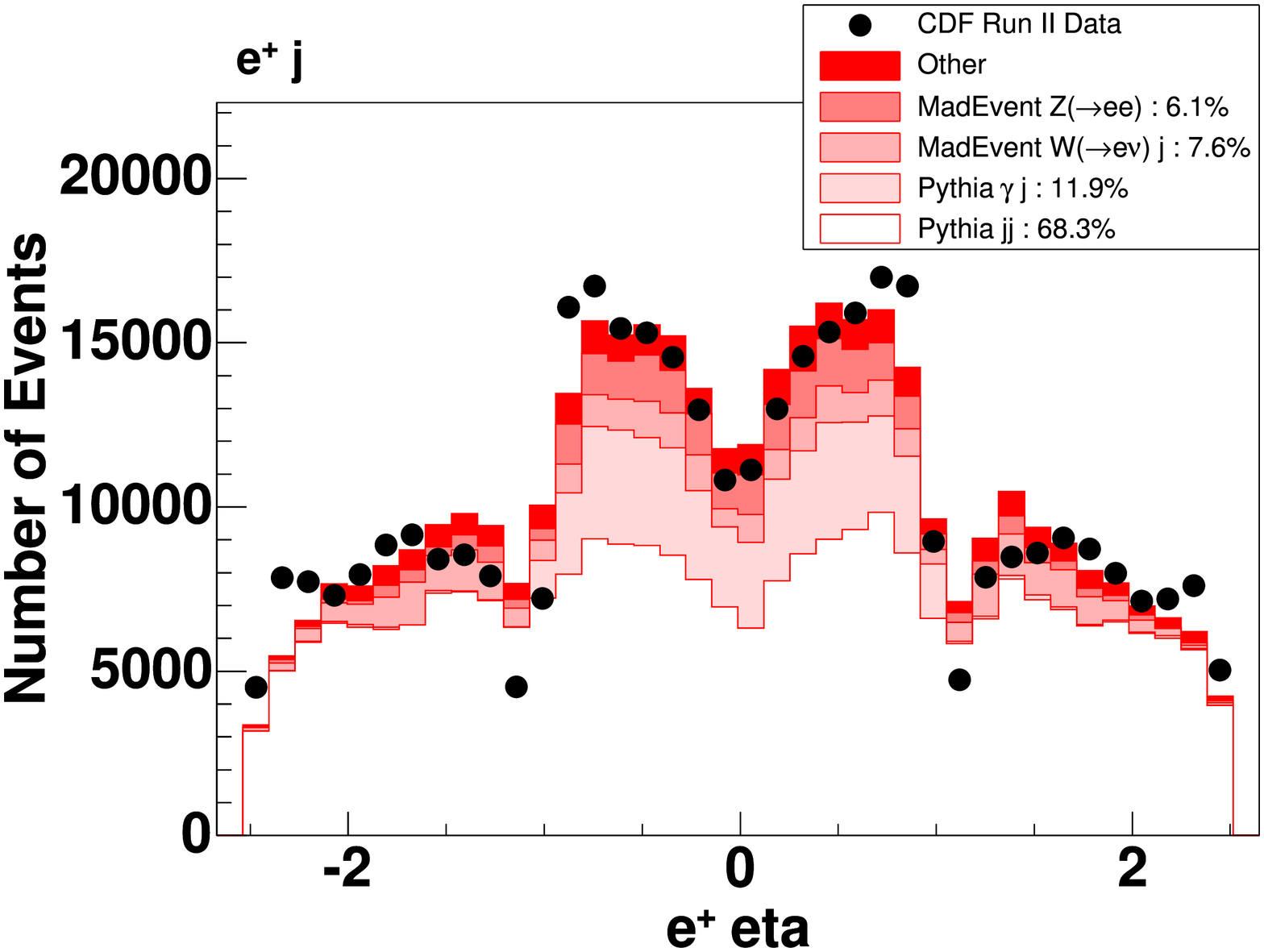} & \includegraphics[width=2.3in,angle=270]{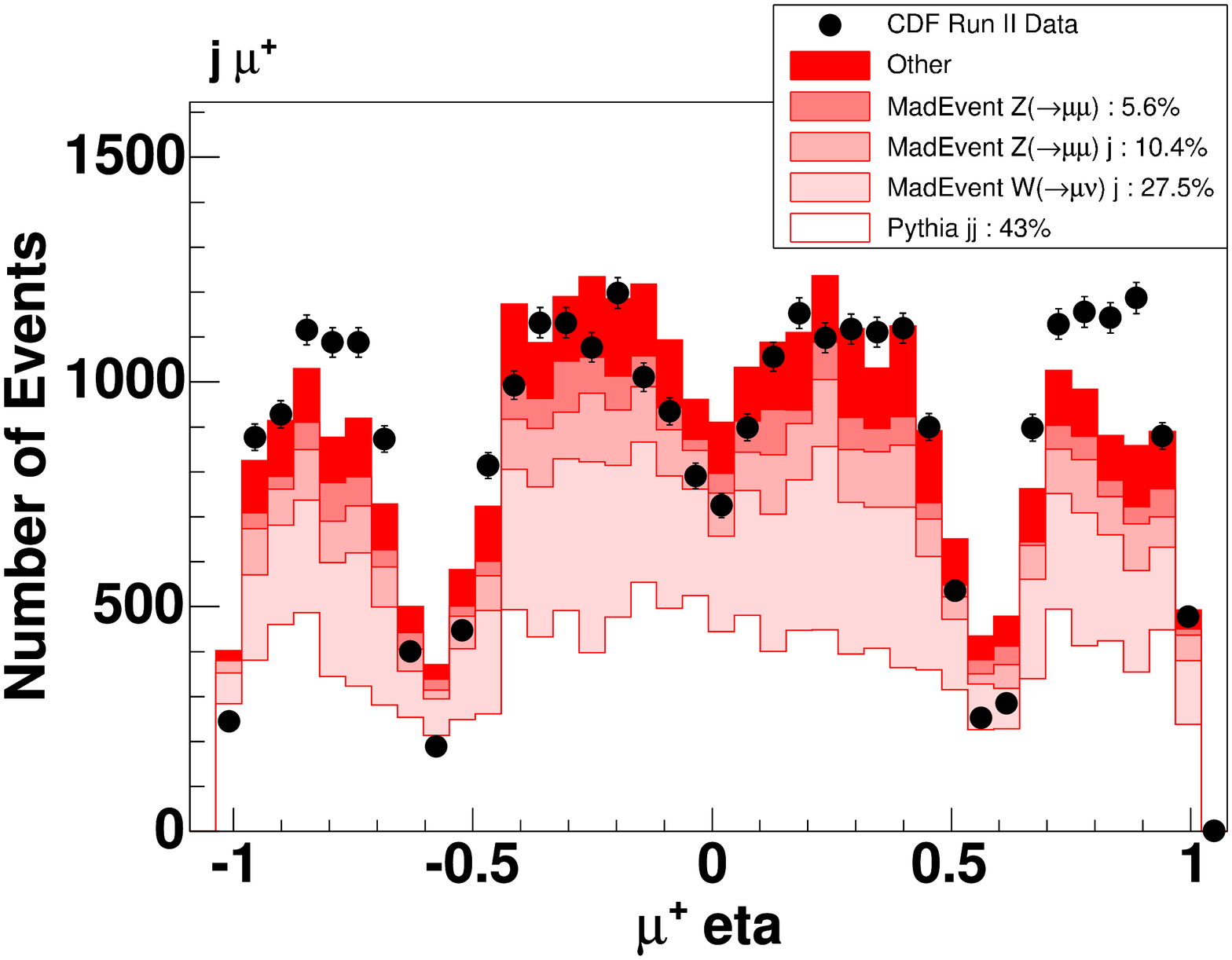} \\
\end{tabular}
\caption[A few of the most discrepant distributions in the final states $ej$ and $j\mu$]{ A few of the most discrepant distributions in the final states $ej$ and $j\mu$, which are greatly affected by the fake rates $\poo{j}{e}$ and $\poo{j}{\mu}$, respectively.  These distributions are among the 13 significantly discrepant distributions identified as resulting from coarseness of the correction model employed. }
\label{fig:1j1fakeLepton_1}
\end{figure*}

\begin{figure*}
\hspace{-1cm}
\begin{tabular}{cc}
\includegraphics[width=2.3in,angle=270]{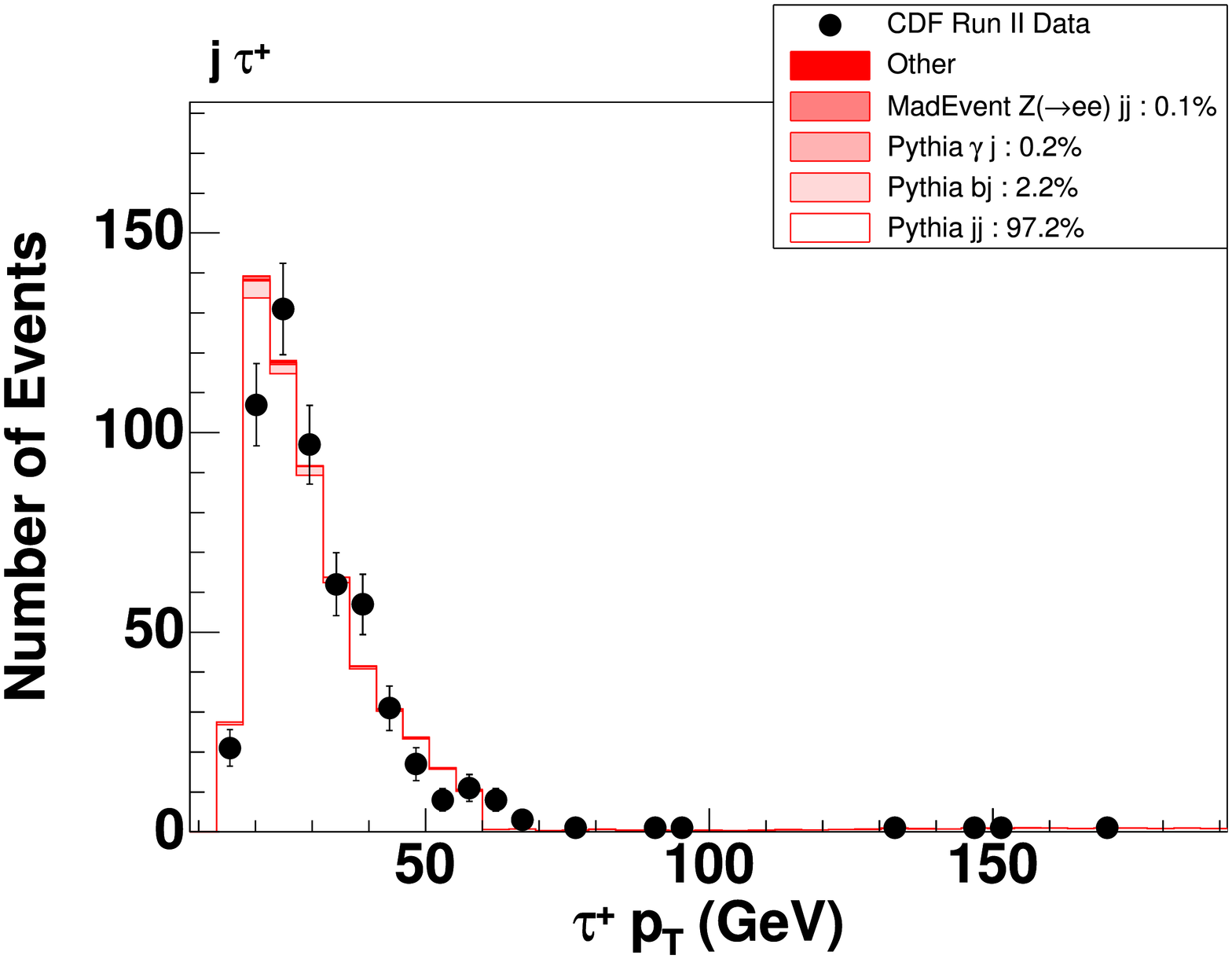} & \includegraphics[width=2.3in,angle=270]{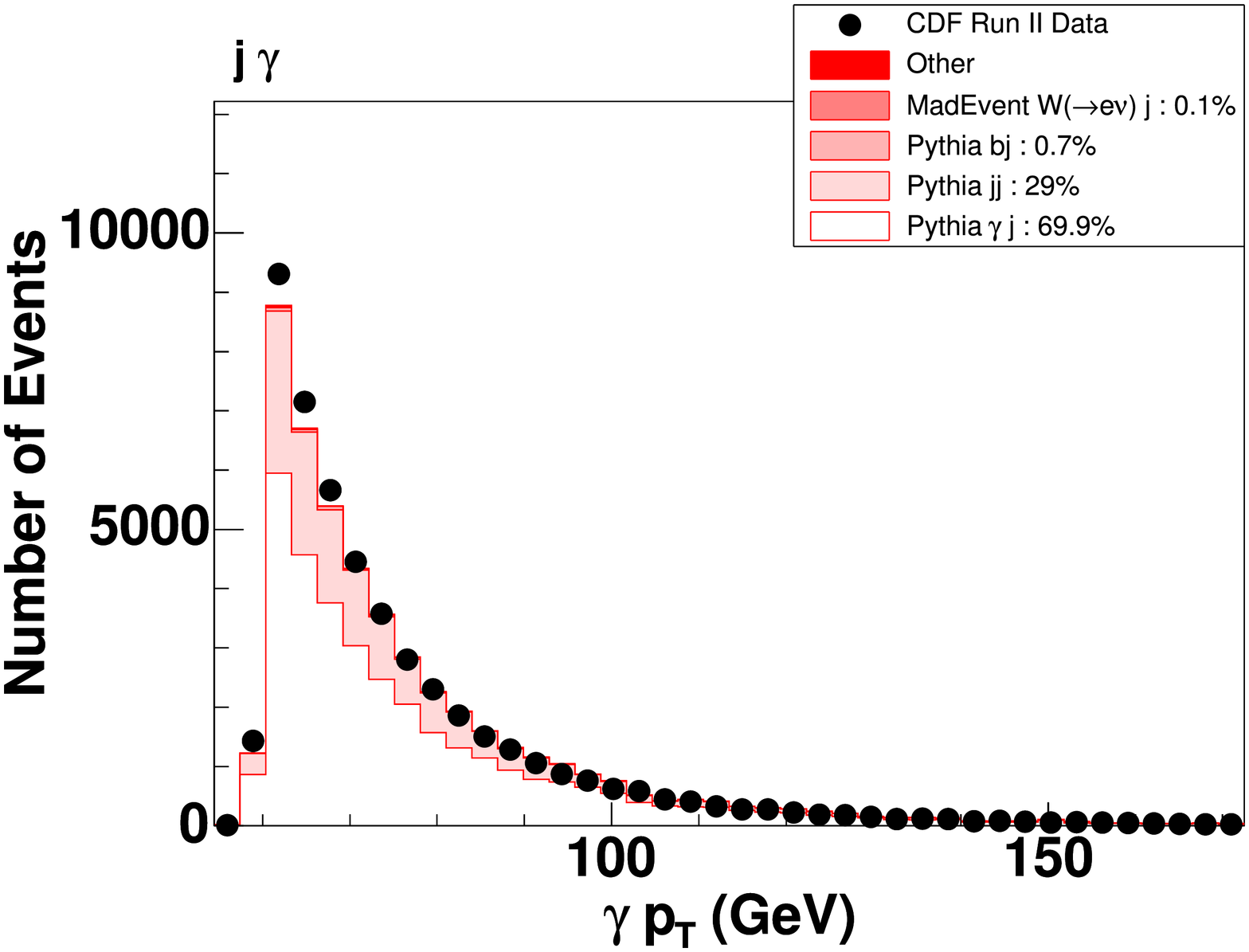} \\
\includegraphics[width=2.3in,angle=270]{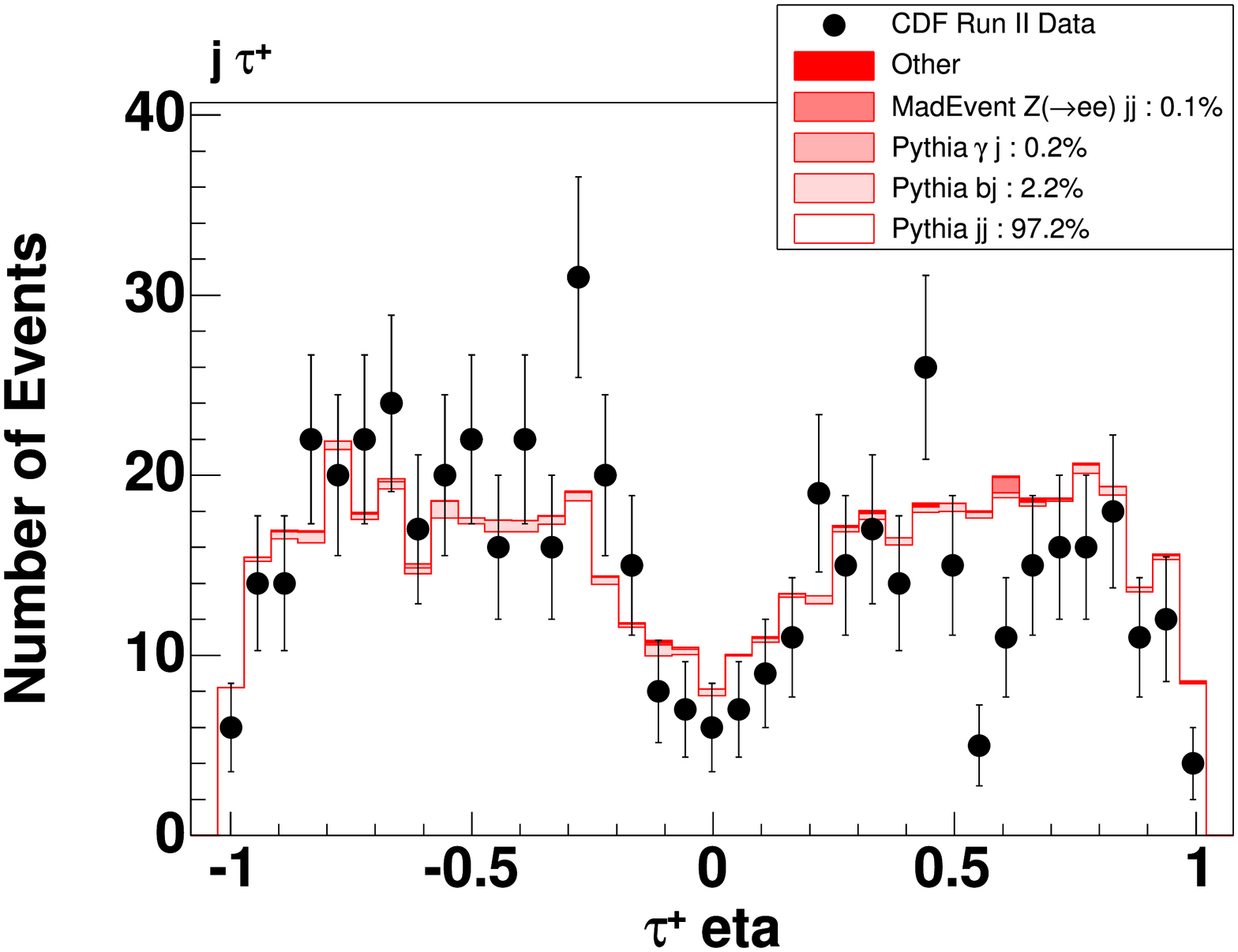} & \includegraphics[width=2.3in,angle=270]{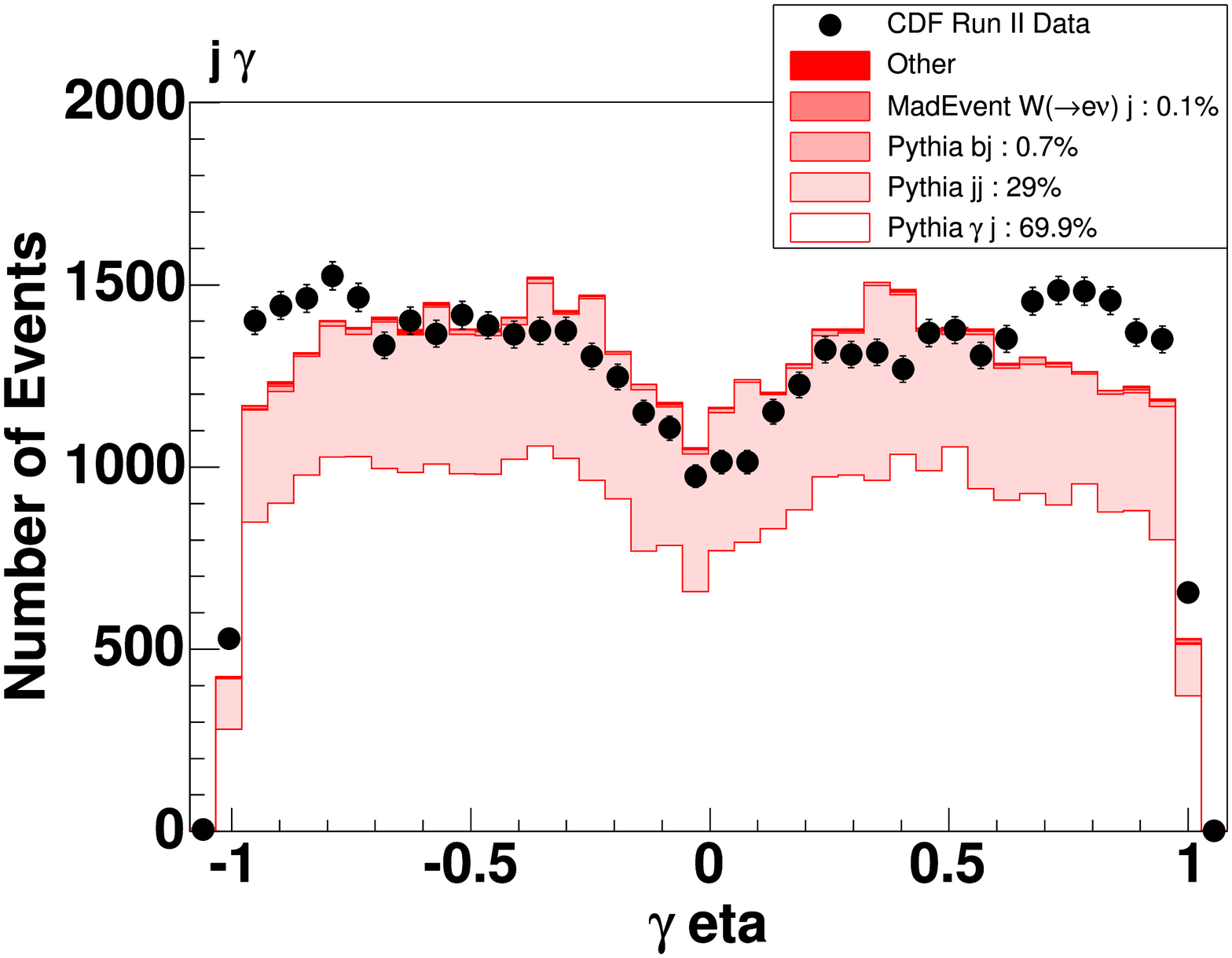}
\end{tabular}
\caption[A few of the most discrepant distributions in the final states $j\tau$ and $j\gamma$]{A few of the most discrepant distributions in the final states $j\tau$ and $j\gamma$, which are greatly affected by the fake rates $\poo{j}{\tau}$ and $\poo{j}{\gamma}$, respectively.  
  The distributions in the $j\gamma$ final state are among the 13 significantly discrepant distributions identified as resulting from coarseness of the correction model employed.}
\label{fig:1j1fakeLepton_2}
\end{figure*}

Here and below $\poo{q}{X}$ denotes a quark fragmenting to $X$ carrying nearly all of the parent quark's energy, and $\poo{j}{X}$ denotes a parent quark or gluon being misreconstructed in the detector as $X$.

The probability for a light quark jet to be misreconstructed as an $e^+$ can be written
\begin{eqnarray}
            \poo{j}{e^+} = & \poo{q}{\gamma} \, \poo{\gamma}{e^+} + \nonumber \\
                                   & \poo{q}{\pi^0} \, \poo{\pi^0}{e^+} + \nonumber \\
                                   & \poo{q}{\pi^+} \, \poo{\pi^+ }{e^+} + \nonumber \\ 
                                   & \poo{q}{K^+} \, \poo{K^+ }{e^+}. 
\label{eqn:poo_j_e+}
\end{eqnarray}
A similar equation holds for a light quark jet faking an $e^-$.  

The probability for a light quark jet to be misreconstructed as a $\mu^+$ can be written
\begin{eqnarray}
            \poo{j}{\mu^+} = & \poo{q}{\pi^+} \, \poo{\pi^+}{\mu^+} + \nonumber \\
                             & \poo{q}{K^+} \, \poo{K^+}{\mu^+}.
\label{eqn:poo_j_mu+}
\end{eqnarray}
Here $\poo{\pi}{\mu}$ denotes pion decay-in-flight, and $\poo{K}{\mu}$ denotes kaon decay-in-flight; other processes contribute negligibly.  A similar equation holds for a light quark jet faking a $\mu^-$. 

The only non-negligible underlying physical mechanisms for a jet to fake a photon are for the parent quark or gluon to fragment into a photon or a neutral pion, carrying nearly all the energy of the parent quark or gluon.  Thus
\begin{eqnarray}
            \poo{j}{\gamma} = & \poo{q}{\pi^0} \, \poo{\pi^0}{\gamma} + \nonumber \\
                              & \poo{q}{\gamma} \, \poo{\gamma}{\gamma}.
\label{eqn:poo_j_ph}
\end{eqnarray}


Up and down quarks and gluons fragment nearly equally to each species of pion; hence
\begin{eqnarray}
          \frac{1}{3} \, \poo{q}{\pi} & = \poo{q}{\pi^+} & =  \poo{q}{\pi^-} \nonumber \\
                                      & = \poo{q}{\pi^0},
\label{eqn:poo_q_pi}
\end{eqnarray}
where $\poo{q}{\pi}$ denotes fragmentation into any pion carrying nearly all of the parent quark's energy.
Fragmentation into each type of kaon also occurs with equal probability; hence
\begin{eqnarray}
     \frac{1}{4} \poo{q}{K} & = \poo{q}{K^+} = \poo{q}{K^-}       \nonumber \\
                            & =    \poo{q}{K^0} = \poo{q}{\bar{K^0}},
\label{eqn:poo_q_K}
\end{eqnarray}
where $\poo{q}{K}$ denotes fragmentation into any kaon carrying nearly all of the parent quark's energy.

\Pythia\ contains a parameter that sets the number of string fragmentation kaons relative to the number of fragmentation pions.  The default value of this parameter, which has been tuned to LEP I data, is 0.3; for every 1 up quark and every 1 down quark, 0.3 strange quarks are produced.  Strange particles are produced perturbatively in the hard interaction itself, and in perturbative radiation, at a ratio larger than 0.3:1:1.  This leads to the inequality
\begin{eqnarray}
   0.3 \lesssim \frac{\poo{q}{K}}{\poo{q}{\pi}} < 1,
\label{eqn:poo_q_Kpi}
\end{eqnarray}
where $\poo{q}{K}$ and $\poo{q}{\pi}$ are as defined above.

The probability for a jet to be misreconstructed as a tau lepton can be written
\begin{equation}
    \poo{j}{\tau^+} = \poo{j}{\tau^+_1} + \poo{j}{\tau^+_3},
\label{eqn:poo_j_tau13}
\end{equation}
where $\poo{j}{\tau^+_1}$ denotes the probability for a jet to fake a 1-prong tau, and $\poo{j}{\tau^+_3}$ denotes the probability for a jet to fake a 3-prong tau.  For 1-prong taus,
\begin{eqnarray}
    \poo{j}{\tau^+_1} = & \poo{q}{\pi^+} \, \poo{\pi^+}{\tau^+} + \nonumber \\
                        & \poo{q}{K^+} \, \poo{K^+}{\tau^+}.
\label{eqn:poo_j_tau1}
\end{eqnarray}
Similar equations hold for negatively charged taus.

\begin{figure}
\centering
\includegraphics[width=2.75in,angle=270]{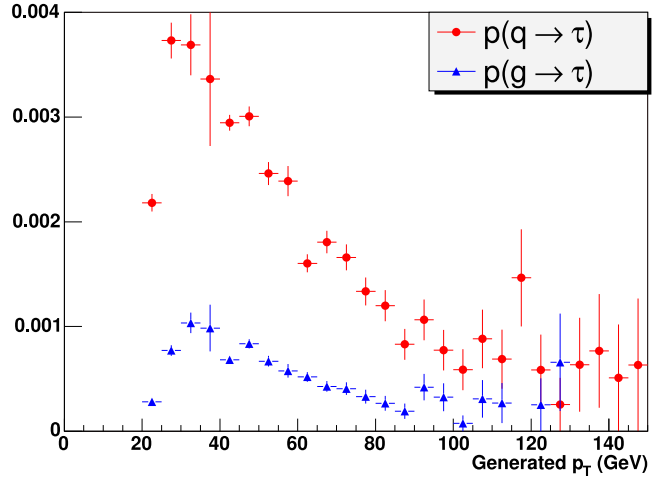}
\caption[The probability for a generated parton to be misreconstructed as a one-prong $\tau$, as a function of the parton's generated $p_T$]{The probability for a generated parton to be misreconstructed as a one-prong $\tau$, as a function of the parton's generated $p_T$.  Red circles show the probability for a jet arising from a parent quark to be misreconstructed as a one-prong tau.  Blue triangles show the probability for a jet arising from a parent gluon to be misreconstructed as a one-prong tau.}
\label{fig:j2tauProbabilities}
\end{figure}

Figure~\ref{fig:j2tauProbabilities} shows the probability for a quark (or gluon) to fake a one-prong tau, as a function of transverse momentum.  Using fragmentation functions tuned on LEP\,1 data, \Pythia\ predicts the probability for a quark jet to fake a one-prong tau to be roughly four times the probability for a gluon jet to fake a one-prong tau.  This difference in fragmentation is incorporated into \Vista's treatment of jets faking electrons, muons, taus, and photons.  The \Vista\ correction model includes such correction factors as the probability for a jet with a parent quark to fake an electron ({\tt 0033} and {\tt 0034}) and the probability for a jet with a parent quark to fake a muon ({\tt 0035}); the probability for a jet with a parent gluon to fake an electron or muon is then obtained by dividing the values of these fitted correction factors by four.

This effect is investigated using fake one-prong taus reconstructed in \Pythia\ dijet samples\cdfSpecific{ ({\tt Pythia\_jj\_018}, {\tt Pythia\_jj\_040} and {\tt Pythia\_jj\_060}, corresponding to lower cuts on $\hat{p}_T$ of 18, 40 and 60~GeV, respectively)}.

\begin{figure}
\centering
\includegraphics[width=3.0in]{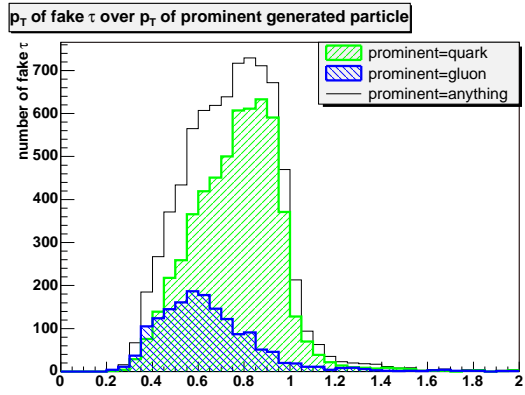}
\caption[Distribution of the $p_T$ of the fake $\tau$ over the $p_T$ of the prominent generated particle]{Distribution of the $p_T$ of the fake $\tau$ over the $p_T$ of the prominent generated particle (pgp), which is defined as the generated particle within $\Delta R < 0.4$ from the reconstructed $\tau$ with the greatest $p_T$. The pgp is almost always a quark or a gluon, and more likely to be a quark by a factor of four.}
\label{fig:j2tauPt}
\end{figure}

Figure~\ref{fig:j2tauPt} shows that the reconstructed fake tau has about $75\pm18\%$ of the $p_T$ of the prominent generated particle, defined to be the generated particle carrying the greatest $p_T$ and being within a cone of $\Delta R < 0.4$ centered on the reconstructed tau.  The $p_T$ of the misreconstructed tau is on average more undermeasured if the generated parton is a gluon than if it is a quark.  This reduction in the $p_T$ of the fake tau is implemented in \Vista\ when a jet is made to fake a $\tau$ during the misreconstruction process.

\begin{figure}
\centering
\begin{tabular}{cc}
\includegraphics[width=3in]{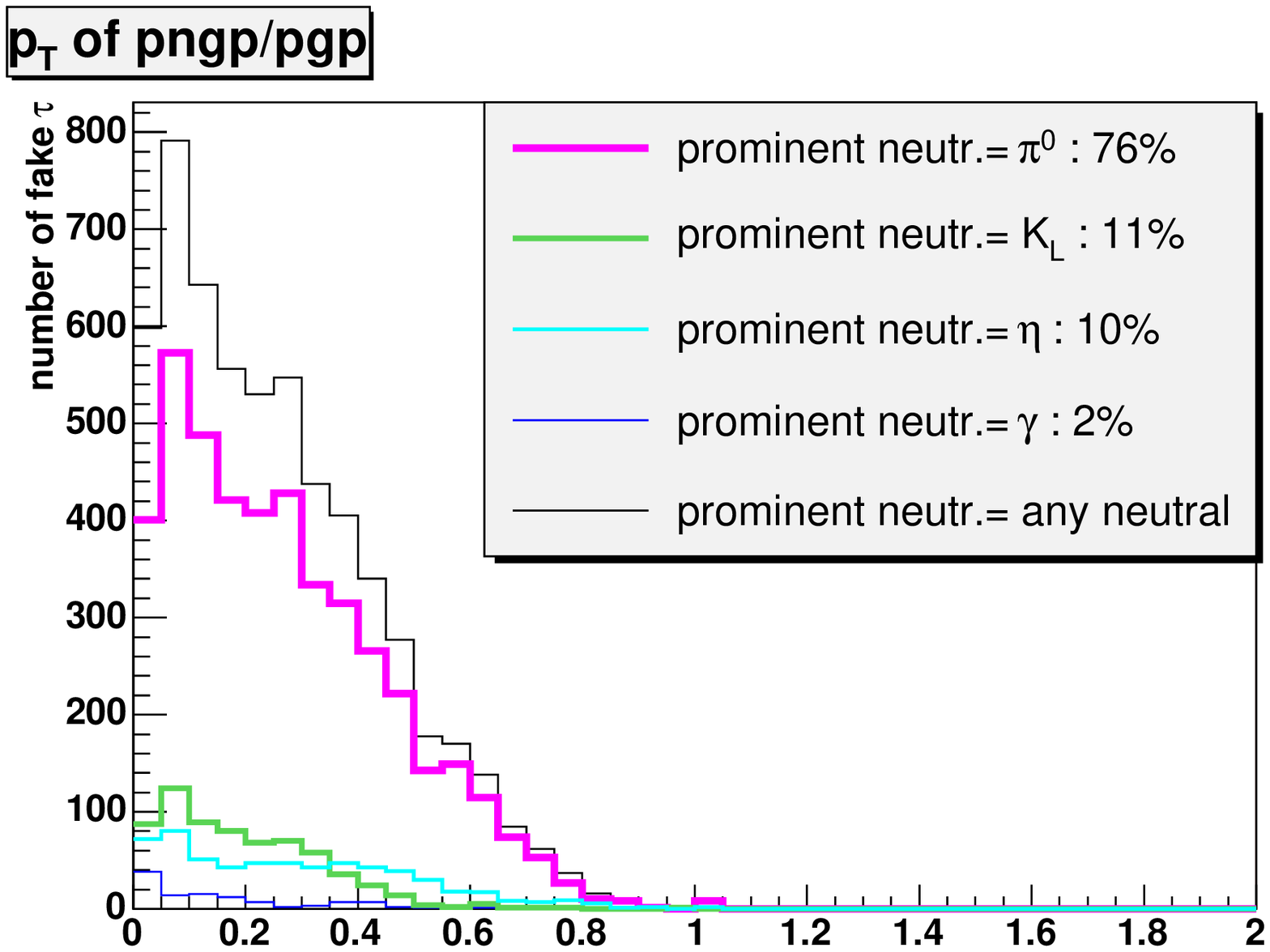} & \includegraphics[width=3in]{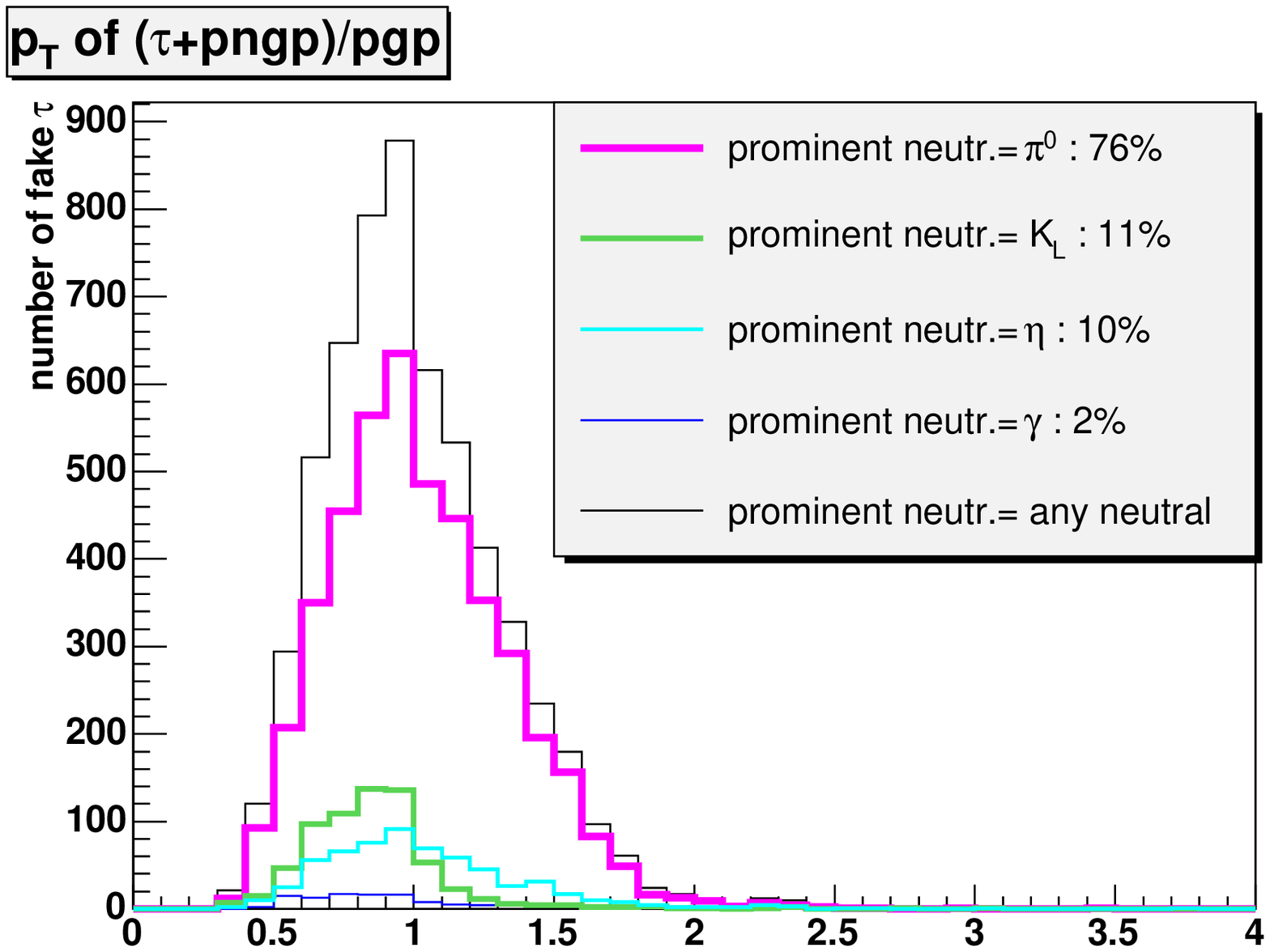} \\
\end{tabular}
\caption[Where the missing $p_T$ in fake $\tau$s goes.]{Upper: The distribution of the $p_T$ of the prominent neutral generated particle (pngp), which is the neutral generated particle with the greatest $p_T$ within a cone of $\Delta R < 0.4$ from the fake one-prong $\tau$, divided by the $p_T$ of the prominent generated particle (pgp), which happens to be either a quark or a gluon. Lower: $p_T$ of the pngp plus the $p_T$ of the reconstructed $\tau$, divided by the $p_T$ of the pgp. The fact that this distribution peaks around 1 shows that the generated $p_T$ that is missing from the fake $\tau$ was carried by the pngp. Most of the times the pngp is a $\pi^0$.}
\label{fig:j2tauNeutrals}
\end{figure}

Figure~\ref{fig:j2tauNeutrals} shows the remaining generated $p_T$ to be carried by neutral particles:  mostly $\pi^0$'s, followed by $K^0_L$'s and $\eta$'s decaying to photons or to three neutral pions.  The $p_T$ of the fake tau is determined by the track and reconstructed $\pi^0$'s.

The physical mechanism underlying the process whereby an incident photon or neutral pion is misreconstructed as an electron is a conversion in the material serving as the support structure of the silicon vertex detector.  This process produces exactly as many $e^+$ as $e^-$, leading to 
\begin{eqnarray}
            \frac{1}{2} \, \poo{\gamma}{e} = \poo{\gamma}{e^+} = \poo{\gamma}{e^-} \nonumber \\
            \frac{1}{2} \, \poo{\pi^0}{e}  = \poo{\pi^0}{e^+}  = \poo{\pi^0}{e^-},
\label{eqn:poo_pi0ph_e}
\end{eqnarray}
where $e$ is an electron or positron.
 
From Fig.~\ref{fig:misId_cdfSim_central_pt}, the average $p_T$ of electrons reconstructed from 25~GeV incident photons is $23.9\pm1.4$~GeV.  The average $p_T$ of electrons reconstructed from incident 25~GeV neutral pions is $23.7\pm1.3$~GeV.

The charge asymmetry between $\poo{K^+}{e^+}$ and $\poo{K^-}{e^-}$ in Table~\ref{tbl:misId_cdfSim_central} arises because $K^-$ can capture on a nucleon, producing a hyperon ($\Sigma^\pm$), which $K^+$ does not produce, due to baryon number and strangeness conservation.  Among the products of the hyperon decay are neutral pions, which decay electromagnetically and deposit in the electromagnetic calorimeter the energy needed to have a fake $e^-$.  The absense of this process in $K^++N$ interaction reduces the $poo{K^+}{e^+}$ relative $p{K^-}{e^-}$ by roughly a factor of two.

The physical process primarily responsible for $\pi^\pm \rightarrow e^\pm$ is inelastic charge exchange
\begin{eqnarray}
            \pi^- p  \rightarrow  \pi^0 n  \nonumber \\
            \pi^+ n  \rightarrow  \pi^0 p
\label{eqn:chargeExchangeProcesses}
\end{eqnarray}
occurring within the electromagnetic calorimeter.  The charged pion leaves the ``electron's'' track in the CDF tracking chamber, and the $\pi^0$ produces the ``electron's'' electromagnetic shower.  No true electron appears at all in this process, except as secondaries in the electromagnetic shower originating from the $\pi^0$.

The average $p_T$ of reconstructed ``electrons'' originating from a single charged pion is $18.8\pm2.2$~GeV, indicating that the misreconstructed ``electron'' in this case is measured to have on average only 75\% of the total energy of the parent quark or gluon.  This is expected, since the recoiling nucleon from the charge exchange process carries some of the incident pion's momentum.

 An additional small loss in energy for a jet misreconstructed as an electron, photon, or muon is expected since the leading $\pi^+$, $K^+$, $\pi^0$, or $\gamma$ takes only some fraction of the parent quark's energy.

The cross sections for $\pi^- p  \rightarrow  \pi^0 n$ and $\pi^+ n  \rightarrow  \pi^0 p$, proceeding through the isospin $I$ conserving and $I_3$ independent strong interaction, are roughly equal.  The corresponding particles in the two reactions are related by interchanging the signs of their $z$-components of isospin.  
\cdfSpecific{
\begin{eqnarray}
  \sqrt{1/3} \, \ket{3/2, -1/2 } - \sqrt{2/3} \, \ket{ 1/2, -1/2 }  \nonumber \\
  \rightarrow   \sqrt{2/3} \, \ket{ 3/2, -1/2 } + \sqrt{1/3} \, \ket{ 1/2, -1/2 },
\end{eqnarray}
the cross section is proportional to ${2/9} \abs{M_3 - M_1}^2$.  Kets $\ket{I,I_3}$ are in isospin space; $M_3$ denotes $\bra{3/2} H \ket{3/2}$ for the Hamiltonian $H$ responsible for the strong interaction, which is assumed to conserve total isospin $I$ and to be independent of $I_3$; and $M_1$ similarly denotes $\bra{1/2} H \ket{1/2}$.
  For $\pi^+ n \rightarrow  \pi^0 p$,
\begin{eqnarray}
  \sqrt{1/3} \, \ket{ 3/2, 1/2 } + \sqrt{2/3} \, \ket{ 1/2, 1/2 }  \nonumber  \\
   \rightarrow   \sqrt{2/3} \, \ket{ 3/2, 1/2 } - \sqrt{1/3} \, \ket{ 1/2, 1/2 },
\end{eqnarray}
the cross section is again proportional to ${2/9} \abs{M_3 - M_1}^2$.  
}

The probability for a 25 GeV $\pi^+$ to decay to a $\mu^+$ can be written
\begin{eqnarray}
            \poo{\pi^+}{\mu^+} = & p(\text{decays within tracker}) +    \nonumber \\
                                & p(\text{decays within calorimeter}).
\label{eqn:poo_pi_mu_decayInFlight}
\end{eqnarray}
The probability for the pion to decay within the tracking volume is
\begin{equation}
            p(\text{decays within tracker}) = 1-e^{-R_{\text{tracker}}/\gamma (c \tau)},
\label{eqn:poo_pi_mu_decayInFlight_cot}
\end{equation}
where $\gamma=25$~GeV~/~140~MeV~$=180$ is the pion's Lorentz boost, the proper decay length of the charged pion is $(c \tau) = 7.8$~meters, and the radius of the CDF tracking volume is $R_{\text{tracker}} = 1.5$~meters, giving $p(\text{decays within tracker}) = 0.001$.  The probability for the pion to decay within the calorimeter volume is
\begin{equation}
            p(\text{decays within calorimeter}) \approx \lambda_I / \gamma (c \tau),
\label{eqn:poo_pi_mu_decayInFlight_calorimeter}
\end{equation}
where $\lambda_I \approx 0.4$~meters is the nuclear interaction length for charged pions on lead or iron and the path length through the calorimeter is $L_{\text{cal}} \approx 2$~meters, leading to $p(\text{decays within calorimeter}) \approx 0.00025$.  Summing the contributions from decay within the tracking volume and decay within the calorimeter volume, $\poo{\pi^+ }{ \mu^+} \approx 0.00125$.

The primary physical mechanism by which a jet fakes a photon is for the parent quark or gluon to fragment into a leading $\pi^0$ carrying nearly all the momentum.  The highly boosted $\pi^0$ decays within the beam pipe to two photons that are sufficiently collinear to appear in the preshower, electromagnetic calorimeter, and shower maximum detector as a single photon.  Thus
\begin{equation}
\poo{j}{\gamma} = \poo{q}{\pi^0} \, \poo{\pi^0}{\gamma}.
\end{equation}
An immediate corollary is that the misreconstructed ``photon'' carries the energy of the parent quark or gluon, and is well measured.


 Since $\poo{q}{\pi^0} \gg \poo{q}{\gamma}$, it follows from Eq.~\ref{eqn:poo_q_pi} and Table~\ref{tbl:misId_cdfSim_central} that the conversion contribution to $\poo{j}{e}$ is $\approx 75\%$, and the charge exchange contribution is $\approx 25\%$:
\begin{eqnarray}
\frac{0.75}{0.25} =  ( & \poo{q}{\gamma} \, \poo{\gamma}{e^+} + & \, \nonumber \\
                       & \poo{q}{\pi^0} \, \poo{\pi^0}{e^+}     & ) \, / \nonumber \\
                     ( & \poo{q}{\pi^+} \, \poo{\pi^+ }{e^+} +  & \, \nonumber \\
                       & \poo{q}{K^+} \, \poo{K^+ }{e^+}        & ). 
\label{eqn:poo_conversion_e}
\end{eqnarray}
\cdfSpecific{This is consistent with the conclusion obtained in Ref.~\cite{BackgroundToElectronsCdfNote}.}

The number of $e^+\,j$ events in data is 0.9 times the number of $e^-\,j$ events.  This charge asymmetry arises from $\poo{K^+}{e^+}$ and $\poo{K^-}{e^-}$ in Table~\ref{tbl:misId_cdfSim_central}.  Quantitatively,
\begin{eqnarray}
    \frac{\poo{j}{e^+}}{\poo{j}{e^-}} = \frac{0.9+ 0.2 \, \poo{K^+}{e^+} / \poo{K}{e}}{0.9+0.2 \, \poo{K^-}{e^-} / \poo{K}{e}},
\label{eqn:poo_j_e_chargeAsymmetry}
\end{eqnarray}
where 0.9 is the sum of 0.75 from Eq.~\ref{eqn:poo_conversion_e} and $0.15\approx0.25\times 0.6$ from Eq.~\ref{eqn:poo_q_Kpi}, and 0.2 is twice $1-0.9$.  From $\poo{K^+}{e^+}$ and $\poo{K^-}{e^-}$ in Table~\ref{tbl:misId_cdfSim_central}, $\poo{K^+}{e^+} / \poo{K}{e} = 1/3$ and $\poo{K^-}{e^-} / \poo{K}{e} = 2/3$, predicting $\poo{j}{e^+}/\poo{j}{e^-}=0.935$, in reasonable agreement with the ratio of the observed number of events in the $e^+\,j$ and $e^-\,j$ final states.

The number of $j\,\mu^+$ events observed in CDF Run II is 1.1 times the number of $j\,\mu^-$ events observed.  This charge asymmetry arises from $\poo{K^+}{\mu^+}$ and $\poo{K^-}{\mu^-}$ in Table~\ref{tbl:misId_cdfSim_central}.

The physical mechanism by which a prompt photon fakes a tau lepton is for the photon to convert, producing an electron or positron carrying most of the photon's energy, which is then misreconstructed as a tau.  The probability for this to occur is equal for positively and negatively charged taus,
\begin{equation}
       \frac{1}{2} \poo{\gamma}{\tau} = \poo{\gamma}{\tau^+} = \poo{\gamma}{\tau^-},
\end{equation}
and is related to previously defined quantities by 
\begin{equation}
       \poo{\gamma}{\tau} = \poo{\gamma}{e} \, \frac{1}{\poo{e}{e}} \, \poo{e}{\tau},
\end{equation}
where $\poo{\gamma}{e}$ denotes the fraction of produced photons that are reconstructed as electrons, $\poo{e}{e}$ denotes the fraction of produced electrons that are reconstructed as electrons, and hence $\poo{\gamma}{e} / \poo{e}{e}$ is the fraction of produced photons that pair produce a single leading electron.

 Note $\poo{e}{\gamma}\approx\poo{\gamma}{e}$ from Table~\ref{tbl:misId_cdfSim_central}, as expected, with value of $\approx 0.03$ determined by the amount of material in the inner detectors and the tightness of isolation criteria.  A hard bremsstrahlung followed by a conversion is responsible for electrons to be reconstructed with opposite sign; hence
\begin{eqnarray}
            \poo{e^\pm}{e^\mp} =       & \poo{e^+}{e^-} = \poo{e^-}{e^+} \nonumber \\
                               \approx & \frac{1}{2} \, \poo{e^\pm}{\gamma}\poo{\gamma}{e^\mp},
\end{eqnarray}
where the factor of $1/2$ comes because the material already traversed by the $e^\pm$ will not be traversed again by the $\gamma$. In particular, track curvature mismeasurement is not responsible for erroneous sign determination in the central region of the CDF detector.

From knowledge of the underlying physical mechanisms by which jets fake electrons, muons, taus, and photons, the simple use of a reconstructed jet as a lepton or photon with an appropriate fake rate applied to the weight of the event needs slight modification to correctly handle the fact that a jet that has faked a lepton or photon generally is measured more accurately than a hadronic jet.  Rather than using the momentum of the reconstructed jet, the momentum of the parent quark or gluon is determined by adding up all Monte Carlo particle level objects within a cone of $\Delta R = 0.4$ about the reconstructed jet.  In misreconstructing a jet in an event, the momentum of the corresponding parent quark or gluon is used rather than the momentum of the reconstructed jet.  A jet that fakes a photon then has momentum equal to the momentum of the parent quark or gluon plus a fractional correction equal to $0.01\times(\text{parent}\,p_T - 25~\text{GeV})/(25~\text{GeV})$ to account for leakage out of the cone of $\Delta R = 0.4$, and a further smearing of $0.2~\sqrt{\text{GeV}} \times \sqrt{\text{parent}\,p_T}$, reflecting the electromagnetic resolution of the CDF detector.  The momenta of jets that fake photons are multiplied by an overall factor of 1.12, and jets that fake electrons, muons, or taus are multiplied by an overall factor of $0.95$.  These numbers are determined by the $\ell\pmiss$, $\ell j$, and $\gamma j$ final states.  The distributions most sensitive to these numbers are the missing energy and the jet $p_T$.

A $b$ quark fragmenting into a leading $b$ hadron that then decays leptonically or semileptonically results in an electron or muon that shares the $p_T$ of the parent $b$ quark with the associated neutrino.  If all hadronic decay products are soft, the distribution of the momentum fraction carried by the charged lepton can be obtained by considering the decay of a scalar to two massless fermions.  Isolated and energetic electrons and muons arising from parent $b$ quarks in this way are modeled as having $p_T$ equal to the parent $b$ quark $p_T$, multiplied by a random number uniformly distributed between 0 and 1.

\section{Additional background sources}
\label{sec:AdditionalBackgroundSources}

This appendix provides additional details on the estimation of the Standard Model prediction.

\subsection{Cosmic ray and beam halo muons}
\label{sec:CorrectionModelDetails:CosmicRays}

\begin{figure*}
\begin{tabular}{cc}
\includegraphics[width=2.3in,angle=270]{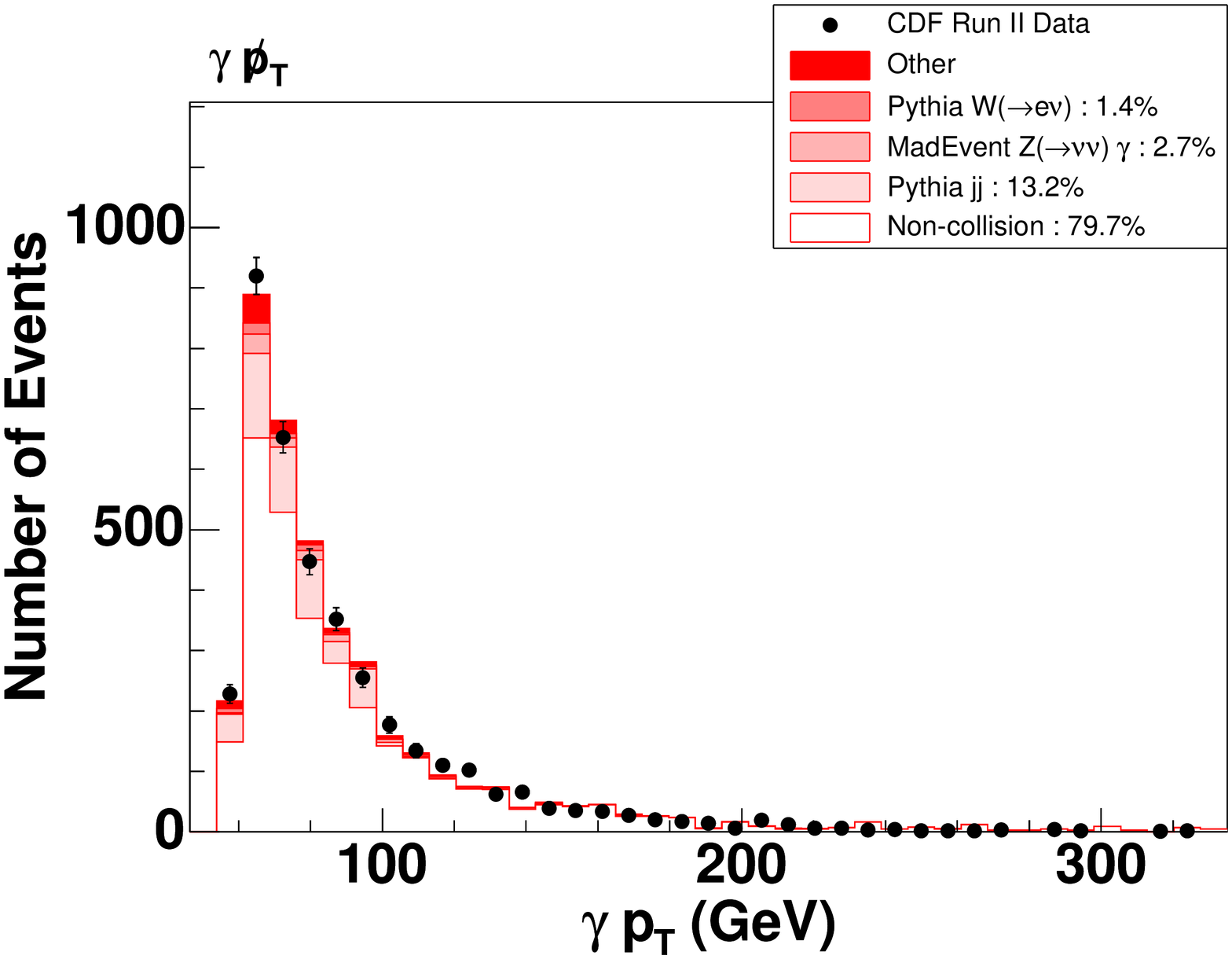} & 
\includegraphics[width=2.3in,angle=270]{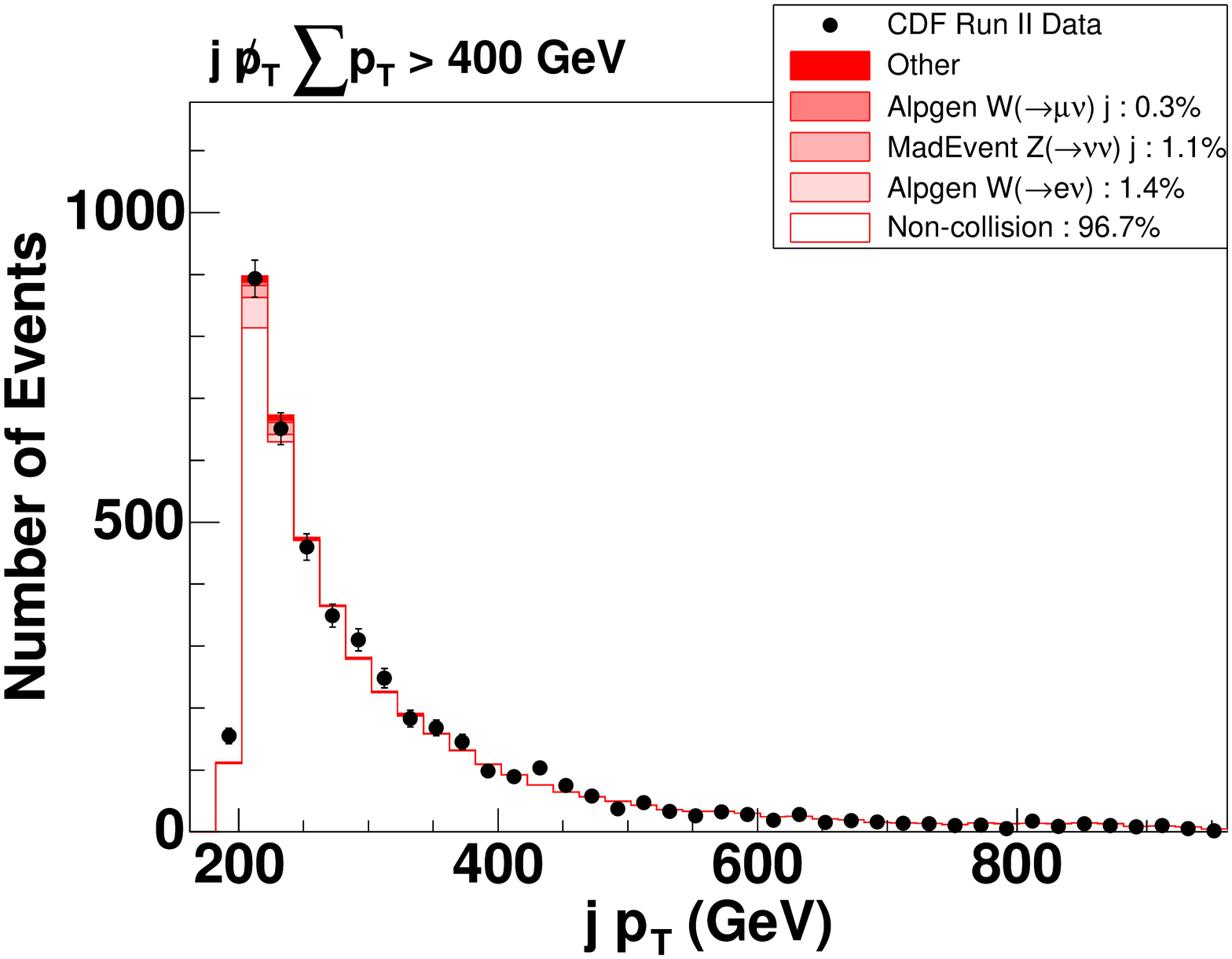} \\
\includegraphics[width=2.3in,angle=270]{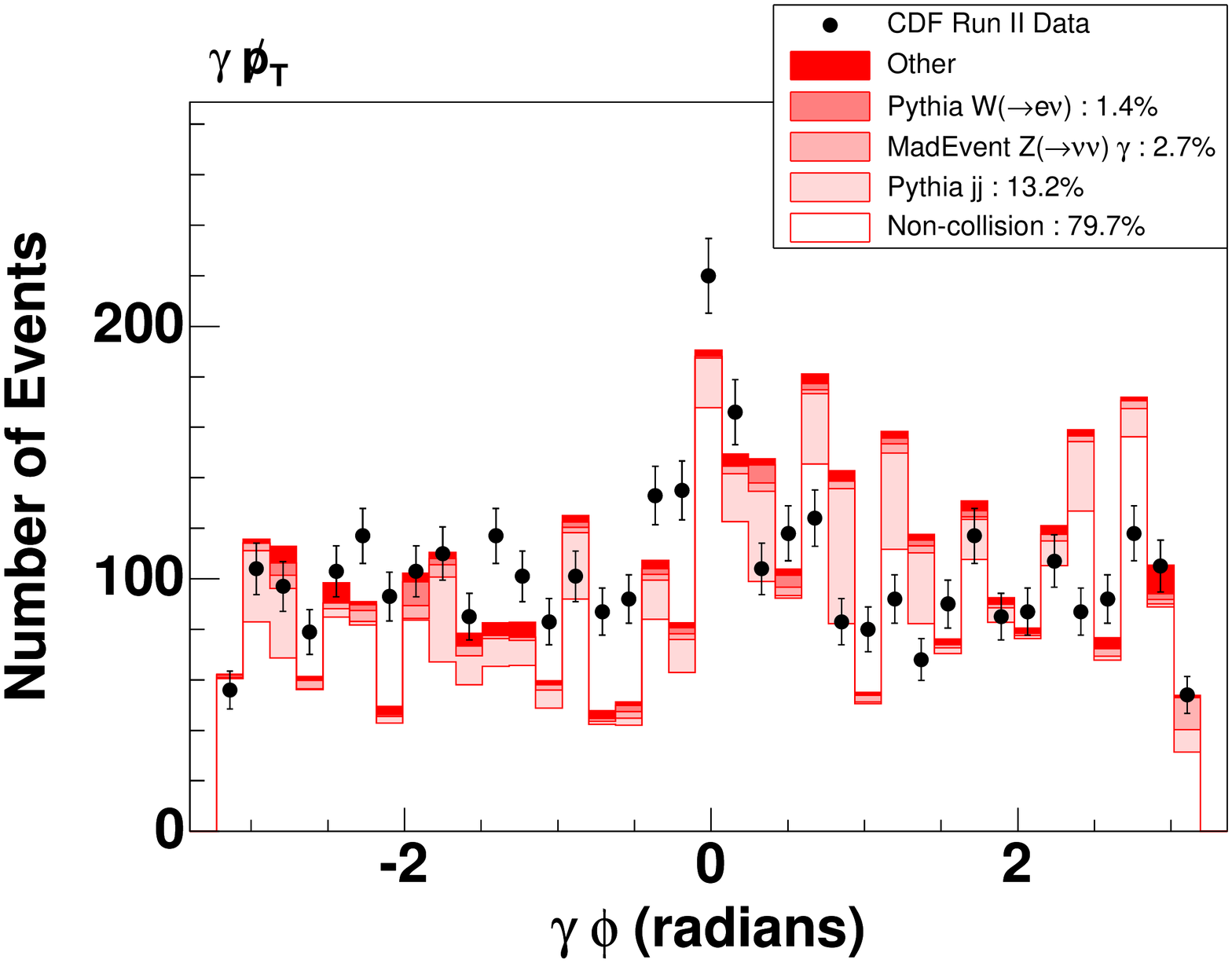} & 
\includegraphics[width=2.3in,angle=270]{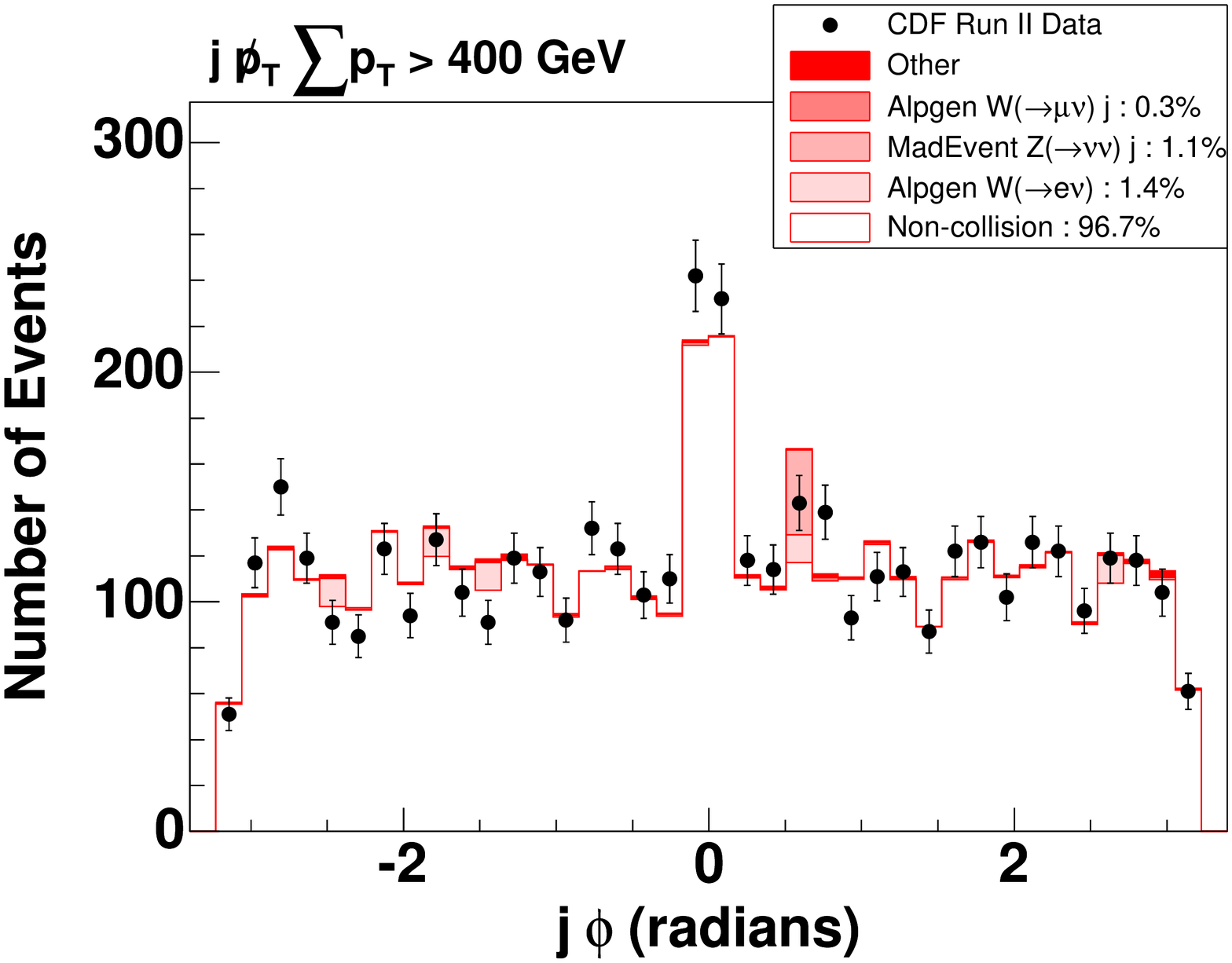} \\
\end{tabular}
\caption[The distribution of transverse momentum and azimuthal angle for photons and jets in the $\gamma\pmiss$ and $j\pmiss$ final states]{The distribution of transverse momentum and azimuthal angle for photons and jets in the $\gamma\pmiss$ and $j\pmiss$ final states, dominated by cosmic ray and beam halo muons.  The vertical axis shows the number of events in each bin.  Data are shown as filled (black) circles; the SM prediction is shown as the shaded (red) histogram.  The prediction includes contributions from cosmic ray and beam halo muons, estimated using events containing fewer than three reconstructed tracks.  The contribution from cosmic ray muons is flat in $\phi$, while the contribution from beam halo is localized to $\phi=0$.  The only degrees of freedom for the background to these final states are the {\tt{cosmic}}~$\gamma$ and {\tt{cosmic}}~$j$ correction factors, whose values are determined from the global fit (Table~\ref{tbl:CorrectionFactorDescriptionValuesSigmas}).}
\label{fig:1ph1j}
\end{figure*}

There are four dominant categories of events caused by cosmic ray muons penetrating the detector: $\mu\pmiss$, $\mu^+\mu^-$, $\gamma\pmiss$, and $j\pmiss$.  There is negligible contribution from cosmic ray secondaries of any particle type other than muons.

A cosmic ray muon penetrating the CDF detector whose trajectory passes within 1~mm of the beam line and within $-60<z<60$~cm of the origin may be reconstructed as two outgoing muons.  In this case the cosmic ray event is partitioned into the final state $\mu^+\mu^-$.  If one of the tracks is missed, the cosmic ray event is partitioned into the final state $\mu\pmiss$.  The standard CDF cosmic ray filter, which makes use of drift time information in the central tracking chamber, is used to reduce these two categories of cosmic ray events.

CDF data events with exactly one track (corresponding to one muon) and events with exactly two tracks (corresponding to two muons) are used to estimate the cosmic ray muon contribution to the final states $\mu\pmiss$ and $\mu^+\mu^-$ after the cosmic ray filter.  This sample of events is used as the SM background process {\tt{cosmic}}~$\mu$.  The {\tt{cosmic}}~$\mu$ sample does not contribute to the events passing the analysis offline trigger, whose cleanup cuts require the presence of three or more tracks.  

The remaining two categories are $\gamma\pmiss$ and $j\pmiss$, resulting from a cosmic ray muon that penetrates the CDF electromagnetic or hadronic calorimeter and undergoes a hard bremsstrahlung in one calorimeter cell.  Such an interaction can mimic a single photon or a single jet, respectively.  The reconstruction algorithm infers the presence of significant missing energy balancing the ``photon'' or ``jet.''  If this cosmic ray interaction occurs during a bunch crossing in which there is a $p\bar{p}$ interaction producing three or more tracks, the event will be partitioned into the final state $\gamma\pmiss$ or $j\pmiss$.

CDF data events with fewer than three tracks are used to estimate the cosmic ray muon contribution to the final states $\gamma\pmiss$ and $j\pmiss$.  These samples of events are used as SM background processes {\tt{cosmic}}~$\gamma$ and {\tt{cosmic}}~$j$ for the modeling of this background, corresponding to offline triggers requiring a photon with $p_T>60$~GeV, or a jet with $p_T>40$~GeV (prescaled) or $p_T>200$~GeV (unprescaled), respectively.  These samples do not contribute to the events passing the analysis offline trigger, whose cleanup cuts require three or more tracks.  The contribution of these events is adjusted with correction factors that are listed as {\tt{cosmic}}~$\gamma$ and {\tt{cosmic}}~$j$ ``$k$-factors'' in Table~\ref{tbl:CorrectionFactorDescriptionValuesSigmas}, but which are more properly understood as reflecting the number of bunch crossings with zero $p\bar{p}$ interactions (resulting in zero reconstructed tracks) relative to the number of bunch crossings with one or more interactions (resulting in three or more reconstructed tracks).  

The cosmic ray muon contribution to the final states $\gamma\pmiss$ and $j\pmiss$ is uniform as a function of the CDF azimuthal angle $\phi$.  Consider the CDF detector to be a thick cylindrical shell, and consider two arbitrary infinitesimal volume elements at different locations in the material of the shell.  Since the two volume elements have similar overburdens, the number of cosmic ray muons with $E\gtrsim20$~GeV penetrating the first volume element is very nearly the same as the number of cosmic ray muons with $E\gtrsim20$~GeV penetrating the second volume element.  Since the material of the CDF calorimeters is uniform as a function of CDF azimuthal angle $\phi$, it follows that the cosmic ray muon contribution to the final states $\gamma\pmiss$ and $j\pmiss$ should also be uniform as a function of $\phi$.  In particular, it is noted that the $\phi$ dependence of this contribution depends solely on the material distribution of CDF calorimeter, which is uniform in $\phi$, and has no dependence on the distribution of the horizon angle of the muons from cosmic rays.

The final states $\gamma\pmiss$ and $j\pmiss$ are also populated by beam halo muons, traveling horizontally through the CDF detector in time with a bunch.  A beam halo muon can undergo a hard bremsstrahlung in the electromagnetic or hadronic calorimeters, producing an energy deposition that can be reconstructed as a photon or jet, respectively.  These beam halo muons tend to lie in the horizontal plane and outside of the Tevatron ring, as if centrifugally hurled away from the beam; they horizontally penetrate the CDF detector along $\hat{z}$ at $y=0$ and $x>0$, hence at $\phi=0$.  

Fig.~\ref{fig:1ph1j} shows the $\gamma\pmiss$ and $j\pmiss$ final states, in which events come primarily from cosmic ray and beam halo muons.

\subsection{Multiple interactions}
\label{sec:Overlaps}

In order to estimate event overlaps, consider an interesting event observed in final state C, which looks like an overlap of two events in the final states A and B.  An example is C={\tt e+e-4j}, A={\tt e+e-} and B={\tt 4j}.  It is desired to estimate how many C events are expected from the overlap of A and B events, given the observed frequencies of A and B.

Let ${\cal L}(t)$ be the instantaneous luminosity as a function of time $t$; let 
\begin{equation}
L = \int_{\text{RunII}}{{\cal L}(t) dt} = \VistaApproximateDefiniteLuminosity~\text{pb}^{-1}
\end{equation}
denote the total integrated luminosity; and let
\begin{equation}
\bar{{\cal L}} = \frac{\int_{\text{RunII}}{{\cal L}(t){\cal L}(t) dt}}{\int_{\text{RunII}}{{\cal L}(t) dt}} \approx 10^{32}~\text{cm}^{-2}\text{s}^{-1}
\end{equation}
be the luminosity-averaged instantaneous luminosity.  Denote by $t_0$ the time interval of 396~ns between successive bunch crossings.  The total number of effective bunch crossings $X$ is then 
\begin{equation}
X = \frac{L}{\bar{{\cal L}} t_0} \approx 5\times 10^{13}.
\end{equation}
Letting $A$ and $B$ denote the number of observed events in final states A and B, it follows that the number of events in the final state C expected from overlap of A and B is
\begin{equation}
C=\frac{A}{X}\frac{B}{X}X=\frac{AB}{X}.
\end{equation}
Overlap events are included in the SM background estimate, although their contribution is generally negligible.

\subsection{Intrinsic $k_T$}
\label{sec:CorrectionModelDetails:IntrinsicKt}

Significant discrepancy is observed in many final states containing two objects {\tt o1} and {\tt o2} in the variables $\Delta\phi${\tt{(o1,o2)}}, {\tt uncl}~$p_T$, and $\pmiss_T$.  These discrepancies are ascribed to the sum of two effects:  (1) an intrinsic Fermi motion of the colliding partons within the proton and anti-proton, and (2) soft radiation along the beam axis.  The sum of these two effects appears to be larger in Nature than predicted by \Pythia\ with the parameter tunes used for the generation of the samples employed in this analysis.  This discrepancy is well known from previous studies at the Tevatron and elsewhere, and affects this analysis similarly to other Tevatron analyses.  

The $W$ and $Z$ electroweak samples used in this analysis have been generated with an adjusted \Pythia\ parameter that increases the intrinsic $k_T$.  For all other generated Standard Model events, the net effect of the Fermi motion of the colliding partons and the soft non-perturbative radiation is hypothesized to be described by an overall ``effective intrinsic $k_T$,'' and the center of mass of each event is given a transverse kick.  Specifically, for every event of invariant mass $m$ and generated summed transverse momentum \SumPt, a random number $k_T$ is pulled from the probability distribution 
\begin{eqnarray}
p(k_T) \varpropto (k_T < m/5) \times & [\frac{4}{5} g(k_T;\mu=0,\sigma_1) + \nonumber \\
                                     &  \frac{1}{5} g(k_T;\mu=0,\sigma_2)],
\end{eqnarray}
where $(k_T < m/5)$ evaluates to unity if true and zero if false; $g(k_T; \mu,\sigma)$ is a Gaussian function of $k_T$ with center at $\mu$ and width $\sigma$;  $\sigma_1=2.55\,\text{GeV}+0.0085\, \SumPt$ is the width of the core of the double Gaussian; and $\sigma_2=5.25\,\text{GeV}+0.0175 \, \SumPt$ is the width of the second, wider Gaussian. 
The event is then boosted to an inertial frame traveling with speed $\abs{\vec{\beta}}=k_T/m$ with respect to the lab frame, in a direction transverse to the beam axis, where $m$ is the invariant mass of all reconstructed objects in the event, along an azimuthal angle pulled randomly from a uniform distribution between 0 and $2\pi$.  The momenta of identified objects are recalculated in the lab frame.  Sixty percent of the recoil kick is assigned to unclustered momentum in the event.  The remaining forty percent of the recoil kick is assumed to disappear down the beam pipe, and contributes to the missing transverse momentum in the event.  This picture, and the particular parameter values that accompany this story, are determined primarily by the {\tt{uncl}}~$p_T$ and $\pmiss_T$ distributions in highly populated two-object final states, including the low-$p_T$ $2j$ final state, the high-$p_T$ $2j$ final state, and the final states $j\gamma$, $e^+e^-$, and $\mu^+\mu^-$.

Under the hypothesis described, reasonable although imperfect agreement with observation is obtained.  The result of this analysis supports the conclusions of previous studies indicating that the effective intrinsic $k_T$ needed to match observation is quite large relative to naive expectation.  That the data appear to require such a large effective intrinsic $k_T$ may be pointing out the need for some basic improvement to our understanding of this physics.

\section{Global fit}
\label{sec:CorrectionFactorFitDetails}

This section describes the construction of the global $\chi^2$ used in the \Vista\ global fit.

\subsection{The $\chi^2_k$}
\label{sec:CorrectionFactorFitDetails:chi_k}
The bins in the CDF high-$p_T$ data sample are labeled by the index $k=(k_1,k_2)$, where each value of $k_1$ represents a phrase such as ``this bin contains events with three objects: one with $17<p_T<25$~GeV and $\abs{\eta}<0.6$, one with $40<p_T<60$~GeV and $0.6<\abs{\eta}<1.0$, and one with $25<p_T<40$~GeV and $1.0<\abs{\eta}$,'' and each value of $k_2$ represents a phrase such as ``this bin contains events with three objects: an electron, muon, and jet, respectively.''  The reason for splitting $k$ into $k_1$ and $k_2$ is that a jet can fake an electron (mixing the contents of $k_2$), but an object with $\abs{\eta}<0.6$ cannot fake an object with $0.6<\abs{\eta}<1.0$ (no mixing of $k_1$).  The term corresponding to the $k^\text{th}$ bin takes the form of Eq.~\ref{eq:chi_k}, where $\text{Data}[k]$ is the number of data events observed in the $k^\text{th}$ bin, $\text{SM}[k]$ is the number of events predicted by the Standard Model in the $k^\text{th}$ bin, $\delta\text{SM}[k]$ is the Monte Carlo statistical uncertainty on the Standard Model prediction in the $k^\text{th}$ bin, and $\sqrt{\text{SM}[k]}$ is the statistical uncertainty on the prediction in the $k^\text{th}$ bin. To legitimize the use of Gaussian errors, only bins containing eight or more data events are considered.  The Standard Model prediction $\text{SM}[k]$ for the $k^\text{th}$ bin can be written in terms of the introduced correction factors as
\begin{eqnarray}
\lefteqn{\text{SM}[k] = \text{SM}[(k_1,k_2)] =} & \nonumber \\
                & \sum_{{k_2}'\in\text{objectLists}}\sum_{l\in\text{processes}} \nonumber \\
                &  (\int{{\cal L}\,dt}) \cdot (\text{kFactor}[l])\cdot (\text{SM}_0[(k_1,{k_2}')][l])\cdot \nonumber \\
                & (\text{probabilityToBeSoMisreconstructed}[(k_1,{k_2}')][k_2])\cdot   \nonumber \\
                &  (\text{probabilityPassesTrigger}[(k_1,k_2)]),
\end{eqnarray}
where $\text{SM}[k]$ is the Standard Model prediction for the $k^\text{th}$ bin; the index $k$ is the Cartesian product of the two indices $k_1$ and $k_2$ introduced above, labeling the regions of the detector in which there are energy clusters and the identified objects corresponding to those clusters, respectively; the index ${k_2}'$ is a dummy summation index; the index $l$ labels Standard Model background processes, such as dijet production or $W$+1~jet production; $\text{SM}_0[(k_1,{k_2}')][l]$ is the initial number of Standard Model events predicted in bin $(k_1,{k_2}')$ from the process labeled by the index $l$; $\text{probabilityToBeMisreconstructedThus}[(k_1,{k_2}')][k_2]$ is the probability that an event produced with energy clusters in the detector regions labeled by $k_1$ that are identified as objects labeled by ${k_2}'$ would be mistaken as having objects labeled by $k_2$; and $\text{probabilityPassesTrigger}[(k_1,k_2)]$ represents the probability that an event produced with energy clusters in the detector regions labeled by $k_1$ that are identified as objects labeled by $k_2$ would pass the trigger.

The quantity $\text{SM}_0[(k_1,{k_2}')][l]$ is obtained by generating some number $n_l$ (say $10^4$) of Monte Carlo events corresponding to the process $l$.  The event generator provides a cross section $\sigma_l$ for this process $l$.  The weight of each of these Monte Carlo events is equal to $\sigma_l /n_l$.  Passing these events through the CDF simulation and reconstruction, the sum of the weights of these events falling into the bin $(k_1,{k_2}')$ is $\text{SM}_0[(k_1,{k_2}')][l]$.

\subsection{$\chi^2_{\text{constraints}}$}
\label{sec:CorrectionFactorFitDetails:chi_constraints}

The term $\chi^2_{\text{constraints}}(\vec{s})$ in Eq.~\ref{eqn:chiSqd} reflects constraints on the values of the correction factors determined by data other than those in the global high-$p_T$ sample.  These constraints include $k$-factors taken from theoretical calculations and numbers from the CDF literature when use is made of CDF data external to the \Vista\ high-$p_T$ sample.  The constraints imposed are:
\begin{itemize}
\item The luminosity ({\tt 0001}) is constrained to be within 6\% of the value measured by the CDF \v{C}erenkov luminosity counters.
\item The fake rate $\poo{q}{\gamma}$  ({\tt 0039}) is constrained to be $2.6\times10^{-4} \pm 1.5\times10^{-5}$, from the single particle gun study of Appendix~\ref{sec:MisidentificationMatrix}.
\item The fake rate $\poo{e}{\gamma}$ ({\tt 0032}) plus the efficiency $\poo{e}{e}$ ({\tt 0026}) for electrons in the plug is constrained to be within 1\% of unity.
\item Noting $\poo{q}{\gamma}$ corresponds to correction factor {\tt 0039}, $\poo{q}{\pi^\pm}=2\, \poo{q}{\pi^0}$, and $\poo{q}{\pi^0}=\poo{q}{\gamma} / \poo{\pi^0}{\gamma}$, and taking $\poo{\pi^0}{\gamma}=0.6$ and $\poo{\pi^\pm}{\tau}=0.415$ from the single particle gun study of Appendix~\ref{sec:MisidentificationMatrix}, the fake rate $\poo{q}{\tau}$ ({\tt 0038}) is constrained to $\poo{q}{\tau}=\poo{q}{\pi^\pm}\poo{\pi^\pm}{\tau} \, \pm 10\%$.
\item The $k$-factors for dijet production ({\tt 0018} and {\tt 0019}) are constrained to $1.10 \pm 0.05$ and $1.33\pm 0.05$ in the kinematic regions $\hat{p}_T<150$~GeV and $\hat{p}_T>150$~GeV, respectively, where $\hat{p}_T$ is the transverse momentum of the scattered partons in the $2\rightarrow2$ process in the colliding parton center of momentum frame\cdfSpecific{~\cite{dijetNLO}}.
\item The inclusive $k$-factor for $\gamma + N$jets ({\tt 0004}--{\tt 0007}) is constrained to $1.25\pm 0.15$~\cite{NLOJET:Nagy:2003tz,NLOJET:Nagy:2001xb}.
\item The inclusive $k$-factor for $\gamma\gamma + N$jets ({\tt 0008}--{\tt 0010}) is constrained to $2.0 \pm 0.15$~\cite{Diphox:Binoth:1999qq}.
\item The inclusive $k$-factors for $W$ and $Z$ production ({\tt 0011}--{\tt 0014} and {\tt 0015}--{\tt 0017}) are subject to a 2-dimensional Gaussian constraint, with mean at the NNLO/LO theoretical values~\cite{Stirling:Sutton:1991ay}, and a covariance matrix that encapsulates the highly correlated theoretical uncertainties, as discussed in Appendix~\ref{sec:VistaCorrectionModel:CorrectionFactorValues}.
\item Trigger efficiency correction factors are constrained to be less than unity.
\item All correction factors are constrained to be positive.
\end{itemize}

\subsection{Covariance matrix}
\label{sec:CorrectionFactorCovarianceMatrix}

\begin{sidewaystable*}
\hspace{+0in}
\tiny
\begin{minipage}{8.1in}
\begin{verbatim}
       0001 0002 0003 0004 0005 0006 0007 0008 0009 0010 0011 0012 0013 0014 0015 0016 0017 0018 0019 0020 0021 0022 0023 0024 0025 0026 0027 0028 0029 0030 0031 0032 0033 0034 0035 0036 0037 0038 0039 0040 0041 0042 0043 0044 
0001   1    -.32 -.7  -.56 -.53 -.45 -.26 -.36 -.21 -.14 -.87 -.77 -.51 -.28 -.82 -.55 -.31 -.95 -.96 -.94 -.94 -.88 -.88 -.62 -.54 -.17 -.46 -.37 -.09 -.1  0    -.24 +.08 +.17 +.08 -.01 -.04 +.01 +.02 -.02 -.22 -.21 -.13 -.11 
0002   -.32 1    +.21 +.37 +.38 +.33 +.2  +.34 +.18 +.12 +.28 +.25 +.17 +.09 +.27 +.18 +.1  +.3  +.31 +.31 +.3  +.28 +.28 +.2  +.18 +.06 +.15 +.12 -.31 +.02 0    +.08 -.14 -.06 -.03 0    +.01 -.03 -.07 -.07 +.07 +.07 +.04 +.04 
0003   -.7  +.21 1    +.39 +.37 +.31 +.18 +.25 +.14 +.1  +.61 +.53 +.35 +.2  +.57 +.38 +.21 +.66 +.66 +.66 +.65 +.61 +.61 +.43 +.38 +.12 +.32 +.26 +.06 +.07 -.01 +.17 -.05 -.12 -.06 +.01 +.03 -.01 -.01 +.01 +.15 +.14 +.09 +.08 
0004   -.56 +.37 +.39 1    +.9  +.77 +.48 +.61 +.33 +.23 +.49 +.43 +.29 +.16 +.46 +.32 +.18 +.5  +.53 +.53 +.52 +.49 +.49 +.35 +.3  +.1  +.25 +.2  -.46 +.03 0    +.13 -.44 -.09 -.03 -.01 +.02 -.32 -.62 -.17 +.11 +.11 +.07 +.06 
0005   -.53 +.38 +.37 +.9  1    +.75 +.46 +.62 +.31 +.21 +.46 +.41 +.27 +.15 +.44 +.3  +.17 +.5  +.51 +.48 +.5  +.47 +.46 +.33 +.29 +.1  +.24 +.19 -.49 +.03 -.02 +.12 -.43 -.09 -.04 0    +.02 -.29 -.57 -.16 +.11 +.1  +.07 +.06 
0006   -.45 +.33 +.31 +.77 +.75 1    +.4  +.54 +.29 +.13 +.4  +.35 +.24 +.13 +.38 +.26 +.14 +.43 +.44 +.42 +.42 +.35 +.4  +.28 +.25 +.09 +.2  +.16 -.45 +.02 -.01 +.1  -.36 -.07 -.04 0    +.01 -.24 -.46 -.14 +.09 +.09 +.06 +.05 
0007   -.26 +.2  +.18 +.48 +.46 +.4  1    +.34 +.18 +.09 +.23 +.2  +.13 +.08 +.22 +.15 +.08 +.24 +.25 +.25 +.24 +.21 +.22 +.02 +.14 +.05 +.12 +.09 -.29 +.01 -.02 +.06 -.23 -.04 -.02 +.01 +.01 -.15 -.3  -.09 +.05 +.05 +.03 +.03 
0008   -.36 +.34 +.25 +.61 +.62 +.54 +.34 1    +.37 +.28 +.32 +.28 +.19 +.1  +.3  +.22 +.12 +.34 +.34 +.34 +.33 +.31 +.31 +.22 +.18 +.06 +.16 +.12 -.61 -.03 0    +.11 -.29 -.06 -.03 0    +.01 -.09 -.17 -.28 +.07 +.07 +.04 +.04 
0009   -.21 +.18 +.14 +.33 +.31 +.29 +.18 +.37 1    +.06 +.19 +.17 +.11 +.06 +.2  +.06 +.11 +.2  +.2  +.19 +.19 +.18 +.18 +.13 +.08 +.03 +.07 +.06 -.31 +.05 0    +.06 -.15 -.03 -.01 0    +.01 -.04 -.08 -.29 +.04 +.04 +.03 +.02 
0010   -.14 +.12 +.1  +.23 +.21 +.13 +.09 +.28 +.06 1    +.13 +.11 +.08 +.06 +.13 +.11 -.03 +.13 +.14 +.13 +.13 +.12 +.12 +.09 +.05 -.01 +.05 +.04 -.19 +.06 0    +.07 -.1  -.03 -.01 0    +.01 -.04 -.07 -.26 +.03 +.04 +.01 +.01 
0011   -.87 +.28 +.61 +.49 +.46 +.4  +.23 +.32 +.19 +.13 1    +.85 +.58 +.32 +.89 +.61 +.33 +.83 +.84 +.82 +.82 +.76 +.77 +.54 +.25 +.09 +.16 +.15 +.07 +.07 0    +.12 +.1  +.04 +.02 0    +.01 -.01 -.02 +.01 -.13 -.04 -.11 -.09 
0012   -.77 +.25 +.53 +.43 +.41 +.35 +.2  +.28 +.17 +.11 +.85 1    +.33 +.35 +.79 +.49 +.33 +.72 +.74 +.74 +.72 +.68 +.67 +.47 +.21 +.08 +.15 +.13 +.06 +.06 +.01 +.11 +.1  -.02 -.09 -.01 +.01 -.01 -.01 +.01 -.14 +.01 -.06 -.05 
0013   -.51 +.17 +.35 +.29 +.27 +.24 +.13 +.19 +.11 +.08 +.58 +.33 1    -.21 +.52 +.35 +.15 +.5  +.49 +.46 +.48 +.46 +.45 +.36 +.15 +.06 +.1  +.09 +.04 +.04 -.01 +.07 +.05 +.07 -.07 0    0    -.01 -.01 +.01 -.1  -.07 -.06 -.05 
0014   -.28 +.09 +.2  +.16 +.15 +.13 +.08 +.1  +.06 +.06 +.32 +.35 -.21 1    +.29 +.26 -.04 +.28 +.27 +.28 +.26 +.21 +.26 +.09 +.08 +.03 +.05 +.05 +.02 +.02 0    +.03 +.01 0    -.07 0    0    -.01 -.01 +.01 -.05 -.01 -.03 -.02 
0015   -.82 +.27 +.57 +.46 +.44 +.38 +.22 +.3  +.2  +.13 +.89 +.79 +.52 +.29 1    +.58 +.35 +.77 +.78 +.77 +.76 +.71 +.71 +.5  +.09 +.04 +.06 +.05 +.05 +.02 0    +.03 -.02 -.03 -.06 0    0    -.01 -.02 +.03 +.04 +.03 +.04 +.03 
0016   -.55 +.18 +.38 +.32 +.3  +.26 +.15 +.22 +.06 +.11 +.61 +.49 +.35 +.26 +.58 1    -.09 +.52 +.53 +.52 +.52 +.49 +.48 +.35 +.1  +.03 +.08 +.07 +.03 +.02 0    +.04 -.03 -.01 -.1  0    +.01 -.01 -.02 +.02 0    +.01 -.02 -.01 
0017   -.31 +.1  +.21 +.18 +.17 +.14 +.08 +.12 +.11 -.03 +.33 +.33 +.15 -.04 +.35 -.09 1    +.3  +.3  +.29 +.29 +.25 +.28 +.16 +.03 -.02 +.04 +.04 +.02 +.04 0    +.04 -.03 -.06 -.06 0    +.01 -.01 -.01 -.02 +.03 +.05 +.01 +.01 
0018   -.95 +.3  +.66 +.5  +.5  +.43 +.24 +.34 +.2  +.13 +.83 +.72 +.5  +.28 +.77 +.52 +.3  1    +.91 +.92 +.89 +.85 +.83 +.6  +.51 +.16 +.43 +.35 +.09 +.1  -.07 +.23 -.16 -.23 -.16 +.02 0    -.01 -.03 +.01 +.21 +.18 +.12 +.1  
0019   -.96 +.31 +.66 +.53 +.51 +.44 +.25 +.34 +.2  +.14 +.84 +.74 +.49 +.27 +.78 +.53 +.3  +.91 1    +.91 +.91 +.84 +.85 +.59 +.52 +.16 +.44 +.36 +.09 +.1  +.03 +.23 -.07 -.17 -.08 -.06 +.04 -.01 -.02 +.02 +.21 +.2  +.12 +.11 
0020   -.94 +.31 +.66 +.53 +.48 +.42 +.25 +.34 +.19 +.13 +.82 +.74 +.46 +.28 +.77 +.52 +.29 +.92 +.91 1    +.87 +.84 +.83 +.6  +.51 +.16 +.43 +.35 +.08 +.1  -.05 +.23 -.13 -.24 -.13 +.01 +.01 -.02 -.03 +.01 +.21 +.2  +.12 +.11 
0021   -.94 +.3  +.65 +.52 +.5  +.42 +.24 +.33 +.19 +.13 +.82 +.72 +.48 +.26 +.76 +.52 +.29 +.89 +.91 +.87 1    +.82 +.83 +.57 +.51 +.16 +.43 +.35 +.08 +.1  +.04 +.23 -.07 -.16 -.07 -.08 +.04 -.01 -.02 +.02 +.2  +.19 +.12 +.1  
0022   -.88 +.28 +.61 +.49 +.47 +.35 +.21 +.31 +.18 +.12 +.76 +.68 +.46 +.21 +.71 +.49 +.25 +.85 +.84 +.84 +.82 1    +.73 +.55 +.47 +.15 +.4  +.33 +.08 +.09 -.04 +.21 -.1  -.21 -.1  +.01 +.02 -.01 -.03 +.02 +.19 +.18 +.11 +.1  
0023   -.88 +.28 +.61 +.49 +.46 +.4  +.22 +.31 +.18 +.12 +.77 +.67 +.45 +.26 +.71 +.48 +.28 +.83 +.85 +.83 +.83 +.73 1    +.53 +.48 +.15 +.4  +.33 +.08 +.09 +.01 +.21 -.06 -.15 -.07 -.04 +.03 -.01 -.02 +.02 +.19 +.18 +.11 +.1  
0024   -.62 +.2  +.43 +.35 +.33 +.28 +.02 +.22 +.13 +.09 +.54 +.47 +.36 +.09 +.5  +.35 +.16 +.6  +.59 +.6  +.57 +.55 +.53 1    +.33 +.11 +.28 +.23 +.05 +.06 -.01 +.15 -.09 -.16 -.07 +.01 +.02 -.01 -.02 +.01 +.13 +.13 +.08 +.07 
0025   -.54 +.18 +.38 +.3  +.29 +.25 +.14 +.18 +.08 +.05 +.25 +.21 +.15 +.08 +.09 +.1  +.03 +.51 +.52 +.51 +.51 +.47 +.48 +.33 1    +.23 +.6  +.49 +.05 +.04 -.01 +.25 -.03 -.23 -.05 +.01 +.04 -.01 -.02 +.09 +.12 +.28 +.19 +.17 
0026   -.17 +.06 +.12 +.1  +.1  +.09 +.05 +.06 +.03 -.01 +.09 +.08 +.06 +.03 +.04 +.03 -.02 +.16 +.16 +.16 +.16 +.15 +.15 +.11 +.23 1    +.18 +.15 +.01 +.01 0    -.66 -.03 +.37 -.01 0    -.02 0    -.01 +.19 +.07 -.44 +.05 +.04 
0027   -.46 +.15 +.32 +.25 +.24 +.2  +.12 +.16 +.07 +.05 +.16 +.15 +.1  +.05 +.06 +.08 +.04 +.43 +.44 +.43 +.43 +.4  +.4  +.28 +.6  +.18 1    +.29 +.05 +.1  0    +.27 -.15 -.25 0    0    +.05 -.01 -.01 0    +.35 +.3  -.33 +.33 
0028   -.37 +.12 +.26 +.2  +.19 +.16 +.09 +.12 +.06 +.04 +.15 +.13 +.09 +.05 +.05 +.07 +.04 +.35 +.36 +.35 +.35 +.33 +.33 +.23 +.49 +.15 +.29 1    +.05 +.08 0    +.23 -.1  -.19 +.03 0    +.04 -.01 -.01 0    +.26 +.23 +.32 -.54 
0029   -.09 -.31 +.06 -.46 -.49 -.45 -.29 -.61 -.31 -.19 +.07 +.06 +.04 +.02 +.05 +.03 +.02 +.09 +.09 +.08 +.08 +.08 +.08 +.05 +.05 +.01 +.05 +.05 1    +.06 0    +.03 +.31 -.02 0    +.01 +.01 +.01 +.01 +.21 +.03 +.03 +.01 +.01 
0030   -.1  +.02 +.07 +.03 +.03 +.02 +.01 -.03 +.05 +.06 +.07 +.06 +.04 +.02 +.02 +.02 +.04 +.1  +.1  +.1  +.1  +.09 +.09 +.06 +.04 +.01 +.1  +.08 +.06 1    0    -.13 -.02 -.03 0    0    +.03 0    0    -.76 +.08 +.05 +.01 +.01 
0031   0    0    -.01 0    -.02 -.01 -.02 0    0    0    0    +.01 -.01 0    0    0    0    -.07 +.03 -.05 +.04 -.04 +.01 -.01 -.01 0    0    0    0    0    1    0    +.07 +.04 +.07 -.83 +.03 +.01 +.03 +.01 -.01 0    -.01 -.01 
0032   -.24 +.08 +.17 +.13 +.12 +.1  +.06 +.11 +.06 +.07 +.12 +.11 +.07 +.03 +.03 +.04 +.04 +.23 +.23 +.23 +.23 +.21 +.21 +.15 +.25 -.66 +.27 +.23 +.03 -.13 0    1    -.06 -.48 -.02 0    +.05 0    0    -.08 +.17 +.57 +.06 +.05 
0033   +.08 -.14 -.05 -.44 -.43 -.36 -.23 -.29 -.15 -.1  +.1  +.1  +.05 +.01 -.02 -.03 -.03 -.16 -.07 -.13 -.07 -.1  -.06 -.09 -.03 -.03 -.15 -.1  +.31 -.02 +.07 -.06 1    +.23 +.17 -.02 -.01 +.2  +.39 +.14 -.55 -.18 -.21 -.18 
0034   +.17 -.06 -.12 -.09 -.09 -.07 -.04 -.06 -.03 -.03 +.04 -.02 +.07 0    -.03 -.01 -.06 -.23 -.17 -.24 -.16 -.21 -.15 -.16 -.23 +.37 -.25 -.19 -.02 -.03 +.04 -.48 +.23 1    +.16 -.01 -.04 +.01 +.02 +.09 -.31 -.89 -.22 -.19 
0035   +.08 -.03 -.06 -.03 -.04 -.04 -.02 -.03 -.01 -.01 +.02 -.09 -.07 -.07 -.06 -.1  -.06 -.16 -.08 -.13 -.07 -.1  -.07 -.07 -.05 -.01 0    +.03 0    0    +.07 -.02 +.17 +.16 1    -.02 +.02 +.01 +.01 0    -.12 -.1  -.26 -.23 
0036   -.01 0    +.01 -.01 0    0    +.01 0    0    0    0    -.01 0    0    0    0    0    +.02 -.06 +.01 -.08 +.01 -.04 +.01 +.01 0    0    0    +.01 0    -.83 0    -.02 -.01 -.02 1    0    0    +.01 0    +.01 +.01 +.02 +.01 
0037   -.04 +.01 +.03 +.02 +.02 +.01 +.01 +.01 +.01 +.01 +.01 +.01 0    0    0    +.01 +.01 0    +.04 +.01 +.04 +.02 +.03 +.02 +.04 -.02 +.05 +.04 +.01 +.03 +.03 +.05 -.01 -.04 +.02 0    1    +.01 +.01 -.03 +.06 +.07 +.03 +.02 
0038   +.01 -.03 -.01 -.32 -.29 -.24 -.15 -.09 -.04 -.04 -.01 -.01 -.01 -.01 -.01 -.01 -.01 -.01 -.01 -.02 -.01 -.01 -.01 -.01 -.01 0    -.01 -.01 +.01 0    +.01 0    +.2  +.01 +.01 0    +.01 1    +.51 +.06 0    0    0    0    
0039   +.02 -.07 -.01 -.62 -.57 -.46 -.3  -.17 -.08 -.07 -.02 -.01 -.01 -.01 -.02 -.02 -.01 -.03 -.02 -.03 -.02 -.03 -.02 -.02 -.02 -.01 -.01 -.01 +.01 0    +.03 0    +.39 +.02 +.01 +.01 +.01 +.51 1    +.12 0    -.01 0    0    
0040   -.02 -.07 +.01 -.17 -.16 -.14 -.09 -.28 -.29 -.26 +.01 +.01 +.01 +.01 +.03 +.02 -.02 +.01 +.02 +.01 +.02 +.02 +.02 +.01 +.09 +.19 0    0    +.21 -.76 +.01 -.08 +.14 +.09 0    0    -.03 +.06 +.12 1    -.04 -.11 +.01 +.01 
0041   -.22 +.07 +.15 +.11 +.11 +.09 +.05 +.07 +.04 +.03 -.13 -.14 -.1  -.05 +.04 0    +.03 +.21 +.21 +.21 +.2  +.19 +.19 +.13 +.12 +.07 +.35 +.26 +.03 +.08 -.01 +.17 -.55 -.31 -.12 +.01 +.06 0    0    -.04 1    +.37 +.39 +.33 
0042   -.21 +.07 +.14 +.11 +.1  +.09 +.05 +.07 +.04 +.04 -.04 +.01 -.07 -.01 +.03 +.01 +.05 +.18 +.2  +.2  +.19 +.18 +.18 +.13 +.28 -.44 +.3  +.23 +.03 +.05 0    +.57 -.18 -.89 -.1  +.01 +.07 0    -.01 -.11 +.37 1    +.25 +.22 
0043   -.13 +.04 +.09 +.07 +.07 +.06 +.03 +.04 +.03 +.01 -.11 -.06 -.06 -.03 +.04 -.02 +.01 +.12 +.12 +.12 +.12 +.11 +.11 +.08 +.19 +.05 -.33 +.32 +.01 +.01 -.01 +.06 -.21 -.22 -.26 +.02 +.03 0    0    +.01 +.39 +.25 1    +.07 
0044   -.11 +.04 +.08 +.06 +.06 +.05 +.03 +.04 +.02 +.01 -.09 -.05 -.05 -.02 +.03 -.01 +.01 +.1  +.11 +.11 +.1  +.1  +.1  +.07 +.17 +.04 +.33 -.54 +.01 +.01 -.01 +.05 -.18 -.19 -.23 +.01 +.02 0    0    +.01 +.33 +.22 +.07 1    
\end{verbatim}
\end{minipage}
\caption[Correction factor correlation matrix.]{Correction factor correlation matrix.  The top row and left column show correction factor codes.  Each element of the matrix shows the correlation between the correction factors corresponding to the column and row.  Each matrix element is dimensionless; the elements along the diagonal are unity; the matrix is symmetric; positive elements indicate positive correlation, and negative elements anti-correlation.}
\label{tbl:CorrectionFactorCorrelationMatrix}
\end{sidewaystable*}

This section describes the correction factor covariance matrix $\Sigma$.  The inverse of the covariance matrix is obtained from 
\begin{equation}
\Sigma^{-1}_{ij} = \frac{1}{2} \frac{\partial^2 \chi^2(\vec s)}{\partial s_i \partial s_j} \bigg| _{\vec s_0},
\label{eqn:logL4}
\end{equation}
where $\chi^2(\vec{s})$ is defined by Eq.~\ref{eqn:chiSqd} as a function of the correction factor vector $\vec{s}$, vector elements $s_i$ and $s_j$ are the $i^{\text{th}}$ and $j^{\text{th}}$ correction factors, and $\vec{s}_0$ is the vector of correction factors that minimizes $\chi^2(\vec{s})$.  Numerical estimation of the right hand side of Eq.~\ref{eqn:logL4} is achieved by calculating $\chi^2$ at $\vec s_0$ and at positions slightly displaced from $\vec s_0$ in the direction of the $i^{\text{th}}$ and $j^{\text{th}}$ correction factors, denoted by the unit vectors $\hat i$ and $\hat j$.  Approximating the second partial derivative
\begin{eqnarray}
  \label{eq:secondDerivative}
  \frac{\partial^2 \chi^2}{\partial s_j \partial s_i} \bigg|_{\vec s_0} &=& \frac{\chi^2(\vec s_0 + \hat i \, \delta s_i + \hat j \, \delta s_j) - \chi^2(\vec s_0 + \hat j \, \delta s_j)}{\delta s_j \delta s_i}  - \nonumber \\
                                                                        & & \frac{\chi^2(\vec s_0 + \hat i \, \delta s_i) - \chi^2(\vec s_0)}{\delta s_j \delta s_i} \nonumber
\end{eqnarray}
leads to
\begin{eqnarray}
\Sigma^{-1}_{ij} & = & [ \chi^2(\vec{s}_0+\delta s_i \, \hat{i}+\delta s_j \, \hat{j}) \nonumber \\
            &   &  - \chi^2(\vec{s}_0+\delta s_i \, \hat{i}) \nonumber \\
            &   &  - \chi^2(\vec{s}_0+\delta s_j \, \hat{j}) \nonumber \\
            &   & + \chi^2(\vec{s}_0) ] /  (2 \delta s_i \, \delta s_j),
\label{eqn:CorrectionFactorSigmaInverse}
\end{eqnarray}
for appropriately small steps $\delta s_i$ and $\delta s_j$ away from the minimum.  The covariance matrix $\Sigma$ is calculated by inverting $\Sigma^{-1}$.  The diagonal element $\Sigma_{ii}$ is the variance $\sigma_i^2$ of the $i^{\text{th}}$ correction factor, and the correlation $\rho_{ij}$ between the $i^{\text{th}}$ and $j^{\text{th}}$ correction factors is $\rho_{ij}=\Sigma_{ij}/\sigma_i\sigma_j$.  The variances of each correction factor, corresponding to the diagonal elements of the covariance matrix, are shown in Table~\ref{tbl:CorrectionFactorDescriptionValuesSigmas}.  The correlation matrix obtained is shown in Table~\ref{tbl:CorrectionFactorCorrelationMatrix}.

\section{Correction factor values}
\label{sec:VistaCorrectionModel:CorrectionFactorValues}

This section provides notes on the values of the \Vista\ correction factors obtained from a global fit of Standard Model prediction to data.  The correction factors considered are numbers that can in principle be calculated {\em{a priori}}, but whose calculation is in practice not feasible.  These correction factors divide naturally into two classes, the first of which reflects the difficulty of calculating the Standard Model prediction to all orders, and the second of which reflects the difficulty of understanding from first principles the response of the experimental apparatus.

The theoretical correction factors considered are of two types.  The difficulty of calculating the Standard Model prediction for many processes to all orders in perturbation theory is handled through the introduction of $k$-factors, representing the ratio of the true all orders prediction to the prediction at lowest order in perturbation theory.  Uncertainties in the distribution of partons inside the colliding proton and anti-proton as a function of parton momentum are in principle handled through the introduction of correction factors associated with parton distribution functions, but there are currently no discrepancies to motivate this.

Experimental correction factors correspond to numbers describing the response of the CDF detector that are precisely calculable in principle, but that are in practice best constrained by the high-$p_T$ data themselves.  These correction factors take the form of the integrated luminosity, object identification efficiencies, object misidentification probabilities, trigger efficiencies, and energy scales.  

\subsection{$k$-factors}

For nearly all Standard Model processes, $k$-factors are used as an overall multiplicative constant, rather than being considered to be a function of one or more kinematic variables.  The spirit of the approach is to introduce as few correction factors as possible, and to only introduce correction factors motivated by specific discrepancies.

\begin{figure}
\centering
\includegraphics[width=3.5in]{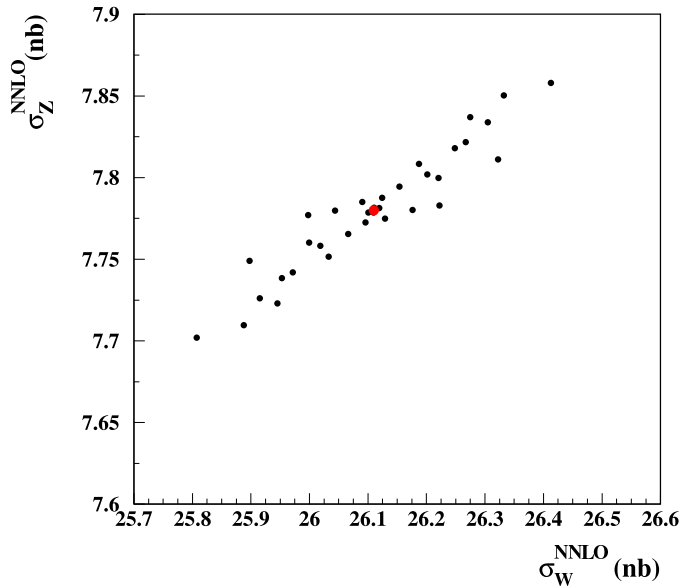} \\
\caption[Variation of the $k$-factors for inclusive $W$ and $Z$ production under different choices of parton distribution functions]{Variation of the $k$-factors for inclusive $W$ and $Z$ production under different choices of parton distribution functions, from the Alekhin parton distribution error set~\cite{Alekhin:2005gq}.  The correlation of the uncertainty on these two $k$-factors due to uncertainty in the parton distribution functions is 0.955.}
\label{fig:alekhin_kWkZ_pdfs}
\end{figure}

\paragraph*{\tt{0001}.}
The integrated luminosity of the analysis sample has a close relationship with the theoretically determined values of inclusive $W$ and $Z$ production at the Tevatron.  Figure~\ref{fig:alekhin_kWkZ_pdfs} shows the variation in calculated inclusive $W$ and $Z$ $k$-factors under changes in the assumed parton distribution functions.  Each point represents a different $W$ and $Z$ inclusive cross section determined using modified parton distribution functions.  The use of 16 bases to reflect systematic uncertainties results in 32 black dots in Fig.~\ref{fig:alekhin_kWkZ_pdfs}.  The uncertainties in the $W$ and $Z$ cross sections due to variations in the renormalization and factorization scales are nearly 100\% correlated; varying these scales affects both the $W$ and $Z$ inclusive cross sections in the same way.  The uncertainties in the parton distribution functions and the choice of renormalization and factorization scales represent the dominant contributions to the theoretical uncertainty in the total inclusive $W$ and $Z$ cross section calculations at the Tevatron.  The term in $\chi^2_{\text{constraints}}$ that reflects our knowledge of the theoretical prediction of the inclusive $W$ and $Z$ cross sections explicitly acknowledges this high degree of correlation.

Theoretical constraints on all other $k$-factors are assumed to be uncorrelated with each other, not because the uncertainties of these calculations are indeed uncorrelated, but rather because the correlations among these computations are poorly known.

\begin{figure}
\centering
\includegraphics[width=5in]{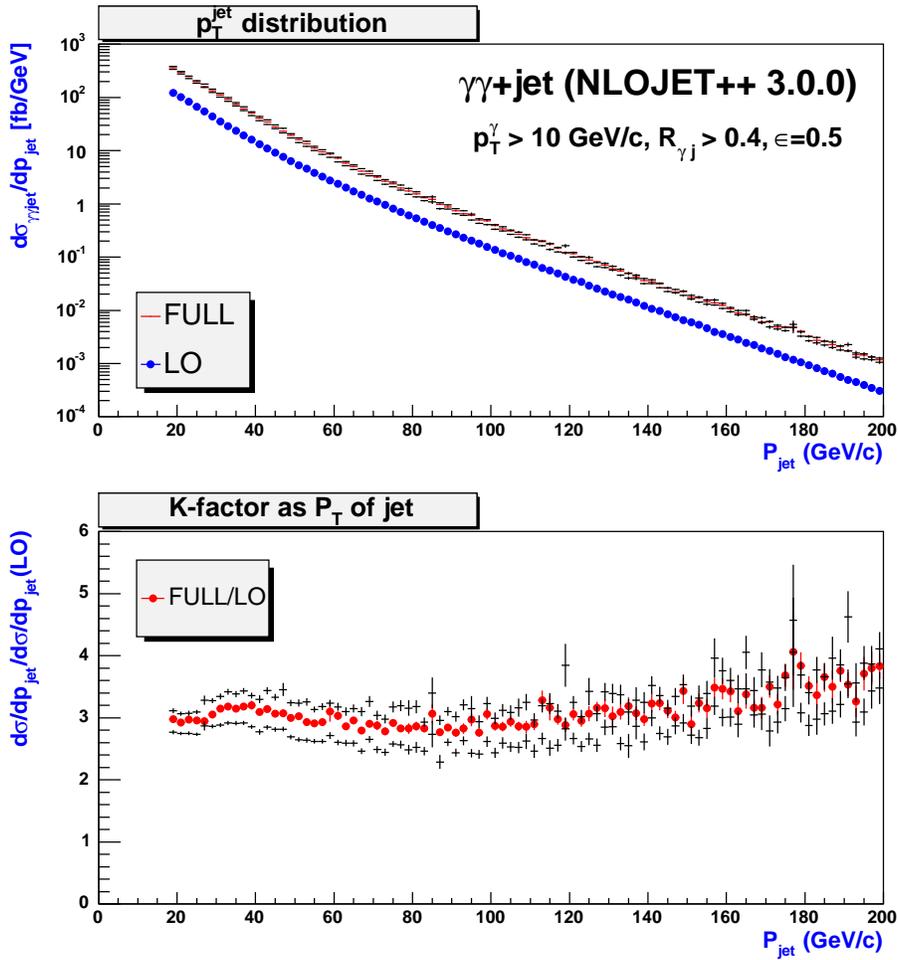}
\caption[Calculation of the $\gamma\gamma j$ $k$-factor, as a function of jet transverse momentum.]{Calculation of the $\gamma\gamma j$ $k$-factor, as a function of jet transverse momentum.  The effect of changing the factorization scale by a factor of two in either direction is also shown (small black points with error bars).}
\label{fig:soonJun_kfactor_2ph1j}
\end{figure}

\paragraph*{\tt{0002}, \tt{0003}.}
The {\tt{cosmic}}~$\gamma$ and {\tt{cosmic}}~$j$ backgrounds are estimated using events recorded in the CDF data with one or more reconstructed photons and with two or fewer reconstructed tracks.  The use of events with two or fewer reconstructed tracks is a new technique for estimating these backgrounds.  These correction factors are primarily constrained by the number of events in the \Vista\ $\gamma\pmiss$ and $j\pmiss$ final states.  The values are related to (and consistent with) the fraction of bunch crossings with one or more inelastic $p\bar{p}$ interactions, complicated slightly by the requirement that any jet falling in the final state $j\pmiss$ has at least 5~GeV of track $p_T$ within a cone of 0.4 relative to the jet axis.
\paragraph*{{\tt{0004}}, {\tt{0005}}, {\tt{0006}}, {\tt{0007}}.}
The NLOJET++ calculation of the $\gamma j$ inclusive $k$-factor constrains the cross section weighted sum of the $\gamma j$, $\gamma 2j$, $\gamma 3j$, and $\gamma 4j$ correction factors to $1.25\pm0.15$~\cite{NLOJET:Nagy:2003tz,NLOJET:Nagy:2001xb}.
\paragraph*{{\tt{0008}}, {\tt{0009}}, {\tt{0010}}.}
The DIPHOX calculation of the inclusive $\gamma\gamma$ cross section at NLO constrains the weighted sum of these correction factors to $2.0\pm0.15$~\cite{Diphox:Binoth:1999qq}.  From Table~\ref{tbl:CorrectionFactorDescriptionValuesSigmas}, the $\gamma\gamma j$ $k$-factor ({\tt{0009}}) appears anomalously large.  Figure~\ref{fig:soonJun_kfactor_2ph1j} shows a calculation of this $\gamma\gamma j$ $k$-factor using NLOJET++~\cite{NLOJET:Nagy:2003tz,NLOJET:Nagy:2001xb} as a function of summed transverse momentum.  The NLO correction to the LO prediction is found to be large, and not manifestly inconsistent with the value for this $k$-factor determined from the \Vista\ fit.  The cross section for $\gamma\gamma 2j$ production has not been calculated at NLO.
\paragraph*{{\tt{0011}}, {\tt{0012}}, {\tt{0013}}, {\tt{0014}}.}
These correction factors correspond to $k$-factors for $W$ production in association with zero, one, two, and three or more jets, respectively.  A linear combination of these correction factors is constrained by the requirement that the inclusive $W$ production cross section is consistent with the NNLO calculation of Ref.~\cite{Alekhin:2005gq}.  The values of these correction factors, and their trend of decreasing as the number of jets increases, depends heavily on the choice of renormalization and factorization scales.  The individual correction factors are not explicitly constrained by a NLO calculation.
\paragraph*{{\tt{0015}}, {\tt{0016}}, {\tt{0017}}.}
These correction factors correspond to $k$-factors for $Z$ production in association with zero, one, and two or more jets, respectively.  A linear combination of these correction factors is constrained by the requirement that the inclusive $Z$ production cross section is consistent with the NNLO calculation of Ref.~\cite{Alekhin:2005gq}.  
\paragraph*{{\tt{0018}}, {\tt{0019}}.}
The two $k$-factors for dijet production correspond to two bins in $\hat{p}_T$, the $p_T$ of the hard two to two scattering in the parton center of mass frame.  These correction factors are constrained by a NLO calculation~\cite{2j_kfactor:Stump:2003yu}, and show expected behavior as a function of $\hat{p}_T$.
\paragraph*{{\tt{0020}}, {\tt{0021}}.}
The two $k$-factors for 3-jet production, corresponding to two bins in $\hat{p}_T$, are unconstrained by any NLO calculation, but show reasonable behavior as a function of $\hat{p}_T$.
\paragraph*{{\tt{0022}}, {\tt{0023}}.}
The $k$-factors for 4-jet production, corresponding to two bins in $\hat{p}_T$, are unconstrained by any NLO calculation, but show reasonable behavior as a function of $\hat{p}_T$.
\paragraph*{\tt{0024}.}
The $k$-factor for the production of five or more jets, constrained primarily by the \Vista\ low-$p_T$ $5j$ final state, is found to be close to unity.

\subsection{Identification efficiencies}

The correction factors in this section, although billed as ``identification efficiencies,'' are truly ratios of the identification efficiency in the data relative to the identification efficiency in \CdfSim.  A correction factor value of unity indicates a proper modeling of the overall identification efficiency by \CdfSim; a correction factor value of 0.5 indicates that \CdfSim\ overestimates the overall identification efficiency by a factor of two.  

\paragraph*{{\tt{0025}}.}
The central electron identification efficiency scale factor is close to unity, indicating the central electron efficiency measured in data is similar (to within 1\%) to the central electron efficiency in the CDF detector simulation.  This reflects an emphasis within CDF on tuning the detector simulation for central electrons.  The determination of this correction factor is dominated by the \Vista\ final states $e\pmiss$ and $e^+e^-$, where one of the electrons has $\abs{\eta}<1$.
\paragraph*{{\tt{0026}}.}
The plug electron identification efficiency scale factor is several percent less than unity, indicating that the CDF detector simulation slightly overestimates the electron identification efficiency in the plug region of the CDF detector.  The determination of this correction factor is dominated by the \Vista\ final states $e\pmiss$ and $e^+e^-$, where one of the electrons has $\abs{\eta}>1$.
\paragraph*{{\tt{0027}}, {\tt{0028}}.}
To reduce backgrounds hypothesized to arise from pion and kaon decays in flight with a substantially mismeasured track, a very good track fit in the CDF tracker is required.  Partially due to this tight track fit requirement, CDF muon identification efficiencies in the regions $\abs{\eta}<0.6$ and $0.6<\abs{\eta}<1.5$ are overestimated in the CDF detector simulation by over 10\%.  The determination of the identification efficiencies $\poo{\mu}{\mu}$ is dominated by the \Vista\ final states $\mu\pmiss$ and $\mu^+\mu^-$.
\paragraph*{\tt{0029}.}
The central photon identification efficiency scale factor is determined primarily by the number of events in the \Vista\ final states $j\gamma$ and $\gamma\gamma$.  The uncertainty on this correction factor is highly correlated with the uncertainties on the $\gamma j$ $k$-factor, the $\poo{j}{\gamma}$ fake rate, and the $\gamma\gamma$ $k$-factor.
\paragraph*{\tt{0030}.}
The plug photon identification efficiency scale factor is determined primarily by the number of events in the \Vista\ final state $\gamma\gamma$.  The uncertainty on this correction factor is highly correlated with the uncertainty on the plug $\poo{j}{\gamma}$ fake rate.
\paragraph*{\tt{0031}.}
The $b$-jet identification efficiency is determined to be consistent with the prediction from \CdfSim.

\subsection{Fake rates}

\paragraph*{\tt{0032}.}
The fake rate $\poo{e}{\gamma}$ for electrons to be misreconstructed as photons in the plug region of the detector is added on top of the significant number of electrons misreconstructed as photons by \CdfSim.
\paragraph*{\tt{0033}.}
In \Vista, the contribution of jets faking electrons is modeled by applying a fake rate $\poo{j}{e}$ to Monte Carlo jets.  \Vista\ represents the first large scale Tevatron analysis in which a completely Monte Carlo based modeling of jets faking electrons is employed.  Significant understanding of the physical mechanisms contributing to this fake rate has been achieved, as summarized in Appendix~\ref{sec:MisidentificationMatrix}.  Consistency with this understanding is required; for example, $\poo{j}{e} \approx \poo{j}{\gamma}\poo{\gamma}{e}$.  The value of this correction factor is determined primarily by the number of events in the \Vista\ final state $ej$, where the electron is identified in the central region of the CDF detector.  It is notable that this fake rate is independent of global event properties, and that a consistent simultaneous understanding of the $ej$, $e2j$, $e3j$, and $e4j$ final states is obtained.
\paragraph*{\tt{0034}.}
The value of the fake rate $\poo{j}{e}$ in the plug region of the CDF detector is roughly one order of magnitude larger than the corresponding fake rate $\poo{j}{e}$ in the central region of the detector, consistent with an understanding of the relative performance of the detector in the central and plug regions for the identification of electrons.  This correction factor is determined primarily by the number of events in the \Vista\ final state $ej$, where the electron is identified in the plug region of the CDF detector.
\paragraph*{\tt{0035}.}
In \Vista, the contribution of jets faking muons is modeled by applying a fake rate $\poo{j}{\mu}$ to Monte Carlo jets.  \Vista\ represents the first large scale Tevatron analysis in which a completely Monte Carlo based modeling of jets faking muons is employed.  The value obtained from the \Vista\ fit is seen to be roughly one order of magnitude smaller than the fake rate $\poo{j}{e}$ in the central region of the detector, consistent with our understanding of the physical mechanisms underlying these fake rates, as described in Appendix~\ref{sec:MisidentificationMatrix}.  The value of this correction factor is determined primarily by the number of events in the \Vista\ final state $j\mu$.
\paragraph*{{\tt{0036}}.}
The fake rate $\poo{j}{b}$ has $p_T$ dependence explicitly imposed.  The number of tracks inside a typical jet, and hence the probability that a secondary vertex is (mis)reconstructed, increases with jet $p_T$.  The values of these correction factors are consistent with the mistag rate determined using secondary vertices reconstructed on the other side of the beam axis with respect to the direction of the tagged jet \cite{bTagging}.  The value of this correction factor is determined primarily by the number of events in the \Vista\ final states $bj$ and $bb$.
\paragraph*{{\tt{0037}}, {\tt{0038}}.}
The fake rate $\poo{j}{\tau}$ decreases with jet $p_T$, since the number of tracks inside a typical jet increases with jet $p_T$.  The values of these correction factors are determined primarily by the number of events in the \Vista\ final state $j\tau$.
\paragraph*{{\tt{0039}}, {\tt{0040}}.}
The fake rate $\poo{j}{\gamma}$ is determined separately in the central and plug regions of the CDF detector.  The values of these correction factors are determined primarily by the number of events in the \Vista\ final states $j\gamma$ and $\gamma\gamma$.  The value obtained for {\tt{0039}} is consistent with the value obtained from a study using detailed information from the central preshower detector.  The fake rate determined in the plug region is noticeably higher than the fake rate determined in the central region, as expected.

\subsection{Trigger efficiencies}

\paragraph*{{\tt{0041}}.}
The central electron trigger inefficiency is dominated by not correctly reconstructing the electron's track at the first online trigger level.
\paragraph*{{\tt{0042}}.}
The plug electron trigger inefficiency is due to inefficiencies in clustering at the second online trigger level.
\paragraph*{{\tt{0043}}, {\tt{0044}}.}
The muon trigger inefficiencies in the regions $\abs{\eta}<0.6$ and $0.6<\abs{\eta}<1.0$ derive partly from tracking inefficiency, and partly from an inefficiency in reconstructing muon stubs in the CDF muon chambers.\\
~\\
The value of these corrections factors are consistent with other trigger efficiency measurements made using additional information~\cite{Messina:2006zz}.

\subsection{Energy scales}

The \Vista\ infrastructure also allows the jet energy scale to be treated as a correction factor.  At present this correction factor is not used, since there is no discrepancy requiring it.

To understand the effect of introducing such a correction factor, a jet energy scale correction factor is added and constrained to $1\pm0.03$, reflecting the jet energy scale determination at CDF~\cite{jetEnergyScale:Bhatti:2005ai}.  The fit returns a value with a very small error, since this correction factor is highly constrained by the low-$p_T$ $2j$, $3j$, $e\,j$, and $e\,2j$ final states.  Assuming perfectly correct modeling of jets faking electrons, as described in Appendix~\ref{sec:MisidentificationMatrix}, this is a correct energy scale error.  The inclusion of additional correction factor degrees of freedom to reflect possible imperfections in this modeling of jets faking electrons increases the energy scale error.  The interesting conclusion is that the jet energy scale (considered as a lone free parameter) is very well constrained by the large number of dijet events; adjustment to the jet energy scale must be accompanied by simultaneous adjustment of other correction factors (such as the dijet $k$-factor) in order to retain agreement with data.


\section{\Sleuth\ details}
\label{sec:Sleuth:MinimumNumberOfEvents}

This appendix elaborates on the \Sleuth\ partitioning rule, and on the minimum number of events required for a final state to be considered by \Sleuth.

\subsection{Partitioning}
\label{sec:Sleuth:Partitioning}

\begin{table*}
\hspace*{-0.0in}\mbox{\tiny\begin{minipage}{7.5in}
\begin{tabular}{p{1.5cm}p{5cm}} 
\Sleuth & \Vista\ Final States   \\ 
\hline 
\hline 
$b \bar{b}$ & $b$$j$, $b$2$j$, 2$b$$j$, 2$b$, 3$b$ \\ 
\hline 
$b \bar{b} \ell^{+}\ell^{-}$ & $e^{+}$$e^{-}$$b$$j$, $\mu^{+}$$\mu^{-}$$b$$j$, $e^{+}$$e^{-}$$b$2$j$, $\mu^{+}$$\mu^{-}$$b$2$j$, $\mu^{+}$$\mu^{-}$2$b$$j$, $e^{+}$$e^{-}$2$b$, $\mu^{+}$$\mu^{-}$2$b$ \\ 
\hline 
$b \bar{b} \ell^{+}\ell^{-} 2j$ & $e^{+}$$e^{-}$$b$3$j$, $\mu^{+}$$\mu^{-}$$b$3$j$ \\ 
\hline 
$b \bar{b} \ell^{+}\ell^{-} 2j \pmiss$ & $\mu^{+}$$\mu^{-}$2$b$2$j$$\pmiss$ \\ 
\hline 
$b \bar{b} \ell^{+}\ell^{-} \pmiss$ & $e^{+}$$e^{-}$$b$$j$$\pmiss$, $\mu^{+}$$\mu^{-}$$b$$j$$\pmiss$, $e^{+}$$e^{-}$$b$2$j$$\pmiss$, $\mu^{+}$$\mu^{-}$$b$2$j$$\pmiss$, $e^{+}$$e^{-}$2$b$$j$$\pmiss$, $e^{+}$$e^{-}$2$b$$\pmiss$, $\mu^{+}$$\mu^{-}$2$b$$\pmiss$ \\ 
\hline 
$b \bar{b} \ell^{+} 2j  \ell'^{-}  \pmiss$ & $e^{+}$$\mu^{-}$$b$3$j$$\pmiss$ \\ 
\hline 
$b \bar{b} \ell^{+} 2j  \gamma \pmiss$ & $\mu^{+}$$\gamma$$b$3$j$$\pmiss$ \\ 
\hline 
$W b\bar{b} jj$ & $e^{+}$$b$3$j$$\pmiss$, $\mu^{+}$$b$3$j$$\pmiss$, $e^{+}$2$b$2$j$$\pmiss$, $\mu^{+}$2$b$2$j$$\pmiss$ \\ 
\hline 
$b \bar{b} \ell^{+} 2j \tau^{+}$ & $\mu^{+}$$\tau^{+}$$b$3$j$ \\ 
\hline 
$b \bar{b} \ell^{+} \ell'^{+}$ & $e^{+}$$\mu^{+}$2$b$ \\ 
\hline 
$b \bar{b} \ell^{+} \ell'^{-}$ & $e^{+}$$\mu^{-}$$b$$j$ \\ 
\hline 
$b \bar{b} \ell^{+} \ell'^{-}  \pmiss$ & $e^{+}$$\mu^{-}$$b$$j$$\pmiss$, $e^{+}$$\mu^{-}$$b$2$j$$\pmiss$, $e^{+}$$\mu^{-}$2$b$$\pmiss$ \\ 
\hline 
$b \bar{b} \ell^{+} \gamma \pmiss$ & $e^{+}$$\gamma$$b$2$j$$\pmiss$, $\mu^{+}$$\gamma$$b$2$j$$\pmiss$ \\ 
\hline 
$W b\bar{b}$ & $e^{+}$$b$$j$$\pmiss$, $\mu^{+}$$b$$j$$\pmiss$, $e^{+}$$b$2$j$$\pmiss$, $\mu^{+}$$b$2$j$$\pmiss$, $e^{+}$2$b$$\pmiss$, $e^{+}$2$b$$j$$\pmiss$, $\mu^{+}$2$b$$j$$\pmiss$, $\mu^{+}$2$b$$\pmiss$, $e^{+}$3$b$$\pmiss$ \\ 
\hline 
$b \bar{b} \ell^{+} \pmiss \tau^{-}$ & $e^{+}$$\tau^{-}$$b$2$j$$\pmiss$, $e^{+}$$\tau^{-}$$b$$j$$\pmiss$, $\mu^{+}$$\tau^{-}$$b$$j$$\pmiss$, $e^{+}$$\tau^{-}$2$b$$\pmiss$ \\ 
\hline 
$b \bar{b} \ell^{+} \tau^{+}$ & $e^{+}$$\tau^{+}$$b$$j$ \\ 
\hline 
$b \bar{b} \ell^{+} \tau^{-}$ & $e^{+}$$\tau^{-}$$b$$j$, $e^{+}$$\tau^{-}$$b$2$j$, $e^{+}$$\tau^{-}$2$b$, $\mu^{+}$$\tau^{-}$$b$$j$ \\ 
\hline 
$b \bar{b} 2j$ & $b$3$j$, 2$b$2$j$ \\ 
\hline 
$b \bar{b} 2j \gamma$ & $\gamma$$b$3$j$, $\gamma$2$b$2$j$ \\ 
\hline 
$b \bar{b} 2j \gamma \pmiss$ & $\gamma$$b$3$j$$\pmiss$ \\ 
\hline 
$b \bar{b} 2j \pmiss$ & $b$3$j$$\pmiss$, 2$b$2$j$$\pmiss$ \\ 
\hline 
$\gamma b \bar{b}$ & $\gamma$$b$$j$, $\gamma$$b$2$j$, $\gamma$2$b$, $\gamma$2$b$$j$, $\gamma$3$b$ \\ 
\hline 
$b \bar{b} \gamma \pmiss$ & $\gamma$$b$$j$$\pmiss$, $\gamma$$b$2$j$$\pmiss$, $\gamma$2$b$$\pmiss$ \\ 
\hline 
$b \bar{b} \pmiss$ & $b$2$j$$\pmiss$, 2$b$$j$$\pmiss$, $b$$j$$\pmiss$, 2$b$$\pmiss$ \\ 
\hline 
$b \bar{b} \pmiss \tau$ & $\tau^{+}$$b$$j$$\pmiss$ \\ 
\hline 
$b \bar{b} \tau^{+} \tau^{-}$ & $\tau^{+}$$\tau^{-}$$b$$j$, $\tau^{+}$$\tau^{-}$$b$2$j$ \\ 
\hline 
$b \bar{b} 4j$ & 2$b$4$j$ \\ 
\hline 
$2b4j\pmiss$ & $b$6$j$$\pmiss$ \\ 
\hline 
$\gamma \gamma b \bar{b}$ & 2$\gamma$$b$$j$, 2$\gamma$$b$2$j$ \\ 
\hline 
$2b6j$ & 2$b$6$j$ \\ 
\hline 
$\ell^{+}\ell^{-}$ & $e^{+}$$e^{-}$, $\mu^{+}$$\mu^{-}$, $e^{+}$$e^{-}$$j$, $\mu^{+}$$\mu^{-}$$j$, $e^{+}$$e^{-}$$b$, $\mu^{+}$$\mu^{-}$$b$ \\ 
\hline 
$\ell^{+}\ell^{-} 2j$ & $e^{+}$$e^{-}$2$j$, $\mu^{+}$$\mu^{-}$2$j$, $e^{+}$$e^{-}$3$j$, $\mu^{+}$$\mu^{-}$3$j$ \\ 
\hline 
$\ell^{+}\ell^{-}\ell'2j\pmiss$ & $e^{+}$$e^{-}$$\mu^{+}$2$j$$\pmiss$ \\ 
\hline 
$\ell^{+}\ell^{-} 2j \gamma$ & $e^{+}$$e^{-}$$\gamma$2$j$, $\mu^{+}$$\mu^{-}$$\gamma$2$j$, $e^{+}$$e^{-}$$\gamma$3$j$, $\mu^{+}$$\mu^{-}$$\gamma$3$j$ \\ 
\hline 
$\ell^{+}\ell^{-} 2j \pmiss$ & $e^{+}$$e^{-}$2$j$$\pmiss$, $\mu^{+}$$\mu^{-}$2$j$$\pmiss$, $e^{+}$$e^{-}$3$j$$\pmiss$, $\mu^{+}$$\mu^{-}$3$j$$\pmiss$ \\ 
\hline 
$\ell^{+}\ell^{-} \tau^{+} 2j \pmiss$ & $e^{+}$$e^{-}$$\tau^{+}$2$j$ \\ 
\hline 
$\ell^{+}\ell^{-} \ell' \gamma \pmiss$ & $e^{+}$$\mu^{+}$$\mu^{-}$$\gamma$$j$ \\ 
\hline 
$\ell^{+}\ell^{-} \ell' \pmiss$ & $e^{+}$$\mu^{+}$$\mu^{-}$, $e^{+}$$e^{-}$$\mu^{+}$$\pmiss$, $e^{+}$$e^{-}$$\mu^{+}$$j$$\pmiss$, $e^{+}$$e^{-}$$\mu^{+}$, $e^{+}$$\mu^{+}$$\mu^{-}$$j$, $e^{+}$$\mu^{+}$$\mu^{-}$$\pmiss$ \\ 
\hline 
$\ell^{+}\ell^{-} \gamma$ & $e^{+}$$e^{-}$$\gamma$, $\mu^{+}$$\mu^{-}$$\gamma$, $e^{+}$$e^{-}$$\gamma$$j$, $\mu^{+}$$\mu^{-}$$\gamma$$j$ \\ 
\hline 
$\ell^{+}\ell^{-} \gamma  \pmiss$ & $e^{+}$$e^{-}$$\gamma$$\pmiss$, $e^{+}$$e^{-}$$\gamma$$j$$\pmiss$ \\ 
\hline 
$\ell^{+}\ell^{-} \pmiss$ & $e^{+}$$e^{-}$$\pmiss$, $\mu^{+}$$\mu^{-}$$\pmiss$, $e^{+}$$e^{-}$$j$$\pmiss$, $\mu^{+}$$\mu^{-}$$j$$\pmiss$, $\mu^{+}$$\mu^{-}$$b$$\pmiss$, $e^{+}$$e^{-}$$b$$\pmiss$ \\ 
\hline 
$\ell^{+}\ell^{-} \pmiss  \tau^{+}$ & $e^{+}$$e^{-}$$\tau^{+}$, $e^{+}$$e^{-}$$\tau^{+}$$j$, $\mu^{+}$$\mu^{-}$$\tau^{+}$ \\ 
\hline 
$\ell^{+}\ell^{-} 4j$ & $e^{+}$$e^{-}$4$j$, $\mu^{+}$$\mu^{-}$4$j$ \\ 
\hline 
$\ell^{+}\ell^{-} 4j \pmiss$ & $e^{+}$$e^{-}$4$j$$\pmiss$, $\mu^{+}$$\mu^{-}$4$j$$\pmiss$ \\ 
\hline 
$\ell^{+}\ell^{-} \tau^{+} 4j \pmiss$ & $e^{+}$$e^{-}$$\tau^{+}$4$j$ \\ 
\hline 
$\ell^{+} \ell'^{+} jj$ & $e^{+}$$\mu^{+}$2$j$, $e^{+}$$\mu^{+}$3$j$ \\ 
\hline 
$\ell^{+} \ell'^{+} \pmiss jj$ & $e^{+}$$\mu^{+}$2$j$$\pmiss$ \\ 
\hline 
$\ell^{+} \ell'^{-} jj$ & $e^{+}$$\mu^{-}$2$j$, $e^{+}$$\mu^{-}$3$j$ \\ 
\hline 
$\ell^{+} \ell'^{-} \pmiss jj$ & $e^{+}$$\mu^{-}$2$j$$\pmiss$, $e^{+}$$\mu^{-}$3$j$$\pmiss$ \\ 
\hline 
$W \gamma jj$ & $\mu^{+}$$\gamma$2$j$$\pmiss$, $e^{+}$$\gamma$2$j$$\pmiss$, $\mu^{+}$$\gamma$3$j$$\pmiss$, $e^{+}$$\gamma$3$j$$\pmiss$ \\ 
\hline 
$W jj$ & $e^{+}$2$j$$\pmiss$, $\mu^{+}$2$j$$\pmiss$, $e^{+}$3$j$$\pmiss$, $\mu^{+}$3$j$$\pmiss$ \\ 
\hline 
$\ell^{+} \tau^{+} \pmiss jj$ & $\mu^{+}$$\tau^{+}$2$j$$\pmiss$ \\ 
\hline 
$\ell^{+} \tau^{-} \pmiss jj$ & $e^{+}$$\tau^{-}$2$j$$\pmiss$, $e^{+}$$\tau^{-}$3$j$$\pmiss$, $\mu^{+}$$\tau^{-}$3$j$$\pmiss$, $\mu^{+}$$\tau^{-}$2$j$$\pmiss$ \\ 
\hline 
$\ell^{+} \tau^{+} jj$ & $e^{+}$$\tau^{+}$2$j$, $e^{+}$$\tau^{+}$3$j$, $\mu^{+}$$\tau^{+}$2$j$, $\mu^{+}$$\tau^{+}$3$j$ \\ 
\hline 
$\ell^{+} \tau^{-} jj$ & $e^{+}$$\tau^{-}$2$j$, $\mu^{+}$$\tau^{-}$2$j$, $e^{+}$$\tau^{-}$3$j$, $\mu^{+}$$\tau^{-}$3$j$ \\ 
\hline 
$\ell^{+} \ell'^{+}$ & $e^{+}$$\mu^{+}$, $e^{+}$$\mu^{+}$$j$ \\ 
\hline 
$\ell^{+} \ell'^{+} \pmiss$ & $e^{+}$$\mu^{+}$$\pmiss$, $e^{+}$$\mu^{+}$$j$$\pmiss$ \\ 
\hline 
$\ell^{+} \ell'^{-}$ & $e^{+}$$\mu^{-}$, $e^{+}$$\mu^{-}$$j$ \\ 
\hline 
$\ell^{+} \ell'^{-} \gamma \pmiss$ & $e^{+}$$\mu^{-}$$\gamma$$j$$\pmiss$ \\ 
\hline
\end{tabular} 
\hspace{0.6cm} 
\begin{tabular}{p{1.5cm}p{5cm}} 
\Sleuth & \Vista\ Final States   \\ 
\hline 
\hline 
$\ell^{+} \ell'^{-} \pmiss$ & $e^{+}$$\mu^{-}$$\pmiss$, $e^{+}$$\mu^{-}$$j$$\pmiss$, $e^{+}$$\mu^{-}$$b$$\pmiss$ \\ 
\hline 
$W \gamma$ & $\mu^{+}$$\gamma$$\pmiss$, $e^{+}$$\gamma$$\pmiss$, $\mu^{+}$$\gamma$$j$$\pmiss$, $e^{+}$$\gamma$$j$$\pmiss$, $e^{+}$$\gamma$$b$$\pmiss$ \\ 
\hline 
$\ell^{+} \tau^{-} \gamma \pmiss$ & $e^{+}$$\tau^{-}$$\gamma$$\pmiss$ \\ 
\hline 
$\ell^{+} \tau^{+} \gamma$ & $e^{+}$$\tau^{+}$$\gamma$ \\ 
\hline 
$\ell^{+} \tau^{-} \gamma$ & $e^{+}$$\tau^{-}$$\gamma$, $\mu^{+}$$\tau^{-}$$\gamma$$j$ \\ 
\hline 
$W             $ & $e^{+}$$\pmiss$, $\mu^{+}$$\pmiss$, $e^{+}$$j$$\pmiss$, $\mu^{+}$$j$$\pmiss$, $e^{+}$$b$$\pmiss$, $\mu^{+}$$b$$\pmiss$ \\ 
\hline 
$\ell^{+} \tau^{+} \pmiss$ & $e^{+}$$\tau^{+}$$\pmiss$, $\mu^{+}$$\tau^{+}$$\pmiss$, $e^{+}$$\tau^{+}$$j$$\pmiss$, $\mu^{+}$$\tau^{+}$$j$$\pmiss$ \\ 
\hline 
$\ell^{+} \tau^{-} \pmiss$ & $e^{+}$$\tau^{-}$$\pmiss$, $\mu^{+}$$\tau^{-}$$\pmiss$, $e^{+}$$\tau^{-}$$j$$\pmiss$, $\mu^{+}$$\tau^{-}$$j$$\pmiss$, $\mu^{+}$$\tau^{-}$$b$$\pmiss$ \\ 
\hline 
$\ell^{+} \tau^{+}$ & $e^{+}$$\tau^{+}$, $e^{+}$$\tau^{+}$$j$, $\mu^{+}$$\tau^{+}$, $\mu^{+}$$\tau^{+}$$j$, $e^{+}$$\tau^{+}$$b$ \\ 
\hline 
$\ell^{+} \tau^{-}$ & $e^{+}$$\tau^{-}$, $e^{+}$$\tau^{-}$$j$, $\mu^{+}$$\tau^{-}$, $\mu^{+}$$\tau^{-}$$j$, $e^{+}$$\tau^{-}$$b$ \\ 
\hline 
$\ell^{+}\ell'^{+}4j\pmiss$ & $e^{+}$$\mu^{+}$4$j$$\pmiss$ \\ 
\hline 
$\ell^{+} \ell'^{-} 4j$ & $e^{+}$$\mu^{-}$4$j$ \\ 
\hline 
$\ell^{+} \ell'^{-} \pmiss 4j$ & $e^{+}$$\mu^{-}$4$j$$\pmiss$ \\ 
\hline 
$W \gamma 4j$ & $\mu^{+}$$\gamma$4$j$$\pmiss$, $e^{+}$$\gamma$4$j$$\pmiss$ \\ 
\hline 
$W 4j$ & $e^{+}$4$j$$\pmiss$, $\mu^{+}$4$j$$\pmiss$ \\ 
\hline 
$\ell^{+} 4j \tau^{+}$ & $e^{+}$$\tau^{+}$4$j$ \\ 
\hline 
$\ell^{+} 4j \tau^{-}$ & $e^{+}$$\tau^{-}$4$j$, $\mu^{+}$$\tau^{-}$4$j$ \\ 
\hline 
$W \gamma \gamma$ & $\mu^{+}$2$\gamma$$\pmiss$, $e^{+}$2$\gamma$$\pmiss$, $e^{+}$2$\gamma$$j$$\pmiss$, $\mu^{+}$2$\gamma$$j$$\pmiss$ \\ 
\hline 
$jj$ & 2$j$, 3$j$ \\ 
\hline 
$\gamma jj$ & $\gamma$2$j$, $\gamma$3$j$ \\ 
\hline 
$\gamma \pmiss jj$ & $\gamma$2$j$$\pmiss$, $\gamma$3$j$$\pmiss$ \\ 
\hline 
$\gamma \pmiss  \tau^{+}  jj$ & $\tau^{+}$$\gamma$2$j$$\pmiss$ \\ 
\hline 
$jj \pmiss$ & 3$j$$\pmiss$, 2$j$$\pmiss$ \\ 
\hline 
$\tau \pmiss jj$ & $\tau^{+}$2$j$$\pmiss$, $\tau^{+}$3$j$$\pmiss$ \\ 
\hline 
$\tau^{+} \tau^{-} 2j$ & $\tau^{+}$$\tau^{-}$2$j$, $\tau^{+}$$\tau^{-}$3$j$ \\ 
\hline 
$\gamma \gamma jj$ & 2$\gamma$2$j$, 2$\gamma$3$j$ \\ 
\hline 
$jj \gamma \gamma \pmiss$ & 2$\gamma$2$j$$\pmiss$, 2$\gamma$3$j$$\pmiss$ \\ 
\hline 
$2\tau^+2j$ & 2$\tau^{+}$2$j$, 2$\tau^{+}$3$j$ \\ 
\hline 
$\gamma \gamma \gamma jj$ & 3$\gamma$2$j$ \\ 
\hline 
$\gamma j$ & $\gamma$$j$, $\gamma$$b$ \\ 
\hline 
$\gamma \pmiss$ & $\gamma$$\pmiss$, $\gamma$$j$$\pmiss$, $\gamma$$b$$\pmiss$ \\ 
\hline 
$\tau \pmiss \gamma$ & $\tau^{+}$$\gamma$$j$$\pmiss$, $\tau^{+}$$\gamma$$\pmiss$ \\ 
\hline 
$\tau^+\tau^-\gamma$ & $\tau^{+}$$\tau^{-}$$\gamma$ \\ 
\hline 
$\tau^{+} \tau^{+} \gamma$ & 2$\tau^{+}$$\gamma$ \\ 
\hline 
$j \pmiss$ & $j$$\pmiss$, $b$$\pmiss$ \\ 
\hline 
$\tau \pmiss$ & $\tau^{+}$$j$$\pmiss$, $\tau^{+}$$b$$\pmiss$ \\ 
\hline 
$\tau^+\tau^-\pmiss$ & $\tau^{+}$$\tau^{-}$$j$$\pmiss$, $\tau^{+}$$\tau^{-}$$\pmiss$ \\ 
\hline 
$\tau^{+} \tau^{-}$ & $\tau^{+}$$\tau^{-}$, $\tau^{+}$$\tau^{-}$$j$, $\tau^{+}$$\tau^{-}$$b$ \\ 
\hline 
$b \bar{b} b \bar{b}$ & 3$b$$j$, 3$b$2$j$, 4$b$ \\ 
\hline 
$W b \bar{b} b \bar{b}$ & $e^{+}$3$b$$j$$\pmiss$, $\mu^{+}$3$b$$j$$\pmiss$ \\ 
\hline 
$4b2j$ & 3$b$3$j$, 3$b$4$j$, 4$b$2$j$ \\ 
\hline 
$\gamma b \bar{b} b \bar{b}$ & $\gamma$3$b$$j$ \\ 
\hline 
$b \bar{b} b \bar{b} \pmiss$ & 3$b$$j$$\pmiss$, 3$b$2$j$$\pmiss$ \\ 
\hline 
$4b4j$ & 4$b$4$j$ \\ 
\hline 
$\ell^{+} \ell^{+}$ & 2$e^{+}$, 2$e^{+}$$j$, 2$\mu^{+}$ \\ 
\hline 
$\ell^{+}\ell^{-}\ell^{+} jj \pmiss$ & 2$e^{+}$$e^{-}$2$j$, 2$e^{+}$$e^{-}$3$j$ \\ 
\hline 
$\ell^{+}\ell^{-}\ell^{+} \pmiss$ & 2$e^{+}$$e^{-}$, 2$e^{+}$$e^{-}$$j$, 2$e^{+}$$e^{-}$$\pmiss$ \\ 
\hline 
$\ell^{+}\ell^{+} jj$ & 2$e^{+}$2$j$ \\ 
\hline 
$\ell^{+}\ell^{+} \ell'^{-} \pmiss$ & $e^{+}$2$\mu^{-}$$\pmiss$ \\ 
\hline 
$\ell^{+}\ell^{+} \gamma$ & 2$e^{+}$$\gamma$ \\ 
\hline 
$\ell^{+}\ell^{+} \gamma  \pmiss$ & 2$e^{+}$$\gamma$$\pmiss$ \\ 
\hline 
$\ell^{+}\ell^{+} \pmiss$ & 2$e^{+}$$\pmiss$, 2$\mu^{+}$$\pmiss$, 2$e^{+}$$j$$\pmiss$, 2$\mu^{+}$$j$$\pmiss$ \\ 
\hline 
$\ell^{+}\ell^{+} 4j$ & 2$e^{+}$4$j$ \\ 
\hline 
$4j$ & 4$j$ \\ 
\hline 
$\gamma 4j$ & $\gamma$4$j$ \\ 
\hline 
$\gamma 4j \pmiss$ & $\gamma$4$j$$\pmiss$ \\ 
\hline 
$4j \pmiss$ & 4$j$$\pmiss$ \\ 
\hline 
$\tau^{+} \pmiss 4j$ & $\tau^{+}$4$j$$\pmiss$ \\ 
\hline 
$\tau^+\tau^-4j$ & $\tau^{+}$$\tau^{-}$4$j$ \\ 
\hline 
$\gamma \gamma 4j$ & 2$\gamma$4$j$ \\ 
\hline 
$\gamma \gamma$ & 2$\gamma$, 2$\gamma$$j$, 2$\gamma$$b$ \\ 
\hline 
$\gamma \gamma \pmiss$ & 2$\gamma$$\pmiss$, 2$\gamma$$j$$\pmiss$ \\ 
\hline 
$\tau^{+} \tau^{+}$ & 2$\tau^{+}$, 2$\tau^{+}$$j$ \\ 
\hline 
$3 \gamma$ & 3$\gamma$ \\ 
\hline 
 \hline 
\end{tabular}

\end{minipage}}
\caption[Correspondence between \Sleuth\ and \Vista\ final states.]{Correspondence between \Sleuth\ and \Vista\ final states.  The first column shows the \Sleuth\ final state formed by merging the populated \Vista\ final states in the second column.  Charge conjugates of each \Vista\ final state are implied.}
\label{tbl:sleuthFinalStateContentIndex}
\end{table*}

Table~\ref{tbl:sleuthFinalStateContentIndex} lists the \Vista\ final states associated with each \Sleuth\ final state.


\subsection{Minimum number of events}

This section expands on a subtle point in the definition of the \Sleuth\ algorithm:  for purely practical considerations, only final states in which three or more events are observed in the data are considered.  

Suppose $\scriptP_{e^+e^-b\bar{b}} = 10^{-6}$; then in computing $\twiddleScriptP$ all final states with $b>10^{-6}$ must be considered and accounted for.  (A final state with $b=10^{-7}$, on the other hand, counts as only $\approx 0.1$ final states, since the fraction of hypothetical similar experiments in which $\scriptP<10^{-6}$ in this final state is equal to the fraction of hypothetical similar experiments in which one or more events is seen in this final state, which is $10^{-7}$.)  This is a large practical problem, since it requires that all final states with $b>10^{-6}$ be enumerated and estimated, and it is difficult to do this believably.

To solve this problem, let \Sleuth\ consider only final states with at least ${d_{\text{min}}}$ events observed in the data.  The goal is to be able to find $\twiddleScriptP < 10^{-3}$.  There will be some number $N_{\text{fs}}(b_{\text{min}})$ of final states with expected number of events $b>b_{\text{min}}$, writing $N_\text{fs}$ explicitly as a function of $b_{\text{min}}$; thus $b_{\text{min}}$ must be chosen to be sufficiently large that all of these $N_{\text{fs}}(b_{\text{min}})$ final states can be enumerated and estimated.  The time cost of simulating events is such that the integrated luminosity of Monte Carlo events is at most 100 times the integrated luminosity of the data; this practical constraint restricts $b_{\text{min}}>0.01$.  The number of \Sleuth\ Tevatron Run II final states with $b>0.01$ is $N_{\text{fs}}(b_{\text{min}}=0.01) \approx 10^3$.  

For small $\scriptP_{\text{min}}$, keeping the first term in a binomial expansion yields $\twiddleScriptP = \scriptP_{\text{min}} N_{\text{fs}}(b_{\text{min}})$, where $\scriptP_{\text{min}}$ is the smallest $\scriptP$ found in any final state.  From the discussion above, the computation of $\twiddleScriptP$ from $\scriptP_{\text{min}}$ can only be justified if $\scriptP_{\text{min}} > ({b_{\text{min}}}^{d_{\text{min}}})$; if otherwise, final states with $b<b_{\text{min}}$ will need to be accounted for.  Thus $\twiddleScriptP$ can be confidently computed only if $\twiddleScriptP > ({b_{\text{min}}}^{d_{\text{min}}}) N_{\text{fs}}(b_{\text{min}})$.

Solving this inequality for $d_{\text{min}}$ and inserting values from above,
\begin{equation}
d_{\text{min}}  \geq  \frac{\log_{10}{\twiddleScriptP} - \log_{10}{N_{\text{fs}}(b_{\text{min}})}}{\log_{10}{b_{\text{min}}}} \approx  \frac{-3 - 3}{-2} = 3.
\end{equation}
A believable trials factor can be computed if $d_{\text{min}} \geq 3$.


At the other end of the scale, computational strength limits the maximum number of events \Sleuth\ is able to consider to $\lesssim 10^4$.  Excesses in which the number of events exceed $10^4$ are expected to be identified by \Vista's normalization statistic.


\subsection{\pvalmin, population and \scriptP}

\Sleuth\ estimates \scriptP\ for a given final state by producing pseudo-data, i.e.\ \sumPt\ values that are distributed according to the Standard Model prediction.  It then scans all \sumPt\ tails, finds the smallest \pval\ and compares it to the \pvalmin\ from the actual data.  That is repeated with many different distributions of pseudo-data, until the fraction of more interesting pseudo-data distributions (which is \scriptP) is determined with 5\% relative uncertainty.

In each pseudo-data distribution that is produced, the population of pseudo-data is randomly distributed according to a Poisson distribution, whose mean is the Standard Model predicted total population ($B$) for the final state.

Each examined \sumPt\ tail has a \pval\ that is not taking into account the statistical uncertainty in the background ($b$) contained in the tail.  The same is true for both data and pseudo-data, therefore the effect in the final \scriptP\ is negligible.

\begin{figure}
\includegraphics[width=4.0in,angle=-90]{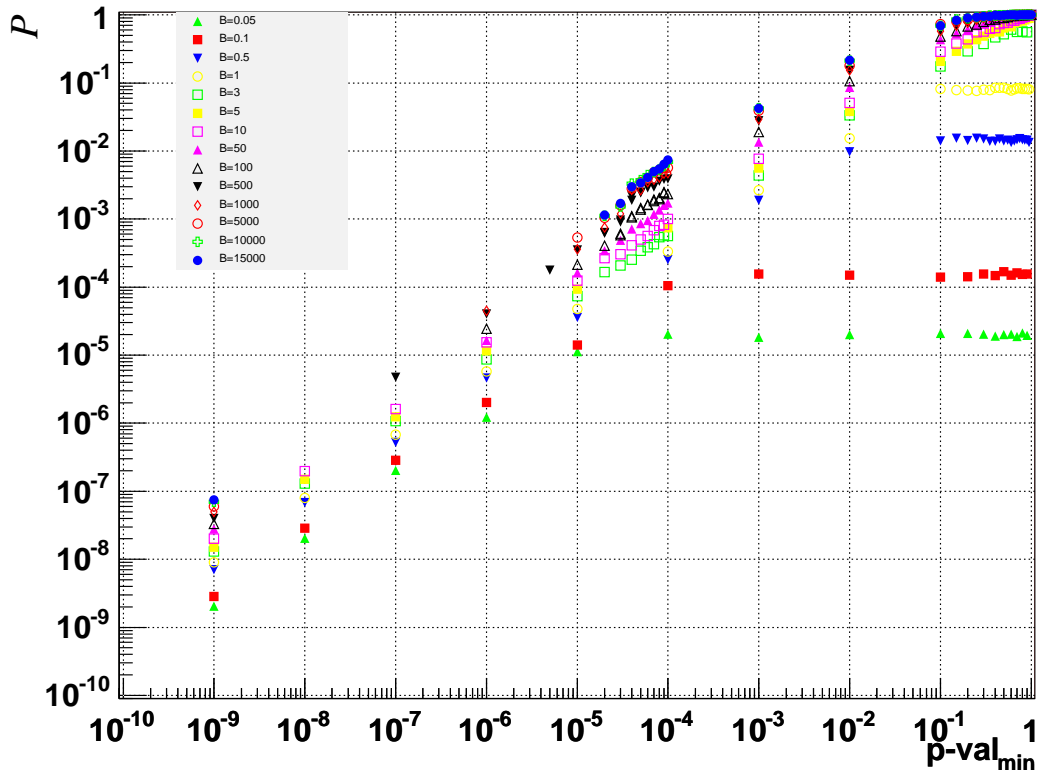}
\caption[\scriptP\ as a function of \pvalmin, for final states of different expected populations]{\scriptP\ as a function of \pvalmin, for final states of different expected populations $B$.  \scriptP\ reaches a plateau at $\pval_{\max}=\sum_{i=3}^{\infty}\frac{B^i}{i!}e^{-B}$, which is visible for small $B$, and reflects the requirement to have at least 3 data events in a \sumPt\ tail to consider it.  \scriptP\ values have been estimated to 5\% relative uncertainty.}
\label{fig:scriptPvsPvalmin}
\end{figure}

Regardless of the particular shape of an expected \sumPt\ distribution, \pvalmin\ in pseudo-data follows the same distribution.  Therefore, \scriptP\ depends only on the \pvalmin\ observed in data, and on the overall expected population; the larger the population, the bigger the average number of considered \sumPt\ tails in pseudo-data, therefore the larger the \scriptP.
The dependence of \scriptP\ on \pvalmin\ and $B$ is shown in Fig.~\ref{fig:scriptPvsPvalmin}.  The advantage of having tabulated this dependence, is that then one does not have to produce pseudo-data repeatedly to estimate \scriptP; he can simply read it from Fig.~\ref{fig:scriptPvsPvalmin}, for a given $B$ and \pvalmin.  This technique makes the execution of \Sleuth\ incredibly fast, allowing for studies such as sensitivity tests, projections to different luminosity, propagation of systematic uncertainties to \tildeScriptP, and frequent assessment of the \sumPt\ excesses in data.  

%

\chapter{Correction Model Details, reflecting the 2 fb$^{-1}$ analysis}
\label{sec:CorrectionModelDetails2}

\section{Details on Event Selection}
\label{sec:trigger2}
Although specific online triggers are not explicitly required, it is still possible to identify the primary online triggers which feed this analysis. These are:
\begin{itemize}
\item {\tt electron\_central\_18} 
\item {\tt muon\_central\_18}
\item {\tt photon\_25\_iso}
\item {\tt jet20} 
\item {\tt jet100}
\item susy dilepton triggers: 
 {\tt electron\_central\_8\_\&\_track8}
 {\tt cem4\_cmup4}
 {\tt cem4\_cmx4}
 {\tt cem4\_pem8}
 {\tt cmup4\_pem8}
 {\tt cmx4\_pem8}
 {\tt dielectron\_central\_4}
 {\tt dimuon\_cmup4\_cmx4}
 {\tt dimuon\_cmupcmup4}

\item susy dilepton triggers {\tt muon\_cmup8\_\&\_track8} and {\tt muon\_cmx8\_\&\_track8} (introduced in run number 200274, roughly 600~pb$^{-1}$ into Run II)
\item hadronic ditau trigger (introduced roughly 300~pb$^{-1}$ into Run II)
\end{itemize}

The following datasets were used:
\begin{itemize}
\item HighPt Central Electron stream: bhel0d, bhel0h, bhel0i, bhel0j
\item HighPt CMUP and CMX muon stream: bhmu0d, bhmu0h, bhmu0i, bhmu0j
\item HighPt Photon stream: cph10d, cph10h, cph10i, cph10j
\item SUSY dilepton stream: edil0d, edil0h, edil0i, edil0j
\item Ditau stream: etau0d, etau0h, etau0i, etau0j
\item Jet20 stream: gjt10d, gjt10h, gjt10i, gjt10j
\item Jet100 stream: gjt40d, gjt40h, gjt40i, gjt40j
\end{itemize}

\section{Details on Particle Identification}
\label{sec:particle_id_detail}

This section contains tables of information related to particle identification.  Electron identification is described in Tables~\ref{tbl:central_electron_id} and~\ref{tbl:plug_electron_id}; muon identification in Tables~\ref{tbl:common_muon_id},~\ref{tbl:cmup_muon_id},~\ref{tbl:cmx_muon_id}, and~\ref{tbl:bmu_muon_id}; tau identification in Table~\ref{tbl:tau_id}; and photon identification in Tables~\ref{tbl:central_photon_id} and~\ref{tbl:plug_photon_id}.  Standard fiducial criteria apply.  Standard CDF SecVtx algorithm is used to identify $b$-jets.  

Jets are identified using the JetClu~\cite{JetClu:Abe:1991ui} clustering algorithm with cone size $\Delta R = 0.4$, unless the event contains one or more jets with $p_T>200$~GeV and no leptons or photons, in which case cones of $\Delta R = 0.7$ are used.~\cdfSpecific{\footnote{Jet energies are corrected to level 7, using {\tt jetCorr04b}.}}
Jets with $p_T>150$~GeV are required to have at least 5~GeV of track $p_T$ within the cone.




\begin{table}
\centering
\begin{tabular}{l|l}
Variable & Cut \\
\hline
Region & Fiducial CES \\
Track $Z_0$ & $\le 60$~cm \\
COT Ax. Seg. & $\ge 3$ \\
COT St. Seg. & $\ge 2$ \\ 
Signed CES $\Delta X$ & $-3.0 < q \Delta X < 1.5$\\
CES $\Delta Z$ & $< 3.0$~cm \\
Track $p_T$ & $>10$~GeV/c \\
$p_T/E_T$ (if $p_T < 50$) & $>0.5$ \\
Had/Em & $< 0.055 + 0.00045 \times E$\\
Isolation & $< 0.1 \times E_T$\\
LShrTrk & $< 0.2$\\
CES StripChi2 & $< 10.0$\\
Conversion & FALSE \\
\end{tabular}
\caption[Central electron identification criteria]{Central electron identification criteria used in \Vista\ and \Sleuth.
These correspond to TightCEM electrons as defined in \cite{CDF7950}, for Gen5 and Gen6.  The conversion finder flags a second track with $\abs{\Delta XY} < 0.2$~cm, $\abs{\Delta\text{cot}\theta}<0.015$, and $p_T>0$~GeV.}
\label{tbl:central_electron_id}
\end{table}

\begin{table}
\centering
\begin{tabular}{l|l}
Variable & Cut \\
\hline
Region & $|\eta| < 2.6$\\
Had/Em & $<0.05$ \\
Isolation & $< 0.1 \times E_T$ \\
PEM Chi2 & $< 10$ \\
PES U & $>0.65$  \\
PES V & $>0.65$  \\
PHX Track & TRUE \\
N SVX hits & $\ge 3$ \\
deltaR(PHX Track,EM cluster) & $< 0.01$  \\
\end{tabular}
\caption[Plug electron identification criteria]{Plug electron identification criteria used in \Vista\ and \Sleuth.
These correspond to Tight Phoenix electrons as defined in \cite{CDF7950}, except for the cut on $\Delta R$.}
\label{tbl:plug_electron_id}
\end{table}

\begin{table}
\centering
\begin{tabular}{l|l}
Variable & Cut \\
\hline
Larry curvature correction & Applied \\
Track $Z_0$ & $\le 60$~cm \\
COT Ax. Seg. & $\ge 3$ \\
COT St. Seg. & $\ge 3$ \\ 
Iso/$p_T$ & $< 0.1$ \\
EM + Had Energy & $>0.1$~GeV\\
EM Energy & $< 2.0 + 0.0115(p-100)\times(p>100)$\\
Had Energy & $< 6.0 + 0.0280(p-100)\times(p>100)$\\
COT $\chi^2$ (gen5) & $<3$ for $p_T<60$; $<2$ for $p_T\ge60$ \\
COT $\chi^2$ (gen6) & $<2$ \\
Track With Si hits & $|d_0| < 0.02$\\
Track Without Si hits & $|d_0| < 0.2$\\
\end{tabular}
\caption[Common muon identification criteria]{Common muon identification criteria used in \Vista\ and \Sleuth.}
\label{tbl:common_muon_id}
\end{table}

\begin{table}
\centering
\begin{tabular}{l|l}
Variable & Cut \\
\hline
CMU $\Delta X$ & $<7$~cm\\
CMP $\Delta X$ & $<5$~cm\\
CMUP Fiducial & TRUE \\
No bluebeam muons & For Runs $< 154449$\\
\end{tabular}
\caption[CMUP muon identification criteria]{CMUP muon identification criteria used in \Vista\ and \Sleuth. These are in addition to the criteria common to all muons.}
\label{tbl:cmup_muon_id}
\end{table}

\begin{table}
\centering
\begin{tabular}{l|l}
Variable & Cut \\
\hline
CMX $\Delta X$ & $<6$~cm\\
COT exit radius & $>140$~cm \\
CMX Fiducial & TRUE \\
Run & $> 150144$ \\
Keystone and Miniskirt good & $run \ge 190697$\\
Exclude wedge 14, west & For $190697 \le run \le 209760$ \\
\end{tabular}
\caption[CMX Muon identification criteria]{CMX Muon identification criteria used in \Vista\ and \Sleuth. These are in addition to the criteria common to all muons.}
\label{tbl:cmx_muon_id}
\end{table}

\begin{table}
\centering
\begin{tabular}{l|l}
Variable & Cut \\
\hline
BMU $\Delta X$ & $<10$~cm\\
BMU Fiducial & TRUE \\
\end{tabular}
\caption[BMU Muon identification criteria ]{BMU Muon identification criteria used in \Vista\ and \Sleuth. These are in addition to the criteria common to all muons.}
\label{tbl:bmu_muon_id}
\end{table}

\begin{table}
\centering
\begin{tabular}{l|l}
Variable & Cut \\
\hline
Seed Track $p_T$ & $> 10.5$ \\
Track $|z_0|$ & $<60$~cm \\
Fiducial ShowerMax & $9.0 < |Z_\text{CES}| < 230.0$ \\
$\pi^0$ isolation (gen5) & No $\pi^0$ ($p_T>0.5$)\\
                         & in annulus 10-30$^o$\\
$\pi^0$ isolation (gen6) & SumPt of $\pi^0$ with $p_T>0.5$ in \\
                         & annulus 0.15-0.4 rad $< 0.6$\\

Track isolation (gen5) & No track ($p_T>1$) in \\
                       & annulus 10-30$^o$\\
Track isolation (gen6) & SumPt of tracks in annulus \\
                       & 0.2-0.4 rad $< 1.0$ \\
Calorimeter Isolation & Iso/$E_T < 0.1$\\
Calorimeter $E_T$ & Cal $E_T <$ $\text{VisPt} + 1.5\sqrt{\text{VisPt}}$ \\
Visible Mass (tracks+$\pi^0$s) & $<1.8$~GeV \\ 
Track $|d_0|$ & $< 0.2$~cm \\
Seed Track Vertex Consistency & Abs(seedTrack z - \\
Consistency & Primary Vertex z) $< 5$~cm \\
One Prong Tau & N tracks in 10$^o$ cone =1 \\
Electron removal & $\xi > 0.1$ and \\
                 & EMfraction $<0.925$\\
Not a Muon & No matching muon stubs and \\
           & Cal $E_T / $ Seed Track $p_T > 0.5$ \\
\end{tabular}
\caption[$\tau$ identification criteria]{Table of $\tau$ identification criteria used in \Vista\ and \Sleuth.}
\label{tbl:tau_id}
\end{table}

\begin{table*}
\centering
\begin{tabular}{l|l}
Variable & Cut \\
\hline
Fiducial Region (X) & CES $|X|< 21$~cm \\
Fiducial Region (Z) & $9 <$ CES Z $< 230$~cm\\
Had/Em & $<0.125$ $||$ $<0.055+0.00045 \times E$\\
Isolation ($E_T \le 20$) & $<0.1 \times E_T$\\
Isolation ($E_T > 20$) & $<2.0 + 0.02 \times (E_T -20)$\\
Track isolation, cone 0.4 & SumPt $< 2 + 0.005\times E_T$\\
Ntrack (N3D) & N3D $\le$ 1 \\ 
Track $p_T$ (if N3D=1) & $< 1.0 + 0.005 \times E_T$ \\
Chi2 (Strips+Wires)/2.0 & $< 20$ \\
2nd CES clus. $E \sin{\theta}$ ($E_T \le 18$)& strip+wire $<0.14 \times E_T$\\ 
2nd CES clus. $E \sin{\theta}$ ($E_T > 18$)& strip+wire $<2.52 + 0.01 \times E_T$\\ 
\end{tabular}
\caption[Central photon identification criteria]{Central photon identification criteria used in \Vista\ and \Sleuth.
Here $E_T$ refers to corrected photon $E_T$. The ``2nd CES Cluster'' cut is tighter than the standard photon cut.}
\label{tbl:central_photon_id}
\end{table*}

\begin{table*}
\centering
\begin{tabular}{l|l}
Variable & Cut \\
\hline
Region & $1.2 < |\eta| < 2.6$ \\
Had/Em ($E_T \le 100$) & $<0.05$ \\
Had/Em ($E_T > 100$) & $<0.05 + 0.026 \times Log(E_T / 100)$ \\
Isolation ($E_T \le 20$) & $<0.1 \times E_T$ \\
Isolation ($E_T > 20$) & $<2.0 + 0.02 \times E_T$ \\
Track Isolation (in a cone of $dR < 0.4$) & $<2.0 + 0.005 \times E_T$ \\
PEM Chi2 & $< 10$ \\
PES U & $>0.65$ \\
PES V & $>0.65$  \\
\end{tabular}
\caption[Plug photon identification criteria]{Plug photon identification criteria used in \Vista\ and \Sleuth.  Here $E_T$ refers to corrected photon $E_T$. These are the standard Joint Physics cuts.}
\label{tbl:plug_photon_id}
\end{table*}

\section{\Vista: Single Particle Gun Results}
\label{sec:MisidentificationMatrix2}

Tables~\ref{tbl:misId_cdfSim_central2} and \ref{tbl:misId_cdfSim_plug2} show the response of the CDF detector simulation, reconstruction, and particle identification algorithms to single particles in the central and plug regions respectively, with all changes to particle identification criteria discussed in section \ref{sec:particleID2}.  We use a single particle gun to shoot $10^5$ particles of each type, with $p_T=25$~GeV, uniformly distributed in $\theta$ and $\phi$.  The types of generated particles label the rows, while the resulting reconstructed objects label the columns of each table.  Table~\ref{tbl:misId_cdfSim_50GeV_central2} shows a similar study with $10^4$ particles at $p_T=50$~GeV.  These results are not directly used in the analysis, but provide a sensible cross-check for the used fake rates and identification efficiencies.

It should be noted that the number of photons reconstructed as electrons decreased compared to the last round of this analysis.  As expected, the number of electrons which were identified with the wrong charge has decreased proportionately, as well as the number of $\pi^0$ reconstructed as electrons.  All these are results of making the conversion filter tighter, by removing the lower $p_T$ threshold that was previously required when looking for sibling tracks coming from conversion.

\begin{figure}
\vspace{-1cm}
\centering
\includegraphics[width=3.0in]{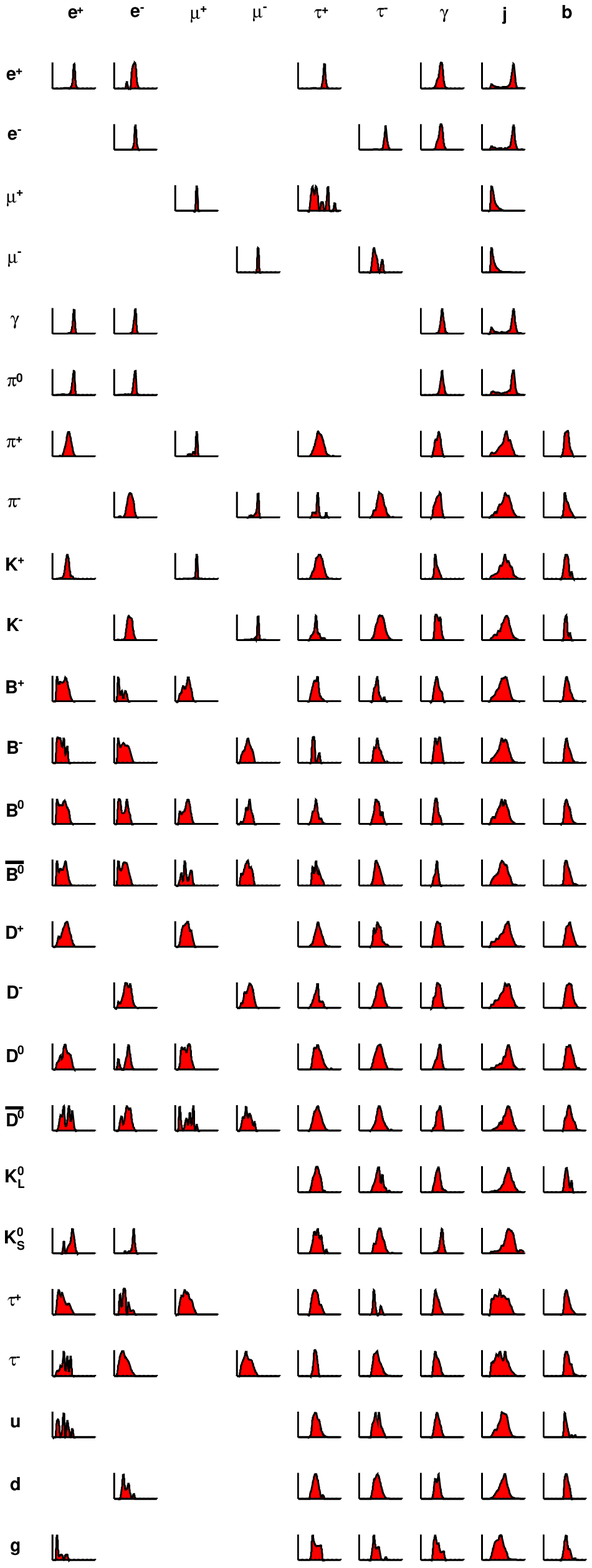}
\label{fig:singleParticleFakePt_central2}
\caption[Transverse momentum distribution of reconstructed objects from single particles shot into the central CDF detector]{Transverse momentum distribution of reconstructed objects (labelling columns) arising from single particles (labelling rows) with $p_T=25$~GeV shot from a single particle gun into the central CDF detector.  The area under each histogram is equal to the number of events in the corresponding misidentification matrix element of Table~\ref{tbl:misId_cdfSim_central2}; histograms with fewer than ten events are not shown.  The horizontal axis ranges from 0 to 50~GeV, with one tick mark each 5~GeV.  The incident single particle distribution is a delta function at the center of each plot, at $p_T=25$~GeV.\label{tbl:misId_cdfSim_central_pt2}}
\end{figure}

\begin{figure}
\vspace{-1cm}
\centering
\includegraphics[width=3.0in]{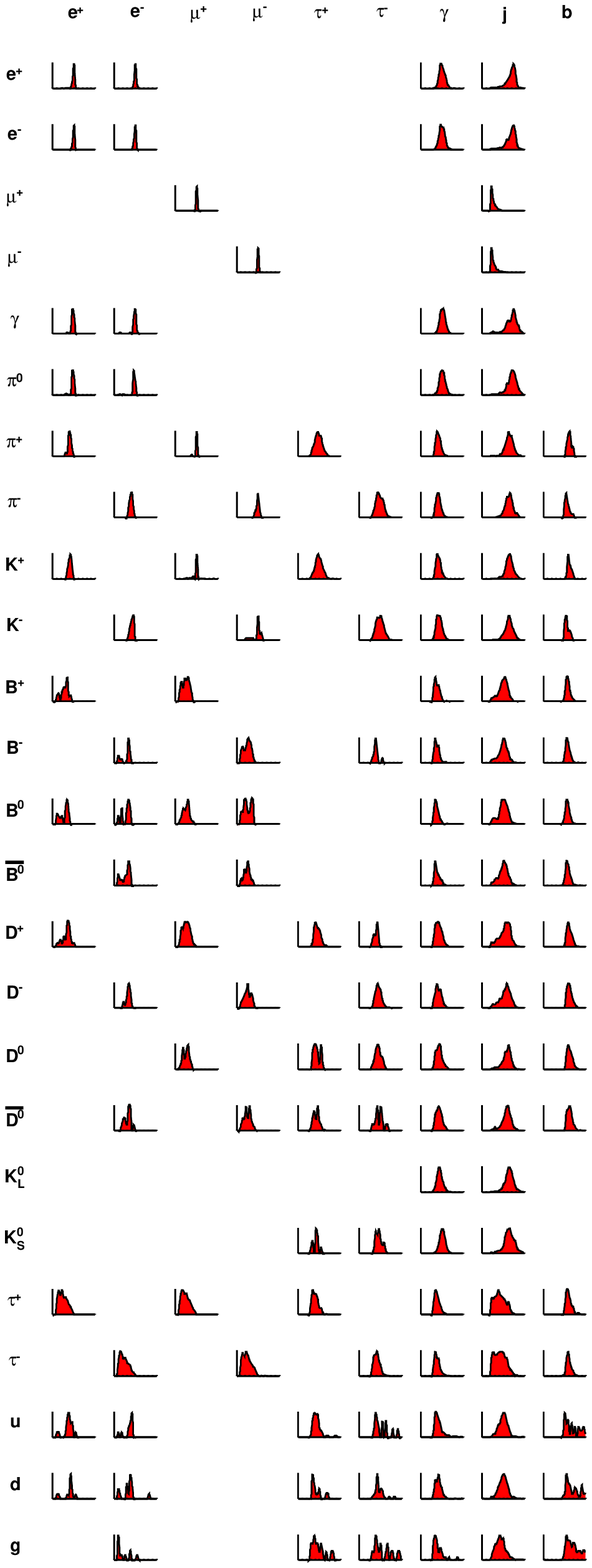}
\label{fig:singleParticleFakePt_plug2}
\caption[Transverse momentum distribution of reconstructed objects from single particles shot into the plug CDF detector]{Transverse momentum distribution of reconstructed objects (labelling columns) arising from single particles (labelling rows) with $p_T=25$~GeV shot from a single particle gun into the plug CDF detector.  The area under each histogram is equal to the number of events in the corresponding misidentification matrix element of Table~\ref{tbl:misId_cdfSim_plug2}; histograms with fewer than ten events are not shown.  The horizontal axis ranges from 0 to 50~GeV, with one tick mark each 5~GeV.  The incident single particle distribution is a delta function at the center of each plot, at $p_T=25$~GeV.\label{tbl:misId_cdfSim_plug_pt2}}
\end{figure}

Figures \ref{fig:singleParticleFakePt_central2} and \ref{fig:singleParticleFakePt_plug2} show the $p_T$ distributions of the reconstructed object (column label), resulting from the initial particle (row label), for the central and plug region of the detector respectively.  We note that the $p_T$ resolution of reconstructed $\tau$s has worsened, consistently with obtaining $p_T$ from calorimeter $E_T$ rather than visible momentum.

\begin{table}
\centering
\hspace*{-0.3in}\mbox{
\begin{tabular}{c|rrrrrrrrr}
 & $e^+$ & $e^-$ & $\mu^+$ & $\mu^-$ & $\tau^+$ & $\tau^-$ & $\gamma$ & $j$ & $b$ \\ \hline 
$e^+$  & 60940 & 13 & 0 & 0 & 265 & 0 & 2009 & 33140 & 0  \\
$e^-$  & 8 & 61124 & 0 & 0 & 0 & 263 & 1988 & 33021 & 1  \\
$\mu^+$  & 0 & 0 & 53142 & 0 & 19 & 0 & 0 & 2079 & 0  \\
$\mu^-$  & 0 & 1 & 0 & 53202 & 0 & 17 & 0 & 2067 & 0  \\
$\gamma$  & 592 & 571 & 0 & 0 & 1 & 2 & 67197 & 27110 & 0  \\
$\pi^0$  & 502 & 499 & 0 & 0 & 2 & 5 & 57655 & 38244 & 0  \\
$\pi^+$  & 306 & 0 & 122 & 0 & 68254 & 5 & 86 & 29195 & 37  \\
$\pi^-$  & 3 & 388 & 0 & 94 & 10 & 67327 & 133 & 30100 & 43  \\
$K^+$  & 183 & 2 & 289 & 0 & 69690 & 8 & 25 & 26930 & 24  \\
$K^-$  & 0 & 282 & 0 & 178 & 32 & 68036 & 98 & 28846 & 23  \\
$B^+$  & 326 & 18 & 231 & 9 & 132 & 17 & 40 & 71606 & 25893  \\
$B^-$  & 17 & 312 & 6 & 232 & 13 & 117 & 42 & 71921 & 25611  \\
$B^0$  & 343 & 97 & 267 & 43 & 103 & 63 & 22 & 72021 & 24990  \\
$\bar{B^0}$  & 83 & 351 & 34 & 262 & 58 & 97 & 20 & 72031 & 25094  \\
$D^+$  & 269 & 4 & 198 & 2 & 2221 & 56 & 209 & 83659 & 11606  \\
$D^-$  & 4 & 275 & 5 & 171 & 80 & 2292 & 229 & 83673 & 11595  \\
$D^0$  & 119 & 20 & 93 & 5 & 310 & 1070 & 250 & 91599 & 5473  \\
$\bar{D^0}$  & 22 & 95 & 15 & 99 & 1170 & 270 & 225 & 91589 & 5466  \\
$K^0_L$  & 0 & 2 & 0 & 0 & 68 & 67 & 194 & 97538 & 27  \\
$K^0_S$  & 17 & 18 & 2 & 1 & 83 & 445 & 9647 & 78364 & 0  \\
$\tau^+$  & 6750 & 20 & 4919 & 0 & 6336 & 10 & 652 & 55677 & 613  \\
$\tau^-$  & 17 & 6623 & 0 & 4907 & 9 & 6064 & 615 & 56201 & 580  \\
$u$  & 12 & 8 & 2 & 0 & 658 & 29 & 247 & 98645 & 24  \\
$d$  & 4 & 16 & 1 & 1 & 55 & 428 & 181 & 98916 & 21  \\
$g$  & 10 & 8 & 0 & 4 & 29 & 31 & 12 & 98190 & 99  \\
\end{tabular}}
\caption[Central single particle misidentification matrix.]{Central single particle misidentification matrix.  Using a single particle gun, $10^5$ particles of each type shown at the left of the table are shot with $p_T=25$~GeV into the central CDF detector, uniformly distributed in $\theta$ and in $\phi$.  The resulting reconstructed object types are shown at the top of the table, labeling the table columns.  Thus the rightmost element of this matrix in the fourth row from the bottom shows $\poo{\tau^-}{j}$, the number of negatively charged tau leptons (out of $10^5$) reconstructed as a jet.\label{tbl:misId_cdfSim_central2}}
\end{table}

\begin{table}
\centering
\begin{tabular}{c|rrrrrrrrr}
 & $e^+$ & $e^-$ & $\mu^+$ & $\mu^-$ & $\tau^+$ & $\tau^-$ & $\gamma$ & $j$ & $b$ \\ \hline 
$e^+$  & 3737 & 26 & 0 & 0 & 1 & 0 & 71307 & 24597 & 0  \\
$e^-$  & 24 & 3834 & 0 & 0 & 0 & 4 & 71003 & 24789 & 0  \\
$\mu^+$  & 0 & 0 & 10661 & 0 & 3 & 0 & 1 & 1061 & 0  \\
$\mu^-$  & 0 & 0 & 0 & 10678 & 0 & 2 & 3 & 1127 & 0  \\
$\gamma$  & 55 & 65 & 0 & 0 & 0 & 0 & 76374 & 23064 & 0  \\
$\pi^0$  & 46 & 53 & 0 & 0 & 0 & 1 & 74111 & 25522 & 0  \\
$\pi^+$  & 17 & 0 & 16 & 0 & 4395 & 2 & 554 & 93462 & 25  \\
$\pi^-$  & 1 & 24 & 0 & 10 & 3 & 4206 & 673 & 93570 & 20  \\
$K^+$  & 13 & 0 & 59 & 0 & 4658 & 0 & 421 & 92807 & 12  \\
$K^-$  & 0 & 36 & 0 & 38 & 1 & 4301 & 834 & 92958 & 23  \\
$B^+$  & 50 & 2 & 102 & 3 & 4 & 0 & 186 & 90077 & 7389  \\
$B^-$  & 3 & 18 & 2 & 81 & 0 & 13 & 160 & 90178 & 7347  \\
$B^0$  & 52 & 12 & 96 & 15 & 3 & 8 & 148 & 90241 & 7016  \\
$\bar{B^0}$  & 10 & 52 & 8 & 107 & 4 & 5 & 126 & 90149 & 7095  \\
$D^+$  & 32 & 4 & 90 & 0 & 136 & 11 & 738 & 94326 & 2148  \\
$D^-$  & 2 & 22 & 1 & 57 & 6 & 127 & 817 & 94367 & 2100  \\
$D^0$  & 9 & 7 & 38 & 2 & 20 & 74 & 671 & 96983 & 1027  \\
$\bar{D^0}$  & 2 & 15 & 3 & 37 & 74 & 17 & 628 & 96928 & 1121  \\
$K^0_L$  & 1 & 0 & 0 & 0 & 6 & 8 & 1089 & 97411 & 6  \\
$K^0_S$  & 2 & 3 & 0 & 0 & 11 & 39 & 9532 & 56689 & 0  \\
$\tau^+$  & 339 & 8 & 1249 & 1 & 341 & 2 & 3198 & 66243 & 104  \\
$\tau^-$  & 5 & 346 & 0 & 1226 & 0 & 336 & 3208 & 66111 & 108  \\
$u$  & 19 & 12 & 2 & 0 & 73 & 13 & 423 & 99359 & 47  \\
$d$  & 13 & 17 & 0 & 0 & 15 & 36 & 359 & 99357 & 60  \\
$g$  & 7 & 11 & 8 & 1 & 19 & 18 & 41 & 98937 & 426  \\
\end{tabular}
\caption[Plug single particle misidentification matrix.]{Plug single particle misidentification matrix.  Using a single particle gun, $10^5$ particles of each type shown at the left of the table are shot with $p_T=25$~GeV into the plug CDF detector, uniformly distributed in $\theta$ and in $\phi$.  The resulting reconstructed object types are shown at the top of the table, labeling the table columns.  Thus the rightmost element of this matrix in the fourth row from the bottom shows $\poo{\tau^-}{j}$, the number of negatively charged tau leptons (out of $10^5$) reconstructed as a jet.\label{tbl:misId_cdfSim_plug2}}
\end{table}

\begin{table}
\centering
\hspace*{-0.3in}\mbox{
\begin{tabular}{c|rrrrrrrrr}
 & $e^+$ & $e^-$ & $\mu^+$ & $\mu^-$ & $\tau^+$ & $\tau^-$ & $\gamma$ & $j$ & $b$ \\ \hline 
$e^+$  & 6060 & 2 & 0 & 0 & 38 & 0 & 139 & 3576 & 1  \\
$e^-$  & 0 & 6103 & 0 & 0 & 0 & 28 & 128 & 3574 & 0  \\
$\mu^+$  & 1 & 0 & 5217 & 0 & 2 & 0 & 0 & 289 & 0  \\
$\mu^-$  & 0 & 1 & 0 & 5201 & 0 & 2 & 1 & 301 & 0  \\
$\gamma$  & 55 & 75 & 0 & 0 & 0 & 0 & 6554 & 3118 & 0  \\
$\pi^0$  & 42 & 38 & 0 & 0 & 0 & 0 & 5751 & 3991 & 0  \\
$\pi^+$  & 19 & 0 & 9 & 0 & 7721 & 0 & 2 & 2089 & 19  \\
$\pi^-$  & 0 & 15 & 0 & 4 & 3 & 7761 & 2 & 2100 & 9  \\
$K^+$  & 10 & 0 & 20 & 0 & 7662 & 2 & 2 & 2109 & 5  \\
$K^-$  & 0 & 25 & 0 & 11 & 5 & 7682 & 3 & 2119 & 10  \\
$B^+$  & 18 & 2 & 6 & 0 & 13 & 1 & 0 & 5160 & 4792  \\
$B^-$  & 1 & 9 & 0 & 6 & 1 & 10 & 1 & 5186 & 4776  \\
$B^0$  & 13 & 3 & 5 & 0 & 13 & 9 & 0 & 5111 & 4836  \\
$\bar{B^0}$  & 0 & 11 & 0 & 3 & 7 & 7 & 0 & 5095 & 4868  \\
$D^+$  & 41 & 1 & 20 & 0 & 726 & 10 & 14 & 7861 & 1241  \\
$D^-$  & 2 & 52 & 1 & 11 & 10 & 696 & 10 & 7898 & 1247  \\
$D^0$  & 9 & 4 & 7 & 2 & 31 & 93 & 25 & 9036 & 767  \\
$\bar{D^0}$  & 3 & 11 & 0 & 9 & 87 & 39 & 24 & 9081 & 693  \\
$K^0_L$  & 0 & 0 & 0 & 0 & 7 & 7 & 13 & 9849 & 16  \\
$K^0_S$  & 0 & 1 & 0 & 0 & 18 & 69 & 810 & 5855 & 1  \\
$\tau^+$  & 889 & 2 & 711 & 0 & 1506 & 1 & 63 & 5008 & 191  \\
$\tau^-$  & 3 & 816 & 0 & 672 & 1 & 1496 & 107 & 5088 & 188  \\
$u$  & 1 & 1 & 0 & 0 & 46 & 2 & 19 & 9923 & 31  \\
$d$  & 1 & 0 & 0 & 0 & 3 & 35 & 7 & 9944 & 24  \\
$g$  & 1 & 1 & 0 & 2 & 3 & 3 & 0 & 9897 & 174  \\
\end{tabular}}
\caption[Central single particle misidentification matrix.]{Central single particle misidentification matrix.  Using a single particle gun, $10^4$ particles of each type shown at the left of the table are shot with $p_T=50$~GeV into the central CDF detector, uniformly distributed in $\theta$ and in $\phi$.  The resulting reconstructed object types are shown at the top of the table, labeling the table columns.  \label{tbl:misId_cdfSim_50GeV_central2}}
\end{table}

\clearpage

\section{Fake Rates}
\label{sec:fake_rates}
It would take too many Monte Carlo events to acquire enough statistics of rare fake processes.  To overcome this difficulty, we apply our own multiplicative fake rates on reconstructed objects, when they are reconstructed more often than the objects thay may fake.  Specifically, we apply fake rates for jets or b-tagged jets faking electrons, muons, photons, $\tau$s, jets faking b-tagged jets, and photons faking electrons.  
Note that other fake processes are not neglected -- they are handled by CDFSim. 
In the interest of simplicity, we try to keep our fake rates as simple as possible. There is generally one overall coefficient for the fake rate, and this value is usually obtained from the \Vista\ fit to the data.
In some cases however, to better model the true fake process, we need to introduce additional modulations as a function of $p_T$ or location within the detector ($\eta$ or $\phi$).
This section details all the special modulations applied for \Vista\ fake rates. Generally, we show a modulating function, which multiplies the appropriate correction factor value to obtain the true fake rate applied.
If not shown here, the fake rate is treated as being constant.

\begin{figure}
\centering
\includegraphics[angle=-90,width=0.5\columnwidth]{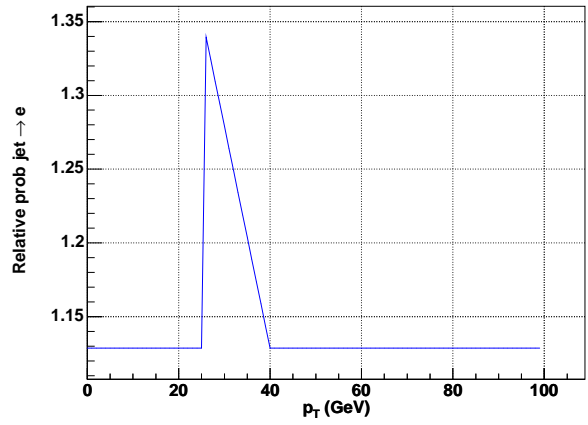}
\caption{The relative fake rate for jets to fake electrons in the plug as a function of the $p_T$ of the jet}
\label{fig:pj2e_relative}
\end{figure}

\begin{figure}
\centering
\includegraphics[angle=-90,width=0.5\columnwidth]{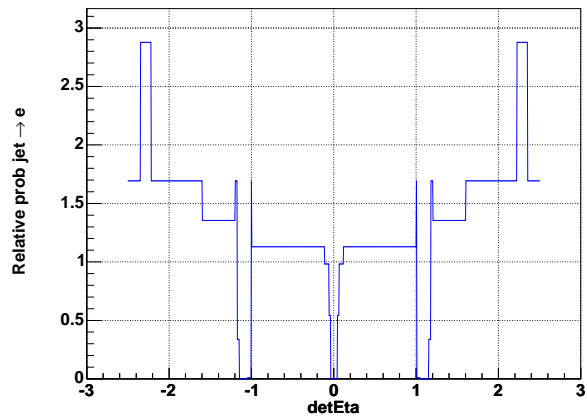}
\caption{The relative fake rate for jets to fake electrons as a function of detEta.}
\label{fig:pj2e_relative_vs_eta}
\end{figure}

\begin{figure}
\centering
\includegraphics[angle=-90,width=0.5\columnwidth]{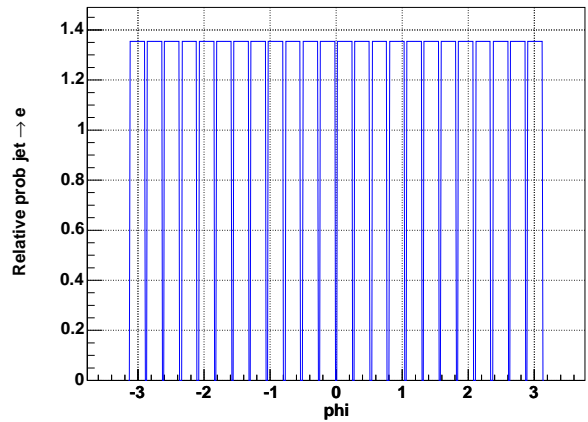}
\caption{The relative fake rate for jets to fake electrons as a function of phi.}
\label{fig:pj2e_relative_vs_phi}
\end{figure}

\begin{figure}
\centering
\includegraphics[angle=-90,width=0.5\columnwidth]{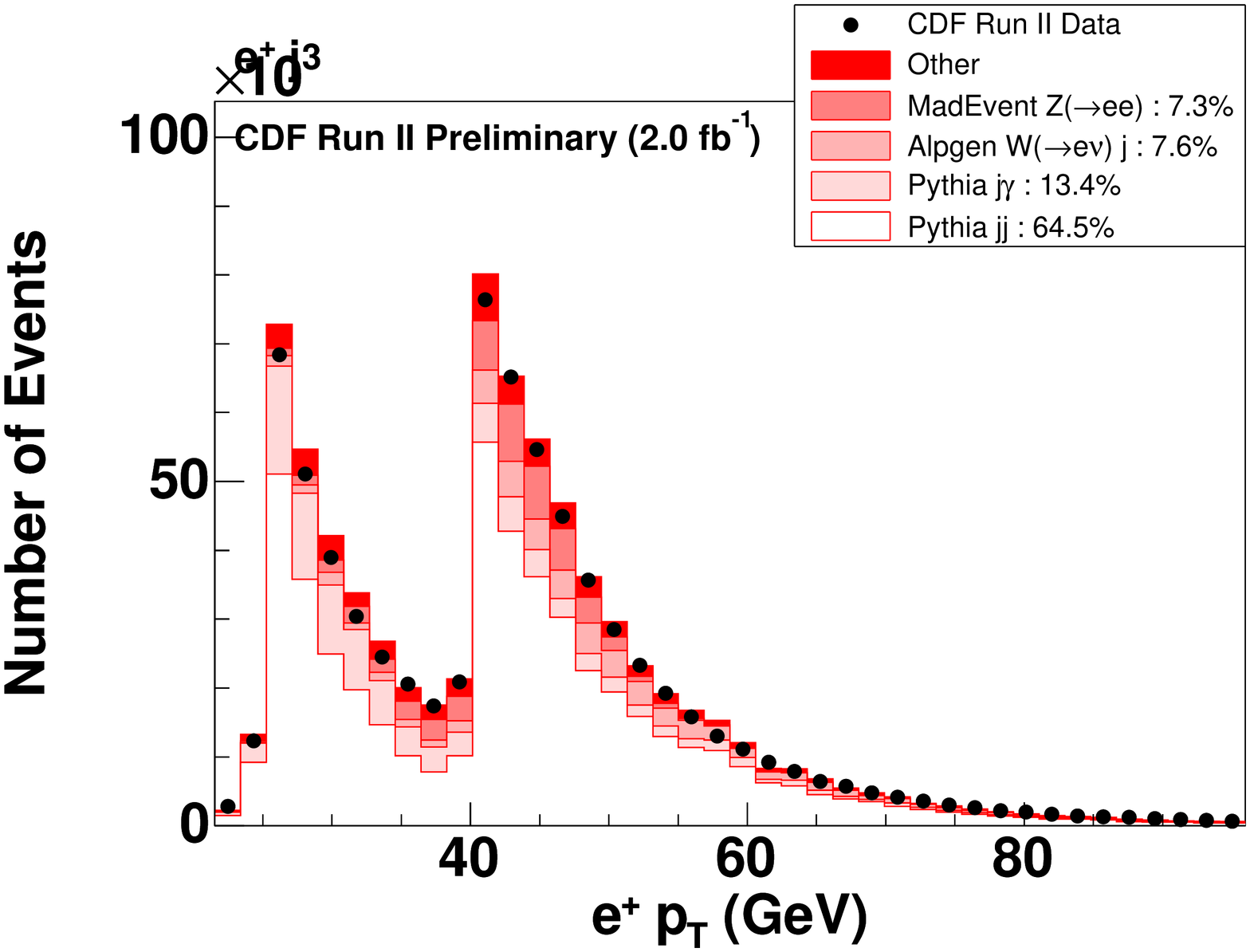}
\caption{Electron $p_T$ distribution in the {\tt 1e+1j} final state.}
\label{fig:plots_1e+1j_e+pt}
\end{figure}

\begin{figure}
\centering
\includegraphics[angle=-90,width=0.5\columnwidth]{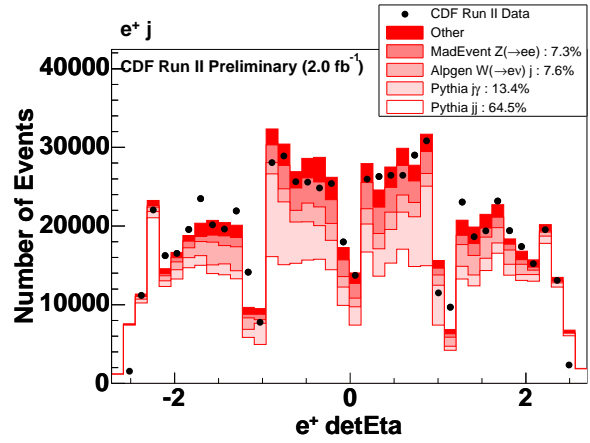}
\caption{Electron detector eta distribution in the {\tt 1e+1j} final state.}
\label{fig:plots_1e+1j_e+deteta}
\end{figure}

\begin{figure}
\centering
\includegraphics[angle=-90,width=0.5\columnwidth]{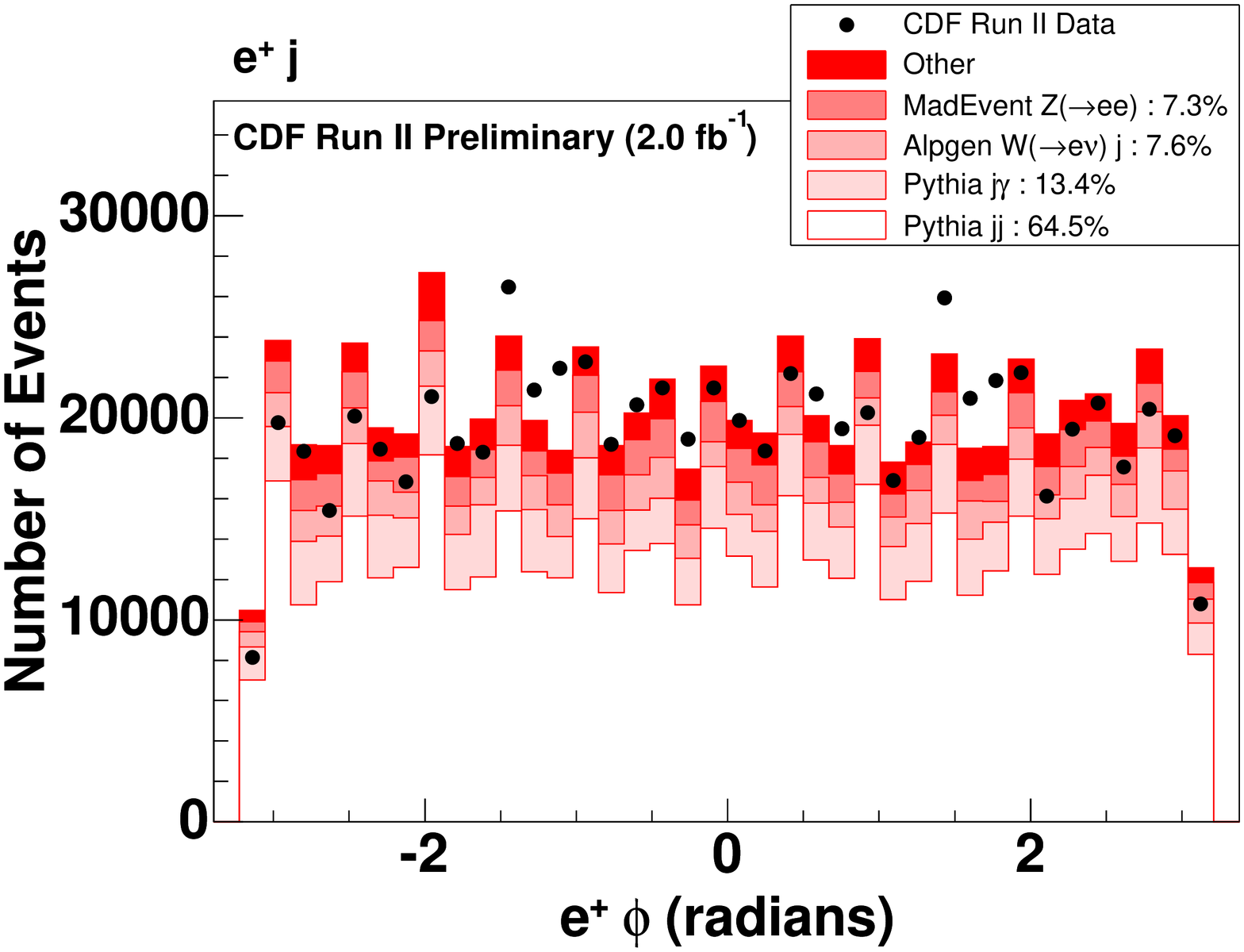}
\caption{Electron phi distribution in the {\tt 1e+1j} final state.}
\label{fig:plots_1e+1j_e+phi}
\end{figure}

Figures \ref{fig:pj2e_relative_vs_eta} and \ref{fig:pj2e_relative_vs_phi} show the relative fake rate for jets to fake electrons as a function of \detEta\ and $\phi$.  These functions of \detEta\ and $\phi$ are multiplied by overall correction factors which represent a crude average fake rate over the appropriate region.  These shaped functions are meant to model more fine details in fake rates than the overall average can contain.  In addition to \detEta\ and $\phi$ dependence, for plug electrons there is a dependance on the $p_T$, shown in Figure \ref{fig:pj2e_relative}.  Figures \ref{fig:plots_1e+1j_e+pt}, \ref{fig:plots_1e+1j_e+deteta}, and \ref{fig:plots_1e+1j_e+phi} show the electron $p_T$, electron \detEta\ and $\phi$ distribution from data in the {\tt 1e+1j} final state, where almost all events come from QCD dijet production where one of the jets fakes an electron. This serves as the dominant control region for determining variations in jet to electron fake rate.

\begin{figure}
\centering
\includegraphics[angle=-90,width=0.5\columnwidth]{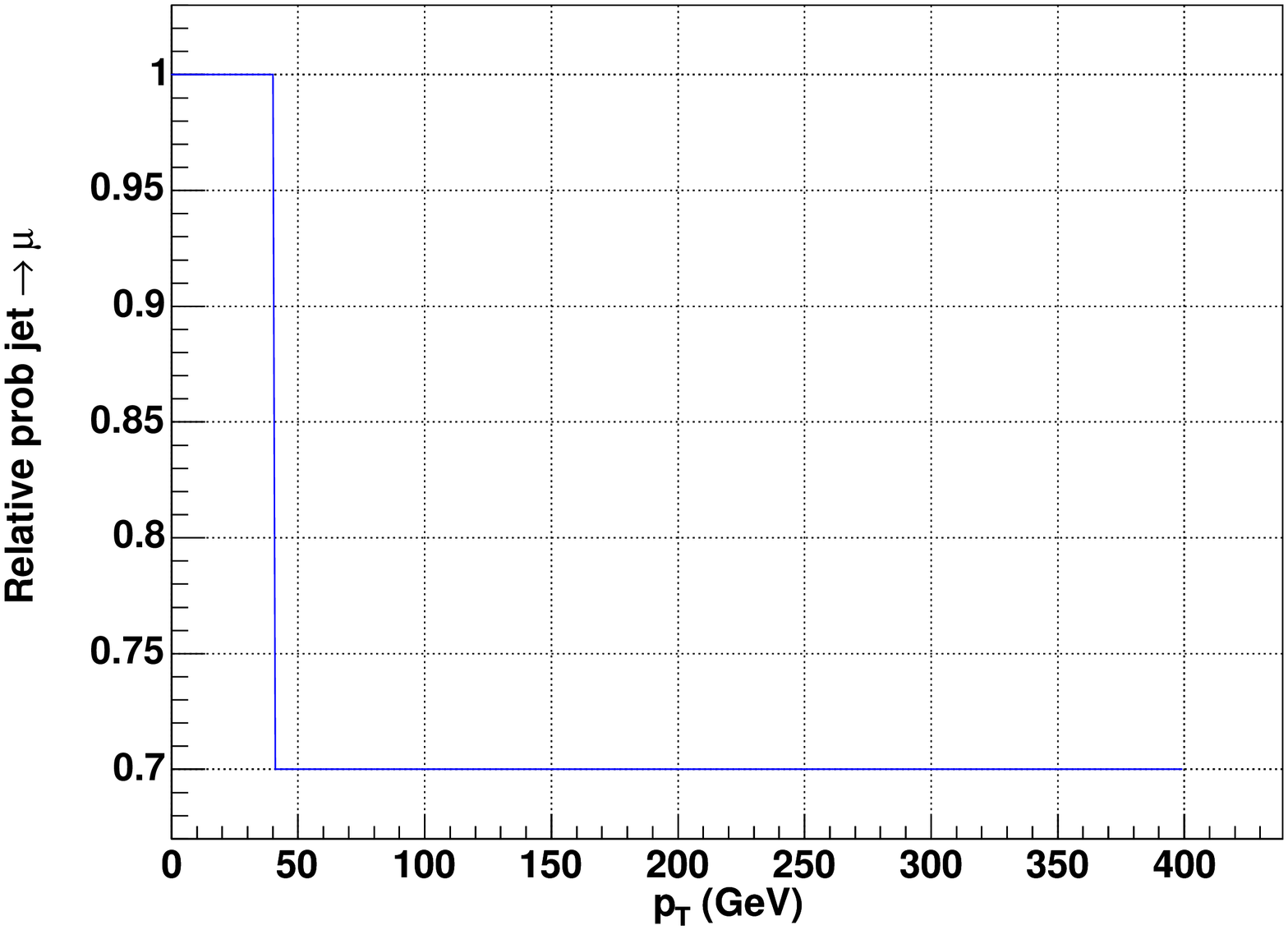}
\caption{The relative fake rate for jets to fake muons as a function of $p_T$.}
\label{fig:pj2mu_relative}
\end{figure}

\begin{figure}
\centering
\includegraphics[angle=-90,width=0.5\columnwidth]{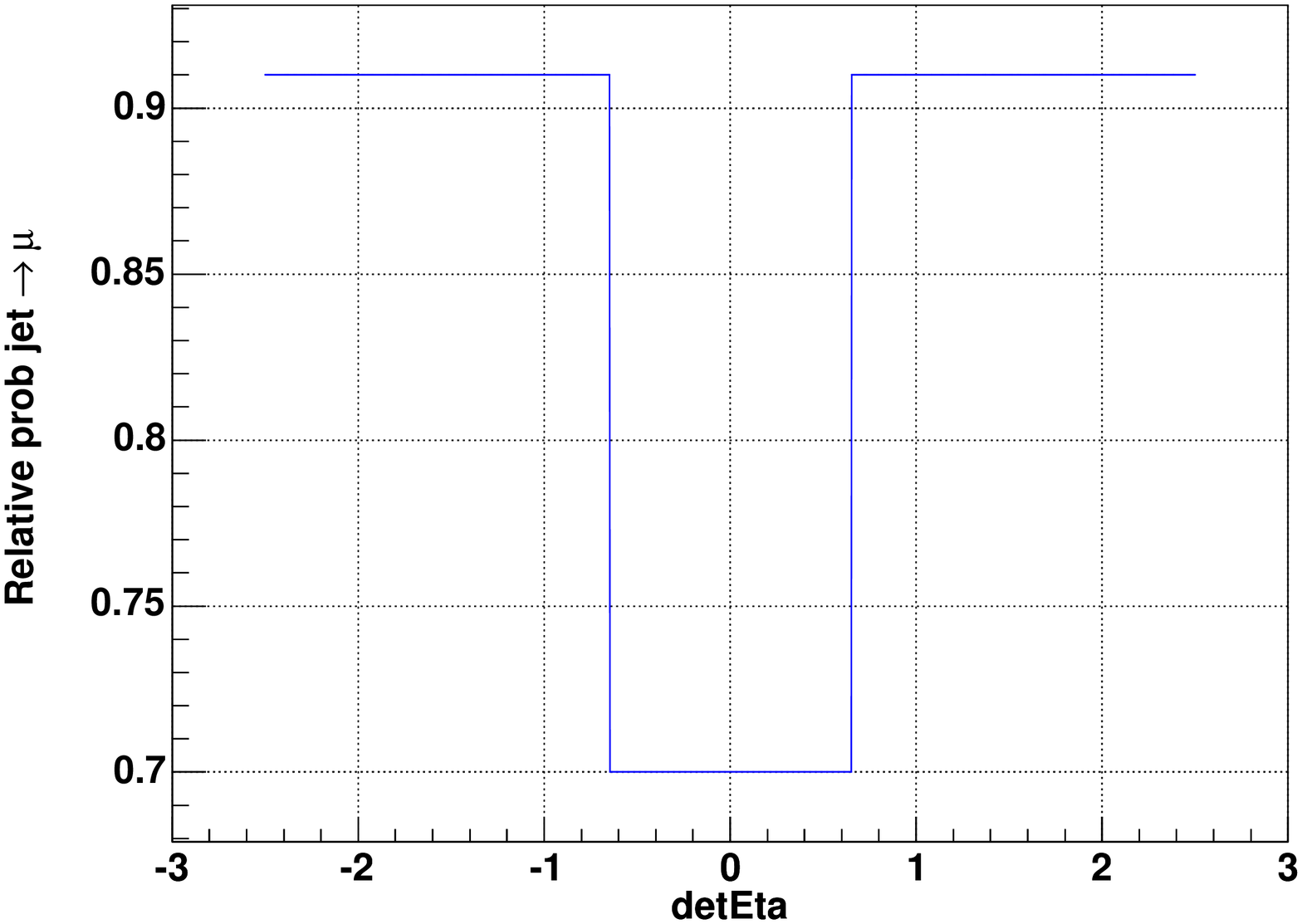}
\caption{The relative fake rate for jets to fake muons as a function of \detEta.}
\label{fig:pj2mu_relative_vs_eta}
\end{figure}

\begin{figure}
\centering
\includegraphics[angle=-90,width=0.5\columnwidth]{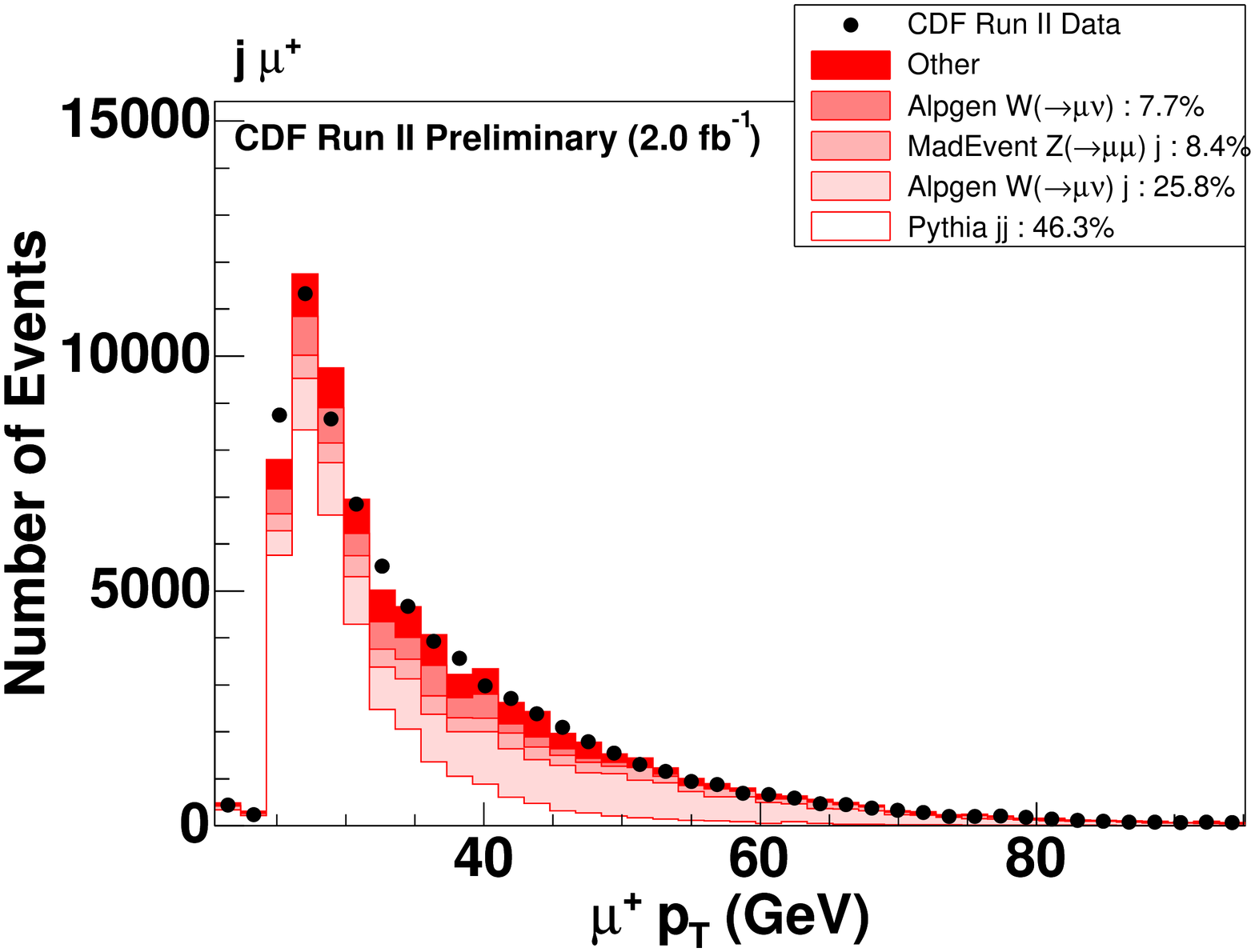}
\caption{Muon $p_T$ distribution in the {\tt 1j1mu+} final state.}
\label{fig:plots_1j1mu+_mu+pt}
\end{figure}

\begin{figure}
\centering
\includegraphics[angle=-90,width=0.5\columnwidth]{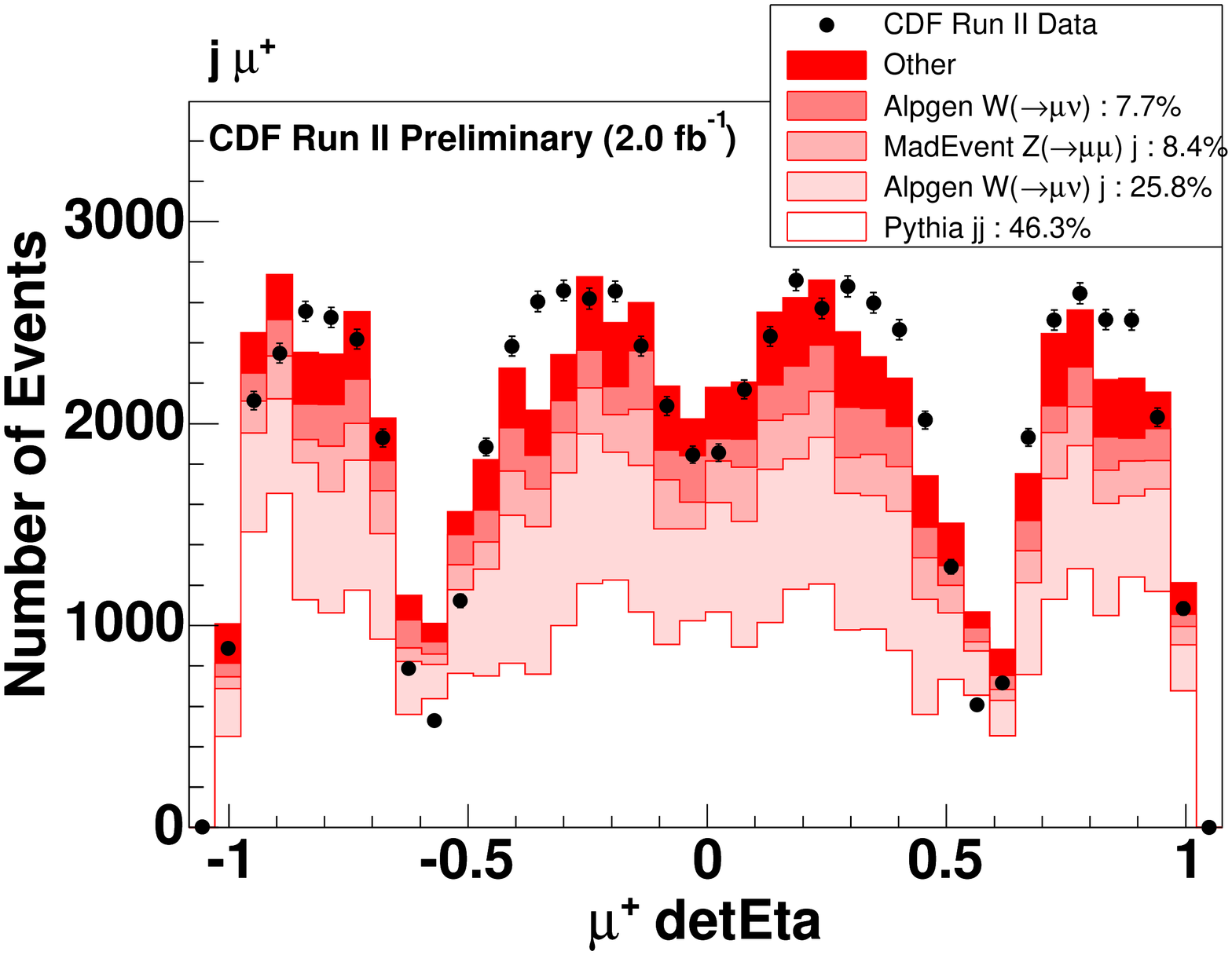}
\caption{Muon \detEta\ distribution in the {\tt 1j1mu+} final state.}
\label{fig:plots_1j1mu+_mu+deteta}
\end{figure}

\begin{figure}
\centering
\includegraphics[angle=-90,width=0.5\columnwidth]{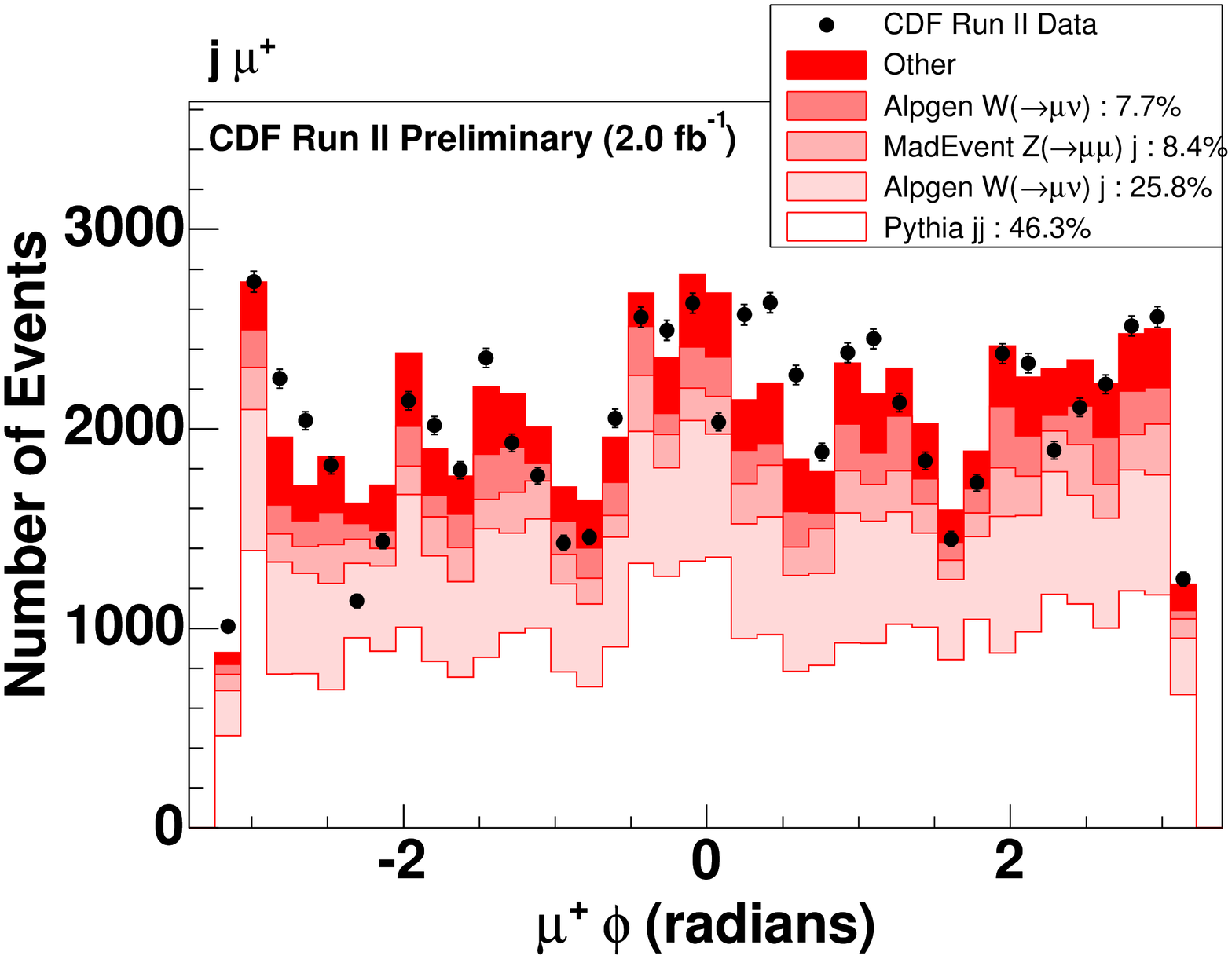}
\caption{Muon $\phi$ distribution in the {\tt 1j1mu+} final state.}
\label{fig:plots_1j1mu+_mu+phi}
\end{figure}

Figures \ref{fig:pj2mu_relative} and \ref{fig:pj2mu_relative_vs_eta} show the fake rate variation for jets to fake muons as a function of $p_T$ and \detEta. The fake rate is higher in CMX than in CMU and CMP. The muon $p_T$, \detEta, and $\phi$ distributions in the {\tt 1j1mu+} final state are shown in Fig.~\ref{fig:plots_1j1mu+_mu+pt}, \ref{fig:plots_1j1mu+_mu+deteta}, and \ref{fig:plots_1j1mu+_mu+phi}. These serve as the dominant control regions determining these fake rates.

\begin{figure}
\centering
\includegraphics[angle=-90,width=0.5\columnwidth]{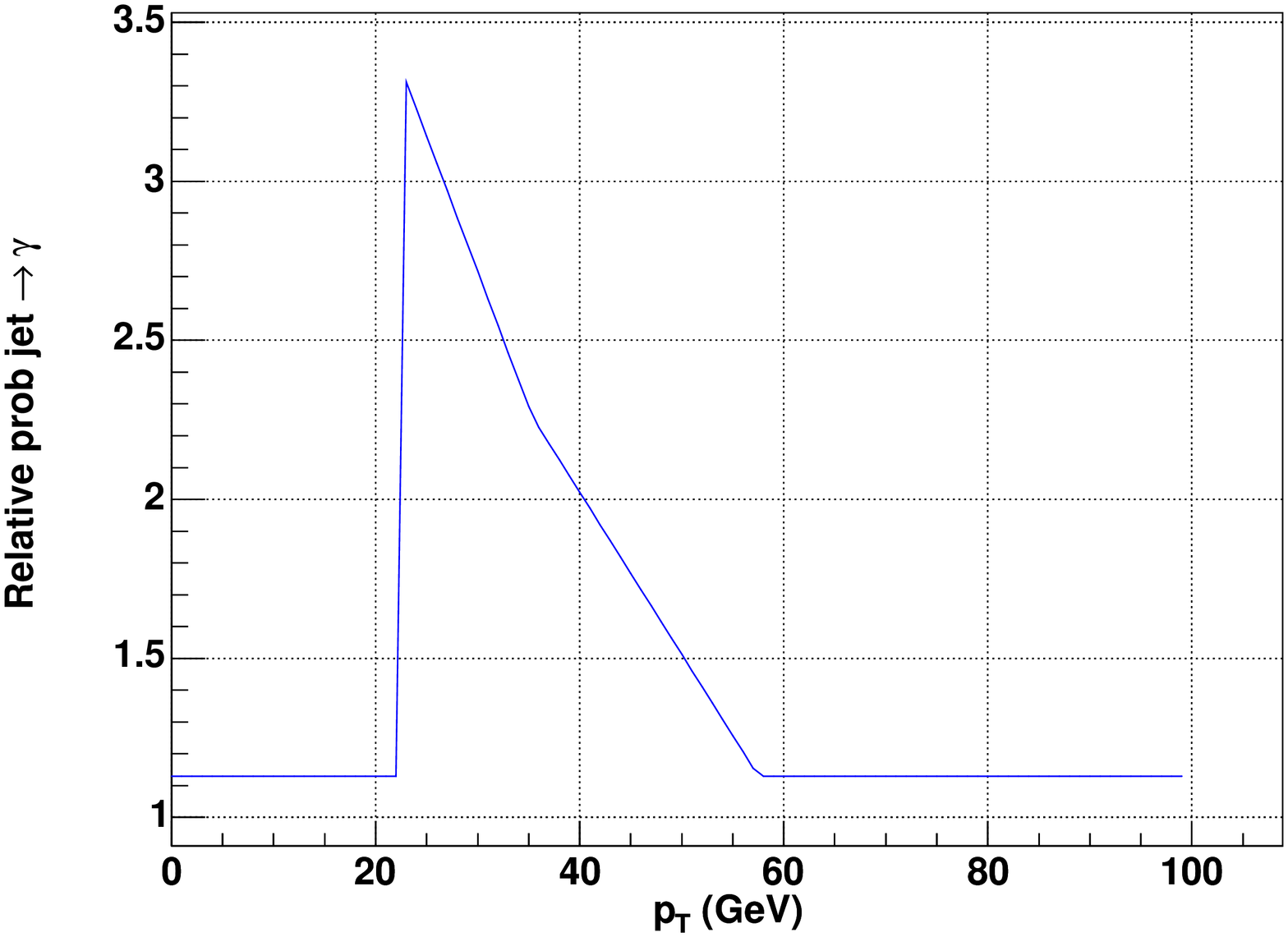}
\caption{The relative fake rate for jets to fake photons as a function of $p_T$.}
\label{fig:pj2ph_relative_pt}
\end{figure}

\begin{figure}
\centering
\includegraphics[angle=-90,width=0.5\columnwidth]{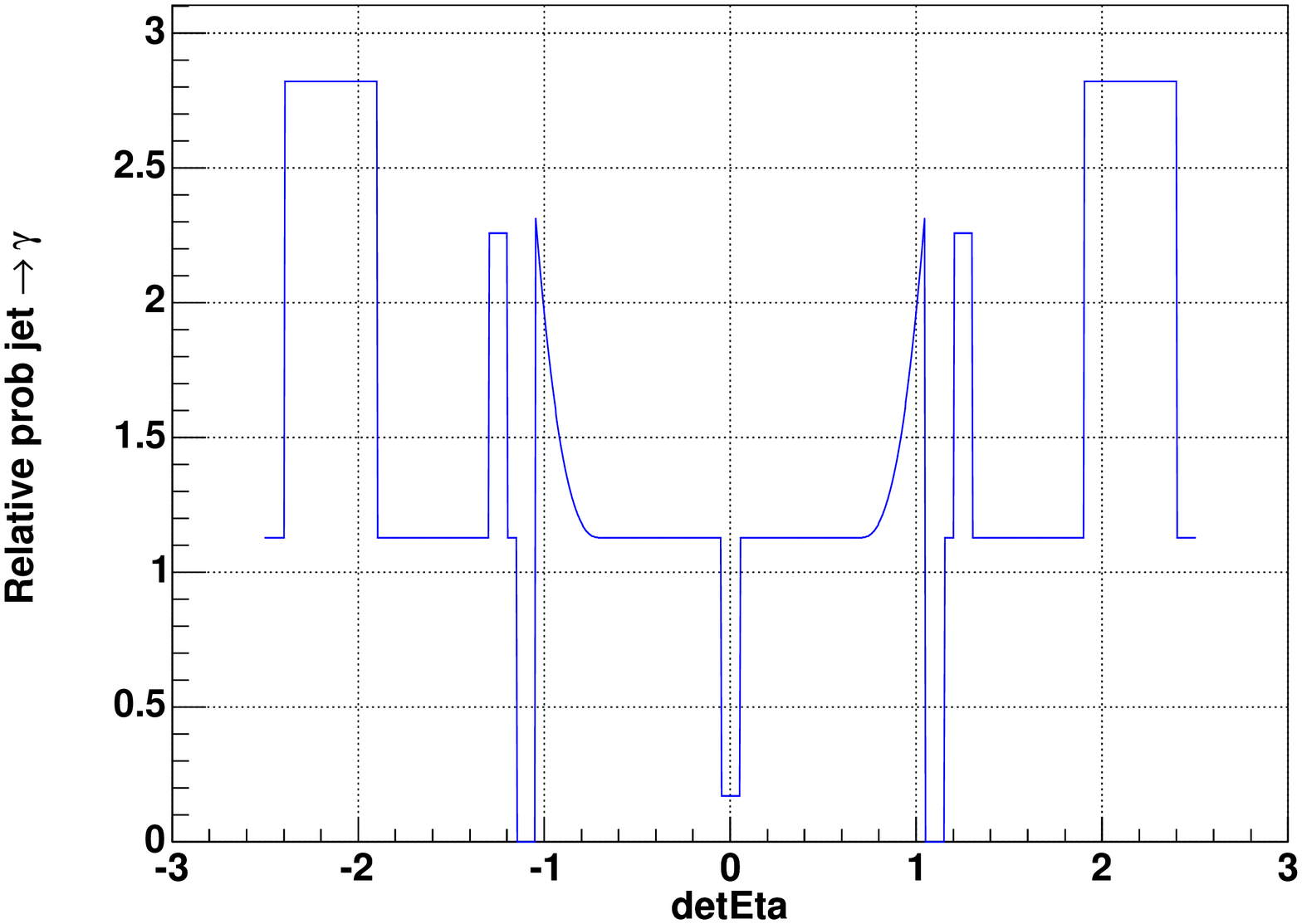}
\caption{The relative fake rate for jets to fake photons as a function of \detEta.}
\label{fig:pj2ph_relative_vs_eta}
\end{figure}

\begin{figure}
\centering
\includegraphics[angle=-90,width=0.5\columnwidth]{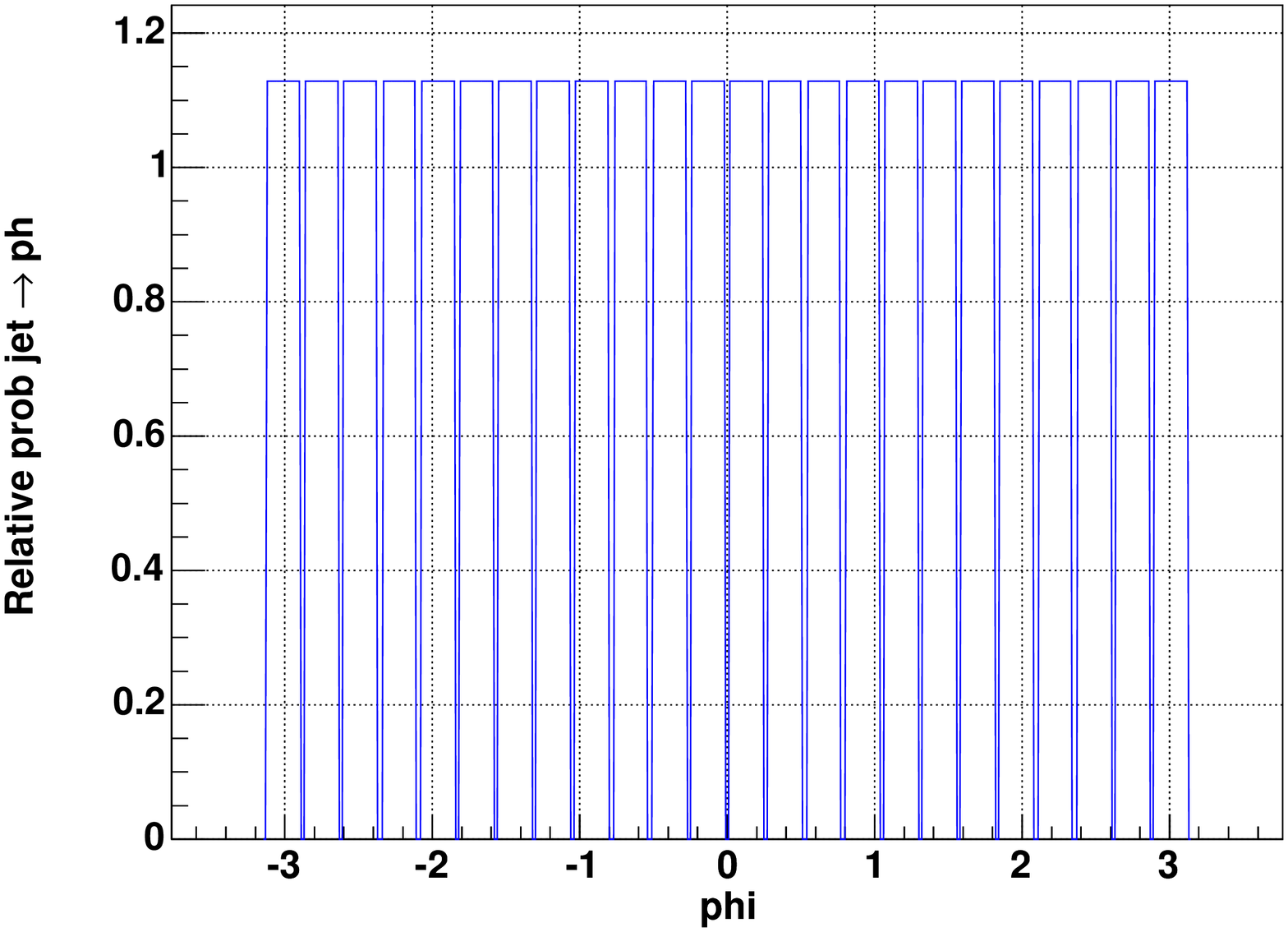}
\caption{The relative fake rate for jets to fake photons as a function of $\phi$.}
\label{fig:pj2ph_relative_vs_phi}
\end{figure}

\begin{figure}
\centering
\includegraphics[angle=-90,width=0.5\columnwidth]{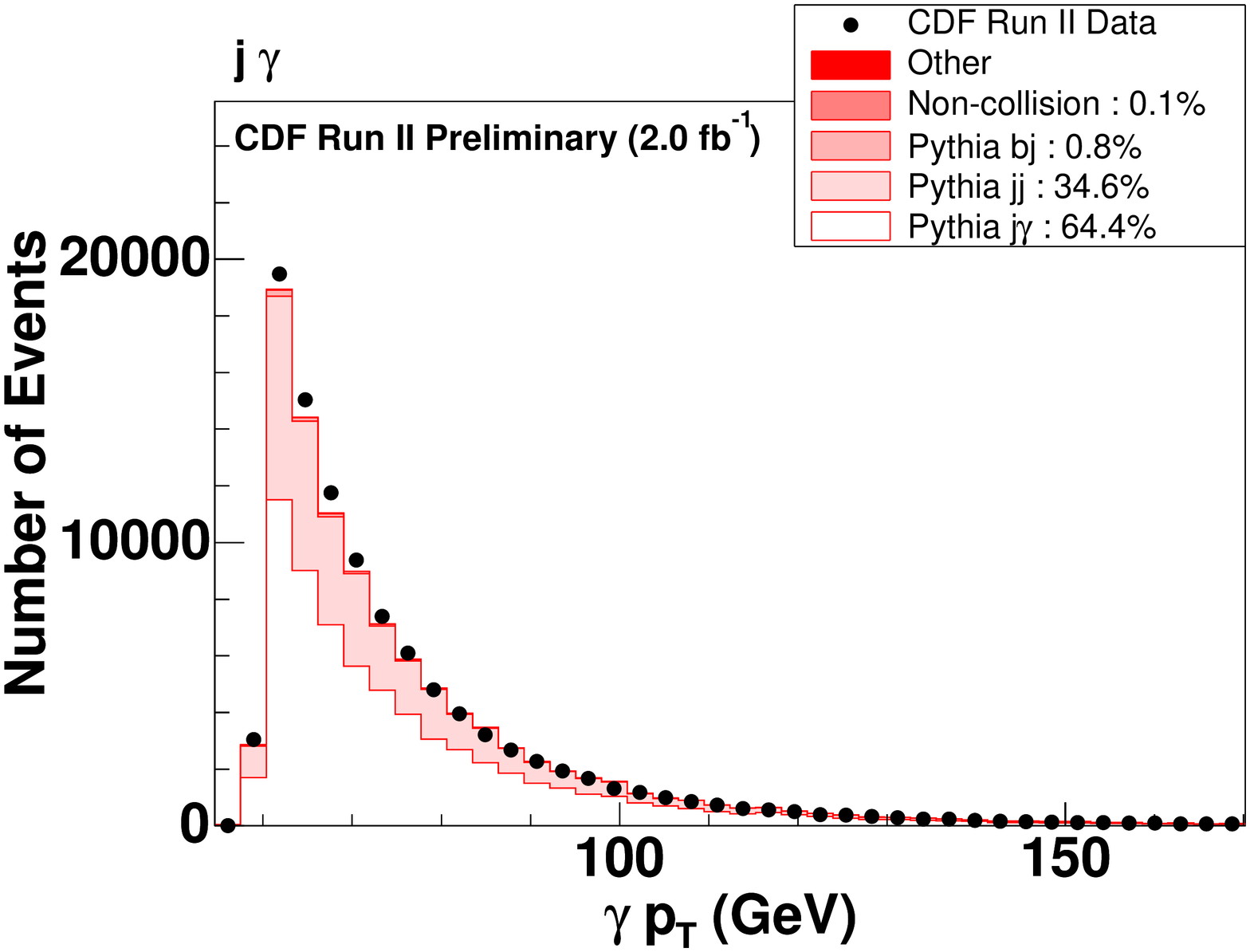}
\caption{Photon $p_T$ distribution in the {\tt 1j1ph} final state.}
\label{fig:plots_1j1ph_phpt}
\end{figure}

\begin{figure}
\centering
\includegraphics[angle=-90,width=0.5\columnwidth]{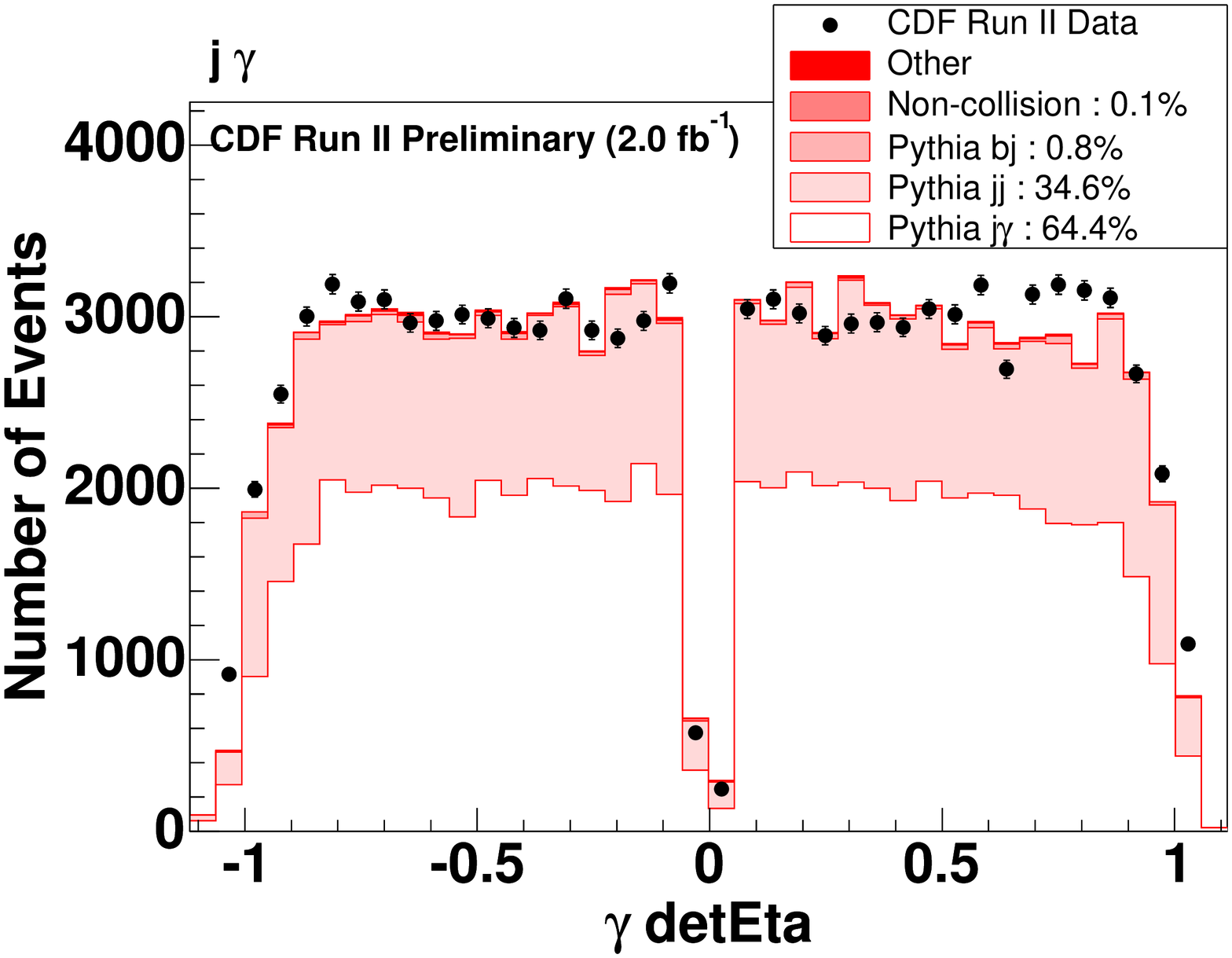}
\caption{Photon \detEta\ distribution in the {\tt 1j1ph} final state.}
\label{fig:plots_1j1ph_phdeteta}
\end{figure}

\begin{figure}
\centering
\includegraphics[angle=-90,width=0.5\columnwidth]{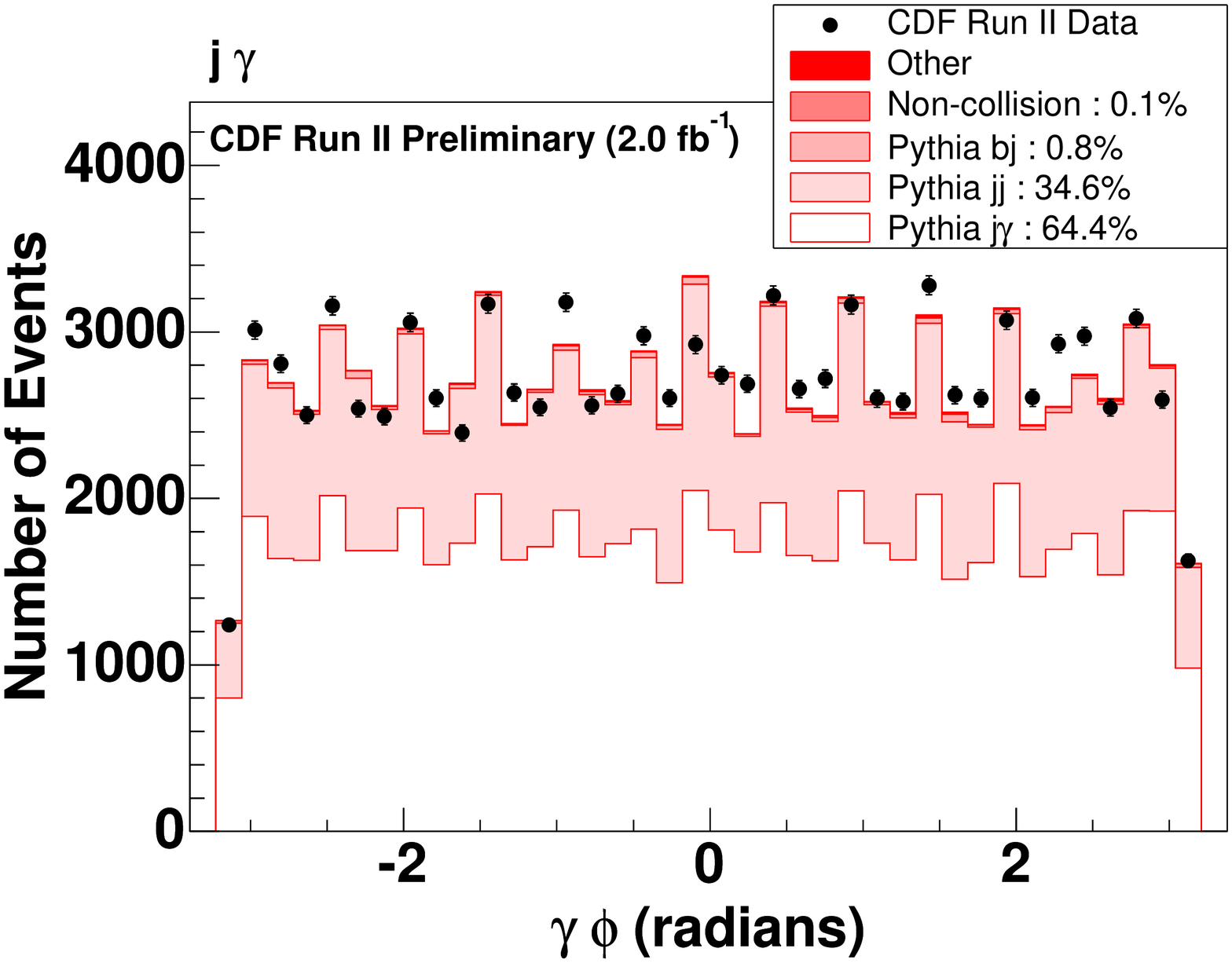}
\caption{Photon $\phi$ distribution in the {\tt 1j1ph} final state.}
\label{fig:plots_1j1ph_phphi}
\end{figure}

Figures \ref{fig:pj2ph_relative_pt}, \ref{fig:pj2ph_relative_vs_eta}, and \ref{fig:pj2ph_relative_vs_phi} show the jet to photon fake rates as functions of $p_T$, \detEta, and $\phi$.  Detector geometry features are analogous to those exhibited in the jet to electron fake rate.  The photon $p_T$, \detEta, and $\phi$ distributions in the {\tt 1j1ph} final state are shown in Fig.~\ref{fig:plots_1j1ph_phpt}, \ref{fig:plots_1j1ph_phdeteta}, and \ref{fig:plots_1j1ph_phphi}. This is one of the dominant control regions determining the jet to photon fake rates. Unlike the previous two cases, this final state is dominated by real $\gamma$+jet production, rather than the fake process, which contributes about 35\% to this final state. 

\begin{figure}
\centering
\includegraphics[angle=-90,width=0.5\columnwidth]{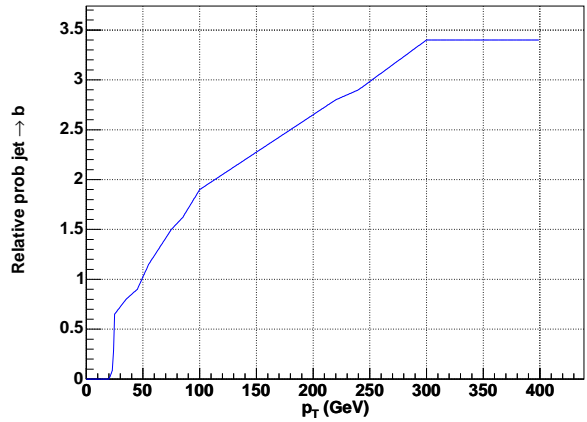}
\caption[The relative fake rate for jets to fake $b-tagged$ jets as a function of $p_T$.]{The relative fake rate for jets to fake $b-tagged$ jets as a function of $p_T$. It is essentially the mistag rate.}
\label{fig:pj2b_relative}
\end{figure}

\begin{figure}
\centering
\includegraphics[angle=-90,width=0.5\columnwidth]{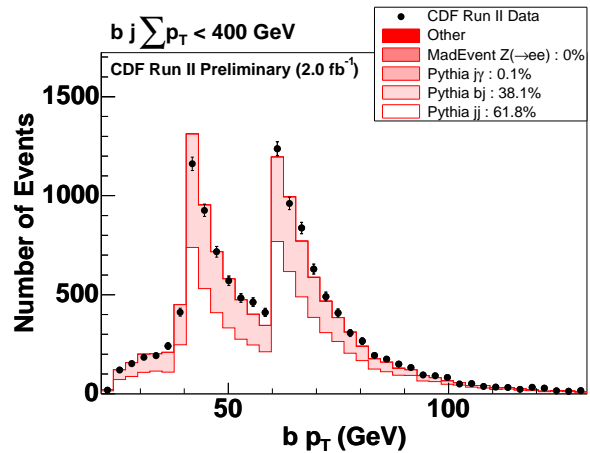}
\caption{The $b$-jet $p_T$ distribution in the {\tt 1b1j} low \sumPt\ final state.}
\label{fig:plots_1b1j_sumPt0-400_bpt}
\end{figure}

\begin{figure}
\centering
\includegraphics[angle=-90,width=0.5\columnwidth]{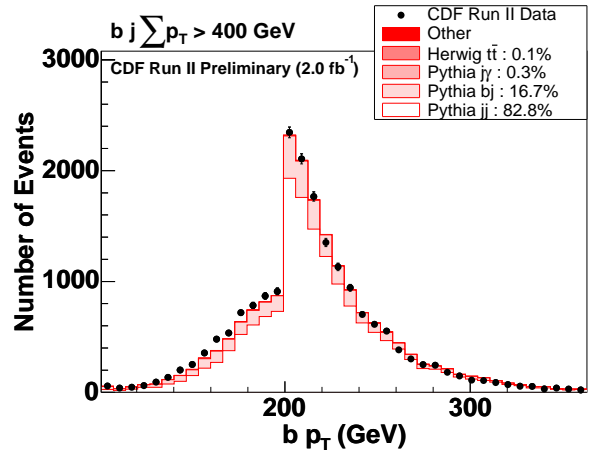}
\caption{The $b$-jet $p_T$ distribution in the {\tt 1b1j} high-\sumPt\ final state.}
\label{fig:plots_1b1j_sumPt400+_bpt}
\end{figure}

The variation in jet faking b-jet rate is shown in \ref{fig:pj2b_relative}, as a function of $p_T$.  This shape is consistent with the one measured by the b-tagging group.  Before comparing absolute values, however, it should be noted that this \Vista\ fake rate includes contributions from charm quarks to fake b, which is not usually included in the b-tagging mistag rate. When we accounted for the expected relative contribution of charmed quarks in our 'denominator jets', we found values consistent with the mistag rates.
The b jet $p_T$ distribution is shown in Fig.~\ref{fig:plots_1b1j_sumPt0-400_bpt} and \ref{fig:plots_1b1j_sumPt400+_bpt}, for the {\tt 1b1j} high \SumPt\ and {\tt 1b1j} low \SumPt\ final states. These are the dominant control regions determining the mistag rates.

\begin{figure}
\centering
\includegraphics[angle=-90,width=0.5\columnwidth]{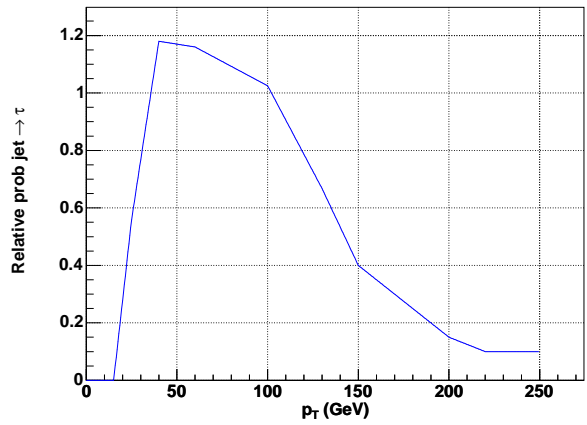}
\caption{The relative fake rate for jets to fake $\tau$s as a function of $p_T$. }
\label{fig:pj2tau_relative}
\end{figure}

\begin{figure}
\centering
\includegraphics[angle=-90,width=0.5\columnwidth]{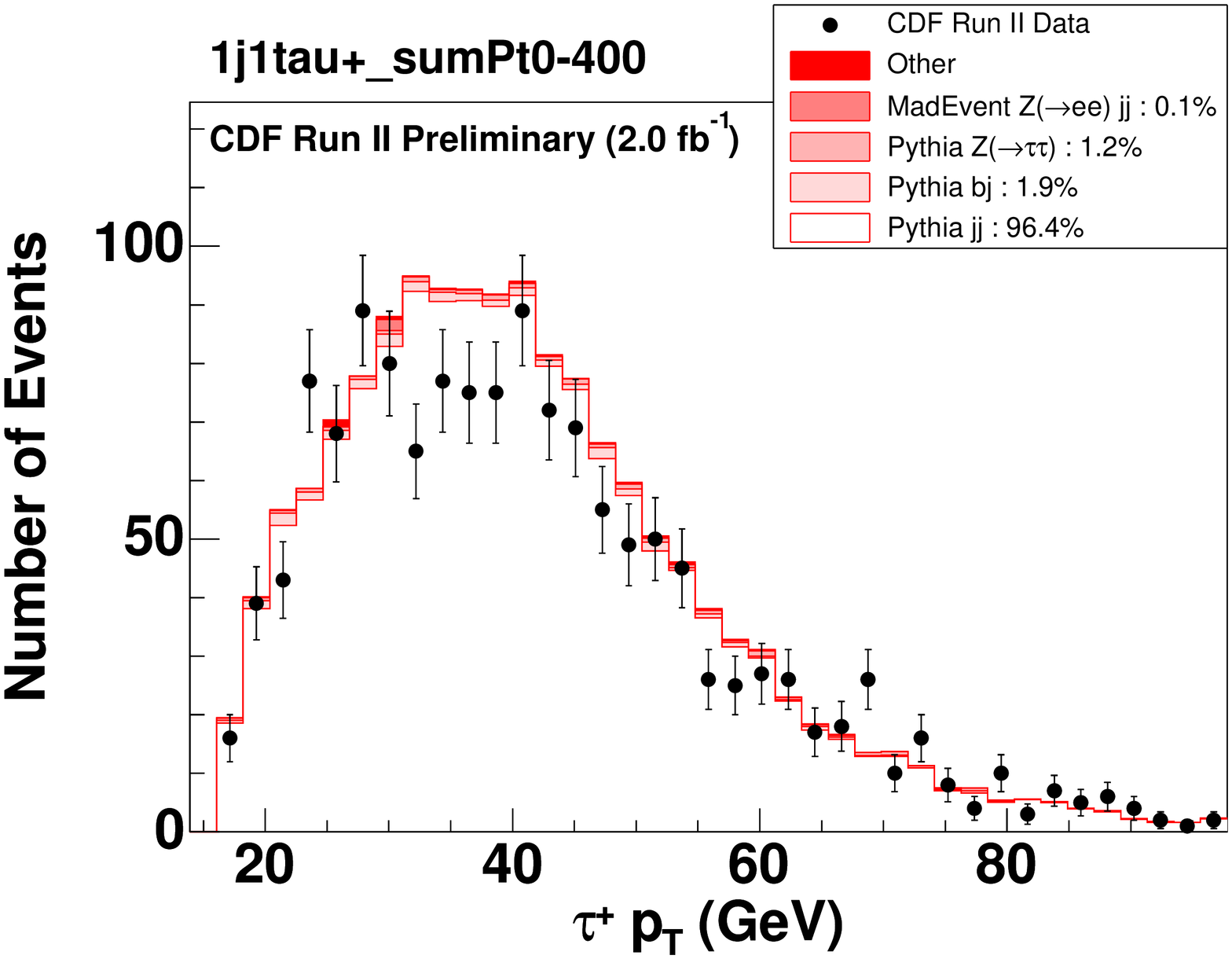}
\caption{The $\tau$ $p_T$ distribution in the {\tt 1j1tau+} low-\sumPt\ final state.}
\label{fig:plots_1j1tau+_sumPt0-400_tau+pt}
\end{figure}

\begin{figure}
\centering
\includegraphics[angle=-90,width=0.5\columnwidth]{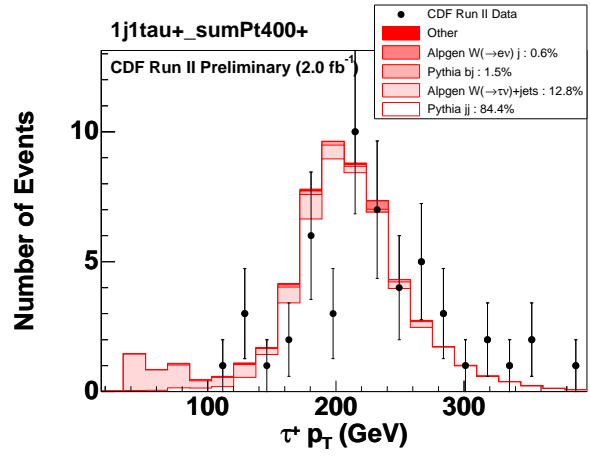}
\caption{The $\tau$ $p_T$ distribution in the {\tt 1j1tau+} high-\sumPt\ final state.}
\label{fig:plots_1j1tau+_sumPt400+_tau+pt}
\end{figure}

\begin{figure}
\centering
\includegraphics[angle=-90,width=0.5\columnwidth]{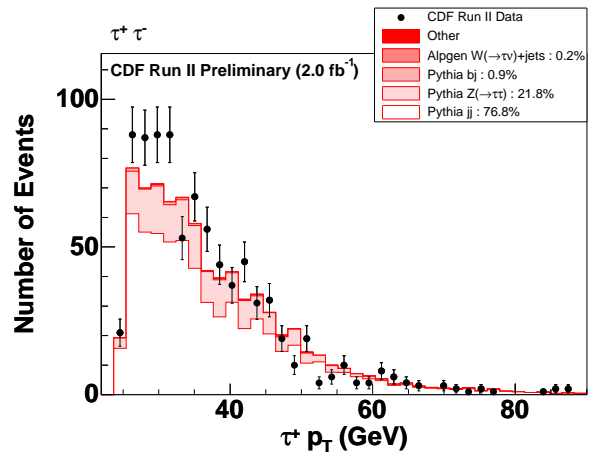}
\caption{The $\tau$ $p_T$ distribution in the {\tt 1tau+1tau-} final state.}
\label{fig:plots_1tau+1tau-_tau+pt}
\end{figure}

The jet to $\tau$ relative fake rate is given in Fig.~\ref{fig:pj2tau_relative}.  This shape is then multiplied by the function $\exp( -{\rm Generated SumPt} / 350\; {\rm GeV})$ and the jet to $\tau$ fake rate correction factor to obtain the final fake rate.  The shape is consistent with previous studies of the jet to $\tau$ fake rate. The $\tau$ $p_T$ distributions in the {\tt 1j1tau+} low-\sumPt, {\tt 1j1tau+} high-\sumPt, and {\tt 1tau+1tau-} final states are shown in Fig.~\ref{fig:plots_1j1tau+_sumPt0-400_tau+pt}, \ref{fig:plots_1j1tau+_sumPt400+_tau+pt}, and \ref{fig:plots_1tau+1tau-_tau+pt}. These serve as the dominant control regions determining the jet to $\tau$ fake rate.

\begin{figure}
\centering
\includegraphics[angle=-90,width=0.5\columnwidth]{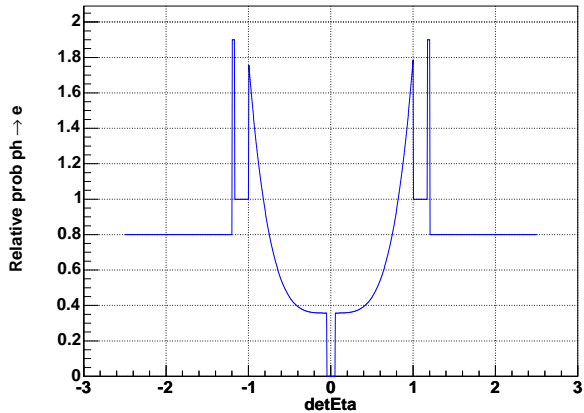}
\caption{The relative fake rate for jets to fake $\tau$s as a function of $p_T$. }
\label{fig:pph2e_relative_vs_eta}
\end{figure}

\begin{figure}
\centering
\includegraphics[angle=-90,width=0.5\columnwidth]{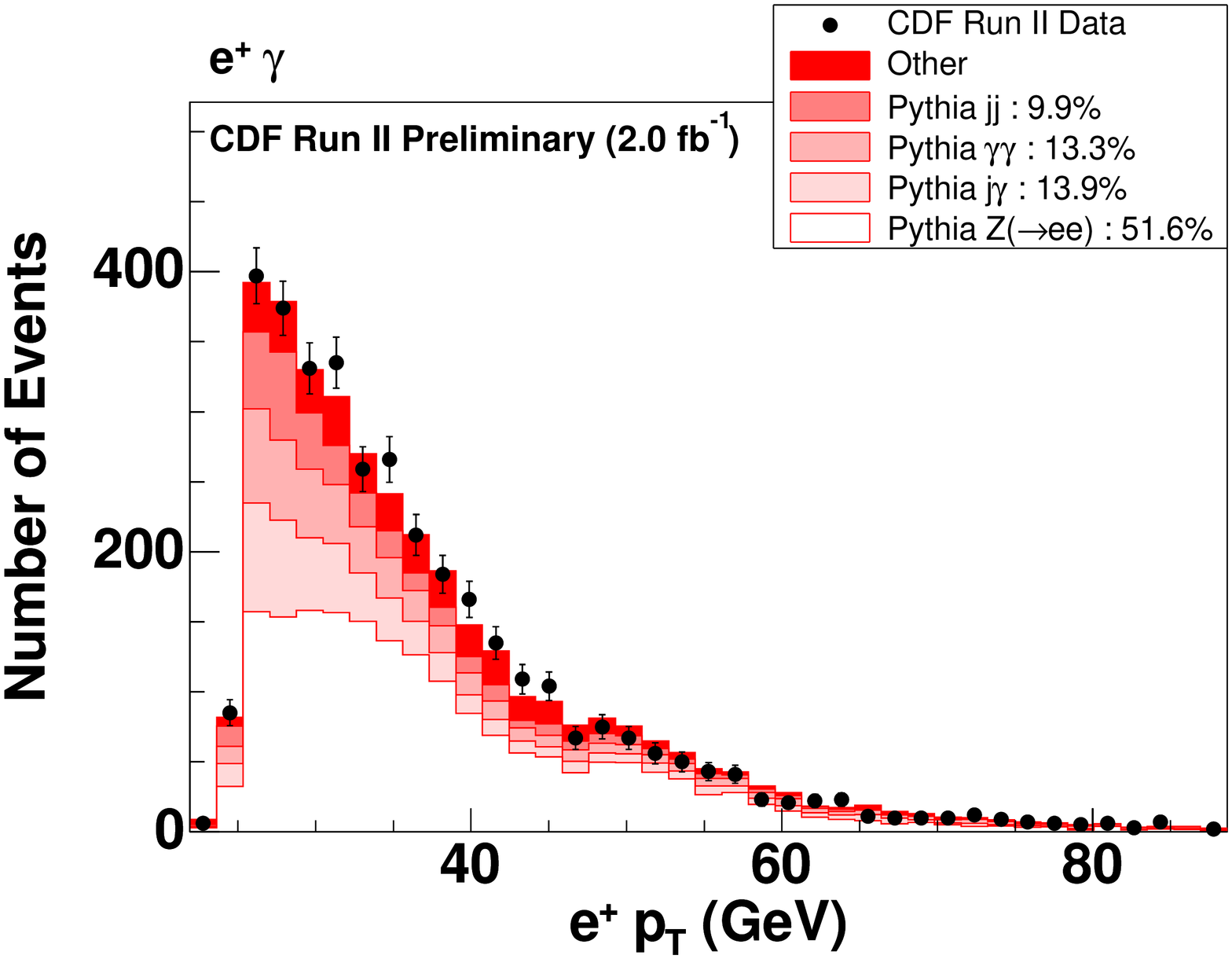}
\caption{The electron $p_T$ distribution in the {\tt 1e+1ph} final state.}
\label{fig:plots_1e+1ph_e+pt}
\end{figure}

\begin{figure}
\centering
\includegraphics[angle=-90,width=0.5\columnwidth]{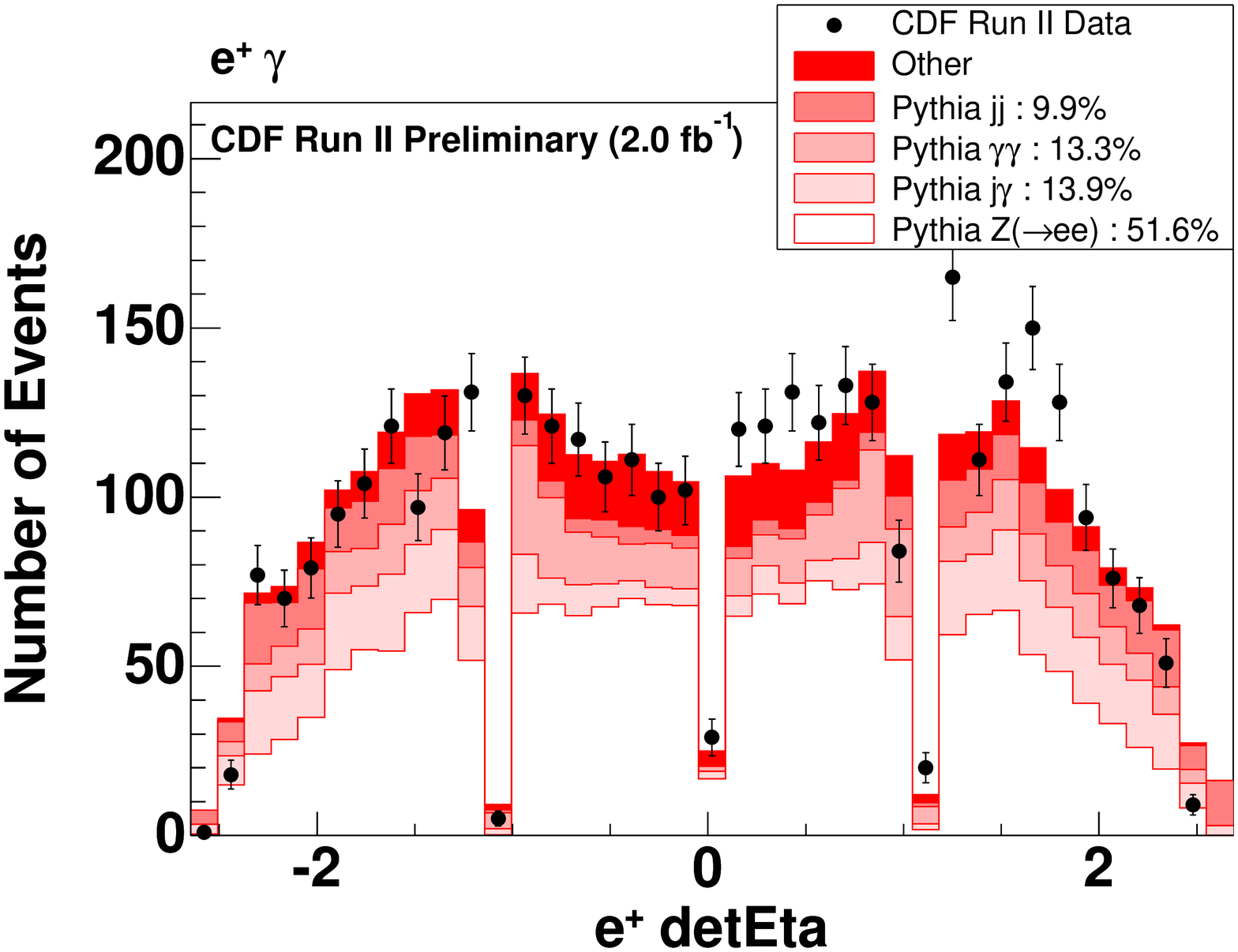}
\caption{The electron \detEta\ distribution in the {\tt 1e+1ph} final state.}
\label{fig:plots_1e+1ph_e+deteta}
\end{figure}

\begin{figure}
\centering
\includegraphics[angle=-90,width=0.5\columnwidth]{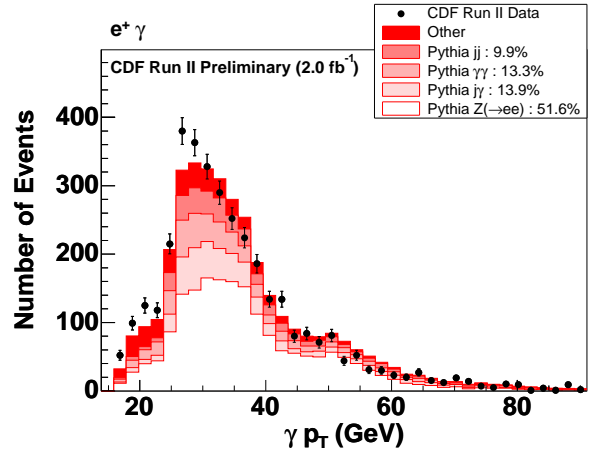}
\caption{The photon $p_T$ distribution in the {\tt 1e+1ph} final state.}
\label{fig:plots_1e+1ph_phpt}
\end{figure}

\begin{figure}
\centering
\includegraphics[angle=-90,width=0.5\columnwidth]{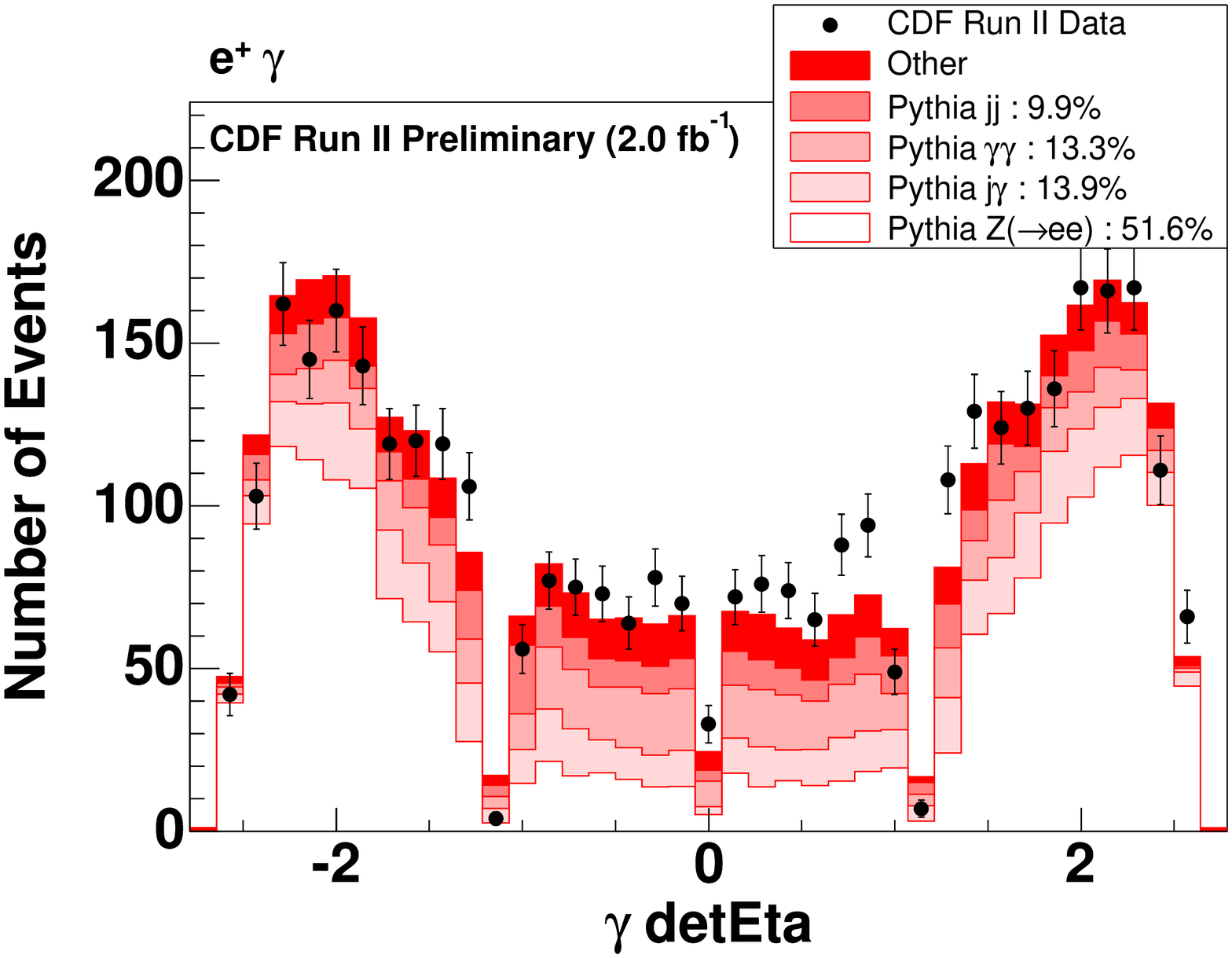}
\caption{The photon \detEta\ distribution in the {\tt 1e+1ph} final state.}
\label{fig:plots_1e+1ph_phdeteta}
\end{figure}

Figure \ref{fig:pph2e_relative_vs_eta} shows the relative fake rate for photons to fake electrons as a function of \detEta. Fig.~\ref{fig:plots_1e+1ph_e+pt} and \ref{fig:plots_1e+1ph_e+deteta} show the electron $p_T$ and \detEta\ distributions in the {\tt 1e+1ph} final state.  This final state is the dominant control region determining the photon to electron fake rate.  However, this underlying process does not contribute very much to the background in this final state and, as a result, the photon to electron fake rate is not as well constrained as other fake rates.  Fig.~\ref{fig:plots_1e+1ph_phpt} and \ref{fig:plots_1e+1ph_phdeteta} show the photon $p_T$ and \detEta\ distributions in this same final state.  As a general comment, this final state is a particularly good example of how well-modelled our fake backgrounds are, since the background contributing to this final state is a mixture of various different fake processes.

\section{Correction Factors}
\label{sec:correctionFactorDetails2}

\begin{table*}
{\hspace{-1cm}
\mbox{
\footnotesize
\begin{minipage}{9in}
\tiny
\begin{tabular}{lll|r@{ $\pm$ }lr@{ $\pm$ }lll}
{\bf Code} & {\bf Category} & {\bf Explanation} & \multicolumn{2}{c}{\bf Before} & \multicolumn{2}{c}{\bf After} & {\bf deviation ($\sigma$)} & {\bf Change(\%)} \\ \hline 
5001 & luminosity & CDF integrated luminosity & 0.927 & 0.02 & 2 & 0.0608 & 53.4 & 115.3 \\ 
5102 & k-factor & cosmic\_ph & 0.69 & 0.05 & 0.81 & 0.05 & 2.5 & 18.4 \\ 
5103 & k-factor & cosmic\_j & 0.45 & 0.014 & 0.19 & 0.014 & -18.2 & -57.1 \\ 
5121 & k-factor & 1$\gamma$1j photon+jet(s) & 0.95 & 0.04 & 0.92 & 0.04 & -0.7 & -2.9 \\ 
5122 & k-factor & 1$\gamma$2j & 1.2 & 0.05 & 1.3 & 0.05 & 1.3 & 5.3 \\ 
5123 & k-factor & 1$\gamma$3j & 1.5 & 0.07 & 1.6 & 0.07 & 1.5 & 7.0 \\ 
5124 & k-factor & 1$\gamma$4j+ & 2 & 0.16 & 1.9 & 0.14 & -0.5 & -3.9 \\ 
5130 & k-factor & 2$\gamma$0j diphoton(+jets) & 1.8 & 0.08 & 1.6 & 0.07 & -2.4 & -10.5 \\ 
5131 & k-factor & 2$\gamma$1j & 3.4 & 0.24 & 3 & 0.17 & -1.8 & -12.8 \\ 
5132 & k-factor & 2$\gamma$2j+ & 1.3 & 0.16 & 1.2 & 0.09 & -0.6 & -7.7 \\ 
5141 & k-factor & W0j W (+jets) & 1.5 & 0.027 & 1.4 & 0.04 & -2.8 & -5.2 \\ 
5142 & k-factor & W1j & 1.1 & 0.03 & 1.3 & 0.04 & 9.1 & 25.7 \\ 
5143 & k-factor & W2j & 1 & 0.03 & 2 & 0.06 & 32.0 & 94.1 \\ 
5144 & k-factor & W3j+ & 0.76 & 0.05 & 2.1 & 0.07 & 26.9 & 177.4 \\ 
5151 & k-factor & Z0j Z (+jets) & 1.4 & 0.024 & 1.4 & 0.03 & -1.3 & -2.2 \\ 
5152 & k-factor & Z1j & 1.2 & 0.04 & 1.2 & 0.04 & 1.3 & 4.5 \\ 
5153 & k-factor & Z2j+ & 1 & 0.05 & 1 & 0.04 & -0.3 & -1.5 \\ 
5161 & k-factor & 2j $\hat{p}_T$$<$150 dijet & 0.96 & 0.022 & 1 & 0.031 & 1.9 & 4.4 \\ 
5162 & k-factor & 2j 150$<$$\hat{p}_T$ & 1.3 & 0.028 & 1.3 & 0.04 & 2.9 & 6.5 \\ 
5164 & k-factor & 3j $\hat{p}_T$$<$150 multijet & 0.92 & 0.021 & 0.94 & 0.03 & 1.0 & 2.3 \\ 
5165 & k-factor & 3j 150$<$$\hat{p}_T$ & 1.4 & 0.032 & 1.5 & 0.05 & 3.7 & 8.7 \\ 
5167 & k-factor & 4j $\hat{p}_T$$<$150 & 0.99 & 0.025 & 1.1 & 0.04 & 3.0 & 7.7 \\ 
5168 & k-factor & 4j 150$<$$\hat{p}_T$ & 1.7 & 0.04 & 1.9 & 0.07 & 5.5 & 12.8 \\ 
5169 & k-factor & 5j low & 1.3 & 0.05 & 1.3 & 0.06 & 1.7 & 6.8 \\ 
5170 & k-factor & 1b2j 150$<$$\hat{p}_T$ heavyflavor multijet & NA & NA & 2.2 & 0.12 & NA & NA \\ 
5171 & k-factor & 1b3j 150$<$$\hat{p}_T$ & NA & NA & 3 & 0.16 & NA & NA \\ 
5211 & misId & p(e$\rightarrow$e) central & 0.99 & 0.006 & 0.98 & 0.007 & -1.5 & -0.9 \\ 
5212 & misId & p(e$\rightarrow$e) plug & 0.93 & 0.009 & 0.97 & 0.007 & 3.6 & 3.5 \\ 
5213 & misId & p($\mu$$\rightarrow$$\mu$) CMUP+CMX & 0.85 & 0.008 & 0.89 & 0.007 & 5.3 & 5.0 \\ 
5216 & misId & p($\gamma$$\rightarrow$$\gamma$) central & 0.97 & 0.018 & 0.95 & 0.013 & -1.6 & -2.9 \\ 
5217 & misId & p($\gamma$$\rightarrow$$\gamma$) plug & 0.91 & 0.018 & 0.85 & 0.007 & -3.2 & -6.4 \\ 
5219 & misId & p(b$\rightarrow$b) central & 1 & 0.04 & 0.97 & 0.02 & -0.8 & -3.2 \\ 
5246 & misId & p($\gamma$$\rightarrow$e) plug & NA & NA & 0.062 & 0.0021 & NA & NA \\ 
5256 & misId & p(q$\rightarrow$e) central & 9.71$\times 10^{-5}$ & 1.9$\times 10^{-6}$ & 7.077$\times 10^{-5}$ & 1$\times 10^{-6}$ & -13.9 & -27.1 \\ 
5257 & misId & p(q$\rightarrow$e) plug & 0.0008761 & 1.8$\times 10^{-5}$ & 0.0007611 & 5$\times 10^{-6}$ & -6.4 & -13.1 \\ 
5261 & misId & p(q$\rightarrow$$\mu$) & 1.157$\times 10^{-5}$ & 2.7$\times 10^{-7}$ & 1.235$\times 10^{-5}$ & 5$\times 10^{-7}$ & 2.9 & 6.7 \\ 
5266 & misId & p(b$\rightarrow$$\mu$) & NA & NA & 3.522$\times 10^{-5}$ & 1.1$\times 10^{-5}$ & NA & NA \\ 
5273 & misId & p(j$\rightarrow$b) 25$<$$p_T$ & 0.0168 & 0.00027 & 0.0183 & 0.00018 & 5.4 & 8.7 \\ 
5285 & misId & p(q$\rightarrow$$\tau$) & 0.0034 & 0.00012 & 0.0052 & 8$\times 10^{-5}$ & 14.9 & 52.5 \\ 
5292 & misId & p(q$\rightarrow$$\gamma$) central & 0.0002651 & 1.5$\times 10^{-5}$ & 0.0002611 & 1.2$\times 10^{-5}$ & -0.3 & -1.5 \\ 
5293 & misId & p(q$\rightarrow$$\gamma$) plug & 0.00159 & 0.00013 & 0.000478 & 4$\times 10^{-5}$ & -8.6 & -70.0 \\ 
5402 & trigger & p(e$\rightarrow$trig) plug, $p_T$$>$25 & 0.83 & 0.015 & 0.86 & 7$\times 10^{-5}$ & 1.8 & 3.2 \\ 
5403 & trigger & p($\mu$$\rightarrow$trig) CMUP+CMX, $p_T$$>$25 & 0.917 & 0.007 & 0.918 & 0.004 & 0.2 & 0.1 \\ 
\end{tabular}
\end{minipage}
}}
\caption[Comparison of correction factors that were used also in the first 0.927~fb$^{-1}$]{Comparison of correction factors that were used also in the first 0.927~fb$^{-1}$. The Luminosity is in units of fb$^{-1}$.}
\label{tbl:correctionFactorComparison2}
\end{table*}

\begin{sidewaysfigure*}
\centering
\includegraphics[width=9.0in]{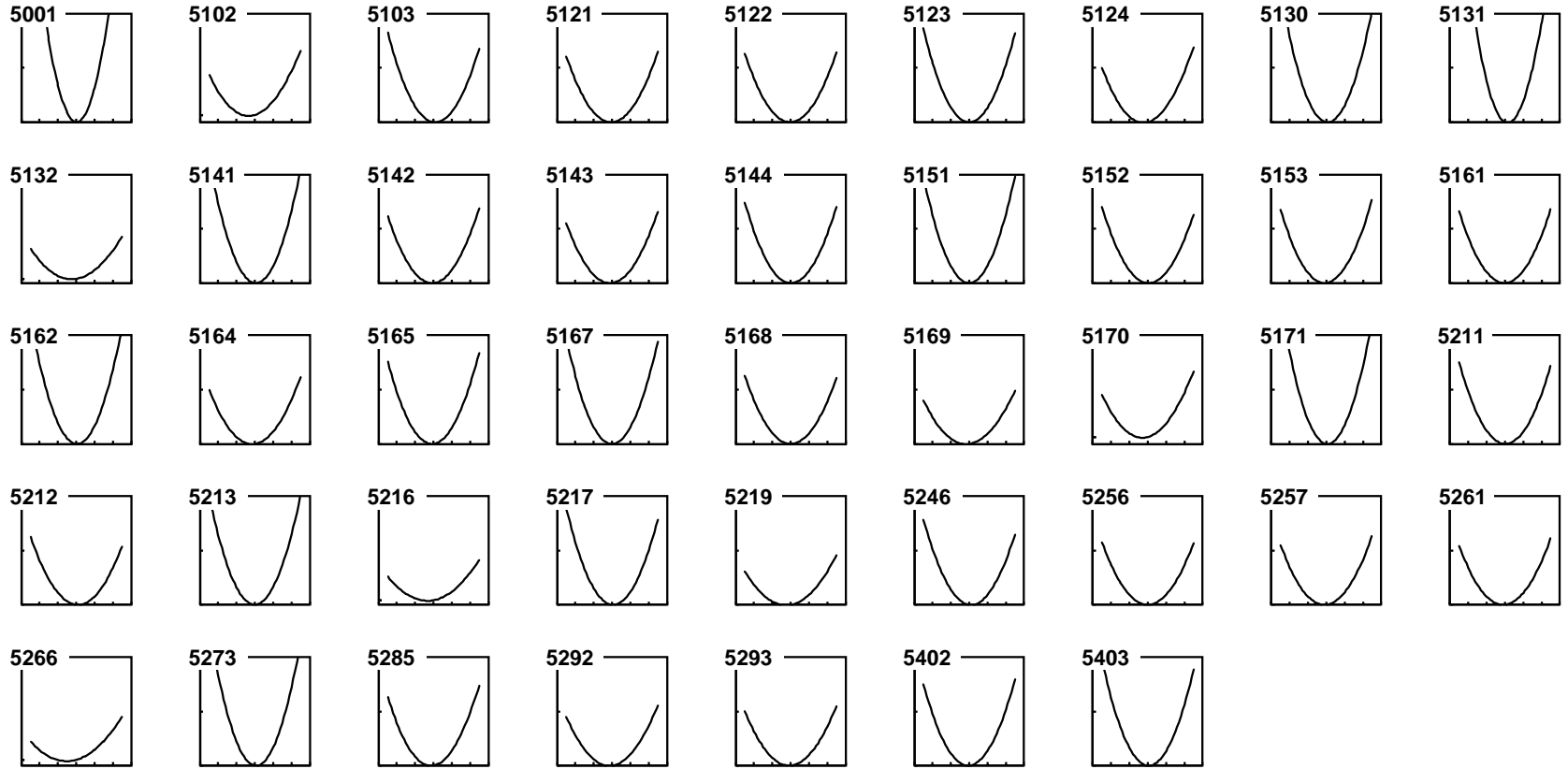}
\caption[Profiles of the $\chi^2$ function at its minimum]{Profiles of the $\chi^2$ function at its minimum along each correction factor axis.  This array of figures is used as a debugging tool to validate the parabolic form of the minimum and the calculation of the error matrix.  The top left pane shows $\chi^2$ as a function of integrated luminosity (correction factor code {\tt 0001}), holding all remaining correction factors fixed; remaining panes show $\chi^2$ profiles along each of the other correction factors.  One tick on the horizontal axis of the $i^\text{th}$ pane corresponds to $\delta s_i$, the obtained error on the correction factor value.  One tick on the vertical axis corresponds to one unit of $\chi^2$.\label{fig:CorrectionFactorMinimumProfiles2} }
\end{sidewaysfigure*}

\begin{table*}
\hspace*{-0.0in}
\centering
\mbox{
\tiny
\centering
\begin{tabular}{lcl}
Code &   Pull Apart &   Contributions \\
\hline \\
0042 &           10.3 &     ( e6pmiss = -3.2 , e6jjpmiss = 2 , e6jj = 1.3 , e6j = -1.1 ) \\
0037 &           10.2 &     ( jtau2 = -1 , ph0tau = 0.9 , jtau1 = -0.9 , ph6tau = 0.8 ) \\
0036 &            9.2 &     ( bj5 = 1.5 , b5j = -1.3 , be0 = 0.8 , b5jj = -0.5 ) \\
0013 &            8.9 &     ( e6jjpmiss = 2.9 , e0jjpmiss = -1.1 , jjmu0pmiss = -0.7 , jjmu0 = -0.5 ) \\
0031 &            8.8 &     ( bbj = -0.9 , bj5 = 0.7 , be0 = 0.5 , bbjj = -0.5 ) \\
0034 &            8.5 &     ( e6ph6 = -2.6 , e6jj = 0.8 , e6ph0 = 0.8 , e6j = -0.7 ) \\
0001 &            8.4 &     ( e0j = -0.5 , e6pmiss = -0.4 , e0pmiss = 0.4 , e6jjpmiss = 0.4 ) \\
0033 &            8.2 &     ( e0j = -4.7 , e0jj = 1.3 , be0 = 1 , e0jjj = 0.4 ) \\
0018 &            7.7 &     ( jj = 1.4 , e0j = -1.3 , e6j = -0.9 , bj = 0.6 ) \\
0034 &            7.2 &     ( e6jj = 2.3 , e6j = -1.6 , be6 = -0.9 , bej = 0.4 ) \\
0014 &            6.9 &     ( e0jjjpmiss = -1 , jjjmu0 = -1 , e6jjjpmiss = 1 , jjjmu0pmiss = -0.6 ) \\
0020 &            6.2 &     ( jjj = -2.1 , e6jj = 1.3 , e0jj = 0.5 , bej = 0.3 ) \\
0004 &            6   &     ( jph0 = 2.2 , e0j = -1.4 , bph0 = -0.6 , be0 = 0.5 ) \\
0026 &            5.7 &     ( e6pmiss = -1.3 , e6e6 = 1.3 , e6jjpmiss = 1 , e6e6j = -0.6 ) \\
0012 &            5.3 &     ( e6jjpmiss = 0.7 , jmu0pmiss = 0.6 , be0pmiss = -0.6 , e0jpmiss = 0.5 ) \\
0016 &            5.1 &     ( e6e6j = -1.5 , jmu0mu0 = 0.5 , constraints = 0.4 , e0jj = 0.27 ) \\
0030 &            5   &     ( constraints = 1.4 , e6ph6 = -1 , mu0ph6pmiss = -0.4 , ph6tau = 0.27 ) \\
0017 &            4.9 &     ( e6e6jj = 0.5 , jjjmu0 = -0.4 , e6e6jjj = -0.27 , e6e6j = -0.24 ) \\
0029 &            4.8 &     ( jph0 = 0.8 , constraints = 0.5 , jjph0 = -0.5 , bph0 = -0.4 ) \\
0005 &            4.5 &     ( jjph0 = -2.3 , e0jj = 0.7 , bjph0 = -0.3 , e6jj = 0.24 ) \\
0040 &            4   &     ( ph6tau = 1.3 , e6ph6 = -0.9 , ph0ph6 = -0.3 , j5ph6 = -0.28 ) \\
0038 &            4   &     ( bmu0 = 0.9 , jjmu0 = -0.8 , jjjmu0 = -0.7 , bjjmu = -0.5 ) \\
0039 &            3.7 &     ( jph0 = 1.5 , jjph0 = -0.5 , bph0 = -0.4 , e6ph0 = 0.22 ) \\
0025 &            3.5 &     ( e0pmiss = 0.9 , e0e0 = -0.8 , e0j = -0.2 , e0jjpmiss = -0.19 ) \\
0015 &            3.3 &     ( e0e0 = -0.7 , e6e6 = 0.6 , e0e6 = -0.4 , constraints = 0.3 ) \\
0006 &            3.1 &     ( jjjph0 = -1.9 , e0jjj = 0.6 , bjjph0 = 0.16 ) \\
0007 &            3.1 &     ( jjjjph0 = -2.1 , e0jjjj = 0.6 , e6jjjj = -0.13 ) \\
0022 &            3   &     ( bjjj = 0.6 , e6jjj = -0.4 , e0jjj = 0.3 , jjjmu0 = -0.3 ) \\
0019 &            2.9 &     ( bj5 = 1.3 , b5j = -1 , bb5 = -0.14 , jj5 = -0.13 ) \\
0035 &            2.9 &     ( jmu0 = 1.1 , jjmu0 = -0.6 , jjjmu0 = -0.5 , bmu0 = 0.4 ) \\
0010 &            2.6 &     ( jjjphph = -0.9 , jjphph = 0.4 , e0jjph6 = 0.23 , e6jjjph6 = 0.17 ) \\
0021 &            2.2 &     ( jjj5 = 1.2 , b5jj = -0.6 , jjj5ph0 = -0.12 ) \\
0024 &            2.1 &     ( e6jjjj = -0.5 , e0jjjj = 0.5 , jjjjj = -0.4 , bjjjj = 0.24 ) \\
0011 &            2   &     ( e0pmiss = 0.6 , e6pmiss = -0.6 , mu0pmiss = -0.25 , constraints = -0.19 ) \\
0027 &            2   &     ( mu0pmiss = -0.5 , jmu0pmiss = 0.16 , jjmu0pmiss = -0.13 , jjjmu0 = -0.12 ) \\
0043 &            1.8 &     ( mu0pmiss = -0.7 , constraints = 0.17 , jmu0pmiss = 0.16 , jjjmu0 = -0.16 ) \\
0025 &            1.7 &     ( b5jj = -0.7 , bjj5 = 0.31 , bb5j = 0.29 , jjj5 = 0.15 ) \\
0008 &            1.6 &     ( ph0ph6 = -0.5 , e6ph0 = 0.3 , constraints = 0.24 , ph0ph0 = -0.22 ) \\
0026 &            1.2 &     ( bbjj5 = -0.5 , b5jjj = 0.29 , bb5jj = 0.25 ) \\
0023 &            1.1 &     ( jjjj5 = -0.6 , b5jjj = 0.3 ) \\
0002 &            0.7 &     ( j5ph0 = -0.11 ) \\
0009 &            0.6 &     ( e6jph0 = 0.19 , constraints = 0.16 , e0jph0 = -0.11 ) \\
\end{tabular}
}
\caption[Correction factor pull apart table]{Correction factor pull apart table, intended to show which correction factors are being pulled in different directions.  Letting $\chi^2_k$ denote the $k^\text{th}$ term in the $\chi^2$ sum, and $s_i$ the $i^\text{th}$ correction factor, the \emph{pull} of the $k^\text{th}$ bin on the $i^\text{th}$ correction factor is denoted $\text{pull}_{ki}$.  Intuitively, bin $k$ ``pulls'' on the $i^\text{th}$ correction factor with a strength of $\text{pull}_{ki}$.  More precisely, the value obtained by the $i^{\text{th}}$ correction factor is $\text{pull}_{ki}$ standard deviations away from where it would be in the absence of the $k^{\text{th}}$ bin.  If bin $k$ pulls the $i^\text{th}$ correction factor toward larger values, $\text{pull}_{ki}$ is positive; if bin $k$ favors smaller values of the $i^\text{th}$ correction factor, $\text{pull}_{ki}$ is negative.  The units of $\text{pull}_{ki}$ are units of $\chi^2$.  The correction factors are sorted in order of decreasing \emph{pull apart}, where the pull apart of the $i^\text{th}$ correction factor is defined as $\text{pullApart}_{i} \equiv \sum_k \abs{\text{pull}_{ki}}$, provided in the second column.  Intuitively, a correction factor has large pull apart if some bins strongly favor a larger value, and some bins strongly favor a smaller value.  In the third column between parentheses are the bins $k$ that contribute most to the pull apart of each correction factor, along with each individual contribution $\text{pull}_{ki}$.  In each line, only the four largest contributions with $\text{pull}\geq 0.1$ are listed.  In the bin labels, a {\tt 0} following a particle specifies its centrality; a {\tt 4} following a particle indicates it has $p_T>200$~GeV; a {\tt 5} following {\tt{mu}} indicates it is a CMX muon in the region $0.6<\abs{\eta}<1.0$; a {\tt 10} following a particle indicates it lies in the plug region $1<\abs{\eta}<2.5$; {\tt{constraints}} specifies the contribution from $\chi^2_{\text{constraints}}$.}
\label{tbl:CorrectionFactorPullApartTable2}
\end{table*}

\begin{table*}
\hspace*{-0.0in}
\centering
\mbox{
\tiny
\centering
\begin{tabular}{lcl}
Bin     &      Total Influence    &   IndividualInfluence \\
\hline \\
e0j &                    8.9          &  ( 0033 = -4.7 , 0004 = -1.4 , 0018 = -1.3 , 0001 = -0.5 ) \\
e6jjpmiss &              8.1          &  ( 0013 = 2.9 , 0042 = 2 , 0026 = 1 , 0012 = 0.7 ) \\
e6jj &                   6.6          &  ( 0034 = 2.3 , 0042 = 1.3 , 0020 = 1.3 , 0034 = 0.8 ) \\
e6pmiss &                6.1          &  ( 0042 = -3.2 , 0026 = -1.3 , 0011 = -0.6 , 0034 = -0.5 ) \\
e6ph6 &                  5.5          &  ( 0034 = -2.6 , 0030 = -1 , 0040 = -0.9 , 0034 = -0.21 ) \\
jph0 &                   5.2          &  ( 0004 = 2.2 , 0039 = 1.5 , 0029 = 0.8 , 0018 = 0.4 ) \\
e6j &                    5            &  ( 0034 = -1.6 , 0042 = -1.1 , 0018 = -0.9 , 0034 = -0.7 ) \\
constraints &            4.7          &  ( 0030 = 1.4 , 0029 = 0.5 , 0016 = 0.4 , 0015 = 0.3 ) \\
bj5 &                    3.9          &  ( 0036 = 1.5 , 0019 = 1.3 , 0031 = 0.7 , 0001 = 0.28 ) \\
be0 &                    3.8          &  ( 0033 = 1 , 0036 = 0.8 , 0031 = 0.5 , 0004 = 0.5 ) \\
jjph0 &                  3.7          &  ( 0005 = -2.3 , 0039 = -0.5 , 0029 = -0.5 , 0020 = -0.2 ) \\
jjjmu0 &                 3.7          &  ( 0014 = -1 , 0038 = -0.7 , 0035 = -0.5 , 0017 = -0.4 ) \\
e0jj &                   3.6          &  ( 0033 = 1.3 , 0005 = 0.7 , 0020 = 0.5 , 0016 = 0.27 ) \\
be6 &                    3.3          &  ( 0034 = -0.9 , 0042 = -0.6 , 0018 = -0.5 , 0036 = -0.4 ) \\
jjmu0 &                  2.9          &  ( 0038 = -0.8 , 0035 = -0.6 , 0013 = -0.5 , 0020 = -0.2 ) \\
b5j &                    2.8          &  ( 0036 = -1.3 , 0019 = -1 , 0031 = -0.3 , 0001 = -0.2 ) \\
ph6tau &                 2.8          &  ( 0040 = 1.3 , 0037 = 0.8 , 0030 = 0.27 , 0042 = 0.2 ) \\
e6e6j &                  2.5          &  ( 0016 = -1.5 , 0026 = -0.6 , 0017 = -0.24 , 0001 = -0.1 ) \\
jjjph0 &                 2.5          &  ( 0006 = -1.9 , 0029 = -0.2 , 0039 = -0.17 , 0022 = -0.12 ) \\
bph0 &                   2.4          &  ( 0004 = -0.6 , 0036 = -0.5 , 0039 = -0.4 , 0029 = -0.4 ) \\
jjjjph0 &                2.4          &  ( 0007 = -2.1 ) \\
jjj &                    2.4          &  ( 0020 = -2.1 , 0001 = -0.24 ) \\
jmu0 &                   2.3          &  ( 0035 = 1.1 , 0038 = 0.5 , 0012 = 0.2 , 0018 = 0.17 ) \\
e6jjjpmiss &             2.3          &  ( 0014 = 1 , 0042 = 0.4 , 0013 = 0.4 , 0026 = 0.21 ) \\
bej &                    2.3          &  ( 0034 = 0.4 , 0020 = 0.3 , 0031 = 0.32 , 0036 = 0.3 ) \\
e0jjjj &                 2.2          &  ( 0007 = 0.6 , 0024 = 0.5 , 0014 = 0.29 , 0033 = 0.28 ) \\
bmu0 &                   2.2          &  ( 0038 = 0.9 , 0035 = 0.4 , 0036 = 0.23 , 0031 = 0.16 ) \\
b5jj &                   2.2          &  ( 0025 = -0.7 , 0021 = -0.6 , 0036 = -0.5 , 0031 = -0.21 ) \\
e6e6 &                   2.1          &  ( 0026 = 1.3 , 0015 = 0.6 , 0001 = 0.18 ) \\
e6ph0 &                  2            &  ( 0034 = 0.8 , 0008 = 0.3 , 0039 = 0.22 , 0029 = 0.21 ) \\
e0jjj &                  2            &  ( 0006 = 0.6 , 0033 = 0.4 , 0022 = 0.3 , 0014 = 0.21 ) \\
e0pmiss &                2            &  ( 0025 = 0.9 , 0011 = 0.6 , 0001 = 0.4 ) \\
e0jjpmiss &              1.9          &  ( 0013 = -1.1 , 0012 = -0.25 , 0025 = -0.19 , 0014 = -0.12 ) \\
be0pmiss &               1.8          &  ( 0012 = -0.6 , 0036 = -0.5 , 0013 = -0.25 , 0031 = -0.19 ) \\
e0e0 &                   1.8          &  ( 0025 = -0.8 , 0015 = -0.7 , 0001 = -0.21 ) \\
e0jjjpmiss &             1.7          &  ( 0014 = -1 , 0013 = -0.4 , 0025 = -0.12 ) \\
ph0tau &                 1.7          &  ( 0037 = 0.9 , 0004 = 0.31 , 0029 = 0.2 , 0039 = 0.15 ) \\
mu0pmiss &               1.7          &  ( 0043 = -0.7 , 0027 = -0.5 , 0011 = -0.25 , 0001 = -0.2 ) \\
jjmu0pmiss &             1.6          &  ( 0013 = -0.7 , 0017 = -0.2 , 0012 = -0.17 , 0043 = -0.14 ) \\
bj &                     1.6          &  ( 0018 = 0.6 , 0031 = 0.5 , 0036 = 0.4 ) \\
\end{tabular}
}
\caption[Correction factor influence table]{Correction factor influence table.  Letting $\chi^2_k$ denote the $k^\text{th}$ term in the $\chi^2$ sum and $s_i$ the $i^\text{th}$ correction factor, the \emph{pull} of the $i^\text{th}$ bin on the $k^\text{th}$ correction factor is denoted $\text{pull}_{ki}$.  The \emph{total influence} of a bin $k$ is defined as $\text{totalInfluence}_{k} \equiv \sum_i \abs{\text{pull}_{ki}}$.  Intuitively, bins with large total influence are ``important'' in influencing the position of the $\chi^2$ minimum.  Bins with large total influence tend to be big (containing many data events), pull on many correction factors, and prefer correction factors values significantly different from the values they would otherwise assume.  Bins in this table are sorted in order of decreasing total influence, provided in the second column.  In the third column between parentheses are the correction factors $s_i$ that are most influenced by the bin.  The extent to which these correction factors are influenced is also shown in the third column, with an entry such as {\tt 0001 = -0.65} indicating correction factor code {\tt 0001} feels a pull of $-0.65$.  In each line, only the four largest contributions with $\text{pull}\geq 0.1$ are listed.  Due to the multiplicative nature of the correction factors, the pull on each correction factor from bin $k$ is typically negative if the Standard Model prediction exceeds the number of data events in bin $k$, and positive if the Standard Model prediction falls short of the data in bin $k$.}
\label{tbl:CorrectionFactorInfluenceTable2}
\end{table*}

\begin{sidewaystable*}
\hspace{-2cm}
\tiny
\begin{minipage}{8.1in}
\begin{verbatim}
       5001 5102 5103 5121 5122 5123 5124 5130 5131 5132 5141 5142 5143 5144 5151 5152 5153 5161 5162 5164 5165 5167 5168 5169 5170 5171 5211 5212 5213 5216 5217 5219 5246 5256 5257 5261 5266 5273 5285 5292 5293 5402 5403 
5001   1    -.45 -.84 -.6  -.6  -.53 -.34 -.47 -.4  -.33 -.97 -.9  -.79 -.64 -.88 -.68 -.48 -.96 -.99 -.96 -.94 -.92 -.86 -.75 -.49 -.53 -.63 -.61 -.58 -.07 -.17 -.07 -.15 -.08 +.01 0    -.02 +.03 -.04 +.01 +.01 +.02 -.03 
5102   -.45 1    +.37 +.46 +.49 +.44 +.28 +.48 +.4  +.3  +.43 +.4  +.36 +.25 +.39 +.31 +.2  +.43 +.44 +.43 +.42 +.41 +.39 +.34 +.22 +.24 +.3  +.28 +.28 -.36 +.02 +.04 +.09 -.18 -.06 -.01 +.02 -.01 +.02 -.04 -.12 0    +.03 
5103   -.84 +.37 1    +.5  +.5  +.44 +.28 +.39 +.34 +.27 +.81 +.75 +.66 +.53 +.73 +.57 +.4  +.8  +.82 +.8  +.78 +.77 +.72 +.63 +.41 +.44 +.53 +.51 +.49 +.06 +.14 +.06 +.13 +.07 -.01 0    +.01 -.02 +.03 -.01 -.01 -.01 +.03 
5121   -.6  +.46 +.5  1    +.92 +.8  +.54 +.75 +.57 +.43 +.58 +.53 +.47 +.38 +.52 +.41 +.28 +.56 +.59 +.58 +.57 +.55 +.52 +.46 +.29 +.31 +.39 +.38 +.37 -.4  +.02 +.07 +.12 -.6  -.13 +.01 0    -.05 -.05 -.61 -.29 -.02 +.02 
5122   -.6  +.49 +.5  +.92 1    +.79 +.54 +.77 +.58 +.44 +.58 +.53 +.47 +.38 +.52 +.41 +.28 +.58 +.59 +.56 +.57 +.55 +.52 +.46 +.3  +.32 +.39 +.38 +.37 -.46 +.01 +.06 +.13 -.57 -.13 -.01 +.02 -.04 -.04 -.51 -.28 -.02 +.02 
5123   -.53 +.44 +.44 +.8  +.79 1    +.48 +.69 +.53 +.37 +.51 +.47 +.41 +.32 +.46 +.36 +.24 +.51 +.52 +.51 +.49 +.44 +.46 +.4  +.26 +.28 +.35 +.33 +.33 -.43 +.01 +.04 +.11 -.5  -.12 -.02 +.02 -.03 -.03 -.43 -.24 -.02 +.02 
5124   -.34 +.28 +.28 +.54 +.54 +.48 1    +.47 +.36 +.26 +.32 +.3  +.27 +.2  +.29 +.24 +.14 +.33 +.33 +.33 +.32 +.3  +.29 +.15 +.17 +.18 +.22 +.21 +.21 -.31 0    +.04 +.07 -.35 -.08 -.01 +.01 -.02 -.02 -.3  -.17 -.01 +.01 
5130   -.47 +.48 +.39 +.75 +.77 +.69 +.47 1    +.63 +.49 +.44 +.41 +.36 +.28 +.4  +.32 +.21 +.45 +.46 +.45 +.44 +.43 +.4  +.35 +.23 +.25 +.32 +.29 +.3  -.63 -.14 +.04 +.18 -.48 -.17 -.01 +.02 -.02 +.05 -.27 -.38 +.01 +.02 
5131   -.4  +.4  +.34 +.57 +.58 +.53 +.36 +.63 1    +.3  +.39 +.36 +.31 +.25 +.35 +.27 +.19 +.39 +.4  +.39 +.38 +.37 +.35 +.3  +.2  +.21 +.26 +.24 +.25 -.51 -.13 +.04 +.15 -.33 -.12 -.01 +.01 -.02 +.04 -.13 -.24 0    +.02 
5132   -.33 +.3  +.27 +.43 +.44 +.37 +.26 +.49 +.3  1    +.31 +.29 +.25 +.2  +.28 +.22 +.13 +.31 +.32 +.31 +.31 +.3  +.28 +.24 +.16 +.17 +.21 +.2  +.2  -.4  -.16 +.03 +.13 -.23 -.09 -.01 +.01 -.01 +.02 -.06 -.12 0    +.02 
5141   -.97 +.43 +.81 +.58 +.58 +.51 +.32 +.44 +.39 +.31 1    +.92 +.81 +.65 +.95 +.72 +.52 +.93 +.96 +.94 +.91 +.89 +.84 +.73 +.48 +.52 +.43 +.47 +.43 +.08 +.16 +.07 +.11 +.08 +.04 +.01 +.01 -.03 +.04 -.01 0    -.08 -.05 
5142   -.9  +.4  +.75 +.53 +.53 +.47 +.3  +.41 +.36 +.29 +.92 1    +.68 +.62 +.88 +.64 +.5  +.86 +.89 +.87 +.84 +.83 +.78 +.68 +.45 +.48 +.4  +.43 +.39 +.07 +.15 +.08 +.11 +.06 +.02 -.07 +.05 -.04 +.03 0    0    -.05 -.02 
5143   -.79 +.36 +.66 +.47 +.47 +.41 +.27 +.36 +.31 +.25 +.81 +.68 1    +.37 +.77 +.59 +.38 +.76 +.78 +.75 +.74 +.73 +.68 +.6  +.39 +.42 +.35 +.39 +.35 +.06 +.13 +.06 +.09 +.05 +.07 +.07 -.09 -.03 +.03 -.01 0    -.12 -.04 
5144   -.64 +.25 +.53 +.38 +.38 +.32 +.2  +.28 +.25 +.2  +.65 +.62 +.37 1    +.62 +.51 +.23 +.62 +.63 +.62 +.6  +.57 +.55 +.45 +.32 +.35 +.29 +.32 +.3  +.06 +.11 +.03 +.08 +.04 +.01 +.05 -.06 0    +.02 -.01 0    -.05 -.04 
5151   -.88 +.39 +.73 +.52 +.52 +.46 +.29 +.4  +.35 +.28 +.95 +.88 +.77 +.62 1    +.71 +.54 +.85 +.87 +.85 +.83 +.81 +.76 +.66 +.43 +.47 +.23 +.22 +.21 +.07 +.13 +.07 +.08 +.06 0    -.01 +.01 -.03 +.02 0    -.01 -.01 +.01 
5152   -.68 +.31 +.57 +.41 +.41 +.36 +.24 +.32 +.27 +.22 +.72 +.64 +.59 +.51 +.71 1    +.17 +.66 +.67 +.66 +.64 +.63 +.59 +.52 +.34 +.36 +.23 +.23 +.23 +.05 +.1  +.05 +.07 +.01 +.01 -.04 +.03 -.03 +.02 -.01 -.01 -.04 -.02 
5153   -.48 +.2  +.4  +.28 +.28 +.24 +.14 +.21 +.19 +.13 +.52 +.5  +.38 +.23 +.54 +.17 1    +.46 +.47 +.46 +.45 +.43 +.42 +.35 +.23 +.25 +.11 +.11 +.11 +.05 +.07 +.04 +.04 +.02 0    +.03 -.05 -.01 +.01 -.01 -.01 0    -.01 
5161   -.96 +.43 +.8  +.56 +.58 +.51 +.33 +.45 +.39 +.31 +.93 +.86 +.76 +.62 +.85 +.66 +.46 1    +.95 +.94 +.9  +.89 +.82 +.73 +.52 +.55 +.6  +.58 +.56 +.06 +.16 +.01 +.15 +.04 -.06 -.07 +.05 -.04 0    -.01 0    -.02 +.03 
5162   -.99 +.44 +.82 +.59 +.59 +.52 +.33 +.46 +.4  +.32 +.96 +.89 +.78 +.63 +.87 +.67 +.47 +.95 1    +.95 +.93 +.91 +.85 +.74 +.5  +.53 +.62 +.6  +.57 +.07 +.17 +.09 +.15 +.08 -.01 -.01 +.02 -.08 +.04 -.01 -.01 -.02 +.03 
5164   -.96 +.43 +.8  +.58 +.56 +.51 +.33 +.45 +.39 +.31 +.94 +.87 +.75 +.62 +.85 +.66 +.46 +.94 +.95 1    +.89 +.9  +.83 +.73 +.5  +.54 +.6  +.59 +.56 +.07 +.16 +.03 +.15 +.06 -.08 -.01 +.01 -.03 +.02 -.02 -.01 -.02 +.03 
5165   -.94 +.42 +.78 +.57 +.57 +.49 +.32 +.44 +.38 +.31 +.91 +.84 +.74 +.6  +.83 +.64 +.45 +.9  +.93 +.89 1    +.86 +.82 +.7  +.29 +.47 +.59 +.57 +.55 +.06 +.16 +.14 +.14 +.08 0    0    +.01 -.06 +.04 0    -.01 -.02 +.03 
5167   -.92 +.41 +.77 +.55 +.55 +.44 +.3  +.43 +.37 +.3  +.89 +.83 +.73 +.57 +.81 +.63 +.43 +.89 +.91 +.9  +.86 1    +.76 +.7  +.48 +.51 +.58 +.56 +.54 +.07 +.15 +.03 +.15 +.06 -.05 -.01 +.01 -.02 +.03 -.02 -.01 -.02 +.03 
5168   -.86 +.39 +.72 +.52 +.52 +.46 +.29 +.4  +.35 +.28 +.84 +.78 +.68 +.55 +.76 +.59 +.42 +.82 +.85 +.83 +.82 +.76 1    +.64 +.4  +.24 +.54 +.52 +.5  +.06 +.14 +.16 +.13 +.08 0    0    +.01 -.09 +.04 0    -.01 -.02 +.03 
5169   -.75 +.34 +.63 +.46 +.46 +.4  +.15 +.35 +.3  +.24 +.73 +.68 +.6  +.45 +.66 +.52 +.35 +.73 +.74 +.73 +.7  +.7  +.64 1    +.39 +.42 +.47 +.46 +.44 +.05 +.13 +.01 +.12 +.05 -.03 -.02 +.02 -.01 +.02 -.02 -.01 -.02 +.03 
5170   -.49 +.22 +.41 +.29 +.3  +.26 +.17 +.23 +.2  +.16 +.48 +.45 +.39 +.32 +.43 +.34 +.23 +.52 +.5  +.5  +.29 +.48 +.4  +.39 1    +.43 +.31 +.3  +.29 +.04 +.08 -.23 +.06 +.02 -.01 -.08 +.07 -.07 -.01 -.02 0    -.01 +.01 
5171   -.53 +.24 +.44 +.31 +.32 +.28 +.18 +.25 +.21 +.17 +.52 +.48 +.42 +.35 +.47 +.36 +.25 +.55 +.53 +.54 +.47 +.51 +.24 +.42 +.43 1    +.34 +.33 +.31 +.04 +.09 -.35 +.07 +.03 -.01 -.07 +.06 +.12 0    -.01 0    -.01 +.01 
5211   -.63 +.3  +.53 +.39 +.39 +.35 +.22 +.32 +.26 +.21 +.43 +.4  +.35 +.29 +.23 +.23 +.11 +.6  +.62 +.6  +.59 +.58 +.54 +.47 +.31 +.34 1    +.77 +.81 +.02 +.12 +.03 +.21 +.06 -.13 -.03 +.03 -.01 +.04 -.01 -.03 +.13 +.21 
5212   -.61 +.28 +.51 +.38 +.38 +.33 +.21 +.29 +.24 +.2  +.47 +.43 +.39 +.32 +.22 +.23 +.11 +.58 +.6  +.59 +.57 +.56 +.52 +.46 +.3  +.33 +.77 1    +.77 +.03 +.14 +.04 +.1  +.08 +.11 +.01 0    -.01 +.06 -.01 +.01 -.2  -.01 
5213   -.58 +.28 +.49 +.37 +.37 +.33 +.21 +.3  +.25 +.2  +.43 +.39 +.35 +.3  +.21 +.23 +.11 +.56 +.57 +.56 +.55 +.54 +.5  +.44 +.29 +.31 +.81 +.77 1    +.01 +.12 +.03 +.17 +.06 -.04 +.07 -.03 -.01 +.04 -.02 -.02 0    -.33 
5216   -.07 -.36 +.06 -.4  -.46 -.43 -.31 -.63 -.51 -.4  +.08 +.07 +.06 +.06 +.07 +.05 +.05 +.06 +.07 +.07 +.06 +.07 +.06 +.05 +.04 +.04 +.02 +.03 +.01 1    +.16 0    -.04 +.51 +.13 +.01 -.01 0    -.01 -.03 +.27 -.01 0    
5217   -.17 +.02 +.14 +.02 +.01 +.01 0    -.14 -.13 -.16 +.16 +.15 +.13 +.11 +.13 +.1  +.07 +.16 +.17 +.16 +.16 +.15 +.14 +.13 +.08 +.09 +.12 +.14 +.12 +.16 1    +.01 -.24 +.1  +.08 +.01 0    0    +.12 +.03 -.47 +.05 0    
5219   -.07 +.04 +.06 +.07 +.06 +.04 +.04 +.04 +.04 +.03 +.07 +.08 +.06 +.03 +.07 +.05 +.04 +.01 +.09 +.03 +.14 +.03 +.16 +.01 -.23 -.35 +.03 +.04 +.03 0    +.01 1    +.02 +.03 +.02 +.09 -.07 -.68 +.04 0    -.02 0    +.01 
5246   -.15 +.09 +.13 +.12 +.13 +.11 +.07 +.18 +.15 +.13 +.11 +.11 +.09 +.08 +.08 +.07 +.04 +.15 +.15 +.15 +.14 +.15 +.13 +.12 +.06 +.07 +.21 +.1  +.17 -.04 -.24 +.02 1    -.02 -.43 -.01 0    +.01 +.07 -.03 -.13 +.02 +.06 
5256   -.08 -.18 +.07 -.6  -.57 -.5  -.35 -.48 -.33 -.23 +.08 +.06 +.05 +.04 +.06 +.01 +.02 +.04 +.08 +.06 +.08 +.06 +.08 +.05 +.02 +.03 +.06 +.08 +.06 +.51 +.1  +.03 -.02 1    +.18 +.03 0    +.01 +.1  +.63 +.3  +.01 0    
5257   +.01 -.06 -.01 -.13 -.13 -.12 -.08 -.17 -.12 -.09 +.04 +.02 +.07 +.01 0    +.01 0    -.06 -.01 -.08 0    -.05 0    -.03 -.01 -.01 -.13 +.11 -.04 +.13 +.08 +.02 -.43 +.18 1    +.07 -.02 -.02 +.04 +.13 +.16 -.61 -.16 
5261   0    -.01 0    +.01 -.01 -.02 -.01 -.01 -.01 -.01 +.01 -.07 +.07 +.05 -.01 -.04 +.03 -.07 -.01 -.01 0    -.01 0    -.02 -.08 -.07 -.03 +.01 +.07 +.01 +.01 +.09 -.01 +.03 +.07 1    -.89 +.05 +.04 0    0    -.06 -.21 
5266   -.02 +.02 +.01 0    +.02 +.02 +.01 +.02 +.01 +.01 +.01 +.05 -.09 -.06 +.01 +.03 -.05 +.05 +.02 +.01 +.01 +.01 +.01 +.02 +.07 +.06 +.03 0    -.03 -.01 0    -.07 0    0    -.02 -.89 1    -.05 -.02 0    0    +.05 +.11 
5273   +.03 -.01 -.02 -.05 -.04 -.03 -.02 -.02 -.02 -.01 -.03 -.04 -.03 0    -.03 -.03 -.01 -.04 -.08 -.03 -.06 -.02 -.09 -.01 -.07 +.12 -.01 -.01 -.01 0    0    -.68 +.01 +.01 -.02 +.05 -.05 1    +.02 +.04 +.01 0    +.01 
5285   -.04 +.02 +.03 -.05 -.04 -.03 -.02 +.05 +.04 +.02 +.04 +.03 +.03 +.02 +.02 +.02 +.01 0    +.04 +.02 +.04 +.03 +.04 +.02 -.01 0    +.04 +.06 +.04 -.01 +.12 +.04 +.07 +.1  +.04 +.04 -.02 +.02 1    +.12 -.26 -.05 -.01 
5292   +.01 -.04 -.01 -.61 -.51 -.43 -.3  -.27 -.13 -.06 -.01 0    -.01 -.01 0    -.01 -.01 -.01 -.01 -.02 0    -.02 0    -.02 -.02 -.01 -.01 -.01 -.02 -.03 +.03 0    -.03 +.63 +.13 0    0    +.04 +.12 1    +.23 +.02 0    
5293   +.01 -.12 -.01 -.29 -.28 -.24 -.17 -.38 -.24 -.12 0    0    0    0    -.01 -.01 -.01 0    -.01 -.01 -.01 -.01 -.01 -.01 0    0    -.03 +.01 -.02 +.27 -.47 -.02 -.13 +.3  +.16 0    0    +.01 -.26 +.23 1    -.09 -.02 
5402   +.02 0    -.01 -.02 -.02 -.02 -.01 +.01 0    0    -.08 -.05 -.12 -.05 -.01 -.04 0    -.02 -.02 -.02 -.02 -.02 -.02 -.02 -.01 -.01 +.13 -.2  0    -.01 +.05 0    +.02 +.01 -.61 -.06 +.05 0    -.05 +.02 -.09 1    +.23 
5403   -.03 +.03 +.03 +.02 +.02 +.02 +.01 +.02 +.02 +.02 -.05 -.02 -.04 -.04 +.01 -.02 -.01 +.03 +.03 +.03 +.03 +.03 +.03 +.03 +.01 +.01 +.21 -.01 -.33 0    0    +.01 +.06 0    -.16 -.21 +.11 +.01 -.01 0    -.02 +.23 1    
\end{verbatim}
\end{minipage}
\caption[Correction factor correlation matrix]{Correction factor correlation matrix.  The topmost row and leftmost column show correction factor codes. Each element of the matrix shows the correlation between the correction factor labeling the element's column and the correction factor labeling the element's row.  Each matrix element is dimensionless; the elements along the diagonal are unity; the matrix is symmetric; positive elements indicate positive correlation, and negative elements anti-correlation.}
\label{tbl:CorrectionFactorCorrelationMatrix2}
\end{sidewaystable*}

\subsection{Comparison with first round}
\label{sec:comparisonWithv01}

The correction factor values obtained in the second round (v02) (corresponding to 2~fb$^{-1}$) are here compared with the correction factor values obtained in the first round (v01) (corresponding to 927~pbb$^{-1}$).
The numerical values can be found in Table~\ref{tbl:correctionFactorComparison2}; analysis of the changes is provided below.

\paragraph*{\tt{5001}.}
The integrated luminosity of the sample has of course increased with respect to v01.  The present integrated luminosity obtained from the fit is again consistent with the luminosity obtained from the CLC measurement.
\paragraph*{\tt{5102}.}
This cosmic photon ``$k$-factor'' has increased due to requiring that this background satisfies the same good run list that are required for the data and by requiring that these events contain at least one reconstructed photon. As a result the number of events in this background has been decreased prompting this $k$-factor to increase accordingly. 
\paragraph*{\tt{5103}.}
This cosmic jet ``$k$-factor'' has decreased due to the cut on the second jet in jet final states, as described in Sec.~\ref{sec:eventGeneration2}. The cut removes events in which the leading jet is due to a cosmic ray, and the other jets are due to the underlying event. As a result of this removal, the kfactor for this background has been reduced.
\paragraph*{\tt{5121}--\tt{5132}.}
The $k$-factors for photon + jet production and diphoton production is consistent with values obtained in v01.
\paragraph*{\tt{5151}--\tt{5153}.}
The $k$-factors for Z + jet production is consistent with values obtained in v01.
\paragraph*{\tt{5141}--\tt{5144}.}
Motivated by a mistake in the modelling of the inoperational period of the keystone and miniskirt portions of the muon detector, we switched from the MadEvent W+jets Monte Carlo sample to the standard Top Group Alpgen W+jets sample. These $k$-factors were changed to correspond to Alpgen cross sections.
\paragraph*{\tt{5161}--\tt{5169}.}
In v01 of this analysis we used $\poo{j}{j}=1$, despite the fact that $\poo{j}{b} \gtrsim 0.01$.  It is logically more consistent to chose $\poo{j}{j}=1-\poo{j}{b}$, so this is what is done in v02.  The result of this modification is that $k$-factors for processes with one or more jets have increased.
\paragraph*{\tt{5170},\tt{5171}.}
These two $k$-factors for heavy flavor multijet production have been introduced.
\paragraph*{\tt{5211},\tt{5212}.}The central electron identification efficiency is consistent with value obtained in v01. The phoenix electron identification efficiency scale factor has changed reflecting our change to the phoenix electron identification criteria. 
\paragraph*{\tt{5213}.}The muon identification efficiency scale factor has changed due to our change to the muon identification criteria, and the correction to the modelling of the inoperational period of the keystone and miniskirt portions of the muon detector. 
\paragraph*{\tt{5216},\tt{5217},\tt{5219}.}
The identification efficiencies $\poo{\gamma}{\gamma}$ in the central and plug regions, and $\poo{b}{b}$ in the central region are consistent with values obtained in v01.
\paragraph*{\tt{5245}.}
The fake rate $\poo{e}{\gamma}$ has been removed after the change to the plug electron and photon identification. It was found to be unnecessary. This vanished correction factor is not listed in Table~\ref{tbl:correctionFactorComparison2}.
\paragraph*{\tt{5246}.}
The fake rate $\poo{\gamma}{e}$ in the plug has been promoted to a correction factor from a fixed value of 0.005. This value increased significantly due to a redefinition of plug photons into electrons in the 1e+1ph final state. This was motivated by the fact that this plug photon was much more likely to have been an electron. We have removed this renaming procedure for the current version of the analysis.
\paragraph*{\tt{5256},\tt{5257}.}
The fake rates $\poo{q}{e}$ in the central and plug regions have decreased by roughly 13\% and 6\%, respectively, due to our improved conversion removal.  In v01 we required a candidate conversion track to have $p_T>2$~GeV; in v02 we make no transverse momentum requirement on the candidate converstion track. The change to the fake rate in the plug region is also affected by our change to the phoenix electron identification.
\paragraph*{\tt{5261}.}
The fake rate $\poo{q}{\mu}$ is consistent with the value obtained in v01.
\paragraph*{\tt{5273}.}
The fake rate $\poo{j}{b}$ is consistent with the value obtained in v01.
\paragraph*{\tt{5285}.}
A different $p_T$ dependence has been imposed for the fake rate $\poo{q}{\tau}$ in v02 than applied in v01 and a dependence on the generated sumPt has also been applied. As a result of not being careful about proper normalizations of those functions, this number is not directly comparable to the one from v01.
\paragraph*{\tt{5292}.}
The value obtained for the fake rate $\poo{q}{\gamma}$ in the central region is consistent with the value obtained in v01.
\paragraph*{\tt{5293}.}
The fake rate $\poo{q}{\gamma}$ in the plug has decreased to due our correction to the plug photon identification criteria.
\paragraph*{\tt{5401}.}
The central electron trigger efficiency has been found to increase to unity in the current version of the analysis, because we now allow an event to pass on any online trigger. As a consequence, it is no longer appropriate to constrain this trigger efficiency to the Joint Physics value for the CEM trigger. We now simply fix the central electron trigger efficiency to $1.0$ and it is no longer a correction factor.
This vanished correction factor is not listed in Table~\ref{tbl:correctionFactorComparison2}.
\paragraph*{\tt{5402}.}
The plug electron trigger efficiency is consistent with the value from v01.
\paragraph*{\tt{5403}.}
We have combined the CMUP and CMX trigger efficiencies due to the fact that they were very close to each other from v01 of the analysis. The value in v02 of the analysis is consistent with the values from v01.


\chapter{Risk of Being {\em Ad Hoc}}
\label{sec:blindOrNot}

\section{Introduction}

Here follows a general discussion, not so much about the actual SM implementation in this analysis, but about concerns such as bias, not being ``blind'', and how these factors affect the meaning of a null result.

In a search for new physics, especially a model-independent one, it is necessary to construct the Standard Model (SM) prediction.  Then, one can test whether the data ($D$) are consistent with it.  

By definition, the data follow the true law of nature.  Denote the true theory by $T$.  If there is physics beyond the SM, then $T \neq SM$.  If new physics is to be observed, the p.d.f.\ of at least one observable quantity needs to differ adequately from that predicted by the SM.

Having the data events distributed according to $T$, one has the freedom to test their consistency with any conceivable theory.  However, what is really interesting, is how well the data agrees with the SM, rather than some arbitrary model, not necessarily well motivated.  We could, for example, construct a model agreeing bin by bin with the data.  Imagine for instance having a dedicated $k$-factor\footnote{$k$-factors are corrections to the cross sections of processes.  Typically, cross sections are calculated to leading-order, or next-to-leading-order, and rarely to an even higher order.  $k$-factors are meant to correct such approximate calculations to the infinite-order cross section, which is incalculable, therefore $k$-factors are inferred from the data.} per final state; then we would be able to adjust this elastic pseudo-theory to match any combination of populations across final states.  That data-obeying model would be consistent with $T$.  Then, by construction, testing the quality of the fit would confirm the null hypothesis, namely that data agree with the constructed model.  The hypothesis test itself would be perfectly legitimate, and its outcome would be correct, yet completely uninteresting, since nobody is interested in that absurd model anyway. 

The problem then begins with the realization that the truly interesting hypothesis, the SM, is itself not known exactly; one needs information about correction factors, such as fake rates, $k$-factors, efficiencies etc. Different values of such parameters result in different ``SM'' predictions. 

\begin{figure}
\centering
\includegraphics[width=13cm,angle=0]{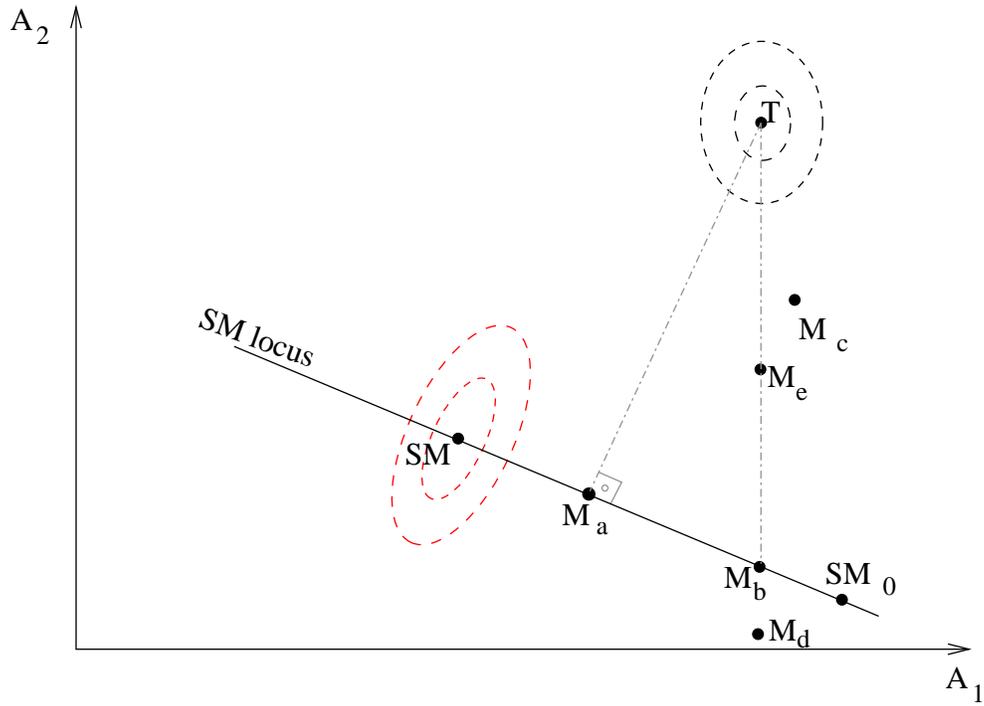} 
\caption{Simplified picture of the p.d.f.s of the true theory and several possibilities for the SM implementation.}
\label{fig:blindOrNot1}
\end{figure}

Let's assume there are only two observable quantities, $A_1$ and $A_2$ (Fig.~\ref{fig:blindOrNot1}).  For example, $A_1$ and $A_2$ could be the populations of events in two final states.  Depending on the values of some correction factors (like $k$-factors etc.), the prediction of the SM implementation can be centered anywhere in some locus.  In this case, the allowed locus is represented by a one-dimensional solid line; in general, the locus may be higher-dimensional. 

The correction factors have some true values, which may be unknown.  The true Standard Model prediction is located at the ``SM'' point, which corresponds to the true values of the correction factors.  Ideally, {\em that} is the SM we would like to compare to the data.

When the work to construct the SM prediction begins, one has no adjustments made yet, which results in some prediction centered on, say, point $SM_0$.  One sees then the data\footnote{Whether he sees all, or part, or only some aspect of them will be discussed later.}, which are by definition near point $T$, and notices the discrepancies in $A_1$ and $A_2$.  Since he has applied no corrections yet, he can not be confident that the current prediction is the real SM. The SM has been successful so far, therefore to rule it out one needs convincing evidence.  To be convincing, he needs to be conservative; he must exploit any source of systematic uncertainty that he can identify in order to correct the prediction in a direction that brings it closer to the data.  
Unfortunately, there is no prescription how to do that correctly.

There are some obvious sources of uncertainty: $k$-factors reflecting the fact that it is not possible to calculate the infinite-order cross section of SM processes, uncertainties in the exact probability by which a particle may be misidentified, uncertainty in the integrated luminosity etc.  For specific discrepancies that are not accounted for by such obvious uncertainties, one needs to become more imaginative to identify what may be causing them, but it is important to not invent false corrections.  It requires judgment to make well motivated adjustments instead of {\em ad hoc} corrections that hide the signal of potentially new physics.  The locus, represented by the solid line in Fig.~\ref{fig:blindOrNot1}, is meant to represent the possible predictions that can be derived by making well motivated corrections, whereas points out of the locus represent the results of poorly motivated corrections.

Suppose that throughout the process one makes well motivated corrections.  Then his prediction should drift along the locus from point $SM_0$ to point $M_a$, which gives the best agreement with the observed data in $A_1$ and $A_2$ simultaneously.  Even though $M_a \neq SM$,  he will need to stop at $M_a$ and not proceed towards the actual SM point.  That is because he has no way to know if he has reached the actual SM to stop there; his only guidance is the data and his judgment.  To be conservative, he would have to bring the prediction as close to point $T$ as allowed, but not closer -- that is point $M_a$.  
The wrong thing to do would be to introduce extraneous, poorly motivated corrections that would drive one from $SM_0$ to a prediction like $M_c$, namely out of the locus.  That would be the result of {\em ad hoc} treatment of discrepancies, which in its extreme limit would result in a model as uninteresting as the data-obeying model mentioned earlier.


What can safeguard one from constructing the prediction of some poorly motivated model?  Only prudence and an over-constrained system that limits systematic uncertainties, making it harder to deviate from the SM locus.  
The risk of implementing an {\em ad hoc} model remains, unless all systematic uncertainties shrunk to zero, in which ideal case the locus would shrink into just the true SM point.
However, there are some ``blind'' approaches that, as will be argued, create the illusion of safety against erring, or the sensation that information is generated out of nothing, by using the data in ``clever'' ways, i.e.\ by not seeing all of them at the same time.

\section{Blind to signal region}

In some cases (not in this analysis) one may presume that the new physics will be affecting $A_2$ but not $A_1$.  That is clearly an assumption, which in many cases can be motivated.  $A_2$ is then treated as ``signal region'', and $A_1$ as ``control region''.  Adjusting the correction model to achieve maximal agreement with the data in $A_1$ is legitimate, since the premise is that the SM should distribute $A_1$ as $T$ does. That leads (if everything is done correctly) to a SM implementation with p.d.f.\ centered on $M_b$.

There is nothing wrong in defining control and signal regions.  Clearly, when interpreting the result of the comparison of the data with $M_b$ one needs to remember that $M_b$ is not the globally best fitting model (that would be $M_a$).  Furthermore, $M_b$ is not necessarily the true SM, but is the model that best fits the control region.  Indicative of the value of such results is the fact what is ``signal'' region in one analysis can be ``control'' in another. Depending on what one defines as ``signal'' and ``control'', the result may vary from agreement to disagreement with the data. Although these results can be valid, they are convincing only if the initial premise is accepted.

Unfortunately, staying ``blind'' in $A_2$ does not guarantee that the final model will not be an absurd and {\em ad hoc} one.  For example, a human error may lead one from $SM_0$ to $M_d$ or $M_e$. Apart from a human error that may occur during the development of the correction model, opening the box (e.g.\ looking at the measured $A_2$) often makes people question the correctness of their implemented model, especially in the event of a discrepancy with the data.  In that phase of reconsideration, one may even accidentally change his background model from $M_b$ to $M_e$, so the notion of ``blindness'' is questionable, unless no discrepancy is seen.  Therefore, as in the non-blind analysis case, prudence and an over-constrained system that limits systematic uncertainties, making it harder to deviate from the SM locus, can prevent testing the goodness of a worthless model (like $M_e$, $M_c$ or $M_d$).

\section{Blind to part of the data}

Another approach considered ``blind'' is to split the whole data set ($D$) in two parts ($D_{\rm control}$, $D_{\rm signal}$), assigning for example every third event to $D_{\rm control}$ and the rest to $D_{\rm signal}$.  Then, $D_{\rm control}$ can be used to develop the correction model, and $D_{\rm signal}$ is only revealed in the end, to check how well it is fitted by the derived background model. 

The supposed advantage of this approach is that $D_{\rm signal}$ is independent from $D_{\rm control}$. So, if agreement is observed between $D_{\rm signal}$ and the background model, that supposedly can not be due to a biased model, as the background model was developed knowing nothing about $D_{\rm signal}$.
Though psychologically reassuring, this impression of safety is false.

Obviously, all data come from the same distribution $T$, therefore there is no reason why $D_{\rm signal}$ would be distributed any differently than $D_{\rm control}$, apart from random statistical fluctuations, which actually become bigger when $D_{\rm control}$ and $D_{\rm signal}$ have smaller populations.

If one makes wrong judgments in the way he uses $D_{\rm control}$, then there are two possibilities: If one observes agreement between the background model and $D_{\rm signal}$, that only means that $D_{\rm signal}$ didn't fluctuate too differently than $D_{\rm control}$. On the other hand, if one observes disagreement, that would only be due to (rare) statistical fluctuation of $D_{\rm signal}$ with respect to $D_{\rm control}$.  In other words, if one makes the wrong use of $D_{\rm control}$ the result is as uninformative as it would be if he had used the whole $D$ in a wrong way.

Furthermore, even if one is very prudent and has an over-constrained system with small systematic uncertainties, still splitting the data makes the situation worse.  Having less data in $D_{\rm control}$ to constrain the correction factors makes the locus where SM could be larger, therefore it is more likely to end up with a correction model farther away from the actual SM, simply due to larger systematic uncertainties.  Furthermore, having a smaller number of data in $D_{\rm signal}$ reduces statistical power, making it harder to observe a real effect that may appear in the measured $A_1$ and $A_2$. 

In summary, splitting $D$ in two does not secure one from implementing wrongly his theoretical prediction.  If one can make proper use of $D_{\rm control}$, then he can also make proper use of the whole $D$, which would offer the advantage of smaller uncertainties.

\section{Summary}

To summarize, there is no way to be sure that the null hypothesis compared to the data is the SM, rather than some other uninteresting one.  However, there is reason to hope that what was tested in this analysis is the agreement of the data with a model that at least is possible to be the SM, namely belongs to the SM locus determined by well motivated systematic uncertainties.  Certainly, the tested model is biased to agree with the data more than the SM may actually agree\footnote{Think of the analogy given by points ``SM'' and $M_a$ in Fig.~\ref{fig:blindOrNot1}.}, since the best fitting choice of correction parameters was made, but that is inevitable, since the SM is assumed correct until proof of the contrary.  The hope that the implemented background model is not far from the actual SM is based on the fact that the correction model is significantly over-constrained by examining not just a couple of observables, but thousands.  After all, human errors are always possible, but the best effort was made to eliminate them.  Well motivated corrections usually fix several problems at once, while mistaken adjustments tend to fix one problem but cause other.  Our global approach allowed us to distinguish the former from the latter, by monitoring simultaneously so many observables before and after the adjustments.

\twocolumn
\chapter{Nomenclature}
\label{chapter:nomenclature}
\renewcommand{\baselinestretch}{1}
{
\footnotesize
\begin{itemize}
\nomenclature{BMU}{Barrel Muon system. Often synonymous to IMU}
\nomenclature{CDF}{Collider Detector at Fermilab}
\nomenclature{CEI}{Charge Exchange Injection}
\nomenclature{CEM}{Central Electromagnetic calorimeter}
\nomenclature{CERN}{Conseil Europ\'een pour la Recherche Nucl\'eaire}
\nomenclature{CES}{Central Electromagnetic Showermax detector}
\nomenclature{CHA}{Central Hadronic calorimeter}
\nomenclature{CKM}{Cabibbo Kobayashi Maskawa}
\nomenclature{CLC}{Cerenkov Luminosity Counter}
\nomenclature{CMP}{Central Muon Upgrade}
\nomenclature{CMUP}{A muon that has both CMU and a CMP hits}
\nomenclature{CMU}{Central Muon Detector}
\nomenclature{CMX}{Central Muon Extension}
\nomenclature{COT}{Central Outer Tracker}
\nomenclature{CPR}{Central Preshower detector}
\nomenclature{CPU}{Central Processor Unit}
\nomenclature{CP}{Charge Parity}
\nomenclature{CSL}{Consumer Server Logger}
\nomenclature{DAQ}{Data Acquisition}
\nomenclature{DIS}{Deep Inelastic Scattering}
\nomenclature{EM}{Electromagnetic}
\nomenclature{EVB}{Event Builder}
\nomenclature{EWK}{Electroweak}
\nomenclature{EWSB}{Electroweak Symmetry Breaking}
\nomenclature{FCC}{Feynman Computing Center}
\nomenclature{FIFO}{First In First Out}
\nomenclature{FNAL}{Fermi National Accelerator Laboratory}
\nomenclature{GMSB}{Gauge Mediated Supersymmetry Breaking}
\nomenclature{GUT}{Grand Unification Theory}
\nomenclature{ID}{Identification}
\nomenclature{IMU}{Intermediate Muon system}
\nomenclature{ISL}{Intermediate Silicon Layer}
\nomenclature{KS}{Kolmogorov Smirnov}
\nomenclature{L00}{Layer 0 of the Silicon Detector}
\nomenclature{LHC}{Large Hadron Collider}
\nomenclature{LO}{leading order}
\nomenclature{MC}{Monte Carlo}
\nomenclature{MET}{Missing Transverse Energy}
\nomenclature{MIP}{Minimum Ionizing Particle}
\nomenclature{MI}{Main Injector}
\nomenclature{PDF}{Parton Distribution Function}
\nomenclature{p.d.f.}{Probability Density Function}
\nomenclature{PEM}{Plug Electromagnetic calorimeter}
\nomenclature{PES}{Plug Electromagnetic Showermax detector}
\nomenclature{PHA}{Plug Hadronic calorimeter}
\nomenclature{PHX}{``Phoenix'', referring to forward tracks reconstructed from silicon hits}
\nomenclature{PMNS}{Pontecorvo Maki Nakagawa Sakata}
\nomenclature{PMT}{Photomultiplier}
\nomenclature{QCD}{Quantum Chromodynamics}
\nomenclature{RF}{Radio Frequency}
\nomenclature{SCPU}{Scanner CPU}
\nomenclature{SM}{Standard Model of elementary particles}
\nomenclature{SUGRA}{Supergravity}
\nomenclature{SUSY}{Supersymmetry}
\nomenclature{SVT}{Silicon Vertex system}
\nomenclature{SVX}{Silicon Vertex Detector}
\nomenclature{TSI}{Trigger Supervisor}
\nomenclature{UV}{Ultraviolet}
\nomenclature{VME}{Virtual Machine Environment, a standard mainframe operating system}
\nomenclature{VRB}{VME Readout Buffer (or Board)}
\nomenclature{WHA}{Endwall Hadronic calorimeter}
\nomenclature{WLS}{Wavelength Shifting (optic fiber)}
\nomenclature{XCES}{Extrapolation to Central Electromagnetic Showermax}
\nomenclature{XFT}{Extremely Fast Tracker}
\nomenclature{XTRP}{Extrapolation Unit}
\end{itemize}
}
\onecolumn

\bibliography{main}
\bibliographystyle{hunsrt} 
\end{document}